\documentclass[twoside,openright,a4paper,11pt,parskip=false,abstract=on,makeidx,cleardoublepage=plain]{scrreprt}
\usepackage[margin=2cm,bindingoffset=1cm,includeheadfoot]{geometry}
\usepackage[utf8]{inputenc}	
\usepackage[T1]{fontenc}
\usepackage{yfonts}
\usepackage[english]{babel}

\usepackage[colorlinks,pagebackref]{hyperref}
\renewcommand*{\backrefalt}[4]{%% alternative interface
\ifnum#1=1 %
\ifnum#3=1 %
(Cited on page %
\else
(Cited on page %
\fi
\else (Cited on pages %
\fi#2.)
}

\usepackage{soul}
\usepackage[sc]{mathpazo}
\usepackage{amsmath, amsfonts, amssymb, cancel, amsthm}
\usepackage{bbm}
\usepackage{xfrac}
\usepackage{bbold}
\usepackage{mathrsfs}
\usepackage{abraces}
\usepackage{slashed}
\usepackage{relsize}
\usepackage{braket}
\usepackage{tensor}
\usepackage{esint}
\usepackage{mathtools}
\usepackage{siunitx}
\usepackage{graphicx}
\usepackage{array}
\usepackage{booktabs}
\usepackage[font=small,labelfont=sc]{caption}
\usepackage[labelfont=normalfont,subrefformat=simple]{subcaption}
\usepackage{ulem}
\usepackage{shadethm}
\usepackage{float}
\floatstyle{plaintop}
\restylefloat{table}
\usepackage{enumerate}
\usepackage{babelbib}
\usepackage{epstopdf}

\usepackage{setspace}

\usepackage[]{tocbibind}

\usepackage{verbatim}

\usepackage{tikz}
\usetikzlibrary{decorations.markings,decorations.pathreplacing,arrows,arrows.meta,calc,shapes.arrows,shapes.geometric}

\usepackage{cite}

\usepackage{makeidx}

\usepackage{chngcntr}
\DeclareRobustCommand{\d}{\mathrm{d}}

\DeclareRobustCommand{\e}{\mathrm{e}}
\DeclareRobustCommand{\i}{\mathrm{i}}
\DeclareRobustCommand{\1}{\mathbbm{1}}

\DeclareMathOperator{\sgn}{sgn}
\DeclareMathOperator{\sech}{sech}

\DeclareMathOperator{\Tr}{Tr}

\DeclareRobustCommand{\op}[2]{\tensor*{\hat{#1}}{_{#2}}}
\DeclareRobustCommand{\opd}[2]{\tensor*{\hat{#1}}{^{\dagger}_{#2}}}

\DeclareRobustCommand{\hc}{\textsc{h.}\text{c.}}

\DeclareRobustCommand{\Green}[2]{\,\vcenter{\hbox{\begin{tikzpicture} \newcommand\h{0.21}; \draw (0,-\h)--(-\h,0)--(0,\h); \draw ($(0,-\h)-(0.1,0)$)--($(-\h,0)-(0.1,0)$)--($(0,\h)-(0.1,0)$); \end{tikzpicture}}} #1 ; #2 \vcenter{\hbox{\begin{tikzpicture} \newcommand\h{0.21}; \draw (0,-\h)--(\h,0)--(0,\h); \draw ($(0,-\h)+(0.1,0)$)--($(\h,0)+(0.1,0)$)--($(0,\h)+(0.1,0)$); \end{tikzpicture}}}}

\DeclareRobustCommand{\Greenline}[2]{\,\vcenter{\hbox{\begin{tikzpicture} \newcommand\h{0.175}; \draw (0,-\h)--(-\h,0)--(0,\h); \draw ($(0,-\h)-(0.1,0)$)--($(-\h,0)-(0.1,0)$)--($(0,\h)-(0.1,0)$); \end{tikzpicture}}} #1 ; #2 \vcenter{\hbox{\begin{tikzpicture} \newcommand\h{0.175}; \draw (0,-\h)--(\h,0)--(0,\h); \draw ($(0,-\h)+(0.1,0)$)--($(\h,0)+(0.1,0)$)--($(0,\h)+(0.1,0)$); \end{tikzpicture}}}}

\DeclareFontFamily{U}{MnSymbolC}{}
\DeclareSymbolFont{MnSyC}{U}{MnSymbolC}{m}{n}
\DeclareFontShape{U}{MnSymbolC}{m}{n}{
    <-6>  MnSymbolC5
   <6-7>  MnSymbolC6
   <7-8>  MnSymbolC7
   <8-9>  MnSymbolC8
   <9-10> MnSymbolC9
  <10-12> MnSymbolC10
  <12->   MnSymbolC12}{}
\DeclareMathSymbol{\intp}{\mathbin}{MnSyC}{'270}
\DeclareMathSymbol{\extp}{\mathbin}{MnSyC}{'54}
\DeclareMathSymbol{\vecp}{\mathbin}{MnSyC}{'25}

\newcommand{\adjvec}[1]
{
	\rotatebox[origin=c]{90}{$\begin{pmatrix} \rotatebox[origin=c]{-90}{$\begin{matrix} #1 \end{matrix}$} \end{pmatrix}$}
}

\newcommand{\subfiglabel}[3]{
\begin{subfigure}{0.49\linewidth}
\begin{tikzpicture}
	\node at (0,0) {#1};
	\node at (#2) {\footnotesize\subref*{#3}};
\end{tikzpicture}
\phantomsubcaption{\label{#3}\vspace{-\baselineskip}}
\end{subfigure}
}

\makeatletter
\newcounter{parentsubcaption}
\newenvironment{subsubcaption}
 {\refstepcounter{sub\@captype}%
  \protected@edef\theparentsubcaption{\@nameuse{thesub\@captype}}%
  \setcounter{parentsubcaption}{\value{sub\@captype}}%
  \setcounter{sub\@captype}{0}%
  \@namedef{thesub\@captype}{\theparentsubcaption.\arabic{sub\@captype}}%
  \ignorespaces
}{%
  \setcounter{sub\@captype}{\value{parentsubcaption}}%
  \ignorespacesafterend
}
\makeatother

\usepackage{fancyhdr}
\makeatletter
\def\headrule
{{
	\if@fancyplain\let\headrulewidth\plainheadrulewidth\fi
	\hrule\@height\headrulewidth\@width\headwidth
}}
\makeatother
\renewcommand{\headrulewidth}{0.5pt}

\usepackage[tocindentauto]{tocstyle}
\usetocstyle{KOMAlike}

\definecolor{pale}{cmyk}{0.04,0.03,0.02,0}
\definecolor{ocean}{cmyk}{0.94,0.63,0.04,0.07}
\definecolor{clover}{cmyk}{0.71,0.05,0.99,0.27}
\definecolor{plum}{cmyk}{0.49,0.91,0.0,0.1}
\definecolor{maraschino}{cmyk}{0,0.80,0.94,0.00}
\definecolor{red}{cmyk}{0.00,0.88,0.85,0.07}
\definecolor{blue}{cmyk}{1.00,0.43,0.00,0.30}

\definecolor{ucdlb}{cmyk}{0.75,0.31,0.01,0.01}
\definecolor{ucddb}{cmyk}{0.95,0.75,0.09,0.23}
\definecolor{ucdg}{cmyk}{0.81,0.06,0.81,0.07}

\makeindex
\usepackage{upgreek}

\usepackage{multirow}

\usepackage{diagbox}

\newtheorem{theorem}{Theorem}
\newtheorem{corollary}{Corollary}

\DeclareRobustCommand{\twiddle}[1]{\widetilde{#1}}
\DeclareRobustCommand{\M}[2]{\tensor*{M}{^{(#1)}_{#2}}}
\DeclareRobustCommand{\tM}[2]{\tensor*{\twiddle{M}}{^{(#1)}_{#2}}}

\usepackage{soul}
\usepackage{xcolor}
\begin{document}

\thispagestyle{empty}
\begin{titlepage}
{
\setstretch{1.25}

\begin{center}
	%\vspace*{4cm}
	%Title			
	%{%\fontsize{21}{24}\selectfont
	\textsc{{
	{\LARGE Interactions and Topology in Quantum Matter}}}
	
	{\Large Auxiliary Field Approach \& Generalized SSH Models}
	%}
	
	\vspace{\baselineskip}
	\textcolor{maraschino}{\Large--- arXiv version ---}
	%}
			
	\vspace{\baselineskip}
	
	\includegraphics[height=5\baselineskip]{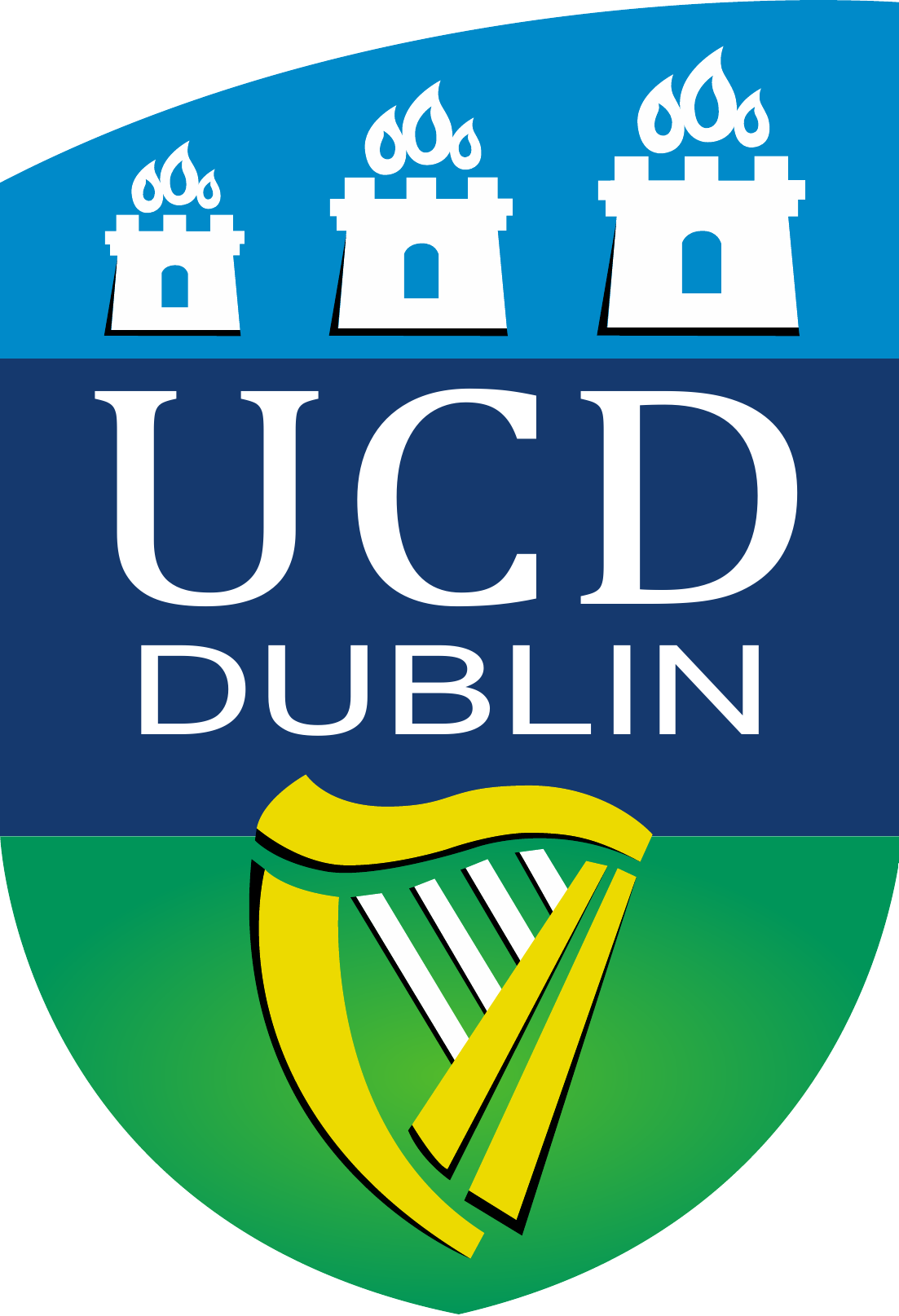}\\
%	\vspace{-\baselineskip}
%	\qquad
%	\includegraphics[width=0.3\textwidth]{Chapters/Titlepage/Images/Logo_THP.png}\\
	\vspace{\baselineskip}

	% Author
	\textsc{\Large Patrick J. Wong} \\ {\large 17209149}%5761263

	%\Large
	\vspace{\baselineskip}
	The thesis is submitted to University College Dublin in fulfillment of the requirements for the degree of Doctor of Philosophy in Physics

	\vspace{2\baselineskip}	
	{\large
	\begin{tabular}{p{0.45\linewidth}p{0.55\linewidth}}
%	\multicolumn{2}{l}{Centre for Quantum Engineering, Science, and Technology}\\\\
	\textcolor{ucdlb}{School of Physics} & \textcolor{ucdlb}{Scoil na Fisice}\\
	\textcolor{ucddb}{University College Dublin} & \textcolor{ucddb}{An Col\'aiste Ollscoile, Baile \'Atha Cliath}\\
	\textcolor{ucdg}{Dublin, Ireland} & \textcolor{ucdg}{Baile \'Atha Cliath, \'Eire}
	\\
	\multirow{4}{*}{\hspace{-1em}\includegraphics[width=0.75\linewidth]{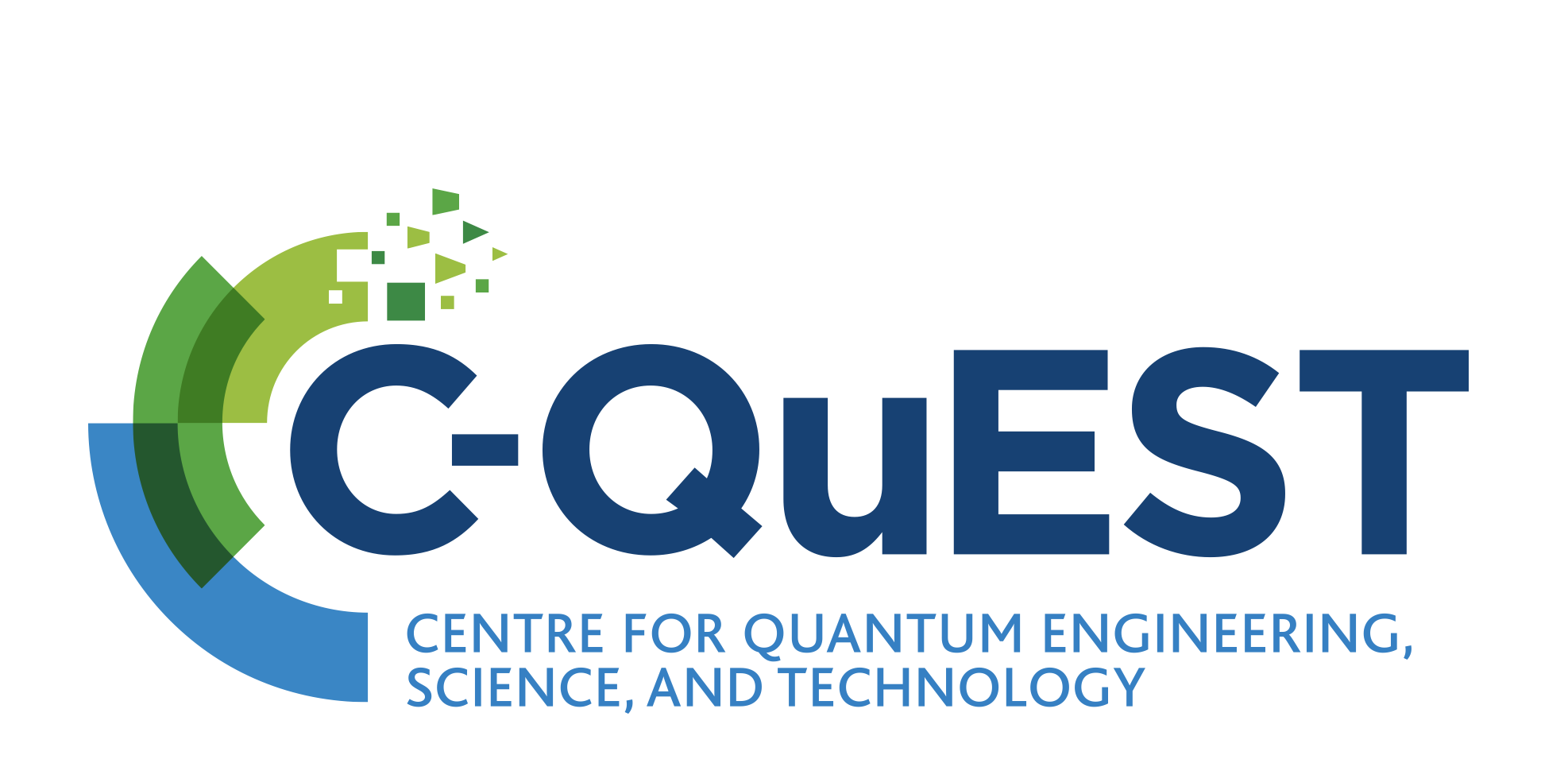}} \\ & \rule{0pt}{18pt} \\ & \textcolor{ucddb}{UCD Quantum Centre} \\ & \textcolor{ucdlb}{Ionad um Quantum UCD}
	\end{tabular}
	}
	
%	{%\footnotesize
%	\begin{tabular}{lll}
%	Centre for & \textcolor{ucdlb}{School of Physics} & \textcolor{ucdlb}{Scoil na Fisice}\\
%	Quantum Engineering, & \textcolor{ucddb}{University College Dublin} & \textcolor{ucddb}{An Col\'aiste Ollscoile, Baile \'Atha Cliath}\\
%	Science and Technology &\textcolor{ucdg}{Dublin, Ireland} & \textcolor{ucdg}{Baile \'Atha Cliath, \'Eire}
%	\end{tabular}
%	}

	\vspace{2\baselineskip}

	\begin{tabular}{ll}
	Head of School:& Associate Prof. Emma Sokell
	\\
	Principal Supervisor:& Assistant Prof. Andrew K. Mitchell
	\\
	Research Studies Panel:& Prof. Peter Duffy \\& Associate Prof. Nicolae-Viorel Buchete
	\end{tabular}
	
	%
	
	% Bottom of the page
	\vfill
	
	{\large May 2022}

\end{center}
}
\end{titlepage}

\cleardoublepage

\pagenumbering{roman}
\setcounter{page}{1}

\tableofcontents
\cleardoublepage

\listoffigures
%\addcontentsline{toc}{chapter}{List of Figures}

%\listoftables
%\addcontentsline{toc}{chapter}{List of Tables}

%\thispagestyle{plain}
\chapter*{Abstract}
\addcontentsline{toc}{chapter}{Abstract}

Condensed matter systems in the solid state owe much of their properties to the quantum behavior of their electronic degrees of freedom. Despite these dynamical degrees of freedom being relatively simple, the emergent phenomena which appear from collective behavior of the many constituent particles can lead to highly non-trivial characteristics. Two such non-trivial characteristics are topological phases, and the phenomena which result from many strongly correlated degrees of freedom.

Presented in this thesis are a set of projects which lie at the intersection between strong correlations and topological phases of matter.
The first of these projects is a treatment of an infinite dimensional generalization of the Su-Schrieffer-Heeger model with local Coulomb interactions which is treated exactly using the technique of dynamical mean-field theory, with the numerical renormalization group as the impurity solver. Observed in the solution is power-law augmentation of the non-interacting density of states. The topological spectral pole becomes broadened into a power-law diverging spectrum, and the trivial gapped spectrum becomes a power-law vanishing pseudogap. At stronger interaction strengths we have a first-order transition to a fully gapped Mott insulator. This calculation represents an exact solution to an interacting topological insulator in the strongly correlated regime at zero temperature.

The second set of projects involves the development of methods for formulating non-interacting auxiliary models for strongly correlated systems.
These auxiliary models are able to capture the full dynamics of the original strongly correlated model, but with only completely non-interacting degrees of freedom, defined in an enlarged Hilbert space. We motivate the discussion by performing the mapping analytically for simple interacting systems using non-linear canonical transformations via a Majorana decomposition. For the nontrivial class of interacting quantum impurity models,  the auxiliary mapping is established numerically exactly for finite-size systems using exact diagonalization, and for impurity models in the thermodynamic limit using the numerical renormalization group, both at zero and finite temperature. We find that the auxiliary systems take the form of generalized Su-Schriefeer-Heeger models, which inherit the topological characteristics of those models. These generalized Su-Schrieffer-Heeger models are also formalized and investigated in their own right as novel systems. 

Finally, we apply the auxiliary field methodology to study the Mott transition in the Hubbard model. In terms of the auxiliary system, we find that the Mott transition can be understood as a topological phase transition, which manifests as the formation and dissociation of topological domain walls.

\cleardoublepage
\chapter*{Statement of Original Authorship}
\addcontentsline{toc}{chapter}{Statement of Original Authorship}

I hereby certify that the submitted work is my own work, was completed while registered as a candidate for the degree stated on the Title Page, and I have not obtained a degree elsewhere on the basis of the research presented in this submitted work.

%\vspace{3\baselineskip}
\vfill

This thesis is based on the following publications:
\begin{itemize}
	\item[\cite{motttopology}] S. Sen, P. J. Wong, A. K. Mitchell, ``The Mott transition as a topological phase transition,'' \textit{Phys. Rev. B} \textbf{102} 081110(R) \textbf{Editors' Suggestion} (2020). \href{https://www.arxiv.org/abs/2001.10526}{arXiv:2001.10526 [cond-mat.str-el]}
	\item[\cite{bethessh}] P. J. Wong, A. K. Mitchell, ``Topological phases of the interacting SSH model on the Bethe lattice,'' (\textit{In Progress}).
	\item[\cite{generalizedssh}] P. J. Wong, A. K. Mitchell, ``Extended SSH tight-binding models,'' (\textit{In Progress}).
	\item[\cite{auximp}] P. J. Wong, S. Sen, A. K. Mitchell, ``Effective theories for quantum impurity models'' (\textit{In Progress}).
	\item[\cite{phasymmotttopology}] P. J. Wong, S. Sen, A. K. Mitchell, ``Effective Topology in the $ph$-asymmetric Hubbard Model,'' (\textit{In Progress}).
%	\item \fullcite{motttopology}
\end{itemize}

\vfill

\phantom{.}

\chapter*{Acknowledgements}
\addcontentsline{toc}{chapter}{Acknowledgements}

I would first of all thank my supervisor Andrew Mitchell for inviting me to Dublin and into his burgeoning group and into his first batch of PhD students. His support and strong investment in my education over the past few years is what made this thesis possible.

Among the group members, I collaborated most with Dr. Sudeshna Sen, and without her expertise and contributions, several aspects of the work in this thesis would not be as high quality as it is.

I am grateful to the other group members, Emma Minarelli
and
Jonas Rigo,
for providing camaraderie in our shared situation.

Similarly, I have appreciated the company of Eoin, Eoin, and George in our shared office.

I must also thank the School of Physics management staff, particularly Bairbre Fox and John Brennan
for all their help with bureaucratic and administrative affairs, as well as the acquisition of IKEA furniture.

I would like to thank my Research Studies Panel members Profs. Vio Buchete and Peter Duffy for their additional guidance and support over the course of my studies here.

I would also like to thank my examiners, Prof. Vladimir Loboskin and Prof. Nuala Caffrey, and particularly my special external examiner Prof. C\'edric Weber, for their time in evaluating this work.

Outside of the academic sphere, I am grateful to my band mates from the Jazz Society, Ben, Marcie, Michael, Ois\'in, and Ted, for the many gigs we managed to perform together.

Finally I must acknowledge the support of my parents for innumerable aspects of life, but more directly as it pertains to this thesis, for harboring me in their shelter during a global pandemic which occurred during the progress of this work.

\cleardoublepage

\pagenumbering{arabic}
\setcounter{page}{1}
%\linespread{1.3}
{
\onehalfspacing
%\doublespacing
%\input{Chapters/prologue.tex}

\chapter{Introduction}

Condensed matter physics is the area of physics concerned with aggregates of many constituents. 
Aggregates which exist on macroscopic scales are comprised of electrons and atomic nuclei at the nanoscale. Their phenomenology is therefore dictated by quantum mechanical behavior.

A major area of study within condensed matter physics is then describing the macroscopic electronic behavior of solid materials in terms of their microscopic properties.
One of the basic classifications within this realm is that of electronic phases of matter. The primary phases describe whether or not these substances do or do not conduct electrical current, termed to be metallic or insulating phases.

In substances which are metallic, the electrons in the outer shells of the atomic constituents are free to propagate through the material.

However, there are various mechanisms which might lead a system to exhibit insulating behavior.
The most basic is that of the atomic insulator, where the local atomic potential is greater than the kinetic energy of the outermost valence electrons. An example of such an insulator is solid Ar.
A type of system for which the valence electrons do possess meaningful kinetic energy, but nevertheless exhibit insulating behavior is the band insulator, where there exists a large energy gap at the Fermi level. This means that there exist no electronic states which have sufficient energy to propagate through the material.

Another manner in which electrons may be prohibited from propagating is through the presence of disorder which takes the form of randomized local potentials on the lattice sites. This disorder can be the result of a high density of impurities or defects in the system. This type of insulator is known as an Anderson insulator.

There also exist insulators which might otherwise be concluded to be metals from elementary band theory. An example of such an insulator is that of a Mott insulator. In a Mott insulator the insulating behavior arises as a result of a strong Coulomb repulsion between electrons. In a system where electronic interactions play a meaningful role, an otherwise metallic phase may become insulating if
the interaction strength is sufficiently high enough to inhibit propagation of the charge carriers.

The Mott insulating phenomenon is an example of the non-trivial phenomenology which can arise from strongly correlated electrons. The level of correlation between electrons in a material can dramatically affect its macroscopic characteristics. As such, these correlations cannot generally be ignored in the analysis of these systems as their inherent properties can be dramatically altered by them, and completely new phenomena may arise due to them as well.

%impurity models

A special type of band insulator which has been discovered in recent decades is that of the
topological insulator~\cite{hk}. A topological insulator is a band insulator in which the momentum space energy bands possess non-trivial topology. Mathematically, topology refers to the global properties of a manifold. These properties are robust under small deformations of its geometry. A sphere and an ellipsoid are considered topologically equivalent, but topologically distinct from the torus. Quantitatively, this difference can be seen from the `hairy ball' theorem, which says that the tangent vector field on the sphere or ellipsoid would possess a singularity, whereas the tangent vector field on the torus would not. 
Phenomenologically, topological insulators are systems whose bulk possesses an insulating spectral gap, but on the boundary possess localized gapless electronic states.

Early notions of this sort of topology in condensed matter were in semi-conducting wells~\cite{volkovpankratov} and the integer quantum Hall effect~\cite{originalqhe}. In the integer quantum Hall effect, a $2d$ insulating material was observed to exhibit quantized conductance on its boundary in the presence of a strong external magnetic field. 
A later discovery was the
quantum spin Hall effect~\cite{kanemele,bernevigzhang}
which exhibited similar properties, but did not require the external magnetic field.

\begin{comment}
quantum anomalous Hall effect

topological insulators~\cite{hk} and topological superconductors~\cite{kitaev}

MOSFET\footnote{metal-oxide-semiconductor field-effect transistor} device

$\mathrm{Ga}\mathrm{As}\mathrm{-}\mathrm{Al}_{x}\mathrm{Ga}_{1-x}\mathrm{As}$ heterostructure

\end{comment}

When developing a theory describing a particular system, it is tempting to start with the most foundational elements of that system from which to base the theory on.
Nearly all known physical phenomena can be described, at a reductive level, by the quantum field theory arising from the action functional\footnote{The dynamics of the spacetime background is given by the \textit{classical} action $S_{\text{GR}}[\vartheta] = \frac{1}{2c\varkappa}\int \left[ \star (\vartheta \extp \vartheta) \extp R \right]$ which at time of writing does not yet have a satisfactory quantum description. Some elements of this open question are addressed in~\cite{mscthesis}.}
\begin{equation}
	S_{\text{SM}}[\psi,A,\Phi] = \int \left[ -\frac14 F \extp \star F + \overline{\psi} \left( \i \slashed{\mathrm{D}} - \star m \right) \psi + \mathrm{D} \Phi \extp \star \mathrm{D} \Phi - V(|\Phi|^2) + \mathcal{L}(\psi,\Phi) \right]
\end{equation}
where the integrand is the Standard Model of particle physics Lagrangian 4-form.
While this quantum field theory is considered to be fundamental to nearly all observed physical phenomena, it is not possible to directly reproduce all such phenomena from it. Even for relatively high-energy subnuclear physics at energy scales $\sim \SI{1}{\giga\electronvolt/c^2}$ it is often necessary to work with effective theories rather than the underlying fundamental one. 
One of the main reasons for evaluating effective models is the computational intractability of the more fundamental theory. 

Additionally, the fundamental products of reductionism cannot in turn capture emergent phenomena which appear at different parameter scales~\cite{moreisdifferent}.
This leads to another reason for developing effective theories. Only at this higher effective level is it possible to construct a theory which correctly describes the relevant phenomena. 

In terms of constructing effective theories, the fundamental constituents of a system are not necessarily the appropriate objects to model in order to produce a theory which describes a particular phenomenon.
It is even not uncommon for models of particular phenomena to appear drastically different than the actual systems they intend to represent.
An example of a system whose dynamical degrees of freedom are non-trivially related their elementary constituents is that of those exhibiting the fractional quantum Hall effect~\cite{fqhexp}. In the fractional quantum Hall effect, the system's charge carrying excitations have an electric charge which is a fraction of the unit charge of the electron~\cite{laughlinfqh,fracchargef,fracchargei}. In terms of its elementary constituents, systems exhibiting the fractional quantum Hall effect are comprised of only elementary electrons and atomic nuclei. These objects possess a precisely quantized integer electric charge. None of the elementary constituents of these systems possess fractional charge. The macroscopic phenomenon of fractional charge carriers then is an effect of emergent behavior arising from non-trivial interplay between the system's elementary constituents. 

This provides an example of a quantum quasiparticle. Quasiparticles are emergent degrees of freedom which arise from the collective dynamics of the elementary constituents of a system, but nonetheless may behave as the primary dynamical degrees of freedom of their system.
While quasiparticles are typically only thought of in quantum systems, the notion of a classical quasiparticle also exists~\cite{mattuck}. Such a classical quasiparticle may be thought of as a particle existing at the center of mass of a two-body problem, such as two orbiting asteroids in space, whose mass is the sum of the masses of the two bodies. There is no actual physical object with the total mass at the center of mass, but the system can be modeled as if there were such a body there.

A particularly notable type of quasiparticle in recent condensed matter studies are Majorana degrees of freedom~\cite{kitaev}. Majorana fermions were first derived as a consequence of relativistic quantum electrodynamics which a primary characteristic being that they are their own anti-particle. While there are no present experimental signatures of elementary particles which are Majorana fermions~\cite{pdg}, Majorana quasiparticles may appear in condensed matter systems, the classic example of manifesting as zero bias conductance peaks in the spectra of electrons tunneling across Josephson junctions~\cite{kitaev,majoranaforqc}.

%Other examples of exotic quasiparticles that arise from many-body dynamics which may appear in condensed matter contexts are skyrmions~\cite{altlandsimons} , or even axions~\cite{pdg,mscthesis}. The latter of which may exist from astrophysical sources

The first way in which the notion of effective theories enter into the work of this thesis is in the construction of basic models of condensed matter systems.
These are toy models that capture the basic physics of interest in a simplified setting. One such example is the Kitaev superconducting wire\index{Kitaev superconductor}, which is a model which is able to realize Majorana degrees of freedom~\cite{kitaev}.
The second manner in which the notion of effective theories enter is in the construction of auxiliary models, particularly in the second half of the thesis.
These auxiliary models are on an additional layer of effective theory in that their relevant degrees of freedom do not correspond to the actual physical degrees of freedom of the system. Nevertheless, these auxiliary models with their non-physical auxiliary degrees of freedom enable manipulation of the model into a form in which the physical properties of the model are still derivable.

\section{Outline of the Thesis}

The methods of quantum condensed matter physics employed in the work of this thesis are reviewed in \S\ref{ch:methods}.
In particular the theory of Green functions is discussed. Green functions in many-body quantum theory are quantities which are much more convenient to work with than the many particle wavefunctions of the dynamical degrees of freedom and numerous physically meaningful quantities can be derived from them.

Also covered in this chapter is more detailed quantitative analysis of the models described in the above introduction. Surveyed here is a collection of models which appropriately illustrate certain classes of physically interesting systems, and form the platform of the studies undertaken in this text.

A particularly relevant model discussed is the Su-Schrieffer-Heeger (SSH) model which describes a $1d$ topological insulator. In \S\ref{ch:genssh}, this model is generalized in various manners to other $1d$ models which exhibit topological features. The types of generalizations developed here serve as reference models for the auxiliary systems which appear in the later chapters of this thesis.

The first half of the thesis cumulates in \S\ref{ch:bethessh}, which analyzes a system exhibiting the topological features of the SSH model while incorporating strong electronic correlations. The system is solved numerically exactly in the limit of infinite dimensions using the DMFT-NRG method established in \S\ref{ch:methods}.

The second half of the thesis begins with \S\ref{ch:aux} and emphasizes the second theme of this work. It presents a set of novel auxiliary field mappings for strongly correlated electron systems which can map fully interacting systems to completely non-interacting ones.
The first of these methods is based on nonlinear canonical transformations of interacting fermionic systems based on their decomposition into Majorana degrees of freedom. This method makes extensive use of the Clifford algebra structure obeyed by Majorana degrees of freedom and the geometric structures which appear. While there are some complete calculations of specific applications given in \S\ref{sec:majorana}, the primary function of this section is a proof of principle that an analytic transformation of an interacting system to a non-interacting system exists. For the type of systems considered in this thesis this method of nonlinear canonical transformations has limited applicability due to its algebraic complexity for all but the simplest systems. However, the method in general may be able to achieve goals different than those of this thesis or have interesting applications to other types of systems not considered here. As such, the method is presented with some generality with numerous points of outlook provided to serve as a starting point for future work.

The second auxiliary field mapping presented in \S\ref{ch:aux} is the method which sees most application in the products of this thesis. 
The mapping is first developed for systems of finite Hilbert space before being adapted for systems with continuous spectrum. In both cases the mapping is applied to impurity models both at zero temperature as well as at finite temperature. In particular the effects of temperature on the parameters of the auxiliary system are discussed.

While the general methodology of the mapping is presented in \S\ref{ch:aux}, a more detailed study is reserved for the next chapter \S\ref{ch:motttopology} where a richer spectrum of results are obtained, as well as a connection involving the work of the first part of the thesis on topological phases of matter. This chapter represents an in-depth case study of the auxiliary field mapping for the Hubbard model across the Mott metal-insulator transition.

The thesis concludes in \S\ref{ch:conclusion} where the primary novel results of the thesis are recapitulated and elements of potential future work are highlighted.

\subsubsection{Notation \& Conventions}
%Chapters are labelled with capital Roman numerals, chapter sections are labelled with Arabic numerals, and chapter subsections are labelled with lowercase Roman numerals. Equations and floats are labelled with Arabic numerals continuously within each chapter, with the format chapter numeral.section numeral.equation/float number.

Except where noted, \textit{e.g.} in \S\ref{sec:transport}, units are chosen such that $\hslash = 1$, $k_{\text{B}} = 1$, and any lattice spacing constants are taken to be unitary and dimensionless,  meaning that the dimension of momentum $k$ is similarly $[k] = 1$. The inverse thermodynamic temperature is notated as $\beta = 1/T$. 
With this choice of units, the remaining free dimension is energy. Energy scales here will generally be with respect to the system under consideration's bandwidth, $D$, with arbitrary units.

%$\mathbbm{Z}^+ = \mathbbm{N}$ $0 \notin \mathbbm{Z}^+$

Matrix objects are notated in bold and $\mathbbm{1}$ denotes the identity matrix with its dimension only prescribed where context alone is insufficient.
The abbreviation $\hc$ stands for Hermitian conjugate.
Additional conventions and notation are defined as introduced as well as in the index.

\chapter{Elements of Condensed Matter Physics\label{ch:methods}}

This chapter introduces the technical apparatuses utilized throughout this thesis.
We begin by reviewing the basic theoretical framework employed to study many-body quantum systems.
A central object introduced is that of Green functions, from which a variety of physical quantities can be derived. The primary quantity examined in this thesis is the local spectral density of states.
The Green functions also play a central role in numerical computational schemes employed throughout this thesis.
Two in particular which are described in the following are the numerical renormalization group (NRG)~\cite{nrg} and dynamical mean-field theory (DMFT)~\cite{dmft}.

This chapter ends with an introduction to topology as it relates to condensed matter systems, and the Su-Schrieffer-Heeger model~\cite{ssh,shortcourse} is presented as an example system which exhibits these topological features.

%%%%%%%%%%%%%%%%%%%%
\section{Many-Body Quantum Theory}
%%%%%%%%%%%%%%%%%%%%

A many-body system consisting of $Q$ electrons, with positions $r_q$, and $P$ nuclei, with atomic numbers $Z_p$ and positions $R_p$, in the absence of external electromagnetic fields and ignoring nuclear and relativistic effects can be described by the Hamiltonian
\begin{equation}
\begin{aligned}
	\hat{H}	=
		&-\sum_{q=1}^{Q} \frac{\hslash^2}{2 m_{e}} \triangle_{q}
		+ \frac12 \sum_{\substack{q,q'=1\\ q'\neq q}}^{Q} \frac{e^2}{|r_{q} - r_{q'}|}
		-\sum_{p=1}^{P} \sum_{q=1}^{Q} \frac{e^2 Z_p}{|R_p - r_q|}
		\\
		&-\sum_{p=1}^{P} \frac{\hslash^2}{2 M_{p}} \triangle_{p}
		+ \frac12 \sum_{\substack{p,p'=1 \\ p'\neq p}}^{P} \frac{e^2 Z_{p} Z_{p'}}{|R_{p} - R_{p'}|}
\end{aligned}
\label{eq:toe}
\end{equation}
in Gau{\ss}ian units where $e$ is the unit electric charge, $m_e$ the mass of the electron and $M_p$ the mass of the $p^{\text{th}}$ nuclei.

The many-body wavefunction for this system $\mathnormal{\Psi} = \mathnormal{\Psi}(t, \{r\}, \{R\})$ obeys the Schr\"odinger equation
\begin{equation}
	-\frac{\hslash}{\i} \frac{\partial}{\partial t} \mathnormal{\Psi} = \hat{H} \mathnormal{\Psi} \,.
\label{eq:toeschroedinger}
\end{equation}
Such a wavefunction in principle describes most electronic condensed matter phenomena. This system however is in general not solvable, so approximations need to be made in order to obtain useful results for physical systems.

The physical systems modeled in this work are those which are in the solid state where the atomic nuclei are arranged in the structure of a crystal lattice with negligible kinetic energy. The rigidity of the nuclei also implies the negligibility of the potential between the nuclei as well, rendering both terms in the second line of \eqref{eq:toe} irrelevant. This case can be described more formally by the Born-Oppenheimer approximation~\cite{bornoppenheimer}. Nonetheless, the Schr\"odinger equation \eqref{eq:toeschroedinger} is still typically computationally intractable as $Q \sim \mathcal{O}(10^{23})$ in the solid state.

A many electron wavefunction is anti-symmetric under exchange of electrons. This means that the algebra of single electron wavefunctions is an exterior algebra $\bigwedge(V)$.
A many body wavefunction of $Q$ electrons $\mathnormal{\Psi}_Q$ can be  described by a linear combination of the exterior products of the single electron wavefunctions $\psi_{q_i}$ as
\begin{equation}
	\mathnormal\Psi_Q = \sum_{\pi} a_{q_1,q_2,\cdots,q_Q} \psi_{q_1} \extp \psi_{q_2} \extp \cdots \extp \psi_{q_Q}
\label{eq:antisymmetricstatistics}
\end{equation}
with the sum over all permutations $\pi$.
The total $Q$-body wavefunction may also be written more traditionally in terms of the Slater determinant
\begin{equation}
	\mathnormal\Psi(r_1,\ldots,r_Q) = \frac{1}{\sqrt{Q!}} \begin{Vmatrix} \psi_{1}(r_1) & \psi_{2}(r_1) & \cdots & \psi_{Q}(r_1) \\ \psi_{1}(r_2) & \psi_{2}(r_2) & \cdots & \psi_{Q}(r_2) \\ \vdots & \vdots & \ddots & \vdots \\ \psi_{1}(r_Q) & \psi_{2}(r_Q) & \cdots & \psi_{Q}(r_Q) \end{Vmatrix} \,.
\end{equation}
The alternating nature of the determinant automatically satisfies the sign change of exchanging a pair of fermionic particles.
Dealing with such a wavefunction for $Q>2$ quickly becomes highly computationally intensive. It is therefore desirable to use an alternative computational strategy which is more computationally tractable. Such a computational strategy may be found in the formalism of field quantization.

In statistical mechanics, systems with dynamical particle number are treated with the grand canonical ensemble, with partition function $\mathcal{Z}_{\textsc{GC}} = \Tr \e^{-\beta (\hat{H}-\mu\hat{N})}$, rather than the canonical ensemble, whose partition function is $\mathcal{Z}_{\textsc{C}} = \Tr \e^{-\beta \hat{H}}$. The relevant Hamiltonian is then $\hat{\mathcal{H}} = \hat{H} - \mu \hat{N}$ rather than simply $\hat{H}$. In this work the Fermi level is normalized such that $\mu = 0$, with the understanding that there exist filled states at energies $\varepsilon < 0$. The grand canonical Hamiltonian will then be notated simply as $\hat{H}$.

%%%%%%%%%%%%%%%%
\subsection{Field Quantization}
%%%%%%%%%%%%%%%%

The methods of ordinary quantum mechanics are suitable for treating systems comprised of a small number of particles with the particle number fixed.
When dealing with many particles in which the particle number may not be conserved, it is useful to treat the quantum particles instead as excitations of a quantum field.
Many-body degrees of freedom may be quantized according to field quantization~\cite{fradkinqft,altlandsimons,feynmanstatmech}. 
Field quantization also goes by the descriptive name of ``occupation number formalism'' of quantum mechanics, and for historical reasons it also goes by the less descriptive nomenclature of ``second quantization''\index{second quantization}.

Historically the quantum field approach was developed in the context of quantizing the electromagnetic field and to solve paradoxes associated with early approaches to relativistic quantum mechanics, but it is now understood more generally as a method for treating many identical particles, which inherently includes the case where particle number is a dynamical quantity, unlike ordinary quantum mechanics where particle number is fixed. As previously mentioned, this amounts to using the grand canonical ensemble versus the canonical ensemble in statistical mechanics.

A difference between the relativistic case and the non-relativistic many-body case is that the many-body case can always, in principle, be treated with a many-body Schr\"odinger equation, whereas such a framework is inherently unavailable for relativistic theories.  
The necessary conclusion that relativistic quantum mechanics is inherently a many-body quantum theory may be seen from the Dirac equation of relativistic fermions~\cite{feynmanqed}: The Dirac equation possesses an unbounded negative energy spectrum. This would imply that there exists finite amplitude for a positive energy Dirac fermion to transition to an arbitrary negative energy state releasing an arbitrary amount of radiation energy. This issue can be cured by postulating that all negative energy states are already filled, thereby preventing positive energy states from transitioning to those states. Therefore, in order to consider a single relativistic fermion, it is actually necessary to consider the infinite body case and the field formalism is inevitable.
In the non-relativistic condensed matter scenario the number of states is finite but typically rather large, $\mathcal{O}(10^{23})$.
Another distinguishing feature of relativistic Dirac fermions in vacuum and fermions in condensed matter is that the vacuum is defined by the Fermi level, so that the `negative energy' states are populated by actual fermions whose energies are below the Fermi level.

%for example due the non-zero amplitude of a particle to propagate outside its forward light cone.

The space of states of a quantum system with dynamical particle number is given by the direct sum of fixed number Hilbert spaces. This algebraic space is known as a Fock space\index{Fock space},
\begin{equation}
	\mathcal{F} = \mathcal{H}_{0} \oplus \mathcal{H}_{1} \oplus \cdots \oplus \mathcal{H}_{N} \oplus \cdots
\end{equation}
where $\mathcal{H}_j$ is the $j$-particle Hilbert space.
An element of Fock space takes the form of a state vector
\begin{equation}
	\left\lvert n_0 , n_1 , \ldots, n_N , \ldots \right\rangle = \lvert n_0 \rangle \oplus \lvert n_1 \rangle \oplus \cdots \oplus \lvert n_N \rangle \oplus \cdots
\label{eq:fockvector}
\end{equation}
where $\lvert n_i \rangle$ denotes the occupation of a single-particle state $n_i$. In the fermionic case, the occupation is either 0 or 1 for each state with unique quantum numbers.

Operators acting on Fock space vectors take the form of $\opd{a}{i}$ and $\op{a}{i}$ which are maps ${\hat{a}}{^{(\dagger)}_i} : \mathcal{F} \to \mathcal{F}$, but with $\opd{a}{i} : \mathcal{H}_{i} \to \mathcal{H}_{i+1}$ and $\op{a}{i} : \mathcal{H}_{i} \to \mathcal{H}_{i-1}$. They function as creating or annihilating a state $\lvert n_i \rangle$ by increasing or decreasing that state's occupation. Fock space operators are formally taken to be
operator valued distributions 
where the the operators are smeared by a localized function, such as an
$f(k) \in L^2$.
An operator valued distribution takes the form of~\cite{macroqft}
\begin{equation}
	\opd{a}{j,s} = \frac{1}{(2\pi)^{d/2}} \int \d^dk\, f_j(k) \opd{a}{k,s}
\end{equation}
with spatially localized wave packet $f_j(x)$
\begin{equation}
	f_j(x) = \frac{1}{(2\pi)^{d/2}} \int \d^dk\, f_j(k) \e^{\i k \cdot x}
\end{equation}
which belongs to an orthonormal set,
\begin{equation}
	( f_i , f_j ) = \int \d^dk\, f^*_i(k) f_j(k) = \delta_{ij}
\end{equation}
The action of the operator valued distribution $\opd{a}{j,s}$ on the vacuum state is the creation of a state localized at $j$ with quantum number(s) $s$ and wavefunction $f_j$.
The necessity of employing operator valued distributions rather than Hilbert space operators directly is due to the fact that position is not a well-defined quantum number. It is also necessary as plane wave states are non-normalizable.

The operators arising in field quantization respect the exchange statistics of their respective fields. For fermionic fields, the operators inherit the exterior algebra of the single particle fermionic wavefunctions \eqref{eq:antisymmetricstatistics}.

The space of physical states can be constructed by first defining a vacuum state $\lvert 0 \rangle$ which satisfies
\begin{equation}
	\op{a}{j} \lvert 0 \rangle = 0 \,.
\end{equation}
Finitely occupied states can be constructed by application of the $\opd{a}{j}$ operator to this vacuum state as
\begin{equation}
	\lvert j \rangle \vcentcolon= \opd{a}{j} \lvert 0 \rangle = \int \d^dk\, f_{j}(k) \opd{a}{k} \lvert 0 \rangle
\end{equation}
where the second equality emphasizes the distributional nature of the operator.
These states then form the basis for constructing the Fock space.
%In terms of distributions, the energy operator $\hat{H}$ can be thought of appropriately as
%\begin{equation}
%	\hat{H} \lvert j \rangle = \int \d^dk\, E_k f_j(k) \opd{a}{j} \lvert 0 \rangle
%\end{equation}
%field expansion
%\begin{equation}
%	\varphi(x) = \int \d^dk \left[ u(k) \op{a}{k} \e^{\i k\cdot x - \i E_k t} + v(k) \opd{b}{k} \e^{-\i k\cdot x + \i E_k t} \right]
%\end{equation}

The action of a non-relativistic quantum field is given by
\begin{equation}
	S[\hat{\psi}^\dagger,\hat{\psi}]
	=
	\int \d^d r \left[ -\frac\hslash\i \opd{\psi}{}(r) \frac{\partial}{\partial t} \op{\psi}{}(r) - \frac{\hslash^2}{2m} \nabla \opd{\psi}{}(r) \cdot \nabla \op{\psi}{}(r) - V(r) \opd{\psi}{}(r) \op{\psi}{}(r) \right]
\end{equation}
where $\hat{\psi}^{(\dagger)}(r)$ are field operators which create or annihilate a field excitation at point $r$. The canonical momenta for $\op{\psi}{}$ and $\opd{\psi}{}$ are 
\begin{align}
	\op{\Pi}{\psi} &= \i \hslash \opd{\psi}{}
	&
	\op{\Pi}{\psi^\dagger} &= -\i \hslash \op{\psi}{}
\end{align}
which obey the canonical commutation relations.
For fermionic fields, the canonical commutation relation involves the anticommutator,
\begin{equation}
	\{ \op{\psi}{}(r) , \op{\Pi}{\psi}(r') \} = \i\hslash \delta(r-r') \,.
\end{equation}
It follows from this that the field $\hat{\psi}$ obeys the commutation algebra
\begin{align}
	\{ \op{\psi}{a}(r) , \opd{\psi}{b}(r') \} &= \delta_{ab} \delta(r-r')
	&
	\{ \op{\psi}{a}(r) , \op{\psi}{b}(r') \} &= 0 = \{ \op{\psi}{a}(r) , \opd{\psi}{b}(r') \}
\label{eq:car}
\end{align}
for general quantum numbers $a$ and $b$. The resulting equations of motion of this action yield the field equivalent of the Schr\"odinger equation.
The corresponding Hamiltonian may be derived from the action as
\begin{equation}
	\hat{H} = \int \d^d r \left[ -\frac{\hslash^2}{2m} \opd{\psi}{}(r) \triangle \op{\psi}{}(r) + \opd{\psi}{}(r) \hat{V}(r) \op{\psi}{}(r) \right] \,.
\end{equation}

The discussion here only involved fermionic fields, which obey antisymmetric exchange statistics, \eqref{eq:car}. Bosonic fields on the other hand obey symmetric exchange statistics, and therefore require a slightly different quantization treatment. Since bosonic fields do not feature in this thesis their treatment will be omitted here, but can be found any standard field theory text, such as~\cite{altlandsimons,fetterwalecka,abrikosov}.

The physical systems which will be discussed in the following will be models of solids whose atomic nuclei are arranged into crystalline lattices with the localized orbitals around these atomic nuclei being the relevant dynamical degrees of freedom. 
The interpretation of the quantum many-body system in condensed matter on a lattice is not of many independent indistinguishable electrons orbiting atoms at various lattice sites, but rather that of a single electron quantum field permeating the entire system with excitations of this field occurring at the various lattice sites.

%%%%%%%%%%%%%%%%%%
\subsection{Tight-Binding Models}
%%%%%%%%%%%%%%%%%%

A large class of materials in the solid state at the microscopic level take the form of a periodic crystalline arrangement of atoms. These periodic structures are the unit cells of the system. Basic phenomenological models often involve unit cells with just one, or only a few, atomic sites. Real materials, particularly elaborate compounds, can consist of unit cells containing a dozen or more atoms.
While any real material is of course of finite size, the bulk of a material sample contains such a large number of unit cells that the bulk can be said to consist of an infinite number of them. The atomic potential can then be described as possessing translational symmetry which manifests in the relation
\begin{equation}
	V(r + \ell) = V(r)
\end{equation}
where $\ell$ is a displacement vector between the position $r$ in a unit cell with the corresponding position in a neighboring unit cell.
This means that the electronic wavefunction $\psi(r)$ must also obey this symmetry such that $\psi(r+\ell) = \psi(r)$.
This symmetry means that the wavefunctions may be written in terms of Bloch functions
\begin{equation}
	\psi_{k}(r) = u_k(r) \e^{\i k \cdot r}
\end{equation}
where $u_k(r+\ell) = u_k(r)$ and $k$ is $2\pi$ periodic.

Wavefunctions which are localized on each lattice site are Wannier functions, which are formed from the Fourier transform of the Bloch functions as
\begin{equation}
	\phi(r-R_j) = \frac{1}{\sqrt{\mathcal{V}}}\int \d^dk\, \e^{\i k \cdot R_j} \psi_k(r)
\end{equation}
where $\mathcal{V}$ is the volume of the unit cell.
Wannier functions form an orthogonal basis of wavefunctions,
\begin{equation}
	\int \d^dr\, \tensor*{\phi}{^*_a}(r-R_i) \tensor*{\phi}{_b}(r-R_j) = \delta_{ij} \delta_{ab} \,,
\end{equation}
taking into account quantum numbers $a$ and $b$.

\begin{figure}[h]
\centering
\begin{tikzpicture}[yscale=0.075,xscale=1.25]
\begin{scope}
    \clip (-0.6,2) rectangle (6.6,-35);
	\draw[blue] (-2,-32)--++(11,0);
	\draw[blue] (-2,-19)--++(11,0);
	\draw[blue] (-2,-11)--++(11,0);
	\draw[blue] (-2,-6)--++(11,0);
	\def\r{0.4}
	\foreach \a in {-1,0,1,2,3,4,5,6,7,8}
		\draw[scale=1,domain=((\a*10+1)/10):((\a*10+9)/10),smooth,variable=\x,purple,fill=white]  plot ({\x},{
			-\r/((\x+2)*(\x+2))
			-\r/((\x+1)*(\x+1))
			-\r/((\x)*(\x))
			-\r/((\x-1)*(\x-1))
			-\r/((\x-2)*(\x-2))
			-\r/((\x-3)*(\x-3))
			-\r/((\x-4)*(\x-4))
			-\r/((\x-5)*(\x-5))
			-\r/((\x-6)*(\x-6))
			-\r/((\x-7)*(\x-7))
			-\r/((\x-8)*(\x-8))
			-\r/((\x-9)*(\x-9))
			});
\end{scope} 
\end{tikzpicture}
\caption[Schematic of potential well of condensed matter lattice]{Schematic of potential well of condensed matter lattice with energy levels shown. In the tight-binding paradigm all energy states except the highest are ``tightly bound'' to the local atomic well, meaning that only an electron occupying the highest energy bound state possesses enough kinetic energy to have finite amplitude to tunnel into a neighboring well.}
\end{figure}
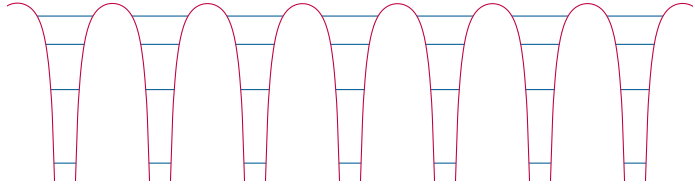
The approximation of the tight-binding model is that only the highest bound state has finite amplitude to tunnel through the potential to an available bound state on a neighboring site. The tight-binding model can also be used to treat the case where multiple orbitals are dynamical, or where there exist the possibility of long range tunneling to next-nearest neighbor (NNN) sites, or N$^n$NN sites in general. Systems in which the orbitals are highly localized,  which is the situation idealized by the tight-binding approximation, include the cases of partially filled $d$- or $f$-shell orbitals in transition metals and rare earth elements.

In the tight-binding approximation, the kinetic term of the electrons in Eq.~\eqref{eq:toe} may be obtained from analyzing the overlap between the localized wavefunction on neighboring sites. 
\begin{equation}
	t_{ij,\sigma} = \int \d^d r\, \tensor*{\phi}{^*_{\sigma}}(r-R_i) \left( -\frac{\hslash^2}{2m} \triangle + V(r,R) \right) \tensor*{\phi}{_{\sigma}}(r-R_j)
\label{eq:hoppingterm}
\end{equation}
where $V(r,R)$ is the potential produced by the underlying atomic lattice. For Hermitian dynamics, $t_{ij,\sigma}$ is symmetric in $i$ and $j$.

This calculation is necessary for generating a tight-binding model based on a specific material where the particular interatomic potentials $V$ is in principle able to be determined from \textit{ab initio} methods, such as density functional theory~\cite{dft,kohnsham}.
The models considered in this thesis are abstract models designed to capture features of certain broad classes of systems and not intended to obtain numerical data for specific materials. As such, the work which follows defines the $t$'s at the level of the tight-binding model and treats them as free model parameters.

The $t$ parameter describes the kinetics of a quantum field excitation propagating from one lattice site to another and it is conventionally called the hopping parameter, although the `$t$' notation originates from the terminology of these excitations `tunneling' through the lattice potential to different sites.

The inter-electron Coulomb interactions in the tight-binding formalism may be parameterized as
\begin{equation}
	U_{ijkl,\sigma\sigma'} = \frac12 \int \d^d r \int \d^d r' \, \tensor*{\phi}{^*_{\sigma}}(r-R_i) \tensor*{\phi}{^*_{\sigma'}}(r'-R_j) \frac{e^2}{\lvert r-r' \rvert} \tensor*{\phi}{_{\sigma'}}(r'-R_k) \tensor*{\phi}{_{\sigma}}(r-R_l) \,.
\end{equation}
In the cases which will be considered in the following, the Coulomb interaction will be considered to be a local interaction only, so only electrons occupying the same atomic site experience the interaction. This means that the interaction parameter reduces to
\begin{equation}
	U_{ijkl,\sigma\sigma'} \simeq U_{\sigma\sigma'} \delta(R_i - R_j) \delta(R_k - R_k) \delta(R_k - R_l) \equiv U \,.
\label{eq:localU}
\end{equation}
Since electrons obey the Pauli exclusion principle, the electrons involved in the local $U$ interaction must possess different quantum numbers. For electron quasiparticles considered here, the relevant quantum number is the spin projection, $\sigma \in \{\uparrow,\downarrow\}$, such that $U_{\sigma\sigma'} \equiv U_{\uparrow\downarrow}$.

The field theory formulation of quantum theory on a lattice can be constructed as described in the previous section where the smearing functions involved in the definition of the operator valued distributions are the Wannier functions,
%A field operator creating a state at $r$ can be defined as
%\begin{equation}
%	\opd{\psi}{\sigma}(r) = \sum_{i} \phi(r-R_i) \opd{c}{i\sigma}
%\end{equation}
%This relation can be inverted to define
\begin{equation}
	\opd{c}{i\sigma} = \int \d r \, \tensor*{\phi}{_\sigma}(r-R_i) \opd{\psi}{\sigma}(r) \,.
\end{equation}
This operator creates a state with wavefunction $\phi_\sigma(r-R_i)$. These operators obey the fermionic commutation algebra
\begin{align}
	\{ \op{c}{i,r} , \opd{c}{j,s} \} &= \delta_{ij} \delta_{rs}
	&
	\{ \op{c}{i,r} , \op{c}{j,s} \} &= 0 = \{ \opd{c}{i,r} , \opd{c}{j,s} \}
\end{align}
where $i$ and $j$ are lattice sites and $r$ and $s$ are quantum number labels.

In terms of the full microscopic description of solid state materials, the dynamical degrees of freedom may not be individual single electrons, but composite objects consisting of many particles. These composite objects are termed quasiparticles\index{quasiparticle} and can still be treated as individual quantum objects. In terms of the preceding discussion, the wavefunctions $\phi(r)$ now represent the localized wavefunction of the quasiparticle, the kinetic term in Eq.~\eqref{eq:hoppingterm} is the kinetic energy of the quasiparticle, and the potential $V$ in Eq.~\eqref{eq:hoppingterm} is the effective potential affecting the quasiparticle, which may be a composite result from the atomic nuclei as well as the electrons. The overall formalism remains the same as quasiparticles are treated as single quantum particles. These quasiparticles typically can be labelled with the same quantum numbers as elementary electrons, but they can exhibit some differences. An example is in the fractional quantum Hall effect~\cite{fqhexp} where the quasiparticles are observed to possess fractional electric charge~\cite{laughlinfqh,fracchargef,fracchargei}.

The simplest dynamical tight-binding model is that of fermions propagating on a regular lattice $\Gamma$,
\begin{equation}
	\hat{H} = \sum_{\boldsymbol{j}\in\Gamma} \tensor*{\varepsilon}{_{\boldsymbol{j}}} \opd{c}{\boldsymbol{j}} \op{c}{\boldsymbol{j}} + \sum_{\boldsymbol{j}\in\Gamma} \sum_{\boldsymbol{r}} \left( \tensor*{t}{_{\boldsymbol{j},\boldsymbol{j}+\boldsymbol{r}}} \opd{c}{\boldsymbol{j}} \op{c}{\boldsymbol{j}+\boldsymbol{r}} + \hc \right)
\label{eq:tightbindingchain}
\end{equation}
where $\boldsymbol{j}$ labels the lattice sites and $\boldsymbol{j}+\boldsymbol{r}$ parameterizes a Manhattan metric displacement from site $\boldsymbol{j}$ by $\boldsymbol{r}$ sites. $\varepsilon_{\boldsymbol{j}}$ is the energy of the single-particle state $\lvert \boldsymbol{j} \rangle$. Note that these energies are not eigenenergies of the Hamiltonian as single-particle quantum numbers are not good quantum numbers for a many-body Hamiltonian. The models this thesis is concerned with are of nearest neighbor kinetics, where $\boldsymbol{j}+\boldsymbol{r}$ is a displacement of 1 lattice site away from site $\boldsymbol{j}$. In general $\tensor*{t}{_{\boldsymbol{j},\boldsymbol{j}+\boldsymbol{r}}} \in \mathbbm{C}$ with $(\tensor*{t}{_{\boldsymbol{j},\boldsymbol{j}+\boldsymbol{r}}})^\dagger = \tensor*{t}{^*_{\boldsymbol{j}+\boldsymbol{r},\boldsymbol{j}}}$, but for the following discussion and in subsequent chapters it will be assumed that $\tensor*{t}{_{\boldsymbol{j},\boldsymbol{j}+\boldsymbol{r}}} = \tensor*{t}{_{\boldsymbol{j}+\boldsymbol{r},\boldsymbol{j}}} \in \mathbbm{R}$ without loss of generality.

As an example, consider a $d$-dimensional homogeneous system such that $\tensor*{t}{_{\boldsymbol{j},\boldsymbol{j}+1}} = t$ and $\tensor*{\varepsilon}{_{\boldsymbol{j}}} = \varepsilon \,$ $\forall \boldsymbol{j}\in\Gamma$, with $\Gamma$ the hypercubic lattice.
The eigenenergies for this system can be found by diagonalizing the Hamiltonian \eqref{eq:tightbindingchain} by Fourier transformation into momentum space,
\begin{align}
	\op{c}{\boldsymbol{j}} &= \frac{1}{\sqrt{2\pi}} \sum_{\boldsymbol{k}} \op{c}{\boldsymbol{k}} \e^{\i \boldsymbol{k} \cdot \boldsymbol{j}} \,,
	&
	\opd{c}{\boldsymbol{j}} &= \frac{1}{\sqrt{2\pi}} \sum_{\boldsymbol{k}} \opd{c}{\boldsymbol{k}} \e^{-\i \boldsymbol{k} \cdot \boldsymbol{j}} \,,
\end{align}
which results in the Hamiltonian taking the form of a free electron gas
\begin{equation}
	\hat{H} = \sum_{\boldsymbol{k}} \tensor{\varepsilon}{_{\boldsymbol{k}}} \opd{c}{\boldsymbol{k}} \op{c}{\boldsymbol{k}}
\end{equation}
which has dispersion
\begin{equation}
	\varepsilon_{\boldsymbol{k}} = \varepsilon + 2t \sum_{n=1}^{d} \cos(k_n) \,.
\label{eq:squaredispersion}
\end{equation}
This transformation also made use of the identity for the discrete delta function
\begin{equation}
	\sum_{\boldsymbol{j}} \e^{\i ( \boldsymbol{k} - \boldsymbol{k}' ) \boldsymbol{j}} = 2\pi \tensor{\delta}{_{\boldsymbol{k},\boldsymbol{k}'}} \,.
\end{equation}
%
%In higher dimensions the lattice site index $j$ and displacement index $r$ are vectors, $\boldsymbol{j}$ and $\boldsymbol{r}$. For a $d$-dimensional homogeneous hypercubic lattice, 
In the $1d$ case where $\Gamma$ is described as a chain, the dispersion relation is $\tensor{\varepsilon}{_k} = \varepsilon + 2 t \cos(k)$. The bandwidth $D$ can be identified as $D = 2 t$.

It is also convenient to write the Hamiltonian in other representations than in terms of the operators $\hat{c}^{(\dagger)}_{j}$. One representation is in terms of state vectors, or in Dirac notation.
In Dirac notation the Hamiltonian \eqref{eq:tightbindingchain} is expressed as
\begin{equation}
	\hat{H} = \sum_{j} \left( \left\lvert j \right\rangle \varepsilon \left\langle j \right\rvert + \left\lvert j+1 \right\rangle t \left\langle j \right\rvert + \left\lvert j \right\rangle t \left\langle j+1 \right\rvert \right)
\end{equation}
where the state vectors label the occupation of the $j^{\text{th}}$ state of the many-body state vector: $\lvert j \rangle \equiv \lvert \cdots, n_j , \cdots \rangle$ in the notation of Eq.~\eqref{eq:fockvector}. The creation and annihilation operators in this notation are then given by $\opd{c}{j} = \lvert j+1 \rangle \langle j \rvert$ and $\op{c}{j} = \lvert j-1 \rangle \langle j \rvert$.
%are given by $\lvert j \rangle = \opd{c}{j} \lvert \text{vac} \rangle$. 
A Hamiltonian matrix can be formed from the matrix elements of the Hamiltonian as
\begin{equation}
	[\boldsymbol{H}]_{ij} = \langle i \lvert \hat{H} \rvert j \rangle \,.
\end{equation}
For the Hamiltonian \eqref{eq:tightbindingchain} with $\Gamma \simeq \mathbbm{Z}^+$, the Hamiltonian matrix is
\begin{equation}
	\boldsymbol{H} = 
	\begin{pmatrix}
		\varepsilon_{1} & t_{12} &  & O	\\
		t_{21} & \varepsilon_{2} & t_{23} & 	\\
		 & t_{32} & \varepsilon_{3} & \ddots	\\
		O &  & \ddots & \ddots
	\end{pmatrix} \,,
\end{equation}
with the corresponding operator reconstructed as
\begin{equation}
	\hat{H} = \lvert \vec\psi \rangle \boldsymbol{H} \langle \vec\psi \rvert \,.
\end{equation}
This matrix representation is particularly useful for describing systems with internal degrees of freedom, where states with $S$ internal degrees of freedom can be written as
\begin{equation}
	\tensor*{\vec{\psi}}{^\dagger_j} = \adjvec{\opd{c}{j_1} & \opd{c}{j_2} & \cdots & \opd{c}{j_S}}
\end{equation}
with the Hamiltonian taking the form of
\begin{equation}
	\hat{H} = \sum_{j\in\Gamma} \tensor*{\vec{\psi}}{^\dagger_j} \tensor*{\boldsymbol{h}}{_0} \tensor*{\vec{\psi}}{_j} + \tensor*{\vec{\psi}}{^\dagger_{j+r}} \tensor*{\boldsymbol{h}}{_1} \tensor*{\vec{\psi}}{_j} + \tensor*{\vec{\psi}}{^\dagger_j} \tensor*{\boldsymbol{h}}{^\dagger_1} \tensor*{\vec{\psi}}{_{j+r}}
\end{equation}
where $\boldsymbol{h}_0$ and $\boldsymbol{h}_1$ are $S\times S$ matrices describing dynamics of orbitals within unit cells and orbitals between unit cells respectively.

%%%%%%%%%%%%%%%%%%%%%%%%%%%%%%%%%%%%%%%%%%%%%%%%%%%%%%%%%%%%%%%%%%%%%
\section{Physical Models}
%%%%%%%%%%%%%%%%%%%%%%%%%%%%%%%%%%%%%%%%%%%%%%%%%%%%%%%%%%%%%%%%%%%%%

This section presents some models in the tight-binding formalism which serve as base starting points for modeling various electronic phenomena in materials stemming from inter-electron interactions. An example of a non-interacting tight-binding model displaying non-trivial phenomena will be discussed in detail in \S\ref{sec:sshmodel}.

%%%%%%%%%%%%%%%%%%%
\subsection{Anderson Impurity Model}
%%%%%%%%%%%%%%%%%%%

One of the most basic models which incorporate interactions is that of an impurity model, where the system is a model of free non-interacting fermions with the interactions only occurring at a finite number of specific locations. These locations, the impurities, may in general consist of many, but finite number of, internal degrees of freedom, such as a small cluster of sites whose states may possess spin or isospin quantum numbers. 

A basic impurity model is the single impurity Anderson model~\cite{andersonmodel,hewson,kondo}.
This was originally conceived as a model describing localized magnetic moments on magnetic ions dissolved in non-magnetic metals. As the name suggests, the model consists of a single impurity in a non-interacting bath. The Anderson model corresponds to the dilute limit where the density of the magnetic impurities is sufficiently low that each of the impurities can be be treated independently of each other. An example of such a system is Mo-Nb alloys doped with Fe atoms~\cite{andersonmodel}.

%The spherical and plane-wave representations of the Anderson model \cite{kondo}

The Hamiltonain of the single impurity Anderson model is given by
\begin{equation}
	\hat{H}_{\textsc{siam}}
	=
	\underbrace{\sum_{k,\sigma} \tensor*{\varepsilon}{_{k}} \opd{c}{k,\sigma} \op{c}{k,\sigma}}_{\text{bath}}
	+ \underbrace{\sum_{k,\sigma} \left( \tensor*{V}{_{k,\sigma}} \opd{c}{k,\sigma} \op{d}{\sigma} + \tensor*{V}{^*_{k,\sigma}} \opd{d}{\sigma} \op{c}{k,\sigma} \right)}_{\text{hybridization}}
	+ \underbrace{\sum_{\phantom{,}\sigma\phantom{,}} \tensor*{\varepsilon}{_{d}} \opd{d}{\sigma} \op{d}{\sigma} + U \opd{d}{\uparrow} \op{d}{\uparrow} \opd{d}{\downarrow} \op{d}{\downarrow}}_{\text{impurity}}
\label{eq:siam}
\end{equation}
which consists of three parts: the bath, the impurity, and the hybridization between them. The $\op{c}{}$ operators act only on the bath, the $\op{d}{}$ operators act only on the impurity, and $\sigma \in \{\uparrow,\downarrow\}$ labels their spin. 
Since the bath is non-interacting, it can be diagonalized in momentum space. 
From the hybridization amplitude $V_{k}$, it is typical to define a hybridization function $\Delta(z)$ as
\begin{equation}
	\Delta(z) \vcentcolon= \sum_{k} \frac{\left\lvert V_k \right\rvert^2}{z - \varepsilon_k}
\label{eq:siamDelta}
\end{equation}
where $z$ is complex frequency.\footnote{The necessity of using complex frequency and the relation to real frequency $\omega$ will be introduced in \S\ref{sec:complexfrequency}.}
%\index{$0$@\textbf{List of Edits}!201@added missing definition}
This function describes the dynamics between the bath and impurity. In the literature~\cite{nrg,hewson}, the quantity $\pi \sum_{k} \lvert V_{k} \rvert^2 \delta(\omega - \varepsilon_{k}) = -\Im\Delta(\omega)$ is often taken as the definition of the hybridization function as it represents coupling of amplitude $V_{k}$ from the impurity to the bath density of states $\sum_{k} \delta(\omega - \varepsilon_{k})$. $\Delta(z)$ is an analytic function and so can be completely specified by its imaginary part. The form of Eq.~{\eqref{eq:siamDelta}} can also be obtained from the functional integral form of {\eqref{eq:siam}} where the bath degrees of freedom $\op{c}{}$ are integrated out, leaving a term of the form $\sum_k \tensor*{V}{^*_k} \opd{d}{} \frac{1}{z - \varepsilon_k} \tensor*{V}{_k} \op{d}{} = \Delta(z)$, which contributes to the non-interacting propagator for the impurity degrees of freedom $\op{d}{}$. 
%\index{$0$@\textbf{List of Edits}!202@added notes on $\Delta$}
The hybridization function essentially defines the characteristics of the impurity model: both the bath and impurity individually have elementary characteristics (as will be shown below in \S\ref{sec:gfeom} and \S\ref{sec:hubbardatomgf} respectively). It is the coupling between the two which results in non-trivial behavior.
%\cite{kondo}\cite{hewson}Brillouin-Wigner perturbation theory\cite{thouless}
%
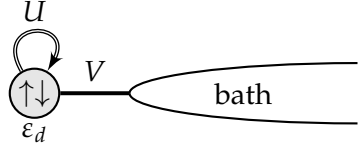
\begin{figure}[h]
\centering
\begin{tikzpicture}[{thick}]
	\node[circle,draw=black,fill=black!10,inner sep=1pt] (1) at (0,0) {$\uparrow\downarrow$};
	\node[below=0.25cm] at (1) {$\varepsilon_d$};
	\def\p{1.25}
	\def\l{3}
	\def\w{0.4}
	\coordinate (s) at (-\p,0);
	\coordinate (d) at (\p,0);
	\draw (d) arc(0:-90:-\l cm and \w cm);
	\draw (d) arc(0:-90:-\l cm and -\w cm);
	 \node at ($(d)+(1.5,0)$) {bath};
	 \draw[-,line width=1.5pt] (d)--(1) node[midway,above] {$V$};
 	\path (1) edge[-latex',line width=0.5pt,double distance=0.5pt,in=60,out=120,looseness=6] node[above] {$U$} (1);
\end{tikzpicture}
\caption[Schematic of the Anderson impurity model]{Schematic of the Anderson impurity model.\label{fig:anderson}}
\end{figure}
More general impurity systems may consist of additional internal degrees of freedom on the impurity, as well as hybridizations onto multiple different baths. In contemporary literature these more general impurity systems are commonly termed quantum dots. They are studied particularly in the non-equilibrium context where the various baths may have external voltage biases applied, thereby driving a current through the dot. Such a system is realized, for example, in semiconductor nanoribbon devices~\cite{emma2cck}.

%%%%%%%%%%%%%%%
\subsection{Hubbard Model}
%%%%%%%%%%%%%%%

The Hubbard model~\cite{gutzwiller,kanamori,hubbard} is a minimal model of a system with interacting fermions on a lattice.
%\cite{fradkin}
Its Hamiltonian is
\begin{equation}
    \hat{H}_{\textsc{h}} = \underbrace{\varepsilon \sum_{j,\sigma} \opd{c}{j,\sigma} \op{c}{j,\sigma} + t \sum_{j,\ell,\sigma} \left( \opd{c}{j,\sigma} \op{c}{j+\ell,\sigma} + \opd{c}{j+\ell,\sigma} \op{c}{j,\sigma} \right)}_{\op{H}{0}} + \underbrace{U \sum_{j} \opd{c}{j,\uparrow} \op{c}{j,\uparrow} \opd{c}{j,\downarrow} \op{c}{j,\downarrow}}_{\op{H}{I}}
\label{eq:hubbard}
\end{equation}
where lattice sites are labeled by $j$, $\ell$ is a displacement between sites on a given lattice $\Gamma$, and $\sigma \in \{\uparrow,\downarrow\}$ labels the spin. The $\op{H}{0}$ and $\op{H}{I}$ are the free (kinetic) and interacting parts of the Hamiltonian respectively. The original motivation of this model was to provide an explanation of the itinerant ferromagnetism of transition metals, such as Fe and Ni, but its use cases far exceed that context. Indeed, the Hamiltonian \eqref{eq:hubbard} is the tight-binding model corresponding to the field quantized Eq.~\eqref{eq:toe} under the approximation \eqref{eq:localU}.
%%%%%%%%%%%%%
A characteristic of the Hubbard model is the Mott metal-insulator transition~\cite{mott,mottreview,mitreview}\index{Mott transition}, where the ground state of the system transitions from being metallic to being insulating upon increase of the interaction strength $U$. A system in this insulating phase is called a Mott insulator. This phase transition can occur without symmetry breaking. Although the ground state of many Mott insulators is magnetically ordered, this is not a requirement, and paramagnetic Mott insulators are known~\cite{qptcusc}.
\begin{figure}[h]
\centering
\includegraphics[width=0.75\linewidth]{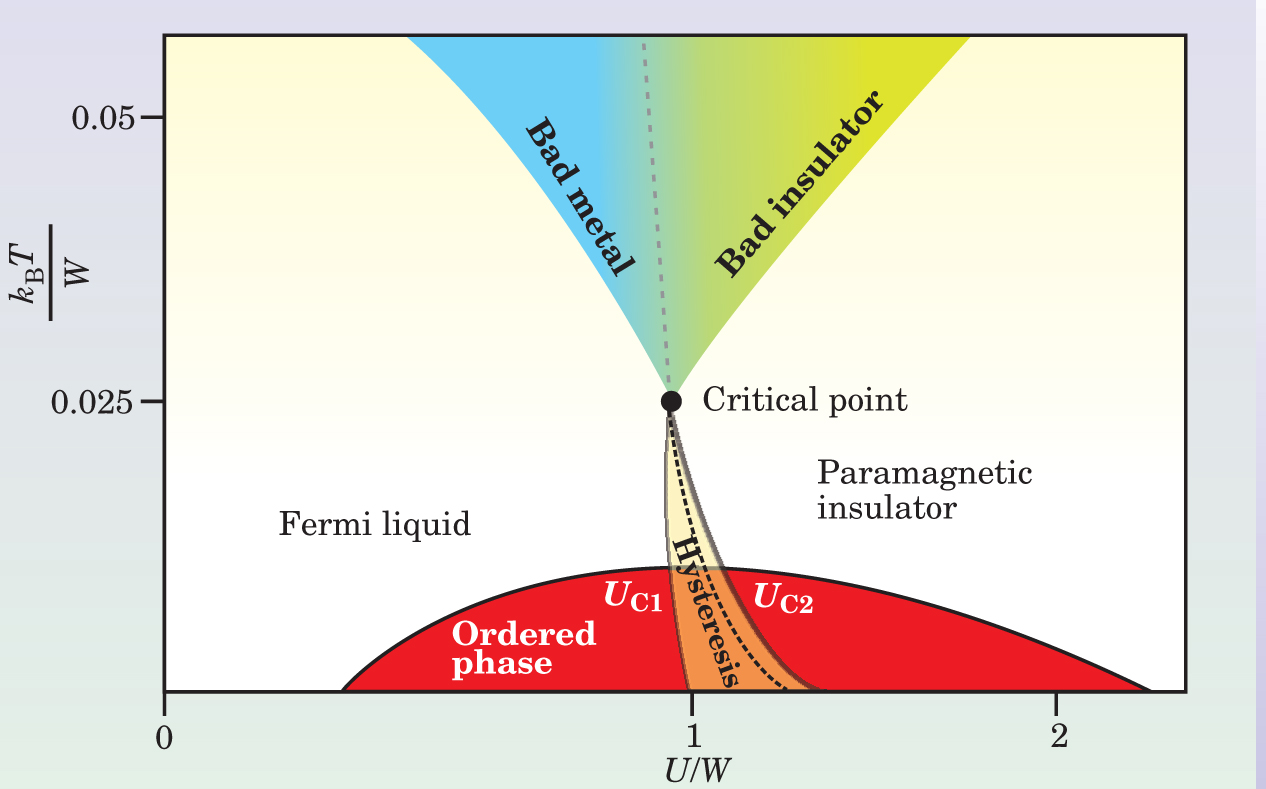}
\caption[Phase diagram of the Hubbard model]{Phase diagram of the Hubbard model (figure reproduced from~\cite{physicstodaydmft}). $U_{\text{c}1}$ marks the critical $U$ for the transition from insulator to metal (decreasing $U$), and $U_{\text{c}2}$ marks the critical $U$ for the metal to insulator transition (increasing $U$). The relation between $W$ in the figure and $D$ in this text is $W = 2D$.\label{fig:hubbardphasediagram}}
\end{figure}
A schematic phase diagram of the Hubbard is shown in Fig.~\ref{fig:hubbardphasediagram}. At low temperatures the Hubbard model undergoes a first-order transition between metallic and insulating phases, which may occur at some critical interaction strength, $U_{c1}$ or $U_{c2}$. The transition occurs at $U_{c2}$ when the system is adiabatically evolved from the metallic phase with increasing $U$. In contrast, the transition occurs at $U_{c1}$ when the system is adiabatically evolved with decreasing $U$ from the insulating phase. The parameter space region $U_{c1} < U < U_{c2}$ is a hysteresis coexistence region. The coexistence region, as well as the discontinuous nature of the transition, is characteristic of a first-order phase transition. Additionally, as mentioned previously the Mott transition is possible without symmetry breaking, which is another characteristic of a first-order transition. At higher temperatures, the $U_{c1}$ and $U_{c2}$ transition curves merge to a second-order critical point. Above this point the metal-insulator transition becomes continuous, and therefore second-order.
%\index{$0$@\textbf{List of Edits}!203@added additional discussion of Hubbard phases}

The Hubbard model \eqref{eq:hubbard} as employed in this thesis is a single band model with local Coulomb interactions. A type of interaction not covered by this model is the Hund interaction, conventionally parameterized by $J$, which is a coupling between spins in different orbitals. As this thesis only deals with the single band Hubbard model, the Hund interaction will not be considered in the following.
%\index{$0$@\textbf{List of Edits}!204@added discussion of magnetism}

The Mott insulating phase is the result of strong Coulomb interactions between electrons on the atomic sites. This is in contrast to conventional band insulators where the electronic energy gap is the result of the lattice. A Mott insulator is also distinguished from a system whose insulating behavior arises from Anderson localization, where the insulating characteristic is due to the presence of disorder~\cite{andersonlocalization}. A Mott insulator still possesses translation invariance on its lattice and is non-disordered. A further distinction between the Mott insulating and insulation arising from Anderson localization is that the Mott phase is inherently a many-body characteristic as it manifests from the interactions between particles. Anderson localization on the other hand is due to the characteristics of the background potential and is therefore not dependent on many dynamical degrees of freedom.
%\index{$0$@\textbf{List of Edits}!205@added additional distinction between Anderson and Mott}

The Hubbard model admits an exact solution in $1d$~\cite{liebwu} which may be obtained from Bethe ansatz methods, more generally known as the quantum inverse scattering method. While this method produces an exact solution of the eigenstates and eigenspectrum, it does not provide the dynamics.

A major avenue of research involving the Hubbard model is its $2d$ form. The $2d$ Hubbard model on a square lattice is model which is representative of CuO$_2$ planes which exist in cuprate superconductors~\cite{qptcusc,htcsc}.
It is suspected that the doped $2d$ Hubbard model on the square lattice is the appropriate model for high-$T_c$ superconductors, although this remains an open question~\cite{qptcusc,htcsc}.

\section{Green Functions\label{sec:greenfunctions}}

The concept of Green functions were introduced by George Green for the solution of linear differential equations arising in classical electrodynamics~\cite{green}. Classical Green functions have since been abstracted as a general method to formally solve inhomogeneous linear differential equations~\cite{mathewswalker, byronfuller}.
An example of such an equation can be given by
\begin{equation}
	\hat{L}[\partial] u(x) = f(x)
\end{equation}
where $\hat{L}[\partial]$ is a linear functional of the differentiation operator $\partial$. This equation admits the formal solution
\begin{equation}
	u(x) = u_0(x) + \int \d^d x'\, G(x,x') f(x')
\end{equation}
where $u_0(x)$ is the homogeneous solution and $G(x,x')$ satisfies the relation
\begin{equation}
	\hat{L} G(x,x') = \delta(x-x') \,.
\label{eq:gislinverse}
\end{equation}
The object $G(x,x')$ is the Green function for the operator $\hat{L}$. The form of Eq.~\eqref{eq:gislinverse} implies that the Green function can be interpreted as the inverse of the operator $\hat{L}$.

%introduced by definition or derived from the Feynman path integral

The notion of Green functions in quantum theory\footnote{The quantum Green function is alternatively referred to as the ``propagator'' or ``kernel'' (of a path integral) in the literature.} originally appeared in the context of the elementary processes of quantum electrodynamics~\cite{feynman,schwingergreen} and were first applied in condensed matter systems in the context of superconductivity~\cite{salam}.
Green function techniques were systematically developed for quantum many body systems in
\cite{galitskiimigdal,martinschwinger} following the formal developments of quantum field theory.

%Schwinger quantum action principle

The systems analyzed in this work are primarily characterized and understood in terms of their Green functions\index{Green function}~\cite{abrikosov,fetterwalecka,bruusflensberg,mattuck,economou}. These are single-particle correlation functions in the many-body quantum system. In particular, the equation of motion approach~\cite{martinschwinger,zubarev} is an important tool. A reason for studying the Green functions of a system is that the elementary excitation spectrum is given by the poles of the Green function. As will be shown, signatures of topological states are also evident in the boundary Green functions.

\begin{comment}
For the eigenvalue equation
\begin{equation}
	\hat{L} u_n = \lambda_n u_n
\end{equation}
it follows that
\begin{equation}
	f(\hat{L}) u_n = f(\lambda_n) u_n
\end{equation}
holds for any well-behaved function $f$.
Proof of this statement follows from a power series expansion of $f(\hat{L})$. From this property it follows that
\begin{equation*}
	\e^{\i \hat{H} t} \lvert\psi_n\rangle = \e^{\i E_n t} |\psi_n\rangle
\end{equation*}
and
\begin{equation*}
	\frac{\1}{E_m - \hat{H}} \lvert \psi_n \rangle = \frac{1}{E_m - E_n} \lvert \psi_n \rangle
\end{equation*}
\end{comment}

\begin{comment}
state $\lvert \Psi(t') \rangle$
insert fermion at time $t'$ $\opd{c}{a}(t') \lvert \Psi(t') \rangle$
evolve in time to $t>t'$ $\hat{U}(t,t') \opd{c}{a}(t') \lvert \Psi(t') \rangle$
calculate overlap with $\opd{c}{a}(t) \lvert \Psi(t) \rangle$
\begin{align}
	\langle \Psi(t) \rvert \op{c}{a}(t) \hat{U}(t,t') \opd{c}{a}(t') \lvert \Psi(t') \rangle
	&=	\langle \Psi \rvert \op{c}{a}(t) \opd{c}{a}(t') \lvert \Psi \rangle
\end{align}
\end{comment}

The concept of the Green function in quantum theory is as follows.
From an initial state $\lvert \mathnormal\Psi \rangle$,
act with operator $\hat{B}(t')$ at time $t'$
and
calculate overlap with state $\hat{A}(t) \lvert \mathnormal\Psi \rangle$ at some time $t$.
The Green function is the
quantum thermal expectation value over all such processes,\footnote{$\displaystyle
	\left\langle \cdots \right\rangle \vcentcolon= \mathcal{Z}^{-1} \sum_{n} \left\langle n \left\lvert \cdots \, \e^{-\beta\hat{H}} \right\rvert n \right\rangle
$ and $\mathcal{Z}$ is the (grand canonical) partition function and $\{\lvert n \rangle\}$ is a complete basis of eigenstates.}
\begin{equation}
	G^c(t,t')
	\vcentcolon=
	-\i \left\langle \mathsf{T} \hat{A}(t) \hat{B}(t') \right\rangle \,.
\end{equation}
A Green function of this form is called
the causal double-time Green function\index{Green function!double-time} where $\mathsf{T}$ denotes causal time ordering of the operators. Such a Green function is distinguished from many-time Green functions that arise, for example, in elementary particle physics calculations which involve many intermediary processes. In condensed matter and statistical mechanics, the double-time Green functions contain sufficiently complete information about the many-body system for most purposes~\cite{zubarev}.

Green functions come in the varieties of causal\index{Green function!causal}, retarded\index{Green function!retarded}, and advanced\index{Green function!advanced}, defined as
\begin{subequations}
\begin{align}
%	\Green{\hat{A}(t)}{\hat{B}(t')}^c	=	
	G^c(t,t')	&\vcentcolon=	-\i \left\langle \mathsf{T} \hat{A}(t) \hat{B}(t') \right\rangle
	\\
%	\Green{\hat{A}(t)}{\hat{B}(t')}^r	=	
	G^r(t,t')	&\vcentcolon=	-\i \theta(t-t') \left\langle \{ \hat{A}(t) , \hat{B}(t') \} \right\rangle
	\\
%	\Green{\hat{A}(t)}{\hat{B}(t')}^a	=	
	G^a(t,t')	&\vcentcolon=	\phantom{-}\i \theta(t'-t) \left\langle \{ \hat{A}(t) , \hat{B}(t') \} \right\rangle
\end{align}
\end{subequations}
where $\{\cdot\,,\cdot\}$ is the anticommutator associated with Fermi statistics.\footnote{While the anticommutator is used exclusively in this work, the Green function may also be defined using the commutator instead. The use of either the commutator or anticommutator in the Green function is not defined \textit{a priori} and is in general given by $[\hat{A},\hat{B}]_{\zeta} = \hat{A} \hat{B} + \zeta \hat{A} \hat{B}$ where $\zeta = \pm$. The choice of $\zeta$ is determined by the properties of $\hat{A}$ and $\hat{B}$, usually the commutator ($-$) for bosonic operators and the anticommutator ($+$) for fermionic operators. The $\hat{A}$ and $\hat{B}$ may also in general be compound operators which potentially obey more complicated statistics than those of single Bose or Fermi operators, and in these cases the choice of commutator is based on convenience for the problem and operators involved~\cite{zubarev}.}
%
\begin{comment}
$\hat{A}(t) = \e^{\i \mathcal{H} t} \hat{A} \e^{-\i \mathcal{H} t}$

time ordering
\begin{equation}
	\mathsf{T} \hat{A}(t) \hat{B}(t') = \theta(t-t') \hat{A}(t) \hat{B}(t') - \theta(t'-t) \hat{B}(t') \hat{A}(t)
\end{equation}

\begin{align}
	G_r(t-t') &= \int_{-\infty}^{\infty} \d E \, \e^{- \i E (t-t')} G_r(E)
	\\
	G_r(E) &= \frac{1}{2\pi} \int_{-\infty}^{\infty} \d t \, \e^{\i E t} G_r(t)
\end{align}

\begin{align}
	G_r(E) &= \frac{1}{2\pi \i} \int_{-\infty}^{\infty} \d t \, \e^{\i E (t-t')} \theta(t'-t) \left\langle \{ \hat{A}(t) , \hat{B}(t') \} \right\rangle
\end{align}
\end{comment}
%
The retarded Green function is analytic in the upper half plane and the
advanced Green function is analytic in the lower half plane.
The retarded Green function is generally the most useful for extracting physically relevant quantities of a system, and will therefore be the only type of Green function considered in the following and notated with the superscript omitted.
%
\begin{comment}
\begin{figure}[h]
\centering
\begin{tikzpicture}[decoration={markings, mark=at position 0.425 with {\arrow[xshift=3pt]{Stealth[length=6pt,width=4pt]}}, mark=at position 0.9 with {\arrow[xshift=3pt]{Stealth[length=6pt,width=4pt]}}},scale=0.75]
\draw[->] (-2.5,0)--(2.5,0);
\draw[->] (0,-2.5)--(0,2.5);
%
\draw[-,line width=1pt,postaction={decorate}] (2,0) arc (0:180:2cm and 2cm)--(-2,0)--(2,0)--cycle;
%
\foreach \a in {0.2,0.4,...,1.99}
{
	\node[scale=0.67] at (-\a,0.15) {$\times$};
}
\end{tikzpicture}
\hspace{5em}
\begin{tikzpicture}[decoration={markings, mark=at position 0.15 with {\arrow[xshift=3pt]{Stealth[length=6pt,width=4pt]}}, mark=at position 0.72 with {\arrow[xshift=3pt]{Stealth[length=6pt,width=4pt]}}},scale=0.75]
\draw[->] (-2.5,0)--(2.5,0);
\draw[->] (0,-2.5)--(0,2.5);
%
\draw[-,line width=1pt,postaction={decorate}] (2,0) arc (0:180:2cm and -2cm)--(-2,0)--(2,0)--cycle;
%
\foreach \a in {0.2,0.4,...,1.99}
{
	\node[scale=0.67] at (\a,-0.15) {$\times$};
}
\end{tikzpicture}
\end{figure}
\end{comment}
%

Using the Heisenberg representation of the operators, $\hat{A}(t) = \e^{\i \hat{H} t} \hat{A} \e^{-\i \hat{H} t}$, a useful property of the Green function for time-independent Hamiltonians is
\begin{equation}
\begin{aligned}[b]
	G(t,t')
	&=	-\i \theta(t-t') \left\langle  \{ \hat{A}(t) , \hat{B}(t') \} \right\rangle
	\\
	&=	-\i \theta(t-t') \left\langle \hat{A} \e^{-\i (\hat{H}-E) (t-t')} \hat{B} + \hat{B} \e^{+\i (\hat{H}-E) (t-t')} \hat{A}  \right\rangle \,,
\end{aligned}
\end{equation}
which means that $G(t,t') \equiv G(t-t')$.
This dependence of the Green function only on the interval means that the Green function can be found in frequency space by taking the Fourier transform as
\begin{equation}
	G(\omega) = \int_{-\infty}^{\infty} \d t \, \e^{\i \omega t} G(t) \,.
\end{equation}
This transform, however, is only well-defined if the integral converges.
This can be ensured by imposing a convergence factor of $\e^{\pm\delta t}$ as
\begin{equation}
	G(\omega) = \int_{-\infty}^{0} \d t \, \e^{\i \omega t + \delta t} G(t) + \int_{0}^{+\infty} \d t \, \e^{\i \omega t - \delta t} G(t)
\end{equation}
with $\delta >0$ and evaluated with the limit $\lim_{\delta\to0^+}$.\label{sec:complexfrequency}

The transform to frequency then involves a complex Fourier transform, or Laplace transform, with complex frequency $z \in \mathbbm{C}$ rather than real frequency $\omega \in \mathbbm{R}$. The frequency space Green function is then notated by $G(z)$.\footnote{The general complex frequency $z$ is also useful as it allows the definition of the single $G(z)$ for both real- and imaginary-time formulations. The imaginary-time Green function can be obtained as the case where $z = \i \omega_n$ where $\omega_n$ are the Matsubara frequencies~\cite{bruusflensberg}.}
The momentum space Green function $G(\boldsymbol{k})$ can be obtained by performing a Fourier transform on position. This can only be defined in the case of translation invariant Hamiltonians, which will not generally hold for the systems under consideration in the following, so the momentum space Green function will not be discussed further.

It is useful to introduce for the Green functions the alternative notation of Ref.~\cite{zubarev}, 
\begin{equation}
	\Greenline{\hat{A}(t)}{\hat{B}(t')}\index{$\Green{\cdot\,}{\cdot}$}\index{Green function!Zubarev notation|see {$\Greenline{\cdot\,}{\cdot}$}}
	\vcentcolon= G(t,t')\,,
\end{equation}
as $\hat{A}$ and $\hat{B}$ may in general be compound operators, which arises in the case of higher-order Green functions.\footnote{Note that this `order' terminology refers to the multiplicity of the operators $\hat{A}$ and/or $\hat{B}$, and not \textit{e.g.} the order of perturbation theory. What is referred to as `higher-order Green functions' still arise in non-perturbative calculations.} This notation additionally allows for transparency of which basis of the Hamiltonian the operators $\hat{A}$ and $\hat{B}$ are elements of.\footnote{An example of such a utility will be demonstrated in \S\ref{sec:domwalls}} The notation is also well suited for the equation of motion method for calculating Green functions, as will be shown in the subsequent section.
With this notation, the complex frequency Green function can be expressed as
\begin{equation}
%	\langle\langle \hat{A} ; \hat{B} \rangle\rangle_z = \int_{0}^{\infty} \d t\ e^{\i z t} \langle\langle \hat{A}(t) ; \hat{B}(0) \rangle\rangle .
	\Green{\hat{A}}{\hat{B}}_z \vcentcolon= \i\int_{0}^{\infty} \d t\, \e^{\i z t} \Green{\hat{A}(t)}{\hat{B}(0)} \,.
\end{equation}

An important quantity which can be obtained from the retarded Green function is the spectral function, or density of states.
%\begin{align}
%	\mathcal{A}(\omega)
%	&= \sum_{n} | n \rangle \langle n | \delta(\omega-\omega_n)
%	\\
%	&= \sum_{n} | n \rangle \langle n |\frac1\pi\lim_{\eta\to0^+}\frac{\eta}{(\omega-\omega_n)^2 + \eta^2}
%	\\
%	&= \sum_{n} | n \rangle \langle n |\frac1\pi\lim_{\eta\to0^+}\Im\frac{-1}{\omega-\omega_n + \i \eta}
%	\\
%	&= -\frac1\pi\sum_{n} \lim_{\eta\to0^+} \Im \frac{| n \rangle \langle n |}{\omega-\omega_n + \i \eta}
%	\\
%	&= -\frac1\pi \Im G^R(\omega)
%\end{align}
%
%\begin{align*}
%	G^R(\omega)
%	&= \int \d E \left[ \frac{A}{E - \omega + \i \delta} + \frac{B}{E + \omega + \i \delta} \right]
%	\\
%	&=	- \i \pi \int \d E A \delta(E - \omega) + \pi \Re G^R(\omega)
%\end{align*}
This can be obtained from the Lehmann representation\index{Lehmann representation} of the Green function,
%\begin{equation}
\begin{align}
	\Green{\op{c}{i}(t)}{\opd{c}{j}(0)}
	&=	\frac{1}{\mathcal{Z}} \sum_{n,m} \e^{-\beta E_n} \left( \langle n \lvert \op{c}{i} \rvert m \rangle \langle m \lvert \opd{c}{j} \rvert n \rangle \e^{\i (E_n - E_m) t} + \langle n \lvert \opd{c}{j} \rvert m \rangle \langle m \lvert \op{c}{i} \rvert n \rangle \e^{-\i (E_n - E_m) t} \right) \notag
	\intertext{which in the energy representation is}
	\Green{\op{c}{i}}{\opd{c}{j}}_z
	&=	\frac{1}{\mathcal{Z}} \sum_{n,m} \frac{\langle n \lvert \op{c}{i} \rvert m \rangle \langle m \lvert \opd{c}{j} \rvert n \rangle}{z + E_n - E_m} \left( \e^{-\beta E_n} + \e^{-\beta E_m} \right) \,. \label{eq:greenlehmann}
\end{align}
%\end{equation}
Here the states $\lvert n \rangle$ and $\lvert m \rangle$ label a complete basis of many-particle eigenstates which satisfy $\hat{H} \lvert n \rangle = E_n \lvert n \rangle$, and the identity $\sum_m \lvert m \rangle \langle m \rvert = 1$ has been used. The Lehmann representation is a convenient form from which Green functions can be calculated. However, obtaining the exact expression for the Green function requires knowing the complete basis of eigenfunctions and eigenenergies of the system in question. Its utility for exact calculations is therefore limited in many-body systems due to the exponential increase in dimension of the Hilbert space. From the form \eqref{eq:greenlehmann} it is apparent that the Fourier transform of the Green function cannot be with real frequency and the frequency space (causal) Green function must be analytically continued into the complex plane, such that the transform to frequency space then becomes a Laplace transform. The eigenenergy values which appear in the denominator lie on the real axis, since they are the eigenvalues of a Hermitian operator. The contour in the Laplace transform must be shifted to avoid these poles. The retarded Green function is obtained from the analytic continuation to $z = \omega + \i\delta$.
By means of the Sokhotski-Plemelj theorem \eqref{eq:sokhotskiplemelj},
\begin{equation}
	\lim_{\epsilon\to0} \int_{a}^{b} \frac{f(x-x_0)}{x - x_0 \pm \i \epsilon} \d x = \mp \i \pi \int_{a}^{b} f(x) \delta(x-x_0) \d x + \fint_{a}^{b} \frac{f(x-x_0)}{x-x_0} \d x \,,
%	\lim_{\delta\to0} \int_{a}^{b} \frac{\lvert u_{nm;j} \rvert^2}{\omega - E_n + E_m + \i \delta} \d \omega = - \i \pi \int_{a}^{b} \lvert u_{nm;j} \rvert^2 \delta(\omega - E_n + E_m) \d \omega + \fint_{a}^{b} \frac{\lvert u_{nm;j} \rvert^2}{\omega - E_n + E_m} \d \omega
\tag{\ref*{eq:sokhotskiplemelj}}
\end{equation}
the Green function can be decomposed into real and imaginary parts.
As the retarded Green function is analytic, it follows that the part involving the Cauchy principal value is the real part
\begin{equation}
	\Re G(\omega) = \frac1\pi \fint \frac{\Im G(\omega')}{\omega' - \omega} \d\omega' \,.
\end{equation}
The remaining imaginary part of the Green function can be identified with the spectral function, or local density of states. The spectral function\index{spectral function} for site $j$ can be defined as
\begin{equation}
	\mathcal{A}_{j}(\omega)
	=	\frac{1}{\mathcal{Z}} \sum_{m,n} \lvert u_{nm;j} \rvert^2 \delta\left( \omega - ( E_m - E_n ) \right) \left( \e^{-\beta E_n} + \e^{-\beta E_m} \right)
\end{equation}
which consists of a set of delta poles with weights given by $\lvert u_{nm;j} \rvert^2$ at positions $\omega_{j} = E_m-E_n$, where $u_{nm;j} = \langle n \lvert \op{c}{j} \rvert m \rangle$ and $u_{mn;j}^*= \langle m \lvert \opd{c}{j} \rvert n \rangle$. The spectral function measures the number of states in the many-body spectrum at a given energy $\omega$. In this thesis the normalization of spectral functions $\mathcal{A}(\omega)$ is chosen such that
\begin{equation}
	\int_{-\infty}^{\infty} \d\omega\, \mathcal{A}(\omega) = 1
\end{equation}
although other normalization choices exist in the literature.\footnote{A common alternative definition of the spectral function from the retarded Green function is $\mathcal{A}(\omega) = - 2 \Im \Greenline{\op{c}{j}}{\opd{c}{j}}_{\omega+\i0^+}$ \cite{altlandsimons,bruusflensberg} such that the integration measure is $\displaystyle \frac{\d\omega}{2\pi}$ rather than $\d\omega$, with $\displaystyle \int \frac{\d\omega}{2\pi} \mathcal{A}(\omega) = 1$.}
It follows from Eq.~\eqref{eq:sokhotskiplemelj} that the spectral function can be obtained from the local retarded Green function on site $j$ as
\begin{equation}
	\mathcal{A}_{j}(\omega) = -\frac1\pi \lim_{\delta\to0^+} \Im \Green{\op{c}{j}}{\opd{c}{j}}_{\omega+\i\delta} \,.
\end{equation}
%\begin{equation}
%	z \Green{\hat{A}}{\hat{B}}_z = \left\langle [ \hat{A} , \hat{B} ]_\zeta \right\rangle + \Green{[\hat{A} , \hat{H}]_-}{\hat{B}}_z
%\end{equation}
%Resolvant operator
%\begin{equation}
%	G_{i,j} = \left\langle \psi_i \left| \frac{1}{z\mathbbm1 - \hat{H}} \right| \psi_j \right\rangle
%\end{equation}
The spectral function is a scalar function of a real variable $\omega$ with $\mathcal{A}(\omega)$ being obvious notation. For complex analytic functions, such as the retarded single-particle Green function $G(z)$, it is useful to adopt the notation that for functions $f$ of complex variable $z$,
\begin{equation}
	f(\omega) \equiv f(\omega^+) \equiv f(\omega+\i0^+) \vcentcolon= \lim_{\delta\to0^+} f(\omega+\i \delta)
\label{eq:omeganotation}
\end{equation}
so that the notation $G(\omega)$ has a well defined meaning.

For systems with repeating unit cells or internal degrees of freedom, it is useful to express the Green function in terms of a larger matrix, with each matrix element being the Green function over specific elements of these internal degrees of freedom.
The matrix Green function for systems with $S$ internal degrees of freedom is
\begin{equation}
	\boldsymbol{G}_{\mu,\nu}(z) = \begin{pmatrix} \Green{\op{c}{\mu,n_1}}{\opd{c}{\nu,n_1}}_z & \cdots & \Green{\op{c}{\mu,n_1}}{\opd{c}{\nu,n_S}}_z \\ \vdots & \ddots & \vdots \\ \Green{\op{c}{\mu,n_S}}{\opd{c}{\nu,n_1}}_z & \cdots & \Green{\op{c}{\mu,n_S}}{\opd{c}{\nu,n_S}}_z \end{pmatrix}
\end{equation}
$\mu$, $\nu$ are indices enumerating the unit cells and the $n_i$ indices enumerate the internal degrees of freedom.

For a finite system, the spectrum consists of a set of discrete delta poles. However, these poles can be broadened under certain circumstances, such as finite lifetime of the quasiparticles or scattering from electron-phonon or electron-electron interactions.
Consider a Green function of the form
\begin{equation}
	G(z) = \frac{1}{z - E - \i \Gamma} \,.
\end{equation}
The spectrum can be shown to be
\begin{equation}
	\mathcal{A}(\omega) = \frac1\pi \frac{\Gamma}{(\omega - E)^2 + \Gamma^2}
\label{eq:broadA}
\end{equation}
which is not a delta function, but rather a Lorentzian peak of finite width. Examples of such a $\Gamma$ are the inverse lifetime of decaying quasiparticles.
In this circumstance the retarded Green function takes the form of $G(t) \approx -\i\theta(t) \e^{-\i E t} \e^{-t/\tau}$ where $\tau$ is the quasiparticle lifetime. This essentially amounts to a shift in the energy $E$ to $E + \i/\tau$. The spectrum then takes the form of Eq.~\eqref{eq:broadA} with $\Gamma = 1/\tau$.
A finite lifetime for quasiparticle states can arise, for example, due to coupling with an external bath, or contributions from the imaginary part of the self-energy, as will be seen later. The similar way in which both processes enter the Green functions suggests that the effect of interactions can be viewed as coupling to external baths. The self-energy arises from particle interactions and the broadening is interpreted as a spread in the energy spectrum due to scattering.
From the Lorenzian form of the delta function \eqref{eq:lorentzian}
\begin{equation}
	\delta(x-x_0) = -\frac1\pi \lim_{\eta\to0^+} \Im \frac{1}{x-x_0+\i\eta} = \frac1\pi \lim_{\eta\to0^+} \frac{\eta}{(x-x_0)^2 + \eta^2} \,,
\end{equation}
it can be seen that for $\Gamma \to 0$ (or $\tau \to \infty$), the spectrum becomes a pole.

\subsection{Calculation Methods\label{sec:calcmeth}}

\subsubsection{Green Function Equations of Motion}\label{sec:gfeom}

The Green function equation of motion can be applied as a method to solve for the Green function algebraically.
For the retarded Green function, the equation of motion can be computed as
\begin{equation}
\begin{aligned}
	\frac{\d}{\d t'} \Green{\hat{A}(t)}{\hat{B}(t')}
	&= \frac{\d}{\d t'} \left( - \i \theta(t-t') \left\langle \{ \hat{A}(t) , \hat{B}(t') \}\right\rangle \right)
	\\
	&= \i \delta(t-t') \left\langle \{ \hat{A}(t) , \hat{B}(t') \}\right\rangle - \i \theta(t-t') \left\langle \{ \hat{A}(t) , \tfrac{\d}{\d t'} \hat{B}(t') \}\right\rangle
	\\
	&= \i \delta(t-t') \left\langle \{ \hat{A}(t) , \hat{B}(t') \}\right\rangle - \theta(t-t') \left\langle \{ \hat{A}(t) , [ \hat{H} , \hat{B}(t') ] \}\right\rangle
	\\
	&= \i \delta(t-t') \left\langle \{ \hat{A}(t) , \hat{B}(t') \}\right\rangle + \Green{ \hat{A}(t) }{ [ \hat{H} , \hat{B}(t') ] } \,.
\end{aligned}
\end{equation}
The Green function equation of motion is particularly useful as it takes the same form for the causal, retarded, and advanced forms, as can be easily shown~\cite{zubarev}.\footnote{
To see this it is also helpful to note the identity $\Greenline{[ \hat{A}(t) , \hat{H} ]}{\hat{B}(t')}\, = \Greenline{ \hat{A}(t) }{ [ \hat{H} , \hat{B}(t') ] }\,$.}

The retarded single particle Green function is given by$\Greenline{\op{c}{i}(t)}{\opd{c}{j}(t')}\,$,
which means that the Green function equation of motion takes the form of%\footnote{Note here that the time derivative has the opposite sign than the Schr\"odinger equation as it is conventionally written. This is due to the time derivative acting on $t'$ rather than $t$.}
\begin{equation}
	\left( \frac{\hslash}{\i} \frac{\d}{\d t'} - \hat{H} \right) \Green{\op{c}{i}(t)}{\opd{c}{j}(t')} = \delta_{i,j} \delta(t-t') \,.
\end{equation}
From this expression it is seen that the single particle quantum Green function is functionally analogous to the classical Green function for the Schr\"odinger operator which supports the nomenclature for this quantity.

It is also illuminating to consider an interacting Hamiltonian of the form $\hat{H} = \hat{H}_0 + \hat{H}_I$ where $\hat{H}_0$ is the free kinetic part and $\hat{H}_I$ contains the interactions. Then the Green function equation of motion can be written in the form
\begin{equation}
	\left( \frac{\hslash}{\i} \frac{\d}{\d t'} - \hat{H}_0 \right) \Green{\op{c}{i}(t)}{\opd{c}{j}(t')}
	=
	\delta_{i,j} \delta(t-t') + \Green{\op{c}{i}(t)}{[\hat{H}_I , \opd{c}{j}(t')]} \,.
\end{equation}
In this form it is clear that unlike the classical Green function, the quantum Green function is in principle a non-linear entity due to the presence of the interaction term.

Performing a Laplace transform of the equation of motion to complex frequency\index{Green function!equation of motion} yields
%\begin{equation}
\begin{align}
	\i \int_{0}^{\infty} \d t\, \e^{\i z (t-t')} \frac{\d}{\d t'} \Green{\hat{A}(t)}{\hat{B}(t')}
	&=	\i \int_{0}^{\infty} \d t\, \e^{\i z (t-t')}
		\begin{multlined}[t][0.125\linewidth]
		\Big[
		\i \delta(t-t') \left\langle \{ \hat{A}(t) , \hat{B}(t') \}\right\rangle \\- \theta(t-t') \left\langle \{ \hat{A}(t) , [ \hat{H} , \hat{B}(t') ] \}\right\rangle
		\Big]
		\end{multlined}\notag
	\\
	z \Green{\hat{A}}{\hat{B}}_z
	&=	\left\langle \{ \hat{A} , \hat{B} \} \right\rangle + \Green{[\hat{A}}{[\hat{H} , \hat{B}]}_z \label{eq:greenzeom}\,.
\end{align}
%\end{equation}
%The resulting final expression for the equations of motion for the retarded Green's function is
%The equation of motion for the Green function is obtained through the Heisenberg equations of motion and is given by
%\begin{equation}
%	z \langle\langle \hat{A} ; \hat{B} \rangle\rangle_z = \langle \{ \hat{A} , \hat{B} \} \rangle + \langle\langle [\hat{A} , \hat{H}] ; \hat{B} \rangle\rangle_z .
%	z \Green{\hat{A}}{\hat{B}}_z = \left\langle \{ \hat{A} , \hat{B} \} \right\rangle + \Green{[\hat{A} , \hat{H}]}{\hat{B}}_z .
%\label{greeneom}
%\end{equation}
As before, this equation of motion holds for causal, advanced, and retarded varieties. The retarded Green function is obtained from $z = \omega + \i \delta$.
This form is a highly convenient algebraic form from which the Green function can be solved for, either analytically or numerically. The Green function equation of motion Eq.~\eqref{eq:greenzeom} will appear as a central calculational tool throughout this thesis.

These equations of motion generally lead to an infinite hierarchy of Green functions of increasing order in the field operators.
From such an infinite hierarchy tower, a solution to the Green function can be found by resumming the tower or truncating with an appropriate approximation at some level in the hierarchy. An approximate solution for the Green function may also be found from applying perturbation theory to the Green function equations of motion.

With the basic formalism of the Green function equations of motion set up, it is now possible to examine a few examples. Consider the tight-binding model of a $1d$ chain with nearest-neighbor kinetics,
\begin{equation}
	\hat{H} = \sum_{j\in\Gamma} \left[ \tensor*{\varepsilon}{_j} \opd{c}{j} \op{c}{j} + \tensor*{t}{_j} \left( \opd{c}{j+1} \op{c}{j} + \opd{c}{j} \op{c}{j+1} \right) \right] .
\label{eq:basickineticham}
\end{equation}
The Green function equation of motion for the correlation between arbitrary sites $i$ and $j$ reads as
\begin{equation}
	( z - \varepsilon_{j} ) G_{i,j}(z) = \delta_{i,j} + t_{j-1} G_{i,j-1}(z) + t_{j} G_{i,j+1}(z)
	\label{eq:simplegreeneom}
\end{equation}
where $G_{i,j}(z) \equiv \Greenline{\op{c}{i}}{\opd{c}{j}}_z$.
A typical quantity of interest is the density of states on a particular site of the system, such as the boundary site. 
The general method for solving the equation of motion in $1d$ is to put the equation into a continued fraction form. For a local Green function such a continued fraction takes the form
\begin{equation}
\begin{aligned}[b]
	( z - \varepsilon_{j} ) G_{j,j}(z)
	&=	1 + t_{j-1} G_{j,j-1}(z) + t_{j} G_{j,j+1}(z)
	\\
	\left( z - \varepsilon_{j} - t_{j-1} \frac{G_{j,j-1}(z)}{G_{j,j}(z)} - t_{j} \frac{G_{j,j+1}(z)}{G_{j,j}(z)} \right) G_{j,j}(z)
	&=	1
	\\
	G_{j,j}(z)
	&=	\cfrac{1}{z - \varepsilon_{j} - t_{j-1} \frac{G_{j,j-1}(z)}{G_{j,j}(z)} - t_{j} \frac{G_{j,j+1}(z)}{G_{j,j}(z)}}
\label{eq:greencontinuedfraction}
\end{aligned}
\end{equation}
The fractions of Green functions in the denominator can be obtained by computing the equation of motion for $G_{i,j\pm1}(z)$, to yield
\begin{equation}
\begin{aligned}[b]
	z G_{i,j+1}(z)
	&=	\varepsilon_{j+1} G_{i,j+1}(z) + t_{j} G_{i,j} + t_{j+2} G_{i,j+2}(z)
	\\
	t_{j} \frac{G_{i,j+1}(z)}{G_{i,j}(z)}
	&=	\cfrac{t_{j}^2}{z - \varepsilon_{j+1} - t_{j+1} \frac{G_{i,j+2}(z)}{G_{i,j+1}(z)}} .
\label{eq:green12}
\end{aligned}
\end{equation}
Iterating the equations of motion yields the continued fraction
\begin{equation}
	G_{j,j}(z)
	=	\cfrac{1}{z - \varepsilon_{j} - \cfrac{t_{j-1}^2}{z - \varepsilon_{j-1} - \cfrac{t_{j-2}^2}{z - \ddots}} - \cfrac{t_{j}^2}{z - \varepsilon_{j+1} - \cfrac{t_{j+1}^2}{z - \ddots}}} .
\end{equation}
In general this continued fraction can only be treated numerically, but
for a homogeneous system where $\varepsilon_{j} = \varepsilon$ and $t_{j} = t$ $\forall j$, the equation of motion for a particular site $j$ can be solved analytically. 
For a semi-infinite system where the chain has a boundary at $j=1$, the continued fraction \eqref{eq:greencontinuedfraction} for the boundary site $j=1$ is
\begin{equation}
	G_{1,1}(z)
	=	\cfrac{1}{z - \varepsilon - t \frac{G_{1,2}(z)}{G_{1,1}(z)}}
\label{eq:green11fraction}
\end{equation}
which now requires the solution of $G_{1,2}(z)$. Making use of the homogeneity of the chain parameters in Eq.~\eqref{eq:green12},
\begin{equation}
	t \frac{G_{1,2}(z)}{G_{1,1}(z)} = t^2 G_{1,1}(z)
\end{equation}
Inserting this expression back into the equation of motion for $G_{1,1}(z)$ produces
\begin{equation}
	G_{1,1}(z)
	=	\cfrac{1}{z - \varepsilon - t^2 G_{1,1}(z)}
\tag{\ref*{eq:green11fraction}$^\prime$}
\end{equation}
which may be resummed to obtain the solution
\begin{equation}
	G_{1,1}(z) = \frac{z - \varepsilon - \sqrt{(z - \varepsilon)^2 - 4 t^2}}{2 t^2}
\label{eq:green11}
\end{equation}
This demonstrates a procedure for computing the Green functions of $1d$ lattice Hamiltonians from their equation of motion. The situation in $d>1$ is generally much more complicated. The Green function for an analogous system on the $2d$ square lattice for instance does not take the form of a continued fraction.

For the bulk site of an infinite homogeneous $1d$ chain, it can be observed that the Green function equation of motion \eqref{eq:greencontinuedfraction} takes the form of an arbitrary site in the bulk coupled to two semi-infinite chains. The Green function for the bulk site is
\begin{equation}
	G_{\text{bulk}}(z) = \cfrac{1}{z - \varepsilon - 2 t^2 G_{1,1}(z)}
\end{equation}
where $G_{1,1}(z)$ is given by \eqref{eq:green11}. The final expression for the bulk Green function is then
\begin{equation}
	G_{\text{bulk}}(z) = \frac{1}{\sqrt{(z - \varepsilon)^2 - 4 t^2}} \,.
\end{equation}

This procedure may also be applied to Hamiltonians defined on higher dimensional lattices. A notable example which is important in this thesis is the Bethe lattice\cite{bethe}\index{Bethe lattice}, a cluster of which is illustrated in Fig.~\ref{bethelatt}.
\begin{figure}[h]
\centering
\begin{tikzpicture}[scale=0.65, line width=2pt, every node/.style={scale=0.75,inner sep=3pt}, every path/.style={scale=0.75}]
\input{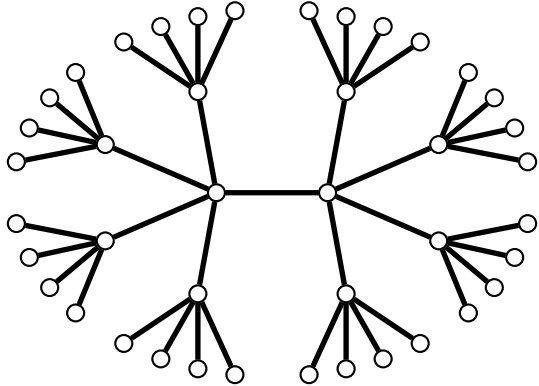}
\end{tikzpicture}
\caption[Schematic of a Bethe lattice]{Schematic of the Bethe lattice with $\kappa = 5$. Shown here is a cluster subset (or Cayley tree). The true Bethe lattice has infinite extent.\label{bethelatt}}
\end{figure}
The Bethe lattice is particularly useful as a basis for finding exact solutions to problems in statistical physics~\cite{baxter}.
The Bethe lattice $\Gamma_{\textsc{bl}}$ is defined as a simple connected undirected regular acyclic graph parameterized only by a coordination number (number of each site's nearest neighbors) $\kappa$. This means that there exists exactly one path between any two lattice sites. The Bethe lattice is defined to have an infinite number of sites. A Bethe lattice with a finite number of sites is known as a Bethe lattice cluster or a Cayley tree.
The infinite $1d$ chain can be interpreted as the minimal example of a Bethe lattice with $\kappa = 2$.
Since the Bethe lattice is infinite in extent with each site having the same coordination number, the Green function on all sites of a homogeneous Hamiltonian  are equivalent, $G_{j,j}(z) \equiv G_{\textsc{bl}}(z)$ $\forall j \in \Gamma_{\textsc{bl}}$. This Green function may be computed from the equation of motion in a similar manner to the previous example to yield
\begin{align}
	G_{\textsc{bl}}(z)
	&= \cfrac{1}{z-\varepsilon - \cfrac{\kappa}{2(\kappa-1)} \left( z-\varepsilon - \sqrt{(z-\varepsilon)^2 - 4(\kappa-1)t^2} \right)} \,.
\end{align}
For the case $\kappa=2$, the Bethe lattice takes the form of the $1d$ chain. Substituting this value into the Bethe lattice Green function readily produces the Green function for the bulk of an infinite $1d$ tight-binding model as to be expected.

A limit of the Bethe lattice which will be important later is the limit of infinite coordination number, $\lim_{\kappa\to\infty}\Gamma_{\textsc{bl}}$. Such a limit often proves useful for producing exact solutions in statistical mechanics~\cite{baxter}.
\label{infinitelimitbethe}
In this limit, the Green function is
\begin{align}
	G_{\textsc{bl}}(z)
	&= \lim_{\kappa\to\infty} \cfrac{1}{z-\varepsilon - \cfrac{\kappa}{2(\kappa-1)} \left( z-\varepsilon - \sqrt{(z-\varepsilon)^2 - 4(\kappa-1)t^2} \right)} \,.
\end{align}
It is seen that the Green function, and therefore the density of states, vanishes in the infinite coordination limit. In order to keep the density of states finite and non-trivial, it is necessary to scale the kinetic term by
\footnote{On the Bethe lattice it is sometimes taken in the literature that
$\displaystyle \tilde{t} = \frac{t}{\sqrt{\kappa-1}}$ to more exactly cancel the $\kappa$-dependent prefactor. The two choices result in the same limit as $\kappa\to\infty$.
}
\begin{equation}
	t \mapsto \frac{\widetilde{t}}{\sqrt{\kappa}}
\end{equation}
With this scaling, the Green function becomes finite
\begin{align}
	G_{\textsc{bl}}(z)
	&= \lim_{\kappa\to\infty} \cfrac{1}{z-\varepsilon - \cfrac{\kappa}{2(\kappa-1)} \left( z-\varepsilon - \sqrt{(z-\varepsilon)^2 - 4(\kappa-1) \left(\frac{\widetilde{t}}{\sqrt{\kappa}}\right)^2} \right)}
	\\
	&= \frac{2}{z-\varepsilon + \sqrt{(z-\varepsilon)^2 - 4 \widetilde{t}^2}}
	\intertext{or more conventionally,}
	&= \frac{z - \varepsilon - \sqrt{(z-\varepsilon)^2 - 4 \widetilde{t}^2}}{2 \widetilde{t}^2} \,,
\end{align}
which is equivalent to the Green function on the boundary of a semi-infinite $1d$ homogeneous chain \eqref{eq:green11} with hopping parameter $\widetilde{t}$.

Another way in which the equation of motion method can be applied to lattices of dimension $d>1$ is via a partial Fourier transformation~\cite{convolutionmethod}. A semi-infinite $3d$ system with a $2d$ boundary is an example of an applicable situation. In this case the degrees of freedom of the infinite (or periodic) $2d$ boundary are Fourier transformed to a diagonal momentum space representation. The system then takes the form of a set of decoupled $1d$ chains, one for each quasimomentum $k_n$, extending into the bulk of the original system. A Green function for each $k_n$ can then be obtained from the equations of motion approach for the $1d$ chains as described above.

\subsubsection{Numerical Fast Recursion Method}

As shown above, for simple non-interacting systems it is possible to obtain an analytic solution for the Green function. However, often the desired end result is not the Green function itself, but rather objects which are obtained from the Green function, such as the spectral function. This involves taking the imaginary component of the analytically continued Green function which may be non-trivial, for example by needing to take the appropriate branch cut in the complex plane. Furthermore, in the case of more involved systems which are periodic but with a large unit cell, the solution for the Green function from the equations of motion would be a high order polynomial which would take a very complicated form.

These issues can be avoided by instead calculating the Green functions by numerically iterating the equations of motion.
Solving the Green function equation of motion for a non-interacting one dimensional system results in a solution in the form of a continued fraction with depth equal to the length of the system~\cite{cfrac}. For high resolution, it is desirable to use a very large system. Directly computing the Green function from such a large continued fraction is, however, computationally inefficient. This necessitates the implementation of a more efficient computational strategy. The strategy adopted here is that of a recursion algorithm where each iteration increases the effective system size exponentially by exploiting self-similarity of the system down the $1d$ chain~\cite{thing,otherthing}.
The typical form of a $1d$ non-interacting tight-binding Hamiltonian in condensed matter physics is
\begin{equation}
	\hat{H} = \sum_{j} \left( | \psi_j \rangle \boldsymbol{h}_0 \langle \psi_j | + | \psi_{j+1} \rangle \boldsymbol{h}_1 \langle \psi_{j} | + | \psi_{j} \rangle \boldsymbol{h}^\dagger_1 \langle \psi_{j+1} | \right)
\end{equation}
%which for finite or semi-infinite systems can be written in a tridiagonal matrix form
%\begin{equation}
%	H	=
%	\begin{pmatrix}
%		h_0 & h_1 & 0 & \cdots \\
%		h_1^\dagger & h_0 & h_1 & \ddots \\
%		0 & h_1^\dagger & h_0 & \ddots \\
%		\vdots & \ddots & \ddots & \ddots
%	\end{pmatrix}
%\end{equation}
where $\boldsymbol{h}_0$ represents dynamics within a unit cell and $\boldsymbol{h}_1$ represents dynamics between unit cells with $| \psi_j \rangle$ an $L$-dimensional vector representing the $j$-th unit cell consisting of $L$ degrees of freedom. The submatrices $\boldsymbol{h}_0$ and $\boldsymbol{h}_1$ are of dimension $L \times L$.
The Green function equation of motion for this Hamiltonian can be adapted as a matrix variant of Eq.~\eqref{eq:simplegreeneom},
%\begin{subequations}
\begin{align}
	\left( z\mathbbm{1} - \tensor*{\boldsymbol{\boldsymbol{h}}}{_0} \right) \boldsymbol{G}_{j,0}(z) &= \tensor*{\boldsymbol{h}}{^\dagger_1} \boldsymbol{G}_{j-1,0}(z) + \tensor*{\boldsymbol{h}}{_1} \boldsymbol{G}_{j+1,0}(z) . \label{eq:greenitereom}
\\
\intertext{This equation of motion can be re-expressed as}
	\boldsymbol{G}_{j,0}(z) &= \boldsymbol{\tau}_0 \boldsymbol{G}_{j-1,0}(z) + \widetilde{\boldsymbol{\tau}}_0 \boldsymbol{G}_{j+1,0}(z) \tag{\ref*{eq:greenitereom}$^\prime$} \label{eq:greeniter}
\end{align}
%\end{subequations}
with the introduction of auxiliary transfer matrices
\begin{subequations}
\begin{align}
	\boldsymbol{\tau}_0 &= \left(z\mathbbm{1} - \boldsymbol{h}_0\right)^{-1} \boldsymbol{h}_1^\dagger
	\\
	\widetilde{\boldsymbol{\tau}}_0 &= \left(z\mathbbm{1} - \boldsymbol{h}_0\right)^{-1} \tensor{\boldsymbol{h}}{_1}
\end{align}
\end{subequations}
The equation of motion Eq.~\eqref{eq:greeniter} can be iterated to produce
\begin{align*}
	\boldsymbol{G}_{j,0}(z)
	&= \boldsymbol{\tau}_0 \left( \boldsymbol{\tau}_0 \boldsymbol{G}_{j-2,0}(z) + \widetilde{\boldsymbol{\tau}}_0 \boldsymbol{G}_{j,0}(z) \right) + \widetilde{\boldsymbol{\tau}}_0 \left( \boldsymbol{\tau}_0 \boldsymbol{G}_{j,0}(z) + \widetilde{\boldsymbol{\tau}}_0 \boldsymbol{G}_{j+2,0}(z) \right)
	\\
	&= (\mathbbm{1} - \boldsymbol{\tau}_0 \widetilde{\boldsymbol{\tau}}_0 - \widetilde{\boldsymbol{\tau}}_0 \boldsymbol{\tau}_0)^{-1} \boldsymbol{\tau}_0^2 \boldsymbol{G}_{j-2,0}(z) + (\mathbbm{1} - \boldsymbol{\tau}_0 \widetilde{\boldsymbol{\tau}}_0 - \widetilde{\boldsymbol{\tau}}_0 \boldsymbol{\tau}_0)^{-1} \widetilde{\boldsymbol{\tau}}_0^2 \boldsymbol{G}_{j+2,0}(z)
	\\
	&\equiv \boldsymbol{\tau}_1 \boldsymbol{G}_{j-2,0}(z) + \widetilde{\boldsymbol{\tau}}_1 \boldsymbol{G}_{j+2,0}(z) \tag{\ref*{eq:greeniter}$^\prime$}\label{eq:greeniterprime}
\end{align*}
with an iterated pair of auxiliary transfer matrices $\boldsymbol{\tau}_1$ and $\widetilde{\boldsymbol{\tau}}_1$. Each additional iteration of Eq.~\eqref{eq:greeniterprime} results in a Green function incorporating a factor of 2 sites more than the previous iteration. The $n$-th iteration is
\begin{align}
	\boldsymbol{G}_{j,0}(z) &= \boldsymbol{\tau}_n \boldsymbol{G}_{j-2^n,0}(z) + \widetilde{\boldsymbol{\tau}}_n \boldsymbol{G}_{j+2^n,0}(z)
\end{align}
where the auxiliary transfer matrices are recursively obtained following Eq.~\eqref{eq:greeniterprime}
\begin{subequations}
\begin{align}
	{\boldsymbol{\tau}}_{n+1} &= \left[ \mathbbm{1} - {\boldsymbol{\tau}}_n \widetilde{\boldsymbol{\tau}}_n - \widetilde{\boldsymbol{\tau}}_n {\boldsymbol{\tau}}_n \right]^{-1} {\boldsymbol{\tau}}_n^2
	\\
	\widetilde{\boldsymbol{\tau}}_{n+1} &= \left[ \mathbbm{1} - {\boldsymbol{\tau}}_n \widetilde{\boldsymbol{\tau}}_n - \widetilde{\boldsymbol{\tau}}_n {\boldsymbol{\tau}}_n \right]^{-1} \widetilde{\boldsymbol{\tau}}_n^2
\end{align}
\end{subequations}
The $\boldsymbol{\tau}$ and $\widetilde{\boldsymbol{\tau}}$ matrices have the interpretation of being a generalization of the interstitial hopping amplitude for cells $2^n$ apart.
In order to relate the non-local Green functions to a local one, the $n$-th order iteration may be taken with $j = 2^n$
\begin{align}
	\boldsymbol{G}_{2^n,0}(z) &= \boldsymbol{\tau}_n \boldsymbol{G}_{0,0}(z) + \widetilde{\boldsymbol{\tau}}_n \boldsymbol{G}_{2^{n+1},0}(z)
	\label{eq:poweriter}
\end{align}
and building a new iteration series now based on the $n$ index and iterating with $n=0,1,2,\ldots,N$ as
\begin{equation}
\begin{aligned}[b]
	\boldsymbol{G}_{1,0}(z)
	&= \boldsymbol{\tau}_0 \boldsymbol{G}_{0,0}(z) + \widetilde{\boldsymbol{\tau}}_0 \boldsymbol{G}_{2,0}(z)
	\\
	&= \left( \boldsymbol{\tau}_0 + \widetilde{\boldsymbol{\tau}}_0 \boldsymbol{\tau}_1 \right) \boldsymbol{G}_{0,0}(z) + \widetilde{\boldsymbol{\tau}}_1 \boldsymbol{G}_{4,0}(z)
	\\
	&= \left( \boldsymbol{\tau}_0 + \widetilde{\boldsymbol{\tau}}_0 \boldsymbol{\tau}_1 + \widetilde{\boldsymbol{\tau}}_0 \widetilde{\boldsymbol{\tau}}_1 \boldsymbol{\tau}_2 \right) \boldsymbol{G}_{0,0}(z) + \widetilde{\boldsymbol{\tau}}_2 \boldsymbol{G}_{8,0}(z)
	\\
	&\vdotswithin{=}
	\\
	&\equiv \boldsymbol{T} \boldsymbol{G}_{0,0}(z) + \widetilde{\boldsymbol{\tau}}_N \boldsymbol{G}_{2^{N+1},0}(z)
\end{aligned}
\end{equation}
where each step is iterated from Eq.~\eqref{eq:poweriter} with each $\boldsymbol{G}_{2^n,0}$ term downfolded to $\boldsymbol{G}_{0,0}$ and $\boldsymbol{G}_{1,0}$ terms from the preceding iterations, which produces each term of the $\boldsymbol{\tau}$, $\widetilde{\boldsymbol{\tau}}$-polynomial $\boldsymbol{T}$. This matrix $\boldsymbol{T}$ takes the form for general $N$
\begin{equation}
	\boldsymbol{T} = \boldsymbol{\tau}_0 + \sum_{n=1}^{N} \left[ \left(\prod_{m=0}^{n-1} \widetilde{\boldsymbol{\tau}}_{m}\right) \boldsymbol{\tau}_{n} \right] \,.
\end{equation}
The recursion is truncated with the approximation $\boldsymbol{G}_{2^{N+1},0}(z) \simeq 0$, leaving $\boldsymbol{G}_{1,0}(z) \simeq \boldsymbol{T} \boldsymbol{G}_{0,0}(z)$.
Returning to the original equation of motion for the boundary unit cell Green function,
\begin{equation}
\begin{aligned}[b]
	\left( z\mathbbm{1} - \tensor*{\boldsymbol{h}}{_0} \right) \boldsymbol{G}_{0,0}(z) &= \mathbbm{1} + \tensor*{\boldsymbol{h}}{_1} \boldsymbol{G}_{1,0}(z)
	\\
	&\simeq \mathbbm{1} + \tensor*{\boldsymbol{h}}{_1} \boldsymbol{T} \boldsymbol{G}_{0,0}(z)
	\\
	\boldsymbol{G}_{0,0}(z) &= \left[ z \mathbbm{1} - \boldsymbol{h}_0 - \boldsymbol{h}_1 \boldsymbol{T} \right]^{-1} \,.\label{eq:recursiongreen}
\end{aligned}
\end{equation}
This recursion scheme resulting in the final expression Eq.~\eqref{eq:recursiongreen} therefore allows for the computation of boundary-cell Green functions for very long systems of length $2^N$ directly from the elementary subblocks of the Hamiltonian which can accurately capture the properties of semi-infinite systems at low computational cost. The algorithm may also be adapted to produce the Green function for the second unit cell from the boundary~\cite{thing,otherthing}.

\subsubsection{Interacting Green functions}

%The Green functions for interacting systems are generally not solvable exactly, except for certain special cases. One such case is the single site Hubbard atom.

%\subsubsection{Example: The Hubbard Atom}
\label{sec:hubbardatomgf}

The Green function for an interacting model generally cannot be computed exactly, but it is possible to obtain exact solutions for models which are sufficiently simple. Such a model is the Hubbard atom, which consists of a single interacting site. This model may be interpreted as the Hubbard model in the limit $U/t \to \infty$. In this limit the dynamics are dominated by the on-site interaction and hybridization between lattice sites becomes negligible. Each site may therefore be analyzed independently from each other.

This system is described by the Hamiltonian
\begin{equation}
	\op{H}{\textsc{ha}} = \varepsilon \left( \opd{c}{\uparrow} \op{c}{\uparrow} + \opd{c}{\downarrow} \op{c}{\downarrow} \right) + U \opd{c}{\uparrow} \op{c}{\uparrow} \opd{c}{\downarrow} \op{c}{\downarrow} \,.
\end{equation}
The Green function is calculated from the equation of motion
\begin{align}
	z \Green{\op{c}{\sigma}}{\opd{c}{\sigma}}_z
	&=	\langle \{ \op{c}{\sigma} , \opd{c}{\sigma} \} \rangle + \Green{\op{c}{\sigma}}{[\op{H}{\textsc{ha}},\opd{c}{\sigma}]}_z
	\intertext{to yield}
	&=	1 + \varepsilon \Green{\op{c}{\sigma}}{\opd{c}{\sigma}}_z + U \Green{\op{c}{\sigma}}{\tensor*{\hat{n}}{_{-\sigma}}\opd{c}{\sigma}}_z
\end{align}
which makes use of the commutator
$
	[ \op{H}{\textsc{ha}} , \opd{c}{\sigma} ] = \varepsilon \opd{c}{\sigma} + U \tensor*{\hat{n}}{_{-\sigma}} \opd{c}{\sigma} \,.
$
It is now necessary to compute the Green function $\Greenline{\op{c}{\sigma}}{\tensor*{\hat{n}}{_{-\sigma}}\opd{c}{\sigma}}_z$. This involves the commutator
\begin{equation}
\begin{aligned}
	[ \op{H}{\textsc{ha}} , \tensor*{\hat{n}}{_{-\sigma}} \opd{c}{\sigma} ] &= \varepsilon \tensor*{\hat{n}}{_{-\sigma}^2} \opd{c}{\sigma} + U \tensor*{\hat{n}}{_{-\sigma}^2} \opd{c}{\sigma} \\&= \varepsilon \tensor*{\hat{n}}{_{-\sigma}} \opd{c}{\sigma} + U \tensor*{\hat{n}}{_{-\sigma}} \opd{c}{\sigma} \,,
\end{aligned}
\end{equation}
where the second equality follows from the fact that the fermion number operator is indempotent, $\tensor*{\hat{n}}{_{-\sigma}^2} = \tensor*{\hat{n}}{_{-\sigma}}$.
This Green function is now calculated to be
\begin{equation}
\begin{aligned}[b]
	z \Green{\op{c}{\sigma}}{\tensor*{\hat{n}}{_{-\sigma}}\opd{c}{\sigma}}_z
	&=	\langle \{ \op{c}{\sigma} , \tensor*{\hat{n}}{_{-\sigma}} \opd{c}{\sigma} \} \rangle + \Green{\op{c}{\sigma}}{[ \op{H}{\textsc{ha}} , \tensor*{\hat{n}}{_{-\sigma}}\opd{c}{\sigma} ]}_z
	\\
	&=	\langle \tensor*{\hat{n}}{_{-\sigma}} \rangle + \varepsilon \Green{\op{c}{\sigma}}{\tensor*{\hat{n}}{_{-\sigma}}\opd{c}{\sigma}}_z + U \Green{\op{c}{\sigma}}{\tensor*{\hat{n}}{_{-\sigma}}\opd{c}{\sigma}}_z
	\\
	\Green{\op{c}{\sigma}}{\tensor*{\hat{n}}{_{-\sigma}}\opd{c}{\sigma}}_z
	&=	\frac{\langle \tensor*{\hat{n}}{_{-\sigma}} \rangle}{z - \varepsilon - U} \,.
\end{aligned}
\end{equation}
%
%\begin{align}
%	\{ \op{c}{\sigma} , \tensor*{\hat{n}}{_{-\sigma}} \opd{c}{\sigma} \} = \tensor*{\hat{n}}{_{-\sigma}}
%\end{align}
%
In this expression there are no more Green functions left to calculate.
It is now possible to write the Green function in closed form as
\begin{equation}
\begin{aligned}[b]
	(z-\varepsilon) \Green{\op{c}{\sigma}}{\opd{c}{\sigma}}_z
	&=	1 + U \Green{\op{c}{\sigma}}{\tensor*{\hat{n}}{_{-\sigma}}\opd{c}{\sigma}}_z
	\\
	\Green{\op{c}{\sigma}}{\opd{c}{\sigma}}_z
	&=	\frac{1}{z - \varepsilon} + \frac{U}{z-\varepsilon} \frac{\langle \tensor*{\hat{n}}{_{-\sigma}} \rangle}{z - \varepsilon - U}
	\\
	&=
	\frac{1 - \langle \tensor*{\hat{n}}{_{-\sigma}} \rangle}{z - \varepsilon} + \frac{\langle \tensor*{\hat{n}}{_{-\sigma}} \rangle}{z - \varepsilon - U} \,.
\end{aligned}
\end{equation}
This leads to a spectral function which consists of two poles at $\omega_{p_1} = \varepsilon$ and $\omega_{p_2} = \varepsilon + U$, with weights $1 - \langle \op{n}{-\sigma} \rangle$ and $\langle \op{n}{-\sigma} \rangle$ respectively. As the Hamiltonian is symmetric under $\sigma \leftrightarrow -\sigma$, the Green functions are similarly symmetric under such an exchange.

Following from the equilibrium fluctuation-dissipation theorem, the general temperature dependent filling of the atom is described by
\begin{equation}
	\langle \op{n}{\sigma} \rangle = \int \d\omega f(\omega) \mathcal{A}_{\sigma}(\omega)
\end{equation}
where $f(\omega) = 1/(1 + \e^{\beta \omega})$ is the Fermi function. Given the form of the spectral function, the filling of the atom can be computed as
\begin{equation}
	\langle \op{n}{\pm\sigma} \rangle
	=
	\frac{f(\varepsilon)}{1 - f(\varepsilon+U) + f(\varepsilon)} \,.
\label{eq:hafilling}
\end{equation}
Provided that at sufficiently low temperature and if $-U < \varepsilon < 0$, it follows that $\langle \op{n}{-\sigma} \rangle \approx \frac12$.
Imposing particle-hole symmetry, where $\op{H}{\textsc{ha}}$ is symmetric under $\opd{c}{\sigma} \leftrightarrow \op{c}{\sigma}$, leads to the condition that $\varepsilon = -U/2$. In this case, $\langle \op{n}{\sigma} \rangle = \frac12$ exactly, by the particle-hole
%\index{$0$@\textbf{List of Edits}!206@discussion of filling}
symmetry, independent of temperature.
The Green function at particle-hole symmetry is then
\begin{equation}
	\Green{\op{c}{\sigma}}{\opd{c}{\sigma}}_z
	=	\frac{\frac12}{z + \frac{U}{2}} + \frac{\frac12}{z - \frac{U}{2}}
\end{equation}
with spectral function
\begin{equation}
	\mathcal{A}(\omega) = \frac12 \delta\left( \omega + \tfrac{U}{2} \right) + \frac12 \delta\left( \omega - \tfrac{U}{2} \right) \,,
\end{equation}
which consists of two poles situated at $\omega = \pm \frac{U}{2}$ each with weight $\frac12$. This is plotted in Fig.~\ref{fig:haspec}.
\begin{figure}[h]
\centering
\includegraphics{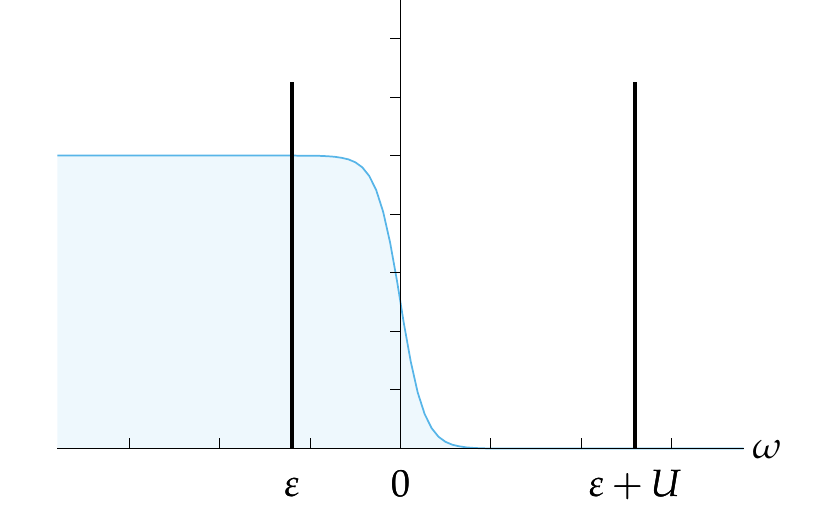}
\caption{Spectrum of the Hubbard atom consisting of two poles at $\omega=-\frac{U}{2}$ and $\omega=+\frac{U}{2}$ The shaded curve is the Fermi function $f(\omega)$.\label{fig:haspec}}
\end{figure}
It is observed that that $\mathcal{A}_{\sigma}(\omega) = \mathcal{A}_{\sigma}(-\omega)$, which corresponds to the system's particle-hole symmetry. For $\varepsilon \neq -U/2$, it still holds that $\langle \op{n}{\sigma} \rangle = \frac12$, meaning the system is still at half-filling, however the system no longer possesses particle-hole symmetry. Parameterizing the particle-hole asymmetry as $\eta \vcentcolon= 1 + 2\varepsilon/U$, the spectrum becomes $\mathcal{A}_{\sigma}(\omega) = \frac12 \delta\left( \omega + \tfrac{U}{2}(1-\eta) \right) + \frac12 \delta\left( \omega - \tfrac{U}{2}(1+\eta) \right)$. It follows that the system is only particle-hole symmetric for $\eta = 0$, or $\varepsilon = -U/2$, as $\mathcal{A}_{\sigma}(\omega) \neq \mathcal{A}_{\sigma}(-\omega)$ for $\eta \neq 0$.
%\index{$0$@\textbf{List of Edits}!206@discussion of filling}

While the single site Hubbard atom is a relatively trivial example of determining an interacting system's Green functions from the equations of motion method, this method can scale exponentially with increasing system size. A Hubbard model consisting of only two sites, a Hubbard dimer, can also be solved from the Green function equations of motion, however this involves the calculation of 105 Green functions in order to close the hierarchy~\cite{hubbarddimereom}. The Hubbard dimer is however more easily solved by exact diagonalization and the spectrum can then be recovered by the Lehmann representation.

For the full Hubbard model on a fully interacting lattice, the local Green Function equation of motion is
\begin{equation}
\begin{aligned}
	z \Green{\op{c}{j,\sigma}}{\opd{c}{j,\sigma}}_z
		&=	\langle \{ \op{c}{j,\sigma} , \opd{c}{j,\sigma} \} \rangle + \Green{\op{c}{j,\sigma}}{[\op{H}{\textsc{h}} , \opd{c}{j,\sigma}]}_z
	\\	&=	\delta_{j,j} + \Green{\op{c}{j,\sigma}}{[\op{H}{0} , \opd{c}{j,\sigma}]}_z + \Green{\op{c}{j,\sigma}}{[\op{H}{I} , \opd{c}{j,\sigma}]}_z
\end{aligned}
\end{equation}
where
\begin{subequations}
\begin{equation}
\begin{aligned}
    \Green{\op{c}{j,\sigma}}{[\op{H}{0} , \opd{c}{j,\sigma}]}_z
        &=  \epsilon \Green{\op{c}{j,\sigma}}{\opd{c}{j,\sigma}}_z + t \Green{\op{c}{j,\sigma}}{\opd{c}{j+1,\sigma}}_z + t \Green{\op{c}{j,\sigma}}{\opd{c}{j-1,\sigma}}_z
\end{aligned}
\end{equation}
and
\begin{equation}
\begin{aligned}[b]
%    \left[ \hat{H}_I , \opd{c}{j,\sigma} \right]
%        &=  U \sum_{i} \left( \opd{c}{i,\uparrow} \op{c}{i,\uparrow} \opd{c}{i,\downarrow} \op{c}{i,\downarrow} \opd{c}{j,\sigma} - \opd{c}{j,\sigma} \opd{c}{i,\uparrow} \op{c}{i,\uparrow} \opd{c}{i,\downarrow} \op{c}{i,\downarrow} \right)
%    \\  &=  U \sum_{i} \left( \opd{c}{i,\uparrow} \op{c}{i,\uparrow} \opd{c}{i,\downarrow} ( \delta_{i,j} \delta_{\downarrow,\sigma} - \opd{c}{j,\sigma} \op{c}{i,\downarrow} ) - \opd{c}{j,\sigma} \opd{c}{i,\uparrow} \op{c}{i,\uparrow} \opd{c}{i,\downarrow} \op{c}{i,\downarrow} \right)
%    \\  &=  U \sum_{i} \left( \opd{c}{i,\uparrow} \op{c}{i,\uparrow} \opd{c}{i,\downarrow} \delta_{i,j} \delta_{\downarrow,\sigma} + \opd{c}{i,\uparrow} \op{c}{i,\uparrow} \opd{c}{j,\sigma} \opd{c}{i,\downarrow} \op{c}{i,\downarrow} - \opd{c}{j,\sigma} \opd{c}{i,\uparrow} \op{c}{i,\uparrow} \opd{c}{i,\downarrow} \op{c}{i,\downarrow} \right)
%    \\  &=  U \sum_{i} \left( \opd{c}{i,\uparrow} \op{c}{i,\uparrow} \opd{c}{i,\downarrow} \delta_{i,j} \delta_{\downarrow,\sigma} + \opd{c}{i,\uparrow} ( \delta_{i,j} \delta_{\uparrow,\sigma} - \opd{c}{j,\sigma} \op{c}{i,\uparrow} ) \opd{c}{i,\downarrow} \op{c}{i,\downarrow} - \opd{c}{j,\sigma} \opd{c}{i,\uparrow} \op{c}{i,\uparrow} \opd{c}{i,\downarrow} \op{c}{i,\downarrow} \right)
%    \\  &=  U \left( \opd{c}{j,\uparrow} \op{c}{j,\uparrow} \opd{c}{j,\downarrow} \delta_{\downarrow,\sigma} + \opd{c}{j,\uparrow} \opd{c}{j,\downarrow} \op{c}{j,\downarrow} \delta_{\uparrow,\sigma} \right)
%    \\
    \Green{\op{c}{j,\sigma}}{[\op{H}{I} , \opd{c}{j,\sigma'}]}_z
        &=  U \Green{\op{c}{j,\sigma}}{\opd{c}{j,\sigma'} \opd{c}{j,-\sigma'} \op{c}{j,-\sigma'}}_z \,.
\end{aligned}
\end{equation}
\end{subequations}
This Green function can then be written in the compact form of
\begin{equation}
\begin{aligned}[b]
	z \Green{\op{c}{j,\sigma}}{\opd{c}{j,\sigma}}_z
%	&=
%	1	+ \epsilon \Green{\op{c}{j,\sigma}}{\opd{c}{j,\sigma}}_z + t \Green{\op{c}{j,\sigma}}{\opd{c}{j+1,\sigma}}_z + t \Green{\op{c}{j,\sigma}}{\opd{c}{j-1,\sigma}}_z
%		+ U \Green{\op{c}{j,\sigma}}{\opd{c}{j,\sigma} \opd{c}{j,-\sigma} \op{c}{j,-\sigma}}_z
%	\\
%	\Green{\op{c}{j,\sigma}}{\opd{c}{j,\sigma}}_z
	&=
	\cfrac{1}{z - \epsilon - K_{\sigma}(z) - \Sigma_{\sigma}(z)}
\end{aligned}
\end{equation}
with
\begin{subequations}
\begin{align}
	&	K_{\sigma}(z) = t \sum_{r} \cfrac{\Green{\op{c}{j,\sigma}}{\opd{c}{j+r,\sigma}}_z}{\Green{\op{c}{j,\sigma}}{\opd{c}{j,\sigma}}_z}
	\\
	&	\Sigma_{\sigma}(z) = U \cfrac{\Green{\op{c}{j,\sigma}}{\opd{c}{j,\sigma} \opd{c}{j,-\sigma} \op{c}{j,-\sigma}}_z}{\Green{\op{c}{j,\sigma}}{\opd{c}{j,\sigma}}_z} \label{eq:foverggreen}
\end{align}
\end{subequations}
where $\Sigma_\sigma(z)$ is the local self-energy. Explicit forms for the higher order Green functions obtained from the equations of motion of the elements in $K_\sigma(z)$ and $\Sigma_\sigma(z)$ are found in the appendix \S\ref{appendixeom}.

%In the limit of infinite dimensions, the Green function $\Green{\op{c}{j,\sigma}}{\opd{c}{j+r,\sigma}}_z$ factors exactly as $\Green{\op{c}{j,\sigma}}{\opd{c}{j+r,\sigma}}_z = \tilde{t} \Green{\op{c}{j,\sigma}}{\opd{c}{j,\sigma}}_z \Green{\op{c}{j,\sigma}}{\opd{c}{j,\sigma}}_z$ \cite{eomhubbardinfinite}

%\begin{align*}
%	z \Green{\op{c}{j,\sigma}}{\opd{c}{j,\sigma} \opd{c}{j,-\sigma} \op{c}{j,-\sigma}}_z
%	&=	\left\langle \{ \op{c}{j,\sigma} , \opd{c}{j,\sigma} \opd{c}{j,-\sigma} \op{c}{j,-\sigma} \} \right\rangle + \Green{\op{c}{j,\sigma}}{ [ \hat{H} , \opd{c}{j,\sigma} \opd{c}{j,-\sigma} \op{c}{j,-\sigma} ] }_z
%	\\
%	&=	\left\langle \op{c}{j,\sigma} \opd{c}{j,\sigma} \opd{c}{j,-\sigma} \op{c}{j,-\sigma} + \opd{c}{j,\sigma} \opd{c}{j,-\sigma} \op{c}{j,-\sigma} \op{c}{j,\sigma} \right\rangle + \Green{\op{c}{j,\sigma}}{ [ \hat{H} , \opd{c}{j,\sigma} \opd{c}{j,-\sigma} \op{c}{j,-\sigma} ] }_z
%	\\
%	&=	\left\langle ( 1 - \opd{c}{j,\sigma} \op{c}{j,\sigma} ) \opd{c}{j,-\sigma} \op{c}{j,-\sigma} + \opd{c}{j,\sigma} \op{c}{j,\sigma} \opd{c}{j,-\sigma} \op{c}{j,-\sigma} \right\rangle + \Green{\op{c}{j,\sigma}}{ [ \hat{H} , \opd{c}{j,\sigma} \opd{c}{j,-\sigma} \op{c}{j,-\sigma} ] }_z
%	\\
%	&=	\left\langle \opd{c}{j,-\sigma} \op{c}{j,-\sigma} \right\rangle + \Green{\op{c}{j,\sigma}}{ [ \hat{H} , \opd{c}{j,\sigma} \opd{c}{j,-\sigma} \op{c}{j,-\sigma} ] }_z
%\end{align*}

It is apparent from the Green functions and their equations of motion that the two particle expectation value does not factor into single particle expectation values
\begin{equation}
	\langle \tensor*{\hat{n}}{_\uparrow} \tensor*{\hat{n}}{_\downarrow} \rangle \neq \langle \tensor*{\hat{n}}{_\uparrow} \rangle \langle \tensor*{\hat{n}}{_\downarrow} \rangle
\end{equation}
which demonstrates the inherent strongly correlated nature of the Hubbard model.

The Green functions for strongly correlated systems is not easily obtained. Unlike for non-interacting models, the hierarchy of the Green functions generally does not close, even in $1d$. For weak coupling, the Green functions can be computed in a perturbative expansion, but for strong coupling, non-perturbative methods or well defined approximations are needed.

\begin{comment}
\cite{dmft}
\cite{infinitehubbard}
\cite{jarrellhubbard}
\cite{bullahubbard}

Kramers-Kronig relations~\cite{mathewswalker}

\begin{align}
	\Re\Sigma(\omega) &= \frac1\pi\fint_{-\infty}^{\infty} \frac{\Im\Sigma(\omega')}{\omega'-\omega} \d\omega'
	&
	\Im\Sigma(\omega) &= -\frac1\pi\fint_{-\infty}^{\infty} \frac{\Im\Sigma(\omega')}{\omega'-\omega} \d\omega'
\end{align}

where $\fint$ notates the Cauchy principal value
\end{comment}

\section{Numerical Renormalization Group}

The numerical renormalization group (NRG) is a fully non-perturbative renormalization group\footnote{The renormalization group is not technically a group in the formal mathematical sense as there is no notion of an inverse operation contained in renormalization group actions. The set of renormalization group actions is rather a monoid. Alternatively it may be thought of as a semigroup as the presence of an identity operation is superfulous.} transformation which can solve quantum impurity models numerically exactly~\cite{wilson,kww1,kww2,nrg}. NRG was originally developed to treat the Kondo model, which similar to the aforementioned Anderson impurity model describes dilute magnetic alloys, an example being single Fe atom magnetic impurities embedded in a non-magnetic material such as Au. NRG has since been generalized for a broader class of physical systems which may have multiple, but still few, interacting degrees of freedom such as quantum dots coupled to one or more non-interacting fermionic baths.

Quantum impurity models are essentially of the form of the Anderson impurity model, with an interacting impurity hybridized to a non-interacting bath. The NRG calculation is first initialized by forming the integral representation of the impurity model as
\begin{subequations}
\begin{align}
	\hat{H}^{\int}_{\text{bath}} &= \sum_{\sigma} \int_{-D}^{D} \d \epsilon \, g(\epsilon) \opd{a}{\epsilon \sigma} \op{a}{\epsilon \sigma}
	\\
	\hat{H}^{\int}_{\text{hyb}} &= \sum_{\sigma} \int_{-D}^{D} \d \epsilon \, h(\epsilon) \left( \opd{d}{\sigma} \op{a}{\epsilon \sigma} + \opd{a}{\epsilon \sigma} \op{d}{\sigma} \right)
\end{align}
\label{eq:nrgintegralrep}
\end{subequations}
where $D$ is the bandwidth of the hybridization and the bath operators $\tensor*{\hat{a}}{^{(\dagger)}_{\epsilon \sigma}}$ satisfy the usual fermionic anticommutation relations. The function $g(\epsilon)$ is the dispersion of the band and $h(\epsilon)$ is the hybridization between the impurity and the band states. These functions are related to the impurity hybridization function $\Delta(z)$, \textit{e.g.} \eqref{eq:siamDelta}, as
\begin{equation}
	\Im \Delta(\omega) = \int_{-D}^{D} \d\epsilon\, h(\epsilon)^2 \delta(\omega - g(\epsilon)) = h[g^{-1}(\omega)]^2 \frac{\d}{\d\omega} g^{-1}(\omega) \,.
\end{equation}
%\index{$0$@\textbf{List of Edits}!207@missing definitions}
 
The key to NRG is the logarithmic separation of energy scales. Quantum impurity models are typically characterized by features at energy scales much lower than that of the bare Hamiltonian due to the Kondo effect~\cite{hewson,kondo}. The use of logarithmic scaling enables the use of exact diagonalization to resolve the low energy features, which would be impossible if the scaling was linear.

The domain of energy support of the bath spectral function is divided into a set of logarithmic intervals $\left[ \Lambda^{-n-1} , \Lambda^{-n} \right]$ with $\Lambda > 1$. The spectral function is then discretized by replacing the continuous spectrum with a discrete pole of the same total weight, as illustrated in Fig.~\ref{fig:nrgdisc}.
\begin{figure}[h!]
\centering
\begin{tikzpicture}[every path/.style = {xscale=6,yscale=4}, every node/.style={scale=1,font=\scriptsize}, lambda/.style={line width = 0.5pt,densely dashed}, pole/.style={line width = 1.5pt,orange}]
\filldraw[fill=blue!10!white,draw=black,line width = 1pt] (-1,0) rectangle (1,0.37);
\node[below] at (-1,0) {$\tensor[^{\phantom{-}}]{-D\phantom{-}}{^{\phantom{-}}}$};
\node[below] at (1,0) {$\tensor[^{\phantom{-}}]{D}{^{\phantom{-}}}$};
\draw[lambda] (-0.5,0.37)--(-0.5,0) node[below] {$-\Lambda^{-1}$};
\draw[lambda] (-0.25,0.37)--(-0.25,0) node[below] {$-\Lambda^{-2}$};
\draw[lambda] (-0.125,0.37)--(-0.125,0);
\draw[lambda] (-0.0625,0.37)--(-0.0625,0) node[below] {$-\Lambda^{-n}$};
\draw[lambda] (-0.03125,0.37)--(-0.03125,0);
\draw[lambda] (0.5,0.37)--(0.5,0) node[below] {$\phantom{-}\Lambda^{-1}$};
\draw[lambda] (0.25,0.37)--(0.25,0) node[below] {$\phantom{-}\Lambda^{-2}$};
\draw[lambda] (0.125,0.37)--(0.125,0);
\draw[lambda] (0.0625,0.37)--+(0,-0.37) node[below] {$\phantom{-}\Lambda^{-n}$};
\draw[lambda] (0.03125,0.37)--+(0,-0.37);
%
%\node[below] at (0,0) {$\tensor[^{\phantom{-}}]{\cdots}{^{\phantom{-}}}$};
%
%
\draw[pole] (-0.75,0.665)--(-0.75,0);
\draw[pole] (-0.375,0.3325)--(-0.375,0);
\draw[pole] (-0.1875,0.16625)--(-0.1875,0);
\draw[pole] (-0.09375,0.083125)--(-0.09375,0);
\draw[pole] (-0.046875,0.041525)--(-0.046875,0);
\draw[pole] (-0.0234375,0.02078125)--(-0.0234375,0);
\draw[pole] (0.75,0.665)--(0.75,0);
\draw[pole] (0.375,0.3325)--(0.375,0);
\draw[pole] (0.1875,0.16625)--(0.1875,0);
\draw[pole] (0.09375,0.083125)--(0.09375,0);
\draw[pole] (0.046875,0.041525)--(0.046875,0);
\draw[pole] (0.0234375,0.02078125)--(0.0234375,0);
\draw[-latex',line width = 1.5pt] (-1.2,0)--(1.2,0) node[right,font=\normalsize] {$\omega$};
\draw[-latex',line width = 1.5pt] (0,0)--(0,0.6) node[above,font=\normalsize] {$\Delta(\omega)$};
\end{tikzpicture}
\caption[Logarithmic discretization of the hybridization function in NRG]{Logarithmic discretization of the hybridization function in NRG. The hybridization function $\Delta(\omega)$ is the shaded region with the discretization bins marked by dashed lines. The orange peaks are poles centered in each bin with weight equal to the spectral weight of that bin.\label{fig:nrgdisc}}
\end{figure}
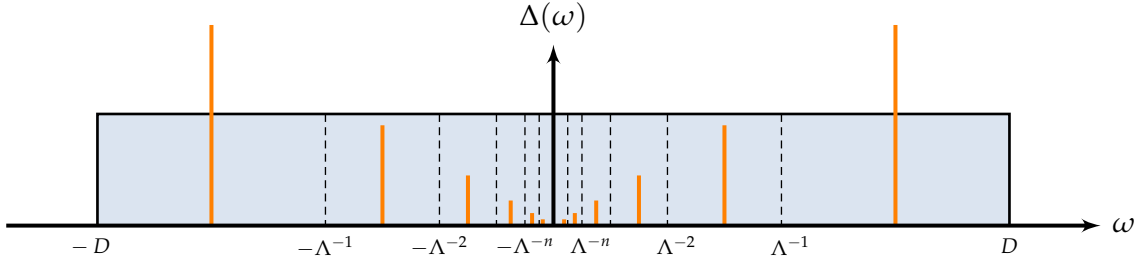
Within each discretization band, an orthonormal set of functions is defined
\begin{equation}
	\tensor*{\psi}{^{\pm}_{np}}(\epsilon) = \begin{cases} \frac{\Lambda^{n/2}}{(1 - \Lambda^{-1})^{1/2}} \e^{\pm \i \omega_n p \epsilon} &  |\epsilon| \in \left( \Lambda^{-n-1} , \Lambda^{-n} \right] \\ \hfil 0 & |\epsilon| \notin \left( \Lambda^{-n-1} , \Lambda^{-n} \right] \end{cases}
\end{equation}
with $p \in \mathbbm{Z}$ and
\begin{equation}
	\omega_n \equiv \frac{2\pi}{\Lambda^{-n} - \Lambda^{-(n+1)}} \,.
\end{equation}
With these basis functions, the bath operators are expressed in a Fourier expansion as
\begin{equation}
	\op{a}{\epsilon \sigma} = \sum_{n,p} \left[ \op{a}{np\sigma} \tensor*{\psi}{^+_{np}}(\epsilon) + \op{b}{np\sigma} \tensor*{\psi}{^-_{np}}(\epsilon) \right]
\end{equation}
with the inverse Fourier expressions being
\begin{align}
	\op{a}{np\sigma} &= \int_{-D}^{D} \d\epsilon \left[ \tensor*{\psi}{^+_{np}}(\epsilon) \right]^* \op{c}{\epsilon\sigma} \,,
	&
	\op{b}{np\sigma} &= \int_{-D}^{D} \d\epsilon \left[ \tensor*{\psi}{^-_{np}}(\epsilon) \right]^* \op{c}{\epsilon\sigma} \,,
\end{align}
and $\op{a}{np\sigma}$ and $\op{b}{np\sigma}$ satisfy the usual fermionic anticommutation relations.

The elements of the Hamiltonian are then re-expressed with these operators lying in these discretization intervals.
%\begin{equation}
%	\hat{H}^{\int}_{\text{hyb}}
%	=	\opd{d}{\sigma} \sum_{n,p}
%\end{equation}
For $h(\epsilon) = h = \text{const.}$, the
impurity couples only to the $p=0$ orbital. For general $\Delta(\omega)$, this can be accomplished by setting $h(\epsilon) = \text{const.}$ within each interval where the constant value is the average of the hybridization within that interval.

The discretized Hamiltonian for the impurity model is now
\begin{equation}
\begin{aligned}
	\frac{\hat{H}}{D} = \hat{H}_{\text{imp}}
		&+ \sum_{n,\sigma} \left( \xi^+_n \opd{a}{n0\sigma} \op{a}{n0\sigma} + \xi^-_n \opd{b}{n0\sigma} \op{b}{n0\sigma} \right)
		\\&+ \frac{1}{\sqrt{\pi}} \sum_{n,\sigma} \left( \opd{d}{\sigma} \left( \gamma^+_n \op{a}{n0\sigma} + \gamma^-_n \op{b}{n0\sigma} \right)
		+ \left( \gamma^+_n \opd{a}{n0\sigma} + \gamma^-_n \opd{b}{n0\sigma} \right) \op{d}{\sigma} \right)
\end{aligned}
\end{equation}
with the coefficients $\xi^\pm_n$ and $\gamma^\pm_n$ defined as
\begin{align}
	\xi^\pm_n	&=	\frac{\int_{\pm\mathcal{I}_n} \d\epsilon\, \epsilon\Delta(\epsilon)}{\int_{\pm\mathcal{I}_n} \d\epsilon\, \Delta(\epsilon)} \,,
	&
	( \gamma^\pm_n )^2	&=	\int_{\pm\mathcal{I}_n} \d\epsilon\, \Delta(\epsilon) \,,
\end{align}
and the integral domains are the intervals $+\mathcal{I}_n = [\Lambda^{-(n+1)},\Lambda^{-n}]$, $-\mathcal{I}_n = [-\Lambda^{-n},-\Lambda^{-(n+1)}]$.

An issue with the discretization is that it systematically underestimates the hybridization function. This can be compensated for by a correction factor $A_\Lambda$, whose exact expression depends on the functional form of the hybridization function.
For quantum impurity models the hybridization function input of NRG often takes the form of a flat band, as illustrated in Fig.~\ref{fig:nrgdisc}. However in other contexts, such as those which will be seen to appear in dynamical mean-field theory, non-trivial initial hybridization functions generally need to be considered. A particularly notable case is where the low energy behavior of the hybridization function exhibits power-law behavior, $\Delta(\omega) \sim |\omega|^r$.
For this case, the
scale factor takes the form of \cite{ingersent.PhysRevB.57.14254}
\begin{equation}
	A_{\Lambda,r} = \left[ \frac{1-\Lambda^{-(2+r)}}{2+r} \right]^{1+r} \left[ \frac{1+r}{1-\Lambda^{-(1+r)}} \right]^{2+r} \ln\Lambda
\end{equation}
which includes the flat-band case, $r=0$.

This discretized Hamiltonian is now mapped onto a semi-infinite chain,
known as the
Wilson chain.
The model will now consist of the impurity on the $-1^{\text{th}}$ site of the semi-infinite chain which is hybridized to the single $0^{\text{th}}$ site of the chain~\cite{kww1}. The conduction operators on this chain for $n\geq0$ are given by
\begin{equation}
	\op{c}{0\sigma} = \frac{1}{\sqrt{\xi_0}} \sum_{n} \left( \gamma^+_n \op{a}{n0\sigma} + \gamma^-_n \op{b}{n0\sigma} \right)
\end{equation}
which results in the Hamiltonian taking the form
\begin{equation}
\begin{multlined}[c][0.75\linewidth]
	\frac{\hat{H}}{D} = \hat{H}_{\text{imp}} + \sqrt{\frac{\xi_0}{\pi}} \sum_{\sigma}\left( \opd{d}{\sigma} \op{c}{0,\sigma} + \opd{c}{0,\sigma} \op{d}{\sigma} \right) \\+ \sum_{\sigma,n=0}^{\infty} \left[ \tensor{\varepsilon}{_{n}} \opd{c}{n,\sigma} \op{c}{n,\sigma} + \tensor{t}{_n} \left( \opd{c}{n,\sigma} \op{c}{n+1,\sigma} + \opd{c}{n+1,\sigma} \op{c}{n,\sigma} \right) \right]
\end{multlined}
\end{equation}
where $\xi_0 = \int_{-D}^D \d\epsilon\, \Delta(\epsilon)$.

The parameters of the chain Hamiltonian are obtained recursively with the initialization
\begin{equation}
\begin{aligned}
	\varepsilon_0	&=	\frac{1}{\xi_0} \int_{-D}^{D} \d\varepsilon \; \varepsilon \Delta(\varepsilon) \,,
	\\
	t_0^2	&=	\frac{1}{\xi_0} \sum_{m} \left[ (\xi^+_m - \varepsilon_0)^2 (\gamma^+_m)^2 + (\xi^-_m - \varepsilon_0)^2 (\gamma^-_m)^2 \right] \,,
	\\
	u_{1,m}	&=	\frac{1}{t_0} (\xi^+_m - \varepsilon_0) \frac{\gamma^+_m}{\sqrt{\xi_0}} \,,
	\\
	v_{1,m}	&=	\frac{1}{t_0} (\xi^-_m - \varepsilon_0) \frac{\gamma^-_m}{\sqrt{\xi_0}} \,,
\end{aligned}
\end{equation}
and the iteration proceeding for $n\geq1$ as
\begin{equation}
\begin{aligned}
	\varepsilon_n	&=	\sum_{m} \left( \xi^+_m u_{n,m}^2 + \xi^-_m v_{n,m}^2 \right) \,,
	\\
	t_n^2	&=	\frac{1}{\xi_0} \sum_{m} \left[ (\xi^+_m)^2 u_{n,m}^2 + (\xi^-_m)^2 v_{n,m}^2 \right] - t_{n-1}^2 - \varepsilon_n^2 \,,
	\\
	u_{n+1,m}	&=	\frac{1}{t_n} \left[ (\xi^+_m - \varepsilon_n) u_{n,m} - t_{n-1} u_{n-1,m} \right] \,,
	\\
	v_{n+1,m}	&=	\frac{1}{t_n} \left[ (\xi^-_m - \varepsilon_n) v_{n,m} - t_{n-1} v_{n-1,m} \right] \,.
\end{aligned}
\end{equation}
It is worth noting that the $1d$ chain form of the discretized bath, the Wilson chain, is independent of the actual physical geometry of the bath. The interpretation of successive sites are as orbitals successively further away from the impurity. Physical details of the actual bath enter only through the effective Wilson chain parameters $t_n$ and $\varepsilon_n$. The Wilson chain is shown schematically in Fig.~\ref{fig:wilsonchain}.

The next procedure in the calculation is an iterative diagonalization.
This is the step in which
the renormalization group character of NRG appears. The Hamiltonian is considered as a limit of Hamiltonians
\begin{equation}
\begin{aligned}
	\hat{H}_N
	= \Lambda^{\frac{N-1}{2}} \Bigg[ \hat{H}_{\text{imp}} &+ \sqrt{\frac{\xi_0}{\pi}} \sum_{\sigma} \left( \opd{d}{\sigma} \op{c}{0,\sigma} + \opd{c}{0,\sigma} \op{d}{\sigma} \right) \\ 
	&+ \left. \sum_{\sigma,n=0}^{N} \tensor{\varepsilon}{_{n}} \opd{c}{n,\sigma} \op{c}{n,\sigma} + \sum_{\sigma,n=0}^{N-1} \tensor{t}{_n} \left( \opd{c}{n,\sigma} \op{c}{n+1,\sigma} + \opd{c}{n+1,\sigma} \op{c}{n,\sigma} \right) \right]
\end{aligned}
\end{equation}
with
\begin{equation}
	\frac{\hat{H}}{D} = \lim_{N\to\infty} \Lambda^{\frac{N-1}{2}} \hat{H}_{N}
\end{equation}
Successive Hamiltonians are constructed iteratively as
\begin{equation}
	\hat{H}_{N+1} = \Lambda^{\frac{1}{2}} \hat{H}_{N} + \Lambda^{\frac{N}{2}} \sum_{\sigma} \tensor{\varepsilon}{_{N+1}} \opd{c}{N+1,\sigma} \op{c}{N+1,\sigma} + \Lambda^{\frac{N}{2}} \sum_{\sigma} \tensor{t}{_{N+1}} \left( \opd{c}{N,\sigma} \op{c}{N+1,\sigma} + \opd{c}{N+1,\sigma} \op{c}{N,\sigma} \right)
\end{equation}
This construction of successive Hamiltonians is the renormalization operation.
At each Hamiltonian $\hat{H}_{N}$ an energy spectrum is found from an eigenbasis. 
The states take the form of
\begin{equation}
	\lvert Q ,\, S_z ;\, r \rangle
\end{equation}
with charge and spin projection quantum numbers, $Q$ and $S_z$; and an index $r$ labelling states with the same $Q$ and $S_z$.
The basis at successive iterations $N+1$ is constructed from the previous iteration. Since this basis would grow exponentially, thereby hindering the exact diagonalization of the Hamiltonian, only a truncated set of states are used to build the basis for the next iteration. The truncation is determined by a set energy cutoff with only the lowest energy states retained at each step. This process is diagrammed schematically in Fig.~\ref{fig:nrgstates}.

A complete basis of eigenstates, the Anders-Schiller basis, can be formed from the total set of discarded states~\cite{andersschiller}. The reason for needing this basis rather than the basis produced from the kept states, is that the kept states possess non-zero overlap with each other. This leads to an over counting of contributions to the Lehmann sum~\cite{fdmnrg}. This can be avoided by using the Anders-Schiller basis, which is complete.

\begin{figure}[ht!]
\centering
\begin{tikzpicture}[every node/.style={line width=1pt,inner sep=4pt,scale=1.},scale=1.5]
\node[circle,draw=black] (o) at (0,0){$\phantom{o}$};
\foreach \n [count=\m] in {0,...,4}
	{
	\node[rectangle,draw=black] (\m) at ($(\n,0)+(1,0)$){$\phantom{o}$};
	\node[below=2pt,inner sep=8pt] at (\m) {$\tensor*[_{\phantom{\n}}]{\varepsilon}{_{\m}}$};
	\ifnum \n>0
	\draw[black,line width=10pt/\m] (\n)--(\m) node[midway,above] {$t_\n$};
	\fi
	}
\node[] (end) at (6,0){};
\draw[-,line width=1pt,double distance = 1pt] (o)--(1) node[midway,above] {$V$};
\draw[black,dashed,line cap=round,line width=1.5pt] (5)--(end);
\end{tikzpicture}
\caption[Schematic of the NRG Wilson chain]{Schematic of the NRG Wilson chain. The amplitude of the $t_n$'s decays logarithmically with increasing $n$.\label{fig:wilsonchain}}
\end{figure}
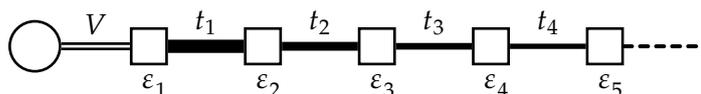
\begin{figure}[ht!]
\centering
\includegraphics[scale=1]{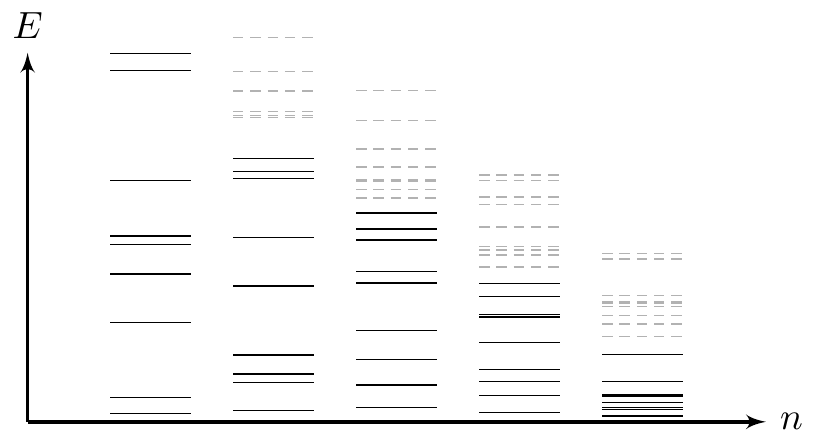}
\caption[Truncation of states in NRG]{Truncation of states in NRG. At each iteration, only a certain number of lowest energy states are kept (black lines). The higher energy states are discarded (gray dashed lines). The discarded states however are kept to form the complete Anders-Schiller basis~\cite{andersschiller} used for calculating spectral quantities.\label{fig:nrgstates}}
\end{figure}

%The exact analytical solution of the Kondo model was derived using Bethe ansatz methods~\cite{andrei,wiegmann}. This solution obtains the eigenspectrum and eigenstates, as well as static quantities such as susceptibility, free energy, and specific heat. This solution however does not include dynamical quantities such as the Green functions.

%An early implementation of NRG as the impurity solver for DMFT was in the solution to the Hubbard model in Ref.~\cite{bullahubbard}.

A particularly relevant quantity which will need to be calculated is the Green function. These are constructed in NRG through their Lehmann representation\index{Lehmann representation} from the eigenstates obtained in the iterative diagonalization~\cite{fdmnrg,nrggreen}. Of particular relevance is the self-energy due to the interactions on the impurity. These are related to the Green function
\begin{equation}
\begin{aligned}[b]
	F_\sigma(z)
	&\vcentcolon=	\Green{\op{d}{\sigma}}{[\hat{H}_{\text{Int}},\opd{d}{\sigma}]}_z
%	\\
%	&= U \Green{\op{c}{\sigma}}{\opd{c}{\sigma} \opd{c}{-\sigma} \op{c}{-\sigma}}_z
\end{aligned}
\end{equation}
which appears from the contribution of the interaction Hamiltonian in the single particle equations of motion.
From the Green function equations of motion the self-energy can be observed to be~\cite{bullaFG}
\begin{equation}
	\Sigma_{\sigma}(z) = \frac{F_\sigma(z)}{G_\sigma(z)} \,.
\label{eq:FoverG}
\end{equation}
For the single impurity Anderson model,
\begin{equation}
	F_{\sigma}(z) = U \Green{\op{d}{\sigma}}{\opd{d}{\sigma} \opd{d}{-\sigma} \op{d}{-\sigma}}_z
\end{equation}
and note the comparison with Eq.~\eqref{eq:foverggreen}.
The imaginary parts of $F_\sigma(z)$ and $G_\sigma(z)$ are calculated in NRG via the Lehmann representation.
Analogously to $G_\sigma(z)$, it is possible to write a spectral function for $F_\sigma(z)$ as
\begin{equation}
	\mathcal{B}_{\sigma}(\omega) = -\frac1\pi \Im F_\sigma(\omega+\i0^+)
\end{equation}
which in the Lehmann representation is
\begin{equation}
	\mathcal{B}_{\sigma}(\omega)
	=	\frac{1}{\mathcal{Z}} \sum_{m,n} \langle n \lvert \op{d}{\sigma} \rvert m \rangle \langle m \lvert \opd{d}{\sigma} \opd{d}{-\sigma} \op{d}{-\sigma} \rvert n \rangle \delta\left( \omega - ( E_m - E_n ) \right) \left( \e^{-\beta E_n} + \e^{-\beta E_m} \right)
\end{equation}
where $m,n$ span the a complete basis.
The discrete spectral poles from the Lehmann representation calculation are broadened to form a continuous spectrum.
This gives the imaginary part of $F_{\sigma}(\omega+\i0^+)$.

In cases where poles are distributed linearly, poles at $\omega_p$ of weight $w_p$ can be broadened using, for example, a Lorentzian distribution
\begin{align}
	w_p \delta(\omega-\omega_p) &\mapsto w_p \frac1\pi \frac{\eta}{(\omega-\omega_p)^2 + \eta^2}
\label{eq:lorentzianpole}
%\end{align}
\intertext{with $0 < \eta \ll 1$.
As the $\{\omega_n\}$ are distributed according to a logarithmic scale in NRG, it is more appropriate to broaden the poles instead using a logarithmic distribution, as~\cite{bullaFG}}
%\begin{equation}
	w_p \delta(\omega-\omega_p) &\mapsto w_p \frac{\e^{-b^2/4}}{b \omega_p \sqrt{\pi}} \e^{-\left(\frac{\ln(\omega/\omega_p)}{b}\right)^2} \,.
	\tag{\ref*{eq:lorentzianpole}$^{\prime}$}
\end{align}
where $b$ is a broadening parameter, with values typically in the range of $0.3 \leq b \leq 0.6$.
This kernel then gives the appropriate broadening of the spectral poles to produce an approximate continuous spectrum.
A Hilbert transform is applied to the continuous imaginary parts to compute the corresponding real part such that the full function satisfies the Kramers-Kronig relations\index{Kramers-Kronig relations}.
The self-energy is then calculated according to Eq.~\eqref{eq:FoverG}

The self-energy may also be calculated directly from the Dyson equation
\begin{equation}
	\tensor*{\Sigma}{_{\sigma}}(z) = \tensor*{G}{^{(0)}_{\sigma}}(z)^{-1} - \tensor*{G}{_{\sigma}}(z)^{-1}
\label{eq:SEfromDyson}
\end{equation}
where
\begin{equation}
	\tensor*{G}{^{(0)}_{\sigma}}(z) = \frac{1}{z - \varepsilon_f - \Delta_{\sigma}(z)}
\end{equation}
is the free impurity Green function and $\Delta_\sigma(z)$ is the hybridization function as before. While this equation is exact analytically, it is numerically it is prone to errors. Conversely in \eqref{eq:FoverG}, since $F_{\sigma}(z)$ and $G_{\sigma}(z)$ are divided, only relative errors propagate to the solution for $\Sigma_{\sigma}(z)$. This means that \eqref{eq:FoverG} produces a more stable result numerically than the inverted Dyson expression \eqref{eq:SEfromDyson}.

\label{sec:seproblems}
Numerical errors can result in the case where $\Im\Sigma(\omega) > 0$ at low temperatures near the Fermi level $\omega=0$. In the context of the NRG-DMFT presented in this thesis this error is corrected by a sign flip, \textit{i.e.} $\Im\Sigma(\omega) \mapsto -\Im\Sigma(\omega)$ when $\Im\Sigma(\omega) > 0$. The errors introduced by this correction do not generally affect the convergence of the DMFT solution as the magnitude and domain of the flipped self-energy is small compared to the self-energy overall. An exception to this is when the self-energy is calculated for very weak interactions. In this case the flipped self-energy is of the same order of magnitude as the rest of the self-energy, and is over a large portion of the self-energy's frequency domain. This causes catastrophic errors to accumulate in the resulting Green functions calculated from NRG and the solution cannot be taken as accurate.

Alternative resolutions of the positive overshoot error include setting $\Im\Sigma(\omega) = 0$ in the overshoot region or shifting the whole imaginary part of the self-energy such that $\Im\Sigma(\omega) < 0$ over its entire range. However, also with this correction scheme numerical errors can ruin the calculation. In some instances there are negative divergences at low frequency of $\mathcal{O}(1)$ in the self-energy. While these divergences may be large in magnitude on the order of the self-energy, they are small compared to the hybridization function and only occur at exponentially low frequencies. These errors therefore do not dramatically affect the Green function as computed from the Dyson equation in the cases encountered in this thesis. Performing the shift procedure however can lead to dramatically increasing the entire self-energy over the whole frequency range, which then does significantly affect the Green function solution.

The non-analytic errors in the self-energy arising from the $F/G$ prescription appear to be a systematic error in NRG. This work does not attempt to solve this issue, but identifies it as an area to be addressed in the NRG community. A recent proposal for a new self energy estimator which does not suffer from the aforementioned issues is found in~\cite{newnrgse}.

\subsection{Solution to the Anderson Model\label{sec:amsolution}}

The numerical renormalization group was originally devised to solve the Kondo model and the resistance minimum problem of dilute magnetic impurities in metals~\cite{wilson}.
The method was subsequently expanded upon to solve the full single impurity Anderson model~\cite{kww1,kww2}. The impurity spectral function takes the form of a three-peak structure, with the central peak at particle-hole symmetry taking a maximum at $1/\pi\Gamma$, where $\Gamma = -\frac1\pi\Im\Delta(0)$. The spectral function and self-energy of the impurity in the Anderson model as obtained from NRG are plotted in Fig.~\ref{fig:siamsolution}.
\begin{figure}[h]
%\begin{subfigure}{\linewidth}
%	\includegraphics{siamGfull.pdf}
%\end{subfigure}
%
\begin{subfigure}{0.49\linewidth}
\begin{tikzpicture}
	\node at (0,0) {\includegraphics{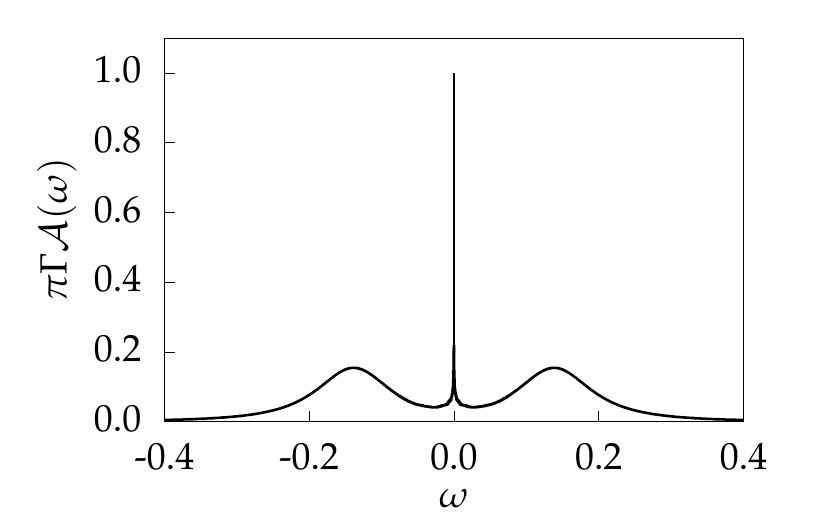}};
	\node at (3.125,2) {\footnotesize\subref*{fig:siamG}};
\end{tikzpicture}
\phantomsubcaption{\label{fig:siamG}}
\end{subfigure}
\hfill
\begin{subfigure}{0.49\linewidth}
\begin{tikzpicture}
	\node at (0,0) {\includegraphics{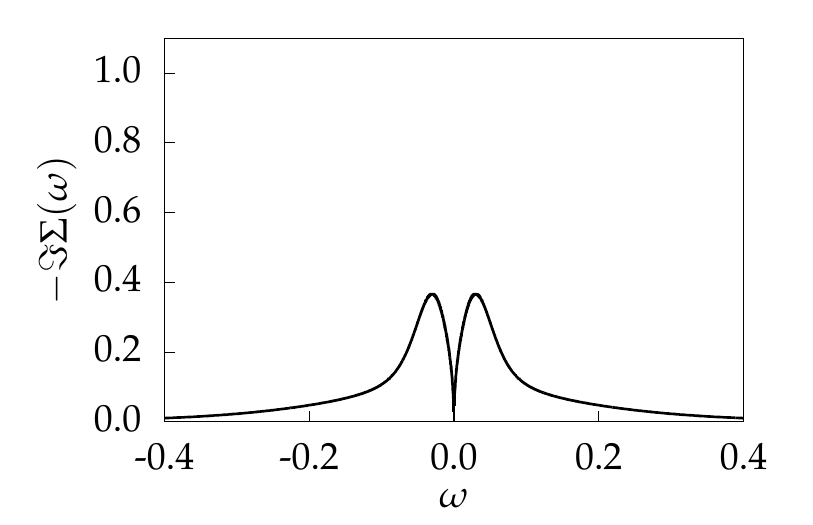}};
	\node at (3.125,2) {\footnotesize\subref*{fig:siamS}};
\end{tikzpicture}
\phantomsubcaption{\label{fig:siamS}}
\end{subfigure}
\vspace{-\baselineskip}
\caption[NRG solution to the single impurity Anderson model]{NRG solution to the single impurity Anderson model parameterized by $V/D = 0.1$, $U/D = 0.3$, and $\epsilon/D = -0.15$. Note that the peak of the spectral function is pinned to $1/\pi\Gamma$ at particle-hole symmetry.\label{fig:siamsolution}}
\end{figure}
The Anderson model has the characteristics of a Fermi liquid, where in particular, the low energy behavior of the self-energy goes as~\cite{hewson,mattuck}
\begin{subequations}
\begin{align}
	\Re\Sigma(\omega) &\sim \omega + \text{const.}
	\\
	\Im\Sigma(\omega) &\sim \omega^2 \,,
\end{align}
\label{eq:lowese}
\end{subequations}
where the constant term in the real part is $\frac{U}{2}$ for system at particle-hole symmetry, but is in general unknown. A Fermi liquid also has the property~\cite{hewson,logangalpin}
\begin{equation}
	\Im\int_{-\infty}^{0} \d\omega\, G(\omega) \frac{\partial \Sigma(\omega)}{\partial \omega} = 0 \,.
\end{equation}
These characteristics will become relevant in later chapters.

Another feature to make note of is the central peak of the spectral function. From the Dyson equation, the impurity Green function is
\begin{align}
	G_{\sigma}(z)
	&= \frac{1}{z - \varepsilon_f - \Sigma_\sigma(z)}
	\\
	&= G^0(z - \Sigma(z))
\end{align}
where $G^0(z)$ is the non-interacting Green function on the impruty with $U=0$. A renormalized energy level can be defined as $\tensor*{\varepsilon}{^*_f} = \tensor*{\varepsilon}{_f} + \Re\Sigma(0)$. At particle-hole symmetry $\tensor*{\varepsilon}{^*_f} = 0$. Evaluating the impurity spectral function at $\omega = 0$ and taking into account the low energy behavior of the self-energy \eqref{eq:lowese} leads to the result that
\begin{equation}
	\mathcal{A}_\sigma(0) = -\frac1\pi \Im G^0(0) \,,
\label{eq:pinned}
\end{equation}
which means that the impurity spectral function at finite $U$ is fixed to the value of the non-interacting density of states at zero energy.

\section{Dynamical Mean-Field Theory\label{sec:dmft}}

While a system with a single interacting site can essentially be solved exactly, as shown in the previous section, a fully interacting system in general cannot. The treatment of such systems inevitably involves making approximations or treating effective models.
One computational method of such an effective theory is 
dynamical mean-field theory (DMFT).

The concept of DMFT is that it treats non-local interactions not directly, as in \textit{e.g.} perturbation theory, but rather through auxiliary degrees of freedom which are self-consistently defined. Local interactions are however treated directly. The restriction from non-local interactions to local ones only is the core of the DMFT approximation.

Ordinary mean-field theory models a system by treating its dynamics only locally, with the behavior of the rest of the system modeled by a static background, the mean-field. A classic example of such a mean-field theory is that of the Ising model of magnetic spins where the mean-field represents a background overall magnetization of the entire system. In the Ising model, correlations are not taken between independent dynamical spins, but rather between single dynamical spins and a static mean-field average.
Dynamical mean-field theory on the other hand employs a similar concept, but allows for the implementation of a non-static mean-field, \textit{i.e.} a mean-field which in principle has energy and momentum dependence (although momentum dependence is suppressed in DMFT, as explained below).

A central concept which facilitates DMFT is that in infinite spatial dimensions the self energy becomes a purely local quantity~\cite{metznervollhardt,infinitehubbard,muellerhartmann}. This allows interactions to be treated only locally and non-local correlations can be ignored.
This can be seen from the diagrammatic perturbative expansion of the self-energy. Here considered are the self-energy
skeleton diagrams with internal lines consisting of the full interacting Green function~\cite{jarrellhubbard,schweitzerczycholl}.
Diagrammatically, the self-energy can be expressed as~\cite{mattuck,dmft}
\begin{equation}
\boldsymbol{\Sigma}_{\sigma}(z) =
\;
\begin{tikzpicture}[baseline={(current bounding box.center)},decoration={markings, mark= at position 0.5 with {\arrow[xshift=3pt]{Stealth[length=6pt,width=4pt]}}}, scale=0.75]
	\coordinate (i0) at (0,0) node[below] {$i$};
	\coordinate (i) at (0,1);
	\draw[postaction={decorate}] (i) arc (90:450:0.5cm and -0.5cm) node[midway,below=0.125cm] {$-\sigma$};
	\draw[dashed] (i)--(i0);
	%\node[below] at (-0.5,0) {$i$};
%	\node[below] at (0.5,0) {$\phantom{j}$};
	\draw[-] (-0.5,0)--(0.5,0);
\end{tikzpicture}
\;+\;
\begin{tikzpicture}[baseline={(current bounding box.center)}, decoration={markings, mark= at position 0.5 with {\arrow[xshift=3pt]{Stealth[length=6pt,width=4pt]}}}, scale=0.75]
	\def\y{1.5}
	\coordinate (i1) at (-1,\y);
	\draw[postaction={decorate}] (i1) arc (30:150:-2cm and 1cm) coordinate (j1) node[midway,below=0.125cm] {$-\sigma\phantom{-}$};
	\draw[postaction={decorate}] (j1) arc (30:150:2cm and -1cm);
	\draw[dashed] (i1)--($(i1)+(0,-\y)$) node[below] {$i$};
	\draw[dashed] (j1)--($(j1)+(0,-\y)$) node[below] {$j$};
	\draw[postaction={decorate}] ($(i1)+(-0.5,-\y)$)--($(j1)+(0.5,-\y)$) node[midway,below] {$\sigma$};
\end{tikzpicture}
\;+\;
\cdots
\label{eq:sediagrams}
\end{equation}
where only the first terms are shown.
The non-local Green functions describing propagation from site $i$ to site $j$ scale with $\sim t \mapsto \widetilde{t}/\sqrt{d}$.
The multiplicity of internal lines in the non-local self-energy diagrams scales as $d$.
As shown in \eqref{eq:sediagrams}, the
diagrams for the non-local contributions to the self-energy scale at least as $(\widetilde{t}/\sqrt{d})^3$.
The overall scaling behavior of the non-local self-energy contributions is then at least as $1/\sqrt{d}$ which vanishes in the $d\to\infty$ limit.
The vanishing of these terms in the $d\to\infty$ limit thereby eliminates all non-local contributions to the self-energy.

The locality of the self-energy in the limit of infinite dimensions holds in both real as well as momentum space: 
%\begin{subequations}
%\begin{equation}
%	\lim_{\kappa\to\infty} \Sigma_{i,j}(\omega) = \Sigma(\omega) \delta_{i,j}
%\end{equation}
%\begin{equation}
%	\lim_{d\to\infty} \Sigma(\omega,k) = \Sigma(\omega)
%\end{equation}
%\end{subequations}
\begin{equation}
\begin{aligned}
	\Sigma_{i,j}(z)
		&=	\Sigma(z,r_j - r_i)
		\\
	\Sigma(z) \delta_{i,j}
		&=	\frac{1}{2\pi} \int \d^d k\, \Sigma(z,k) \e^{\i k \cdot (r_j - r_i)}
		\\
	\Sigma(z) \frac{1}{2\pi} \int \d^d k\, \e^{\i k \cdot (r_j - r_i)}
		&=	\frac{1}{2\pi} \int \d^d k\, \Sigma(z,k) \e^{\i k \cdot (r_j - r_i)}
		\\
		\Sigma(z) &= \Sigma(z,k) \,.
\end{aligned}
\end{equation}

The locality of the self-energy implies that the Hubbard model in infinite dimensions can be mapped onto an Anderson Impurity model \eqref{eq:siam}~\cite{infinitehubbard,jarrellhubbard}.
%\begin{equation}
%	\hat{H}_{\textsc{aim}} = \sum_{k,\sigma} \varepsilon_{k} \opd{c}{k,\sigma} \op{c}{k,\sigma} + \sum_{k,\sigma} \left( V_{k,\sigma} \opd{c}{k,\sigma} \op{d}{\sigma} + V^*_{k,\sigma} \opd{d}{\sigma} \op{c}{k,\sigma} \right) + \varepsilon_{d} \opd{d}{\sigma} \op{d}{\sigma} + U \opd{d}{\uparrow} \op{d}{\uparrow} \opd{d}{\downarrow} \op{d}{\downarrow}
%\end{equation}
From the Green function equations of motion, the Hubbard model self-energy can be found to be
\begin{equation}
	\tensor*{\Sigma}{_{ij,\sigma}}(z) = U \frac{\Green{\op{c}{i,\sigma}}{\op{n}{j,-\sigma} \opd{c}{j,\sigma}}_z}{\Green{\op{c}{i,\sigma}}{\opd{c}{j,\sigma}}_z} \,.
\end{equation}
Such terms with $i\neq j$ arise when calculating the non-local $\Greenline{\op{c}{i,\sigma}}{\opd{c}{j,\sigma}}_z$ Green function.
Computing the Green function $\Greenline{\op{c}{i,\sigma}}{\op{n}{j,-\sigma} \opd{c}{j,\sigma}}_z$ involves evaluating the commutator $[ \hat{H}_{I} , \op{n}{j,-\sigma} \opd{c}{j,\sigma} ] = U \sum_{m} [ \op{n}{m,\uparrow} \op{n}{m,\downarrow} , \op{n}{j,-\sigma} \opd{c}{j,\sigma} ] \propto \sum_{m} \delta_{mj}$ where $\sum_{m}$ sums over all sites with the interaction term present.  For the Anderson impurity model this trivially restricts the self-energy to be local as only the impurity site features the interaction term. In the Hubbard model the sum $\sum_{m}$ extends over all sites in the system, but since $\Sigma_{ij,\sigma}(z) = \Sigma_{\sigma}(z) \delta_{ij}$ in infinite dimensions, the Green function for the Hubbard model coincides with the Green function for the single impurity Anderson model. 

In order to obtain finite and non-trivial models in the infinite dimensional limit, it is necessary to ensure that all parameters of the Hamiltonian are extensive.

Recall the form of the dispersion relation for a $d$ dimensional square lattice Eq.~\eqref{eq:squaredispersion}. The $d$-dimensional sum over $\cos(k_i)$ essentially amounts to a sum over random numbers distributed in $[-1,1]$. By the central limit theorem in the $d\to\infty$ limit, this leads to a density of states with Gau{\ss}ian distribution form
\begin{equation}
	\mathcal{A}(\omega) = \frac{1}{2 t \sqrt{\pi d}} \e^{-\left(\frac{\omega - \varepsilon}{2 t \sqrt{d}}\right)^2} \,.
\end{equation}
In order for this quantity to be finite, it is necessary to scale the kinetic hopping parameter $t$ as $t\mapsto\widetilde{t}/\sqrt{d}$. In the Hubbard model \eqref{eq:hubbard}, the interaction term acts only locally, which means it does not scale with $d$. The scaling $t\mapsto\widetilde{t}/\sqrt{d}$ ensures that the relative energy scales of the kinetic and interaction terms remain comparable, even in the infinite dimensional limit, $\displaystyle \lim_{d\to\infty}\mathcal{O}(U/t) \sim 1$.

In the continuum, the density of states for free fermions in $d$ dimensions is proportional to $\varepsilon^{d/2 - 1}$. This expression cannot be scaled in such a way that it remains finite in the limit $d\to\infty$. It is therefore necessary to study the behavior of such systems regularized by a lattice~\cite{muellerhartmanninfinite}.

As mentioned in \S\ref{infinitelimitbethe}, the Bethe lattice is a lattice with a simple well defined infinite dimensional limit. It is therefore an appropriate lattice for building models to be treated by DMFT, and is the lattice of choice used throughout this work.
Other lattices which possess a well-defined limit to infinite coordination number are the $2d$ hexagonal lattice and $3d$ diamond lattice \cite{santoro}.

%For a Hubbard-type Coulomb interaction, the interactions occur only on-site and so the interaction strength $U$ does not scale with the coordination number. The non-local contributions to the self-energy scale as $\frac{1}{\sqrt{\kappa}}$, so therefore vanish in the limit of $\kappa\to\infty$. The self-energy then becomes a purely local quantity in the limit of infinite coordination number.

The coordination number of the Bethe lattice is not directly related to any spatial dimension. To consider the relation of the infinite coordination number to real-space dimension, consider the (hyper)cubic lattice whose unit lattice vectors span the real-space volume. The coordination number $\kappa$ for the hypercubic lattice is directly related to the spatial dimension $d$ by $\kappa = 2d$. In the infinite coordination number limit, the hopping amplitude then scales as
\begin{subequations}
\begin{equation}
	t \to \frac{t}{\sqrt{\kappa}}
\end{equation}
or
\begin{equation}
	t \to \frac{t}{\sqrt{2d}}
\end{equation}
\end{subequations}
which then yields the equivalence between the infinite coordination number limit and the infinite dimensional limit.

The infinite dimensional limit of DMFT represents a spatial mean-field approximation that appears far away from the $d=2$ or $d=3$ dimensionality of real systems. However for lattice models the coordination number may be much higher than the real-space dimensionality. In $d=3$ the face-centered cubic lattice has coordination number $\kappa=12$, which is a relatively ``large'' number.

It is nevertheless necessary for non-local contributions to be taken into account for precise comparison with real systems. Some methods for incorporating non-local contributions into DMFT are reviewed in \cite{nonlocaldmft}.
One example is that of cluster-DMFT~\cite{clusterdmft}, where instead of a single site impurity, the impurity model is taken to be a finite size cluster of interacting sites.
Another example is that of coupling DMFT with the functional renormalization group (FRG), producing the DMF$^2$RG theory~\cite{dmf2rg}. This scheme solves a model using DMFT as usual in the infinite dimensional limit, but then uses this solution to initialize an FRG calculation for the ``true'' finite dimensional system. The spatial dependence is incorporated through the FRG flow. 

There also exist a variety of schemes incorporating DMFT into electronic structure calculations for real materials~\cite{electronicstructuredmft}. 
A common method of such calculations involve coupling DMFT with density functional theory (DFT)~\cite{weberdmftdft}.

%The remainder of this discussion concentrates explicitly on the dimensionality of the lattice coordination number $\kappa$ rather than the real-space dimensionality $d$.

The DMFT algorithm is initiated by choosing an ansatz for the hybridization from impurity to the bath lattice,
\begin{equation}
	\Delta(z) = \sum_k \frac{\left\lvert V_k \right\rvert^2}{z - \varepsilon_k} \,.
\label{eq:initialhyb}
\end{equation}
This hybridization $\Delta(z)$ then specifies a quantum impurity model for which the self-energy of the impurity $\Sigma_{\text{imp}}(z)$ may be calculated using an impurity solver. This self-energy can then be used to calculate the local lattice Green function\footnote{The hybrid notation $\phi[J_m;y_n)$ indicates that $\phi$ is a functional of functions $\{J_m\}$ and a function of variables $\{y_n\}$.}
\begin{equation}
	G_{\text{latt}}[\Sigma_{\text{latt}};z) = \int \d\omega' \frac{\mathcal{A}_0(\omega')}{z - \varepsilon - \omega' - \Sigma_{\text{latt}}(z)}
\end{equation}
where $\mathcal{A}_0(\omega)$ is the non-interacting density of states for the system under consideration.
As an example, for the Hubbard model on an infinite dimensional Bethe lattice the lattice Green function takes the form
\begin{equation}
\begin{aligned}[b]
	G_{\text{latt}}[\Sigma_{\text{latt}};z)
	&=	\cfrac{1}{z - \varepsilon - \Sigma_{\text{latt}}(z) - \cfrac{\kappa t^2}{z - \varepsilon - \Sigma_{\text{latt}}(z) - \cfrac{\kappa t^2}{\ddots}}}
	\\
	&=	\cfrac{1}{z - \varepsilon - \Sigma_{\text{latt}}(z) - G_{\text{latt}}[\Sigma_{\text{latt}};z)} \,.
\end{aligned}
\end{equation}

At this stage of the DMFT cycle, it is necessary to compute the local self-energy. This can be accomplished from a number of techniques, such as iterated perturbation theory (IPT)~\cite{infinitehubbard}, exact diagonalization~\cite{eddmft,webered}, continuous-time--quantum Monte Carlo (CT-QMC)~\cite{ctqmc,weberctqmcdmft}, density matrix renormalization group (DMRG)~\cite{dmrgdmft}, or the numerical renormalization group (NRG)~\cite{bullahubbard,nrg}.
%The present work utilizes the numerical renormalization group, as described in the previous section.

%\begin{equation}
%	G^{-1}_{\text{imp}}[\Delta,\Sigma_{\text{imp}};z) = {z - \varepsilon_{\text{imp}} - \Delta(z) - \Sigma_{\text{imp}}(z)}
%\end{equation}

%\begin{equation}
%	G_{\text{imp}}[\Delta,\Sigma_{\text{imp}};z) = \frac{1}{z - \epsilon_{\text{imp}} - \Sigma_{\text{imp}}(z) - \Delta(z)}
%\end{equation}

The lattice Green function is then used to self-consistently generate a new hybridization function, restarting the DMFT cycle.
%\footnote{The hybrid notation $\phi[J_m;y_n)$ indicates that $\phi$ possesses functional dependence on functions $\{J_m\}$ and function dependence on variables $\{y_n\}$.}
The local lattice self-energy is taken to be the impurity self-energy of the impurity model,
\begin{equation}
	\Sigma_{\text{latt}}(z) \,\hat{=}\, \Sigma_{\text{imp}}(z) \,,
\end{equation}
such that the lattice Green function is a functional of the impurity self-energy
\begin{equation}
	G_{\text{latt}}[\Sigma;z) \,\hat{=}\, G_{\text{latt}}[\Sigma_{\text{imp}};z) \,.
\end{equation}
This then produces a result for the full interacting Green function of the lattice.
%\begin{equation}
%	\Sigma_{\text{latt}}(z) \hat{=} \Sigma_{\text{imp}}(z) \leadsto G_{\text{latt}}[\Sigma_{\text{imp}};z) \hat{=} G_{\text{imp}}[\Sigma_{\text{imp}},\Delta';z)
%\end{equation}
This Green function is then used to construct a new hybridization function $\Delta'(z)$ which defines a new quantum impurity model for the next cycle in the DMFT algorithm by
\begin{equation}
	G_{\text{latt}}[\Sigma_{\text{imp}};z) \,\hat{=}\, G_{\text{imp}}[\Delta',\Sigma_{\text{imp}};z)
\end{equation}
where $G^{-1}_{\text{imp}}[\Delta,\Sigma_{\text{imp}};z) = {z - \varepsilon - \Delta(z) - \Sigma_{\text{imp}}(z)}$. This reinitializes the DMFT calculation starting from \eqref{eq:initialhyb} and the calculation proceedes again to calculate a new lattice Green function.
The cycle is run until the Green function solution reaches convergence, \textit{i.e.} when the $n^{\text{th}}$ cycle's hybridization function $\Delta^{(n)}(\omega)$ can be said to satisfy $\lvert \Delta^{(n)}(\omega) - \Delta^{(n-1)}(\omega) \rvert < \delta$, $\forall \omega$ with $\delta \ll 1$.

The DMFT result is exact in the limit of infinite dimensions, although it can also be applied to finite dimensional cases. In the case of finite dimensions, the DMFT self-consistency condition is an approximation (as its name suggests, a mean-field approximation). A physical interpretation of the DMFT approximation may be obtained from analyzing the Luttinger-Ward functional $\Phi[G]$~\cite{luttingerward,potthofffunctional}. The physical self-energy can be obtained from the stationary point of the Luttinger-Ward functional
\begin{equation}
	\Phi[G] = \Omega[G^{-1}+\Sigma[G]] - \ln G + \Sigma[G] G
\end{equation}
as
\begin{equation}
	\Sigma =  \frac{\delta \Phi[G]}{\delta G} \,.
\end{equation}
$\Omega$ is the grand potential $\Omega = \Omega[G]$ with $\frac{\delta \Omega}{\delta G} = 0$ exact for physical $G$. %This $\Omega[G]$ not known explicitly in general. Instead, take $\Omega = \Omega[\Sigma]$ with $\frac{\delta \Omega}{\delta \Sigma} = 0$ exact for physical $\Sigma$
DMFT approximates the functional $\Phi[G]$ as the sum over skeleton and two particle irreducible (2PI) local diagrams as opposed to all diagrams~\cite{dmft,potthofffunctional}. The skeleton self-energy diagrams are those which only contain undressed Green function internal lines. This then provides some formal justification for applying DMFT to finite-dimensional systems. In this context the purely local self-energy calculated from DMFT is taken as an approximation.
%https://link.springer.com/content/pdf/10.1140/epjb/e2003-00121-8.pdf

%With relation to real materials DMFT can only in practice be understood as an approximation to experimental phenomenology. It can however be used to augment other techniques which do more accurately obtain information for real materials.
%A method density functional theory (DFT)

\subsection{Solution to the Hubbard Model\label{sec:hubbardsolution}}

The previous section introduced a set of powerful numerical methods which can be used to solve systems which are otherwise intractable with analytical tools or less sophisticated computational methods. DMFT coupled with NRG as the impurity solver is able to compute the exact solution to the Hubbard model in the limit of infinite dimensions, with spectral functions evaluated directly on the real axis, down to $T=0$~\cite{bullahubbard}. 

An example of a system which can be easily be solved by DMFT but is intractable by other means is the Hubbard model \eqref{eq:hubbard}.
The DMFT solution to the Hubbard model on the Bethe lattice in infinite dimensions over a range of $U$ is shown in Fig.~\ref{fig:hubbardsolution}.
Without interactions, $U=0$, the spectral function takes on a semi-elliptic form. As interactions are adiabatically increased, the spectrum begins to broaden and develops a characteristic three-peak structure, with the central peak of the spectral function pinned to $\frac1\pi$ following the discussion in \S\ref{sec:amsolution}. The self-energy takes the functional form of~\cite{analyticse}
\begin{equation}
	\Sigma(z) = z - \varepsilon - \frac{1}{G(z)}
\end{equation}
which implies an inverse relationship between the spectral function and the self-energy. As the spectral function develops a three peak structure, the self-energy therefore takes on a two peak form. This is well illustrated in the $U/t=4$ panel of Fig.~\ref{fig:hubbardsolution}. The locations of the peaks are at $\omega \sim \pm \sqrt{Z}$ where $Z$ is the quasiparticle weight\index{quasiparticle weight} $Z = \left( 1 - \left. \partial \Re\Sigma / \partial \omega \right\rvert_{\omega=0} \right)^{-1}$.

A metal-to-insulator phase transition occurs in the vicinity of $U/t \sim 5.9$ as the interaction strength is adiabatically increased from $U=0$. At this point, the density of states at the Fermi level collapses leaving a hard gap. This transition in the spectral function is accompanied by the appearance of a pole at zero energy in the self-energy, which is referred to as the `Mott pole'. The self-energy of the Mott insulating phase can be written as
\begin{equation}
	\Sigma(z) = \frac{\alpha}{z} + \tilde{\Sigma}(z)
\end{equation}
where $\alpha$ is the weight of the Mott pole and $\tilde{\Sigma}(z)$ represents the portion of the self-energy not including the pole. The pole weight is determined by~\cite{bullahubbard}
\begin{equation}
	\frac{1}{\alpha} = \int \d\omega \frac{\mathcal{A}(\omega)}{\omega^2} \,.
\end{equation}
The appearance of the zero energy Mott pole can be interpreted qualitatively as the result of the merging of the two peaks of the self-energy in the metallic phase. As mentioned above the two peaks are located at $\omega \sim \pm \sqrt{Z}$. The quasiparticle weight at the Fermi level $Z$ decreases with increasing $U$, and vanishes at the metal-insulator phase transition. 
\begin{figure}[htp!]
\begin{subfigure}{0.49\linewidth}
	\includegraphics[scale=1]{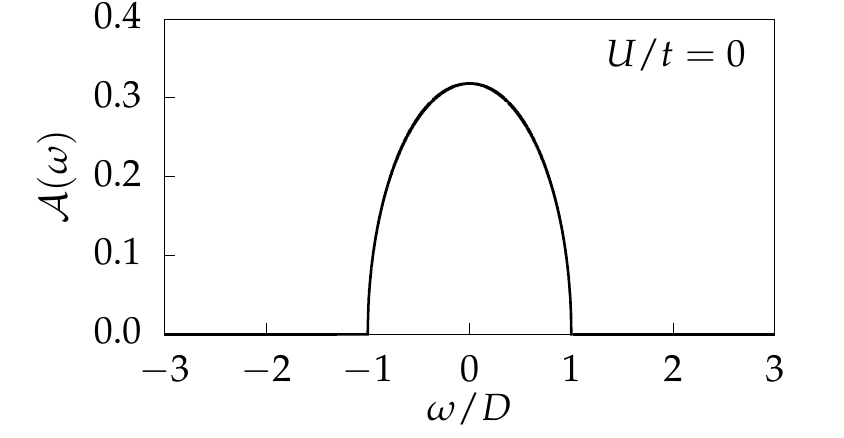}
\end{subfigure}
\hfill
\begin{subfigure}{0.49\linewidth}
	\includegraphics[scale=1]{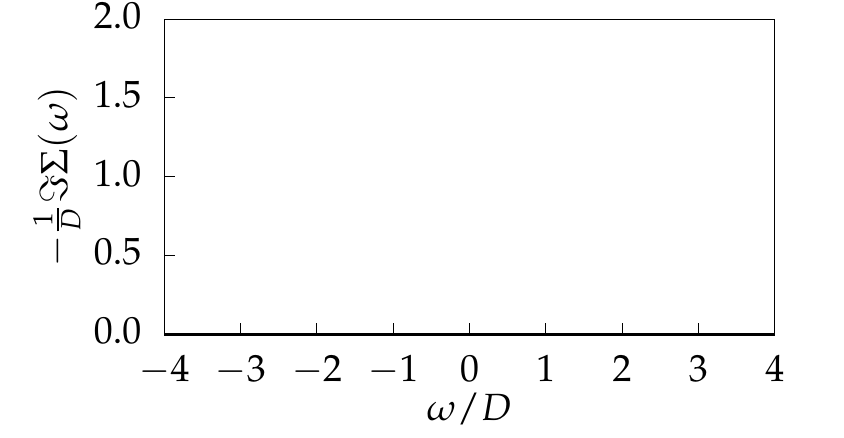}
\end{subfigure}
\begin{subfigure}{0.49\linewidth}
	\includegraphics[scale=1]{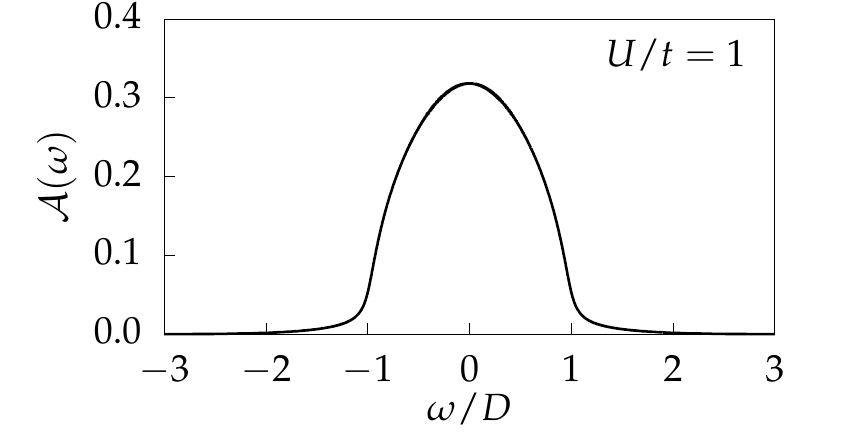}
\end{subfigure}
\hfill
\begin{subfigure}{0.49\linewidth}
	\includegraphics[scale=1]{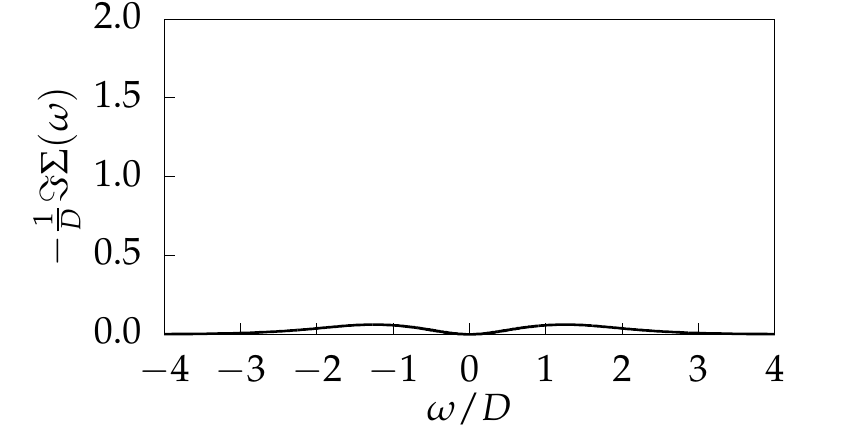}
\end{subfigure}
\begin{subfigure}{0.49\linewidth}
	\includegraphics[scale=1]{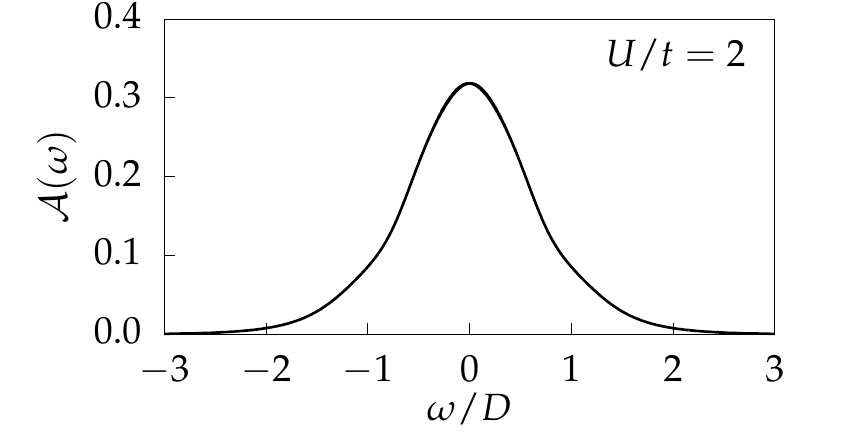}
\end{subfigure}
\hfill
\begin{subfigure}{0.49\linewidth}
	\includegraphics[scale=1]{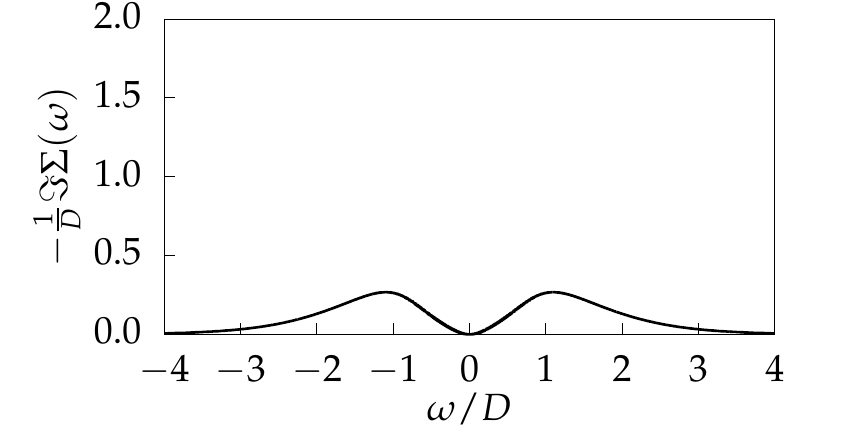}
\end{subfigure}
\begin{subfigure}{0.49\linewidth}
	\includegraphics[scale=1]{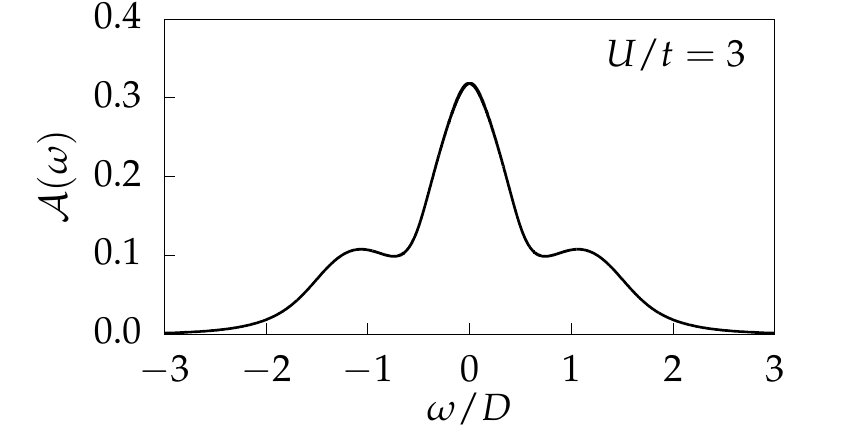}
\end{subfigure}
\hfill
\begin{subfigure}{0.49\linewidth}
	\includegraphics[scale=1]{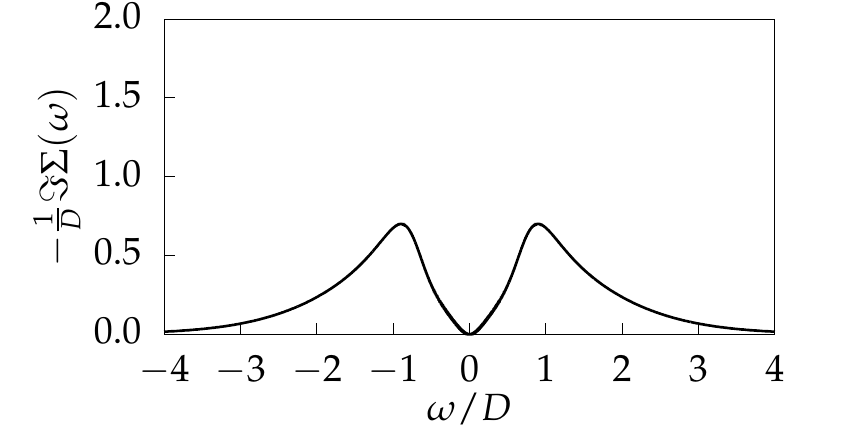}
\end{subfigure}
\begin{subfigure}{0.49\linewidth}
	\includegraphics[scale=1]{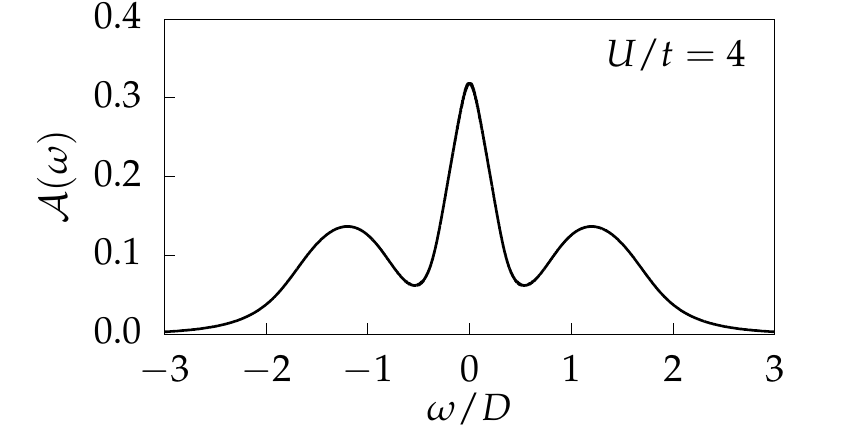}
\end{subfigure}
\hfill
\begin{subfigure}{0.49\linewidth}
	\includegraphics[scale=1]{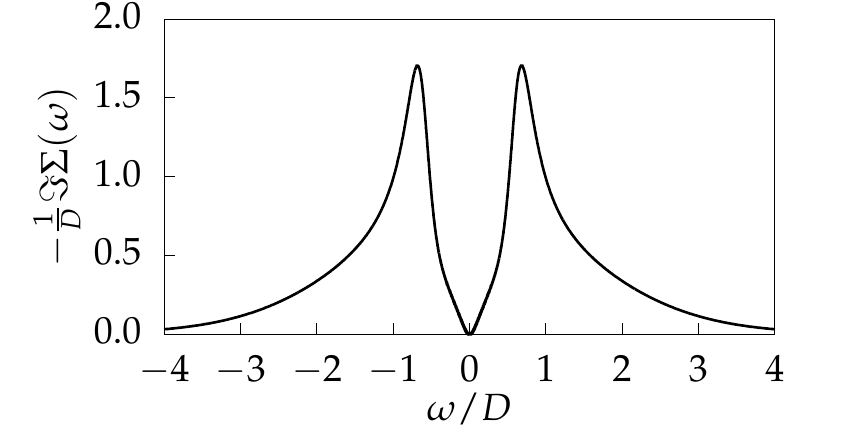}
\end{subfigure}
\caption[DMFT solution to the Hubbard model]{DMFT solution to the Hubbard model on the Bethe lattice with bandwidth $D=2$\label{fig:hubbardsolution}}
\end{figure}
\begin{figure}[htp!]\ContinuedFloat
\begin{subfigure}{0.49\linewidth}
	\includegraphics[scale=1]{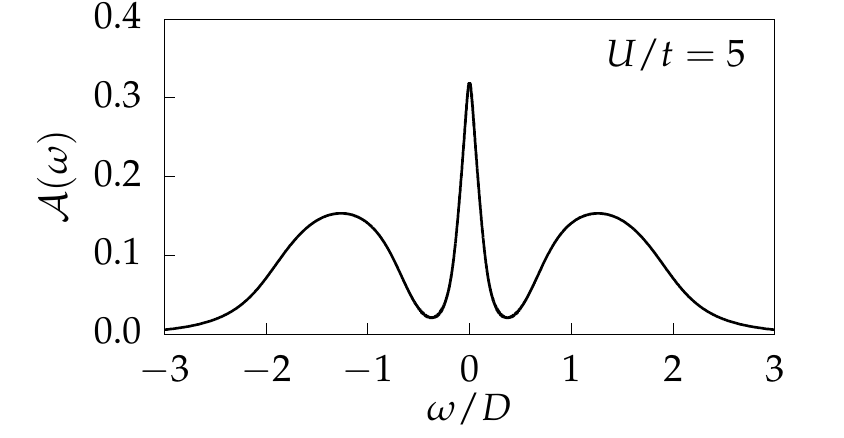}
\end{subfigure}
\hfill
\begin{subfigure}{0.49\linewidth}
	\includegraphics[scale=1]{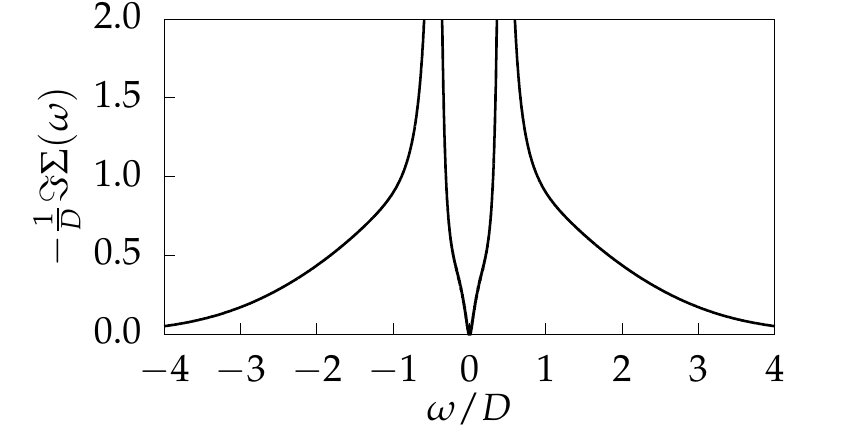}
\end{subfigure}
\begin{subfigure}{0.49\linewidth}
	\includegraphics[scale=1]{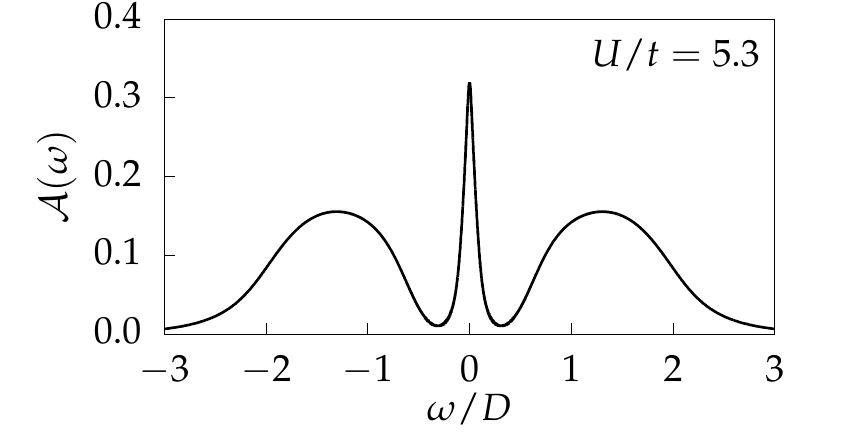}
\end{subfigure}
\hfill
\begin{subfigure}{0.49\linewidth}
	\includegraphics[scale=1]{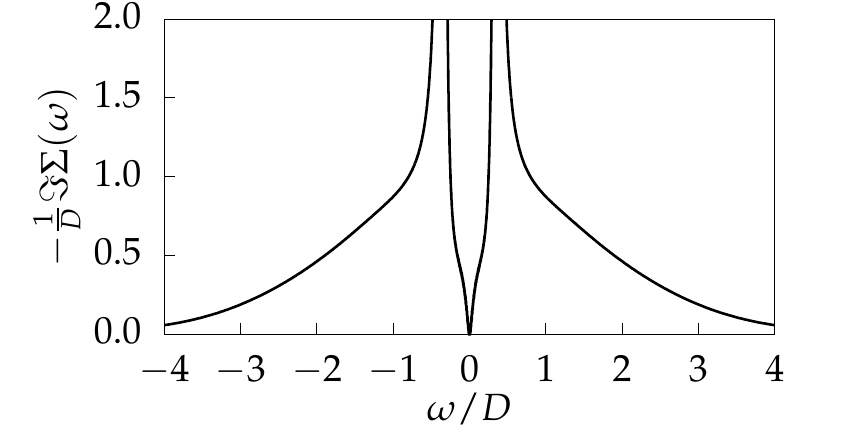}
\end{subfigure}
\begin{subfigure}{0.49\linewidth}
	\includegraphics[scale=1]{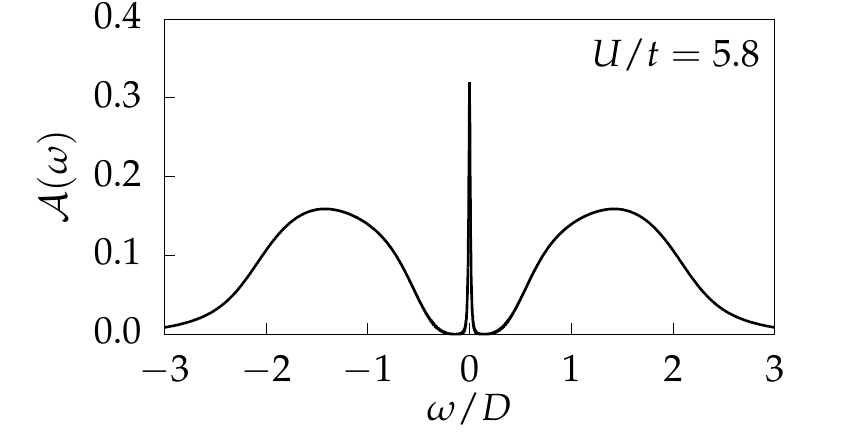}
\end{subfigure}
\hfill
\begin{subfigure}{0.49\linewidth}
	\includegraphics[scale=1]{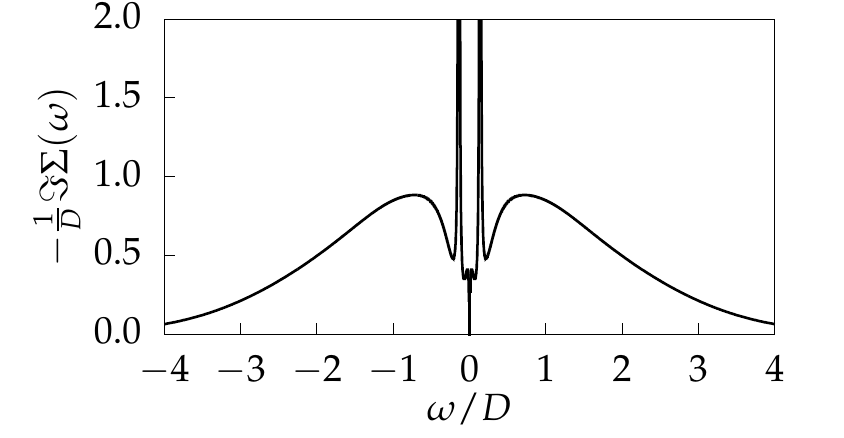}
\end{subfigure}
\begin{subfigure}{0.49\linewidth}
	\includegraphics[scale=1]{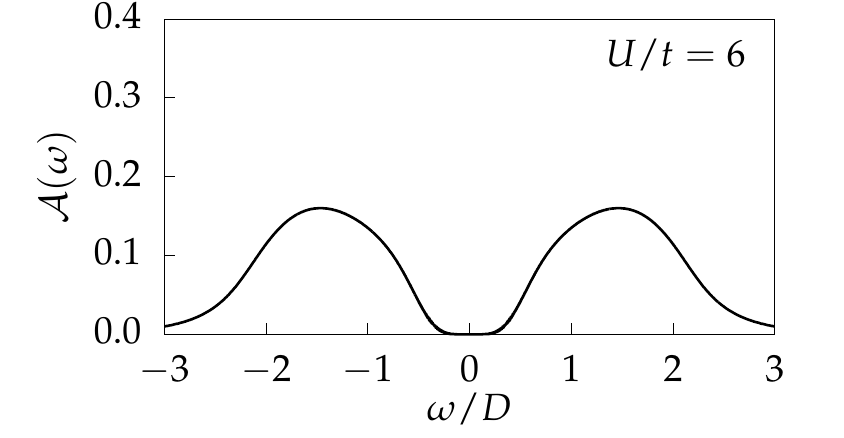}
\end{subfigure}
\hfill
\begin{subfigure}{0.49\linewidth}
	\includegraphics[scale=1]{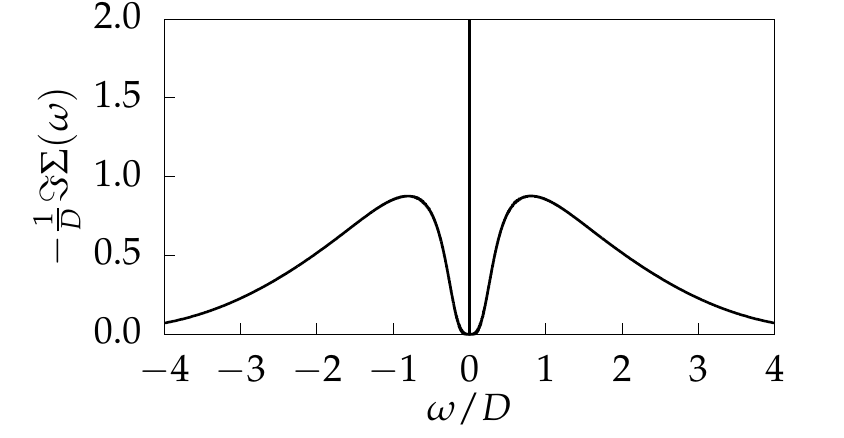}
\end{subfigure}
\begin{subfigure}{0.49\linewidth}
	\includegraphics[scale=1]{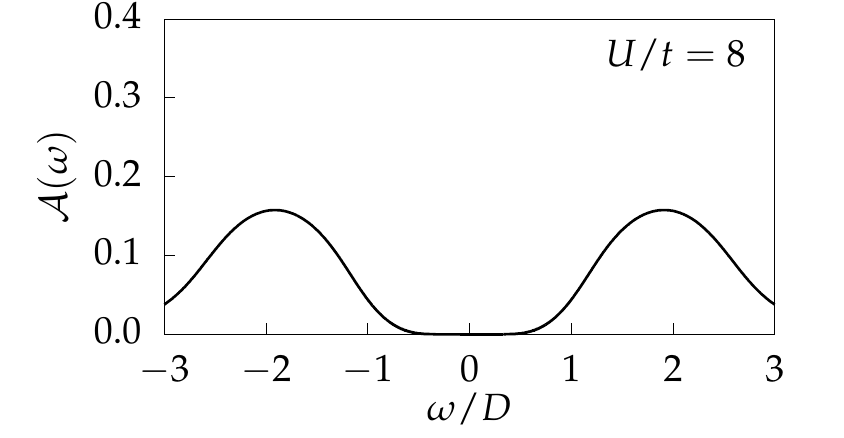}
\end{subfigure}
\hfill
\begin{subfigure}{0.49\linewidth}
	\includegraphics[scale=1]{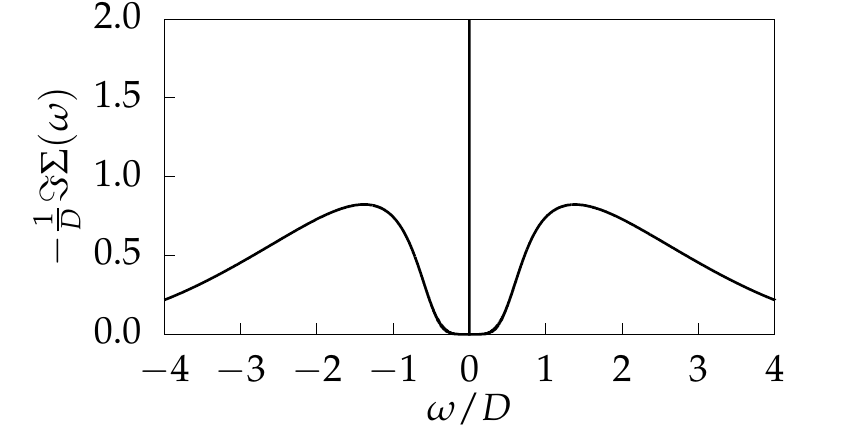}
\end{subfigure}
\caption[DMFT solution to the Hubbard model on the Bethe lattice]{DMFT solution to the Hubbard model on the Bethe lattice with bandwidth $D/t=2$. Note the collapse of the density of states at the Fermi level and the simultaneous appearance of a pole at zero energy in the self-energy at $U/t \sim 5.9$, indicative of the Mott metal-insulator transition.}
\end{figure}

The large $U$ insulating regime where $U \gg t$ can be considered as the many-body extension of the Hubbard atom case analyzed in \S\ref{sec:hubbardatomgf}. Rather than two poles, the spectral function features two bands centered around $\pm\frac{U}{2}$ whose broadened width is due to the many-body interactions generated by weak intersite tunneling $t$.

In addition to the metal-to-insulator phase transition, there also exists an insulator-to-metal phase transition which occurs at a different $U$ than the metal-to-insulator transition. This critical interaction strength is denoted $U_{c1}$ with the metal-to-insulator critical interaction strength termed $U_{c2}$ as $U_{c1} < U_{c2}$. The DMFT solution of the Hubbard model on the Bethe lattice in infinite dimensions reveals that $U_{c1}/t \approx 4.6$ and $U_{c2}/t \approx 5.9$. The region where $U_{c1} < U < U_{c2}$ is the hysteresis region shown in Fig.~\ref{fig:hubbardphasediagram}.
Within the hysteresis region, the system with interaction strength tuned down adiabatically from $U > U_{c2}$ can be described as having interaction strength $U_+$. Conversely, the system with interaction strength tuned up adiabatically from $U < U_{c1}$ can be described as having interaction strength $U_-$.
\begin{figure}[h]
\begin{subfigure}{0.49\linewidth}
	\includegraphics[scale=1]{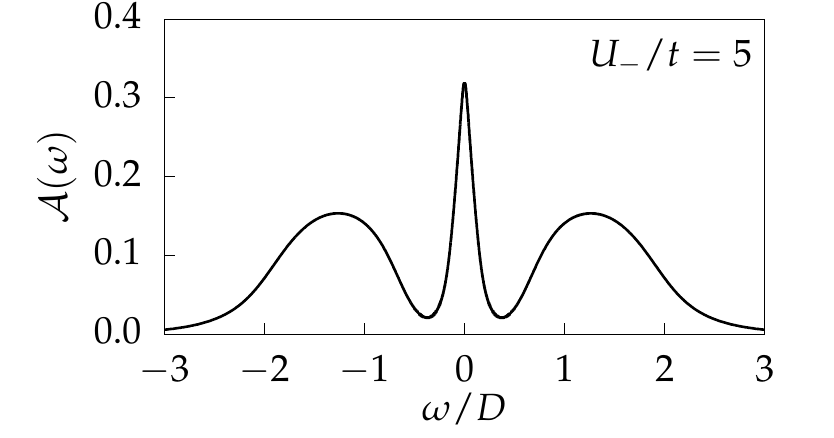}
\end{subfigure}
\hfill
\begin{subfigure}{0.49\linewidth}
	\includegraphics[scale=1]{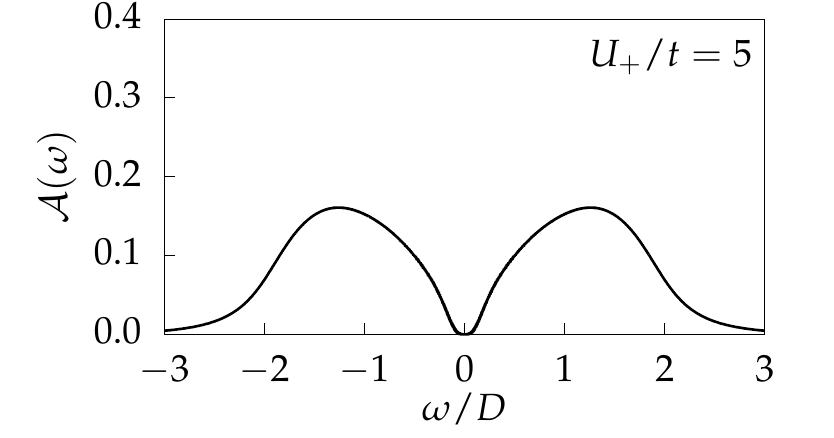}
\end{subfigure}
\caption[Coexistence hysteresis region of the Hubbard model]{Comparison between the spectral function of the Hubbard model at $U/t = 5$ with $D/t=2$ approaching from the metallic phase (left) and from the insulating phase (right). This demonstrates the hysteresis coexistence phase of the Hubbard model.\label{fig:hubbardcoexist}}
\end{figure}
Within this region, when the system is adiabatically tuned to this interaction strength from below ($U_-$), the system is metallic. When the system is adiabatically tuned to this interaction strength from above ($U_+$), it is a Mott insulator. An example of the system in this coexistence region is depicted in Fig.~\ref{fig:hubbardcoexist}. As demonstrated by the sequence in Fig~\ref{fig:hubbardsolution}, as the interaction strength is increased, the outer satellite bands move progressively further away from the Fermi level. Conversely, as the interaction strength is decreased the bands converge towards the Fermi energy. The critical point $U_{c1}$ occurs when the two bands touch.

%%%

Although the DMFT solution is exact only in the infinite dimensional limit, it is reasonably able to capture the physics of some examples of real materials. In particular, the phase diagram of $\mathrm{V}_2\mathrm{O}_3$ is well captured by the DMFT solution to the Hubbard model~\cite{dmft}.

%%%%%%%%%%%%%%%%%%%%%%%%%%%%%%%%%%%%%%%%%%%%%%%%%%%%%%%%%%%%%%%%%%%%%%%%%%%%
%%%%%%%%%%%%%%%%%%%%%%%%%%%%%%%%%%%%%%%%%%%%%%%%%%%%%%%%%%%%%%%%%%%%%%%%%%%%
\section{Topology}\label{sec:topology}

The use of topology in condensed matter physics has been steadily growing over the past few decades~\cite{mermin,avronseilersimon,volovik,horava,hk,fruchartcarpentier,shortcourse,altlandsimons}.
Before turning to its manifestations in physics, presented here first will be a discussion of the mathematical form of topology which appears in the contexts of interest~\cite{nakahara,baezmuniain,eguchi,frankel,schutz}.
The primary element taken into account is whether a certain system is topologically trivial or non-trivial. The triviality or non-triviality of topology can be understood as the difference between the torus $\mathbbm{T}^2$, which is topologically trivial, and the sphere $\mathbbm{S}^2$, which is topologically non-trivial.

The topology of topological materials generally refers to non-trivial topology of their momentum space. For example, a non-topological $2d$ system has a first Brillouin zone which takes the form of a 2-torus $\mathbbm{T}^2$.\footnote{The momenta in the Brillouin zone are $2\pi$ periodic in $k_x$ and $k_y$, which leads to this space taking the form of a torus.} The momentum space is the tangent vector field over this space.

The notion of what is meant by topology can be illustrated in the difference between the M\"obius band and the cylinder, as shown in~Fig.\ref{moebius}.
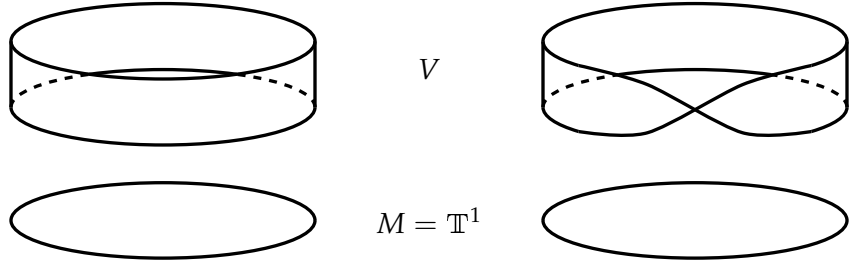
\begin{figure}[h]
\centering
\begin{tikzpicture}[scale=0.5,line width=1.25]
\coordinate (C) at (-3,0);
\coordinate (M) at (3,0);
\node[] at (0,-3) {$M = \mathbbm{T}^1$};
\node[] at (0,1) {$V$};
\node (Cmid) at ($(C)+(-4,1)$) {};
%\draw[dashed] (C) arc (0:180:4cm and 1cm);
\draw (C) arc (180:0:-4cm and -1cm);
\draw[dashed] (C) arc (0:61:4cm and 1cm);
\draw[dashed] ($(C)+(-8,0)$) arc (0:61:-4cm and 1cm);
\draw (Cmid) arc (90:119:4cm and 1cm) coordinate (int);
\draw (Cmid) arc (90:119:-4cm and 1cm);
\draw (C)--($(C)+(0,1.75)$);
\draw ($(C)+(-8,0)$)--($(C)+(-8,1.75)$);
\draw[black] ($(C)+(0,1.75)$) arc (360:0:4cm and -1cm);
\draw ($(C)+(0,-3)$) arc (360:0:4cm and -1cm);
%\node[left] at ($(C)+(-8,-3)$) {$\mathbbm{T}^1$};
%%%
%%%
%%%
\coordinate (Mcent) at ($(M)+(4,0.5)$);
\node (Mmid) at ($(M)+(4,1)$) {};
\draw ($(M)+(8,0)$)--($(M)+(8,1.75)$);
%\draw[dashed] (M) arc (0:180:-4cm and 1cm);
\draw[dashed,black] (M) arc (0:60:-4cm and 1cm);
\draw[dashed] ($(M)+(8,0)$) arc (0:60:4cm and 1cm);
\draw[black] (Mmid) arc (90:120:4cm and 1cm);
\draw (Mmid) arc (90:120:-4cm and 1cm);
\draw[black] (M) arc (180:140:4cm and -1cm) coordinate (bl);
\draw ($(M)+(8,0)$) arc (0:40:4cm and -1cm) coordinate (br);
\draw ($(M)+(0,1.75)$) arc (0:90:-4cm and 1cm);
\draw[black] ($(M)+(8,1.75)$) arc (0:90:4cm and 1cm);
\draw ($(M)+(0,1.75)$) arc (180:140:4cm and -1cm) coordinate (tl);
\draw[black] ($(M)+(8,1.75)$) arc (0:40:4cm and -1cm) coordinate (tr);
\draw[black] plot[smooth] coordinates { (tr) ($(Mcent)+(1.25,0.1)$) ($(Mcent)+(-1.25,-1.175)$) (bl)};
\draw plot[smooth] coordinates { (tl) ($(Mcent)+(-1.25,0.1)$) ($(Mcent)+(1.25,-1.175)$) (br)};
\draw (M)--($(M)+(0,1.75)$);
\draw ($(M)+(0,-3)$) arc (360:0:-4cm and -1cm);
%\node[right] at ($(M)+(8,-3)$) {$\mathbbm{T}^1$};
\end{tikzpicture}
\caption[Cylinder vs. M\"obius band]{Comparison between the topologies of the cylinder (left) and M\"obius band (right). Locally, both have the structure of $\mathbbm{T}^1\times{V}$, but their global topologies differ.\label{moebius}}
\end{figure}
The difference in topologies of the cylinder and M\"obius band may be quantified by analyzing them through the language of fiber bundles\index{fiber bundle}. A fiber bundle is a triplet $\{E,M,\pi\}$ consisting of a total space $E$, base space $M$, and a projection map $\pi : E \to M$. The total space is comprised of fibers $E_p$, $E = \bigcup_p E_p$, defined as $E_p = \{ q \in E | \pi(q) = p ,\ \forall p \in M \}$.
A fiber bundle is termed trivial if the total space takes the form of a direct product of the fiber with the base manifold $E = M \times V$.
In general fiber bundles are only locally trivial, meaning that they are trivial only over a
local coordinate neighborhood $U_{i} \subset M$. In the overlap region of neighborhoods
$U_{\alpha} \cap U_{\beta}$
it is necessary to define a 
transition function $g_{\alpha\beta}$
which appropriately identify local trivializations over each neighborhood with each other.

The cylinder and M\"obius band can be described by a fiber bundle whose fibers are a 1-dimensional vector space $V$ modeled\footnote{Since $x \mapsto -x$ is not a symmetry of $\mathbbm{R}$ the fiber $V$ cannot be taken to be $\mathbbm{R}$ itself, but is instead said to be ``modeled'' on $\mathbbm{R}$~\cite{penrose}.} on the real line $\mathbbm{R}$.
An example of a fiber for the cylinder and M\"obius band is the line segment $V = [-1,1]$. 

An important class of fiber bundle which frequently appears in physics is a $G$-bundle, where the transition functions are elements of a group $G$,
$g_{\alpha\beta} \in G$.
For the M\"obius band $G$ can be taken to be the (multiplicative) group $\mathbbm{Z}_2$ which means that the transition functions take values $g \in \{-1,1\}$.
A $G$-bundle whose fibers are $G$ itself is called a principal-$G$ bundle.
A M\"obius bundle where the fiber is taken to be the boundary of a finite M\"obius band (\textit{i.e.} a fiber consisting of the points $\{-1,1\}$) can be described as a principal-$\mathbbm{Z}_2$ bundle.

%For a point $p \in U_{\alpha} \cap U_{\beta}$

The distinction between the cylinder and the M\"obius band can be obtained quantitatively from the transition functions for a point $p \in U_{\alpha} \cap U_{\beta}$. In the cylinder, the transition functions can map elements of a fiber over overlapping neighborhoods trivially. On the M\"obius band in the presence of the twist, the transition functions perform the map
$U_{\alpha}\times[-1,1] \to U_{\beta}\times[1,-1]$, which indicates the non-trivial topology of the fiber bundle.

A fiber bundle whose fibers have the structure of a Hilbert space is termed a Hilbert bundle.
A fiber bundle which is important in condensed matter physics is the Bloch bundle, which is the direct sum of all momentum space Hilbert bundles belonging to bands below the Fermi level of a given system, $\displaystyle \bigoplus_{n<n_F} \mathcal{H}_n$. Topological band insulators can be identified mathematically from the appearance of non-trivial topology in their Bloch bundle. The total Bloch bundle may be defined as the direct sum over all bands of the system. The total Bloch bundle is always trivial~\cite{fruchartcarpentier}.

An important element in the construction of fiber bundles is the connection. The concept of a connection on a fiber bundle is that it facilitates the comparison between points on different fibers, which is for example needed in the operation of differentiation on sections of a bundle. A section of a fiber bundle over $p \in M$ is defined as $\pi^{-1}(p)$, with the total section being $\bigcup_p \pi^{-1}(p)$.
A classical field in physics can generically be considered mathematically as a section of some vector bundle (a fiber bundle whose fibers are some vector space). Since fiber bundle sections are ubiquitous in physics, the need for a connection becomes apparent. A connection can be understood geometrically as follows: For a vector field $\vec{v} = v^j \hat{e}_j$ with components $v^j$ in basis $\hat{e}_j$ parameterized by the path $\vec{\lambda}$, the derivative of $\vec{v}$ along the component $\lambda^k$ is
\begin{align*}
	\frac{\d \vec{v}}{\d \lambda^k} &= \frac{\d v^j}{\d \lambda^k} \hat{e}_j + v^j \frac{\d \hat{e}_j}{\d \lambda^k} \,, 
\intertext{whose $i^{\text{th}}$ component is}
	\left( \frac{\d \vec{v}}{\d \lambda^k} \right)^i &= \frac{\d v^i}{\d \lambda^k} + v^j \frac{\d \hat{e}_j}{\d \lambda^k} \cdot \hat{e}^i = \frac{\d v^i}{\d \lambda^k} + v^j \tensor*{\Gamma}{^i_{jk}}
\end{align*}
where $\tensor*{\Gamma}{^i_{jk}}$ is the infamous Levi-Civia connection (Christoffel symbol). This illustrates how the connection measures the change of a local coordinate basis along a parameterized path. The non-uniformity of the basis along the path is analogous to the fiber bundle picture where there is no canonical isomorphism between different fibers.
On the Hilbert bundle, a natural connection is the Berry connection, $A$, which is $U(1)$ connection 1-form $A = -\i \left\langle \psi | \d \psi \right\rangle$~\cite{berry,fruchartcarpentier,nakahara}. The $U(1)$ arises from that being the group of transition functions on the Hilbert bundle.

The topology of the M\"obius band can also be determined through its winding number, the closed loop integral of the connection on the band over the circle
\begin{equation}
	\gamma = \oint_{\Gamma} A ,
\end{equation}
which measures the number of twists in the band. In the context of physics, $\gamma$ can be related to the geometric phases of Aharonov and Bohm~\cite{abphase,baezmuniain}, where $A$ is the electromagnetic vector potential, or aforementioned Berry connection.

A measure of topology for a two dimensional manifold $M$ is the first Chern number
\begin{equation}
	\mathrm{Ch}_1 = \frac{1}{2\pi} \iint_M F
\label{chern}
\end{equation}
which is the integral of the curvature 2-form, $F$, on $M$. This number takes on integer values due to the Gau\ss-Bonnet theorem~\cite{frankel,nakahara}.
For a quantum system, the $F$ can be given by the Berry curvature, which is obtained from the Berry connection as
\begin{equation}
\begin{aligned}
	F &= \d A 
	\\&= -\i \left\langle \d \psi | \extp | \d \psi \right\rangle \,,
\end{aligned}
\end{equation}
which follows from applying Stokes' Theorem to the expression for the winding number $\gamma$~\cite{fruchartcarpentier}.
For the connection of a principal $G$-bundle, the curvature takes values in the adjoint representation of the algebra $\mathfrak{g}$ associated to $G$~\cite{frankel,baezmuniain,nakahara}. This discussion of fiber bundles and connections and curvature on bundles has been rather brief and informal; further details may be found in the standard references~\cite{frankel,nakahara,eguchi}, with Ref.~\cite{fruchartcarpentier} of being particular relevance to modern condensed matter physics.

%The first Chern number is equivalent to the Berry phase by means of the generalized Stokes theorem.
In the case of topological insulators, the integral \eqref{chern} is taken over the first Brillouin zone of the valance band. The \index{Chern number}Chern number gives a measure on how topologically non-trivial the valance band Hilbert space is. A valance band with non-trivial topology results in the appearance of topological edge modes in the spectrum.
In condensed matter physics, one of the first recognized instances of a topological state was the quantized Hall conductance of the integer quantum Hall effect~\cite{tknn,niuthoulesswu,avronseilersimon}. In this situation, it was found that the quantized Hall conductance, as computed from the Kubo formula, was proportional to $\mathrm{Ch}_1$. 

%A feature of topological properties is their robustness to perturbations in parameters. 
The winding number and the Chern number are examples of a 
topological invariant, a quantity which captures the global topology of the system and is robust to perturbations in its parameters. For the M\"obius band, the characterizing twist can only be added or removed through a destructive process, such as cutting and re-gluing the band.
For the Chern number, it can be shown that the variation of the functional \eqref{chern} vanishes.

A characteristic of topological insulators is the ``bulk-boundary correspondence''\index{bulk-boundary correspondence} wherein features of the boundary can be determined by properties of the bulk and vice-versa. The presence of topological states on the boundary of a topological insulator can be determined \textit{e.g.} by the Chern number of the bulk. In the case of the integer quantum Hall effect, the Chern number is calculated with respect to the band structure of the (insulating) bulk of the system, but this in turn yields information about the quantized conductance on the boundary of the system.

It should be noted that the aspects of topology discussed here and applied to topological phases and topological states of quantum condensed matter in this thesis are categorically distinct from what are known as topologically \textit{ordered}\index{topological order} phases of matter in the literature~\cite{wen}. A distinguishing characteristic between the two are that topological states exhibit short range ground state entanglement whereas topologically ordered states exhibit long range ground state entanglement. In some cases, such as the integer quantum Hall effect, there is an overlap between the two, but in general they are separate and distinct concepts.

%%%%%%%%%%%%%%%%%%%%%%%%%%%%%%%%%%%%%%%%%%%%
%%%%%%%%%%%%%%%%%%%%%%%%%%%%%%%%%%%%%%%%%%%%
\section{Su-Schrieffer-Heeger Model\label{sec:sshmodel}}
%%%%%%%%%%%%%%%%%%%%%%%%%%%%%%%%%%%%%%%%%%%%
%%%%%%%%%%%%%%%%%%%%%%%%%%%%%%%%%%%%%%%%%%%%

%\cite{hk}
%\cite{tknn,avronseilersimon,niuthoulesswu}
%\cite{horava}

%%%%%%%%%

An elementary condensed matter model which exhibits topological characteristics is the Su--Schrieffer--Heeger (SSH) model~\cite{ssh,solitons,shortcourse}\index{SSH model}. 
The SSH model is a $1d$ non-interacting single-particle tight-binding model of spinless fermions which exhibits the characteristics of a topological insulator.
Originally devised to model \textit{trans}-polyacytlene (\textit{trans}-(CH)$_\text{x}$) as a prototypical system of conducting polymers
and
$\pi$-conjugated polymers in general,
it is now known to be an example of a $1d$ topological insulator~\cite{shortcourse,altlandsimons,marino}. 

In the tight-binding formalism the Hamiltonian of the SSH model is given by
\begin{equation}
	\hat{H}_{\textsc{ssh}}
	=
	\sum_{j\in\mathbbm{Z}^+} \left[ \varepsilon\, \opd{c}{j} \op{c}{j} + \left( \tensor{t}{_A} \opd{c}{2j-1} \op{c}{2j} + \tensor{t}{_B} \opd{c}{2j} \op{c}{2j+1} + \textsc{h.}\text{c.} \right) \right]
\end{equation}
with on-site energies $\varepsilon$ and tunneling amplitudes $t_A$ and $t_B$.
These amplitudes can alternatively be expressed as $t_A = t_0 - \delta t$ and $t_B = t_0 + \delta t$ where $4 t_0$ is the full band width and $4 \delta t$ is the band gap. In the original model of polyacetylene the $\pm\delta t$ originates from the Peierls mechanism of electron-phonon coupling, but here $\delta t$ is taken to be a fundamental parameter of the theory.
The retarded Green function on the boundary is easily obtained from the equation of motion method and takes the continued fraction form
\begin{equation}
	G_{1,1}(z) = \cfrac{1}{z - \varepsilon - \cfrac{t_A^2}{z - \varepsilon - t_B^2 G_{1,1}(z)}}
\end{equation}
from which the explicit expression follows as
\begin{equation}
	G_{1,1}(z) = \frac{(z-\varepsilon)^2 + t_B^2 - t_A^2 + \sqrt{((z-\varepsilon)^2 - t_A^2 + t_B^2)^2 - 4 (z-\varepsilon)^2 t_B^2}}{2 (z-\varepsilon) t_B^2} \,.
\label{eq:sshgreen}
\end{equation}
In the following the local potential $\varepsilon$ is normalized to $\varepsilon = 0$ as the SSH spectrum is symmetric about $\omega = \varepsilon$.
%The Hilbert space of the SSH model is $\mathcal{H} = \mathbbm{C}^{2} \otimes \mathbbm{C}^{\infty} \otimes \ell^2(\mathbbm{Z})$.
%Enforcing chiral symmetry results in the requirement that $\varepsilon_j = 0$, $\forall j$.

The SSH model's simplicity has enabled it to be realized in a variety of experimental setups, such as in
photonic systems~\cite{zakobs},
mechanical acoustic lattices~\cite{acoustic,acousticrev},
and
graphene nanoribbons~\cite{gnano1,gnano2}.
The simplicity of the SSH model also makes it a good starting point for generalizations to examine topology in more sophisticated examples. 
Generalizations of the SSH model previously considered in the literature include the addition of long-range kinetic terms~\cite{extendedrm,longrange2}
and non-Hermitian generalizations~\cite{nonherm}.
The SSH model has also previously been generalized into a tripartite form~\cite{trimer} as well as to a four dimensional unit cell of the SSH$_4$ model~\cite{ssh4}, which has been also realized in ultracold atom experiments~\cite{ssh4exp}. 
Other work has involved generalizing the SSH model to quasi-$1d$ systems, such as SSH ladders~\cite{sshladder}.
Additional generalizations of the SSH model are contained in \S\ref{ch:genssh}.

%Topological insulators typically require the presence of a symmetry which protects the topological state. This symmetry for conventional topological insulators is a subset of time reversal symmetry, charge conjugation symmetry, and chiral symmetry.

A separate type of topological insulator is that of a crystalline topological insulator, which differs from the conventional formalism as the protecting symmetry group is the spatial symmetry group of the system's lattice.
The chiral symmetry of the SSH model can be interpreted as an inversion symmetry on its lattice, and therefore the SSH model can also be used as a prototypical crystalline topological insulator. Similarly, the SSH model can be used as a basis for generalization to higher-order topological insulators~\cite{hoti}.

%The retarded Green function on the boundary site can be calculated via equations of motion techniques
%~\cite{zubarev}
%\begin{equation}
%	z \Green{\hat{A}}{\hat{B}}_z = \big\langle \{\hat{A} , \hat{B}\} \big\rangle + \Green{\hat{A}}{[\hat{H},\hat{B}]}_z
%\end{equation}
%to yield
%\begin{equation}
%	\Green{\op{c}{1}}{\opd{c}{1}}_z
%	\equiv
%	G_{1,1}(z)
%	=	\frac{z^2 + t_B^2 - t_A^2 \pm \sqrt{(z^2 - t_A^2 + t_B^2)^2 - 4 z^2 t_B^2}}{2 z t_B^2}
%\label{eq:sshgreen}
%\end{equation}

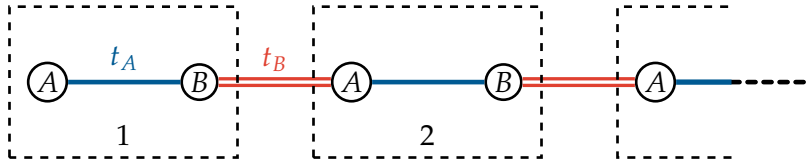
\begin{figure}[h]
\centering
\begin{tikzpicture}[scale=1,site/.style={circle,draw=black,line width=1pt,font=\normalsize,inner sep=1pt}]
	\node[site] (1) at (0,0){$A$};
	\node[site] (2) at ($(1)+(2,0)$){$B$};
	\node[site] (3) at ($(2)+(2,0)$){$A$};
	\node[site] (4) at ($(3)+(2,0)$){$B$};
	\node[site] (5) at ($(4)+(2,0)$){$A$};
	\draw[line width=1.75pt, dashed, line cap=round]($(5)+(1,0)$)--+(1,0);
	\draw[line width=1.75pt,color=blue](1)--(2) node[midway,above,blue] {${t_A}$};
	\draw[line width=1.25pt,color=red,double,double distance=1.25pt](2)--(3) node[midway,above,red] {${t_B}$};
	\draw[line width=1.75pt,color=blue](3)--(4);
	\draw[line width=1.25pt,color=red,double,double distance=1.25pt](4)--(5);
	\draw[line width=2pt,color=blue](5)--+(1,0);
	\draw[dashed,line width=1pt] ($(1)+(-0.5,-1)$)rectangle($(2)+(0.5,1)$) node[midway,below=12pt] {$1$};
	\draw[dashed,line width=1pt] ($(3)+(-0.5,-1)$)rectangle($(4)+(0.5,1)$) node[midway,below=12pt] {$2$};
	\draw[dashed,line width=1pt] ($(5)+(1,-1)$)--($(5)+(-0.5,-1)$)--($(5)+(-0.5,1)$)--($(5)+(1,1)$);
\end{tikzpicture}
\caption[Unit cell grouping of the SSH model]{Unit cell grouping of the SSH model. In this picture $t_A$ represents hopping within the unit cell and $t_B$ represents hopping between unit cells.\label{fig:sshunitcell}}
\end{figure}
It is also instructive to express the Hamiltonian in terms of its unit cells as shown in Fig.~\ref{fig:sshunitcell}. In this form, the Hamiltonian is recast as
\begin{equation}
	\hat{H}_{\textsc{ssh}}
	=	t_A \sum_{\mu\in\mathbbm{Z}^+} \opd{\chi}{\mu} \boldsymbol{\sigma}_1 \op{\chi}{\mu}
		+ t_B \sum_{\mu\in\mathbbm{Z}^+} \left( \opd{\chi}{\mu} \frac{\boldsymbol{\sigma}_1 - \i \boldsymbol{\sigma}_2}{2} \op{\chi}{\mu+1} + \opd{\chi}{\mu+1} \frac{\boldsymbol{\sigma}_1 + \i \boldsymbol{\sigma}_2}{2} \op{\chi}{\mu} \right)
\end{equation}
where $\mu$ labels the unit cell and $\frac12\boldsymbol{\sigma}_a$ are the fundamental representation of $SU(2)$.\footnote{Conventionally also known as the Pauli matrices: $\displaystyle \boldsymbol{\sigma}_1=\begin{pmatrix}0&1\\1&0\end{pmatrix}\,, \boldsymbol{\sigma}_2=\begin{pmatrix}0&-\i\\\i&0\end{pmatrix}\,, \boldsymbol{\sigma}_3=\begin{pmatrix}1&0\\0&-1\end{pmatrix}$.} The operators acting on the unit cells are given by
\begin{equation}
	\op{\chi}{\mu} = \begin{pmatrix} \op{c}{A} \\ \op{c}{B} \end{pmatrix}_{\mu}
\end{equation}
where the $\op{c}{A/B}$ act on $A$ or $B$ sites respectively.
With respect to the matrix fast recursion algorithm described in \S\ref{sec:calcmeth}, the elements of the Hamiltonian representing dynamics within and between unit cells are
\begin{align}
	\boldsymbol{h}_0 &= \begin{pmatrix} 0 & t_A \\ t_A & 0 \end{pmatrix}
	&
	\boldsymbol{h}_1 &= \begin{pmatrix} 0 & 0 \\ t_B & 0 \end{pmatrix}
\end{align}
In matrix form, the Green function equation of motion for the SSH model yields
\begin{equation}
	\boldsymbol{G}_{1,1}(z) = \left[ z \mathbbm{1} - t_A \boldsymbol{\sigma}_1 - t_B^2 \boldsymbol{\sigma}_- \boldsymbol{G}_{1,1}(z) \boldsymbol{\sigma}_+ \right]^{-1}
\label{eq:sshgreenmatrix}
\end{equation}
where
\begin{equation}
	\boldsymbol{G}_{\mu,\nu}(z) = \begin{pmatrix} G_{\mu_A \nu_A}(z) & G_{\mu_B \nu_A}(z) \\ G_{\mu_A \nu_B}(z) & G_{\mu_B \nu_B}(z) \end{pmatrix}
\end{equation}
with $\mu$ and $\nu$ indexing unit cells and $A$ and $B$ labeling sites within the unit cell.
The boundary Green function \eqref{eq:sshgreen} can be recovered from Eq.~\eqref{eq:sshgreenmatrix} by taking the $AA$ component
\begin{equation}
	\left[ \boldsymbol{G}_{1,1}(z) \right]_{A,A} = \Green{\op{\chi}{1_A}}{\opd{\chi}{1_A}}_z = \Green{\op{c}{1}}{\opd{c}{1}}_z \,.
\end{equation}

The unit cell representation of the Hamiltonian is also useful for obtaining the momentum space representation.
Performing a Fourier transformation into momentum space yields the Hamiltonian
\begin{equation}
\begin{aligned}
	\hat{H}_{\textsc{ssh}}(k)
	&=	\sum_k \opd{\chi}{k} \begin{pmatrix} 0 & t_A + t_B e^{\i k} \\ t_A + t_B e^{-\i k} & 0 \end{pmatrix} \op{\chi}{k}
	\\
	&=	\sum_k \opd{\chi}{k} \left[  \left( t_A + t_B \cos(k) \right) \boldsymbol{\sigma}_1 - t_B \sin(k) \boldsymbol{\sigma}_2 \strut \right] \op{\chi}{k}
	\\
	&\equiv	\sum_k \opd{\chi}{k}~\boldsymbol{h}_{\textsc{ssh}}(k)~\op{\chi}{k}
\end{aligned}
\label{eq:sshhamkspace}
\end{equation}
The matrix $\boldsymbol{h}_{\textsc{ssh}}(k)$ is an element of the Gra{\ss}mannian $U(2)/(U(1)\times U(1))$
which puts the SSH model in class $A$III in the classification scheme of topological insulators~\cite{tenfoldclassification}.

The group structure of $U(2)/(U(1)\times U(1))$ can be understood in more concrete terms as follows. As demonstrated by Eq.~\eqref{eq:sshhamkspace}, the Hamiltonian matrix $\boldsymbol{h}(k)$ is a $U(2)$ vector (this is clear from the interpretation of the $SU(2)$ matrices $\boldsymbol{\sigma}_a$ as basis unit vectors). The Hamiltonian is invariant under $U(1)\times U(1)$ gauge transformations, which are transformations that take the form of
\begin{equation}
	\boldsymbol{U} = \begin{pmatrix} \e^{-\i n k} & 0 \\ 0 & 1 \end{pmatrix}
\end{equation}
The Hamiltonian is invariant under this gauge transformation
\begin{equation}
	\boldsymbol{U} \boldsymbol{h}_{\text{ssh}}(k) \boldsymbol{U} = \boldsymbol{h}_{\text{ssh}}(k)
\end{equation}

Models of topological insulators are typically classified based on their spatial dimensions and their symmetries according to the Cartan classification scheme of Riemannian symmetric spaces~\cite{cartan1,cartan2,tenfoldclassification}.
This classification is based on the behavior of a Hamiltonian under chiral, time reversal, and charge conjugation symmetries.
The chiral symmetry generator is
\begin{equation}
	\Gamma = \begin{pmatrix} 1 & 0 \\ 0 & -1 \end{pmatrix}
\end{equation}
which action on the SSH Hamiltonian as
\begin{equation}
	\Gamma \boldsymbol{h}(k) \Gamma^{-1} = -\boldsymbol{h}(k) \,.
\end{equation}
Time reversal symmetry is generated by complex conjugation $T = \mathcal{K}$. The SSH model transforms under this operation as
\begin{equation}
	T \boldsymbol{h}(k) T^{-1} = \boldsymbol{h}(k)^* = \boldsymbol{h}(-k) \,.
\end{equation}
Charge conjugation symmetry can be constructed as a composition of time-reversal and chiral symmetries as $C = T\circ\Gamma$, with the SSH model transforming as
\begin{equation}
	C \boldsymbol{h}(k) C^{-1} = -\boldsymbol{h}(-k) \,.
\end{equation}
%
%The SSH model also possesses an inversion, or isopspin, symmetry, which  $\opd{c}{A} \leftrightarrow \opd{c}{B}$
%\begin{equation}
%	\mathcal{I} = \begin{pmatrix} 0 & 1 \\ 1 & 0 \end{pmatrix}
%\end{equation}
%\begin{equation}
%	\mathcal{I} \boldsymbol{h}(k) \mathcal{I} = \boldsymbol{h}(-k)
%\end{equation}

The momentum space Hamiltonian Eq.~\eqref{eq:sshhamkspace} yields the Bloch energy bands of $\boldsymbol{h}_{\textsc{ssh}}(k)$ as
\begin{equation}
	E_\pm(k)	=	\pm \sqrt{t_A^2 + t_B^2 + 2 t_A t_B \cos(k)} \,.
\end{equation}
Reparameterizing the hopping amplitudes as $t_A = t_0 - \delta t$ and $t_B = t_0 + \delta t$,
the expression under the radical can be rewritten as $2 \left[ t_0^2 + \delta t^2 + (t_0^2 - \delta t^2) \cos(k) \right]$. From here it is clear that the spectrum is gapped unless $\delta t = 0$, where the band crossing occurs at $k = \pi/\mathbbm{Z}$.

The topology of the SSH model can be measured by the winding of the Zak phase~\cite{berry,zak}\index{Zak phase} around the first Brillouin zone
\begin{equation}
	\gamma	=	\frac\i\pi \oint_{\textsc{bz}} \langle \psi(k) \vert \d \psi(k) \rangle
\end{equation}
where
\begin{equation}
	\lvert \psi(k) \rangle = \frac{1}{\sqrt{2}} \begin{pmatrix} 1 \\ \e^{\i\varphi} \end{pmatrix}
\end{equation}
is the eigenvector of the valance band and $\varphi$ is defined from
\begin{equation}
	\tan\varphi = \frac{t_B \sin(k)}{t_A + t_B \cos(k)} \,.
\end{equation}
For the SSH model initialized with a $t_A < t_B$ stagger, the Zak phase is quantized to unity. 
%The Zak phase can be evaluated numerically as
%\begin{equation}
%	\gamma	=	\frac\i\pi \ln \prod_{a} \frac{\psi^\dagger(k_{a}) \psi(k_{a+1})}{| \psi^\dagger(k_{a}) \psi(k_{a+1}) |}
%\label{eq:numzak}
%\end{equation}
%wherein the $k_a$ is a discretization of the momentum in the first Brillouin zone.
The Zak phase of the SSH model has been observed experimentally in cold atom experiments~\cite{zakobs}. Since the Zak phase is technically not gauge invariant, this experiment measured the difference in Zak phases between the trivial and topological phases of the SSH model.

In position space the topology of the SSH model can be seen from its boundary Green function.
This Green function can be calculated numerically using the fast recursion methods described in~\S\ref{sec:calcmeth}.
The boundary site spectral function is obtained from the Green function
\begin{equation}
	\mathcal{A}(\omega) = -\frac1\pi \Im\left[\Green{\op{c}{1}}{\opd{c}{1}}_{\omega + \i0^+}\right]
\end{equation}
\begin{figure}[h]
\centering
\begin{subfigure}{0.49\linewidth}
\centering
\begin{tikzpicture}
\node at (0,0) {\includegraphics[scale=1]{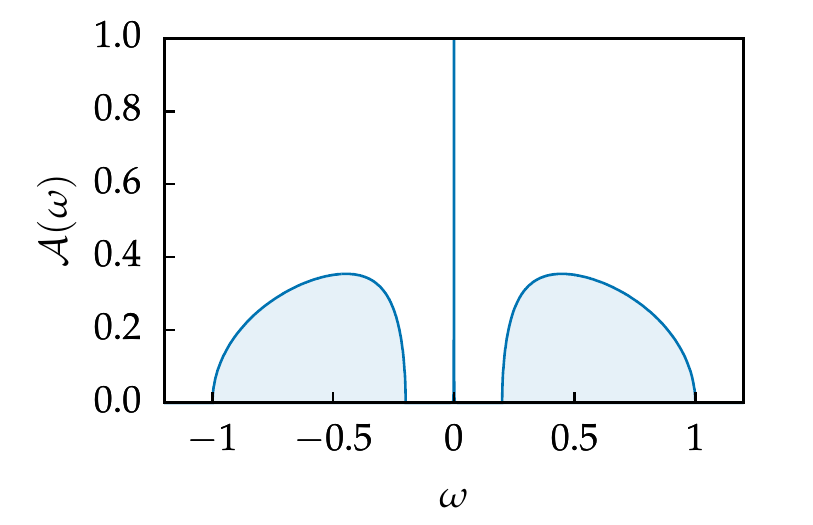}};
\node at (3.125,2.) {\footnotesize (a)};
\end{tikzpicture}
\phantomsubcaption{\label{fig:sshtop}}
\end{subfigure}
\hfill
\begin{subfigure}{0.49\linewidth}
\centering
\begin{tikzpicture}
\node at (0,0) {\includegraphics[scale=1]{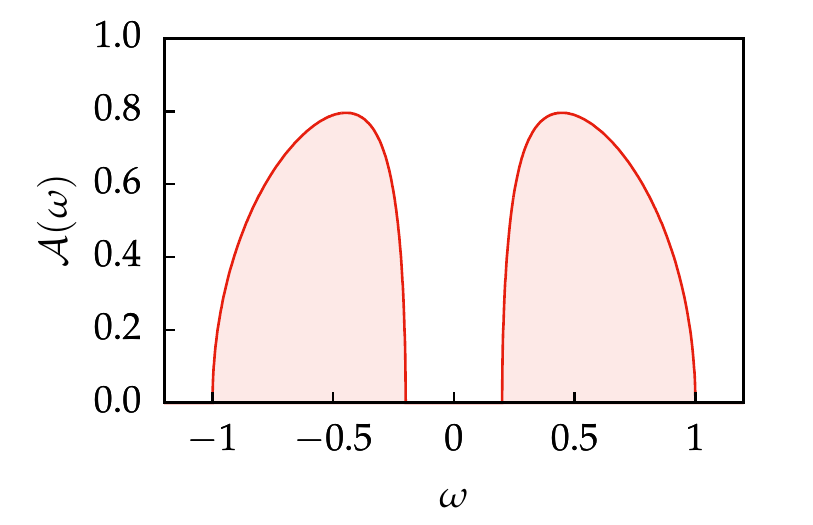}};
\node at (3.125,2.) {\footnotesize (b)};
\end{tikzpicture}
\phantomsubcaption{\label{fig:sshtriv}}
\end{subfigure}
\vspace{-\baselineskip}
\caption[Spectral function of the SSH model]{Spectral function on the boundary site of the SSH model in \subref{fig:sshtop} the topological phase and \subref{fig:sshtriv} the trivial phase. Note that both spectra are normalized to unity, with the difference in magnitude of the valance and conduction bands indicative of the pole weight.\label{fig:sshbands}}
\end{figure}
The primary notable feature of the SSH model spectral function in the topological phase is the presence of a pole at zero energy within a hard gap. This is the topological boundary state. In the trivial phase the spectral function is gapped with no midgap states. The bulk is similarly gapped in both phases.
%\begin{figure}[h]
%	\centering
%	\begin{subfigure}[c]{0.45\linewidth}
%	\centering
%	\includegraphics{sshtop2.pdf}
%	\caption{Topological}
%	\end{subfigure}
%	\hfill
%	\begin{subfigure}[c]{0.45\linewidth}
%	\centering
%	\includegraphics{sshtriv.pdf}
%	\caption{Trivial}
%	\end{subfigure}
%\end{figure}
%The primary characteristic of the SSH model visible in the boundary site spectral function is the presence of a spectral pole at zero energy.
The spectral weight of the topological zero energy pole is
\begin{equation}
	w_p	=	\frac{t_B^2 - t_A^2}{t_B^2} \,.
\end{equation}
Analysis of the zero energy wavefunction reveals that this state is exponentially localized at the boundary. This will be demonstrated in the next section.

The zero energy topological state cannot be removed by local deformations and generally requires a drastic change to the system, such as closing and reopening the band gap or breaking the chiral symmetry. The appearance of a topological characteristic on the system's boundary and a topological characteristic from the bulk, in this case from the winding of the Zak phase, is a manifestation of the bulk-boundary correspondence\index{bulk-boundary correspondence}. 

The spectrum of the bulk of the system is also worth examining.
The Green function on any bulk site can similarly be obtained from the equation of motion as
\begin{equation}
\begin{aligned}[b]
	G_{\text{bulk}}(z)
	&= \cfrac{1}{z - \cfrac{t_A^2}{z - \cfrac{t_B^2}{z - \cfrac{t_A^2}{z - \ddots}}} - \cfrac{t_B^2}{z - \cfrac{t_A^2}{z - \cfrac{t_B^2}{z - \ddots}}}}
	\\
	&= \cfrac{1}{z - t_A^2 G_{t_B,t_A}(z) - t_B^2 G_{t_A,t_B}(z)}
\label{eq:sshbulkgreen}
\end{aligned}
\end{equation}
where the two branches of the continued fraction have been resumed as
\begin{equation}
	G_{t_1,t_2}(z) = \frac{z^2 + t_2^2 - t_1^2 \pm \sqrt{(z^2 - t_1^2 + t_2^2)^2 - 4 z^2 t_2^2}}{2 z t_2^2}
\label{eq:sshgreenfunctionform}
\end{equation}
The bulk SSH Green function is the same regardless of the stagger $t_A < t_B$ or $t_A > t_B$. This change affects the Green function on the boundary site, but for the bulk Green function the change can be seen as swapping the two branches of the continued fraction in \eqref{eq:sshbulkgreen}, which appear on equal footing. Swapping them therefore makes no change to the overall function. Physically, any given site in the bulk is coupled by a strong bond to one neighbor and by a weak bond to its other neighbor.

\begin{figure}[h]
\centering
\begin{tikzpicture}[scale=0.75,site/.style={circle,draw=black,inner sep=4pt,line width=1pt}]
	\node[site,fill=gray] (0) at (0,0){};
	\node[below=3pt] (arrow) at (0) {$\uparrow$};
	\node[below] at (arrow) {$G_{\text{bulk}}(z)$};
	\node[site] (r1) at (2,0){};
	\node[site] (r2) at (4,0){};
	\node[site] (rend) at (6,0){};
	\draw[line width=1.75pt, dashed, line cap=round]($(rend)+(1,0)$)--($(rend)+(2,0)$);
	\draw[line width=1.75pt,color=blue](0)--(r1) node[midway,above,blue] {${t_A}$};
	\draw[line width=1.25pt,color=red,double,double distance=1.25pt](r1)--(r2) node[midway,above,red] {${t_B}$};
	\draw[line width=1.75pt,color=blue](r2)--(rend);
	\draw[line width=1.25pt,color=red,double,double distance=1.25pt](rend)--($(rend)+(1,0)$);
	\node[site] (l1) at (-2,0){};
	\node[site] (l2) at (-4,0){};
	\node[site] (lend) at (-6,0){};
	\draw[line width=1.75pt, dashed, line cap=round]($(lend)+(-1,0)$)--($(lend)+(-2,0)$);
	\draw[line width=1.25pt,color=red,double,double distance=1.25pt](0)--(l1) node[midway,above,red] {${t_B}$};
	\draw[line width=1.75pt,color=blue](l1)--(l2) node[midway,above,blue] {${t_A}$};
	\draw[line width=1.25pt,color=red,double,double distance=1.25pt](l2)--(lend);
	\draw[line width=1.75pt,color=blue](lend)--($(lend)+(-1,0)$);
\end{tikzpicture}
\caption[Bulk Green function of the SSH model]{Bulk Green function}
\end{figure}
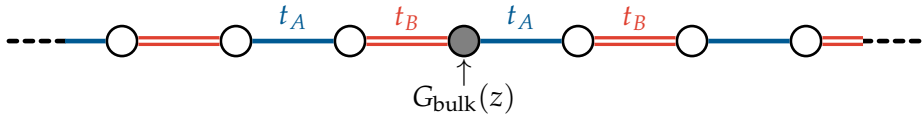

\begin{figure}[htp]
\centering
\includegraphics[scale=1]{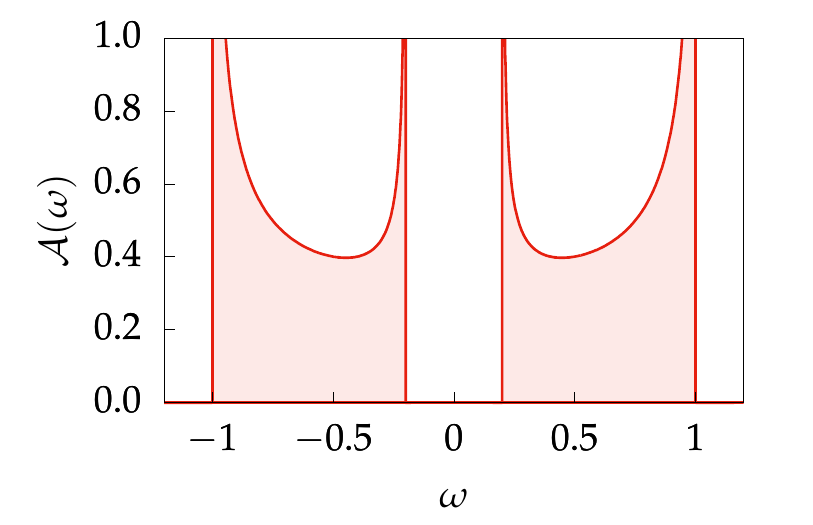}
\caption[Bulk spectrum of the SSH model]{Bulk spectrum of the SSH model given by $\mathcal{A}(\omega)=-\frac1\pi \Im G_{\text{bulk}}(\omega)$ where $G_{\text{bulk}}(z)$ is given by Eq.~\eqref{eq:sshbulkgreen}. The spectrum is always gapped $\forall t_A, t_B$; $t_A \neq t_B$.\label{fig:sshbulk}}
\end{figure}

The appearance of a topological state on the boundary of the system and a bulk which is insulating is the primary defining phenomenological characteristic of a topological insulator.

%%%%%%%%%

\begin{figure}[h]
\centering
\begin{tikzpicture}[scale=0.75,site/.style={circle,draw=black,inner sep=4pt,line width=1pt}]
	\coordinate (t) at (0,1.5);
	\node[site] (t1) at (t){};
	\node[site] (t2) at ($(t1)+(2,0)$){};
	\node[site] (t3) at ($(t2)+(2,0)$){};
	\node[site] (t4) at ($(t3)+(2,0)$){};
	\node[site] (t5) at ($(t4)+(2,0)$){};
	\draw[line width=1.75pt, dashed, line cap=round]($(t5)+(1,0)$)--+(1,0);
	\draw[line width=1.75pt,color=blue](t1)--(t2) node[midway,above,blue] {${t_A}$};
	\draw[line width=1.25pt,color=red,double,double distance=1.25pt](t2)--(t3) node[midway,above,red] {${t_B}$};
	\draw[line width=1.75pt,color=blue](t3)--(t4);
	\draw[line width=1.25pt,color=red,double,double distance=1.25pt](t4)--(t5);
	\draw[line width=2pt,color=blue](t5)--+(1,0);
	\coordinate (b) at ($(0,0)-(t)$);
	\node[site] (b1) at (b){};
	\node[site] (b2) at ($(b1)+(2,0)$){};
	\node[site] (b3) at ($(b2)+(2,0)$){};
	\node[site] (b4) at ($(b3)+(2,0)$){};
	\node[site] (b5) at ($(b4)+(2,0)$){};
	\draw[line width=1.75pt, dashed, line cap=round]($(b5)+(1,0)$)--+(1,0);
	\draw[line width=1.25pt,color=red,double,double distance=1.25pt](b1)--(b2) node[midway,above,red] {${t_A}$};
	\draw[line width=1.75pt,color=blue](b2)--(b3) node[midway,above,blue] {${t_B}$};
	\draw[line width=1.25pt,color=red,double,double distance=1.25pt](b3)--(b4);
	\draw[line width=1.75pt,color=blue](b4)--(b5);
	\draw[line width=1.5pt,color=red,double,double distance=1.5pt](b5)--+(1,0);
	\node (topgraph) at ($(t5)+(7,0)$) {\includegraphics[scale=1]{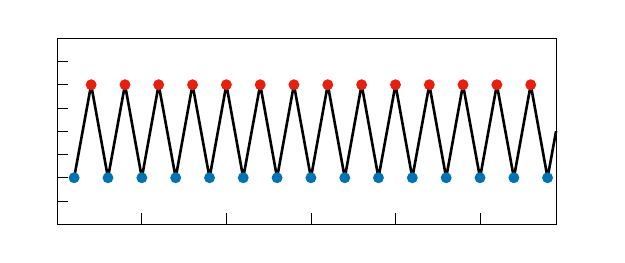}};
	\node at ($(topgraph)+(-3.5,1.5)$) {$t_n$};
	\node (trivgraph) at ($(b5)+(7,0)$) {\includegraphics[scale=1]{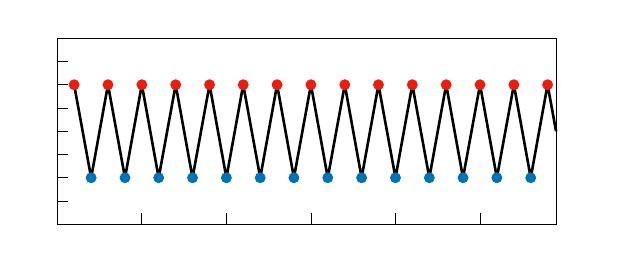}};
	\node at ($(trivgraph)+(3.5,-1.5)$) {$n$};
\end{tikzpicture}
\caption[Parameterizations of the SSH model]{Parameterizations of the SSH model. The two possible configurations are with either $t_A < t_B$ (top) or $t_A > t_B$ (bottom).}
\end{figure}
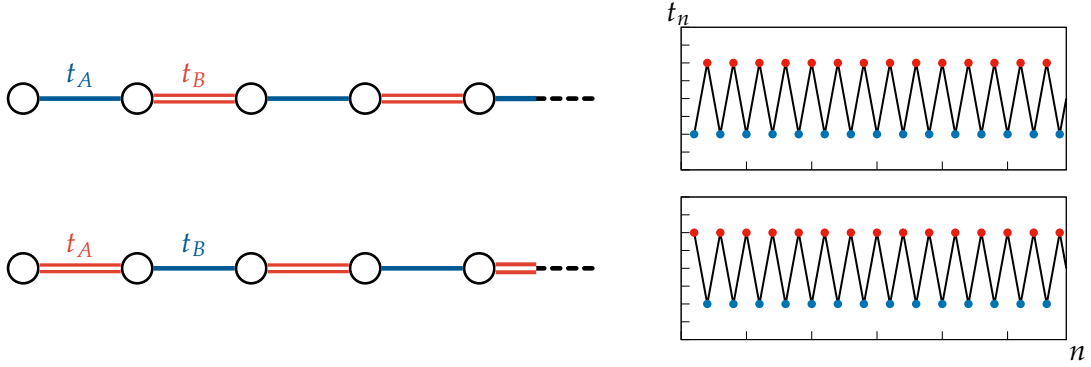

Intuition for the existence of the localized state in the topological phase can be obtained by analyzing the extreme limits of the model's parameterization.
The trivial phase which is without a localized boundary state possesses as an extreme limit $t_B \ll t_A$ such that $t_B \approx 0$. This situation results in a collection of decoupled dimers.
The trivial phase of the SSH model can then be said to be adiabatically connected to an atomic insulator.

The topological phase of the SSH model is adiabatically connected to the opposite limit of $t_A \ll t_B$ such that $t_A \approx 0$. 
A fermion located on the boundary of the system must be completely localized on that site as there is no hybridization to the remainder of the system. As the hybridization to the rest of the system is turned on, tuning $t_A$ finite while keeping $t_A < t_B$, the boundary mode remains localized exponentially, as demonstrated below.

\subsection{Topological State\label{sec:sshtransfer}}
Features of the zero-energy eigenfunction can be obtained from the transfer matrix\index{transfer matrix}~\cite{shortcourse}.
For this analysis it is useful to analyze the SSH model in Dirac notation. The single particle states can be written as
\begin{subequations}
\begin{align}
	\lvert \psi_m^{(A)} \rangle &\vcentcolon= \opd{c}{2m-1} \lvert 0 \rangle
	\\
	\lvert \psi_m^{(B)} \rangle &\vcentcolon= \opd{c}{2m} \lvert 0 \rangle
\end{align}
\end{subequations}
where $\lvert 0 \rangle$ denotes the zero excitation vacuum state. The labels $(A)$ and $(B)$ of the state vectors correspond to states on the $A$ and $B$ sites of the unit cell as shown in Fig.~\ref{fig:sshunitcell}.

In the Dirac bra-ket state vector notation, the SSH Hamiltonian is written as
\begin{equation}
	\hat{H}_{\textsc{ssh}} = \sum_{m=1}^{\infty} \left[ t_A \left( | \psi_{m}^{(A)} \rangle \langle \psi_{m}^{(B)} | + | \psi_{m}^{(B)} \rangle \langle \psi_{m}^{(A)} | \right) + t_B \left( | \psi_{m+1}^{(A)} \rangle \langle \psi_{m}^{(B)} | + | \tensor*{\psi}{_{m}^{(B)}} \rangle \langle \tensor*{\psi}{_{m+1}^{(A)}} | \right) \right]
\end{equation}
The zero-energy eigenstate may be defined with an ansatz of the form
\begin{equation}
	| \mathnormal{\Psi}_0 \rangle = \sum_{n=1}^{\infty} \left[ u_{n}^{(A)} | \psi_{n}^{(A)} \rangle + u_{n}^{(B)} | \psi_{n}^{(B)} \rangle \right]
\end{equation}
which obeys the Schr\"odinger equation
$\hat{H}_{\textsc{ssh}} | \mathnormal{\Psi}_0 \rangle = E_0 | \mathnormal{\Psi}_0 \rangle$. The action of the Hamiltonian on this state is
\begin{align*}
	\hat{H}_{\textsc{ssh}} | \mathnormal{\Psi}_0 \rangle
		&=	\sum_{n=1}^{\infty} \left[ \left( t_A u_{n}^{(B)} \right) | \psi_{n}^{(A)} \rangle + \left( t_A u_{n}^{(A)} \right) | \psi_{n}^{(B)} \rangle + t_B u_{n}^{(A)} | \psi_{n-1}^{(B)} \rangle + t_B u_{n}^{(B)} | \psi_{n+1}^{(A)} \rangle  \right]
\end{align*}
Taking into account the zero eigenenergy, $E_0 = 0$, this leads to a set of algebraic equations for the wavefunction amplitudes
\begin{subequations}
\begin{align}
	t_A u_{n}^{(A)} + t_B u_{n+1}^{(A)} &= 0
	\\
	t_A \tensor*{u}{_{n}^{(B)}} + t_B \tensor*{u}{_{n-1}^{(B)}} &= 0
\end{align}
\end{subequations}
As the system is taken to be semi-infinite, it follows that $u_{n}^{(B)} = 0$ $\forall n$ as $\tensor*{u}{_{n-1}^{(B)}}$ necessarily vanishes due to the non-existence of cells $n\leq0$ from the presence of the boundary.
The equation for $u^{(A)}_{n}$ can be iterated to form a solution in terms of $u^{(A)}_{1}$.

For an SSH model of finite system size with $N$ unit cells ($2N$ sites), edge modes appear on both boundaries. For $t_A < t_B$ an edge mode appears on the left-side boundary with support only on the odd sublattice and the right-side boundary hosts an edge mode with support on the even sublattice only. The right-boundary edge state can be seen to exist following the preceding argument except with state $n=L$. Now it is the $u_{n}^{(A)}$ which all vanish identically due to the absence of states $n>L$ such that $u_{L+1}^{(A)} = 0$. These edge modes decay exponentially into the bulk. A semi-infinite system may be obtained by taking the limit of sending the right-side boundary to infinity. As the edge mode is now infinitely far away, the total wavefunction is zero on the even sublattice as it is suppressed by a factor of $\e^{-\infty}$. A localized state on a boundary site hybridized to the rest of the chain with a strong $t_B$ bond is only present in the case of a SSH model of finite length.

The localization length $\xi$ of the wavefunction can be determined by prescribing the ansatz $u^{(A)}_{n} = e^{-{n}/{\xi}} u^{(A)}_{1}$ for the value of the wavefunction on the $A$ site in unit cell $n$ in terms of the value on the boundary site and iterating the value of the wavefunction from the transfer matrix as
\begin{align*}
	\left| u^{(A)}_{n} \right|^2 &= \left| e^{-{n}/{\xi}} u^{(A)}_{1} \right|^2
	\\
	\left(-\frac{t_A}{t_2}\right)^{2n-2} \left| u^{(A)}_{1} \right|^2 &= e^{-{2n}/{\xi}} \left| u^{(A)}_{1} \right|^2
	\\
	e^{-{2n}/{\xi}} &= \left(\frac{t_A}{t_B}\right)^{2n-2}
	\\
%	-\frac{2n}{\xi} &= (2n-2) \ln\frac{t_A}{t_B}
%	\\
	\Rightarrow \quad \xi &= \frac{n}{1-n} \frac{1}{\ln\frac{t_A}{t_B}}
\end{align*}
For sites deep into the bulk,
\begin{equation}\begin{aligned}
	\xi &\underset{n\gg1}{\approx} \frac{1}{\ln\frac{t_B}{t_A}}
\label{eq:localization}
\end{aligned}\end{equation}
It is apparent that for $t_A < t_B$ the zero-energy state decays exponentially into the bulk. The topological state is therefore exponentially localized on the boundary of the system.

\subsubsection{Disorder}
It is conventionally understood that the topological states of topological insulators are robust to adiabatic perturbations. Strictly speaking, this applies exactly only to perturbations which preserve the symmetry protecting the topological state. The topological state may be stated to be exactly topologically protected against symmetry preserving perturbations. For non-symmetry preserving perturbations, the topological state is only considered to be approximately robust~\cite{sshbulkboundary}.

Disorder which respects the chiral symmetry of the SSH model in momentum space is tantamount to reparameterizing the tunneling amplitudes with $t_A \to t'_A$ and $t_B \to t'_B$ such that $t'_A < t'_B$ still holds.
Disorder applied to the hopping amplitudes can be expressed as $t_B \to t_B + \delta w_j$ and $t_A \to t_A + \delta w_j$, where the $\delta w_j$ are randomly sampled from the interval $[-W,W]$ and $W$ is termed the disorder width.

The spectrum for the disordered SSH model can be obtained directly from the Green function computed from its continued fraction form
\begin{equation}
\begin{aligned}[c]
	G_{1,1}(z)
	=	\cfrac{1}{
		z - \varepsilon_1 - \cfrac{t_{A_1}^2}{
		z - \varepsilon_2 - \cfrac{t_{B_2}^2}{
		z - \varepsilon_3 - \cfrac{t_{A_3}^2}{
		\cfrac{\ddots}{
		z - \varepsilon_{N-1} - \cfrac{t_{B_{N-1}}^2}{
		z - \varepsilon_N
		}
		}
		}
		}
		}
		} \,.
\end{aligned}
\end{equation}
\begin{figure}[h]
\subfiglabel{\includegraphics[scale=1]{sshbandstop.pdf}}{3.125,2}{fig:dtsshspec0}
\subfiglabel{\includegraphics[scale=1]{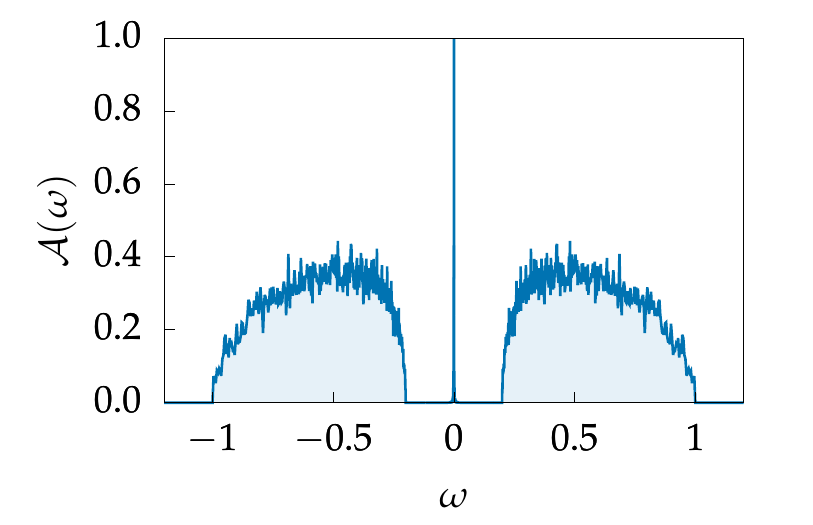}}{3.125,2}{fig:dtsshspec1e-4}
\subfiglabel{\includegraphics[scale=1]{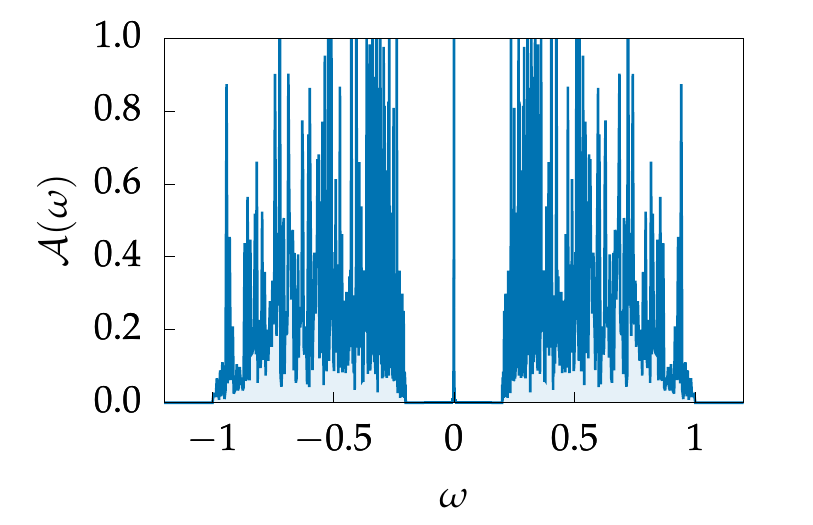}}{3.125,2}{fig:dtsshspec1e-3}
\subfiglabel{\includegraphics[scale=1]{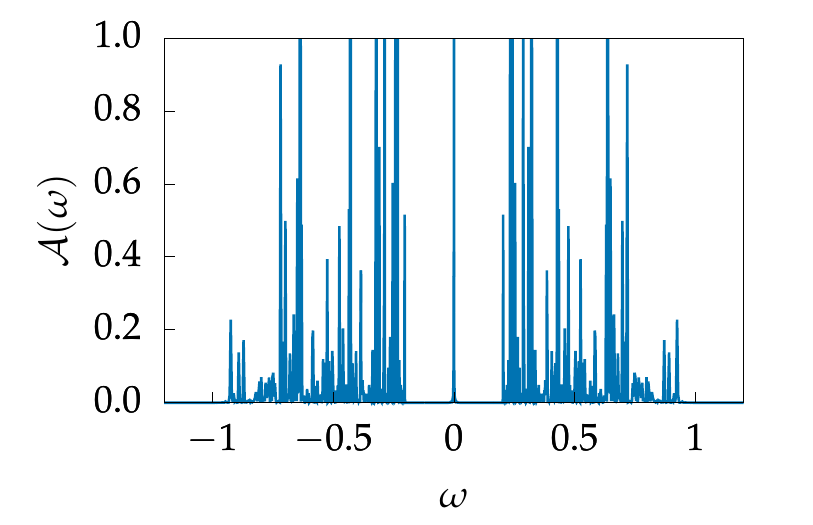}}{3.125,2}{fig:dtsshspec1e-2}
\caption[Spectral function of the SSH model with hopping parameter disorder]{Spectral function of the SSH model in its topological configuration with disorder on the hopping parameters with disorder width of \subref{fig:dtsshspec0} $W = 0$, \subref{fig:dtsshspec1e-4} $W = 10^{-4}$, \subref{fig:dtsshspec1e-3} $W = 10^{-3}$, and \subref{fig:dtsshspec1e-2} $W = 10^{-2}$. Observe that even with heavy distortion, the midgap topological pole remains. Calculation performed with a finite chain of $N=\num{5000000}$ sites.\label{fig:dtsshspec}}
\end{figure}
\begin{figure}[h]
	\includegraphics[scale=1]{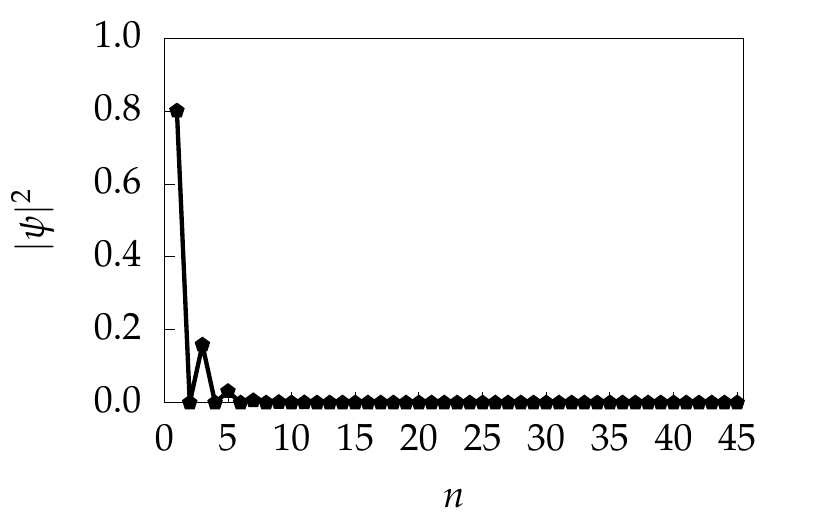}
	\includegraphics[scale=1]{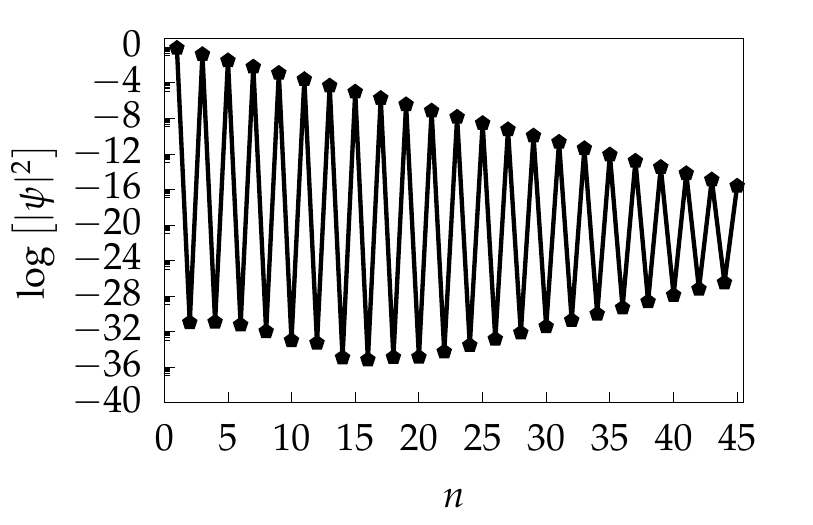}
\caption[Wavefunction of the zero-energy boundary state with hopping parameter disorder]{Wavefunction of the zero-energy boundary state for disorder width of $10^{-3}$ on the hopping parameters. \label{fig:dtsshwfn}}
\end{figure}
In order for the midgap state to collapse, it is necessary to alter the system by more than adding perturbations to the hopping parameters. 
Collapsing the topological state in the midgap generally requires a
drastic change in system parameters,
such as closing and reopening the band gap.

The boundary state also remains localized for disorder in the on-site potential when $\varepsilon \to \delta\varepsilon_j$ and $\delta\varepsilon_j$ are randomly sampled from $[-W,W]$ as in the previous case with disordered hoppings. This effect on the SSH spectrum is shown in Fig.~\ref{fig:desshwfn}.
%\cite{sshsitedisorder}
%
\begin{figure}[h]
\subfiglabel{\includegraphics[scale=1]{sshbandstop.pdf}}{3.125,2}{fig:desshspec0}
\subfiglabel{\includegraphics[scale=1]{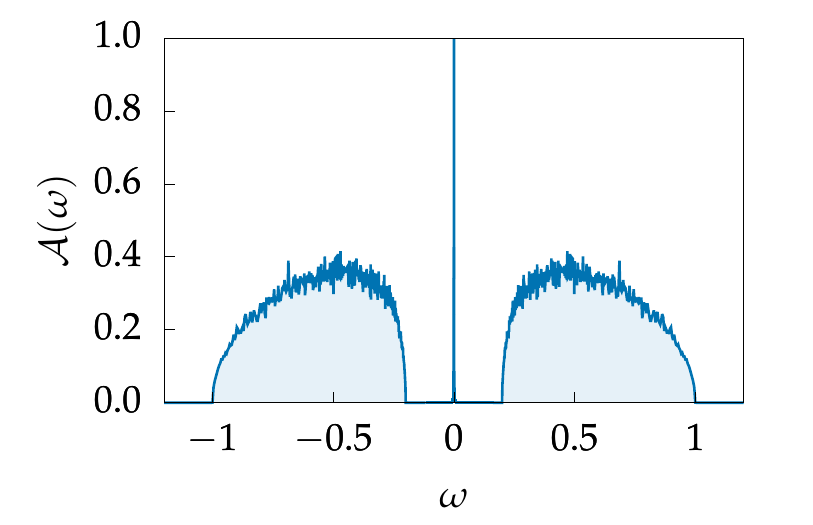}}{3.125,2}{fig:desshspec1e-4}
\subfiglabel{\includegraphics[scale=1]{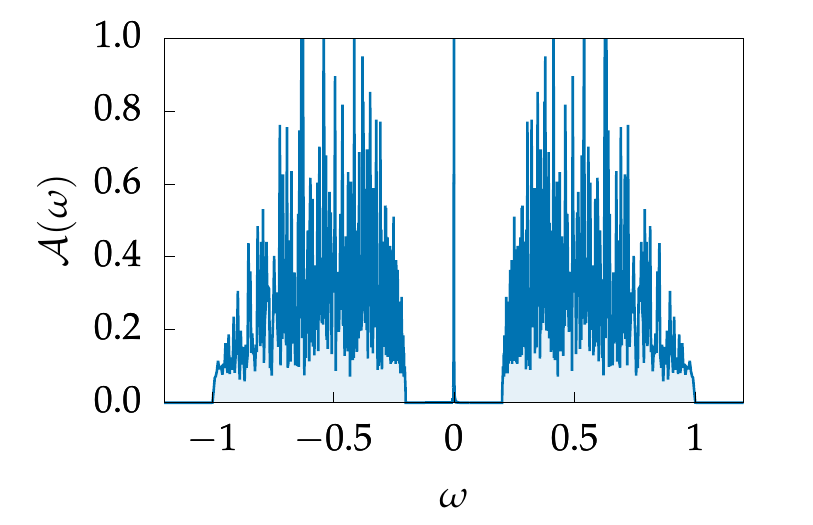}}{3.125,2}{fig:desshspec1e-3}
\subfiglabel{\includegraphics[scale=1]{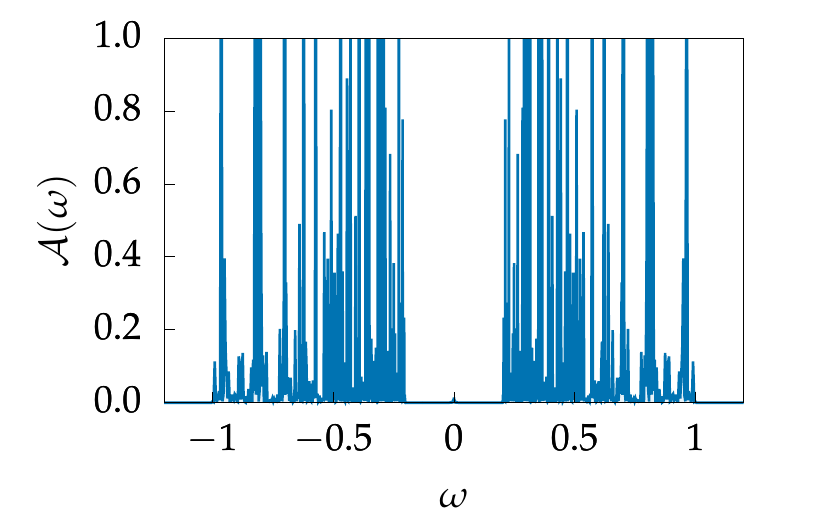}}{3.125,2} {fig:desshspec1e-2}
\caption[Spectral function of the SSH model with on-site potential disorder]{Spectral function of the SSH model in its topological configuration with on-site potential disorder with disorder width of \subref{fig:desshspec0} $W = 0$, \subref{fig:desshspec1e-4} $W = 10^{-4}$, \subref{fig:desshspec1e-3} $W = 10^{-3}$, and \subref{fig:desshspec1e-2} $W = 10^{-2}$. Calculation performed with a finite chain of $N=\num{5000000}$ sites.\label{fig:desshspec}}
\end{figure}
\begin{figure}[h]
	\includegraphics[scale=1]{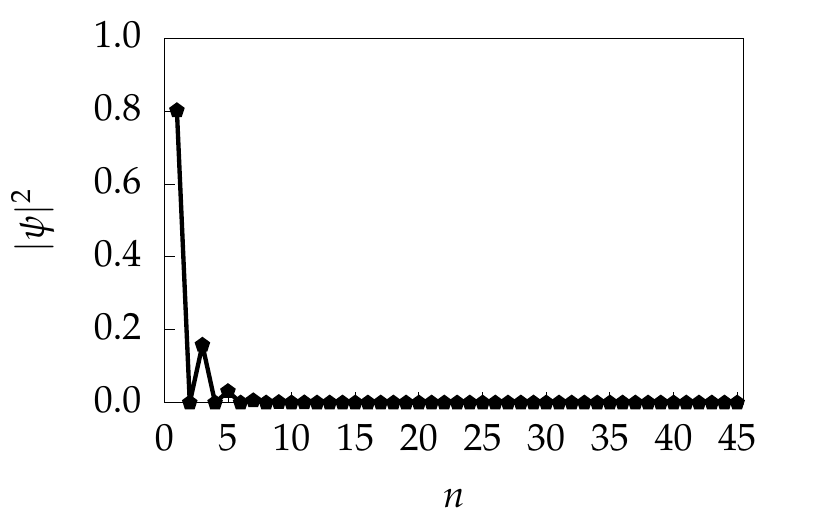}
	\includegraphics[scale=1]{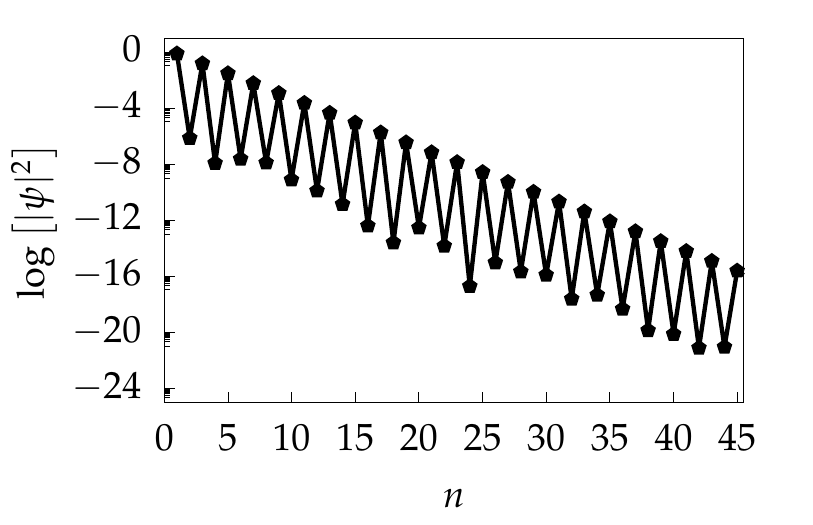}
\caption[Wavefunction of the zero-energy boundary state with on-site potential disorder]{Wavefunction of the zero-energy boundary state for disorder width of $10^{-3}$ on the potentials.\label{fig:desshwfn}}
\end{figure}

Even in the presence of symmetry breaking disorder, the zero energy state remains exponentially localized on the boundary of the system, as shown in Fig.~\ref{fig:desshwfn}. Since the chiral symmetry is broken, the wavefunction now has finite support on the even sublattice, but is still orders of magnitude smaller than the amplitude on the odd sublattice. With potential disorder, the boundary site potential may shift from zero and hence the localized boundary  mode may shift away from the Fermi level. However, the state is still robust and topologically protected, provided that the state remains in the spectral gap.
Sufficiently strong potential disorder will destroy the zero-energy state.
Very strong perturbations are required to destroy such states by moving them into the band continuum. 
This argument regarding robustness in the presence of symmetry-breaking perturbations will become important in the discussion of topology in the Mott transition in the particle-hole asymmetric Hubbard model in \S\ref{sec:phasymmmap}.

\chapter{Generalized SSH Models\label{ch:genssh}}

The SSH model described at the end of the previous chapter is simple enough to provide the basis of a wide variety of extensions which explore realizations of quantum topology in $1d$ systems. Several such extensions are the subject of the present chapter. As mentioned in \S\ref{sec:sshmodel}, many generalizations of the SSH model exist in the literature. The generalizations contained in this chapter are similar in concept to the SSH trimer and SSH$_4$ models\cite{trimer,ssh4} in that they are based on extending the size of the unit cell in $1d$. In contrast to the previous literature, the generalized SSH models here extend the unit cell to arbitrary size with the unit cell hopping parameters given by a particular functional form.

The generalized SSH models presented here also make appearances in the auxiliary systems which are constructed in \S\ref{ch:aux} and \S\ref{ch:motttopology}. This section explores $1d$ models which generalize the SSH model both at the level of the Hamiltonian and at the level of the spectral function.

Discussed first in this chapter are the characteristics of domain walls in the semi-infinite SSH model. The discussion of SSH model domain walls in the literature generally considers only a single domain wall, or a pair of domain walls, in a fully infinite system~\cite{solitons}. These configurations are reviewed below before turning to the semi-infinite case with arbitrary numbers of domain walls, which has not previously been discussed in the literature. Additionally, the method here of reverse engineering the system parameters from the spectrum of topological SSH-type models, performed here via a spectral moment recursion as well as the Lanczos algorithm, is new to the literature.
%\index{$0$@\textbf{List of Edits}!300@new contributions highlighted}

This chapter is based on the paper~\cite{generalizedssh}. Elements of this work also appear in~\cite{motttopology}.

\section{Domain Walls\label{sec:domwalls}}

%Previous studies on domain walls within the SSH model considered finite systems. Here the study considers the semi-infinite SSH model with a focus on the boundary spectrum.

A simple non-trivial generalization of the SSH spectrum is to consider the case where there are two localized mid-gap states. These manifest as poles located at $\pm\omega_p$ due to the system's chiral symmetry.

The number of mid-gap poles is equal to the number of domain walls.
In the absence of domain walls, the bulk of an infinite SSH model possesses an insulating gapped spectral function.
\begin{figure}[h]
\subfiglabel{\includegraphics[scale=1]{sshbandsbulk.pdf}}{3.2,2}{fig:sshspecbulk}
\subfiglabel{\includegraphics[scale=1]{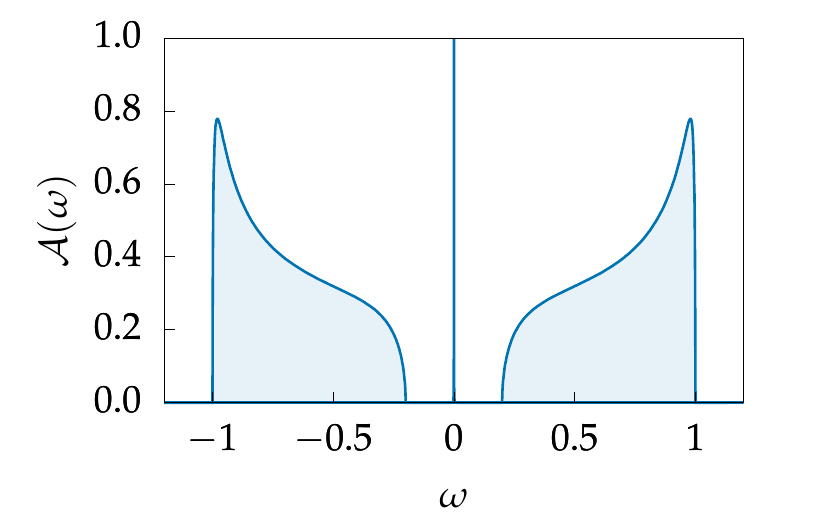}}{3.2,2}{fig:sshspecdw}
\caption[Bulk spectrum in the SSH model without \subref{fig:sshspecbulk} and with \subref{fig:sshspecdw} a domain wall]{Bulk spectrum in the SSH model without \subref{fig:sshspecbulk} and with \subref{fig:sshspecdw} a domain wall. Note that the presence of a domain wall yields the same qualitative spectrum as the boundary in the topological phase.}
\end{figure}
A domain wall may be introduced by reversing the parity of the alternating hopping pattern at one site, demonstrated in Fig.~\ref{fig:simpledwschematic}. This results in the appearance of a domain wall at the site where the hopping parity flips. 
On this site the spectral function is no longer insulating, but takes a form similar to that of the SSH boundary in the topological phase, with a zero energy spectral pole. This result shows that domain walls host localized states in a similar way as the boundary of the SSH model does in its topological configuration. 
In fact, the boundary of a semi-infinite SSH chain may itself be interpreted as being a domain wall between the system and the vacuum, which is topologically trivial.
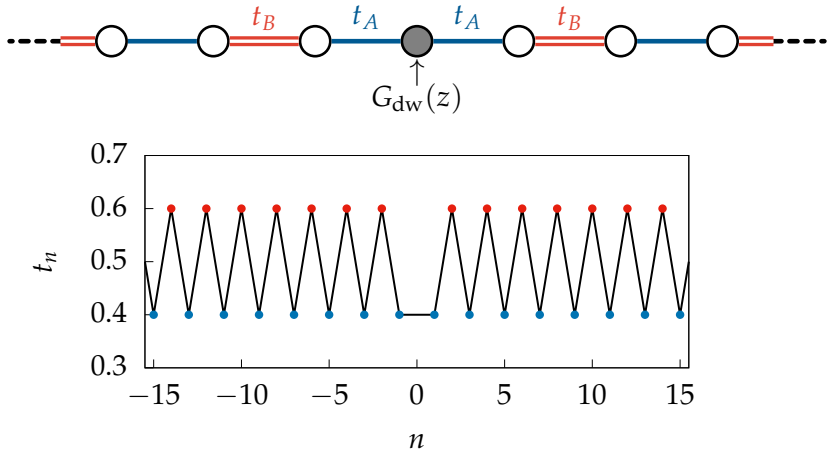
\begin{figure}[h]
\centering
\begin{tikzpicture}[scale=0.67,site/.style={circle,draw=black,inner sep=4pt,line width=1pt}]
	\node[site,fill=gray] (0) at (0,0){};
	\node[below=3pt] (arrow) at (0) {$\uparrow$};
	\node[below] at (arrow) {$G_{\text{dw}}(z)$};
	\node[site] (r1) at (2,0){};
	\node[site] (r2) at (4,0){};
	\node[site] (rend) at (6,0){};
	\draw[line width=1.75pt, dashed, line cap=round]($(rend)+(1,0)$)--($(rend)+(2,0)$);
	\draw[line width=1.75pt,color=blue](0)--(r1) node[midway,above,blue] {${t_A}$};
	\draw[line width=1.25pt,color=red,double,double distance=1.25pt](r1)--(r2) node[midway,above,red] {${t_B}$};
	\draw[line width=1.75pt,color=blue](r2)--(rend);
	\draw[line width=1.25pt,color=red,double,double distance=1.25pt](rend)--($(rend)+(1,0)$);
	\node[site] (l1) at (-2,0){};
	\node[site] (l2) at (-4,0){};
	\node[site] (lend) at (-6,0){};
	\draw[line width=1.75pt, dashed, line cap=round]($(lend)+(-1,0)$)--($(lend)+(-2,0)$);
	\draw[line width=1.75pt,color=blue](0)--(l1) node[midway,above,blue] {${t_A}$};
	\draw[line width=1.25pt,color=red,double,double distance=1.25pt](l1)--(l2) node[midway,above,red] {${t_B}$};
	\draw[line width=1.75pt,color=blue](l2)--(lend);
	\draw[line width=1.25pt,color=red,double,double distance=1.25pt](lend)--($(lend)+(-1,0)$);
	\node at (-0.725,-5) {\includegraphics{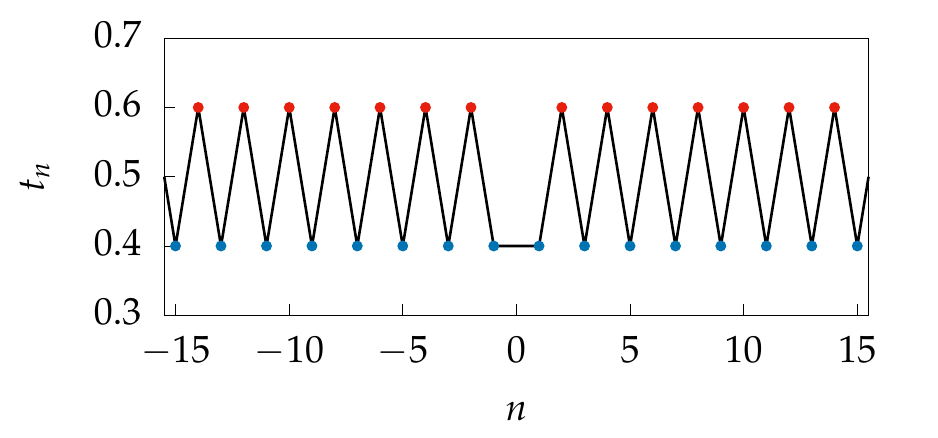}};
\end{tikzpicture}
\caption{Schematic of an elementary domain wall in an infinite SSH model.\label{fig:simpledwschematic}}
\end{figure}
The Green function on an elementary domain wall is given by
\begin{equation}
	G_{\text{dw}}(z) = \cfrac{1}{z - 2 t_A^2 G_{t_B,t_A}(z)}
\end{equation}
where $G_{t_B,t_A}(z)$ is in the notation of Eq.~\eqref{eq:sshgreenfunctionform}. This Green function is produces the spectrum shown in Fig.~\ref{fig:sshspecdw}.

Now consider the case of a semi-infinite SSH model with the boundary in the topological configuration with a single domain wall located a finite distance into the chain, as shown in Fig.~\ref{fig:simpledw}.\footnote{As mentioned above, this set up is equivalent to an infinite SSH model with \textit{two} domain walls in the bulk as the topological boundary can be considered a domain wall with a vacuum.} The spectral function now features instead of a single mid-gap pole, two poles. 
\begin{figure}[h]
\centering
\begin{subfigure}{0.49\linewidth}
\includegraphics[scale=1]{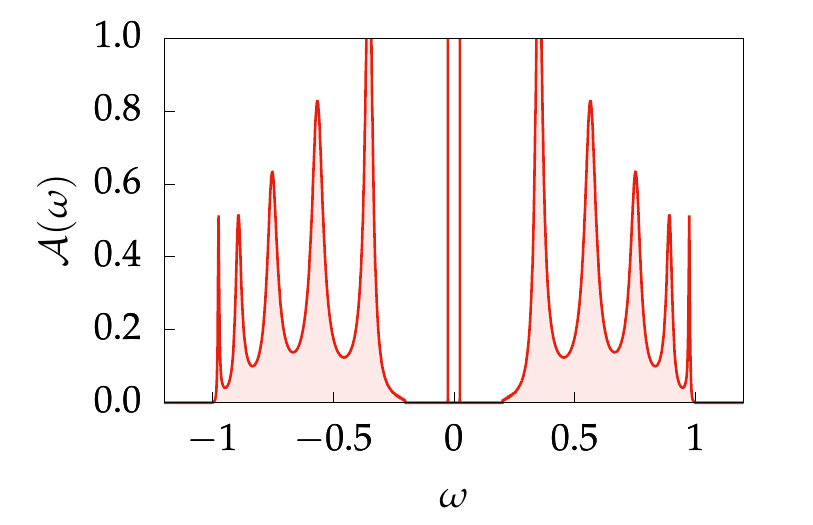}
\end{subfigure}
\begin{subfigure}{0.49\linewidth}
\includegraphics[scale=1]{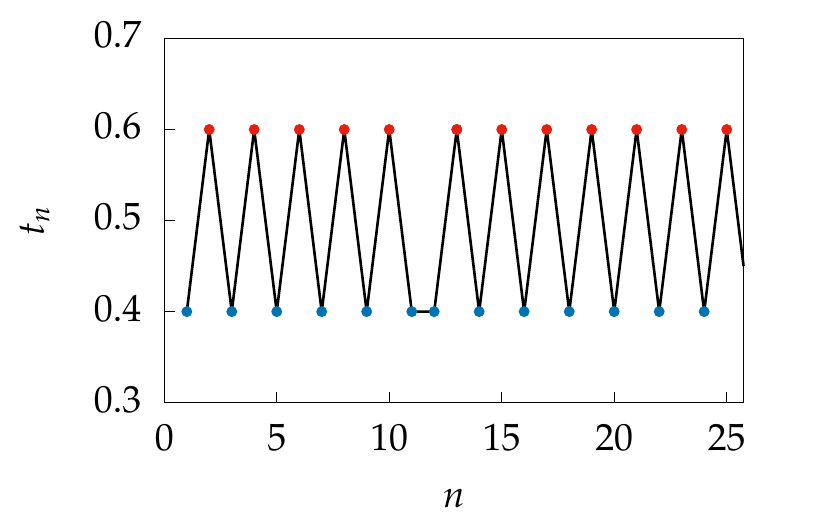}
\end{subfigure}
\caption[An SSH model with a simple domain wall formed by a change in parity of the stagger in the hopping parameters]{An SSH model with a simple domain wall formed by a change in parity of the stagger in the hopping parameters. Note that this produces a pair of gapped poles in the spectrum, and that the outer bands take on a distorted form, but still continuous.\label{fig:simpledw}}
\end{figure}
However, in this case the SSH bands in the spectral function, while still continuous, become distorted, as shown in Fig.~\ref{fig:simpledw}.
It is therefore instructive to investigate under which parameterization of the hopping amplitudes $t_n$ yield multiple mid-gap states as well as preserve the form of the SSH satellite bands, so that the spectrum of an SSH model with a single intragap state with one with multiple intragap states can be directly compared. The motivation here is to produce models whose only difference from the conventional SSH model is the number of localized states in the gap. The objective of preserving the outer bands will become relevant in \S\ref{ch:motttopology} where effective models designed to reproduce specific spectral functions are engineered. 
%\index{$0$@\textbf{List of Edits}!301@added motivation for preserving bands} 
The appropriate parameterization can be found from investigating the continuum quantum field theory associated to the SSH lattice model.

%%%%%%%%%%%%

Rather than focusing on the structure of the of chain parameters and examining the resulting spectrum, it is useful to construct the spectrum first and then reverse engineer the chain parameters.
\label{sec:moments}
A method which can uncover this parameterization numerically is by analyzing the moments of the spectral function. Consider first that a composite spectrum may be written as
\begin{equation}
	\mathcal{A}(\omega) = \frac{1}{\mathcal{N}} \sum_{i} w_{i} \mathcal{A}_{i}(\omega)
\end{equation}
where $w_i$ are the relative weights of the spectral elements $\mathcal{A}_{i}(\omega)$ with normalization $\mathcal{N} = \sum_{i} w_{i}$.
The $k$-th spectral moment of spectral element $i$ is given by
\begin{equation}
	\mu_{i,k} \vcentcolon= \int \omega^k \mathcal{A}_{i}(\omega) \d\omega \,.
\end{equation}
The $k$-th spectral moment for the total spectrum is given by
\begin{equation}
	\mu_{k} = \frac{1}{\mathcal{N}} \sum_{i} w_{i} \mu_{i,j} \,.
\end{equation}
Since moments are additive, it is possible to add the moments of the outer bands of SSH form plus the moments from the spectral poles.

For a particle-hole symmetric system, the spectral function obeys $\mathcal{A}(\omega) = \mathcal{A}(-\omega)$, and therefore only the even moments survive, $\mu_{2n+1} = 0$. Particle-hole symmetry holds for the cases considered in this chapter, although generalizations of this method also exist~\cite{recursionmethod}.
From a set of spectral moments $\{ \mu_0, \mu_2, \ldots, \mu_{2N} \}$, it is possible to construct the first $N$ elements of a chain's hopping parameters $t_n$. The $n$-th chain parameter $t_n$ can be calculated by introducing a set of auxiliary variables $X_k(n)$ obeying the recursion relation~\cite{recursionmethod}
\begin{equation}
	X_{2k}(n) = \frac{X_{2k}(n-1)}{t_{n-1}^2} - \frac{X_{2k-2}(n-2)}{t_{n-2}^2}
\end{equation}
initialized with $X_{2k}(0) = \mu_{2k}$, $X_{2k}(-1) = 0$, and $t_{-1}^2 = 1 = t_{0}^2$. 
The $n$-th chain parameter is recovered from $t_{n}^2 = X_{2n}(n)$. While this moment analysis will give exact results for the $t_n$'s, this scheme is known to be numerically unstable~\cite{gaspard,gautschi}. Using arbitrary precision numerics the maximum number of parameters that can be calculated in practice is about $n_{\text{max}}\sim$120. After this number the calculations deliver unphysical results, such as returning a $t_{n>n_{\text{max}}}^2 < 0$.

An application of the moment analysis is showing that the addition of a zero energy pole of appropriate weight to the spectrum of the SSH model in the trivial phase has the effect of flipping the parity of all the chain parameters of the system.
That is, $t_n = t_0 - (-1)^{n}\delta t \to t_0 - (-1)^{n+1}\delta t$, on adding the zero-energy pole.
%\subsubsection*{Trivial $\boldsymbol+$ Pole $\boldsymbol=$ Topological}
In the trivial phase, the SSH model parameters obey $t_{2j-1} >  t_{2j}$ for $j\in\mathbbm{Z}^+$ and the spectral function features two bands separated by a hard gap without any mid-gap pole.
%	\qquad
%	\qquad
%	\qquad
%Goal:\quad
%\(\begin{matrix}
%	\twiddle{t}_{2n-1} = t-\Delta = t_{2n\phantom{-1}}
%	\\
%	\twiddle{t}_{2n\phantom{-1}} = t+\Delta = t_{2n-1}
%\end{matrix}\)
%\hfill
%$\bullet$ in Topo:
%\(
%	\twiddle{\ \bullet\ }
%\)
In this phase the moments are given by
\begin{equation}
	\mu_{2n} = \int \omega^{2n} \mathcal{A}(\omega) \d\omega
\end{equation}
where the spectral function is 
\begin{equation}
	\mathcal{A}(\omega) = -\frac1\pi \Im\left[ \frac{z^2 + t_1^2 - t_2^2 \pm \sqrt{(z^2 - t_2^2 + t_1^2)^2 - 4 z^2 t_1^2}}{2 z t_1^2} \right]
\end{equation}
and all odd moments vanish by symmetry of the spectral function, $\mathcal{A}(-\omega) = \mathcal{A}(\omega)$.

The spectrum is then altered with the addition of a pole at $\omega=0$ of weight $w_p$ and the weight of the bands is lowered by the same amount to preserve the spectral sum rule. The spectral weight of the added pole is chosen to match the weight of the SSH pole,
%\index{$0$@\textbf{List of Edits}!302@fixed wording}
\begin{equation}
	w_p = \frac{t_1^2 - t_2^2}{t_1^2} \,.
\end{equation}
The moments for this new spectral function are the sum of the moments of the constituent parts, so that the total moment for the new spectrum is obtained from
\begin{equation}
	\twiddle{\mu}_{2n} = \twiddle{\mu}_{2n}^{p} + \twiddle{\mu}_{2n}^{b}
\end{equation}
where the constituent moments are
\begin{subequations}
\begin{align}
	\twiddle{\mu}_{2n}^{p}
	&=	\int \omega^{2n} w_p \delta(\omega) \d\omega
	\intertext{for the pole, and}
	\twiddle{\mu}_{2n}^{b}
	&=	\int \omega^{2n} (1-w_p) \mathcal{A}(\omega) \d\omega\end{align}
\end{subequations}
for the bands, where $\mathcal{A}(\omega)$ is the trivial SSH spectral function. The factor of $(1-w_p)$ comes from the SSH bands decreasing by an amount given by the weight of the zero pole. The bands of the SSH model in the topological phase are the same as those of the bands in the trivial phase scaled by the factor $(1-w_p)$.
Being at $\omega=0$, the pole only possesses a zeroth moment,
\begin{equation}
	\twiddle{\mu}_0^p = w_p \,.
\end{equation}
The zeroth moment of the bands is simply $1-w_p$ due to the normalization of the spectral function. The zeroth moment of the total spectrum is then
\begin{equation}
	\twiddle{\mu}_0 = 1 \,.
\end{equation}
The higher moments of the composite spectrum $\twiddle{\mu}$ are related to the moments of the initial trivial spectrum $\mu$ by
\begin{equation}
\begin{aligned}[b]
	\twiddle{\mu}_{2n} &= (1-w_p) \mu_{2n}
	\\
	&= \frac{t_2^2}{t_1^2} \mu_{2n}
\end{aligned}
\end{equation}
The hopping parameters associated to the composite spectrum moments can now be obtained from the moment recursion as
\begin{subequations}
\begin{align}
&\begin{aligned}[b]
	\twiddle{t}_{1}^2
	&=	\twiddle{X}_{2}(1)
	\\
	&=	\twiddle{\mu}_2
	\\
	&=	\frac{t_2^2}{t_1^2} \mu_{2}
	\\
	&=	\frac{t_2^2}{t_1^2} X_{2}(1)
	\\
	&=	\frac{t_2^2}{t_1^2} t_1^2
	\\
	&=	t_2^2
\end{aligned}
\intertext{and}
&\begin{aligned}[b]
	\twiddle{t}_2^2
	&=	\twiddle{X}_{4}(2)
	\\
	&=	\frac{\twiddle{X}_{4}(1)}{\twiddle{t}_1^2} - \frac{\twiddle{X}_{2}(0)}{\twiddle{t}_0^2}
	\\
	&=	\frac{\twiddle{\mu}_{4}}{\twiddle{t}_1^2} - \twiddle{\mu}_2
	\\
	&=	\frac{\frac{t_2^2}{t_1^2} {\mu}_{4}}{{t}_2^2} - \frac{t_2^2}{t_1^2} {\mu}_2
	\\
	&=	t_2^2 + t_1^2 - \frac{t_2^2}{t_1^2} t_1^2
	\\
	&=	t_1^2 \,.
\end{aligned}
\end{align}
\end{subequations}
As this calculation shows, the parity of the system hopping parameters is reversed, so that the system is now an SSH model in its topological phase.

A similar calculation can be done to insert two poles at $\pm\omega_p$ into the trivial SSH spectrum, as shown later.

\subsection{Continuum Field Theory}\label{sec:continuumfieldtheory}

Information on the nature of domain walls in SSH systems can be gathered from considering the solutions to the continuum quantum field theory corresponding to the SSH lattice model.
%Takayama--Lin-Liu--Maki (TLM) model~\cite{tlm}
%
%Gross--Neveu model~\cite{gn}\label{Gross-Neveu model}
%\begin{align}
%	H	&=	-\i \hslash v_F \psi^\dagger \boldsymbol{\sigma}_3 \partial_x \psi + g \boldsymbol{\sigma}_1 \Delta(x)
%\end{align}

The continuum quantum field theory of the SSH model, also known as the Takayama--Lin-Liu--Maki (TLM) model~\cite{tlm}, can be obtained by linearizing the Hamiltonian around the low-energy near-metallic limit. The metallic phase occurs at the phase space point where $t_B = t_A$, or with the parameterization $t_A = t_0 - t$ and $t_B = t_0 + t$, where $t = 0$. At this point the band crossing occurs at $k = \pi$ in the Brillouin zone. Expanding the Hamiltonian around $k = \pi$ to linear order in $k$ and $t$,
\begin{equation}
\begin{aligned}[b]
	h(k)
		&=	\left( t_A + t_B \cos(k) \right) \boldsymbol{\sigma}_1 - t_B \sin(k) \boldsymbol{\sigma}_2
	\\
	h(k-\pi)
%		&=	- t_B k \boldsymbol{\sigma}_2 + (t_A - t_B) \boldsymbol{\sigma}_1 + \mathcal{O}(k^2)
%	\\	&\equiv	\i \hslash t_B \boldsymbol{\sigma}_2 \partial_x + m(x) \boldsymbol{\sigma}_1
%	\\	&=	\lim_{\epsilon\to0} \frac{\partial}{\partial \epsilon} h_\epsilon(k-\pi)
%	\\	&=	\lim_{\epsilon\to0} \frac{\partial}{\partial \epsilon} \left[ \left( ( t_0 - \epsilon t) + (t_0 + \epsilon t) \cos(k) \right) \boldsymbol{\sigma}_1 - (t_0 + \epsilon t) \sin(k) \boldsymbol{\sigma}_2 \right]
%	\\	&=	\lim_{\epsilon\to0} \frac{\partial}{\partial \epsilon} \left[ t_0 ( (1 + \cos(k)) \boldsymbol{\sigma}_1 - \sin(k) \boldsymbol{\sigma}_2 ) + \epsilon t ( (-1+\cos(k)) \boldsymbol{\sigma}_1 - \sin(k) \boldsymbol{\sigma}_2 ) \right]
%	\\	&=	\adjustlimits\lim_{\epsilon\to0} \lim_{\eta\to0} \frac{\partial}{\partial \epsilon} \frac{\partial}{\partial \eta} \left[ t_0 ( (1 - 1) \boldsymbol{\sigma}_1 + \eta k \boldsymbol{\sigma}_2 ) + \epsilon t ( (-1-1) \boldsymbol{\sigma}_1 + \eta k \boldsymbol{\sigma}_2 ) \right] + \mathcal{O}(\eta^2)
%	\\
	&=	t_0 \boldsymbol{\sigma}_2 k - 2 t \boldsymbol{\sigma}_1 + \mathcal{O}(k^2)
	\\	&\approx	\i \hslash t_0 \boldsymbol{\sigma}_2 \partial_x - m(x) \boldsymbol{\sigma}_1
\end{aligned}
\end{equation}
where $m(x) \vcentcolon= t_B - t_A$.
This is of the form of a Dirac Hamiltonian with position dependent mass. This Hamiltonian may be analyzed considering a domain wall at $x=0$. In this case, the mass term interpolates between the two topologically distinct vacua of $t_B > t_A$ and $t_B < t_A$ parameterized by $m(x\to\pm\infty) = \pm m_0$ where $m_0 = 2 t$.

The Schr\"{o}dinger equation for the zero energy eigenstate reads as
\begin{equation}
\begin{aligned}[b]
	h \psi &= 0\cdot \psi
	\\
	\i \hslash t_0 \frac{\d}{\d x} \psi(x) - \i m(x) \boldsymbol{\sigma}_3 \psi(x) &= 0
	\\
	\i \hslash t_0 \frac{\d}{\d x} \begin{pmatrix} \psi_+(x) \\ \psi_-(x) \end{pmatrix} - \i m(x) \begin{pmatrix} \psi_+(x) \\ -\psi_-(x) \end{pmatrix} &= 0 \,.
\end{aligned}
\end{equation}
The second line follows from multiplying through by $\boldsymbol{\sigma}_2$.
This equation admits the formal solution of
\begin{equation}
	\psi(x) = \psi(0) \e^{-\frac{1}{\hslash t_0} \int_0^x m(x') \d x'}
\end{equation}
In the topological configuration, the boundary of the system may be interpreted as a domain wall with the vacuum. Since the vacuum is regarded to be topologically trivial, the mass profile $m(x)$ then interpolates between topologically trivial and non-trivial phases. A function which smoothly interpolates between two topological configurations is
\begin{equation}
	m(x) \sim \tanh\left( x \right)
\end{equation}
which then yields the formal solution for the wavefunction as
\begin{equation}
	\psi(x) = \psi(0) \sech\left( x \right) \,.
\end{equation}
$\lvert \psi (x)\rvert^2 \sim \sech^2(x)$ decays exponentially, in agreement with the transfer matrix calculation derived in \S\ref{sec:sshmodel}.

The preceding discussion can be made more quantitatively precise by beginning with an alternative formulation of the continuum model.
Rather than starting from the lattice SSH model directly, the starting point of the continuum model can instead be taken to be the Peierls-Fr\"{o}hlich Hamiltonian
\begin{equation}
	\hat{H} = \sum_{n} (t_0 + \alpha(u_{n} + u_{n+1})) ( \opd{c}{n+1} \op{c}{n} + \opd{c}{n} \op{c}{n+1} ) + \frac{K}{2} \sum_{n} (u_{n+1} - u_{n})^2
\end{equation}
which represents a lattice model with electron-phonon coupling. $K$ parameterizes the harmonic potential between the lattice distortions and $u_{n}$ represents the displacement of each atom from its equilibrium position. In this model, the staggered tunneling amplitudes of the SSH model manifest from the Peierls distortion. For physical \textit{trans}-polyacetylene, the values of these parameters are $t_0 = \SI{2.5}{\electronvolt}$, $\alpha = \SI{4.1}{\electronvolt/\angstrom}$, and $K = \SI{21.0}{\electronvolt/\angstrom^2}$. For the present situation the model is taken to be in the adiabatic limit, such that the phonon momentum $p_u$ is negligible, $p_u \approx 0$.
%\index{$0$@\textbf{List of Edits}!303@added missing definition

The continuum model is obtained from linearization around Fermi surface, $k_F = \frac\pi2$ in the Brillouin zone.
The continuum Hamiltonian is
\begin{equation}
	H = -\i \psi^\dagger \boldsymbol{\sigma}_3 \partial_x \psi + g \Delta \psi^\dagger \boldsymbol{\sigma}_1 \psi + \frac12 \Delta^2
\end{equation}
where $\Delta$ represents the phonon field. 
This continuum SSH model is formally equivalent to the static semiclassical field equations of $N_{f}=2$ \index{Gross-Neveu model}Gross-Neveu model~\cite{gn}
\begin{equation}
	\mathcal{L}_{\textsc{gn}} = \overline{\psi} \i \slashed{\partial} \psi + \frac{g}{2} \left( \overline{\psi} \psi \right)^2
\end{equation}
which is a relativistic quantum field theory that is defined in $1+1d$. The Gross-Neveu model can be subjected to
a Hubbard-Stratanovich transformation, which brings it into the form
\begin{equation}
	\mathcal{L}_{\textsc{gn}} = \overline{\psi} \i \slashed{\partial} \psi + g \Delta \overline{\psi} \psi - \frac12 \Delta^2 \,.
\end{equation}
This Lagrangian yields the massive Dirac equation
\begin{align}
	\left( \i \slashed{\partial} - g \Delta \right) \psi &= 0
\end{align}
with the self consistency relation
\begin{equation}
	\Delta = g \overline{\psi} \psi \,.
\end{equation}
The equivalency between the Gross-Neveu model and the previous SSH continuum theory can be seen by changing from the chiral basis to the Dirac basis and performing a chiral rotation of $\theta=\pi/4$.

An effective potential for the $\Delta$ field can be determined through semi-classical methods. 
%An effective action can be obtained from
%\begin{equation}
%	\e^{\i S_{\text{eff}}}
%	=
%	\int \mathcal{D}\psi \mathcal{D}\overline{\psi} \mathcal{D}\Delta~\e^{\i \int\d t \int\d x \left( \overline{\psi} (\i \slashed{\partial} - g \Delta) \psi - \frac12 \Delta^2 \right)}
%\end{equation}
%and integrating out the $\psi$ and $\overline{\psi}$ fields to leading order in $1/N$. This leads to a renormalized effective potential of
%\begin{equation}
%	V_{\text{eff}}(\Delta) = \frac12 \Delta^2 + \frac{g^2 N}{4\pi} \Delta^2 \left( \ln\frac{\Delta^2}{\Delta_0^2} - 1\right)
%\end{equation}
The $\Delta$ field possesses a vacuum expectation value $\langle \Delta \rangle_{0} \neq 0$.
%spontaneously breaks the symmetry
\begin{comment}
\begin{equation}
	\text{Tr}~\e^{-\i H T}
	=
	\int \mathcal{D}\psi \mathcal{D}\overline{\psi} \mathcal{D}\Delta~\e^{\i \int\d t \int\d x \left( \overline{\psi} (\i \slashed{\partial} - g \Delta) \psi - \frac12 \Delta^2 \right)}
\end{equation}

integrate out $\psi$
\begin{equation}
	\text{Tr}~\e^{-\i H T}
	\sim
	\int \mathcal{D}\Delta~\e^{\i \int\d t \int\d x \left( - \frac12 (\Delta^2 - \Delta_0^2) \right) - \i N \sum_i (\omega_i(\Delta) - \omega_i(\Delta_0)) - \i n_0 \omega_0(\Delta)}
\end{equation}

asymptotic boundary conditions
\begin{align}
	\Delta(+\infty) = \pm\Delta_0 = -\Delta(-\infty)
\end{align}

\begin{align}
	( \i \slashed{\partial} - g \Delta) \psi &= 0
	\\
	( \i \slashed{\partial} + g \Delta) ( \i \slashed{\partial} - g \Delta) \psi &= 0
	\\
	\frac{\d^2\psi}{\d x^2} - \left( g^2(\Delta^2 - \Delta_0^2) \pm g \frac{\d\Delta}{\d x} \right) \psi &= -( \omega^2 - g^2 \Delta_0^2) \psi
\end{align}
Eigenenergies
\begin{equation}
	k^2 \equiv \omega^2 - g^2 \Delta_0^2
\end{equation}
Potential
\begin{equation}
	u(x) \equiv g^2(\Delta^2 - \Delta_0^2) \pm g \frac{\d\Delta}{\d x}
\end{equation}

The solution for the $\Delta$ field may be found from inverse scattering methods. 

zero energy bound state

soliton electron bound state wavefunction is finite only on odd sites
\end{comment}
For the soliton with zero energy bound state and the boundary conditions
\begin{align}
	\Delta(+\infty) = \pm\Delta_0 = -\Delta(-\infty) \,,
\end{align}
the semi-classical methods yield the solution~\cite{gnsoliton,gnsemiclassical1,gnsemiclassical2,gnboundstate}
\begin{equation}
	\Delta(x) = \Delta_0 \tanh(x) \,.
\end{equation}
This solution allows the field equations to be solved exactly. The zero energy bound state takes the form
\begin{equation}
	\psi(x) = N_0 \sech(x)
\end{equation}
For $x\gg1$ the analytic soliton solution for $\psi(x)$ exhibits the same asymptotic exponential decay just as was obtained from the transfer matrix calculation.

%From the Atiyah-Singer index theorem the presence of zero energy modes implies that the $\Delta(x)$ takes a topologically non-trivial configuration.

Another admissible solution is that of a polaron, which is essentially a soliton--anti-soliton bound state,
\begin{equation}
	\Delta(x) = \tanh(x-x_0) - \tanh(x+x_0) \,.
\end{equation}
This solution arises from the boundary conditions
\begin{equation}
	\lim_{|x|\to\infty} \Delta(x,t) = |\Delta_0| \,.
\end{equation}
These boundary conditions state the asymptotic form of the $\Delta$ field takes on a value with the same sign in both the $x\to-\infty$ and $x\to+\infty$ asymptotic regimes. In other words, the polaron solution interpolates between a vacuum to another vacuum of the same type.
Unlike the soliton, this is a non-topological state. 
The electronic bound states appear in pairs at energies $\pm\omega_0 \neq 0$.

At a formal level, the relationship between the topological configuration of the phonon field and the presence of fermion zero modes can be obtained from the Atiya-Singer index theorem~\cite{atiyahsinger}. The full technicalities of this theorem and its proof are beyond the scope of this thesis, so here the theorem will be stated in a form relevant to the current discussion and the proof omitted.
The theorem states that for an elliptic differential operator $D$ on a compact oriented differentiable manifold $X$, it follows that
\begin{equation}
	\mathrm{Analytic\ Index\ of\ } D = \mathrm{Topological\ Index\ of\ } X
\label{eq:atiyahsinger}
\end{equation}
where the analytical index is given by $\dim(\ker(D)) - \dim(\ker(D^*))$, \textit{i.e.} the difference in dimensions of the kernel and cokernel of the operator $D$, and the topological index parameterizes the topological (non-)triviality of $X$, where an index of zero implies trivial topology and a non-zero index implies non-triviality. Recall that the kernel of a linear operator $L$ is the set of elements $v$ which satisfy $L v = 0$.

\begin{comment}%%%%%%%%%%%%%%%%%%%%%%%%%%
\begin{theorem}
	For an elliptic pseudodifferential operator $D$ on a compact oriented differentiable manifold $X$
	\begin{equation}
		\mathrm{Analytic\ Index\ of\ } D = \mathrm{Topological\ Index\ of\ } X
	\end{equation}
	The analytic index is given by $\dim(\ker(D)) - \dim(\ker(D^*))$
	\\
	The topological index measures the topology of X
\end{theorem}
\begin{corollary}
	For a Dirac operator, as in \textrm{e.g.} the SSH/Gross-Neveu model with Dirac operator \[D = -\i \hslash t_0 \boldsymbol{\sigma}_2 \partial_x + \boldsymbol{\sigma}_1 \Delta(x), \] zero-energy eigenstates exist if and only if the $\Delta(x)$ is in a non-trivial topological configuration.
\end{corollary}
\end{comment}%%%%%%%%%%%%%%%%%%%%%%%%%%%%
In the case of the SSH or Gross-Neveu models, the elliptic differential operator $D$ and its adjoint $D^*$ are
\begin{align}
	D &= -\i \hslash t_0 \boldsymbol{\sigma}_2 \partial_x + \boldsymbol{\sigma}_1 \Delta(x)
	&
	&\text{and}
	&
	D^* &= \i \hslash t_0 \boldsymbol{\sigma}_2 \partial_x + \boldsymbol{\sigma}_1 \Delta(x)
\end{align}
which act as
\begin{align}
	\left[ -\i \hslash t_0 \boldsymbol{\sigma}_2 \partial_x + \boldsymbol{\sigma}_1 \Delta(x) \right] \psi(x) &= 0
	&
	&\text{and}
	&
	\left[ \i \hslash t_0 \boldsymbol{\sigma}_2 \partial_x + \boldsymbol{\sigma}_1 \Delta(x) \right] \psi(x) &= 0 \,.
\end{align}
For $\displaystyle \psi(x) \sim \begin{pmatrix} 1 \\ 0 \end{pmatrix}$, which represents localization on only one sublattice, it follows that $\boldsymbol{\sigma}_3 \psi = + \psi$ and
\begin{equation}
	\psi(x) = \mathcal{N} \psi(0) \exp\left[-\frac{1}{t_0 \hslash} \int_{0}^{x} \Delta(x') \d x'\right]
	\in
	\ker(D) \,.
\label{eq:psiinkernel}
\end{equation}
For localization on only one sublattice, $\psi$ is unnormalizable for $D^*$, and therefore $\dim(\ker(D^*)) = 0$. From \eqref{eq:psiinkernel} $\dim(\ker(D))$ is finite and therefore the analytic index of $D$ is non-zero. By the Atiyah-Singer index theorem \eqref{eq:atiyahsinger}, the underlying topology is then non-trivial.

%In the case of two domain walls, the localized states have opposite boundary conditions. 
On the other hand, in the case of two domain walls, there is no zero energy state for either $D$ or $D^*$. This means that $\ker(D) = \emptyset = \ker(D^*)$, and so the analytic index $= 0$. This is the situation encountered in the polaron solution as mentioned previously. The Atiyah-Singer index theorem \eqref{eq:atiyahsinger} then concludes that this configuration is topologically trivial.

The topology can also be identified from the topological charge~\cite{marino} as
\begin{align}
	Q	&=	\frac12 \int_{-\infty}^{\infty} \partial_x \Delta(x) \d x
		\\
		&= \frac12 \left[ \Delta(+\infty) - \Delta(-\infty) \right]
\end{align}
which is quantized in terms of the asymptotic vacuum expectation values of the field $\Delta$. 
The topological charge is the integral over the zeroth component of the topological current $j = \star \d \Delta$, which is $1+1d$ in the present case.\footnote{In vector index notation, the topological current is expressed as $j^\mu = \frac12\epsilon^{\mu\nu} \partial_\nu \Delta$. It follows that $j^0 = \frac12 \epsilon^{01} \partial_1 \Delta$.}

%Number of fermion zero modes $=$ topological index

The main result of this section is that the phonon field configuration yielding a domain wall which minimizes the energy takes on a spatial $\tanh$ form. In considering the Peierls-Fr\"ohlich Hamiltonian and interpreting it in terms of a Hamiltonian of SSH-type, it reveals that the width of the $\tanh$ spans over several lattice sites. While the field theoretic derivation of the $\tanh$ form of the domain wall is well-known from the literature~\cite{solitons,marino,gnsoliton,gnsemiclassical1,gnboundstate}, it's employment as the envelope of a domain wall in the lattice SSH model has not previously been reported. 
%\index{$0$@\textbf{List of Edits}!304@clarified literature vs new result}

%%%%%%%%%%%%%%%%%%%%%%%%%%%%%%
\subsection{Single Domain Wall}
\label{sec:singledw}

The derivation that a bulk domain wall within an SSH-type system is parameterized by a $\tanh$ envelope enables the construction of a Hamiltonian for such a system of the form
\begin{equation}
	\hat{H}_{\textsc{dw}} = \sum_{n} \tensor*{t}{_n} \opd{c}{n+1} \op{c}{n} + \hc
\end{equation}
where
\begin{equation}
	t_n	=	t_0 + (-1)^{n} \delta t~\tanh\left(\frac{n + \phi}{\alpha}\right) \,.
\label{eq:singledwtn}
\end{equation}
Here $\phi$ is the parameter which determines the location of the domain wall within the chain. Envelopes of other functional form still result in a localized state on the domain wall, however the nonzero energy bands in the spectral function then become distorted and do not take the form of the SSH bands, whereas their shape is preserved with the $\tanh$ envelope. An envelope which is not a finely-tuned $\tanh$ function produces a `fractal' spectrum, where a multitude of bands form with exponentially small gaps between them and the overall spectrum of these bands forms a distinct shape. Hamiltonians with non-$\tanh$ envelopes without domain walls will be discussed in \S\ref{sec:largeunitcells}.

The node of the $\tanh$ envelope is in general incommensurate with the integer-indexed hopping amplitudes of the lattice, meaning that in general $\phi\notin\mathbbm{Z}$.
%\begin{figure}[h]
%\centering
%\includegraphics[]{tanht.pdf}
%\end{figure}
The location of the domain wall from the boundary of the chain is proportional to $\ln(-\Delta \omega_p)$ where $\Delta \omega_p$
%\index{$0$@\textbf{List of Edits}!305@clarified notation}
is the distance between the two mid-gap poles.

%The $\tanh$ envelope ensures that the bulk bands are of the same functional shape as the standard SSH bands. This is a consequence of the soliton solution from the continuum theory.

%first order $\tanh$
%additional correction terms for large gap

This system can be deformed adiabatically to shift the position of the domain wall by smoothly tuning $\phi$.
By taking the limit $\phi\to\infty$, the domain wall can be propagated to infinity which in effect results in a topological phase transition without bulk gap closing. The system is then left with only a simple domain wall on the boundary with the vacuum. As $\phi\to\infty$, $\Delta \omega_p \to 0$
%\index{$0$@\textbf{List of Edits}!305@clarified notation}
, so that a simple pole at zero emerges.
A different example of a topological phase transition without gap closing involves symmetry breaking~\cite{ezawa}.

\begin{figure}[h]
\centering
\begin{subfigure}{0.49\linewidth}
\includegraphics[scale=1]{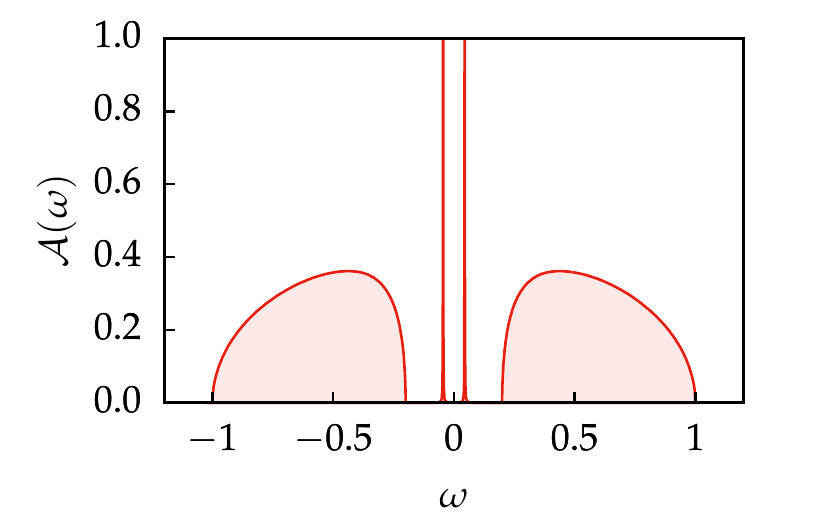}
\end{subfigure}
\begin{subfigure}{0.49\linewidth}
\includegraphics[scale=1]{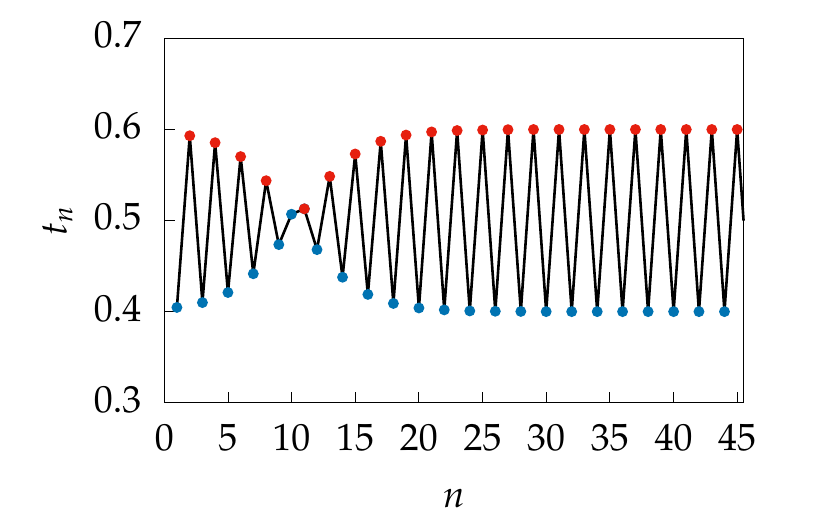}
\end{subfigure}
\caption{Plotted here is the spectral function and hopping parameters for an SSH model with a domain wall parameterized according to Eq.~\eqref{eq:singledwtn} with $\phi=9.35$ and $\alpha=\frac{t_0}{\delta t} = 5.0$.\label{fig:tanhdw}}
\end{figure}

\begin{figure}[h]
\centering
\begin{subfigure}{0.47\linewidth}
\includegraphics{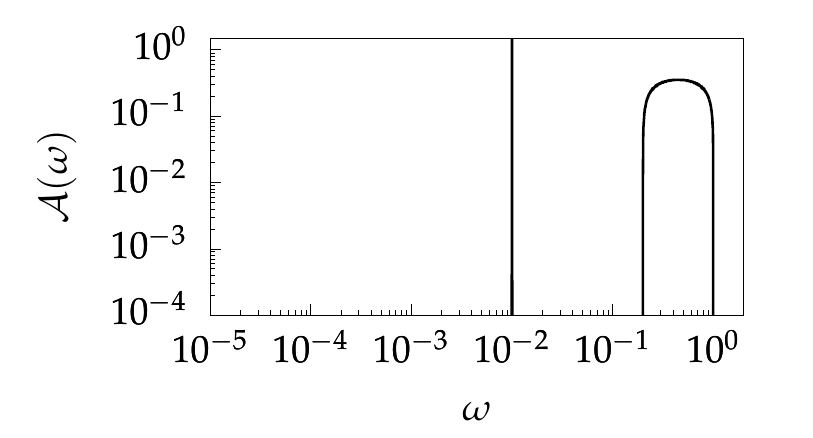}
\end{subfigure}
\begin{subfigure}{0.52\linewidth}
\includegraphics{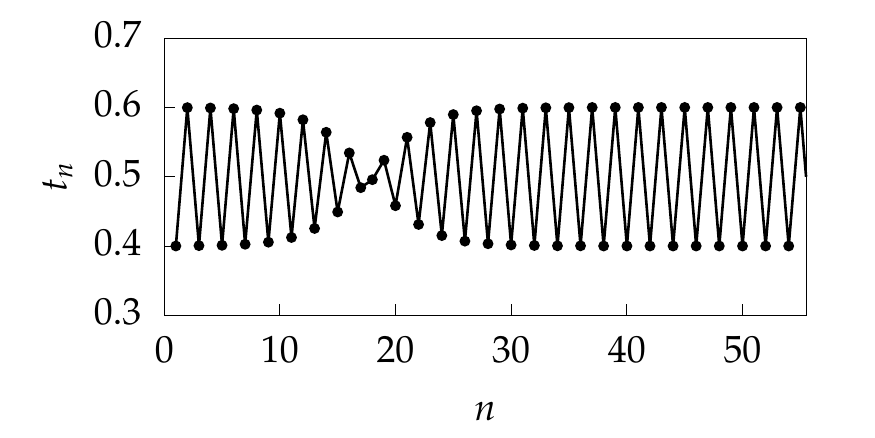}
\end{subfigure}
\begin{subfigure}{0.47\linewidth}
\includegraphics{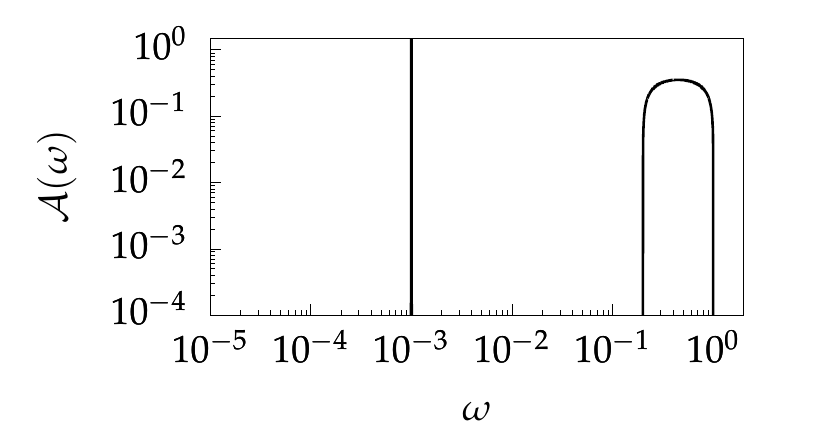}
\end{subfigure}
\begin{subfigure}{0.52\linewidth}
\includegraphics{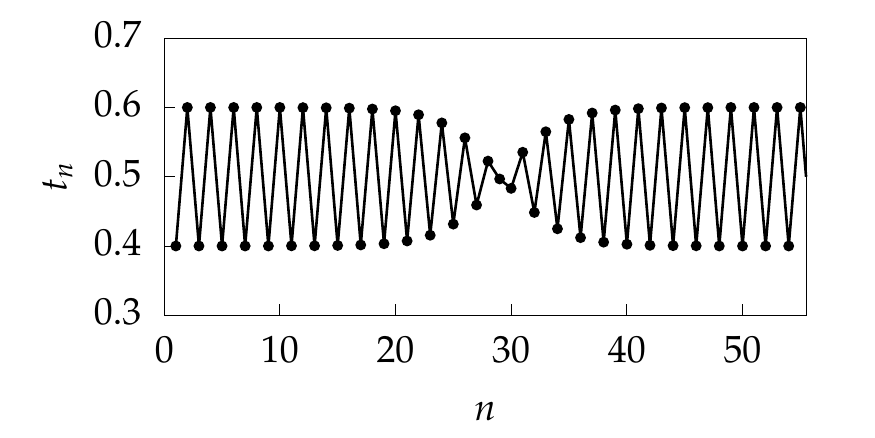}
\end{subfigure}
\begin{subfigure}{0.47\linewidth}
\includegraphics{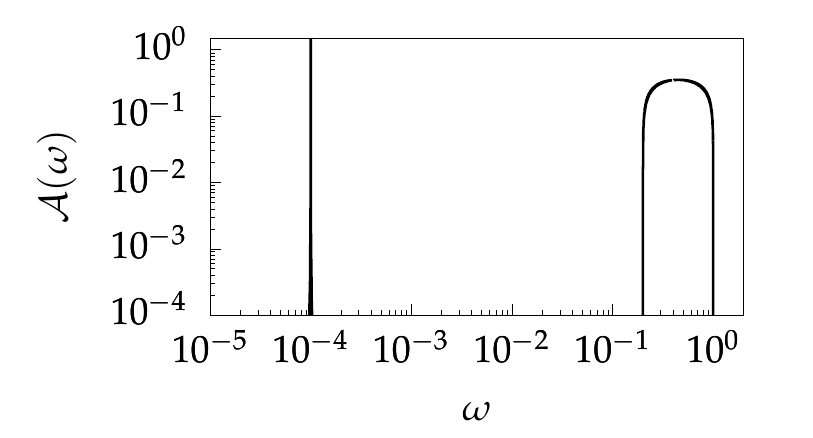}
\end{subfigure}
\begin{subfigure}{0.52\linewidth}
\includegraphics{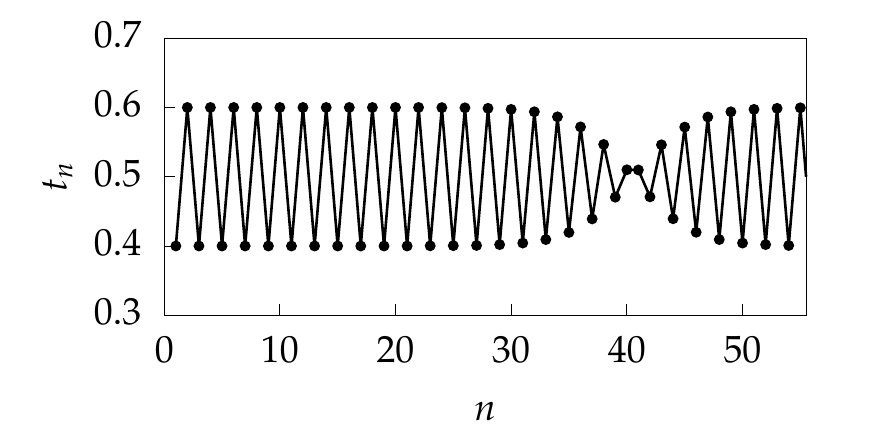}
\end{subfigure}
\caption{Correspondence between the position of the mid-gap poles and the depth from the boundary of the domain wall.\label{fig:dwdistance}}
\end{figure}

\begin{figure}[h]
\includegraphics{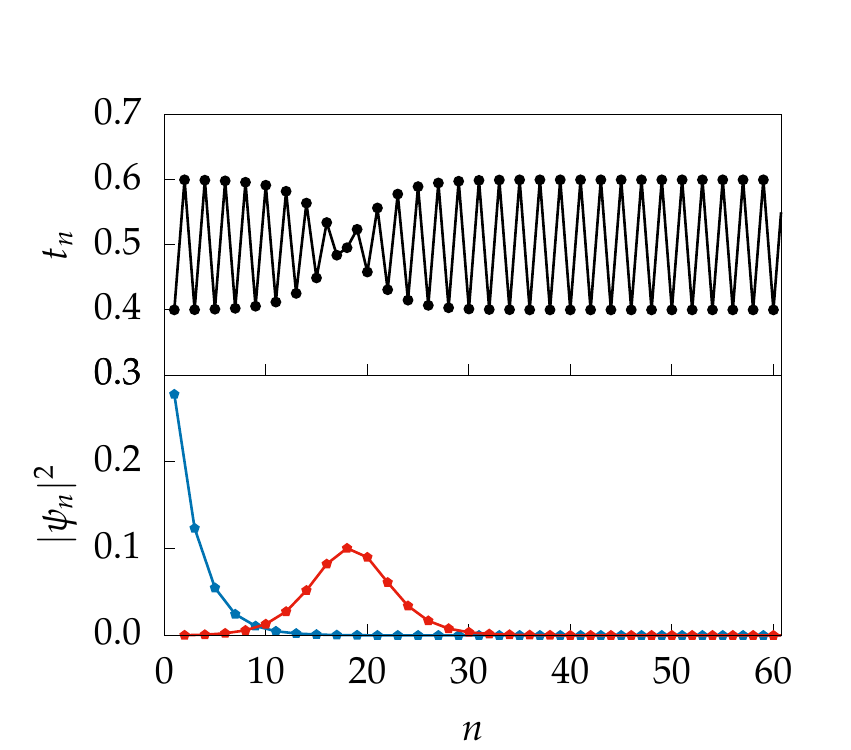}
\includegraphics{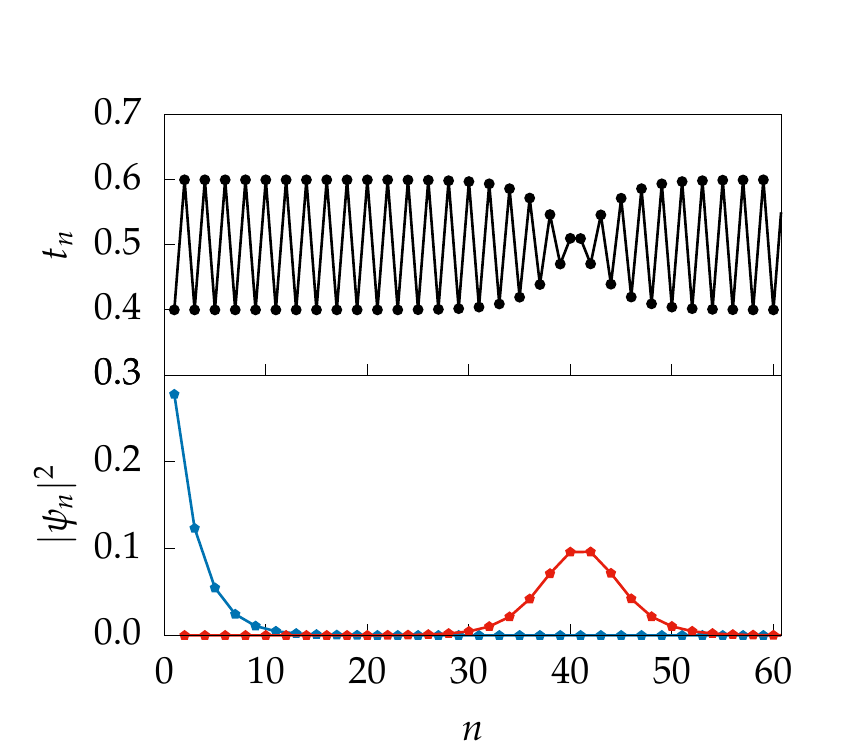}
\caption{Wavefunction of the boundary state localized on the odd sites (blue), and the wavefunction localized on the domain wall with support on the even sites (red).}
\end{figure}

A SSH model in the topological phase with a single domain wall will have two poles in its spectral gap. This configuration is equivalent to the soliton--anti-soliton bound state solution obtained in the continuum field theory.

\subsection{Repeated Domain Walls}\label{sec:repeateddws}

Adding an additional domain wall is equivalent to adding an additional pole in the gap of the spectral function.
It is instructive to analyze this situation from the reverse, by constructing a system which exhibits the gap and satellite bands of the SSH model, but which has an arbitrary configuration of poles lying within the gap.
The task is then to generate a set of parameters $\{\tilde{t}_n\}$ which produces the appropriate tight-binding chain with such a spectral function. 
Using the relationship between the presence of a domain wall in the chain and the mid-gap poles, an ansatz can be chosen for the form of the $\{\tilde{t}_n\}$ with some parameterized dependence on the location of the added spectral poles.

\subsubsection{Effective Hamiltonian}
\label{sec:lanczos}
The Hamiltonian describing the semi-infinite generalized SSH model with multiple mid-gap poles requires knowing the positions of the domain walls. These positions can be obtained by first analyzing an effective Hamiltonian which describes only the mid-gap states.

The mid-gap poles can be described by an effective Hamiltonian whose spectral function consists of only the intragap poles. 
This effective Hamiltonian can be written in a diagonal representation as
\begin{equation}
	\boldsymbol{H}_{\text{D}} = \boldsymbol{H}_{\text{D}}(\{\omega_{p}\}) = \begin{pmatrix} \omega_{p_1} & \phantom{\ddots} & O \\ \phantom{\ddots} & \omega_{p_2} \\ O & & \ddots \end{pmatrix}
\label{eq:diag}
\end{equation}
where the $\omega_{p_j}$ label the position of the intragap poles. To preserve chiral symmetry it is required that $\omega_{p_{2j}} = - \omega_{p_{2j-1}}$.
%\index{$0$@\textbf{List of Edits}!306@added missing minus sign}
The Green function for each of these uncoupled sites is simply
\begin{equation}
	%G_D(z) = 
	\Green{\op{c}{j}}{\opd{c}{j}}_z = \frac{1}{z - \omega_{p_j}}
\end{equation}
with associated spectral function $\mathcal{A}_j(\omega) = \delta(\omega-\omega_{p_j})$.
On the other hand, the spectral function describing the total collection mid-gap poles is given by
\begin{equation}
	\widetilde{\mathcal{A}}(\omega) = \sum_{j=1}^{N} \lvert u_{1p_j} \rvert^2 \delta(\omega-\omega_{p_j})
\end{equation}
where the sum runs over the total number of mid-gap poles, $N = \dim(\boldsymbol{H}_D)$, and $\lvert u_{1p_j} \rvert^2$ determines the weight of the pole at $\omega_{p_j}$.
The weight of these poles sums to the weight of the mid-gap spectral pole of the SSH model in the topological phase, 
\begin{equation}
	\sum_{j=1}^{N} \lvert u_{1p_j} \rvert^2 = \frac{t_B^2 - t_A^2}{t_B^2} \equiv w_p \,.
\label{eq:totalpoleweight}
\end{equation}

The spectral function $\mathcal{A}(\omega)$ can be regarded as being obtained from some Green function as $\mathcal{A}(\omega) = -\frac1\pi \Im \Greenline{\op{f}{1}}{\opd{f}{1}}_{\omega+\i0^+}$. In order to ensure that $\Greenline{\op{f}{1}}{\opd{f}{1}}_{z}$ can properly be considered a Green function of a system, its spectral function must be appropriately normalized, $\int\d\omega \mathcal{A}(\omega) \overset{!}{=} 1$. This requires scaling the Green function by the total weight of the poles, $w_p$, from Eq.~\eqref{eq:totalpoleweight}.
From the diagonal representation \eqref{eq:diag}, this Green function may be written as
\begin{equation}
	\Green{\op{f}{1}}{\opd{f}{1}}_{z} = \frac{1}{w_p} \sum_{j=1}^{N} \frac{\lvert u_{1p_j} \rvert^2}{z - \omega_{p_j}} \,.
\end{equation}
This Green function can also be expressed in terms of a Hamiltonian in tridiagonal form as
\begin{equation}
	\boldsymbol{H}_{\text{T}} = \boldsymbol{H}_{\text{T}}(\{\tilde{t}_n\}) = \begin{pmatrix} 0 & \tilde{t}_1 & \phantom{\ddots} & O \\ \tilde{t}_1 & 0 & \tilde{t}_2 & \phantom{\ddots} \\ & \tilde{t}_2 & 0 & \ddots \\ O & \phantom{\ddots} & \ddots & \ddots \end{pmatrix} \,,
\label{eq:tridiag}
\end{equation}
which corresponds to a $1d$ chain model for the mid-gap states.
The relation between Eq.~\eqref{eq:diag} and Eq.~\eqref{eq:tridiag} is
\begin{equation}
	\boldsymbol{U} \, \boldsymbol{H}_{\text{D}} \, \boldsymbol{U}^\dagger = \boldsymbol{H}_{\text{T}}
\end{equation}
where $\boldsymbol{U}$ is a unitary matrix with the form
\begin{equation}
	\boldsymbol{U} = \begin{pmatrix} u_{1p_1} & u_{1p_2} & \cdots & \\ u_{2p_1} & u_{2p_2} & \cdots & \\ \vdots & \vdots & \ddots \\ \ \end{pmatrix} \,.
\end{equation}
The transformation 
From the equations of motion the Green function in the tridiagonal basis takes the form of a continued fraction
\begin{equation}
	%G_T(z) = 
	\Green{\op{f}{1}}{\opd{f}{1}}_z = \cfrac{1}{z - \cfrac{\tilde{t}_1^2}{z - \cfrac{\tilde{t}_2^2}{z - \ddots}}} \,.
\end{equation}
%
%\begin{equation}
	%G_T(z) = 
%	\Green{\op{f}{1}}{\opd{f}{1}}_z = \mathop{\vcenter{\hbox{\huge K}}}_{p=1}^{N} \frac{-\tilde{t}_{p-1}^2}{z}
%\end{equation}$\tilde{t}_0 = -1$
%
The tridiagonal basis of the Hamiltonian can be constructed from the diagonal basis by means of the Lanczos algorithm~\cite{lanczos}\index{Lanczos algorithm}. 
The Lanczos algorithm is a procedure which iteratively constructs an orthogonal basis $\{|\psi_k\rangle\}$ such that the transition elements $\langle \psi_{m} \rvert \hat{H} \lvert \psi_{n} \rangle$ yield the desired amplitudes.

\begin{comment}
\[Psi] = Table[0, Length[\[Omega]p] + 1];
\[Psi] = Insert[\[Psi], up, 2];
v = Table[0, Length[\[Omega]p] + 2];
t = Table[0, Length[\[Omega]p]];
Do[
 v[[j]] = \[Psi][[j]]/Norm[\[Psi][[j]]];
 \[Psi][[j + 1]] = 
  dH . v[[j]] - v[[j]] (v[[j]] . dH . v[[j]]) - 
   v[[j - 1]] (v[[j - 1]] . dH . v[[j]]);
 t[[j - 1]] = Norm[\[Psi][[j]]];
 , {j, 2, Length[\[Omega]p] + 1}]
\end{comment}

The space of these orthonormal states for a $L$-dimensional Hamiltonian is a Krylov space $\mathcal{K}^L(|\psi_0\rangle) = \mbox{span}\left\{|\psi_0\rangle,\hat{H}|\psi_0\rangle,\hat{H}^2|\psi_0\rangle,\ldots,\hat{H}^L|\psi_0\rangle\right\}$. 
The Krylov space is initialized with the normalized state
\begin{equation}
	|\psi_0\rangle = \frac{1}{w_p} \sum_{i} u_{1p_i} \opd{c}{i} \lvert 0 \rangle \,.
\end{equation}
%which after normalization takes the form
%\begin{equation}
%	|\psi_0\rangle = \begin{pmatrix} u_{1p_1} \\ u_{1p_2} \\ \vdots \end{pmatrix} \,.
%\end{equation}
After this initialization, the first step of the algorithm is to construct a normalized state $\lvert \psi_{1} \rangle$ which is orthogonal to the initial state $\lvert \psi_{0} \rangle$, $\langle \psi_{0} \vert \psi_{1} \rangle \overset{!}{=} 0$. This can be achieved with the ansatz
\begin{equation}
	b_1 \lvert \psi_{1} \rangle = \hat{H} \lvert \psi_{0} \rangle - \lvert \psi_{0} \rangle \langle \psi_{0} \rvert \hat{H} \lvert \psi_{0} \rangle
\label{eq:lanczos1}
\end{equation}
which trivially yields orthogonality as observed from contraction with $\langle \psi_{0} \rvert$ and the normalization of the initial state $\langle \psi_{0} \vert \psi_{0} \rangle = 1$. The coefficient $b_1$ is introduced to ensure that the state $\lvert \psi_1 \rangle$ is normalized, $\langle \psi_1 \vert \psi_1 \rangle = 1$. It can be determined by contraction of \eqref{eq:lanczos1} with $\langle \psi_{1} \rvert$, which yields
\begin{equation}
	b_1 = \langle \psi_{1} \rvert \hat{H} \lvert \psi_{0} \rangle
\end{equation}
due to the orthogonality of $\lvert \psi_{0} \rangle$ and $\lvert \psi_{1} \rangle$.

The second state to be constructed, $\lvert \psi_{2} \rangle$, is required to be orthogonal to both the previous states, $\langle \psi_{0} \vert \psi_{2} \rangle \overset{!}{=} 0$ and $\langle \psi_{1} \vert \psi_{2} \rangle \overset{!}{=} 0$. This state can be constructed in a manner analogous to the previous step by defining $\lvert \psi_{2} \rangle$ as
\begin{equation}
	b_2 \lvert \psi_{2} \rangle = \hat{H} \lvert \psi_{1} \rangle - \lvert \psi_{1} \rangle \langle \psi_{1} \rvert \hat{H} \lvert \psi_{1} \rangle - \lvert \psi_{0} \rangle \langle \psi_{0} \rvert \hat{H} \lvert \psi_{1} \rangle \,.
\end{equation}
Once again the coefficient $b_2$ facilitates $\langle \psi_2 \vert \psi_2 \rangle = 1$ and is determined from the matrix element
\begin{equation}
	b_2 = \langle \psi_{2} \rvert \hat{H} \lvert \psi_{1} \rangle \,.
\end{equation}
The remainder of the Krylov space is produced in a similar fashion. The general form for the production of each state $n\geq1$ is
\begin{equation}
	b_{n} \lvert \psi_{n} \rangle = \hat{H} \lvert \psi_{n-1} \rangle - \sum_{m=0}^{n-1} \lvert \psi_{m} \rangle \langle \psi_{m} \rvert \hat{H} \lvert \psi_{n-1} \rangle \,.
\label{eq:lanczosrecursion}
\end{equation}
The general $b_n$ is given by $b_n = \langle \psi_{n} \rvert \hat{H} \lvert \psi_{n-1} \rangle$. For completeness, a set of parameters $a_n$ can be defined as $a_n = \langle \psi_n \rvert \hat{H} \lvert \psi_n \rangle$. In the present situation only nearest neighbor hopping is considered, so the terms $\langle \psi_{m} \rvert \hat{H} \lvert \psi_{n} \rangle \overset{!}{=} 0$ for $\lvert n - m \rvert > 1$.
The resulting set of parameters $b_{n}$ can be seen to be the hopping amplitudes between nearest neighbor sites.
The tridiagonal Hamiltonian matrix constructed from the Lanczos algorithm is
\begin{equation}
	\boldsymbol{H}_T	=
	\begin{pmatrix}
		a_0 & b_1 &  & & O \\
		b_1 & a_1 & b_2  \\
		& b_2 & a_2 & b_3  \\
		& & b_3 & a_3 & \ddots \\
		O & & & \ddots & \ddots
	\end{pmatrix} \,.
\end{equation}
For the particular case at hand, the Hamiltonian employed in the construction of the Krylov space \eqref{eq:lanczosrecursion} is $\hat{H}_D$. The resulting Hamiltonian constructed from the parameters $\{ a_n , b_n \}$ is the tridiagonal Hamiltonian $\hat{H}_T$.

%The remainder of the Krylov basis is formed from the recursion relation
%\begin{equation}
%	\tilde{t}_{k+1} |\psi_{k+1}\rangle = \boldsymbol{H}_{\text{D}} |\psi_{k}\rangle - |\psi_{k-1}\rangle \langle\psi_{k-1}| \boldsymbol{H}_{\text{D}} |\psi_{k}\rangle - |\psi_{k}\rangle \langle\psi_{k}| \boldsymbol{H}_{\text{D}} |\psi_{k}\rangle
%\end{equation}
%for $k=0$ to $L-1$ given the boundary condition $|\psi_{-1}\rangle \equiv 0$. 

For a Hamiltonian $\hat{H}_D$ of dimension $N$, the Green function on the boundary of the system defined by the corresponding $\hat{H}_T$ produced by the Lanczos algorithm satisfies
\begin{equation}
	\Green{\op{f}{1}}{\opd{f}{1}}_z = \sum_{j=1}^{N} \left| u_{1j} \right|^2 \Green{\op{c}{j}}{\opd{c}{j}}_z \,.
\end{equation}

The Lanczos method is an iterative procedure which crucially depends on the correct normalization and orthogonality of all states at each iteration. As such it potentially suffers from compounding numerical errors with increasing iteration number. In the present context the number of iterations is equal to the number of poles to be included in the spectral gap. For most cases considered in this analysis this number is small, $L<10$. However the Lanczos algorithm was tested for cases considering $L \sim \mathcal{O}(100)$ mid-gap poles and the algorithm was found to be numerically stable and successfully transformed the Hamiltonian $\boldsymbol{H}_D(\omega_1 , \omega_2, \ldots, \omega_L)$ to $\boldsymbol{H}_T(\tilde{t}_1, \tilde{t}_2, \ldots, \tilde{t}_{L-1})$ with their corresponding spectral functions being matching as expected. Prescribing this number of poles within the spectral gap is an extremely high density which verges towards the situation of infinite domain walls producing a full band of states in the spectral gap. The main conclusion of this calculation is that for a modest number of mid-gap poles there are no significant numerical errors in the construction of $\boldsymbol{H}_T$ as the algorithm is stable even when considering iterations orders of magnitude greater than what is needed.

The utility of the Hamiltonian in tridiagonal form $\boldsymbol{H}_{\text{T}}$ is that its parameters $\{\tilde{t}_n\}$ relate to the positions of the domain walls in the full Hamiltonian.

\subsubsection{Construction of Chain Parameters}
For $M$ number of intragap poles, the chain parameters can be obtained by the expression
\begin{equation}
	t_n	=	t_0 + (-1)^{n} \delta t \left[\sum_{m=1}^{M-1} \tanh\left(\frac{n + \phi_m}{(-1)^{m+1} \alpha_m}\right) - \frac{1+(-1)^{M-1}}{2} \right]
\label{eq:tntanh}
\end{equation}
where the $m$ sum produces $M-1$ domain walls. Each domain wall lies at the crossing at the node of a $\tanh$ envelope. The $(-1)^{m+1}$ term in the $\tanh$ envelope is used to ensure that the domain wall flips the parity of the alternating hopping parameters. The last term in \eqref{eq:tntanh} is constant shift necessary to ensure that the alternating parameters are centered about $t_0$.
The parameter $\alpha_m$ is defined as
\begin{equation}
	\alpha_m = \frac{t_0}{\delta t} \,.
%	\beta_m &= \frac{\phantom{-}\xi\phantom{-}}{\frac{t_0}{\delta t}}
\label{eq:tanhparams}
\end{equation}
The specification of the $\alpha_m$ parameter was formulated empirically based on analyzing the numerical values of available input parameters. The index is left explicit as it in principle could be different for each domain wall, as will be seen in Eq.~\eqref{eq:4dwrescale}.

For a chain initialized with a weak bond, there is an additional pole at the end of the chain which therefore results in a final spectral function with $M$ poles. The use of a $\tanh$ envelope is prescribed according to the previous discussion on domain walls in the Gross-Neveu model. 
Domain wall states in SSH chains have localization length $\xi$. Two such states overlap and hybridize. This can be modeled as a tight-binding type tunneling $\tilde{t}$ between domain wall states. With $\tilde{t} \sim \e^{-d/\xi}$ with $d$ the separation between domain walls. Inverting this relationship results in $d \sim \xi \ln \tilde{t}$.
For a single domain wall and a topological boundary, the location of the domain wall in the effective chain goes as $\ln \tilde{t}$
The positions of the domain walls is obtained from the hopping parameters $\tilde{t}_n$ of the effective Hamiltonian describing the intragap poles,
\begin{equation}
	\phi_m = \frac{\alpha_m}{\xi} \sum_{j=1}^m \ln \tilde{t}_j
\label{eq:phifromt}
\end{equation}
which assumes the domain walls are well-separated and their contributions additive.

The expression for the $t_n$ describing the $\tanh$ envelope of domain walls, Eq.~\eqref{eq:tntanh} is only approximate. For a single domain wall, the functional form of the envelope is a $\tanh$. For multiple domain walls, these $\tanh$ functions overlap and the $t_n$'s which lie in the overlap region must be tuned very precisely in order to preserve the smooth continuous form of the high energy bands. This issue is amplified when the domain walls occur close to each other in the chain. The expression Eq.~\eqref{eq:tntanh} should be interpreted as an approximate form which is valid in the case where the bandwidth is small and the domain walls are a reasonable distance from each other.

%\begin{equation}
%	t_n	=	t_0 + (-1)^{n} \delta t \prod_{m=1}^M \tanh\left((-1)^{m+1}\left(\frac{\delta t}{t_0} n + \frac{\phi_m}{\xi}\right)\right)
%\end{equation}

This result is a generalization of the single domain wall case wherein a domain wall in the lattice corresponds to a soliton in the field theory

%\begin{figure}[h]
%\centering
%\includegraphics[]{bubblest.pdf}
%\end{figure}

This $\tanh$ envelope produces an approximate analytical expression for the Hamiltonian parameters $\{t_n\}$. 
For cases where the domain walls appear nearby in the chain, when the distance between $\phi_{m+1}$ and $\phi_m$ is small, care must be taken to ensure that their respective $\tanh$ envelopes do not affect each other. 
An exact set of parameters can be obtained from a moment analysis of the spectral function. Such a calculation will appear in \S\ref{ch:motttopology}.

In addition to constructing SSH models with a multitude of mid-gap poles, it is also possible to construct models whose spectral function exhibits mid-gap bands. 
%An example of this was already seen in \S\ref{sec:} where a repeated unit cell containing a single domain wall results in an SSH-like spectrum but with a mid-gap band rather than a pole. 
Analogously to the cases of many mid-gap poles, a generalized SSH model with infinitely repeating domain walls can result in a spectrum with multiple mid-gap bands. The domain walls host localized states which can be considered to be localized states on a superlattice.
Further discussion of this scenario will be postponed until \S\ref{sec:motttransition}, where systems of this type play a central role.

\subsubsection{Example: 4 Poles}

An example of this calculation strategy is the case where there are $N=4$ poles lying within the SSH band gap. In this example the SSH model parameters $t_A = t_0 - \delta t$ and $t_B = t_0 + \delta t$ will be parameterized with $t_0 = 0.5$ and $\delta t = 0.2$. For this parameterization the SSH pole weight is
\begin{equation}
	w_p = \frac{t_B^2 - t_A^2}{t_B^2} \approx 0.82 \,.
\end{equation}
The intragap poles for this example will be chosen to lie at $\omega_{p_1} = -\omega_{p_2} = 0.001$ and $\omega_{p_3} = -\omega_{p_4} = 0.0001$. The poles will be taken to have equal weight. The Hamiltonian of this system in the diagonal basis is
\begin{equation}
	\boldsymbol{H}_{\text{D}} = \begin{pmatrix} \omega_{p_1} & & & O \\ & \omega_{p_2} & & \\ & & \omega_{p_3} & \\ O & & & \omega_{p_4} \end{pmatrix} \,.
\end{equation}
The system described by this Hamiltonian is that of four decoupled sites with on-site potentials $\omega_{p_j}$. 
Following the notation of \S\ref{sec:lanczos}, the Green function on each site $j=1,\ldots,4$ is given by
\begin{equation}
	\Green{\op{c}{j}}{\opd{c}{j}}_{z} = \cfrac{1}{z - \omega_{p_j}}
\end{equation}
with total spectral function given by $\mathcal{A}(\omega) = \sum_{j=1}^{4} \lvert u_{p_j} \rvert^2 \delta(\omega - \omega_{p_j})$
where $u_{p_j} = \sqrt{\frac{w_p}{4}}$. Applying the Lanczos algorithm yields $N-1=3$ parameters in the tridiagonal $\hat{f}$ basis which have the numerical values $\tilde{t}_1 = 0.000710634$, $\tilde{t}_2 = 0.000696562$, and $\tilde{t}_3 = 0.00014072$. This now represents a system with still with four sites, but now with hybridizations between the sites but with vanishing on-site potentials.
In this basis the Hamiltonian is
\begin{equation}
	\boldsymbol{H}_{\text{T}} = \begin{pmatrix} 0 & \tilde{t}_1 &  & O \\ \tilde{t}_1 & 0 & \tilde{t}_2 &  \\  & \tilde{t}_2 & 0 & \tilde{t}_3 \\ O &  & \tilde{t}_3 & 0 \end{pmatrix} %= \begin{pmatrix} 0 & t_1 & 0 & 0 \\ t_1 & 0 & t_2 & 0 \\ 0 & t_2 & 0 & t_3 \\ 0 & 0 & t_3 & 0 \end{pmatrix}
\end{equation}
and the Green function for the left edge of the system is
\begin{equation}
\begin{aligned}[c]
	\Green{\op{f}{1}}{\opd{f}{1}}_z = \cfrac{1}{z - \cfrac{\tilde{t}_1^2}{z - \cfrac{\tilde{t}_2^2}{z - \cfrac{\tilde{t}_3^2}{z}}}}
\end{aligned}
\end{equation}
which recovers the correct spectral function $\mathcal{A}(\omega)$.
%as required satisfies $\Greenline{\op{f}{1}}{\opd{f}{1}}_z = \sum_{j=1}^{4} \Greenline{\op{c}{j}}{\opd{c}{j}}_{z}$. 

The $\tanh$ envelope of the hopping amplitudes can now be constructed according to Eq.~\eqref{eq:tntanh}. The position of the domain walls is found from Eq.~\eqref{eq:phifromt}, which are
%$\phi_1 = -7.24935$, $\phi_2 = -14.5187$, $\phi_3 = -23.3875$.
$\phi_1 = -15.3559$, $\phi_2 = -30.7542$, $\phi_3 = -49.5403$.

The functional form of the hopping parameters now can be expressed in the form
\begin{equation}
	t_n	=	t_0 + (-1)^{n} \delta t 
	\left[
	\tanh\left(\frac{n + \phi_1}{\alpha_1}\right)
	- \tanh\left(\frac{n + \phi_2}{\alpha_2}\right)
	+ \tanh\left(\frac{n + \phi_3}{\alpha_3}\right)
	\right] \,.
\label{eq:tntanhanalytic}
\end{equation}
The parameters are first chosen such that $\alpha_m$ take the constant form of Eq.~\eqref{eq:tanhparams}.
\begin{figure}[h]
\centering
\includegraphics[scale=1]{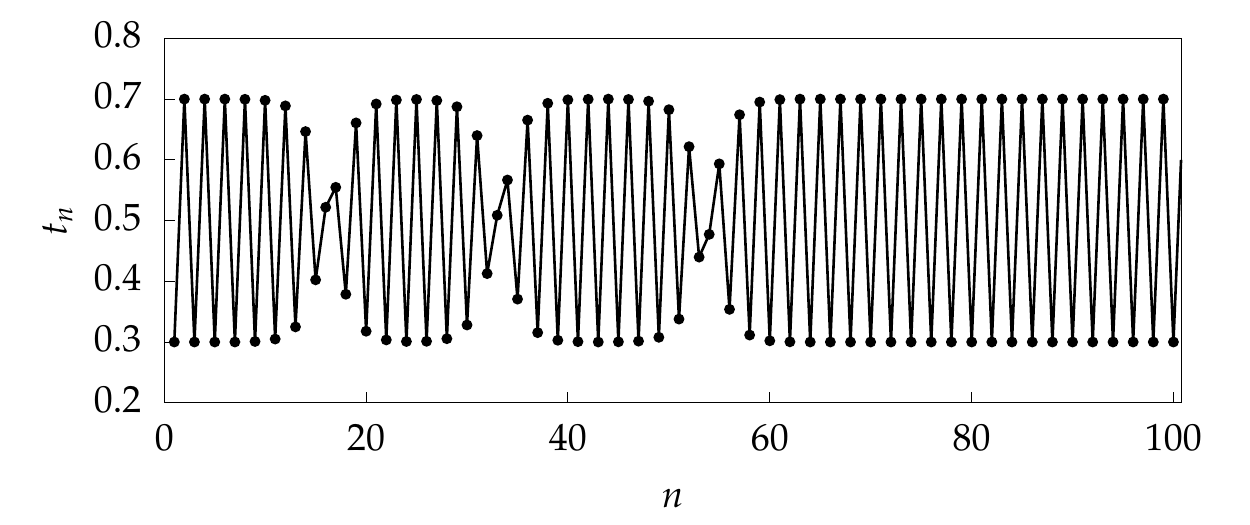}
\caption{Structure of the hopping amplitudes for mid-gap poles situated at $\omega = \pm 0.001, \pm 0.0001$ as derived from the moment expansion, which should be considered the exact result.}
\end{figure}
\begin{figure}[h]
\centering
\includegraphics[scale=1]{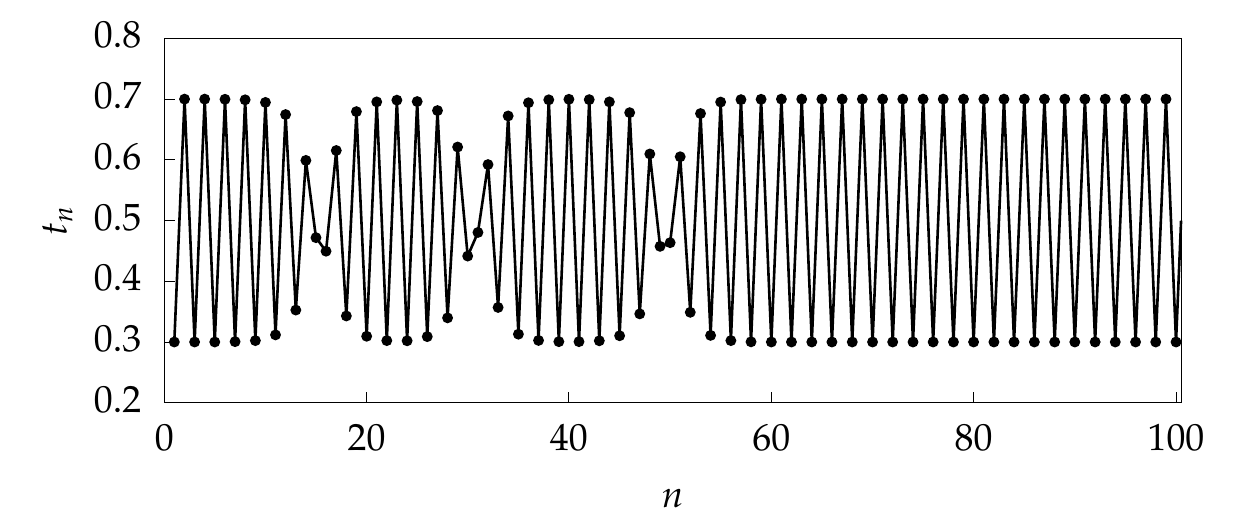}
\caption{Structure of the hopping amplitudes for mid-gap poles situated at $\omega = \pm 0.001, \pm 0.0001$ as prescribed by the toy model Eq.~\eqref{eq:tntanhanalytic}.}
\end{figure}
\begin{figure}[h]
\centering
\begin{subfigure}{0.49\linewidth}
\includegraphics[scale=1]{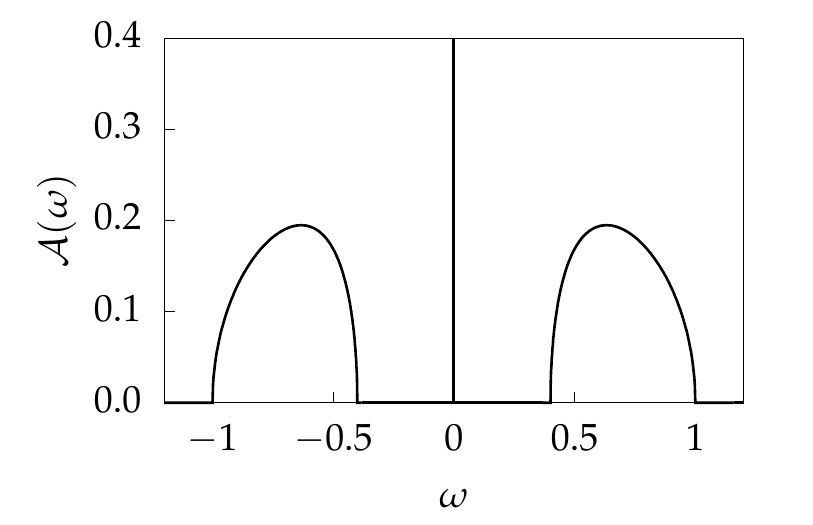}
\end{subfigure}
\begin{subfigure}{0.49\linewidth}
\includegraphics[scale=1]{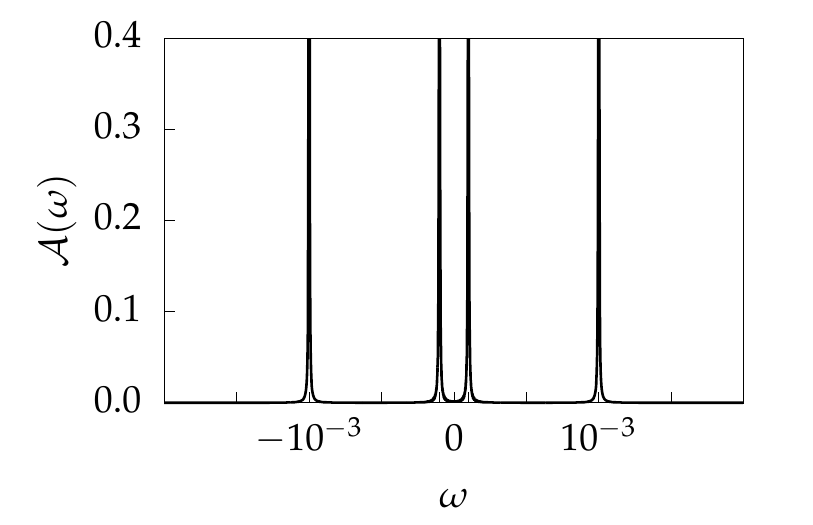}
\end{subfigure}
\caption{Spectrum of a generalized SSH model with a head consisting of three domain walls. The positions of the poles at $\pm 10^{-4}$ and $\pm 10^{-3}$ were prescribed as an initial condition and the chain parameters were calculated using the moment expansion.\label{fig:4polesme}}
\end{figure}
\begin{figure}[h]
\centering
\begin{subfigure}{0.49\linewidth}
\includegraphics[scale=1]{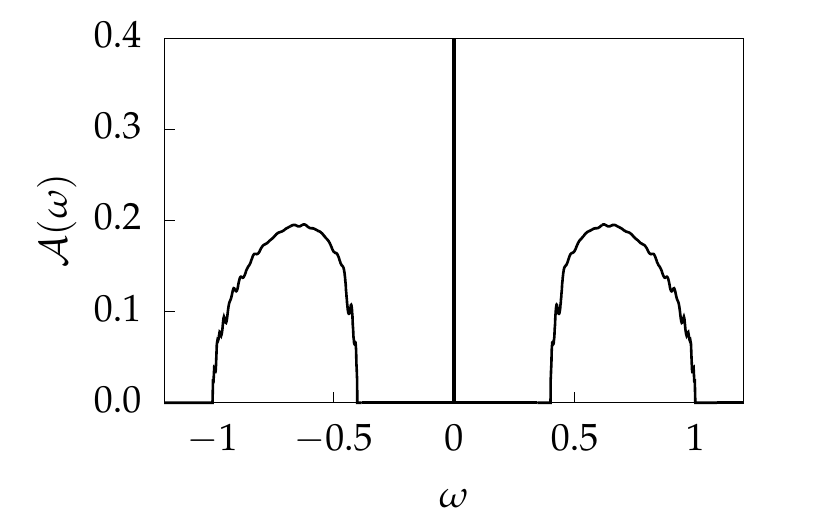}
\phantomsubcaption{\label{fig:4polestoy}}
\end{subfigure}
\begin{subfigure}{0.49\linewidth}
\includegraphics[scale=1]{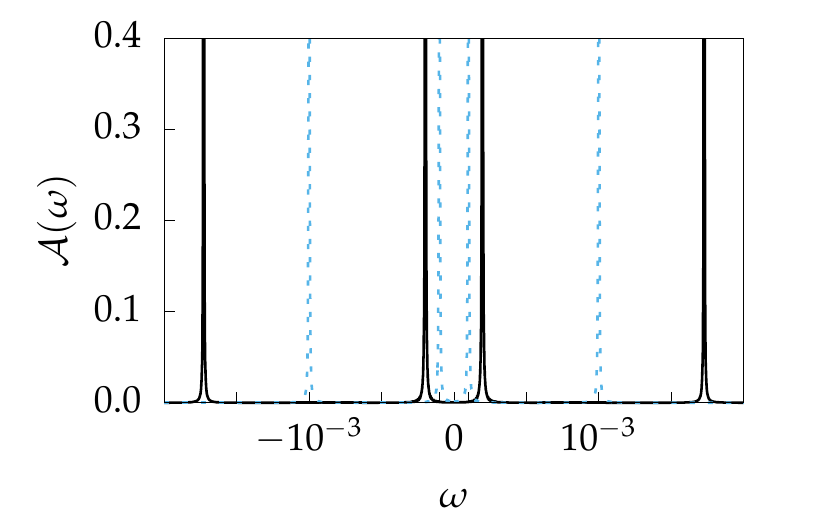}
\phantomsubcaption{\label{fig:4polestoyzoom}}
\end{subfigure}
\vspace{-\baselineskip}
\caption{Spectrum of the generalized SSH model generated from the model \eqref{eq:tntanh}. The model captures the main features of the desired spectrum with reasonable accuracy. Panel \subref{fig:4polestoyzoom} shows the comparison between the positions of the mid-gap poles generated by the toy model (black) and the true positions of the poles (dashed).\label{fig:4polestoyall}}
\end{figure}

As generated by the toy model prescribed in Eq.~\eqref{eq:tntanh}, the poles in the resulting spectrum are approximately located at $\omega = \pm 0.0001975$ and $\omega = \pm 0.001725$.
While the model specified by Eq.~\eqref{eq:tntanh} does not reproduce the location of the mid-gap poles exactly, it does reproduce the correct order of magnitude, with the location of the outer pair of poles being roughly one order of magnitude further out than the inner pair.

The outer bands of the spectrum display some spurious microstructure features coming from the inexactness of the domain wall $\tanh$ profiles, shown in Fig.~\ref{fig:4polestoy}, but otherwise accurately capture the correct shape of the SSH bands.

Some numerical massaging of the parameters yields an essentially exact result:
\begin{align}
	\alpha'_m &= \left\{ \frac{\alpha_1}{1.01} , \frac{\alpha_2}{1.02} , \frac{\alpha_3}{1.03} \right\}
%	&
%	\beta'_m &= \left\{ \frac{\beta_1}{1.06} , \frac{\beta_2}{1.077} , \frac{\beta_3}{1.085} \right\}
\label{eq:4dwrescale}
\end{align}
This rescaling of the $\alpha_m$ is performed by hand. The physically-motivated approximation to the true chain parameters is thus seen to be within $\sim1\%$ error. The advantage over the exact moment expansion is that it is simple and provides physical insight into the mechanism behind the formulation of domain walls and their relationship to mid-gap states.

\subsubsection{Unit Cells With Domain Walls}
\label{sec:ucdw}

A natural generalization of the insertion of domain walls into an SSH lattice is that of 
an extended SSH model where the unit cell spans many sites and there is a domain wall situated within the unit cell.

An example of such a set of $\{t_n\}$ is shown in Fig:~\ref{fig:tn_singledwunitcell}. The corresponding spectral function is shown in Fig.~\ref{fig:G_singledwunitcell}. This spectral function is qualitatively similar to that of the topological phase of the SSH model, with a zero energy feature located within a spectral gap.
Note however that the unit cell is initialized with a strong bond. In the absence of domain walls, this system would be adiabatically connected to the SSH model in its trivial phase.
\begin{figure}[h]
\centering
\begin{subfigure}{0.47\linewidth}
\begin{tikzpicture}
	\node at (0,0) {\includegraphics{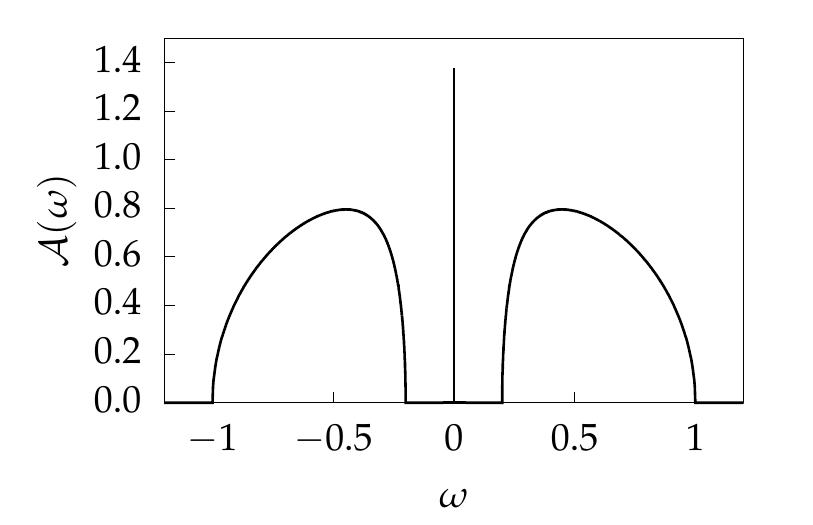}};
	\node at (3.125,2) {\footnotesize\subref*{fig:G_singledwunitcell}};
\end{tikzpicture}
\phantomsubcaption{\label{fig:G_singledwunitcell}}
\end{subfigure}
\begin{subfigure}{0.52\linewidth}
\begin{tikzpicture}
	\node at (0,0) {\includegraphics{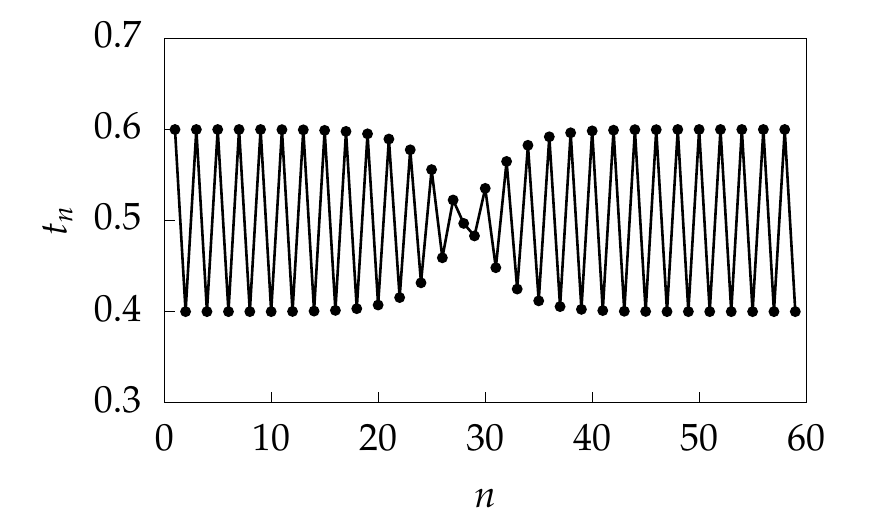}};
	\node at (3.5,2) {\footnotesize\subref*{fig:tn_singledwunitcell}};
\end{tikzpicture}
\phantomsubcaption{\label{fig:tn_singledwunitcell}}
\end{subfigure}
\vspace{-2\baselineskip}
\caption{Spectral function of a generalized SSH model \subref{fig:G_singledwunitcell} whose unit cell contains a domain wall, shown in \subref{fig:tn_singledwunitcell}.\label{fig:singledwunitcell}}
\end{figure}
The spectral function exhibits a mid-gap feature which is not a pole, but rather a band of finite width and spectral height. This characteristic is shown clearly in Fig.~\ref{fig:dwband}. The initial strong bond precludes the presence of a topological zero pole. The band is due to states localized on the domain walls hybridizing to each other.
\begin{figure}[h]
\includegraphics{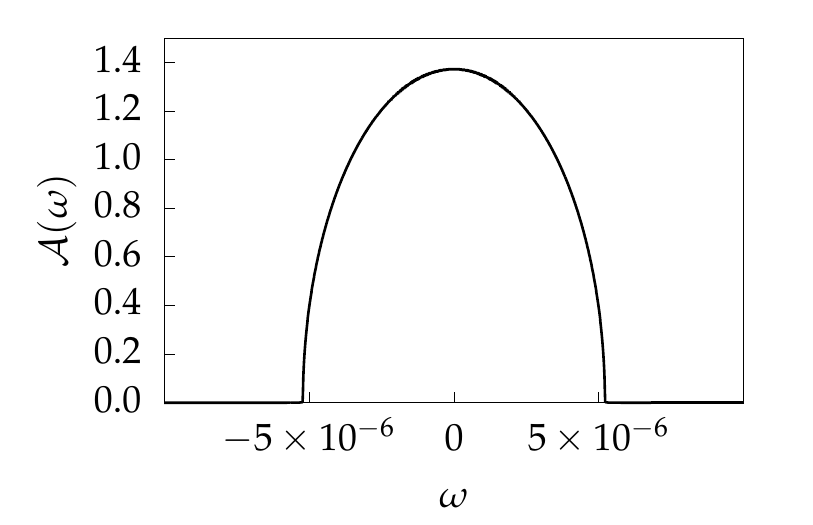}
\includegraphics{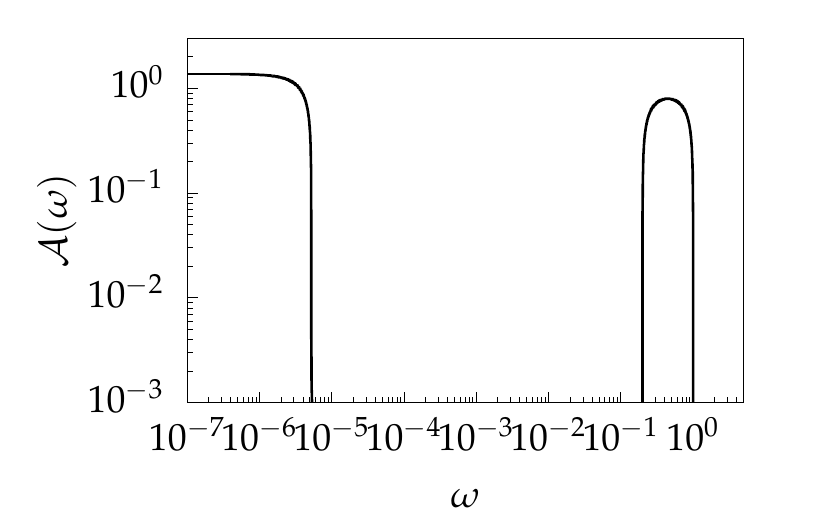}
\caption{Close up of the mid-gap feature in Fig.~\ref{fig:singledwunitcell}. As shown, the mid-gap feature is not a pole, but a well-defined band.\label{fig:dwband}}
\end{figure}
Since the domain walls host localized states, the domain walls can be considered to form a superlattice where the domain walls are the lattice sites and the overlap between wavefunctions localized on each domain wall parameterizes an effective superlattice hopping amplitude. Recall from \S\ref{sec:calcmeth} that for a homogeneous tight binding model the bandwidth is equal to $2t$. The very small width of the mid-gap band is indicative that the hopping amplitude of the superlattice is exponentially small, which is to be expected from the exponentially small overlap between wavefunctions localized on the domain walls.
As with the previous section, the effective tunneling is estimated as $\tilde{t} \sim \delta t \, \e^{-d/\xi}$ with $d$ the separation between domain walls (or equivalently here, the unit cell size). This leads to an effective model for the domain wall states
\begin{equation}
	\op{H}{\textsc{dw}} = \tilde{t} \sum_{j} \opd{f}{j} \op{f}{j+1} + \hc
\end{equation}
where $\op{f}{j}$ is an operator for the $j^{\text{th}}$ domain wall state. This produces a narrow mid-gap band of width $2 \tilde{t} \ll \delta t$.

Further details of these domain wall structures generating spectra with mid-gap bands will be postponed to \S\ref{sec:mottbands} where such models arise within a particularly rich context.

%%%%%%%%%%%%%%%%%%%%%%%%%%%%%%%%%%%%%%%%%%%%%%%%%%%%%%%%
\section{Extended Unit Cells\label{sec:largeunitcells}}

In contrast to the previous sections which have explored generalizations of the SSH model based on modifying its spectrum, in this section extensions of the SSH model are generated by modifying the SSH Hamiltonian directly. Here, the functional form of the SSH model's Hamiltonian is generalized to incorporate unit cells of arbitrary size and structure.
Similar studies considered SSH-type models with three-site unit cells~\cite{trimer} and four-site unit cells~\cite{ssh4}.
Due to the limited size of the unit cells considered there, no functional form of the hopping parameters was prescribed, but rather the analysis was performed with permutations of the relative magnitude of the various hopping parameters with each other. With the extension to unit cells of arbitrary size considered here, a more systematic approach to the strength of the hopping parameters is needed. A variety of parameterizations will be discussed below.

For a unit cell of size $L$, the Hamiltonian of an extended SSH model may be written as
\begin{equation}
	\hat{H} = \sum_{n\in\mathbbm{Z}^+} \sum_{m=0}^{L-1} \tensor{t}{_m}~\opd{c}{L n + m} \op{c}{L n + m+1} + \hc
\end{equation}
with chiral symmetry is enforced such that $\varepsilon_n = 0$.

As in \S\ref{sec:sshmodel} the Hamiltonian can be rewritten using the canonical basis with operators $\hat{\chi}$ which act on an entire unit cell $\alpha$
\begin{equation}
	\hat\chi_{\alpha}	=	\begin{pmatrix} \left\{ \op{c}{m} \right\}_{A} \\ \left\{ \op{c}{m} \right\}_{B} \end{pmatrix}_{\alpha}
\end{equation}
where sublattice $A$ consists of the odd sites within the unit cell, sublattice $B$ consists of the even sites within the unit cell, and $\alpha$ labels the unit cell.
The Hamiltonian then assumes the chiral form of
\begin{equation}
	\hat{H}(k)	=	\hat\chi^\dagger(k) \begin{pmatrix} 0 & \boldsymbol{h}(k) \\ \boldsymbol{h}^\dagger(k) & 0 \end{pmatrix} \hat\chi(k)
\end{equation}
where $\boldsymbol{h}(k)$ is a $\frac{L}{2}\times\frac{L}{2}$ submatrix of the form
\begin{equation}
	\boldsymbol{h}(k)	=	\begin{pmatrix} t_1 & t_2 & 0 & \cdots & t_{L} \e^{\i k} \\ 0 & t_3 & t_4 & \ddots & 0 \\ \vdots & \ddots & \ddots & \ddots & \ddots \\ \vdots & \ddots & \ddots & t_{L-3} & t_{L-2} \\ 0 & \cdots & \cdots & 0 & t_{L-1} \end{pmatrix} \,.
\end{equation}
In this basis the Hamiltonian anti-commutes with the chiral symmetry generator $\Gamma$
\begin{equation}
	\{ \hat{H}(k) , \Gamma \}	=	0
\end{equation}
where the chiral symmetry generator is given by $\Gamma = \boldsymbol{\sigma}_3 \otimes \mathbbm{1}_{N}$
%\begin{equation}
%	\Gamma	=	\begin{pmatrix} \mathbbm1 & 0 \\ 0 & -\mathbbm1 \end{pmatrix}
%\end{equation}

Like the SSH model, a $2N$-dimensional momentum space Hamiltonian of this form falls in the $A$III symmetry class and is an element of $U(2N) / (U(N) \times U(N))$. The standard SSH model can be recovered for $N=1=\frac{L}{2}$.
The $U(N) \times U(N)$ gauge transformations take the form of
\begin{equation}
	\boldsymbol{U} = \begin{pmatrix} \mathbbm{1}_{N} \e^{-\i n k} & 0 \\ 0 & \mathbbm{1}_{N} \end{pmatrix}
\end{equation}

%inversion symmetry
%trimer lattice
%\cite{noncentered}
%non-quantized Zak phase

As an extension of the method of \S\ref{sec:sshtransfer}, the transfer matrix\index{transfer matrix} can be employed to calculate eigenstates of general $1d$ Hamiltonians with unit cells of arbitrary size $N$. The zero-energy eigenstate is then given by the ansatz
\begin{equation}
	| \Psi \rangle = \sum_{n=1}^{N} \sum_{a=1}^{L} u^{(a)}_n | \psi^{(a)}_n \rangle
\end{equation}
with the normalization condition
\begin{equation}
	\sum_{n=1}^{N} \sum_{a=1}^{L} \left\lvert u^{(a)}_n \right\rvert^2 = 1
\end{equation}
and $a$ indexes each site of the unit cell and $n$ indexes each unit cell along the chain.
For zero energy, the Schr\"odinger equation is
\begin{equation}
	\hat{H} | \Psi_0 \rangle = 0 \cdot | \Psi_0 \rangle
\end{equation}
and the above transfer matrix method results in an asymptotic localization length of
\begin{equation}
	\xi \approx \cfrac{1}{\ln\cfrac{\prod_{k\text{ even}} t_k}{\prod_{k\text{ odd}} t_k}}
\end{equation}
which is obtained from calculation by analogous methods to that of \S\ref{sec:sshtransfer}.

In practice, it is often necessary to calculate the Zak phase numerically. This can be accomplished by adapting a method developed in~\cite{numericalchern} for computing the Chern number. Recall that the Zak phase for a particular band $n$ is given by
\begin{equation}
	\gamma_{n\mathcal{C}} = \oint_{\mathcal{C}} A_n
\end{equation}
where
\begin{equation}
	A_n = \i \tensor*{\psi}{^\dagger_n}(k) \tensor{\partial}{_{k}} \tensor*{\psi}{_n}(k) \d k
\end{equation}
is the Berry potential 1-form and $\tensor*{\psi}{_n}(k)$ is the momentum space Bloch eigenfunction of the $n$-th band. As a phase, its exponential is an alternative quantity that makes sense to analyze. The Zak phase can then be written as
\begin{equation}
\begin{aligned}[b]
	\e^{-\i \gamma_{n \mathcal{C}}}
		&=	\e^{-\i \oint_{\mathcal{C}} A_n}
	\\	&=	\e^{-\i \oint_{\mathcal{C}} A_{n}(k) \d k}
	\\	&\approx \e^{-\i \sum_a A_{n}(k_a) \Delta k}
	\\	&= \prod_{a} \e^{-\i A_{n}(k_a) \Delta k}
\end{aligned}
\end{equation}
where the integral has now been approximated by a sum over a discretized Brillouin zone with momentum labelled by $k_a$.
Since $\Delta k$ is small, each factor of the product can be expanded as
\begin{equation}
\begin{aligned}[b]
	\e^{-\i A_{n}(k_a) \Delta k}
	&=	1 - \i A_{n}(k_a) \Delta k + \mathcal{O}(\Delta k^2)
\\	&\approx	1 + \psi^\dagger_n(k_a) \tensor{\partial}{_{k}} \psi_n(k_a) \Delta k
\\	&=	\psi^\dagger_n(k_a) \left[ \psi_n(k_a) + \tensor{\partial}{_{k}} \psi_n(k_a) \Delta k \right]
\\	&\approx	\psi^\dagger_n(k_a) \psi_n(k_a + \Delta k)
\\	&=	\psi^\dagger_n(k_a) \psi_n(k_{a+1})	
\end{aligned}
\end{equation}
where also the discrete derivative has been employed as well as the unitary normalization of the eigenvectors.
The discrete form of the Zak phase for the $n^{\text{th}}$ band can then be obtained as
%\begin{equation}
%	\gamma_{n\mathcal{C}}
%	=	\i \ln \prod_a \frac{\psi^\dagger_n(k_a) \psi_n(k_{a+1})}{\left|\psi^\dagger_n(k_a) \psi_n(k_{a+1})\right|}
%\label{eq:numericalzak}
%\end{equation}
%The division by the modulus is taken so that only the phase of the wavefunction contributes to the calculation.
\begin{equation}
	\gamma_{n\mathcal{C}}	=	\frac\i\pi \ln \left( \prod_a \psi^\dagger_n(k_a) \psi_n(k_{a+1}) \right)
\label{eq:numericalzak}
\end{equation}
where the principal value of the complex logarithm is taken\footnote{The phase of the principal value of the complex logarithm $\ln(z)$ is given by the 2-argument arctangent function, $\tan^{-1}(x,y)$ where $z = x+\i y$.} to ensure that only the phase of the argument contributes to $\gamma$. %The $k$-space integration is taken to be $k\in[0,2\pi)$, or numerically $k_a \in [0 , 2\pi)$
The contour of integration $\mathcal{C}$ is over the whole Brillouin zone, so the product over $a$ is over all $k_a \in [0 , 2\pi]$.
%The total Zak phase is the sum of phases over all bands below the Fermi level.
%\begin{equation}
%	\gamma_{\text{total}\mathcal{C}}	=	\sum_{n < n_F} \frac\i\pi \ln\left( \prod_a \psi^\dagger_n(k_a) \psi_n(k_{a+1}) \right)
%\end{equation}

%%%%%%%%%%%%%%%%%%%%%%%%%%%%%%
\subsection{Periodic Envelopes}
A specific parameterization of the $\{t_n\}$ can be taken to be the form of a periodic envelope.
\begin{equation}
	t_n = t_0 + (-1)^{n} \delta t \left\lvert\cos\left( \tfrac{(n-1) \pi}{q} - \phi \right)\right\rvert
\label{eq:tnenvelope}
\end{equation}
The factor $n-1$ in the argument of the $\cos$ envelope is used such that the first bond in the chain is indexed by $t_1$ and with the displacement initialized by $\cos(-\phi)$ when $n=1$.

In the absence of domain walls, these models are adiabatically connected to the standard SSH model, which is recovered in the limit $q\to\infty$. Taking the absolute value of the envelope in Eq.~\eqref{eq:tnenvelope} ensures there are no domain walls in the parameterization of the $\{t_n\}$.

\begin{figure}[htp!]
\begin{subfigure}{\linewidth}
\begin{subsubcaption}
\subfiglabel{\includegraphics[scale=1]{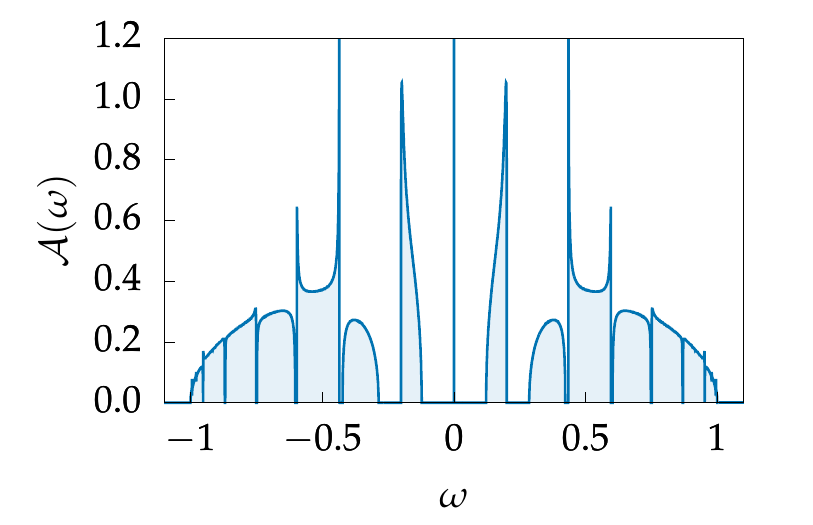}}{3,2}{GenvN30phi0}
\subfiglabel{\includegraphics[scale=1]{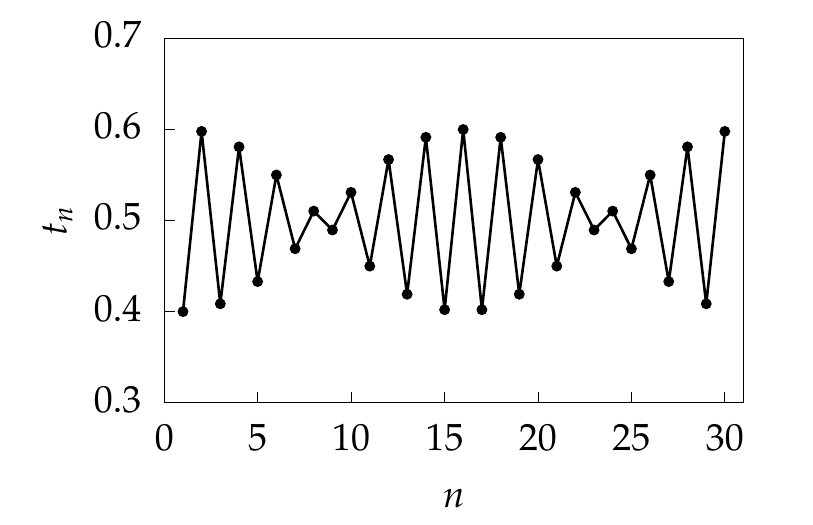}}{3,2}{tnenvN30phi0}
\end{subsubcaption}
\phantomsubcaption{\vspace{-\baselineskip}\addtocounter{subfigure}{-1}\label{fig:envN30phi0}}
\end{subfigure}
\begin{subfigure}{\linewidth}
\begin{subsubcaption}
\subfiglabel{\includegraphics[scale=1]{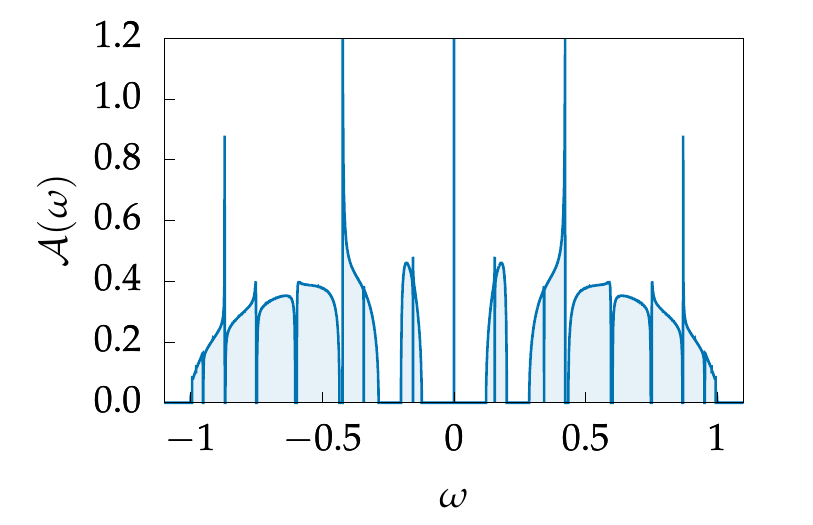}}{3,2}{GenvN30phi2}
\subfiglabel{\includegraphics[scale=1]{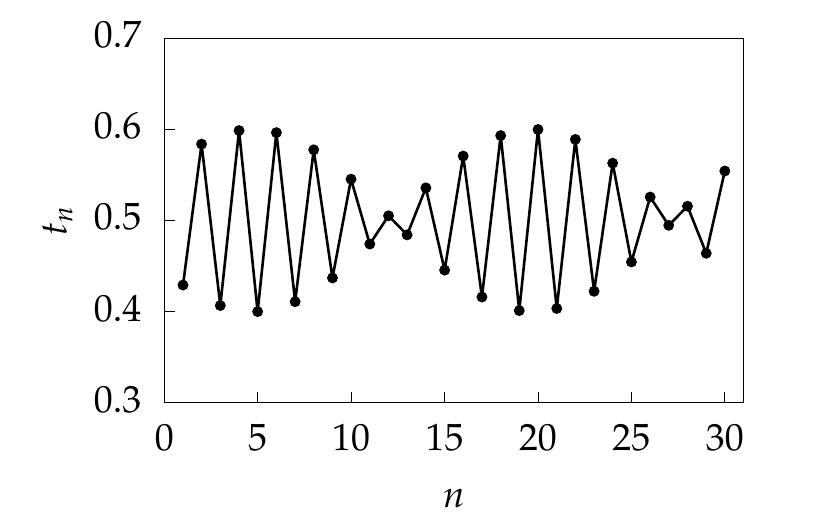}}{3,2}{tnenvN30phi2}
\end{subsubcaption}
\phantomsubcaption{\vspace{-\baselineskip}\addtocounter{subfigure}{-1}\label{fig:envN30phi2}}
\end{subfigure}
\begin{subfigure}{\linewidth}
\begin{subsubcaption}
\subfiglabel{\includegraphics[scale=1]{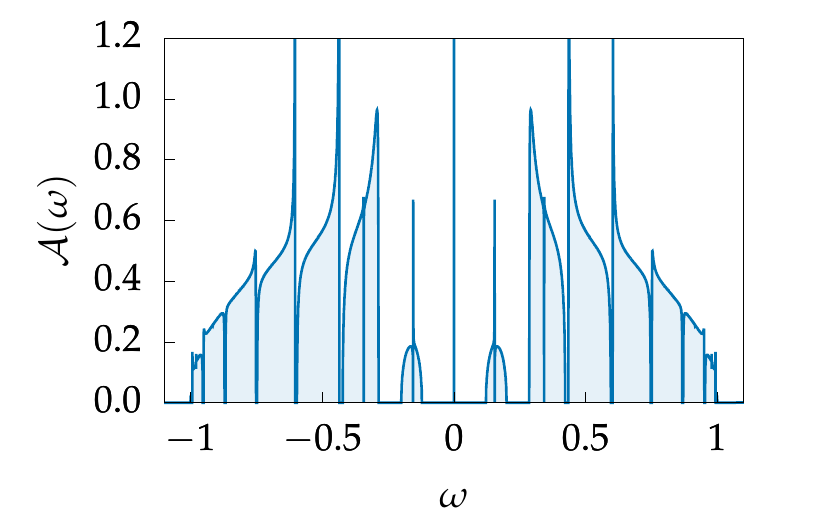}}{3,2}{GenvN30phi4}
\subfiglabel{\includegraphics[scale=1]{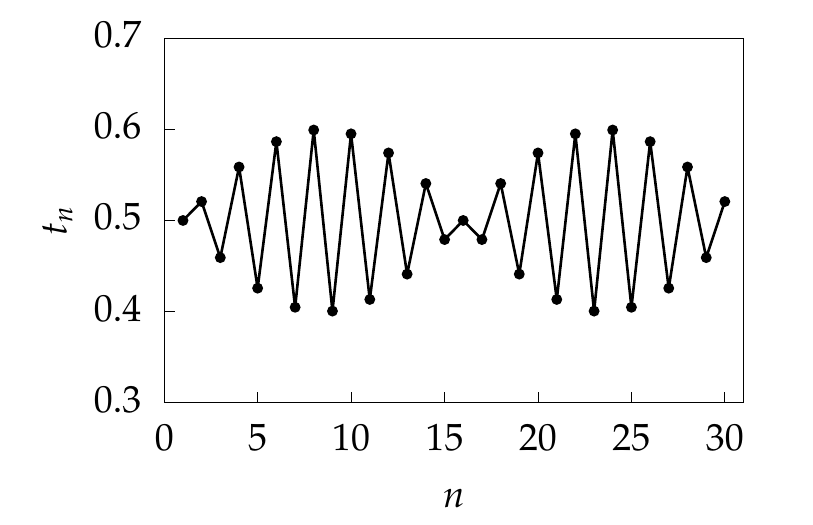}}{3,2}{tnenvN30phi4}
\end{subsubcaption}
\phantomsubcaption{\vspace{-\baselineskip}\addtocounter{subfigure}{-1}\label{fig:envN30phi4}}
\end{subfigure}
\begin{subfigure}{\linewidth}
\begin{subsubcaption}
\subfiglabel{\includegraphics[scale=1]{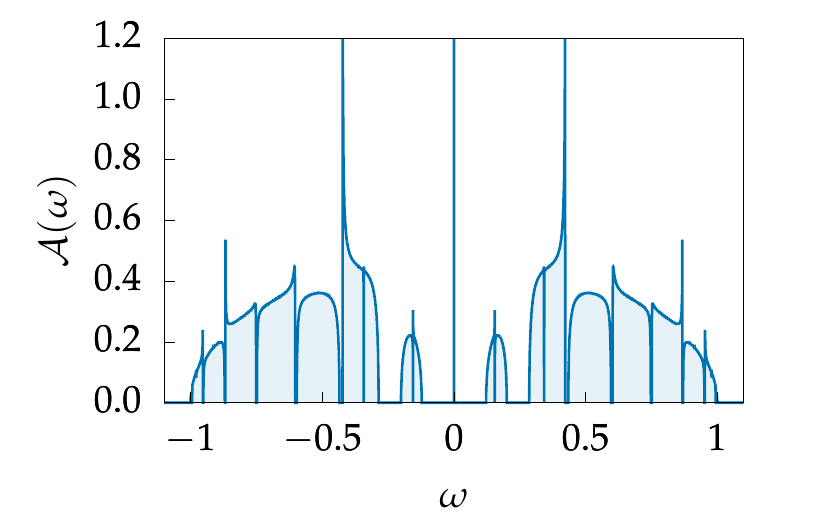}}{3,2}{GenvN30phi6}
\subfiglabel{\includegraphics[scale=1]{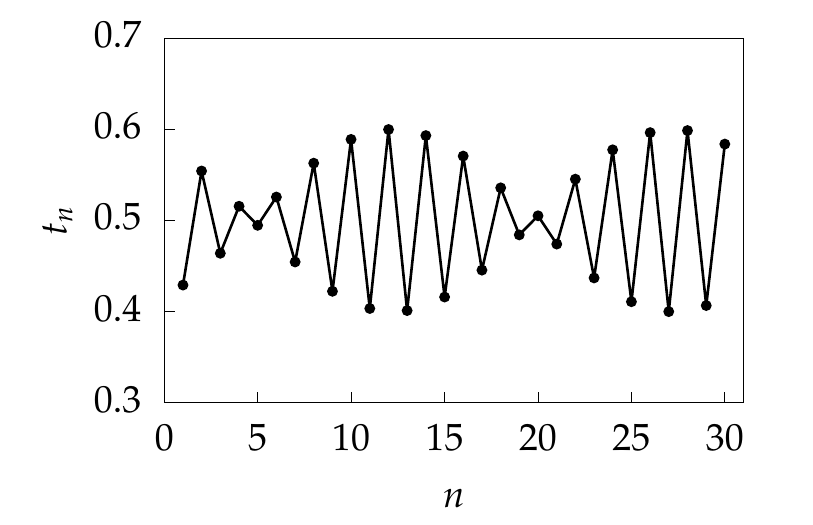}}{3,2}{tnenvN30phi6}
\end{subsubcaption}
\phantomsubcaption{\vspace{-\baselineskip}\addtocounter{subfigure}{-1}\label{fig:envN30phi6}}
\end{subfigure}
\caption{Boundary-site spectral functions (left) corresponding to generalized SSH models with unit cells given by $t_n$'s following Eq.~\eqref{eq:tnenvelope} (right). The chain is initialized with a weak bond with $N=30$ and $\phi=0$ \subref{fig:envN30phi0}, $\phi=\pi/4$ \subref{fig:envN30phi2}, $\phi=\pi/2$ \subref{fig:envN30phi4}, and $\phi=3\pi/4$ \subref{fig:envN30phi6}.\label{fig:envN30}}
\end{figure}

These periodic envelopes can also be interpreted as a kind of periodic spatial disorder which respects the SSH model's chiral symmetry over a larger period than the standard two dimensional symmetry.
\begin{figure}[ht!]
\includegraphics[scale=1]{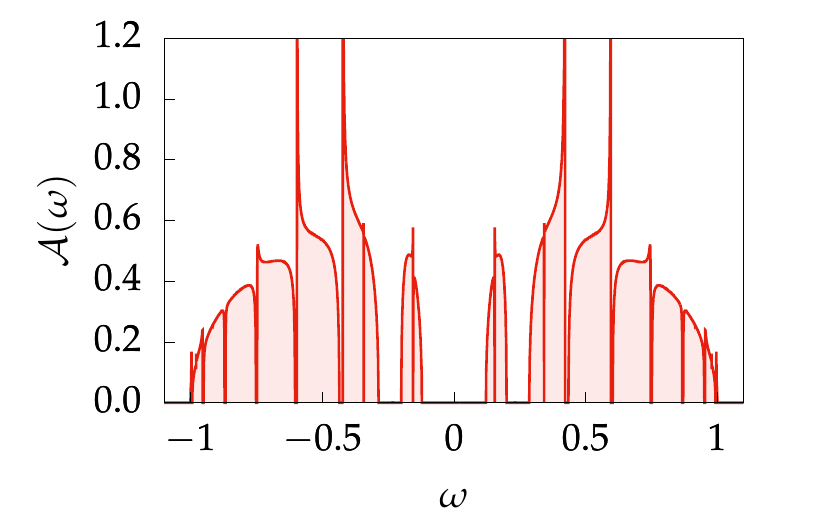}
\includegraphics[scale=1]{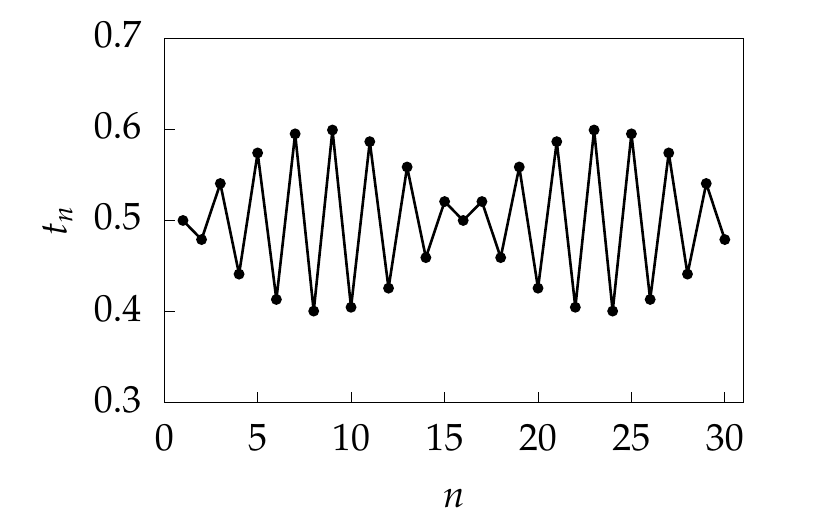}
\caption{Boundary-site spectral function (left) corresponding to a generalized SSH model with unit cells given by $t_n$'s following Eq.~\eqref{eq:tnenvelope} (right). The chain is initialized with a strong bond with $N=30$ and $\phi=\pi/4$. This is the parity reversed case of Fig.~\ref{fig:envN30phi4}.\label{fig:envN30phi4s}}
\end{figure}
An envelope of only half a cosine band can also be implemented. In contrast to the previous envelope, this unit cell repeats its pattern after only half the period of a cosine envelope. 
\begin{figure}[htp!]
\includegraphics[scale=1]{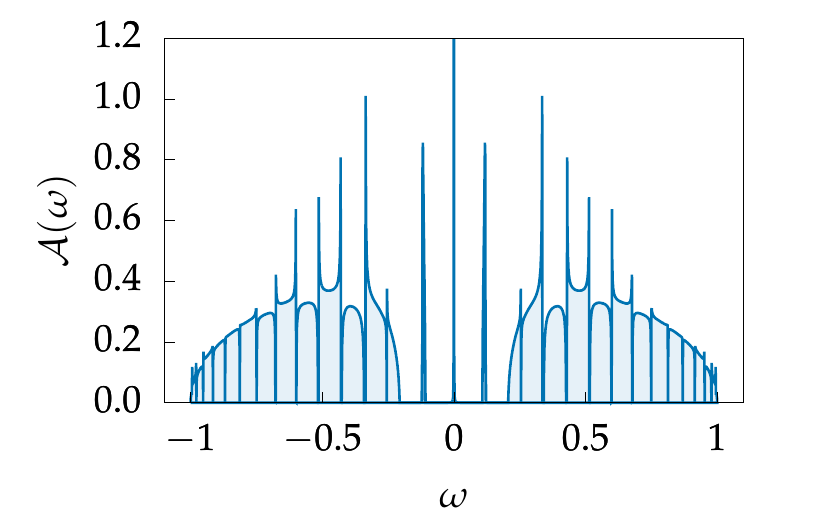}
\includegraphics[scale=1]{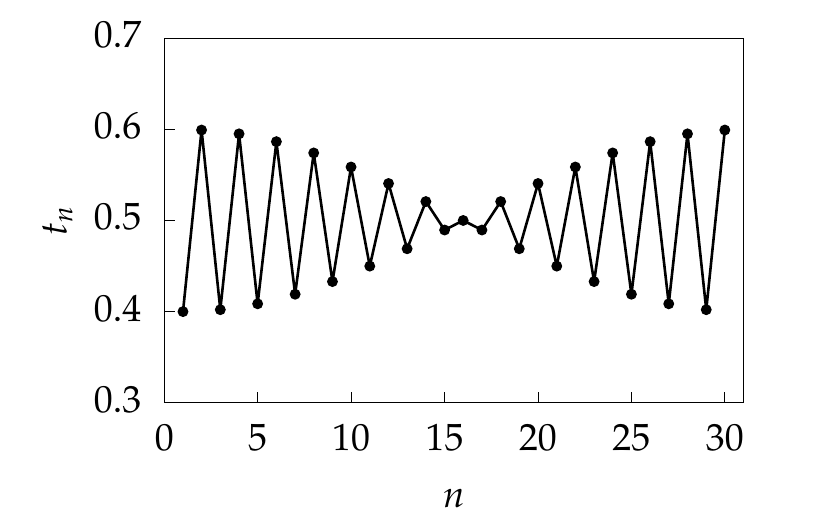}
\caption{Boundary-site spectral function (left) corresponding to a generalized SSH model with unit cells given by $t_n$'s following Eq.~\eqref{eq:tnenvelope} (right). The chain is initialized with a weak bond with $N=60$ and $\phi=0$.\label{fig:envN60phi0}}
\end{figure}
Since these models are adiabatically connected to the SSH model, at low energy they feature a hard gap with or without a spectral pole at zero energy depending on if the chain is initialized with $t_1 < t_2$ or $t_1 > t_2$. 
A difference with the SSH model is that these models are characterized by spectral functions with fractured outer bands. 

For these systems the Zak phase as computed by Eq.~\eqref{eq:numericalzak} is quantized, and their topological classification is clear.

An examples of generalized SSH model with the hopping parameters given by Eq.~\eqref{eq:tnenvelope} is shown in Figs.~\ref{fig:envN30}, \ref{fig:envN30phi4s}, and \ref{fig:envN60phi0}. The shading of the spectral function (blue/red) corresponds to the system being either topological or trivial respectively. The topology is quantified by the Zak phase, as evaluated numerically according to Eq.~\eqref{eq:numericalzak}. For the topological spectra, the Zak phase is integer quantized, and is zero for trivial spectra. The sequence of spectral functions shown in Fig.~\ref{fig:envN30} have unit cells initialized with a weak bond and choses $N=30$, $t_0 = 0.5$, $\delta t = 0.1$, with varying values of $\phi$. The calculation shows that compared to the standard SSH model, the spectrum still features a zero energy mid-gap pole sitting in a hard gap. The SSH bands however take on a fractal form, with many bands separated by exponentially small gaps.

Fig.~\ref{fig:envN30phi4s} shows an example of a generalized SSH model initialized with a strong bond. The parameters are chosen to be $N=30$, $t_0 = 0.5$, $\delta t = 0.1$, and $\phi = 3/\pi/4$. This is the parity reversed case of Fig.~\ref{fig:envN30phi4}. As shown, the spectrum takes on a similar form, but with the topological pole absent, as expected by analogy with the trivial phase of the SSH model.

Shown in Fig.~\ref{fig:envN60phi0} is a model parameterized with $N=60$, $t_0 = 0.5$, $\delta t = 0.1$, and $\phi = 0$. Like the previous examples, this model has a unit cell which is 30 sites in length, but the period of the cosine envelope is now 60 sites, meaning that only half the wavelength is captured by the unit cell.

Tight-binding models parameterized by Eq.~\eqref{eq:tnenvelope} are adiabatically connected to the SSH model since the hopping parameters exhibit a strict alternating behavior. It is therefore expected that the topological phase and presence or absence of a zero energy pole depends on whether the semi-infinite system is initialized with a weak bond or a strong bond.

\subsection{Long Wavelength Modulations}

The models described in the previous section involved generalizations of the SSH model where a long wavelength envelope is superimposed over the hopping amplitudes. The opposite case can also be considered, where the alternating behavior of the hopping amplitudes does not take place over successive bonds, but rather over several bonds.
A parameterization of $\{t_n\}$ which realizes this concept is
\begin{equation}
	t_n = t_0 + (-1)^n \delta t \cos\left( \tfrac{(n-1) 2\pi}{N} - \phi \right)
\label{eq:tnoscillating}
\end{equation}
This parameterization recovers the standard SSH model for a two-site unit cell, $N=2$, and $\phi \neq \pi/2 \mod\pi$.
This parameterization includes as subcases configurations considered in analyses of the SSH$_4$ model~\cite{ssh4}. In contrast to these previous studies, the parameterization \eqref{eq:tnoscillating} specifies a functional form for the $\{t_n\}$. 
%
%Like the SSH model, these models feature a hard gap at the Fermi level with or without a zero-energy pole depending on parameterization. The low energy gap is exponential in the bandwidth and unit cell dimension.
%
The general features of these models can be classified based on their parameterization. This classification is shown in Table~\ref{tab:oscillating}. The primary spectral feature of concern in these models is whether at zero energy the spectrum is gapped ($\mathcal{A}(0)=0$), metallic ($\mathcal{A}(0)=\text{const.}$), or contains a pole ($\mathcal{A}(0) \sim \delta(0)$).
\begin{table}[h]
\centering
\caption[Classification of long wavelength SSH variants]{Classification of long wavelength SSH variants. The parameters are the phase shift $\phi$, the unit cell size $N$, and the qualitative feature of the spectral function at zero frequency $\mathcal{A}(0)$. The notation $X \setminus Y$ denotes the set $X$ with elements of the set $Y$ removed. \textit{E.g.} $2\mathbbm{N} \setminus 4\mathbbm{N}$ denotes cells of even length not including $N$ which are multiples of 4.\label{tab:oscillating}}
\begin{tabular}{ccc}
	$\phi \mod 2\pi$	&	$N$	&	$\mathcal{A}(0)$
	\\\hline
	0	&	$4\mathbbm{N}$	&	Pole	\\
	0	&	$2\mathbbm{N} \setminus 4\mathbbm{N}$	&	Gap	\\
	$\pi$	&	$2\mathbbm{N}$	&	Pole	\\
	\\
	$\pm\pi/2$	&	$8\mathbbm{N}$	&	Pole	\\
	$\pm\pi/2$	&	$4\mathbbm{N} \setminus 8\mathbbm{N}$	&	Gap	\\
	$\pm\pi/2$	&	$2\mathbbm{N} \setminus 4\mathbbm{N}$	&	Metal	\\
\end{tabular}
\end{table}
The Zak phase for models of this type is in general not quantized.
Within this class of generalizations, only systems with an even number of sites in the unit cells were analyzed. Systems with a unit cell size which is odd in dimension cannot possess chiral symmetry.

A sequence of tight-binding chains constructed from the parameterization Eq.~\eqref{eq:tnoscillating} are shown in Fig.~\ref{fig:oscphi4}. The parameters chosen are $t_0 = 0.5$, $\delta t = 0.4$, $\phi = \pi/2$, with $N$ varied as 8, 10, 12, 14 ,16. These values of $N$ demonstrate the variety of spectra which can be generated by the parameterization Eq.~\eqref{eq:tnoscillating}. These choices of $N$ cover all the cases of $\phi = \pi/2~\mod 2\pi$ shown in Table~\ref{tab:oscillating}: $8,16 \in 8\mathbbm{N}$; $12 \in 4\mathbbm{N} \setminus 8\mathbbm{N}$; and $10, 14 \in 2\mathbbm{N} \setminus 4\mathbbm{N}$.

Note that in Fig.~\ref{fig:oscphi4} only the low energy features of the spectral function are plotted. 
So, for example, while the $N=8$ case appears to be identical to the standard SSH configuration, there are bands and gaps at higher energies beyond the plot region; the bandwidth here is $D=1.0$, so there are bands and gaps among the entire bandwidth.
\begin{figure}[htp!]
\includegraphics[scale=1]{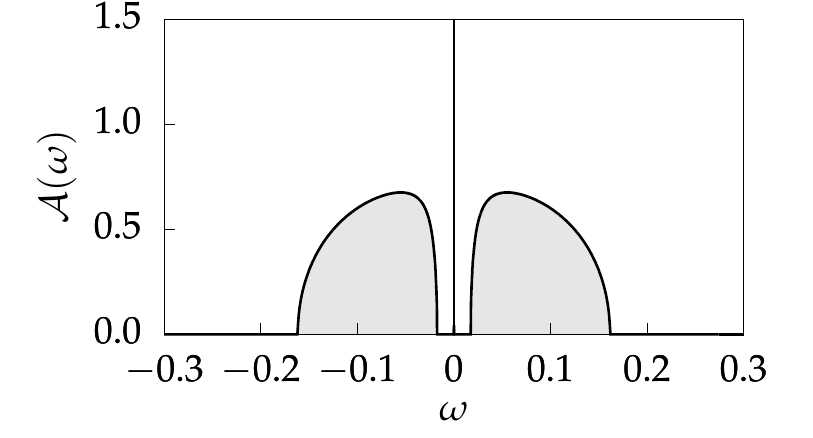}
\includegraphics[scale=1]{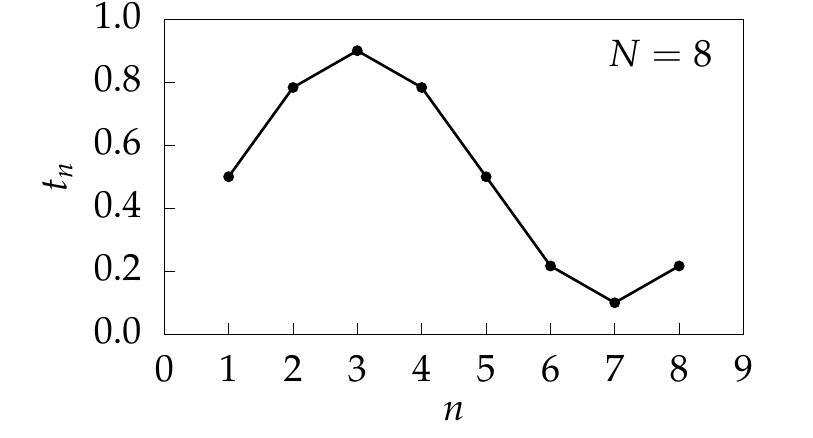}
\includegraphics[scale=1]{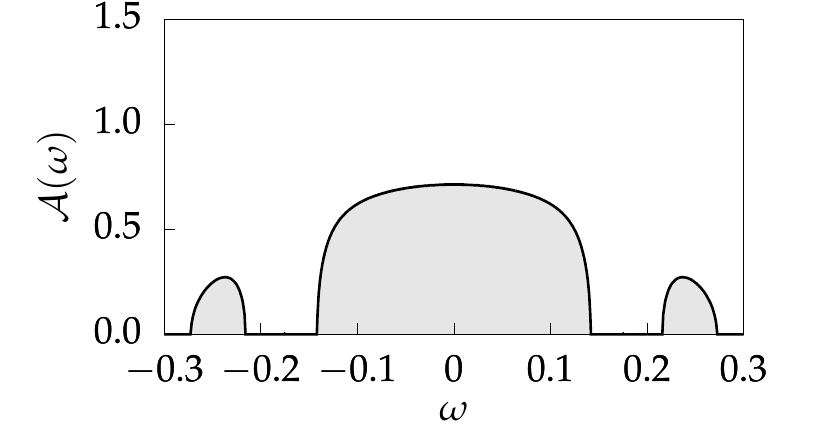}
\includegraphics[scale=1]{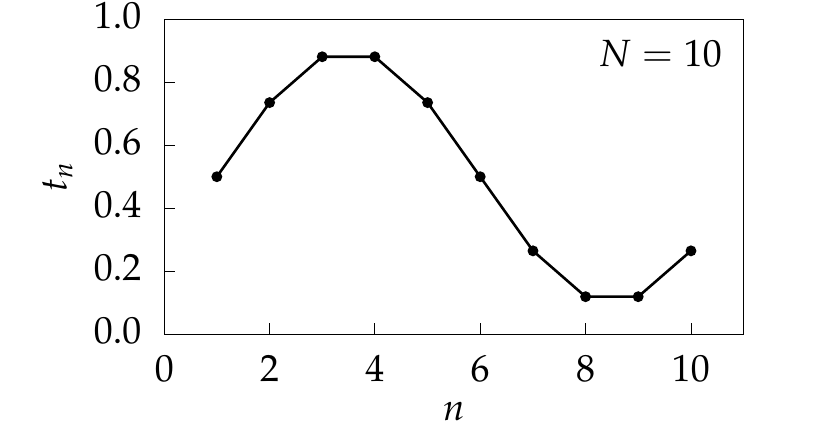}
\includegraphics[scale=1]{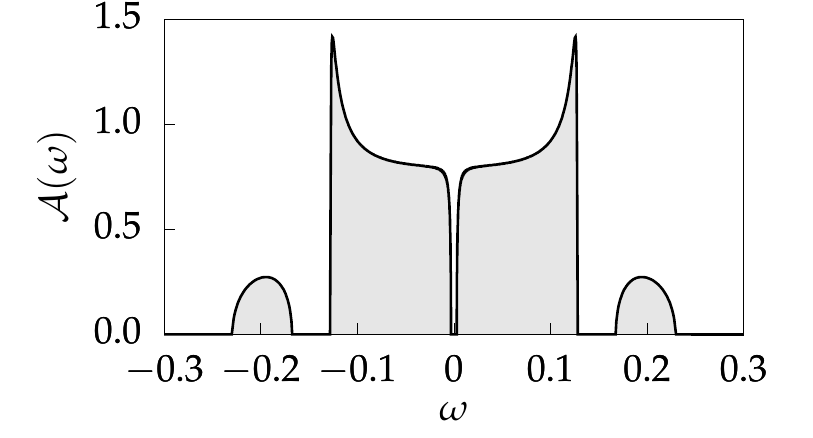}
\includegraphics[scale=1]{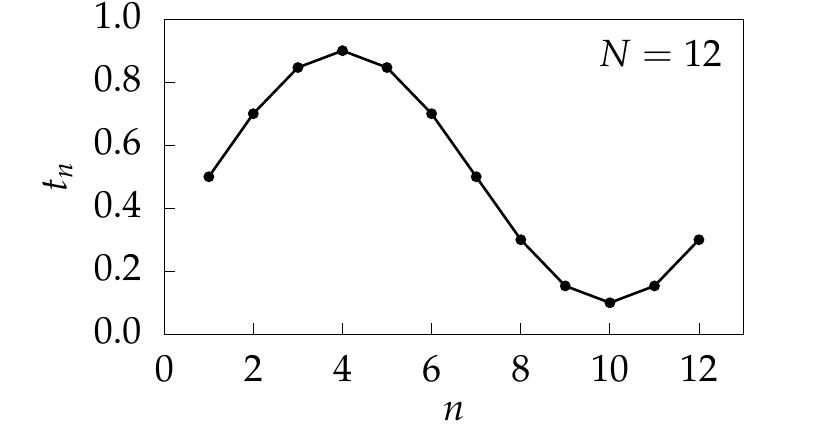}
\includegraphics[scale=1]{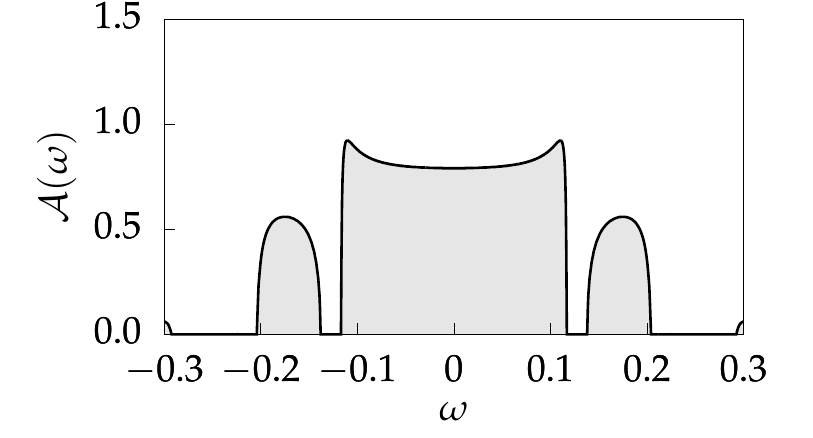}
\includegraphics[scale=1]{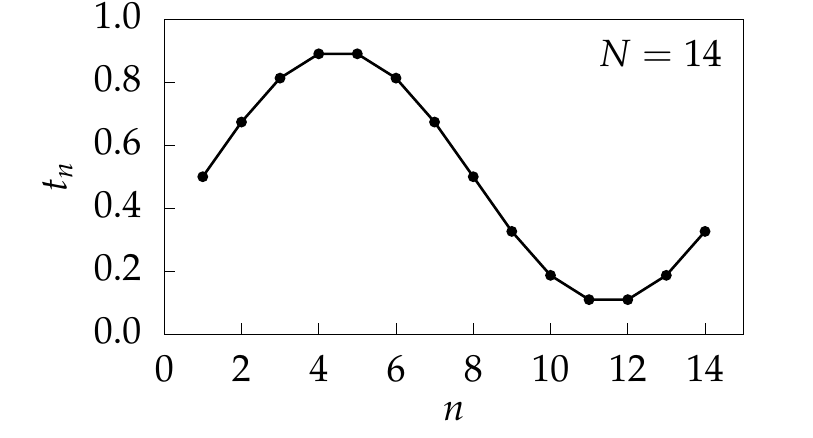}
\includegraphics[scale=1]{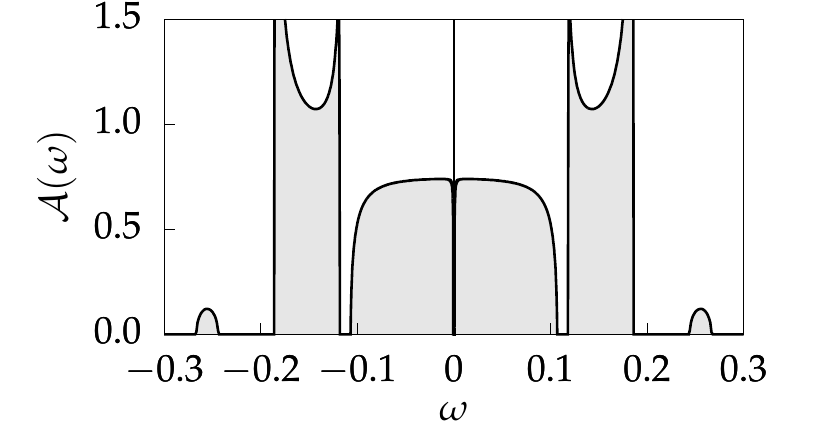}
\includegraphics[scale=1]{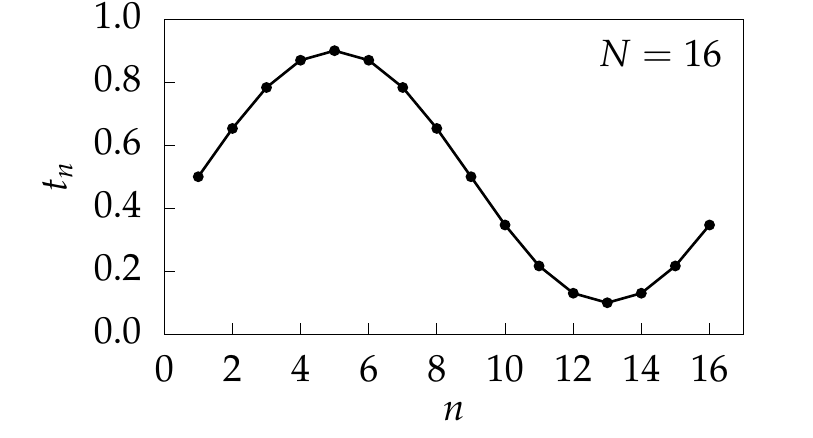}
\caption[Example spectra of long wavelength SSH variants]{Sequence demonstrating the dependence of spectra based on Eq.~\eqref{eq:tnoscillating} on $N$. Note that only the low energy region of the spectral functions is plotted. The spectral functions obey the categorization of Table~\ref{tab:oscillating}.\label{fig:oscphi4}}
\end{figure}

The classification of the long wavelength SSH models presented in Table~\ref{tab:oscillating} was found by analyzing models parameterized by systematic sampling of the parameter space for $\phi$ and $N$. The full theoretical understanding of how this pattern emerges is not yet known and is left for future work.

\section{Power-Law Spectral Functions\label{sec:powerlawssh}}

While one of the main characteristics of the SSH model is the presence of a hard gap about the Fermi energy, the model can be modified to allow cases where the low energy spectrum obeys power-law behavior
\begin{equation}
	\mathcal{A}(\omega) \sim \lvert \omega \rvert^{\pm r}
\label{eq:pmr}
\end{equation}
in the region $\lvert \omega \rvert \ll 1$ and $r>0$.
To obtain a spectrum with power $\pm r$, the chain parameters can be generated by a form of
\begin{equation}
	t_n	=	t_0 \sqrt{ 1 - \zeta (-1)^{n} \frac{r}{n + d} }
\label{eq:powerlawtn}
\end{equation}
where $\zeta = \pm$ and aligns with the sign of the exponent in Eq.~\eqref{eq:pmr}.
The negative sign yields a power-law vanishing spectrum, and the positive sign yields a power-law diverging spectrum, corresponding to the analogous trivial or topological phase of the SSH model. $4 t_0$ is the full band width.

\subsection{Power-Law Vanishing Spectra\label{sec:pseudogapssh}}
A spectrum which is power-law vanishing at low energies is also called a pseudogap.
Pseudogap spectral functions appear in such contexts as cuprate high-$T_c$ superconductors and also in the self-energy of the Hubbard model in infinite dimensions in the metallic phase, as seen in \S\ref{sec:hubbardsolution}.
%\begin{equation}
%	t_n	=	t_0 \sqrt{ 1 + (-1)^{n} \frac{r}{n + d} }
%\label{eq:tnpseudogap}
%\end{equation}
The functional form of the hopping parameters follows Eq.~\eqref{eq:powerlawtn} with $\zeta = +$.
It is clear that $t_1 > t_2$, so this configuration starts on a strong bond, like the trivial phase of the SSH model.
A set of spectra corresponding to models generated from the parameterization Eq.~\eqref{eq:powerlawtn} with $\zeta = +$ for a range of values of $d$ is shown in Figs.~\ref{fig:fraction_r05}, with $r= 0.5$, and \ref{fig:fraction_r2}, with $r = 2$. Note the similarity between the spectra for $r = 2$ in Fig.~\ref{fig:fraction_r05} and the self-energy for the Anderson impurity model, Fig.~\ref{fig:siamS}. In both cases the very low energy features scale as $\omega^2$.

At low energy these spectra exhibit behavior described by $\mathcal{A}(\omega) = \alpha |\omega|^r$. From constructing chains from the ansatz Eq.~\eqref{eq:powerlawtn} and analyzing the resulting spectrum, it is deduced that for fixed $r$, the coefficient $\alpha$ scales as $\alpha \sim d$.
\begin{table}[h]
\centering
\caption{Values of $\alpha$ for varying values of $d$ for fixed $r$.}
\begin{tabular}{|c||c|c|c|c|}\hline
	\diagbox[height=0.67cm]{$r$}{$d$}	&	2	&	5	&	8	&	20	\\\hline\hline
	0.5	&	4.65	&	6.74	&	8.67	&	13.08	\\\hline
	2	&	11.80	&	58.26	&	141.99	&	705.58	\\\hline
\end{tabular}
\end{table}
\begin{figure}[htp!]
%\begin{subfigure}{\linewidth}
\includegraphics[scale=1]{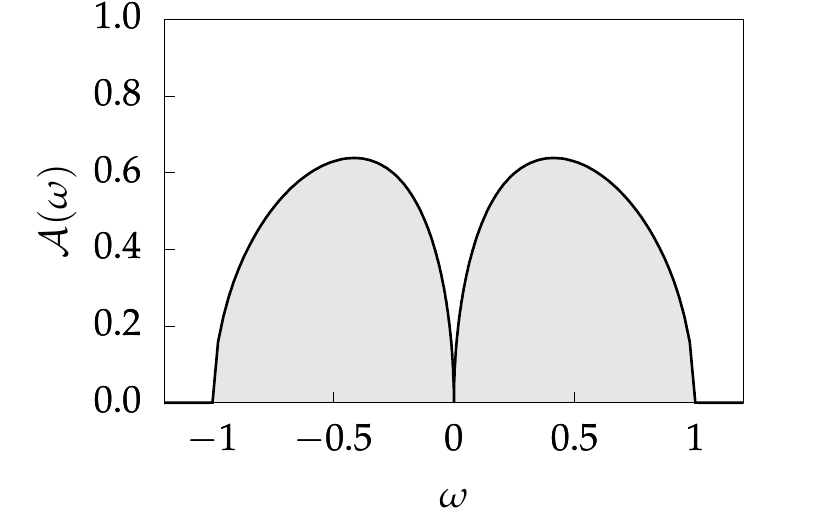}
%\hspace{-1.5em}
\includegraphics[scale=1]{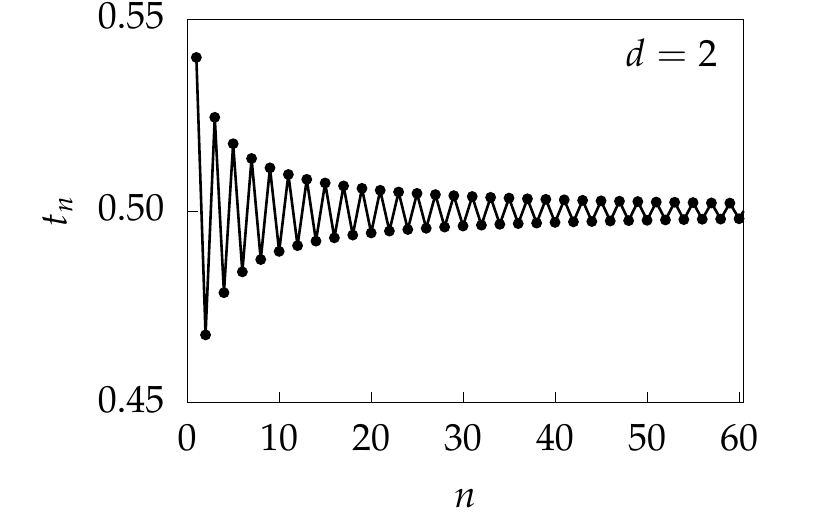}
%\vspace{-1.5\baselineskip}
%\caption{$d=2$}
%\end{subfigure}
%
%\begin{subfigure}{\linewidth}
\includegraphics[scale=1]{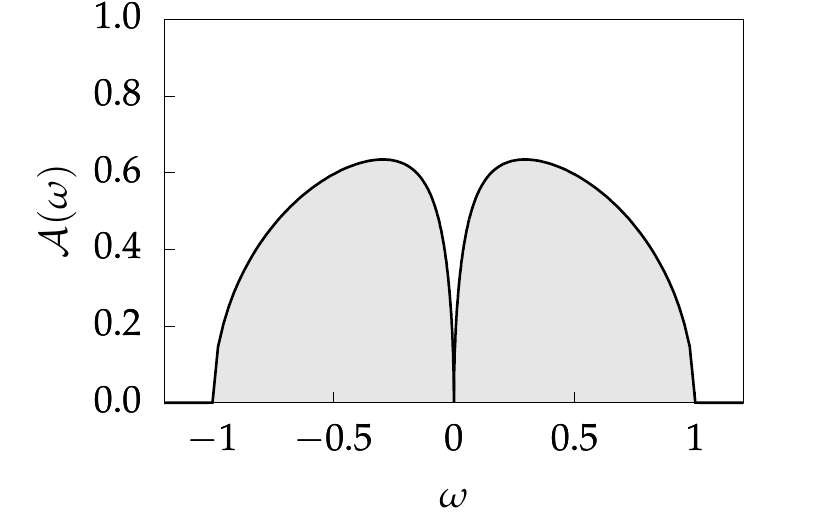}
%\hspace{-1.5em}
\includegraphics[scale=1]{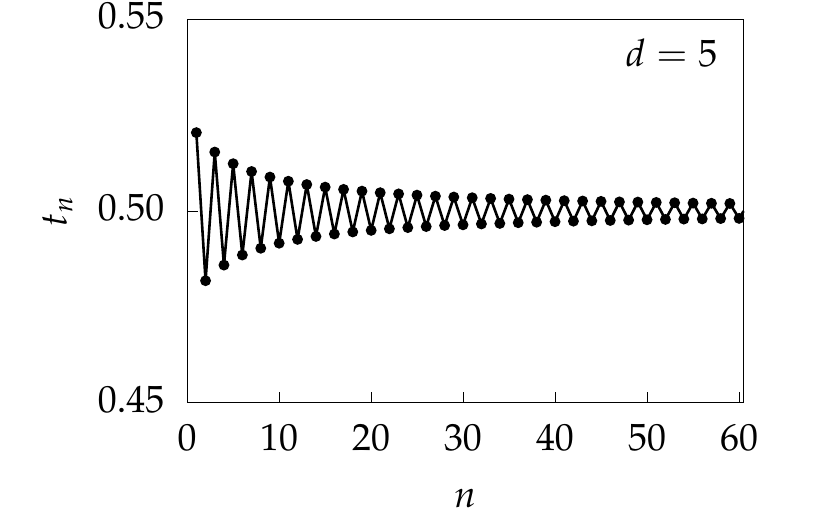}
%\vspace{-1.5\baselineskip}
%\caption{$d=5$}
%\end{subfigure}
%
%\begin{subfigure}{\linewidth}
\includegraphics[scale=1]{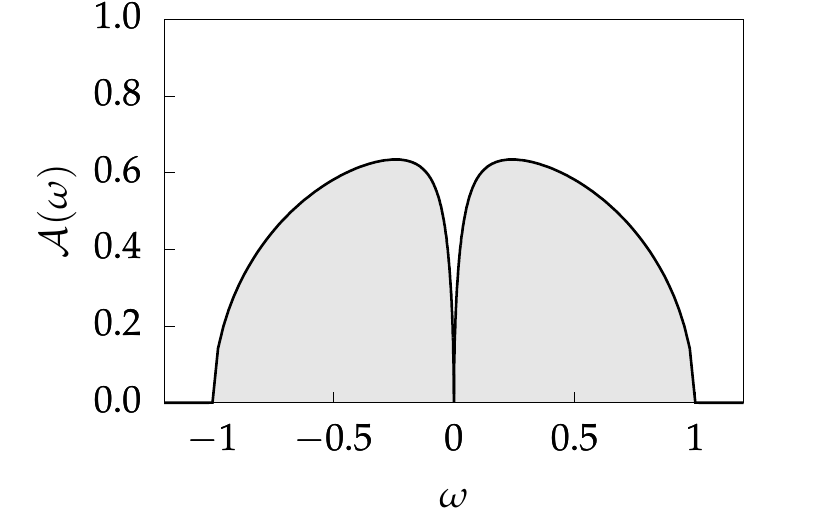}
%\hspace{-1.5em}
\includegraphics[scale=1]{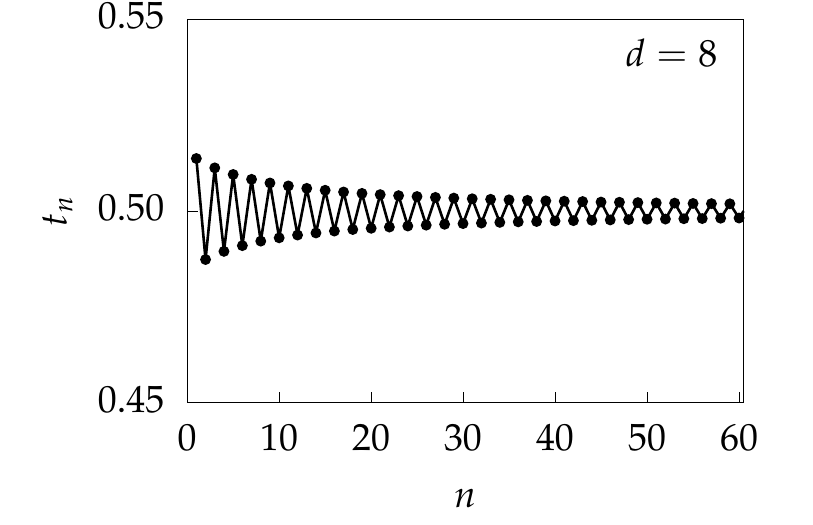}
%\vspace{-1.5\baselineskip}
%\caption{$d=8$}
%\end{subfigure}
%
%\begin{subfigure}{\linewidth}
\includegraphics[scale=1]{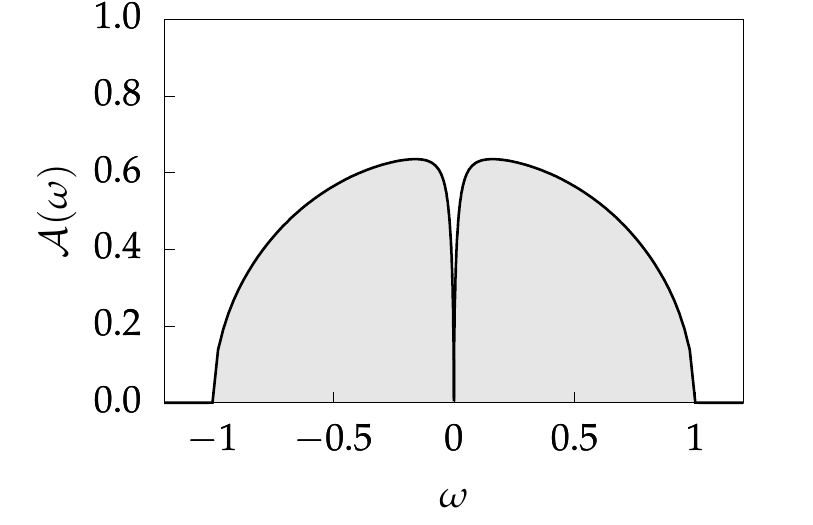}
%\hspace{-1.5em}
\includegraphics[scale=1]{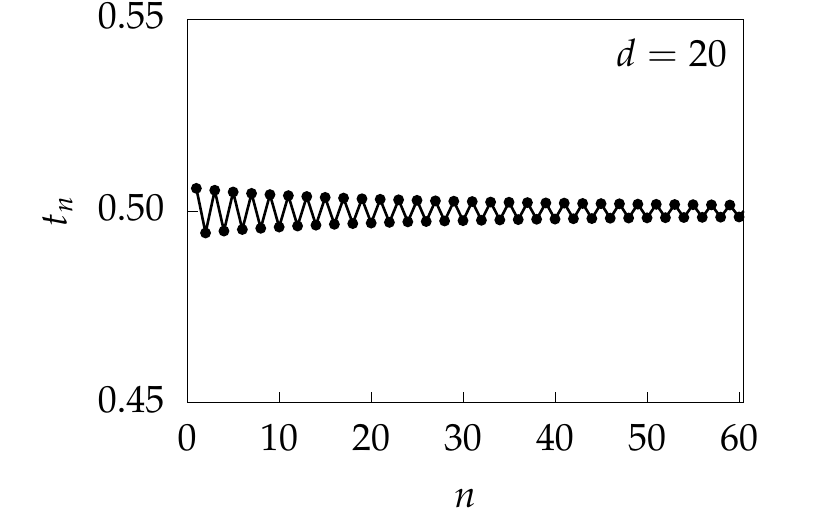}
%\vspace{-1.5\baselineskip}
%\caption{$d=20$}
%\end{subfigure}
%
\caption{Spectral functions (left) corresponding to models with $t_n$'s generated by Eq.~\eqref{eq:powerlawtn} with $\zeta = +$ (right) at various values of $d$ with $r = 0.5$.\label{fig:fraction_r05}}
\end{figure}
\begin{figure}[htp!]
\includegraphics[scale=1]{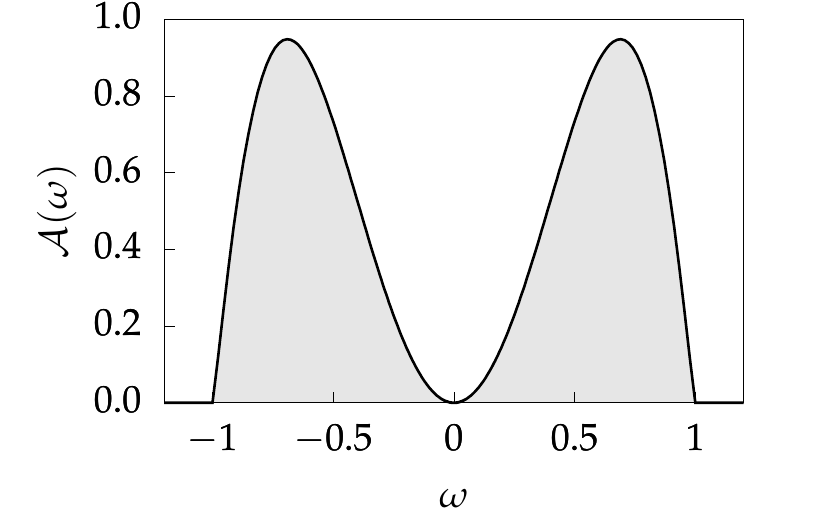}
\includegraphics[scale=1]{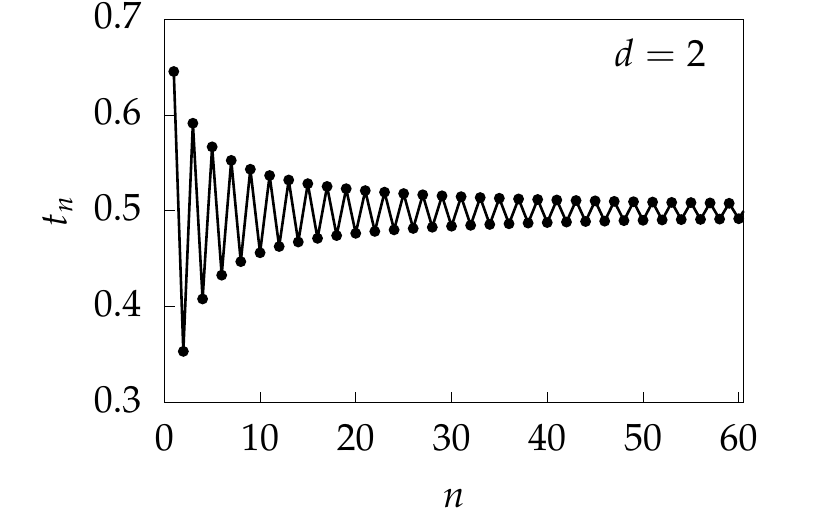}
\includegraphics[scale=1]{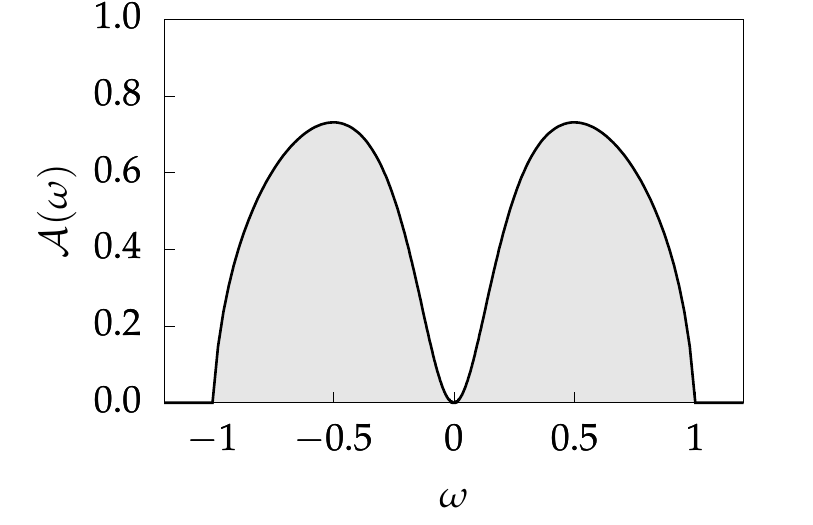}
\includegraphics[scale=1]{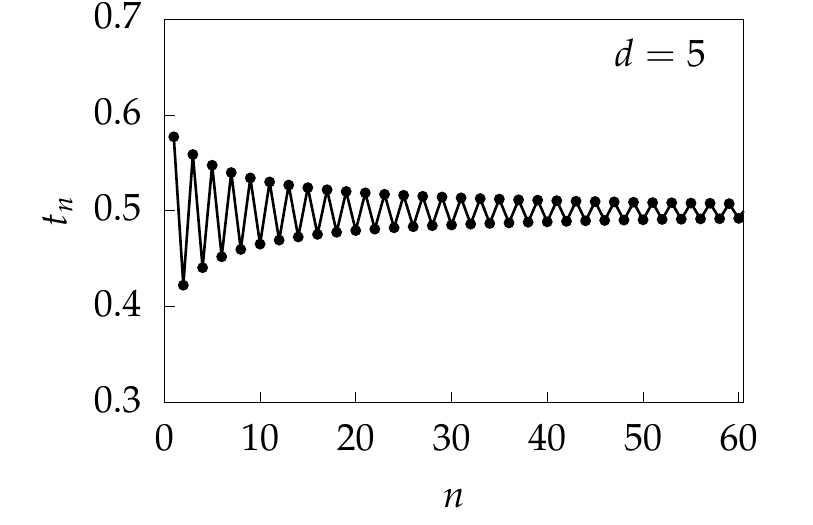}
\includegraphics[scale=1]{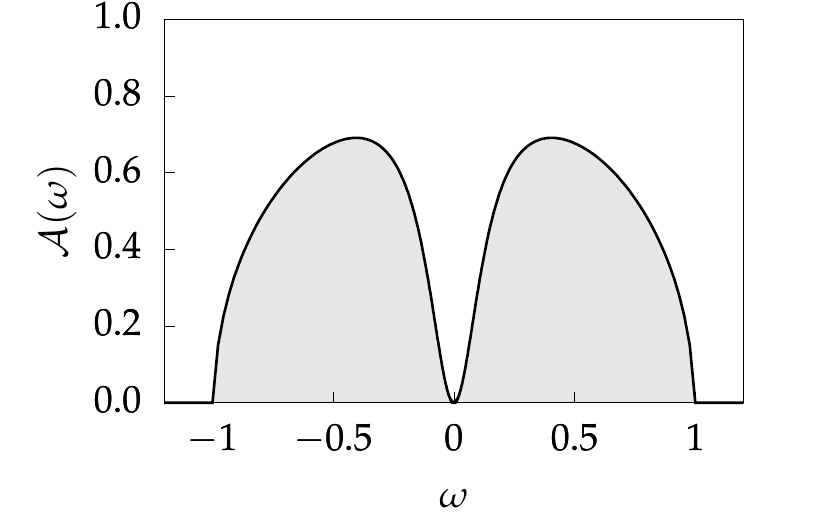}
\includegraphics[scale=1]{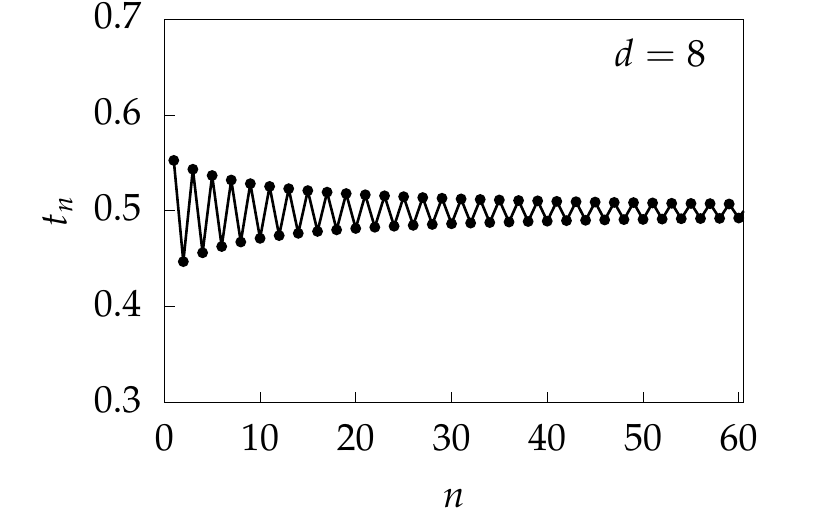}
\includegraphics[scale=1]{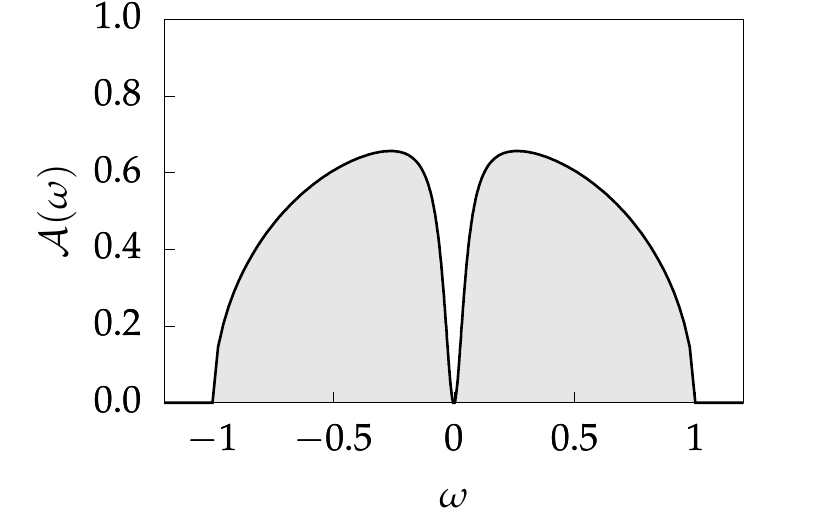}
\includegraphics[scale=1]{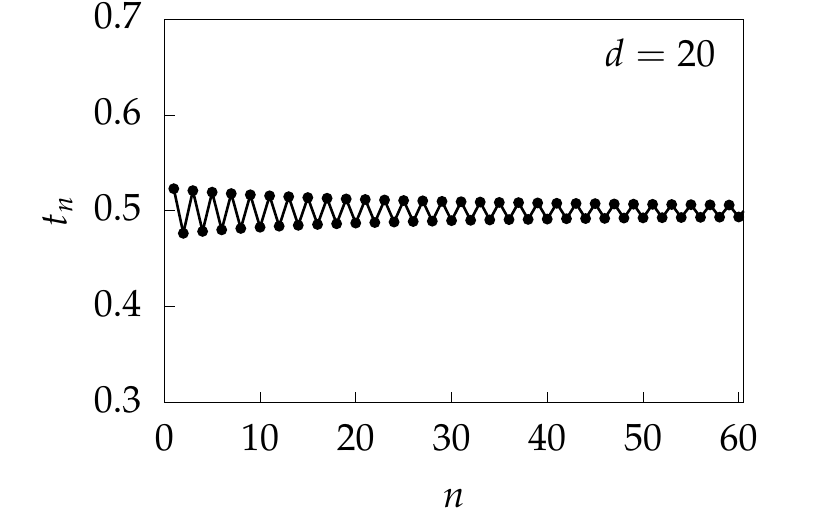}
\caption{Spectral functions (left) corresponding to models with $t_n$'s generated by Eq.~\eqref{eq:powerlawtn} with $\zeta = +$ (right) at various values of $d$ with $r = 2$.\label{fig:fraction_r2}}
\end{figure}
An interpretation of the $1/{n}$ envelope lies in noting that the spectral gap of the SSH model is directly proportional to the difference of neighboring hoppings, and that features at large $n$ correspond to spectral features at low $\omega$. For hoppings parameterized by Eq.~\eqref{eq:powerlawtn}, hopping amplitudes at large $n$ become asymptotically close in magnitude, such that their difference becomes very small. 

In the parameterization Eq.~\eqref{eq:powerlawtn}, the power-law features in the resulting spectral functions typically set in at very low energies, $|\omega| \ll 1$. As such the power-law feature can be difficult to identify on linear scale plots. Spectral functions for $r = 0.5$ and $r = 2$ at $d = 20$ are plotted on a log scale in Fig.~\ref{fig:powerlogd20}. The powers of $r = 0.5$ and $r = 2$ are clearly seen in Figs.~\ref{fig:green_fraction_r05d20_log} and \subref{fig:green_fraction_r2d20_log} respectively.
\begin{figure}[htp!]
\subfiglabel{\includegraphics{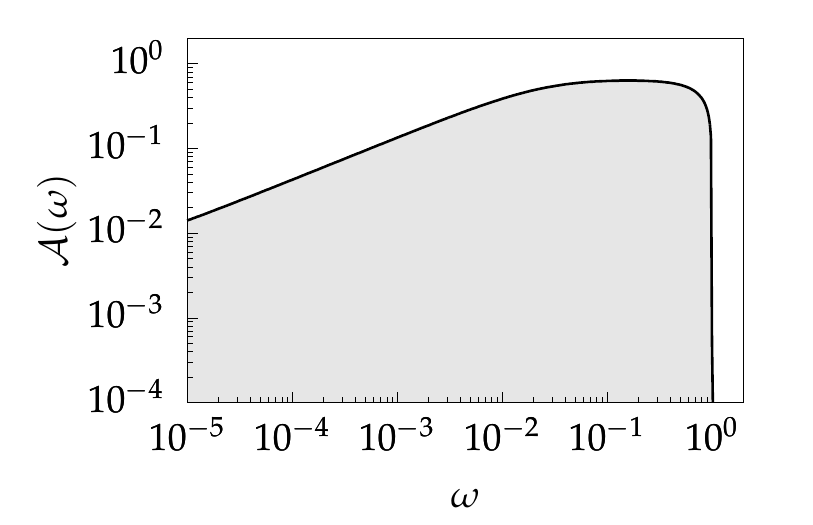}}{3.125,2}{fig:green_fraction_r05d20_log}
\subfiglabel{\includegraphics{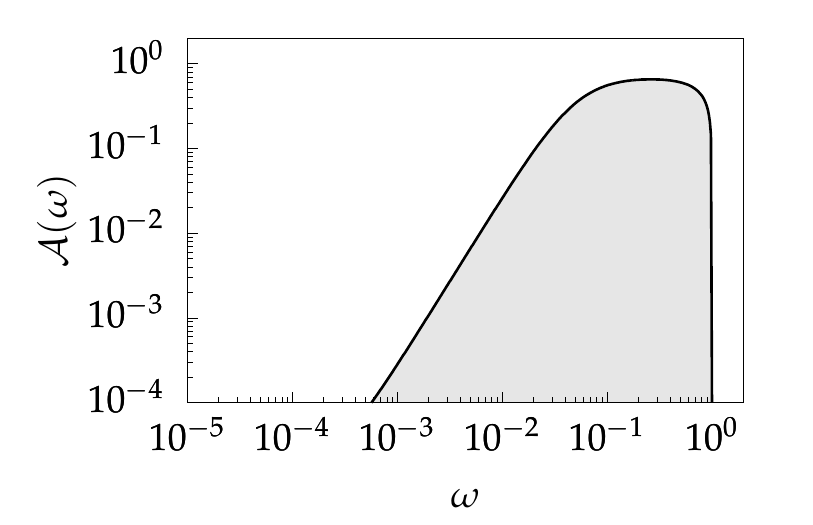}}{3.125,2}{fig:green_fraction_r2d20_log}
\caption{Spectral functions for the power-law vanishing SSH model from Eq.~\eqref{eq:powerlawtn} with $\zeta = +$ and $d=20$ at \subref{fig:green_fraction_r05d20_log} $r=0.5$ and \subref{fig:green_fraction_r2d20_log} $r=2$. Log scale clearly shows the low energy power as $\mathcal{A}(\omega) \sim |\omega|^{r}$\label{fig:powerlogd20}}
\end{figure}

\subsection{Power-Law Diverging Spectra\label{sec:powerdivergencessh}}
The opposite case to a pseudogap is where the spectral function exhibits a power-law divergence at low energy. This case is made possible with the parameterization of the $\{t_n\}$ initialized with a weak bond, following Eq.~\eqref{eq:powerlawtn} with $\zeta = -$.
%\begin{equation}
%	t_n	=	t_0 \sqrt{ 1 + (-1)^{n} \frac{r}{n + d} } \,.
%\label{eq:tnpowerdivergence}
%\end{equation}
This configuration has $t_1 < t_2$, and therefore the chain parity starts with a weak bond like the topological phase of the SSH model.
Similarly, a diverging state at zero energy can be identified. Also as in the SSH model, the wavefunction of this zero energy state is exponentially localized on the boundary site. The localization of the wavefuction is obtained from a transfer matrix calculation.

A set of spectra corresponding to models generated from the parameterization Eq.~\eqref{eq:powerlawtn} with $\zeta = -$ for a range of values of $d$ is shown in Figs.~\ref{fig:fractiond_r05}, with $r = 0.5$, and \ref{fig:fractiond_r2}, with $r = 2$.
\begin{figure}[htp]
\includegraphics[scale=1]{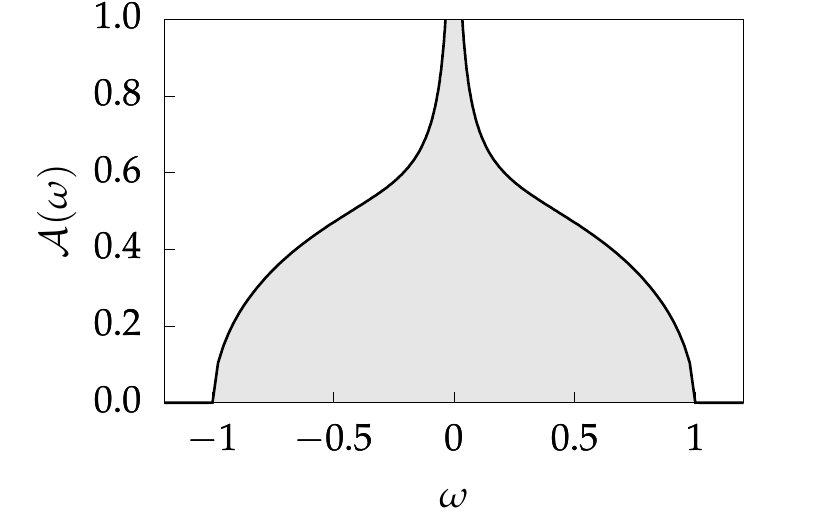}
%\hspace{-1.5em}
\includegraphics[scale=1]{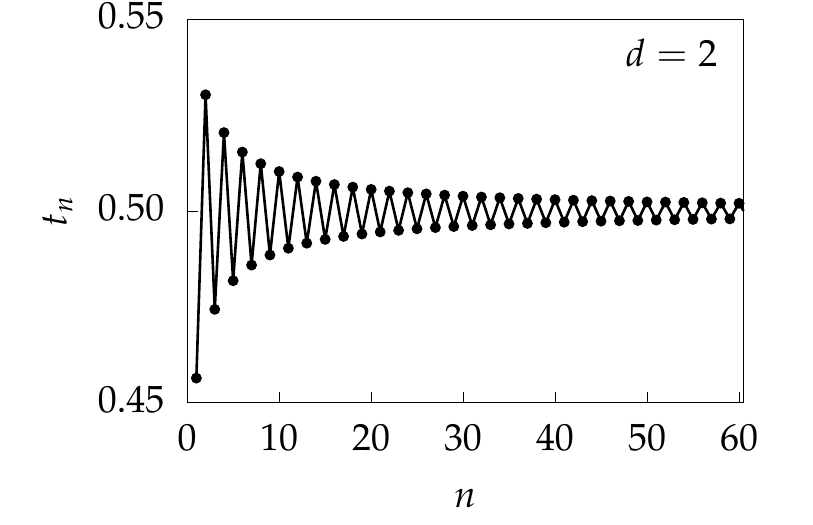}
\includegraphics[scale=1]{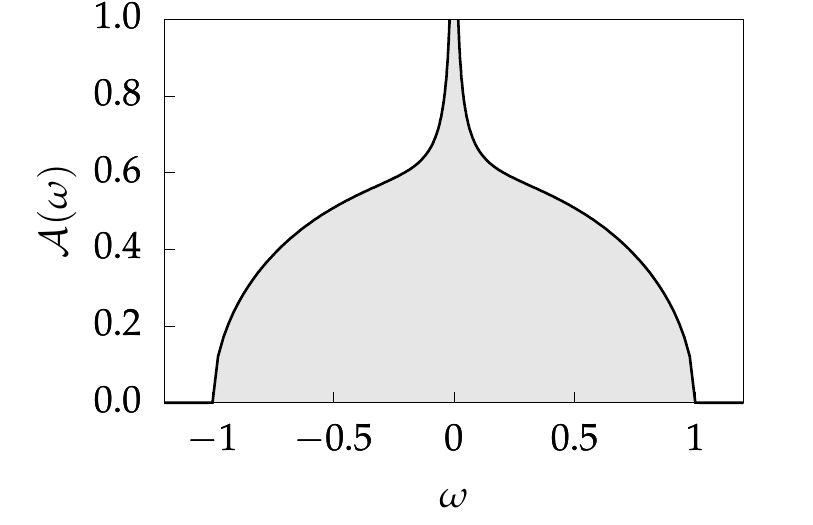}
%\hspace{-1.5em}
\includegraphics[scale=1]{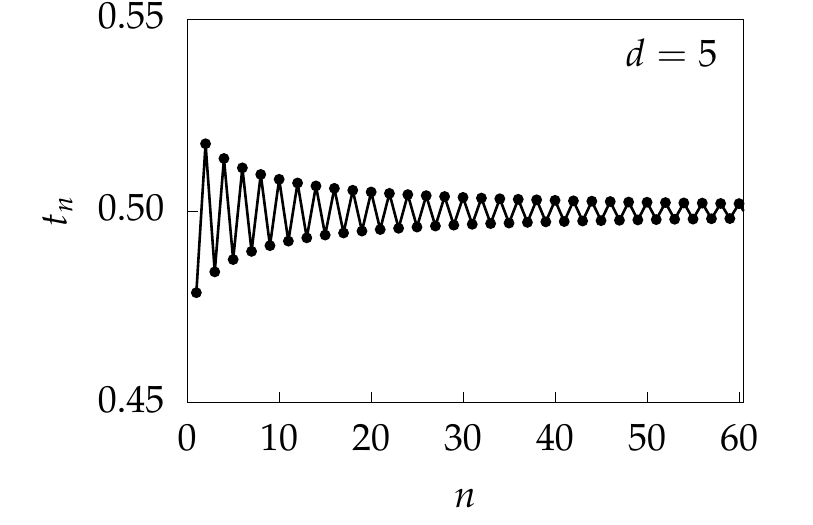}
\includegraphics[scale=1]{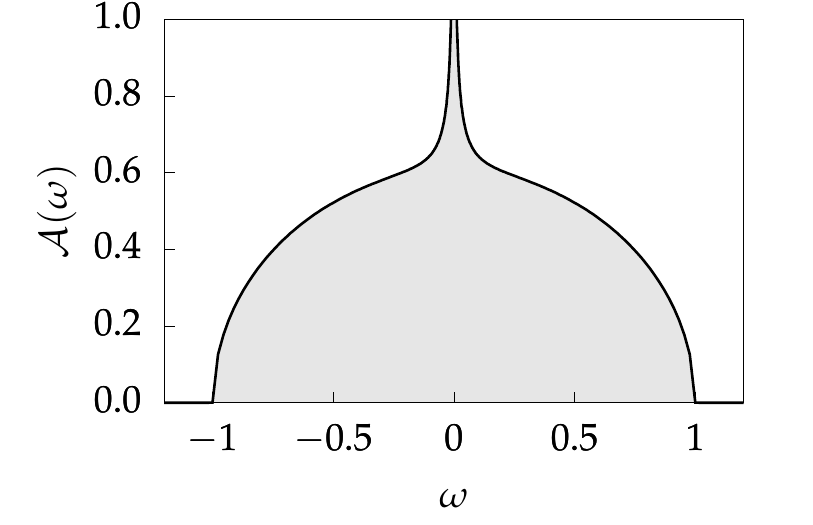}
%\hspace{-1.5em}
\includegraphics[scale=1]{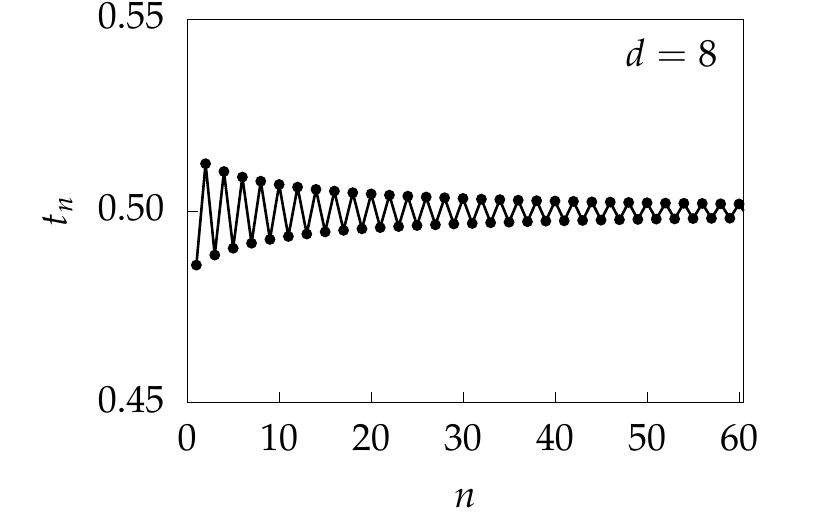}
\includegraphics[scale=1]{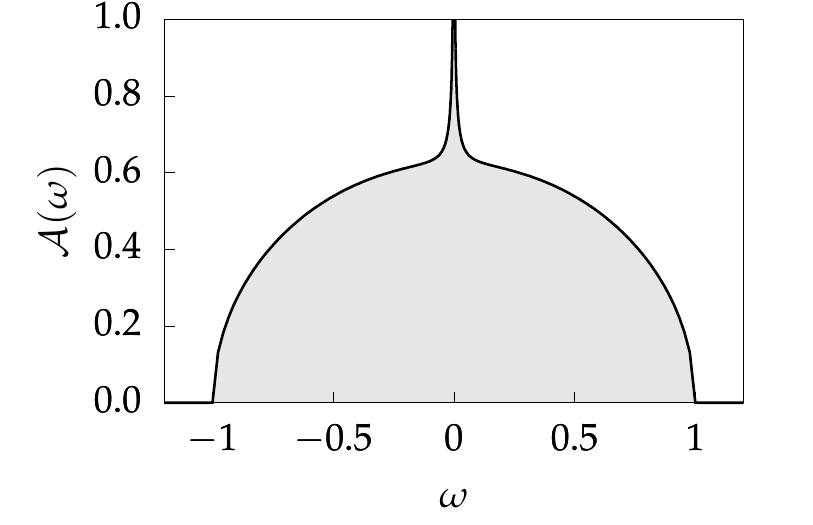}
%\hspace{-1.5em}
\includegraphics[scale=1]{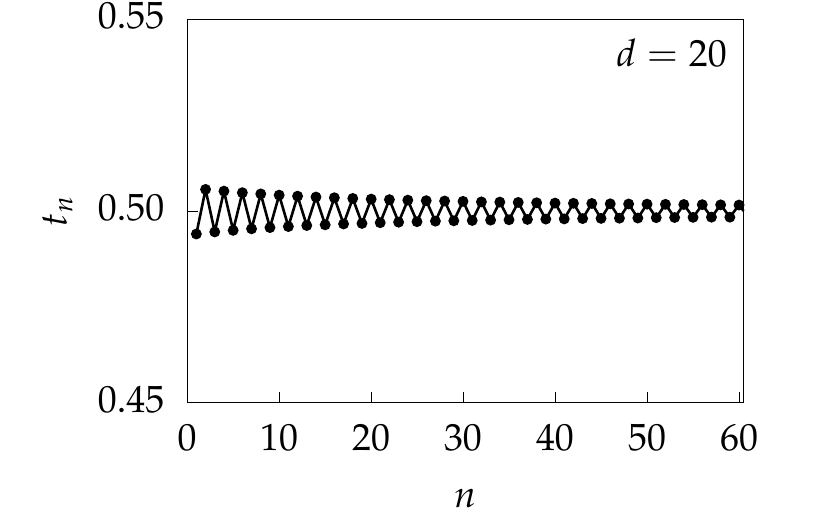}
\caption{Spectral functions (left) corresponding to models with $t_n$'s generated by Eq.~\eqref{eq:powerlawtn} with $\zeta = -$ (right) at various values of $d$ with $r = 0.5$.\label{fig:fractiond_r05}}
\end{figure}
\begin{figure}[htp]
\includegraphics[scale=1]{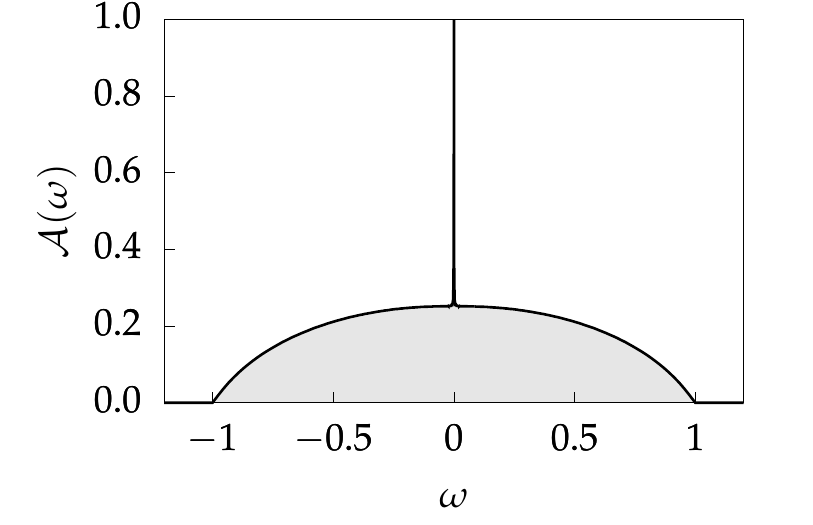}
\includegraphics[scale=1]{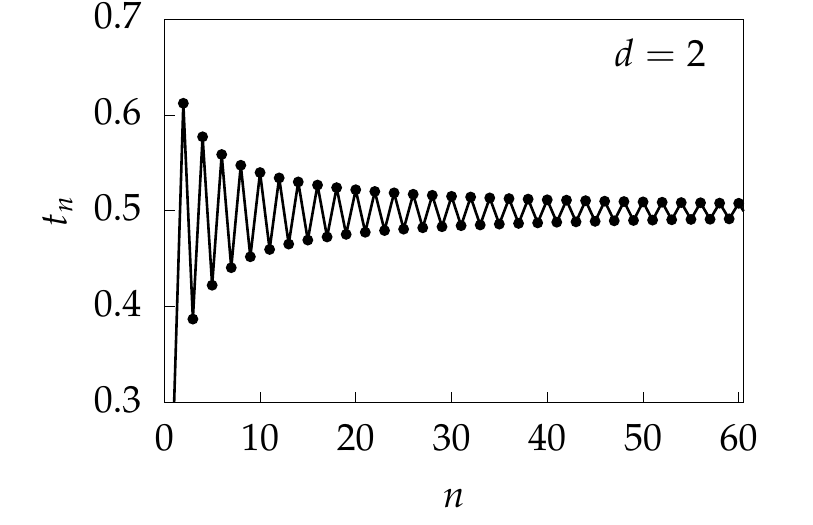}
\includegraphics[scale=1]{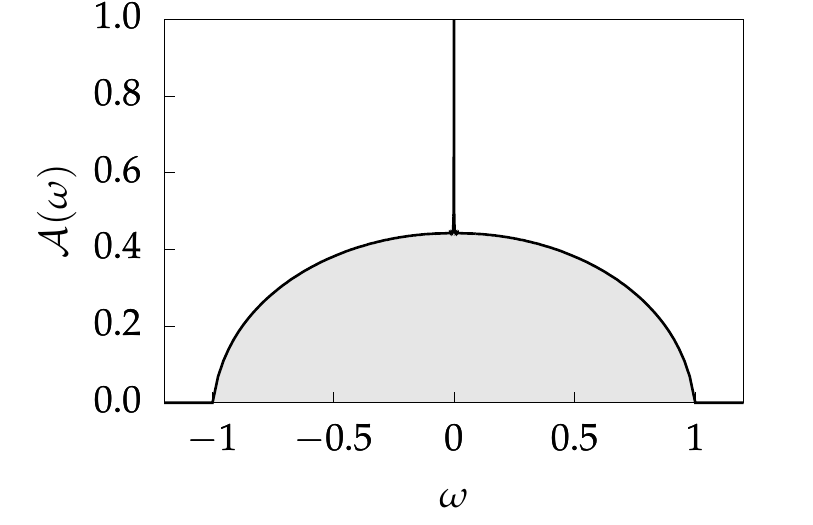}
\includegraphics[scale=1]{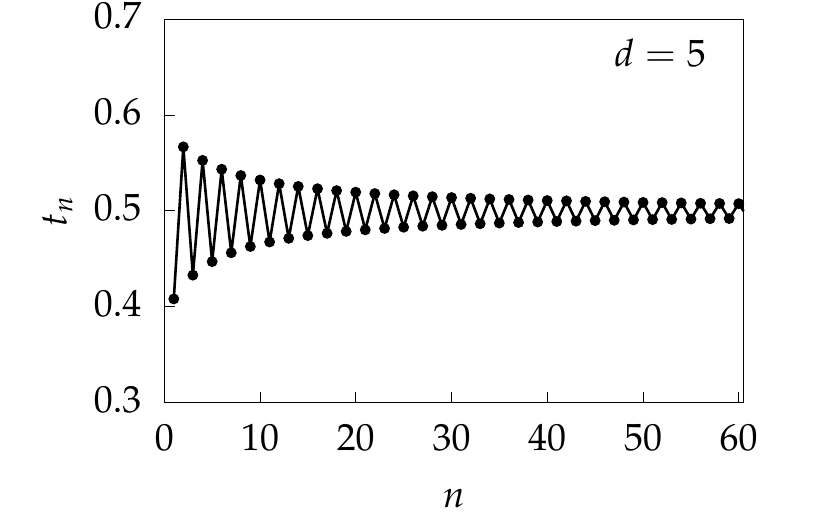}
\includegraphics[scale=1]{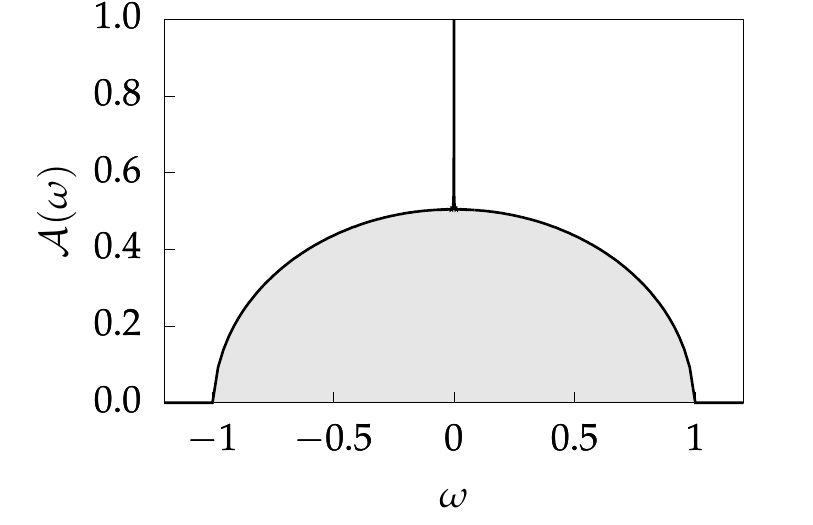}
\includegraphics[scale=1]{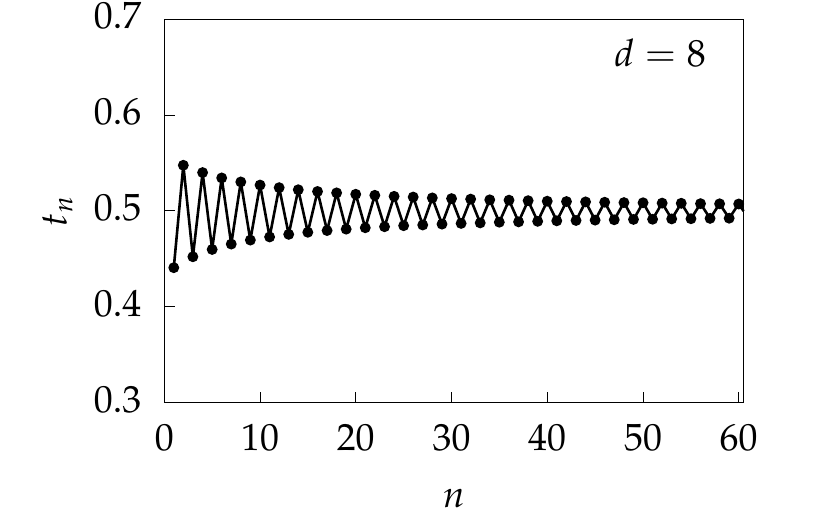}
\includegraphics[scale=1]{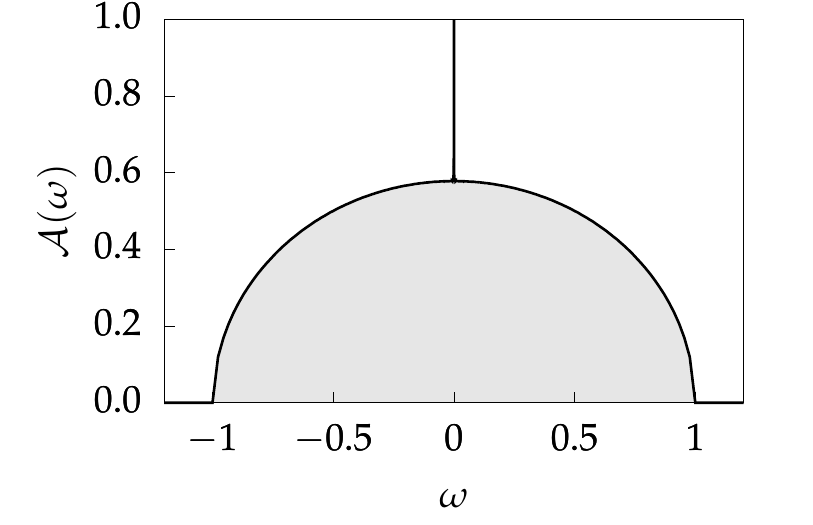}
\includegraphics[scale=1]{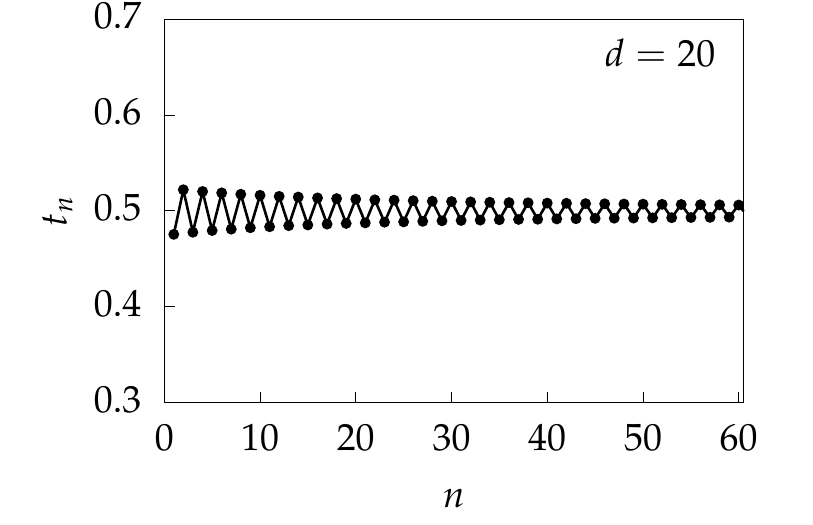}
\caption{Spectral functions (left) corresponding to models with $t_n$'s generated by Eq.~\eqref{eq:powerlawtn} with $\zeta = -$ (right) at various values of $d$ with $r = 2$.\label{fig:fractiond_r2}}
\end{figure}
In the parameterization Eq.~\eqref{eq:powerlawtn}, the power-law features in the resulting spectral functions typically set in at very low energies, $|\omega| \ll 1$. As such the power-law feature can be difficult to identify on linear scale plots. Spectral functions for $r = 0.5$ and $r = 2$ at $d = 20$ are plotted on a log scale in Fig.~\ref{fig:powerlogdd20}.
While on the linear scale plot, the spectrum for $r=2$, $d=20$ in Fig.~\ref{fig:fractiond_r2} appears similar to a pole lying among the continuum, but on the log scale plot shown in Fig.~\ref{fig:green_fractiond_r2d20_log}, the peak is clearly a power-law feature.
\begin{figure}[htp!]
\subfiglabel{\includegraphics{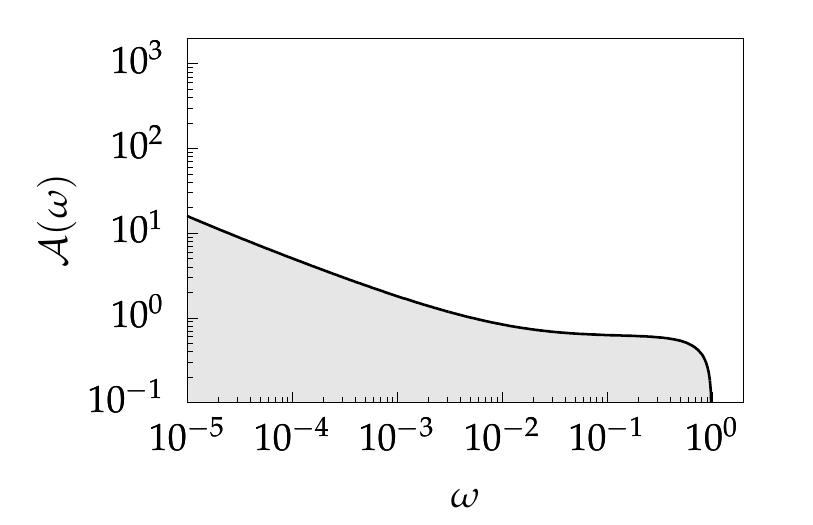}}{3.125,2}{fig:green_fractiond_r05d20_log}
\subfiglabel{\includegraphics{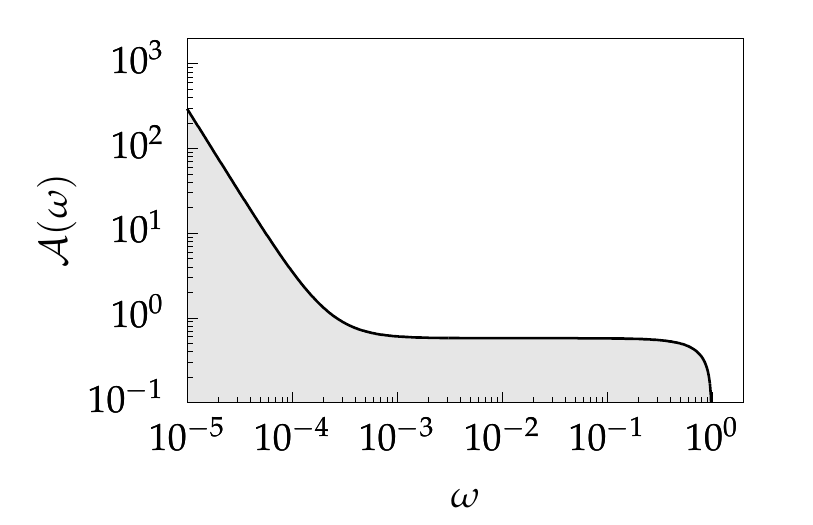}}{3.125,2}{fig:green_fractiond_r2d20_log}
\caption{Spectral functions for the power-law vanishing SSH model from Eq.~\eqref{eq:powerlawtn} with $\zeta = -$ and $d=20$ at \subref{fig:green_fractiond_r05d20_log} $r=0.5$ and \subref{fig:green_fractiond_r2d20_log} $r=2$. Log scale clearly shows the low energy power as $\mathcal{A}(\omega) \sim |\omega|^{-r}$\label{fig:powerlogdd20}}
\end{figure}

The spectrum does not feature a gap and there exist finite energy states infinitesimally close to zero energy. The wavefunctions of these low energy states exhibit a node near to the chain boundary, \textit{i.e.} a point in the chain $n_p$ where $\lvert \psi(n_p) \rvert^2 = 0$, effectively partitioning the wavefunction into separate parts. 
%\index{$0$@\textbf{List of Edits}!307@added clarification}
These wavefunctions exhibit exponential decay between the boundary and the node, and beyond the node the wavefunction is finite and delocalized. The nature of these low energy wavefunctions can be analyzed by considering the entanglement between the two partitions on either side of the node, a portion which can be considered the boundary and a portion which can be considered the bulk. The entanglement entropy is given by
\begin{equation}
	S_{\textsc{e}} = -\text{Tr} \left[\rho_A \ln \rho_A\right]
\end{equation}
where $\rho_A$ is the reduced density matrix of partition $A$. It is found that the portion of the wavefunction near to the boundary has significant entanglement with the bulk state. Because of this, the non-zero energy states should not be interpreted as being topological as they are extended into the bulk.
Only the state precisely at zero energy should be regarded as topological in the power-law diverging case. Note however that it is not separated from the continua by a gap as in the standard SSH case.

\section{Outlook}
Developed in this chapter were several generalizations of the SSH model. Previous generalizations of the SSH model considered unit cells of three and four sites, but the constructions here go further and devise models where the unit cell may possess an arbitrary number of sites. These previous studies exist as special cases of the more general models developed in this chapter.

Another novel generalization constructed here is the non-translation invariant models with power-law suppressed hopping amplitudes. Generalizations of the SSH model in the literature are generally concerned with the system's momentum space representation, and therefore completely miss such non-periodic models.

The SSH model is sometimes taken to be the simplest model of a crystalline topological insulator, or higher-order topological insulator~\cite{hoti}. The lessons learned about the generalized SSH models analyzed in this chapter may provide the basis for generalizing crystalline topological insulators in higher dimensions.

While the generalized SSH models developed here are interesting in their own right, many features of these generalizations will appear again in the context of the auxiliary field mapping in \S\ref{ch:aux} and \S\ref{ch:motttopology}.
%\index{$0$@\textbf{List of Edits}!308@added outlook}

\chapter{Interacting SSH Model on the Bethe Lattice\label{ch:bethessh}}
\chaptermark{Interacting Bethe SSH}

In this chapter a variant of the SSH model is treated with on-site Coulomb interactions. The interactions are treated with DMFT-NRG introduced in \S\ref{sec:dmft}. A prerequisite to performing this calculation involves mapping the original $1d$ SSH model into a form which has a well-defined infinite dimensional limit, suitable for treatment with DMFT~\cite{bethessh}.

Topological phases of matter have been intensely studied in recent years~\cite{hk}, but there still remain a number of open questions regarding the interplay of topology and strong interactions in condensed matter systems~\cite{interacting}, including the role of topological invariants.

The $1d$ SSH model with on-site Coulomb interactions has previously been analyzed perturbatively~\cite{hubbardpeierls}. This analysis showed that the Hubbard interaction does not qualitatively alter the characteristics of the noninteracting SSH model for $U<U_c$. In particular, the soliton solutions of the SSH model persist in the presence of the interaction.
A similar study~\cite{interactingrm} investigated the topology of the $1d$ Rice-Mele model with nearest-neighbor interactions in the low-energy weak-interaction regime of finite system size using functional renormalization group (fRG) and density matrix renormalization group (DMRG) techniques. In this work the topological edge states are modified due to spatial variations in the self-energy close to the system boundary. Such effects are not present in the system considered here due to the local nature of the self-energy within DMFT and the absence of a boundary.
Other previous studies of the SSH model with interactions considered nearest-neighbor density-density interactions~\cite{sirker}.
The SSH model with Hubbard interaction has also been studied using exact diagonalization (ED)~\cite{guo} and density matrix renormalization group (DMRG)~\cite{manmana,barbiero} techniques. These studies were restricted to finite system sizes in $1d$. In contrast, the approach taken below is that of DMFT-NRG and is performed in the limit of infinite dimensions and system size.
This exact solution in the limit of infinite dimensions to date has not been performed in the literature.

A distinguishing feature of the present study in contrast to the aforementioned work is that here the effect of the Mott metal-insulator transition on the topological features is seen. The existing studies mentioned previously have not been able to touch on this element as they have worked exclusively in $1d$, where the Mott transition does not occur in the ground state~\cite{liebwu}.

\section{SSH Model on the Bethe Lattice}

\begin{figure}[ht!]
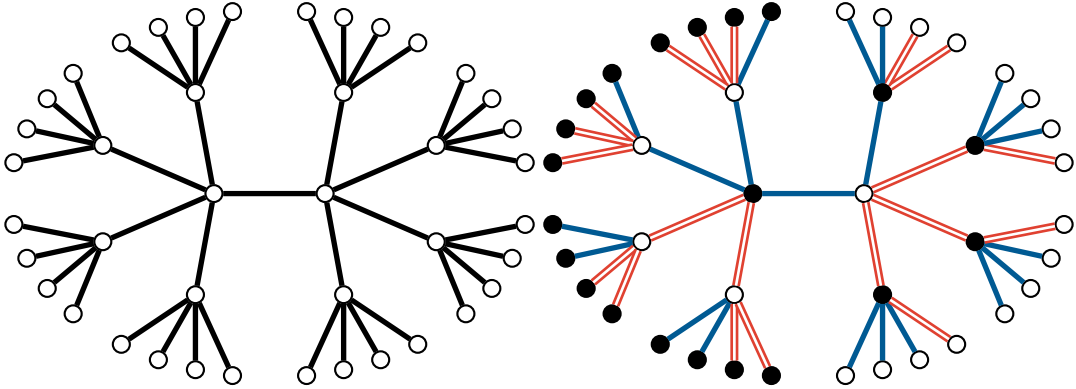

\centering
\begin{tikzpicture}[scale=0.65, line width=2pt, every node/.style={scale=0.75,inner sep=3pt}, every path/.style={scale=0.75}]
\input{bethelatt.tex}
\end{tikzpicture}
\begin{tikzpicture}[scale=0.65, every node/.style={scale=0.75,inner sep=3pt}, every path/.style={scale=0.75}]
\input{bethessh_fig.tex}
\end{tikzpicture}
\caption[Schematic of the SSH model on the Bethe lattice]{Schematic of the higher dimensional SSH model fitted to the Bethe lattice with $\tensor*{\kappa}{_A} = 3$, $\tensor*{\kappa}{_B} = 2$. The single blue links represent links with $t_A$ and the double red links represent links with $t_B$. The $\bullet$- ($\circ$-)sites are those with majority $t_A$ ($t_B$) bonds.\label{fig:sshbethe}}
\end{figure}
This chapter investigates the characteristics of an SSH-type topological insulator with local Coulomb interactions which are handled by DMFT.
DMFT becomes exact in the limit of infinite lattice coordination number. The objective in the present chapter is to exploit this limit to obtain exact results for interacting topologically non-trivial systems. In order to analyze a strongly correlated topological insulator in this framework, it is necessary to construct a model which possesses topological features in this limit. The choice here is to generalize the SSH model to the Bethe lattice, a lattice commonly utilized in DMFT calculations.

To create a model on the Bethe lattice which captures the same properties as the SSH model, it is necessary to not only define two hopping parameters $t_A$ and $t_B$, but also to partition the 
coordination number $\kappa$ into $\tensor*{\kappa}{_A}$ and $\tensor*{\kappa}{_B}$, with $\kappa = \tensor*{\kappa}{_A} + \tensor*{\kappa}{_B}$.
The Hamiltonian for this model can be written as
\begin{equation}
	\op{H}{} =
	\left( \smashoperator[r]{\sum_{\ell\in\{\tensor*{\kappa}{_A}\}}} t_A \, \opd{c}{j+\ell} \op{c}{j} + \smashoperator[r]{\sum_{\ell\in\{\tensor*{\kappa}{_B}\}}} t_B \, \opd{c}{j+\ell} \op{c}{j} + \hc \right)
\end{equation}
where $\ell\in\{\tensor*{\kappa}{_A}\}$ indicates summing over all nearest neighbors in the set $\{\tensor*{\kappa}{_A}\}$ and $\ell\in\{\tensor*{\kappa}{_B}\}$ similarly for $\{\tensor*{\kappa}{_B}\}$.
An example of this lattice is shown schematically in Fig.~\ref{fig:sshbethe}.

%%%
\subsection{The Green functions}

The limit of infinite coordination number is taken in a controlled way in order to ensure that parameters remain finite and the salient non-trivial features of the model are preserved. 
Although the Bethe lattice is infinite in extent and therefore only possesses a bulk and no boundary, it is possible to identify through the Green functions a certain type of bulk-boundary correspondence.

The Green functions for this system may be calculated by considering the Green functions for ``surface'' sites. Since the surface of a $d$-dimensional system is $(d-1)$-dimensional, surface sites of the SSH Bethe lattice are defined as either a $\bullet$- or $\circ$-site with either a $t_A$ or $t_B$ bond removed. 
\begin{subequations}
A surface $\circ$-site with a $t_A$ link removed has the Green function
\begin{align}
	G^\circ_{\text{s}A}(z)
	&=
		\cfrac{1}
		{ z - (\tensor*{\kappa}{_B} - 1) t_A^2 \ G^\bullet_{\text{s}A}(z) - \tensor*{\kappa}{_A} t_B^2 \ G^\bullet_{\text{s}B}(z) }
\end{align}
and the analogous site with a $t_B$ bond removed is
\begin{align}
	G^\circ_{\text{s}B}(z)
	&=
		\cfrac{1}
		{ z - (\tensor*{\kappa}{_A} -1) t_B^2 \ G^\bullet_{\text{s}B}(z) - \tensor*{\kappa}{_B} t_A^2 \ G^\bullet_{\text{s}A}(z) } \,.
\end{align}
The analogous surface Green functions for $\bullet$-sites are
\begin{align}
	G^\bullet_{\text{s}A}(z)
	&=
		\cfrac{1}
		{ z - (\tensor*{\kappa}{_A} - 1) t_A^2 \ G^\circ_{\text{s}A}(z) - \tensor*{\kappa}{_B} t_B^2 \ G^\circ_{\text{s}B}(z) }
	\\
	G^\bullet_{\text{s}B}(z)
	&=
		\cfrac{1}
		{ z - (\tensor*{\kappa}{_B} - 1) t_B^2 \ G^\circ_{\text{s}B}(z) - \tensor*{\kappa}{_A} t_A^2 \ G^\circ_{\text{s}A}(z) } \,.
\end{align}
\end{subequations}
The bulk Green functions can be written in terms of these surface Green functions as
\begin{align}
	G_{\text{b}}^\bullet
	&=
	\cfrac{1}
	{ z - \cfrac{\tensor*{\kappa}{_A} t_A^2}{ z - (\tensor*{\kappa}{_B} - 1) t_A^2 \ G^\bullet_{\text{s}A} - \tensor*{\kappa}{_A} t_B^2 \ G^\bullet_{\text{s}B} } - \cfrac{\tensor*{\kappa}{_B} t_B^2}{ z - (\tensor*{\kappa}{_A} -1) t_B^2 \ G^\bullet_{\text{s}B} - \tensor*{\kappa}{_B} t_A^2 \ G^\bullet_{\text{s}A} }
	}
	\\
	&=
	\cfrac{1}
	{ z - \tensor*{\kappa}{_A} t_A^2 \ G^\circ_{\text{s}A} - \tensor*{\kappa}{_B} t_B^2 \ G^\circ_{\text{s}B}
	}
\intertext{and}
	G_{\text{b}}^\circ
	&=
	\cfrac{1}
	{ z - \cfrac{\tensor*{\kappa}{_A} t_A^2}{ z - (\tensor*{\kappa}{_A} - 1) t_A^2 \ G^\circ_{\text{s}A} - \tensor*{\kappa}{_B} t_B^2 \ G^\circ_{\text{s}B} } - \cfrac{\tensor*{\kappa}{_B} t_B^2}{ z - (\tensor*{\kappa}{_B} - 1) t_B^2 \ G^\circ_{\text{s}B} - \tensor*{\kappa}{_A} t_A^2 \ G^\circ_{\text{s}A} }
	}
	\\
	&=
	\cfrac{1}
	{ z - \tensor*{\kappa}{_A} t_A^2 \ G^\bullet_{\text{s}A} - \tensor*{\kappa}{_B} t_B^2 \ G^\bullet_{\text{s}B}
	} \,.
\end{align}
In the limit of infinite coordination number, $\tensor*{\kappa}{_A} - 1 \approx \tensor*{\kappa}{_A}$ and $\tensor*{\kappa}{_B} - 1 \approx \tensor*{\kappa}{_B}$. In this limit, the $A$ and $B$ surface Green functions on the $\bullet$- and $\circ$-sites become
\begin{equation}
\begin{gathered}
	G^\circ_{\text{s}A}(z)
	\approx
		\cfrac{1}
		{ z - \tensor*{\kappa}{_B} t_A^2 \ G^\bullet_{\text{s}A}(z) - \tensor*{\kappa}{_A} t_B^2 \ G^\bullet_{\text{s}B}(z) }
	\approx
	G^\circ_{\text{s}B}(z)
	\\
	G^\bullet_{\text{s}A}(z)
	\approx
		\cfrac{1}
		{ z - \tensor*{\kappa}{_B} t_B^2 \ G^\circ_{\text{s}B}(z) - \tensor*{\kappa}{_A} t_A^2 \ G^\circ_{\text{s}A}(z) }
	\approx
	G^\bullet_{\text{s}B}(z) \,.
\end{gathered}
\end{equation}
The `surface' sites for both types of sites are equivalent to each their corresponding bulk counterpart, 
\begin{equation}
\begin{aligned}
	G^\circ_{\text{s}*}(z)
	&\approx
	G^\circ_{\text{b}}(z)
	\\
	G^\bullet_{\text{s}*}(z)
	&\approx
	G^\bullet_{\text{b}}(z) \,,
\end{aligned}
\end{equation}
where $* \in \{A,B\}$ and `$\text{b}$' denotes ``bulk.'' The distinction between a $d$-dimensional bulk and its $(d-1)$-dimensional boundary is lost in the limit of infinite dimensions.
The equivalence of the bulk and surface Green functions demonstrates an appearance of the bulk-boundary correspondence in the Bethe SSH model.

In order to have the Green functions be non-trivial under the limit of $\kappa\to\infty$, it is necessary to ensure that the system parameters are extensive. To this end, it is useful to
define the quantities
\begin{align}
	\mathsf{x}^2 &\vcentcolon= \tensor*{\kappa}{_B} t_B^2
	&
	\mathsf{y} &\vcentcolon= \frac{t_B}{t_A}
	&
	\mathsf{z} &\vcentcolon= \frac{\tensor*{\kappa}{_B}}{\tensor*{\kappa}{_A}}
\end{align}
as remaining fixed when limit is taken.
The Green functions in the infinite dimensional limit then take the form of
\begin{subequations}
\begin{align}
	G^\bullet_\infty(z) &= \frac{1}{z - \left( \frac{\mathsf{x}^2}{\mathsf{z}} + \frac{\mathsf{x}^2}{\mathsf{y}^2} \right) G^\circ_\infty(z)}
\intertext{and}
	G^\circ_\infty(z) &= \frac{1}{z - \left( \mathsf{x}^2 + \frac{\mathsf{x}^2}{\mathsf{y}^2 \mathsf{z}} \right) G^\bullet_\infty(z)} \,.
\end{align}
\end{subequations}
These equations recover the Green functions of the standard SSH model for both the trivial and topological phases identifying rescaled hopping parameters as $\tilde{t}_A^2 \vcentcolon= \frac{\mathsf{x}^2}{\mathsf{z}} + \frac{\mathsf{x}^2}{\mathsf{y}^2}$ and $\tilde{t}_B^2 \vcentcolon= \mathsf{x}^2 + \frac{\mathsf{x}^2}{\mathsf{y}^2 \mathsf{z}}$. With these replacements the Green functions for the two sites take the form of
\begin{subequations}
\begin{align}
	G^\circ_\infty(z) &= \frac{1}{z - \cfrac{\tilde{t}_B^2}{z - \tilde{t}_A^2 G^\circ_\infty(z)}}
\intertext{and}
	G^\bullet_\infty(z) &= \cfrac{1}{z - \cfrac{\tilde{t}_A^2}{z - \tilde{t}_B^2 G^\bullet_\infty(z)}} \,,
\end{align}
\end{subequations}
which match the functional form of the edge site Green functions of the $1d$ SSH model in the trivial and topological phases respectively. 

By way of the comparison with the SSH chain Green functions, the $\bullet$-sites exhibit the topological configuration when the compound effective hopping parameters satisfy $\tilde{t}_A < \tilde{t}_B$. In terms of the basic parameters, this enforces the simultaneous requirement that $t_A < t_B$ and $\tensor*{\kappa}{_A} > \tensor*{\kappa}{_B}$. Inverting this requirement results in the $\circ$-sites carrying the topological configuration and the $\bullet$-sites exhibiting the trivial configuration. If either $\mathsf{y}=1$ or $\mathsf{z}=1$, then $\tilde{t}_A = \tilde{t}_B$ and $G^\bullet_\infty(z) = G^\circ_\infty(z)$, which results in the system exhibiting a semi-elliptical metallic local density of states on every site.
The following analysis is performed with the $\bullet$-sites always taking the topological character and the $\circ$-sites always trivial, \textit{i.e.} parameters are chosen such that $\tilde{t}_A < \tilde{t}_B$ from here on.
As with the $1d$ SSH model, the bandwidth is defined as $D = \tilde{t}_A + \tilde{t}_B$. This $D$ sets the energy scale of the system of the interactions in the following.

\begin{figure}[ht!]
	\subfiglabel{\includegraphics[scale=1]{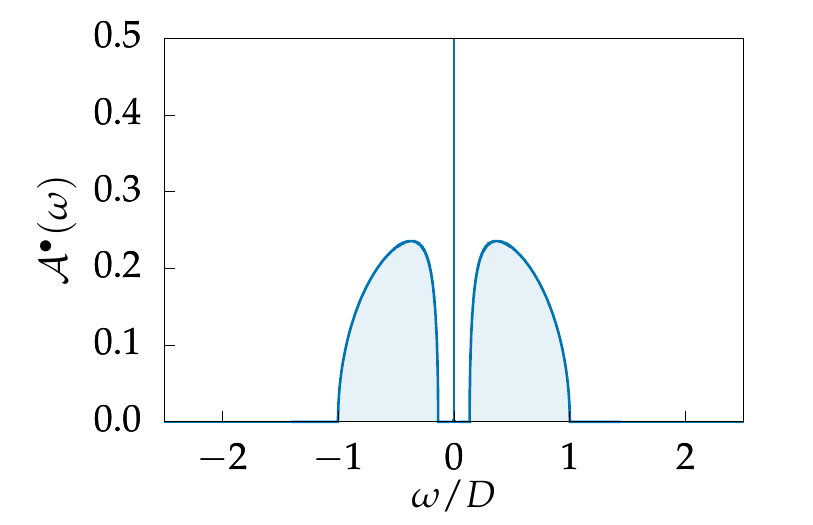}}{3.125,2}{fig:hsshU0a}
	\subfiglabel{\includegraphics[scale=1]{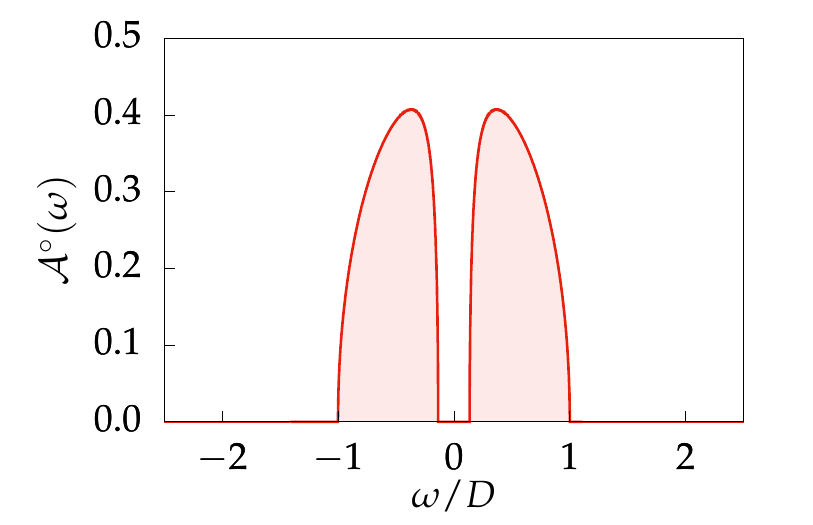}}{3.125,2}{fig:hsshU0b}
	\vspace{-\baselineskip}
\caption{Spectral functions on the $\bullet$-sites \subref{fig:hsshU0a} and $\circ$-sites \subref{fig:hsshU0b} of the non-interacting Bethe SSH model. \label{fig:hsshU0.0}}
\end{figure}

The $\circ$- and $\bullet$-sites then play the roles of the `bulk' and `boundary' sites in the bulk-boundary correspondence\index{bulk-boundary correspondence}. The Bethe SSH lattice contains both types of sites by construction, and enforcing a topological state on one type of site forces the other type to be trivial and insulating. In this way the infinite boundary-less Bethe SSH lattice still realizes the bulk-boundary correspondence.

It is worth noting that these topological sites are still bulk sites, as the Bethe lattice has no real boundary, in contrast to the usual $1d$ model wherein the topological site only occurs on the boundary of the chain.

%\begin{figure}[ht!]
%\centering
%\begin{tikzpicture}[scale=0.65, line width=2pt, every node/.style={scale=0.75,inner sep=3pt}, every path/.style={scale=0.75}]
%\input{bethelatt.tex}
%\end{tikzpicture}
%\caption[Bethe lattice]{Schematic of the Bethe lattice. Shown here is a cluster subset. The true Bethe lattice has infinite extent.\label{blankbethe}}
%\end{figure}

\subsubsection{Choice of Lattice}
The higher dimensional generalization of the SSH model presented here requires a lattice with certain properties. The structure of the lattice must be such that $\mathsf{z}\neq1$ on every site of the lattice. 
There does not exist a lattice exhibiting these properties which also has a reciprocal lattice. A mathematical proof of this is not known to the author at present.

In order to produce an SSH spectral pole, it is not necessary for every site to have the same set of parameters $\{\tensor*{\kappa}{_A},\tensor*{\kappa}{_B},t_A,t_B\}$, but rather it is only necessary that $\tensor*{\kappa}{_A} > \tensor*{\kappa}{_B}$ and $t_A < t_B$ on alternating sites. For randomized parameters restricted to fit the requirements on each site, the spectral function takes a form which is analogous to the $1d$ SSH model with disordered hopping amplitudes, \textit{cf.} Fig.~\ref{fig:dtsshspec}.
%takes the form of Anderson localized states, similar to that of a quasicrystal.

\section{Effect of Interactions}

An interacting version of this model may be constructed by adding a standard on-site Hubbard four-fermion spin-spin interaction term.
\begin{figure}[ht!]
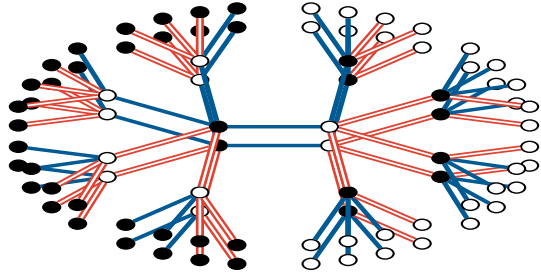

\centering
%\hspace{-1em}
\scalebox{1}[0.65]
{
\begin{tikzpicture}[scale=0.65, every node/.style={scale=0.75,inner sep=3pt}, every path/.style={scale=0.75}]
	\input{bethessh_fig.tex}
\end{tikzpicture}
}\\
\vspace{-7.5\baselineskip}
%\hspace{-2em}
\scalebox{1}[0.65]
{
\begin{tikzpicture}[scale=0.65, every node/.style={scale=0.75,inner sep=3pt}, every path/.style={scale=0.75}]
	\input{bethessh_fig.tex}
\end{tikzpicture}
}\\
\vspace{\baselineskip}
\caption[Schematic of lattice for Hubbard-SSH model]{Schematic of lattice for Hubbard-SSH model, which consists of one copy of the Bethe lattice for each spin and local interactions between corresponding sites.\label{fig:doublebethessh}}
\end{figure}
This involves two copies of the Bethe lattice, one for each spin, with the local Hubbard interaction taking place between corresponding sites, illustrated in Fig.~\ref{fig:doublebethessh}.
The Hamiltonian of this Bethe lattice Hubbard-SSH model is given by
\begin{equation}
%\begin{multlined}[c][0.88\linewidth]
	\op{H}{\textsc{hssh}}
	=
	\smashoperator{\sum_{j\in\Gamma_{\textsc{bl}},\sigma}}\ \left[
		\varepsilon \, \opd{c}{j,\sigma} \op{c}{j,\sigma} + \left( 
		t_A \smashoperator{\sum_{\ell\in\{\tensor*{\kappa}{_A}\}}} \opd{c}{j+\ell,\sigma} \op{c}{j,\sigma} + 
		t_B \smashoperator{\sum_{\ell\in\{\tensor*{\kappa}{_B}\}}} \opd{c}{j+\ell,\sigma} \op{c}{j,\sigma} + \hc\! \right) \right]
	+ U \smashoperator{\sum_{j\in\Gamma_{\textsc{bl}}}} \opd{c}{j,\uparrow} \op{c}{j,\uparrow} \opd{c}{j,\downarrow} \op{c}{j,\downarrow}
%\end{multlined}
\end{equation}
where $j$ indexes the sites on the Bethe lattice, $\ell$ runs over nearest neighbors either in the set $\{\tensor*{\kappa}{_A}\}$ or $\{\tensor*{\kappa}{_B}\}$, and $\sigma \in \{\uparrow,\downarrow\}$ is the spin index.

The system is solved using DMFT-NRG as described in \S\ref{sec:dmft}.
The full interacting Green functions for the two types of sites are given by
\begin{subequations}
\begin{equation}
\begin{aligned}[b]
	G^\bullet_\sigma(z)
	&= \cfrac{1}{z-\varepsilon - \cfrac{\tilde{t}_A^2}{z-\varepsilon - \tilde{t}_B^2 G^\bullet_\sigma(z) - \Sigma^\circ_\sigma(z)} - \Sigma^\bullet_\sigma(z)}
	\\
	&= \frac{1}{z-\varepsilon - K^\circ_\sigma(z) - \Sigma^\bullet_\sigma(z)}
\end{aligned}
\end{equation}
and
\begin{equation}
\begin{aligned}[b]
	G^\circ_\sigma(z)
	&= \cfrac{1}{z-\varepsilon - \cfrac{\tilde{t}_B^2}{z-\varepsilon - \tilde{t}_A^2 G^\circ_\sigma(z) - \Sigma^\bullet_\sigma(z)} - \Sigma^\circ_\sigma(z)}
	\\
	&= \frac{1}{z-\varepsilon - K^\bullet_\sigma(z) - \Sigma^\circ_\sigma(z)} \,.
\end{aligned}
\end{equation}
\label{eq:hsshbethegreenfunctions}
\end{subequations}

%\subsection{Topology}

In the following analysis, parameters are chosen with
\begin{equation}
	( \mathsf{x} , \mathsf{y} , \mathsf{z} ) = ( 1.0,\, 3.0,\, 2 )
\label{eq:hsshparam}
\end{equation}
such that $\tilde{t}_A^2 = \frac{11}{18} \approx 0.61$ and $\tilde{t}_B^2 = \frac{19}{18} \approx 1.06$. Since $\tilde{t}_A < \tilde{t}_B$, the $\bullet$-sites express the topological feature and the $\circ$-sites express the trivial spectrum.
With this parameterization choice the effective band gap is $2\lvert \tilde{t}_B - \tilde{t}_A \vert \approx 0.49$ and the effective full band width is $2D \equiv 2\lvert \tilde{t}_B + \tilde{t}_A \rvert \approx 3.62$.

As the calculation takes place in the infinite dimensional limit, the solution presented here is numerically exact and not an approximation.
The NRG is performed with $\Lambda=2.0$ keeping 3000 states in each iteration at a temperature of $T/D = 10^{-6}$, which is considered to be effectively zero temperature.

The system is now first analyzed in the case with small, but finite, interaction strength, $0<U \ll D$. This is in the perturbative regime, however the solution presented here is obtained from the full non-perturbative DMFT-NRG calculation. The reason for studying this regime is to ascertain how the interactions affect the different elements of the SSH spectrum without being concerned with whether the effects are interfering with each other.
At higher interaction strengths the spectral features become heavily broadened, and it is difficult to determine exactly which spectral features are augmented in which way from the interactions.
While these results are in the perturbative regime, they have been obtained using the full non-perturbative DMFT-NRG.

The spectral functions for the $\bullet$- and $\circ$-sites at $U/D = 0.01$ on a linear scale are plotted in Fig.~\ref{fig:hsshU0.01linear}. The contrasting features of these spectra compared to the non-interacting case ($U=0.0$, \textit{cf.} Fig.~\ref{fig:hsshU0.0}) are not clearly seen as they occur at logarithmically small scales. It is therefore advantageous to examine the spectral functions on a logarithmic scale, as shown in Fig.~\ref{fig:hsshUsmalllog}, where the spectral functions for the $\bullet$- and $\circ$-sites is plotted for $U/D=0.001$ and $U/D=0.01$.
The DMFT-NRG solution shows that the zero-energy pole on the topological $\bullet$-sites is broadened to exhibit power-law behavior which appears at $\omega \sim \mathcal{O}(U)$ and decays exponentially into the preformed gap. On the logarithmic scale plots in Fig.~\ref{fig:hsshUsmalllog}, the low energy power-law appears linear. At higher energy the power-law feature develops a shoulder which decays into the gap. This shoulder is the broadening. The continuum bands outside the gap are not meaningfully affected by the interactions at this strength. This similar to the case of the Hubbard model as solved in \S\ref{sec:hubbardsolution}. In Fig.~\ref{fig:hubbardsolution}, the difference in the continuum band of the Hubbard model spectrum between $U/D = 0$ and $U/D = 1.0$ is minimal. The interaction strength considered for the HSSH model at present is orders of magnitude less than that.
For $\omega < U$, the spectral function exhibits the behavior $\mathcal{A}^{\bullet/\circ}(\omega)\sim|\omega|^{\mp r}$ where the exponent is $-r$ for $\bullet$-sites and $+r$ for $\circ$-sites ($r>0$).
The value of the power-law scaling $r$ is dependent upon the system parameters $\tilde{t}_A$ and $\tilde{t}_B$.
For the parameterization~\eqref{eq:hsshparam}, the exponent is $r\approx0.41$.
This value changes based on the choice of parameters $(\mathsf{x},\mathsf{y},\mathsf{z})$. 
For $(\mathsf{x},\mathsf{y},\mathsf{z}) = ( 1.0, 2.0, 2 )$, the non-interacting band gap is $2 |\tilde{t}_B - \tilde{t}_A| \approx 0.39$ and $r \approx 0.43$.
For $(\mathsf{x},\mathsf{y},\mathsf{z}) = ( 1.0, 3.0, 3 )$, the non-interacting band gap is $2 |\tilde{t}_B - \tilde{t}_A| \approx 0.70$ and $r \approx 0.37$.
From this comparison it is seen that it is coincidental that the power-law scale factor $r$ is approximately the same as the band gap for the parameterization~\eqref{eq:hsshparam} as this is not the case for different choices of parameterization.
\begin{figure}[ht!]
	\includegraphics[scale=1]{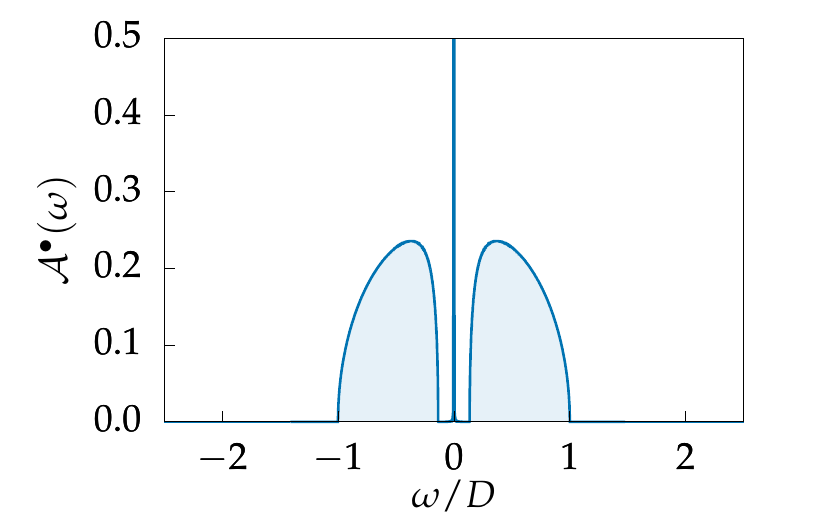}
	\includegraphics[scale=1]{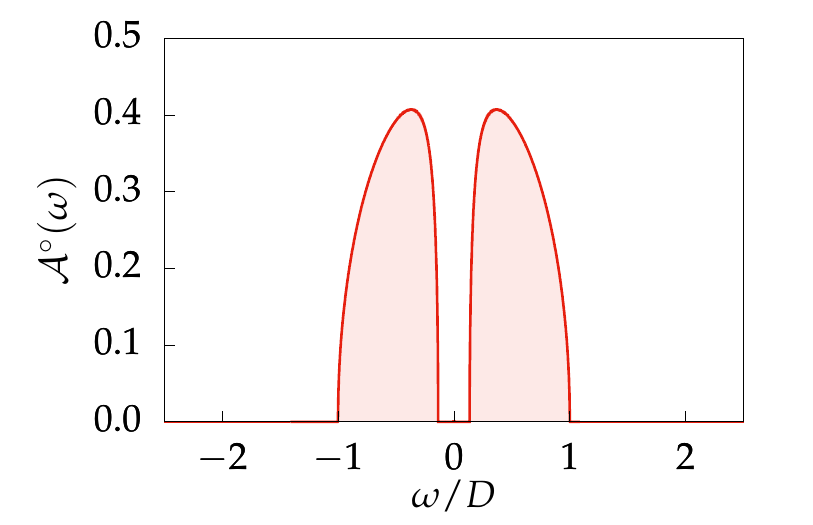}
\caption[Spectral functions for the $\bullet$- and $\circ$-sites at $U/D = 0.01$ on a linear scale]{Spectral functions for the $\bullet$- and $\circ$-sites at $U/D = 0.01$ on a linear scale. On this scale the effects of the interactions are difficult to visualize. See the logarithmic scale plots in Fig.~\ref{fig:hsshUsmalllog} for a better visualization.\label{fig:hsshU0.01linear}}
\end{figure}
\begin{figure}[ht!]
\begin{subfigure}[c]{0.5\linewidth}
\begin{tikzpicture}
	\node at (0,0) {\includegraphics[width=\linewidth]{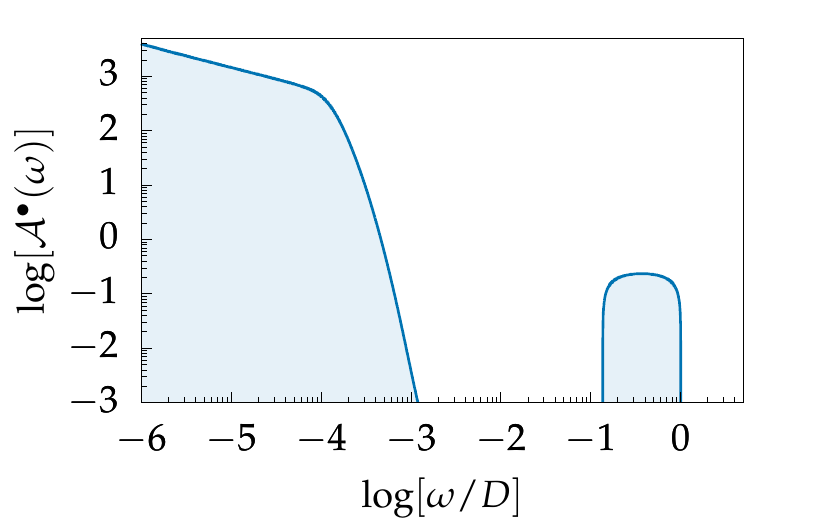}};
	\node at (3,1.875) {\subref*{fig:hsshU0_001alog}};
\end{tikzpicture}
\phantomsubcaption{\vspace{-\baselineskip}\label{fig:hsshU0_001alog}}
\end{subfigure}
\begin{subfigure}[c]{0.5\linewidth}
\begin{tikzpicture}
	\node at (0,0) {\includegraphics[width=\linewidth]{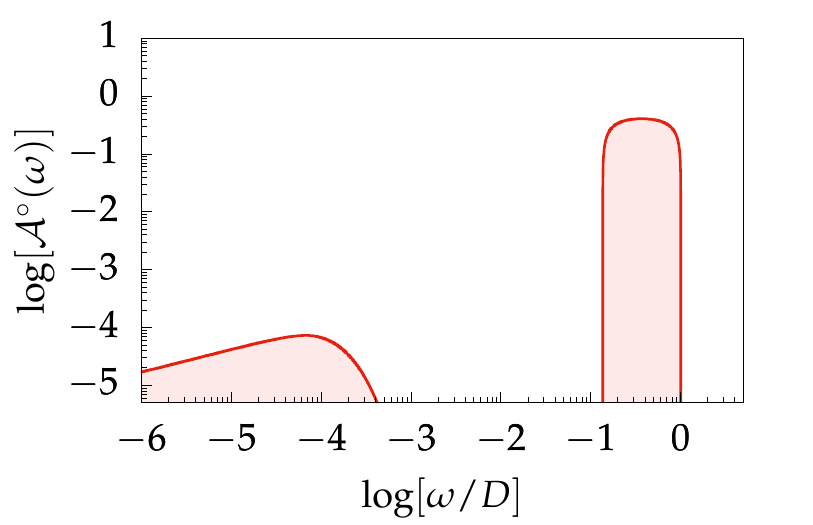}};
	\node at (3,1.875) {\subref*{fig:hsshU0_001blog}};
\end{tikzpicture}
\phantomsubcaption{\vspace{-\baselineskip}\label{fig:hsshU0_001blog}}
\end{subfigure}
\begin{subfigure}[c]{0.5\linewidth}
\begin{tikzpicture}
	\node at (0,0) {\includegraphics[width=\linewidth]{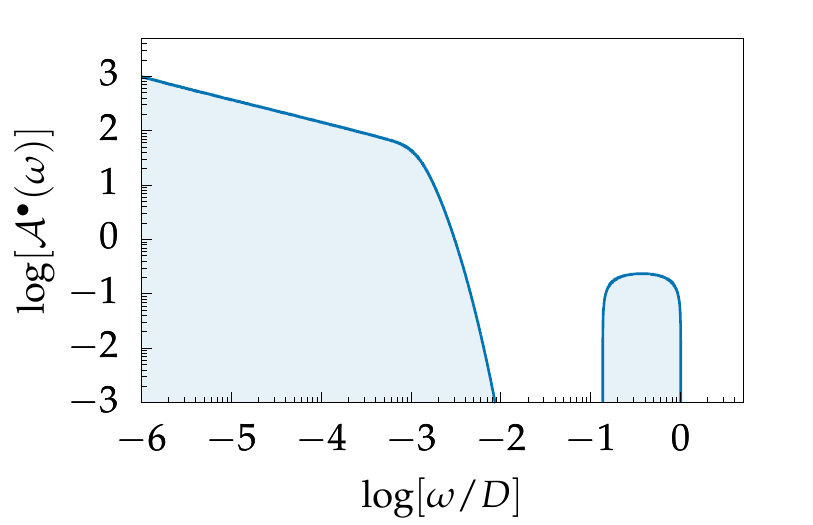}};
	\node at (3,1.875) {\subref*{fig:hsshU0_01alog}};
\end{tikzpicture}
\phantomsubcaption{\vspace{-\baselineskip}\label{fig:hsshU0_01alog}}
\end{subfigure}
\begin{subfigure}[c]{0.5\linewidth}
\begin{tikzpicture}
	\node at (0,0) {\includegraphics[width=\linewidth]{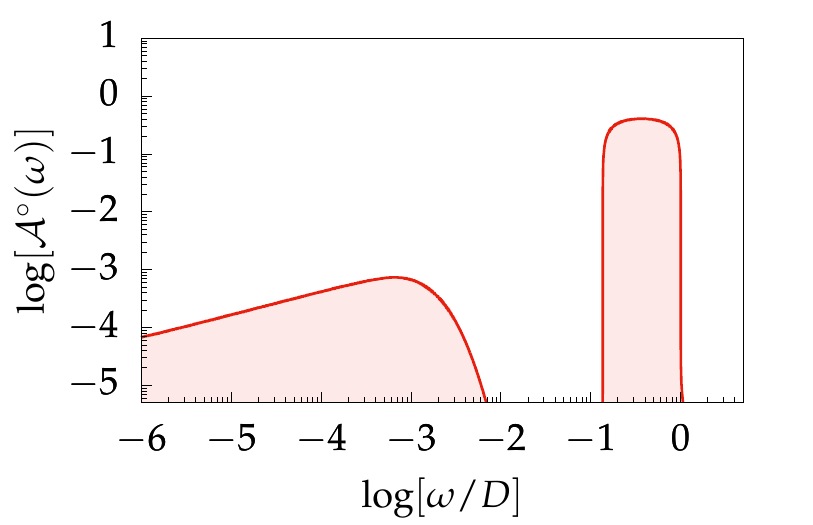}};
	\node at (3,1.875) {\subref*{fig:hsshU0_01blog}};
\end{tikzpicture}
\phantomsubcaption{\vspace{-\baselineskip}\label{fig:hsshU0_01blog}}
\end{subfigure}
\caption[Appearance of power-law features at low $U$ on the interacting Bethe-SSH model]{Appearance of power-law features at low $U$ on the interacting Bethe-SSH model. The feature on the right in all plots are the continuum SSH bands. The space between the bands and the low energy feature is a hard gap. The low energy spectral functions scale as $|\omega|^{\pm r}$ where the power is $-r$ in \subref{fig:hsshU0_001alog} and \subref{fig:hsshU0_01alog}, and $+r$ in \subref{fig:hsshU0_001blog} and \subref{fig:hsshU0_01blog}. Note that the power-law feature onsets at $\omega \sim \mathcal{O}(U/D)$.\label{fig:hsshUsmalllog}}
\end{figure}

As interactions are adiabatically increased into the non-perturbative regime, the low energy spectral features of both $\bullet$- and $\circ$-sites retain their power-law behavior with the same constant power $r$.
Both the low energy feature and the continuum bands exhibit exponential decay into the hard gap until the broadened low energy feature encounters the band. The power-law feature continues to onset at $\omega \sim \mathcal{O}(U)$, such that the broadened low energy feature encounters the band at $U \sim \lvert \tilde{t}_B - \tilde{t}_A \rvert/2$. Once this point is reached, the power-law feature occupies the entire low energy region, but does not extend beyond the former gap width. This can be seen by comparing the spectra at $U/D = 3.0$ in Fig.~\ref{fig:hsshU3log} with the spectra at $U \ll \lvert \tilde{t}_B - \tilde{t}_A \rvert$ in Fig.~\ref{fig:hsshUsmalllog}. The spectra in Fig.~\ref{fig:hsshUsmalllog} show that the inner band edge appears at $\omega/D \sim 10^{-1}$. The spectra at $U/D = 3.0$ in Fig.~\ref{fig:hsshU3} are heavily broadened by the interactions, but the power-law feature is still confined to $|\omega| \lesssim \lvert \tilde{t}_B - \tilde{t}_A \rvert$.
As in the Hubbard model, as interactions are further increased the central features of the spectrum become compressed towards the Fermi level and similarly here the power-law features are pushed away from the bands.
\begin{figure}[ht!]
\includegraphics[scale=1]{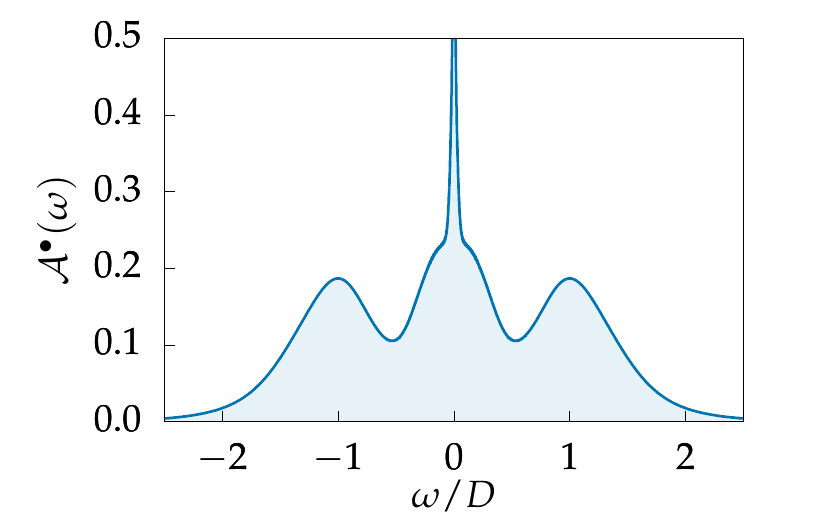}
\includegraphics[scale=1]{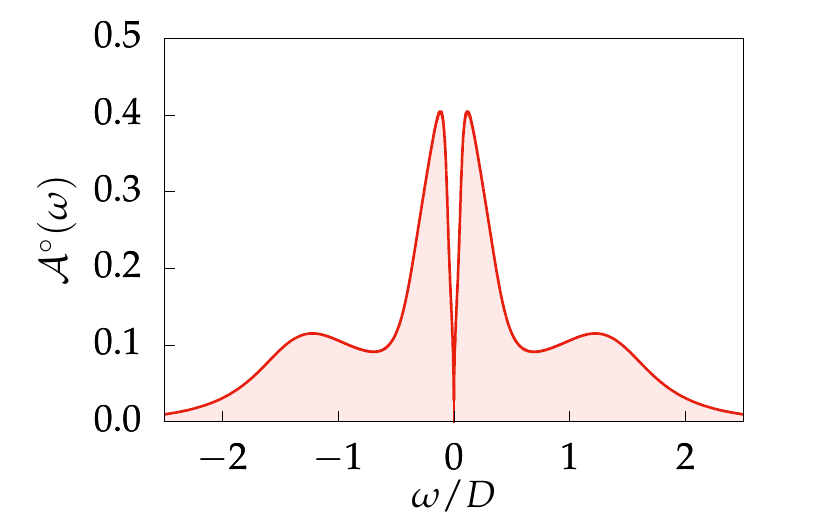}
\caption{Spectral functions for the $\bullet$- and $\circ$-sites at $U/D=3.0$. The original spectral features are now heavily broadened.\label{fig:hsshU3}}
\end{figure}
\begin{figure}[ht!]
\includegraphics[scale=1]{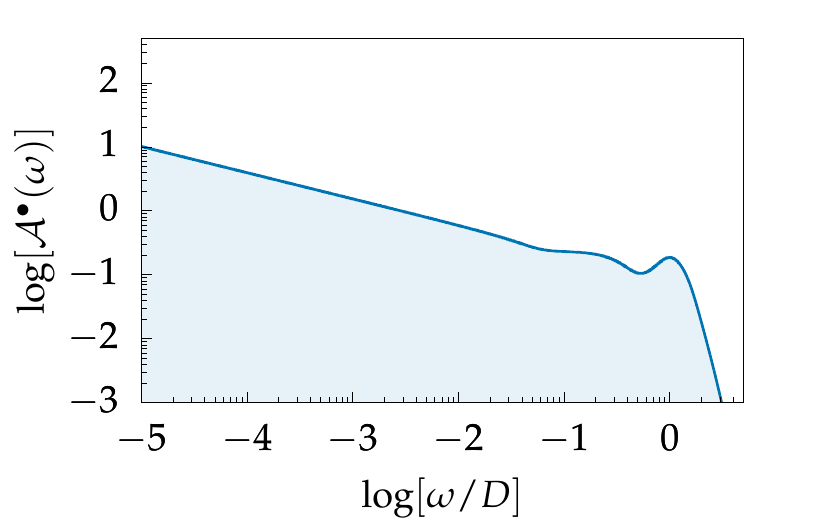}
\includegraphics[scale=1]{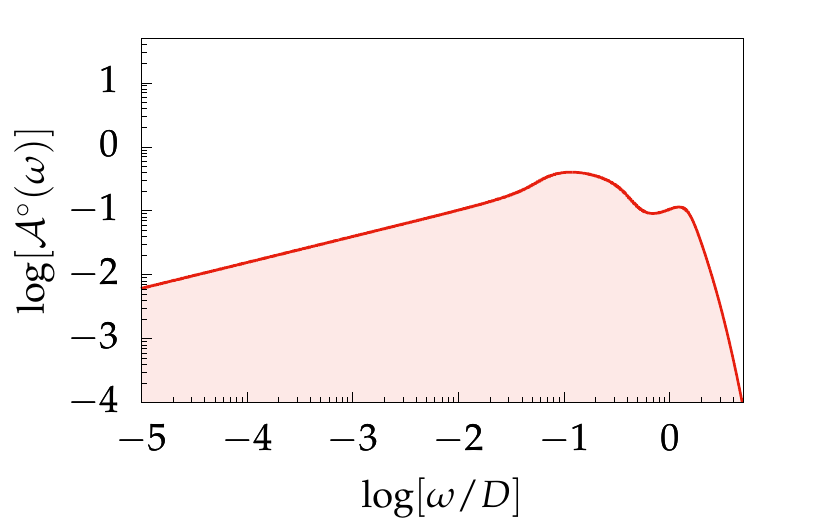}
\caption[Spectral functions for the $\bullet$- and $\circ$-sites at $U/D=3.0$ on a logarithmic scale]{Spectral functions for the $\bullet$- and $\circ$-sites at $U/D=3.0$ on a logarithmic scale. Observed here is the low energy power-law feature which is retained from the perturbative regime. The region of the power-law is confined to the width of the band gap of the non-interacting system.\label{fig:hsshU3log}}
\end{figure}
\begin{figure}[ht!]
\includegraphics[scale=1]{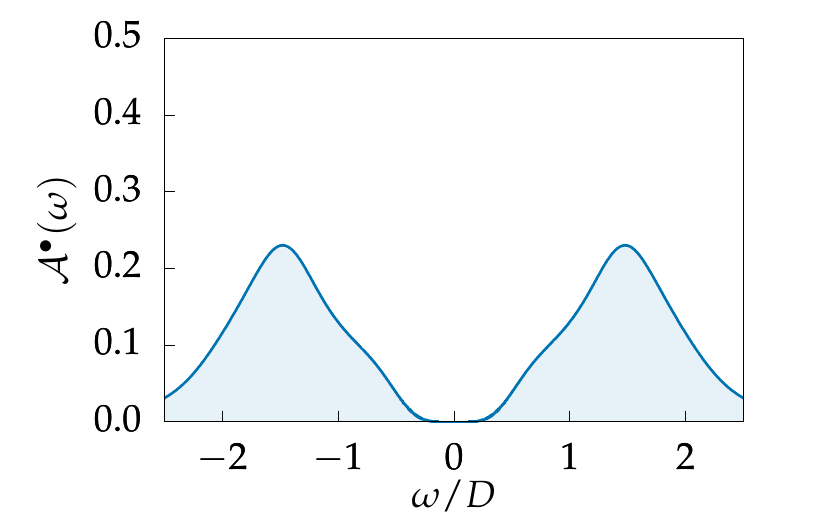}
\includegraphics[scale=1]{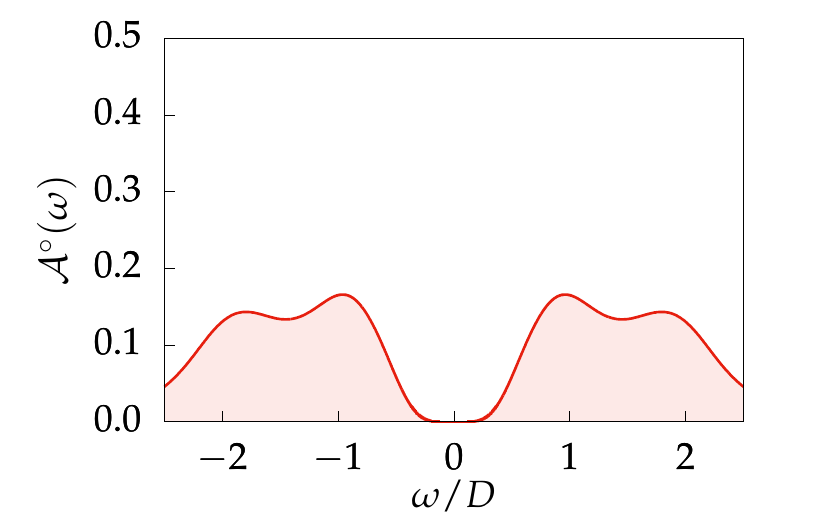}
\caption[Spectral functions for the $\bullet$- and $\circ$-sites at $U/D=5.4$. The spectral functions here are hard gapped and are in a Mott insulating phase.]{Spectral functions for the $\bullet$- and $\circ$-sites at $U/D=5.4$. The spectral functions here are hard gapped and are in a Mott insulating phase. The characteristic Mott pole in their self-energies can be seen in Fig.~\ref{fig:hsshmottse}.\label{fig:hsshMottphase}}
\end{figure}
The main result here is that for the Bethe-SSH model in the presence of local Coulomb interactions, sites in the topological configuration exhibit a Kondo resonance situated at zero-energy and the sites in the trivial configuration exhibit a pseudogap with power-law behavior. The solution is presented in Fig.~\ref{fig:hsshsolution}.
\begin{figure}[htp!]
\begin{subfigure}{0.49\linewidth}
	\includegraphics[scale=1]{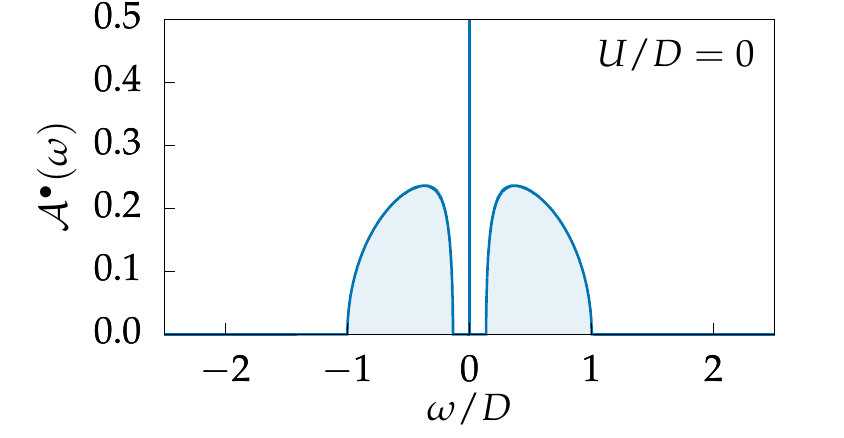}
\end{subfigure}
\hfill
\begin{subfigure}{0.49\linewidth}
	\includegraphics[scale=1]{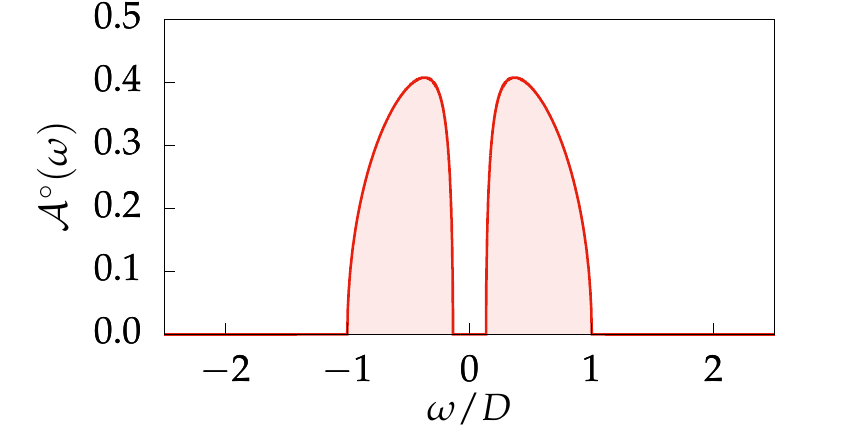}
\end{subfigure}
\begin{subfigure}{0.49\linewidth}
	\includegraphics[scale=1]{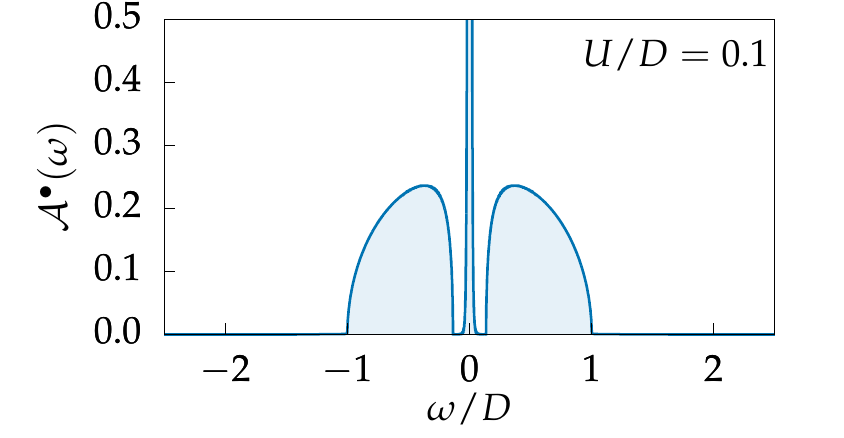}
\end{subfigure}
\hfill
\begin{subfigure}{0.49\linewidth}
	\includegraphics[scale=1]{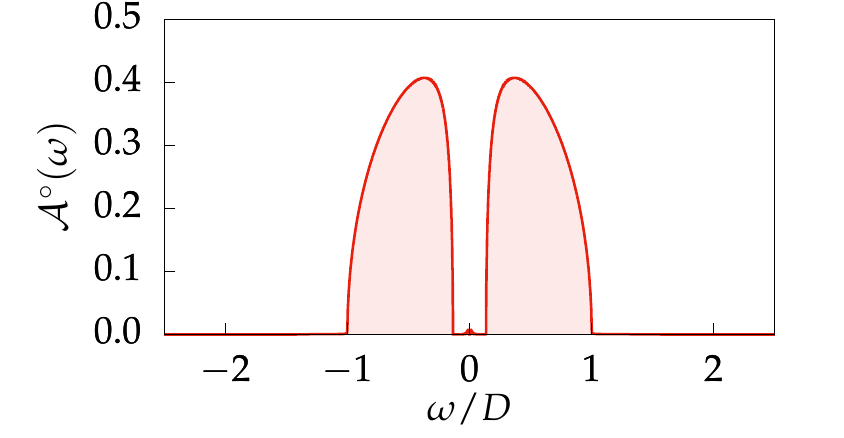}
\end{subfigure}
\begin{subfigure}{0.49\linewidth}
	\includegraphics[scale=1]{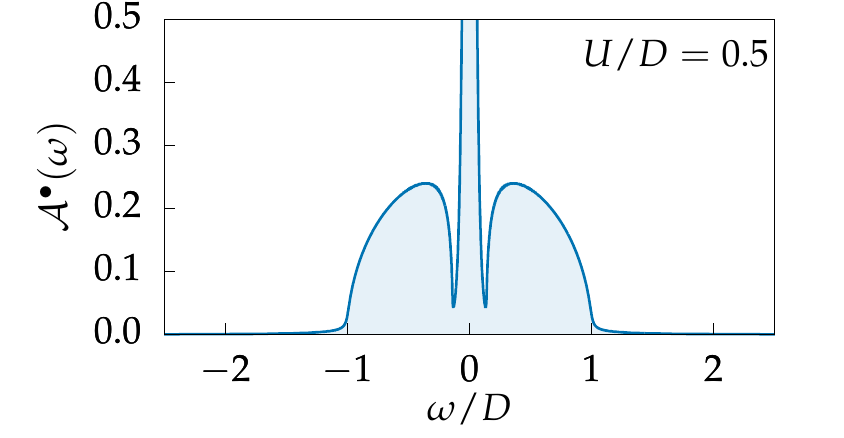}
\end{subfigure}
\hfill
\begin{subfigure}{0.49\linewidth}
	\includegraphics[scale=1]{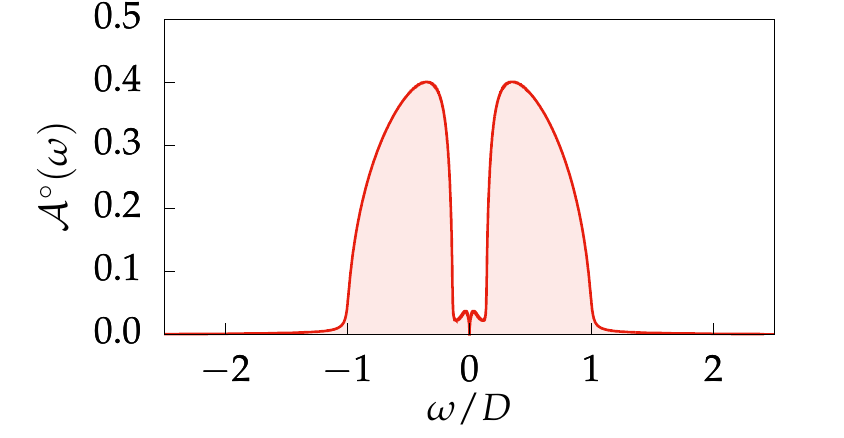}
\end{subfigure}
\begin{subfigure}{0.49\linewidth}
	\includegraphics[scale=1]{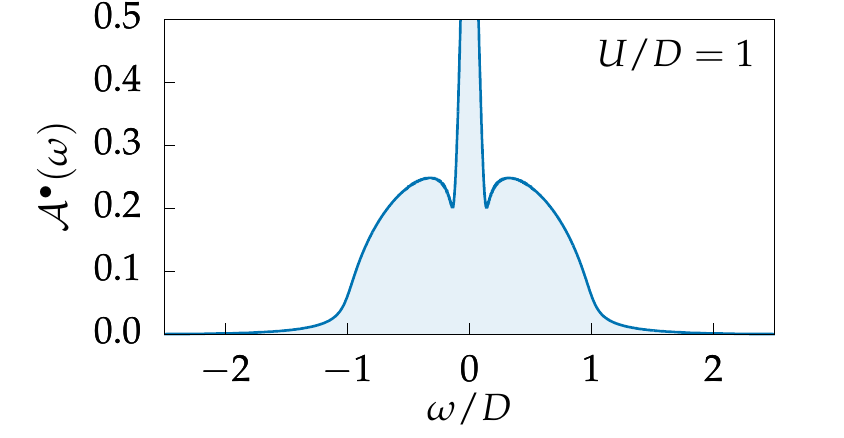}
\end{subfigure}
\hfill
\begin{subfigure}{0.49\linewidth}
	\includegraphics[scale=1]{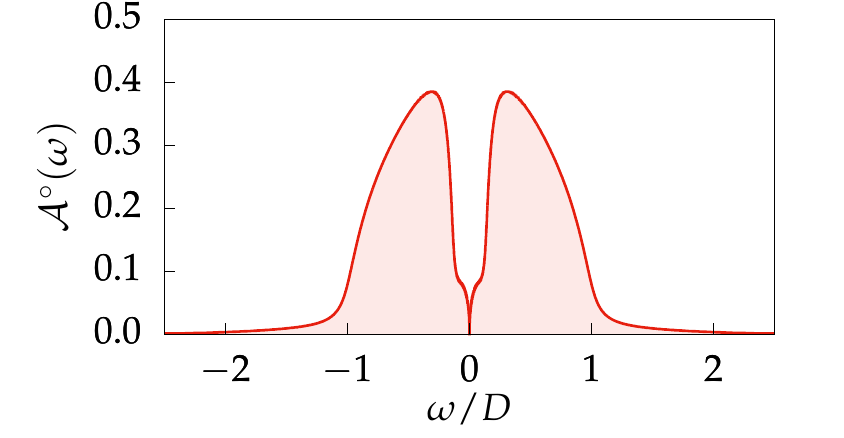}
\end{subfigure}
\begin{subfigure}{0.49\linewidth}
	\includegraphics[scale=1]{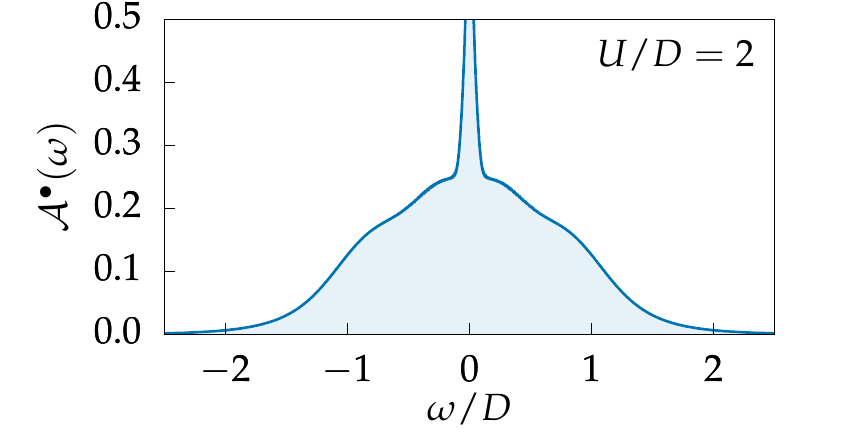}
\end{subfigure}
\hfill
\begin{subfigure}{0.49\linewidth}
	\includegraphics[scale=1]{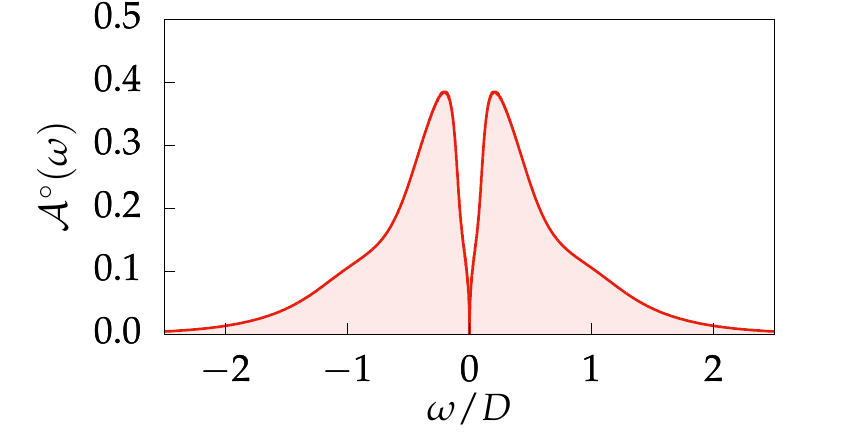}
\end{subfigure}
\caption[DMFT solution to the Bethe lattice Hubbard-SSH model]{DMFT solution to the Bethe lattice Hubbard SSH model for the $\bullet$ (left) $\circ$ (right) sites.\label{fig:hsshsolution}}
\end{figure}
\begin{figure}[htp!]\ContinuedFloat
\begin{subfigure}{0.49\linewidth}
	\includegraphics[scale=1]{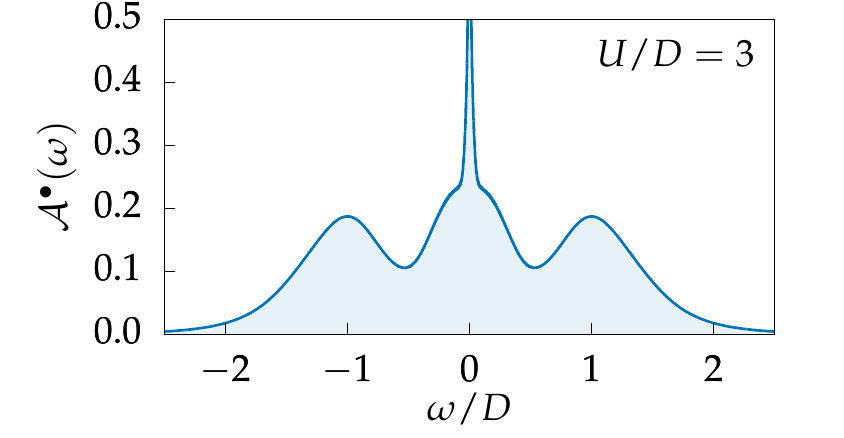}
\end{subfigure}
\hfill
\begin{subfigure}{0.49\linewidth}
	\includegraphics[scale=1]{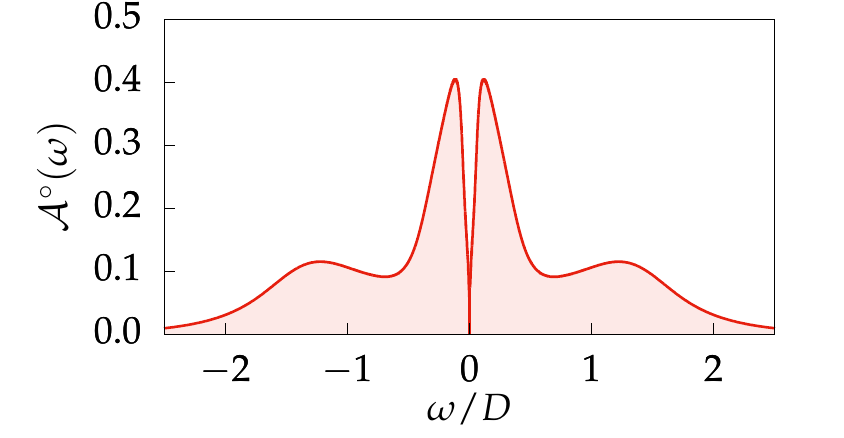}
\end{subfigure}
\begin{subfigure}{0.49\linewidth}
	\includegraphics[scale=1]{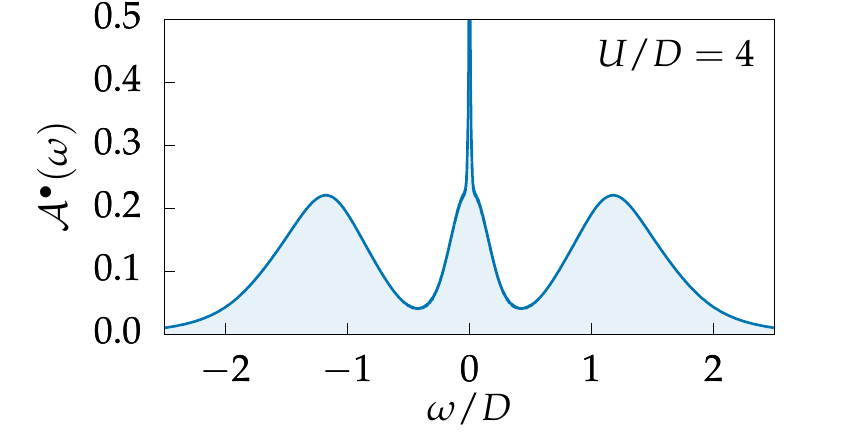}
\end{subfigure}
\hfill
\begin{subfigure}{0.49\linewidth}
	\includegraphics[scale=1]{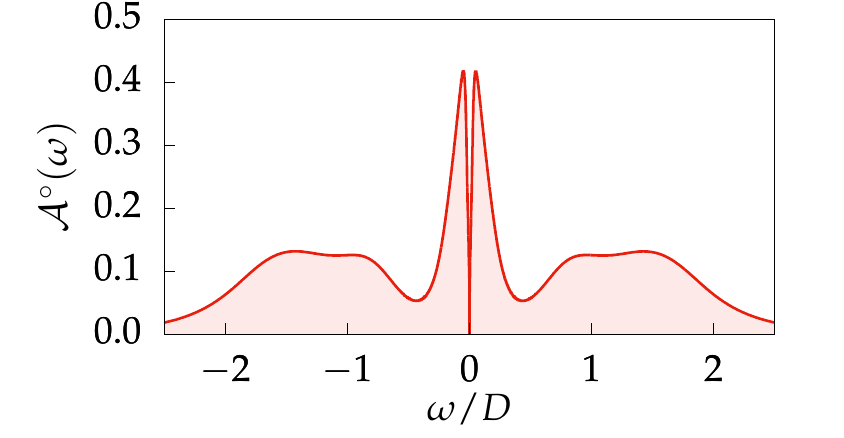}
\end{subfigure}
\begin{subfigure}{0.49\linewidth}
	\includegraphics[scale=1]{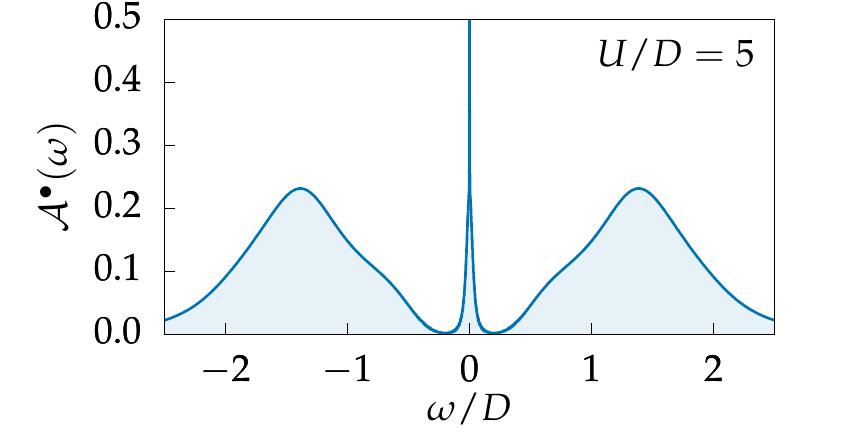}
\end{subfigure}
\hfill
\begin{subfigure}{0.49\linewidth}
	\includegraphics[scale=1]{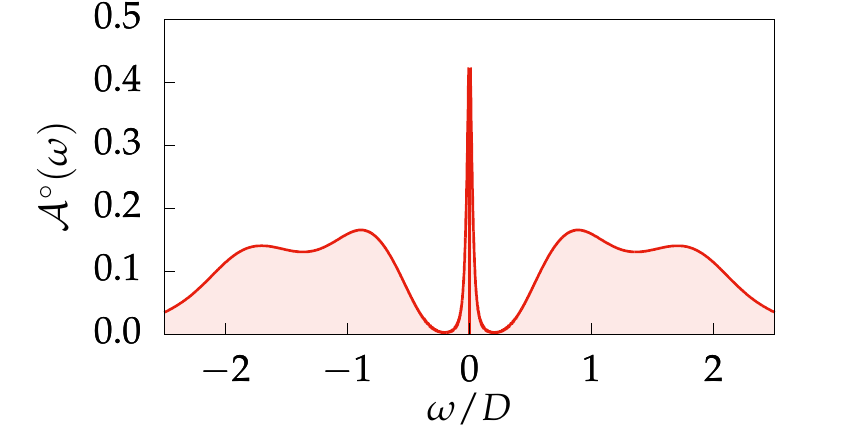}
\end{subfigure}
\begin{subfigure}{0.49\linewidth}
	\includegraphics[scale=1]{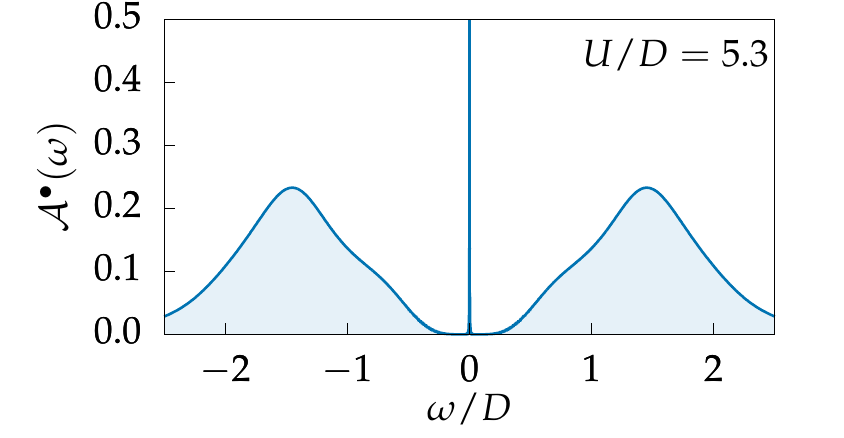}
\end{subfigure}
\hfill
\begin{subfigure}{0.49\linewidth}
	\includegraphics[scale=1]{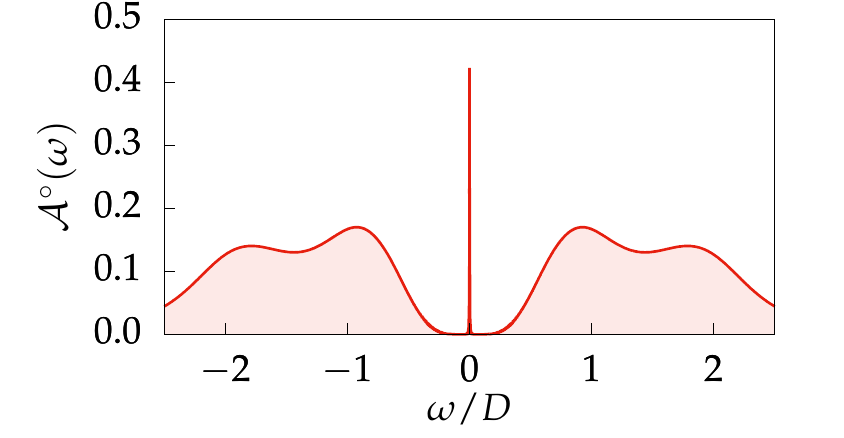}
\end{subfigure}
\begin{subfigure}{0.49\linewidth}
	\includegraphics[scale=1]{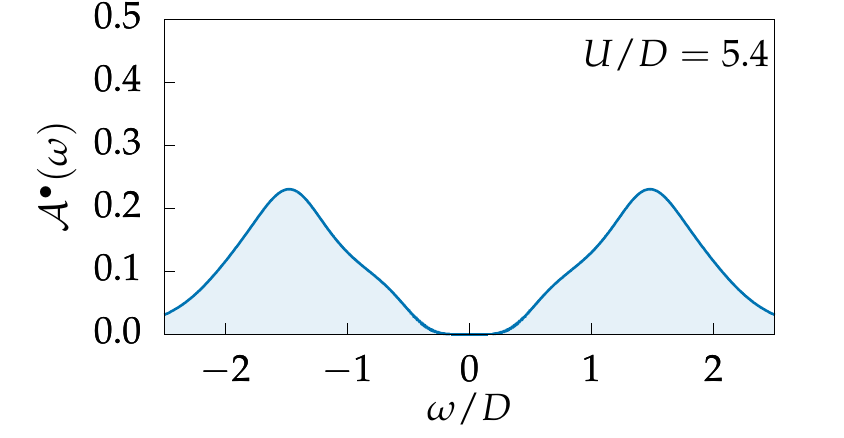}
\end{subfigure}
\hfill
\begin{subfigure}{0.49\linewidth}
	\includegraphics[scale=1]{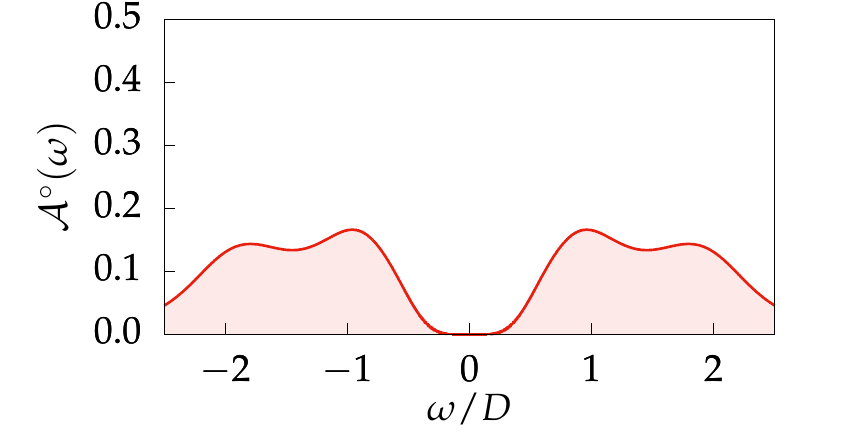}
\end{subfigure}
\caption[DMFT solution to the Bethe lattice Hubbard-SSH model]{DMFT solution to the Bethe lattice Hubbard SSH model for the $\bullet$ (left) $\circ$ (right) sites.}
\end{figure}
%

%For the parameters $(\mathsf{x},\mathsf{y},\mathsf{z}) = (\frac{3\sqrt{5}}{7},\frac{\sqrt{41}}{21},4.0)$, the exponent $r\approx0.33$.

Above a critical interaction strength $U_{c}$, both types of sites exhibit a Mott insulating phase with a hard gap.
Under the parameterization \eqref{eq:hsshparam} the Mott transition\index{Mott transition} occurs at $5.3 < U_{c2}/D < 5.4$. The spectrum on both sites in the Mott insulating phase is shown in Fig.~\ref{fig:hsshMottphase}. The phase with $U>U_{c}$ can unambiguously be called Mott insulating, but the phase with $U<U_{c}$ cannot uniformly be called \textit{e.g.} topological, as the topological states only exist on one sublattice. With respect to the Mott phase transition, it may be informally called the `metallic' phase by analogy with the Hubbard phase diagram, but the system does not take on a uniform phase in this region, with one sublattice having a spectrum adiabatically connected to the boundary spectrum of a topological insulator and the other sublattice being adiabatically connected to a conventional band insulator.

As in the basic Hubbard model~\eqref{eq:hubbard}, this system exhibits a hysteresis region in the phase diagram (\textit{cf.} Fig.~\ref{fig:hubbardcoexist} and the surrounding discussion).
Initializing the DMFT algorithm with a Green function of the Mott insulating phase, such as $U/D > 5.4$, and iterating the calculation with progressively decreasing $U$ it is found that that $U_{c1}$ is $4.2 < U_{c1}/D < 4.3$. This means that the hysteresis region lies in the parameter region $4.3 \lesssim U/D \lesssim 5.4$. The comparison of spectral functions within the coexistence regime is plotted in Fig.~\ref{fig:hsshcoexistence}.
Within the hysteresis region, the system with interaction strength tuned down adiabatically from $U > U_{c2}$ can be described as having interaction strength $U_+$. Conversely, the system with interaction strength tuned up adiabatically from $U < U_{c1}$ can be described as having interaction strength $U_-$.
%
\begin{comment}
\begin{subequations}
\begin{align}
	G^\bullet(z)
	&= \cfrac{1}{z-\varepsilon - \Sigma_b(z) - \cfrac{t_A^2}{z-\varepsilon - \Sigma_a(z) - t_B^2 G^\bullet(z)}}
	\\
	&= \cfrac{1}{z-\varepsilon - \Sigma_b(z) - K_b(z)}
\end{align}
%
\begin{align*}
	\Delta_b(z) = t_A^2 G^\circ(z)
\end{align*}
%
\begin{align}
	G^\circ(z)
	&= \cfrac{1}{z-\varepsilon - \Sigma_a(z) - \cfrac{t_A^2}{z-\varepsilon - \Sigma_b(z) - t_A^2 G^\circ(z)}}
	\\
	&= \cfrac{1}{z-\varepsilon - \Sigma_a(z) - K_a(z)}
\end{align}
%
\begin{align*}
	\Delta_a(z) = t_B^2 G^\bullet(z)
\end{align*}
\end{subequations}
\end{comment}
%
%\begin{figure}[ht!]
%\includegraphics[scale=1]{hsshU0a.pdf}
%\includegraphics[scale=1]{hsshU0b.pdf}
%\caption{Spectral functions for the $\bullet$- and $\circ$-sites at $U/D=0.0$. The spectral functions take the form of the SSH model in the topological and trivial phases respectively.}
%\end{figure}
%
\begin{figure}[ht!]
	\subfiglabel{\includegraphics[scale=1]{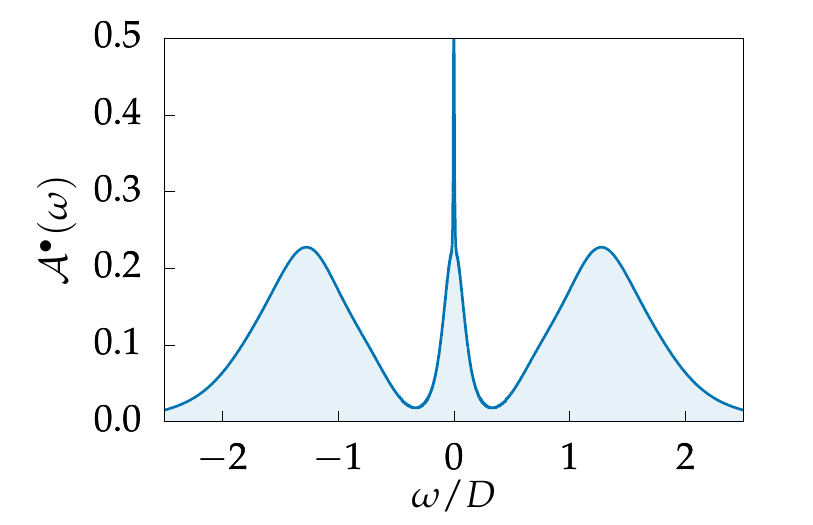}}{3,1.875}{fig:hsshU4_5a}
	\subfiglabel{\includegraphics[scale=1]{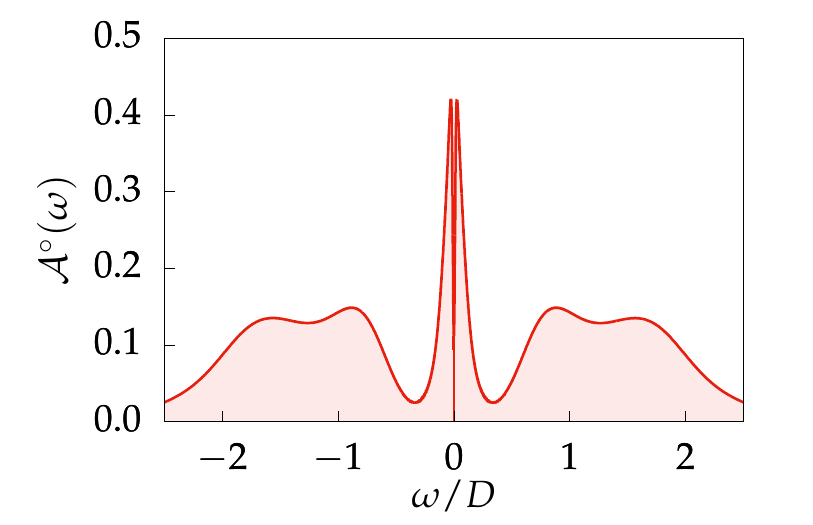}}{3,1.875}{fig:hsshU4_5b}
	\subfiglabel{\includegraphics[scale=1]{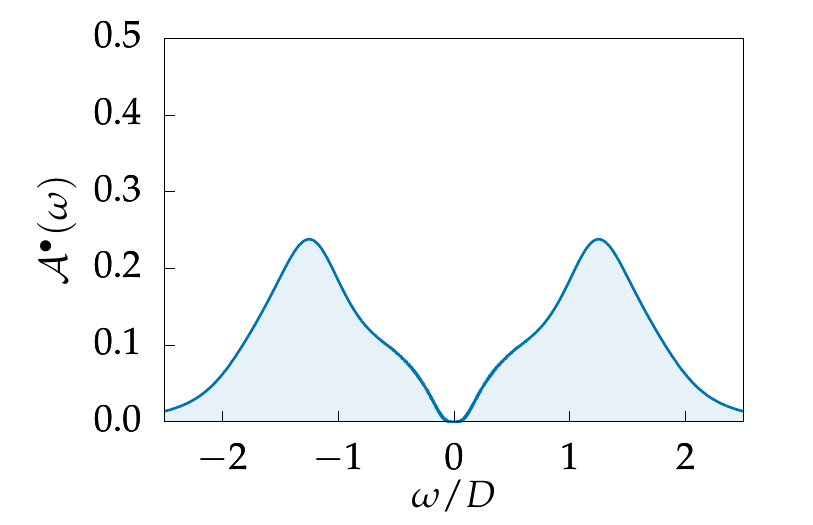}}{3,1.875}{fig:hsshUc1U4_5a}
	\subfiglabel{\includegraphics[scale=1]{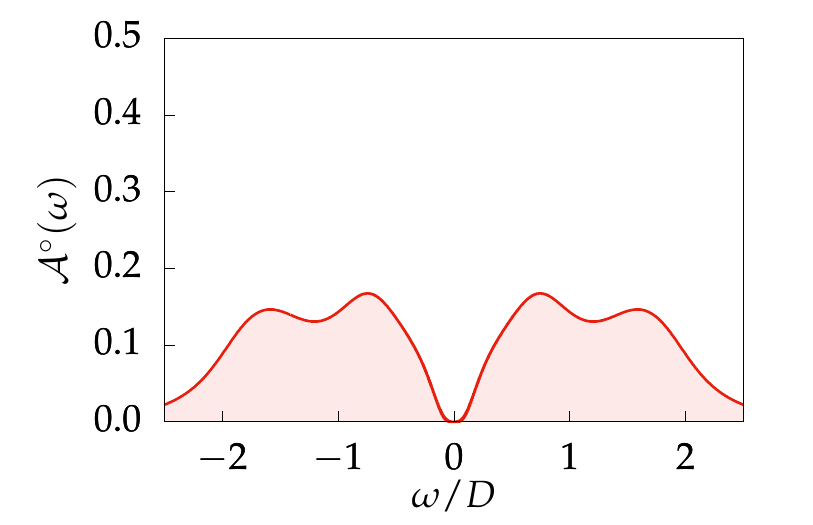}}{3,1.875}{fig:hsshUc1U4_5b}
\caption[Spectral functions on the $\bullet$- and $\circ$-sites at $U/D=4.5$ demonstrating the coexistence hysteresis region.]{Spectral functions on the $\bullet$- and $\circ$-sites at $U/D=4.5$. Panels \subref{fig:hsshU4_5a},\subref{fig:hsshU4_5b} are evaluated at $U_{-}/D = 4.5$, and panels \subref{fig:hsshUc1U4_5a},\subref{fig:hsshUc1U4_5b} are evaluated at $U_{+}/D = 4.5$, where $U_{-}$($U_{+}$) refers to approaching the interaction strength adiabatically from metallic (insulating) phase respectively. This demonstrates the coexistence hysteresis region $U_{c1} < U < U_{c2}$.\label{fig:hsshcoexistence}}
\end{figure}

A remark worth making is that the situation described by the Hubbard-SSH model presented here is different than the class of interacting topological insulators known as `topological Mott insulators'\index{topological Mott insulator} in the literature~\cite{interacting}. Topological Mott insulators are analogous to ordinary non-interacting topological insulators in the sense that they host topological states on their boundaries with an insulating bulk. In topological Mott insulators the bulk is a Mott insulator.
The Hubbard-SSH model here exhibits Mott insulating behavior on both of its sublattices in the $U>U_c$ regime, and in the $U < U_c$ regime where sites of one sublattice exhibit topological features, the sites of the other sublattice are not Mott insulating. As such, it does not present itself as a model of a topological Mott insulator.

\subsection{Self-Energy Analysis}\label{sec:hsshse}
\begin{figure}[ht!]
\subfiglabel{\includegraphics{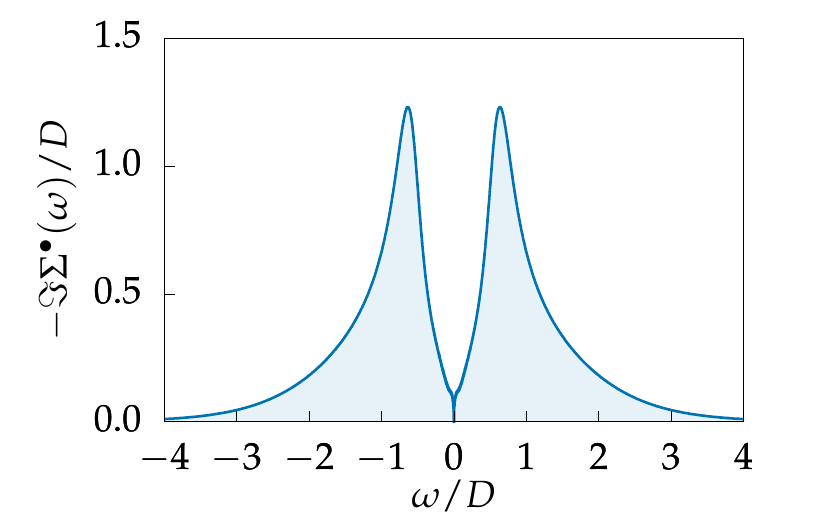}}{3.125,2}{fig:hsshSU3a}
\subfiglabel{\includegraphics{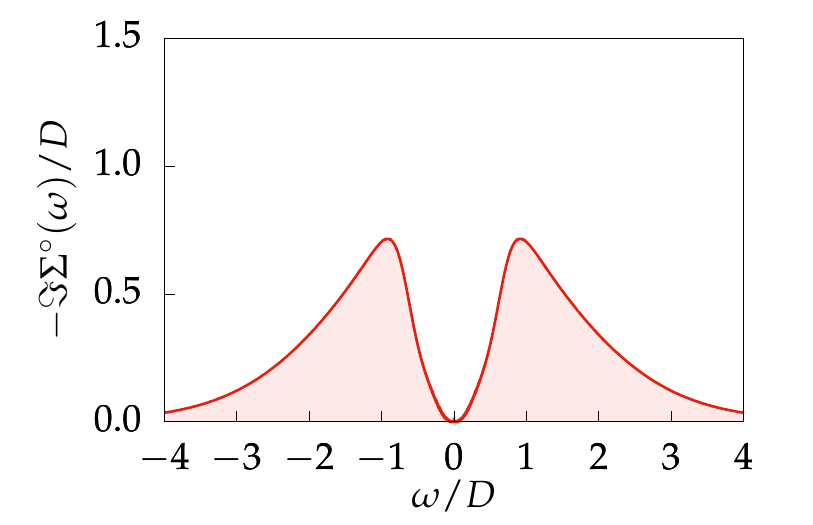}}{3.125,2}{fig:hsshSU3b}
\caption{$-\Im\Sigma(\omega)$ of the HSSH model at $U/D = 3.0$. At low energy the self-energy scales as $\omega^2$, which is indicative of a Fermi liquid.\label{fig:hsshse}}
\end{figure}
As in the full Green functions, the distinguishing characteristics of the self-energies occurs at low energy. In the $U<U_c$ phase, the self-energy on the $\circ$-sites scales as $\omega^2$, shown in Fig.~\ref{fig:hsshSU3b}, which is indicative of a Fermi liquid. This shown explicitly in Fig.~\ref{fig:hsshtrivse}. On the $\bullet$-sites, the self-energy possesses power-law behavior $|\omega|^{r}$, shown in Figs.~\ref{fig:hsshSU3a} and \ref{fig:hsshtopse}. The power $r$ is the same as the power which appears in the Green functions. From the Dyson equation
\begin{equation}
	G^\bullet(z) = \frac{1}{z - \varepsilon - \tilde{t}^2_A G^\circ(z) - \Sigma^\bullet(z) \strut}
\end{equation}
it can be observed that the power-law features of $\Im\Sigma^\bullet(\omega)$ and $G^\circ(\omega)$ contribute constructively to the opposite sign power-law which appears in $G^\bullet(\omega)$. The analogous Dyson equation for $G^\circ(\omega)$ shows that at low energy only the power-law from $G^\bullet(\omega)$ affects $G^\circ(\omega)$ since $\omega^2 \ll |\omega|^{-r}$.
\begin{figure}[ht!]
\includegraphics{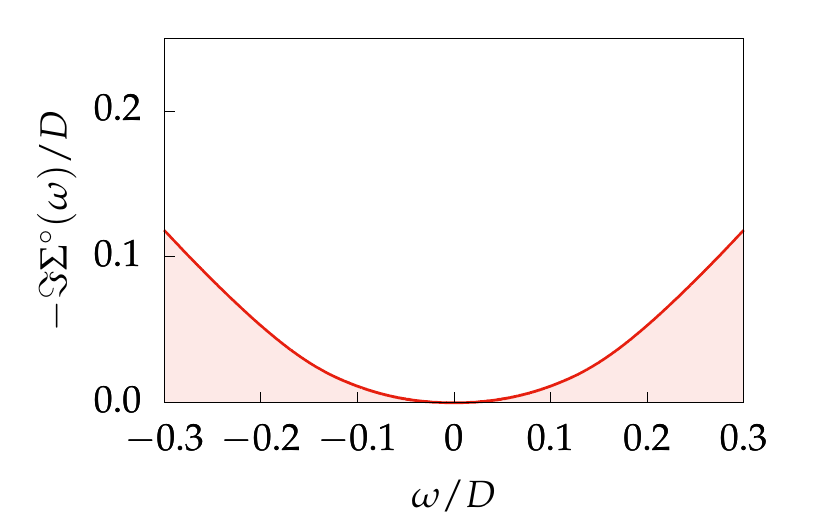}
\includegraphics{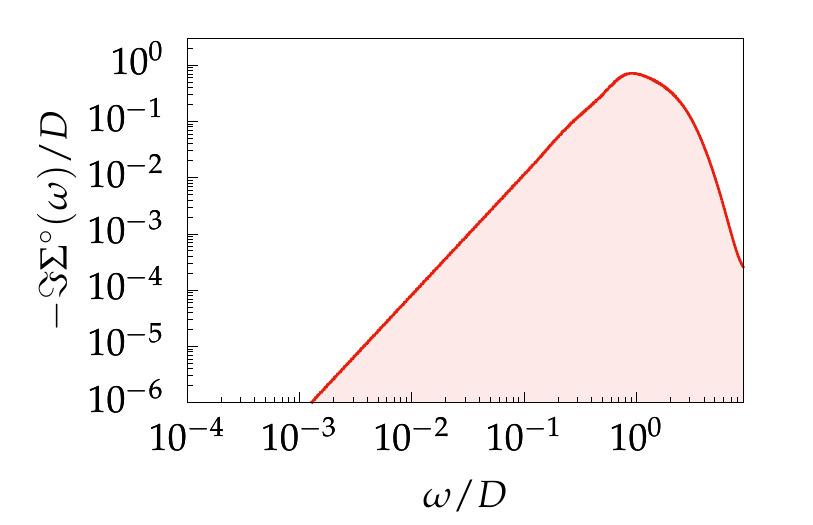}
\vspace{-\baselineskip}
\caption{The low energy feature of $-\Im\Sigma^{\circ}(\omega)$ at $U/D = 3.0 < U_c$. The low energy feature goes as $-\Im\Sigma^{\circ}(\omega) \sim \omega^2$, which is indicative of a Fermi liquid state.\label{fig:hsshtrivse}}
\end{figure}
\begin{figure}[ht!]
\includegraphics{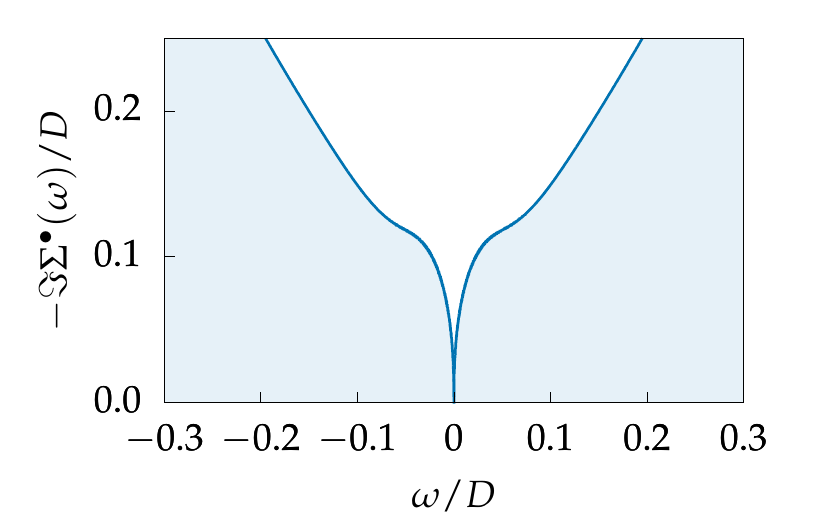}
\includegraphics{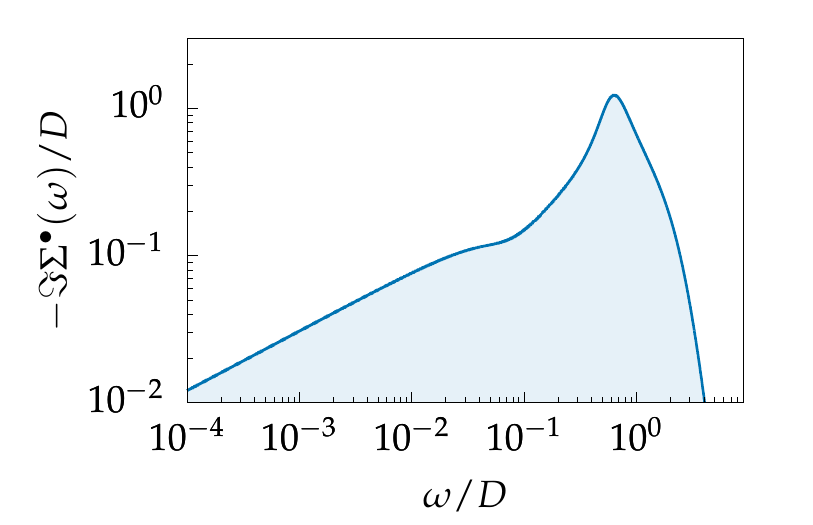}
\vspace{-\baselineskip}
\caption{The low energy feature of $-\Im\Sigma^{\bullet}(\omega)$ at $U/D = 3.0 < U_c$. The power-law feature onsets at $|\omega| \sim 10^{-1}$. This power-law behavior is indicative that the $\bullet$-site is a non-Fermi liquid.\label{fig:hsshtopse}}
\end{figure}
This low energy behavior persists throughout the $U<U_c$ phase, even though higher energy features change with increasing $U$, with the self-energies of both sites following similar behavior as the Hubbard model self-energy as $U\to U_c^-$.
The power $r$ does not depend on the strength of interaction $U$, but the scale at which $\mathcal{A}^{\bullet/\circ}(\omega) \sim |\omega|^{\pm r}$ does.
\begin{figure}[ht!]
\subfiglabel{\includegraphics{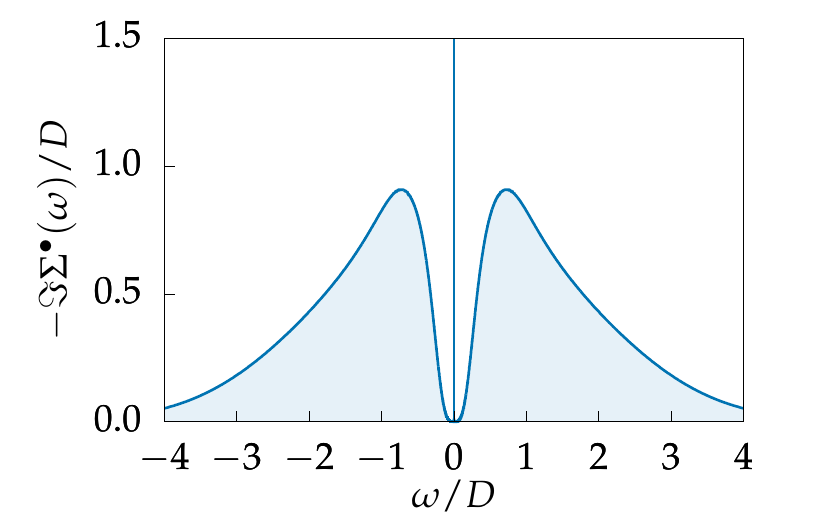}}{3.125,2}{fig:hsshSU5_4a}
\subfiglabel{\includegraphics{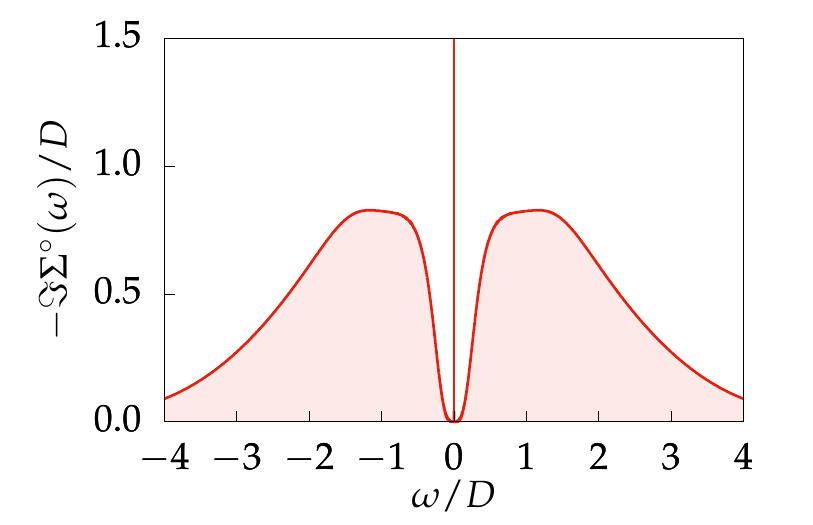}}{3.125,2}{fig:hsshSU5_4b}
\subfiglabel{\includegraphics{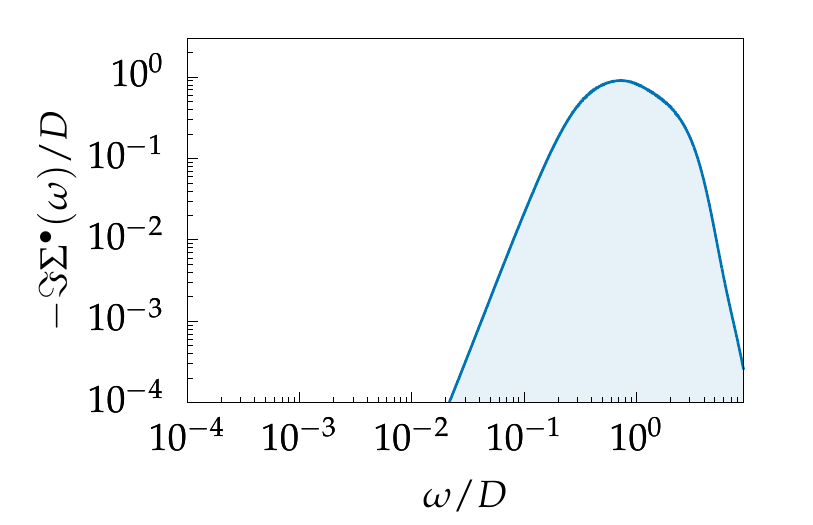}}{3.125,2}{fig:hsshSU5_4loga}
\subfiglabel{\includegraphics{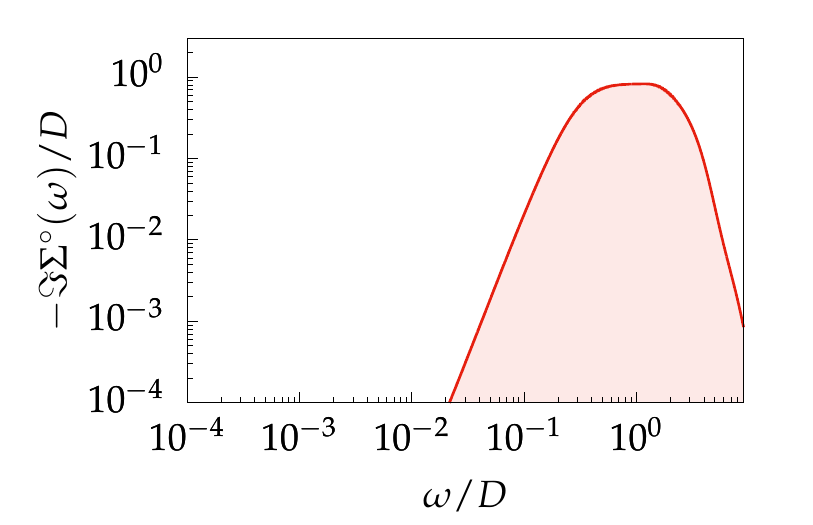}}{3.125,2}{fig:hsshSU5_4logb}
\caption{$-\Im\Sigma(\omega)$ on the $\bullet$- and $\circ$-sites at $U/D = 5.4 > U_c$. The continuum bands decay exponentially into the gap \subref{fig:hsshSU5_4loga},\subref{fig:hsshSU5_4logb}. The Mott pole is not visible on the log-scale plots.\label{fig:hsshmottse}}
\end{figure}

Above the critical interaction strength, the self-energy on both sites features a Mott pole located in a gap. The gap can be identified as a hard gap as the bands of the imaginary part of the self-energy decay exponentially into the gap. This is shown in Fig.~\ref{fig:hsshmottse}.

Qualitative information about the two sites in the two phases can be gathered from examining the Luttinger integral\index{Luttinger integral}. The Luttinger integral is given by
\begin{equation}
	I^{\bullet/\circ}_{L} = \frac{2}{\pi} \Im \int_{-\infty}^{0} \text{d}\omega\, G^{\bullet/\circ}(\omega) \frac{\partial \Sigma^{\bullet/\circ}(\omega)}{\partial \omega}
\end{equation}
for each of the $\bullet$- and $\circ$-sites.
The form of the Luttinger integral has similar functional form to the Volovik-Essen-Gurarie invariant~\eqref{eq:volovikessengurarie} as $\Sigma(\omega) \sim G^{-1}(\omega)$.

For $U> U_c$, the Luttinger integral takes the value $I_{L} = 1$ on both sites. This is indicative of both sites existing in a Mott insulating phase which is corroborated with the observation that the self energy on both sites features a Mott pole.

On the $\circ$-sites with $U<U_c$, the Luttinger integral evaluates to $I^\circ_{L} = 0$ which is implies that the sites which are adiabatically connected to the topologically trivial configuration take the form of a Fermi liquid in the presence of interactions. The behavior of the self energy of these sites in this regime is $\Im\Sigma(\omega) \overset{|\omega| \ll 1}{\sim} \omega^2$ which is consistent with the Fermi liquid picture.

For the parameters used in the above analysis, the $\bullet$-sites with $U<U_c$ feature a Luttinger integral which evaluates to $I^\bullet_{L} \approx 0.17$, which is a non-trivial value. This indicates that the $\bullet$-sites in the $U<U_c$ phase cannot be considered to be Fermi liquids, but also are not simple local moments as in the Mott insulating case.

The Luttinger integral depends on the effective non-interacting bandwidth $\lvert \tilde{t}_B - \tilde{t}_A \rvert$, which controls the power-law $r$, and its value is therefore not a unique identifier of the system's phase. It does however indicate non-Fermi liquid correlations in the system.

%%%%%%%%%%%%%%%%
\section{Classification of the Topological Phases}

A standard method of classifying the topological phases of a system is calculating its Chern number. For the SSH model, this is accomplished by calculating the Zak phase\index{Zak phase}
\begin{equation}
	\gamma_{\textsc{z}} = \smashoperator{\oint_{\textsc{bz}}} {A}(k) = \begin{cases} 0 & t_A \geq t_B \\ \pi & t_A < t_B \end{cases}
\end{equation}
where ${A}(k) = -\text{i}\langle \psi(k) | \partial_k \psi(k) \rangle\text{d}k$ is the $U(1)$ connection 1-form on the Bloch bundle of the first Brillouin zone and the $|\psi(k)\rangle$ is the eigenvector of the lower band. Unlike the higher dimensional Berry phase which is the integral of the curvature 2-form, the Zak phase is not gauge invariant.

An alternative method, which is particularly suited to interacting systems, is given by the Volovik-Essin-Gurarie invariant~\cite{volovik,gurarie,essingurarie} which is defined in terms of the momentum space Green's functions as
\begin{equation}
	\mathcal{N} = \int \bigwedge_{k}^{d} G(\omega,k) \, \d_{k} G^{-1}(\omega,k)
\label{eq:volovikessengurarie}
\end{equation}
where $\bigwedge$ denotes the exterior products of the differential 1-form of the momentum space Green functions $G \d G^{-1}$ over the $d$-dimensions of the system.

A further additional method is by using the topological Hamiltonian~\cite{simpinv,topham}\index{topological Hamiltonian}
\begin{equation}
	H_{\text{T}}(k) = H_0(k) + \Sigma(\omega=0,k) ,
\label{eq:topham}
\end{equation}
which is comprised of the non-interacting part of the Hamiltonian $H_0(k)$, and the self-energy of the interacting system evaluated at zero frequency $\Sigma(0,k)$. Topological invariants are then obtained in the usual manner as if $H_{\text{T}}(k)$ were the Hamiltonian of a free fermion system, for example by the Chern number or the Volovik-Essen-Gurarie functional \eqref{eq:volovikessengurarie}.

While the Bethe lattice does not have a defined dual, a non-interacting tight-binding model on the Bethe lattice (\textit{e.g.} of the form Eq.~\eqref{eq:basickineticham}) can still be diagonalized in terms of a polarization $\theta$ which plays the role of an effective momentum~\cite{diagonalbethe}. 

Since all sites on the Bethe lattice for this model are identical, an arbitrary site is chosen as a `center' and a state on this site is labelled $\lvert0\rangle$. Additional states are labelled $\lvert1\rangle$, $\lvert2\rangle$, $\lvert3\rangle$, $\ldots$, $\lvert n \rangle$, $\ldots$ with each state containing the total number of sites of degree $n$ away from $\lvert0\rangle$.
The action of the Hamiltonian on a given state $n>1$ is
\begin{equation}
	\hat{H} \lvert n \rangle = \sqrt{\kappa-1} \left( \lvert n-1 \rangle + \lvert n+1 \rangle \right) \,.
\end{equation}
The nature of state $\lvert0\rangle$ as a reference state means that transitioning between states $\lvert0\rangle$ and $\lvert1\rangle$ involves $\kappa$ bonds as opposed to $\kappa-1$ bonds for transitioning between states $n>1$.
The dispersion relation is
\begin{equation}
	\varepsilon(\theta) = 2 \sqrt{\kappa-1}\, t \cos\theta
\end{equation}
for finite $\kappa$ and
\begin{equation}
	\varepsilon(\theta) = 2 \tilde{t} \cos\theta
\end{equation}
in the limit of $\kappa\to\infty$.
In the infinite dimensional limit, this scheme recovers the same dispersion relation as for a $1d$ homogeneous chain.

An analogous procedure may be implemented here to the case of the Bethe lattice SSH model. The difference in this case is the inhomogeneity in the hoppings from states $\lvert n \rangle$ to $\lvert n+1 \rangle$. This inhomogeneity can be cured by taking at each successive step the average over all hoppings between states $\lvert n \rangle$ and $\lvert n+1 \rangle$.
%The dispersion of the infinite dimensional Bethe-SSH model can be found analogously as
%\begin{equation}
%	E_\pm(\theta)	=	\pm \sqrt{t_A^2 + t_B^2 + 2 t_A t_B \cos\theta}
%\end{equation}
Taking a $\bullet$-site for $\lvert0\rangle$, the average of all hoppings to state $\lvert1\rangle$ is less than $t_0 \equiv \frac12(t_A + t_B)$ as by definition there are more $t_A$ bonds than $t_B$ bonds connecting to a $\bullet$-site. Similarly, the hoppings involved in hopping from state $\lvert1\rangle$ to state $\lvert2\rangle$ is always greater than $t_0$. For the hoppings involving states $\lvert n \rangle$, $n>0$, the hoppings involve $\tensor*{\kappa}{_A}-1$ or $\tensor*{\kappa}{_B}-1$ as only sites further away from $\lvert0\rangle$ are taken into account. This means that there exists the possibility that some sites involved in the hopping $\lvert n \rangle$ to $\lvert n+1 \rangle$ for which $\tensor*{\kappa}{_B}-1 = \tensor*{\kappa}{_A}$ or $\tensor*{\kappa}{_A}-1 = \tensor*{\kappa}{_B}$. These links however only constitute a subset of all links between the sites of $\lvert n \rangle$ and $\lvert n+1 \rangle$, the remainder of which will have a majority of either $t_B$ or $t_A$, thereby skewing the average in the appropriate way.

The hopping amplitude for the transition $\lvert0\rangle \to \lvert1\rangle$ is unique as all bonds from the 0-site are involved in the transition. For all other transitions $\lvert n \rangle \to \lvert n+1 \rangle$, some bonds do not participate as only sites further away from $\lvert0\rangle$ are taken into account. However, the multiplicity of each bond involved in the transition is the same.
For the finite dimensional example given in Fig.~\ref{fig:sshbethe}, the alternating hopping amplitudes of the diagonalized model are $\tilde{t}_A \approx 0.87$ and $\tilde{t}_B \approx 0.94$.

Since the self-energy of the DMFT solution is purely local it possesses no momentum, or effective momentum, dependence. 
However, defining a topological invariant from the topological Hamiltonian in terms of the Green functions requires the calculation of the momentum space Green function, $G(\theta)$ ($\theta$ being the effective momentum described above). The Green function obtained from the DMFT solution is not this quantity, so additional work must be done in order to employ the topological Hamiltonian formalism to the interacting SSH system investigated in the above.
%In the $U<U_c$ regime, the self-energy vanishes at $\omega=0$, so it does not contribute to the topological Hamiltonian. The topological Hamiltonian is therefore equal to the non-interacting Hamiltonian in the diagonal basis. 
%\begin{equation}
%	H_{\text{T}}(\theta) = H_{0}(\theta) + \cancelto{0}{\Sigma(0,\theta)} \quad = H_{0}(\theta)
%\tag{\ref*{eq:topham}$^\prime$}
%\end{equation}
%Ordinarily the topological Hamiltonian can be expressed in terms of the inverse momentum-space Green function as ${H}_T(k) = {G}^{-1}(0,k)$. However, this is not the Green function which is produced in the above DMFT solution. This Green function however can be written in the form
%\begin{align*}
%	G(\i\omega,k) &= \frac{1}{\i\omega - \boldsymbol{h}(k)\cdot\boldsymbol{\sigma} - \Sigma(\i\omega,k)}
%	\\
%	G_{\text{loc}}(\omega)
%\end{align*}
%\cite{}

%The topological invariant for the two sites when $U<U_c$ is therefore the same as their adiabatically connected counterparts, with $\bullet$-sites classified as topological and $\circ$-sites classified as trivial.
A conjecture is that states which are adiabatically connected to non-interacting topological phases retain their topological classification. While reasonable to make, based on the general notion that topological phases are robust to adiabatic deformations, this conjecture has not been proven to extend to the interacting regime~\cite{interacting}.
It is however known that topological invariants break down under the phase transition to $U>U_c$~\cite{breakdownii} since this phase is not adiabatically connected to states for which a well-defined topological invariant can be calculated. The topological invariant given by the topological Hamiltonian can therefore not be used in the Mott insulating regime. This is particularly relevant for topological Mott insulators.

%%%%%%%%%%%%%%%%%%%%%%%%%%%%%%%%%%%%%%%%%%%%%%%%%%%%%%%%%%%%%%%%%%%%%%%%%%%%%%%%

%These approaches, however, are not available in the present case as they all rely on calculations performed in momentum space. The Bethe lattice does not admit a Fourier dual, which means that a momentum space cannot be defined. An alternative method for classifying the topological phases of this system is thus needed.

%One possibility of classifying the phases is by analyzing the characteristics of the 

%%%%%%%%%%%%%%%%%%
\section{The Particle-Hole Asymmetric Case}

%The Case of $\boldsymbol{\varepsilon \neq -\frac{U}{2}}$

The preceding analysis was performed in the regime of particle-hole symmetry at half-filling, where $\varepsilon = -U/2$. 
The DMFT analysis can easily be extended to the situation away from particle-hole symmetry, as in the case of a doped system.
The asymmetry can be parameterized as $\varepsilon \to \varepsilon = \varepsilon_0 - \mu$ where $\varepsilon_0 = -U/2$ and $\mu$ is the doping parameter.%$\eta=1-2\mu/U$

While relatively straightforward to implement in the DMFT-NRG framework, this calculation is computationally intensive. For high resolution the calculation was performed at a temperature of $T/D = 10^{-12}$ requiring $\sim90$ iterations of NRG at $\Lambda = 2.0$ keeping 6000 states.

The main result of this calculation is that there exists in this model a doping induced quantum phase transition. A sequence of spectral functions on both sublattices depicting this phase transition as $\mu$ is increased is shown in Fig.~\ref{fig:phasymmsequence}.
Near $\varepsilon = -U$, the spectrum develops a hard gap near $\varepsilon = 0$ on both sublattices. This gap is depicted in detail in Fig.~\ref{fig:dopingpt}.
Past this point, on the $\bullet$-sublattice the Kondo resonance reenters the upper band and the system again becomes metallic. 
For $U/D = 3.6$ this reentrant phase transition occurs in the vicinity of $\mu/D \approx 1.5$.
On the $\circ$-sites, at $\mu_c$ the doping destroys the pseudogap state and produces a hard gap in the spectrum. As the doping is increased further, the gap fills in and the spectrum obtains finite value between the satellite bands. This is seen in the $\mu/D = 1.5$ panel of Fig.~\ref{fig:phasymmsequence}.

The phase transition to the hard gap is difficult to capture accurately with high resolution due to the strong competition between the Kondo resonance and the topological state which no longer coincide for $\mu \neq 0$. Due to the intertwined nature of the Green functions, \textit{cf.} Eq.~\eqref{eq:hsshbethegreenfunctions}, this affects the calculation for the pseudogapped spectra as well. For values of $\mu$ near to the phase transition, the DMFT-NRG calculation does not converge, with each DMFT iteration producing a drastically different solution than the previous iteration, even after very many DMFT iterations ($>200$) and small changes in parameters from the initialization ($\Delta \mu/D < 0.05$). Past the transition, $\mu > \mu_c$, the DMFT-NRG calculation does converge.
% \mu 0.0 0.4 0.6 0.9 1.1 1.3 1.5
% \eta 0., -0.222222, -0.333333, -0.5, -0.611111, -0.722222, -0.833333
% 1.2 -0.666667	12/18
%	0	4/18	6/18	9/18	11/18	13/18	15/18
%
%\begin{figure}[htp]
%\centering
%\input{30-33}
%\caption{$U/D=3.0$, $\varepsilon/D = -3.3$}
%\end{figure}
%%%%%%%dU3.6e-2.4 dU3.6e-2.7 dU3.6e-2.9  dU3.6e-3.1 dU3.6e-3.3
\begin{figure}[htp!]
\centering
\begin{subfigure}{\linewidth}
\caption*{\vspace{-\baselineskip} \qquad\qquad $\mu/D = 0.0$}
\includegraphics[scale=1]{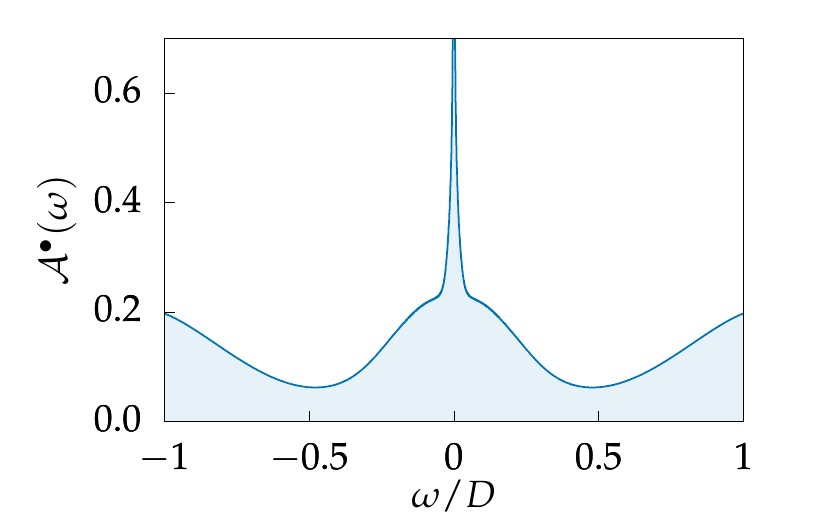}
\includegraphics[scale=1]{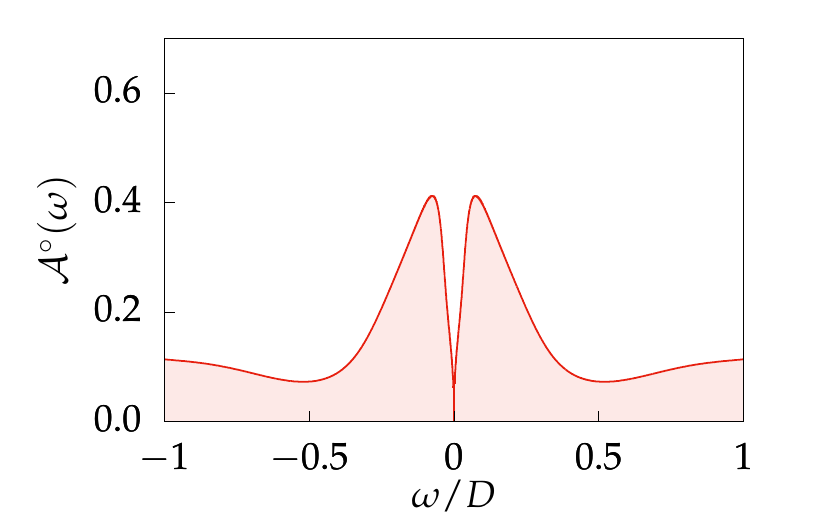}
\vspace{-\baselineskip}
\end{subfigure}
\begin{subfigure}{\linewidth}
\caption*{\vspace{-\baselineskip} \qquad\qquad $\mu/D = 0.4$}
\includegraphics[scale=1]{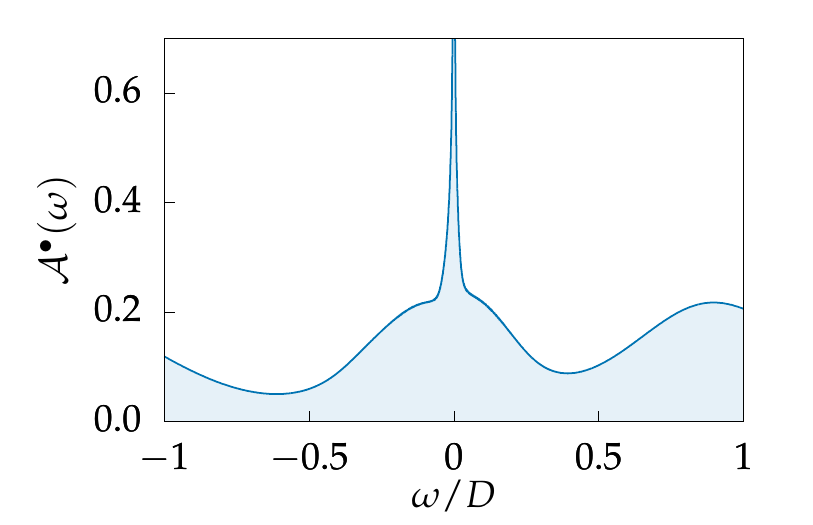}
\includegraphics[scale=1]{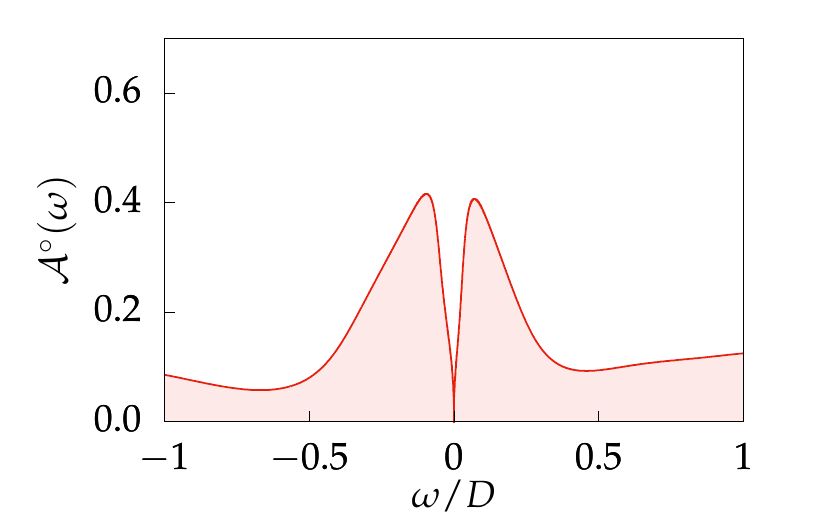}
\vspace{-\baselineskip}
\end{subfigure}
\begin{subfigure}{\linewidth}
\caption*{\vspace{-\baselineskip} \qquad\qquad $\mu/D = 0.6$}
\includegraphics[scale=1]{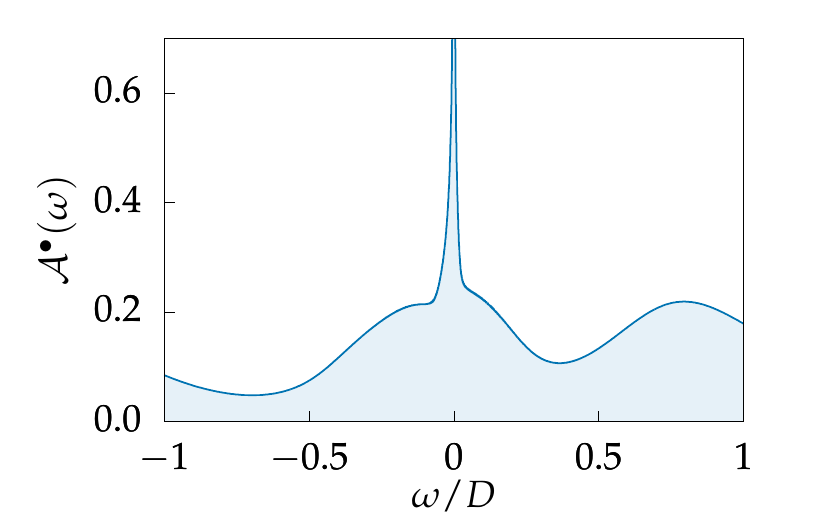}
\includegraphics[scale=1]{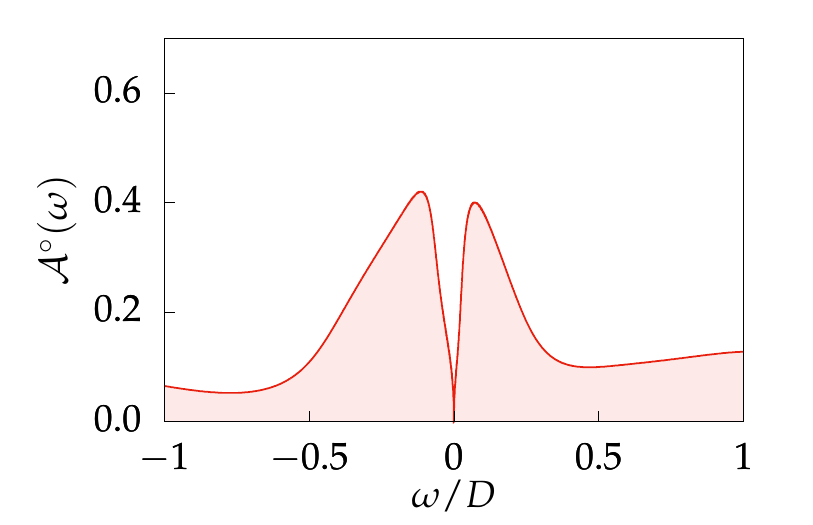}
\vspace{-\baselineskip}
\end{subfigure}
\begin{subfigure}{\linewidth}
\caption*{\vspace{-\baselineskip} \qquad\qquad $\mu/D = 0.9$}
\includegraphics[scale=1]{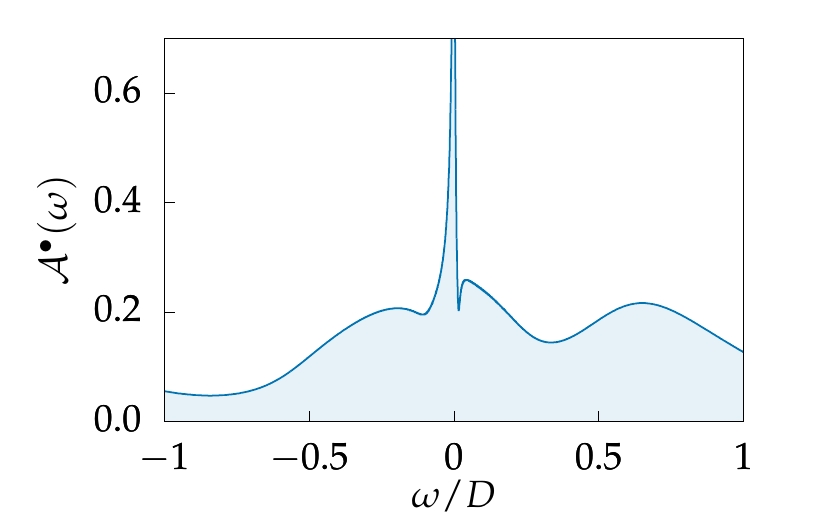}
\includegraphics[scale=1]{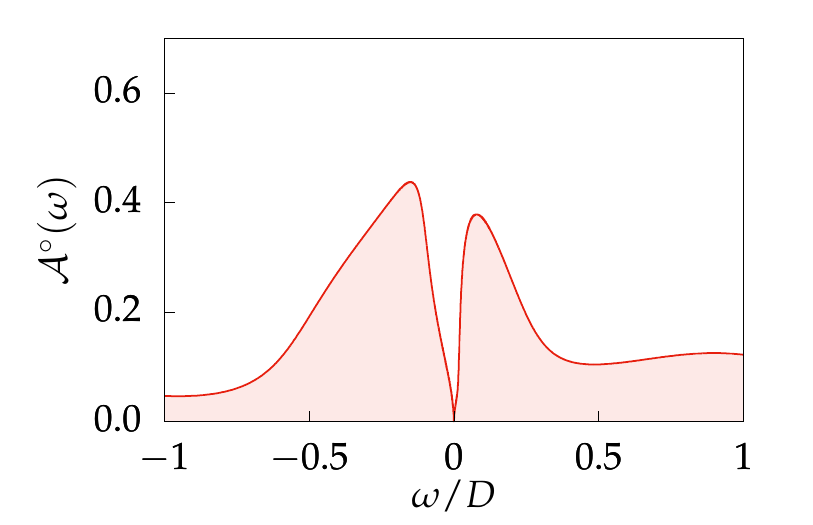}
\vspace{-\baselineskip}
\end{subfigure}
\caption[Spectral functions on the $\bullet$- and $\circ$-sites of the HSSH model at $U/D=3.6$ with doping $\varepsilon = \varepsilon_0 - \mu$ where $\varepsilon_0/D = -1.8$]{Spectral functions on the $\bullet$- and $\circ$-sites of the HSSH model at $U/D=3.6$ with doping $\varepsilon = \varepsilon_0 - \mu$ where $\varepsilon_0/D = -1.8$}
\end{figure}
%%%%%%%%%%%%%%%%%%%%%%%%%%%%%%%%%%%%%%%%%%%%%%%%%%%%%%%
\begin{figure}[htp!]\ContinuedFloat
\centering
\begin{subfigure}{\linewidth}
\caption*{\vspace{-\baselineskip} \qquad\qquad $\mu/D = 1.0$}
\includegraphics[scale=1]{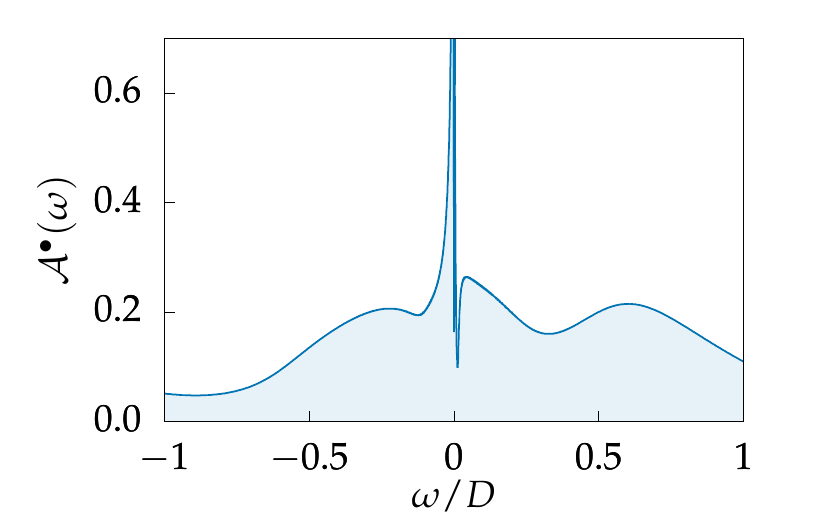}
\includegraphics[scale=1]{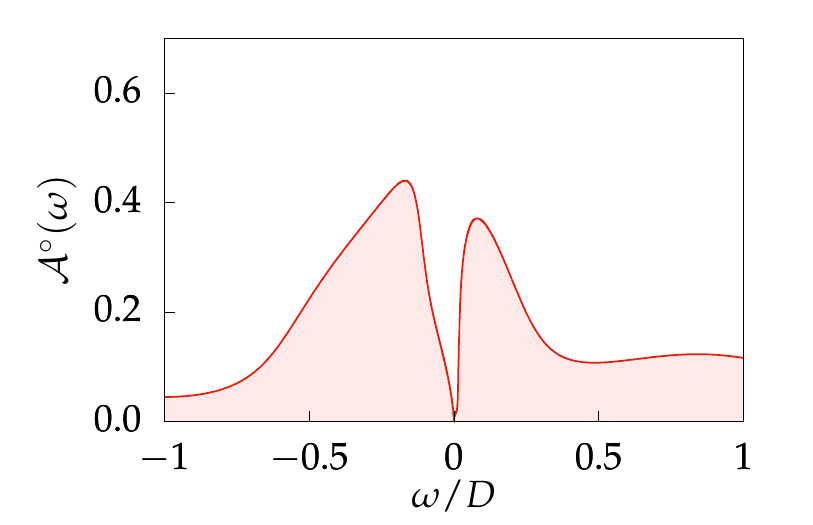}
\vspace{-\baselineskip}
\end{subfigure}
\begin{subfigure}{\linewidth}
\caption*{\vspace{-\baselineskip} \qquad\qquad $\mu/D = 1.1$}
\includegraphics[scale=1]{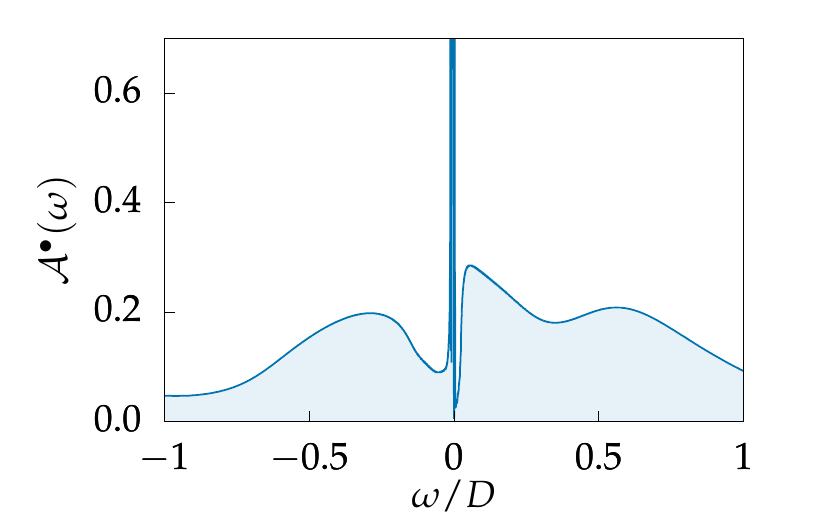}
\includegraphics[scale=1]{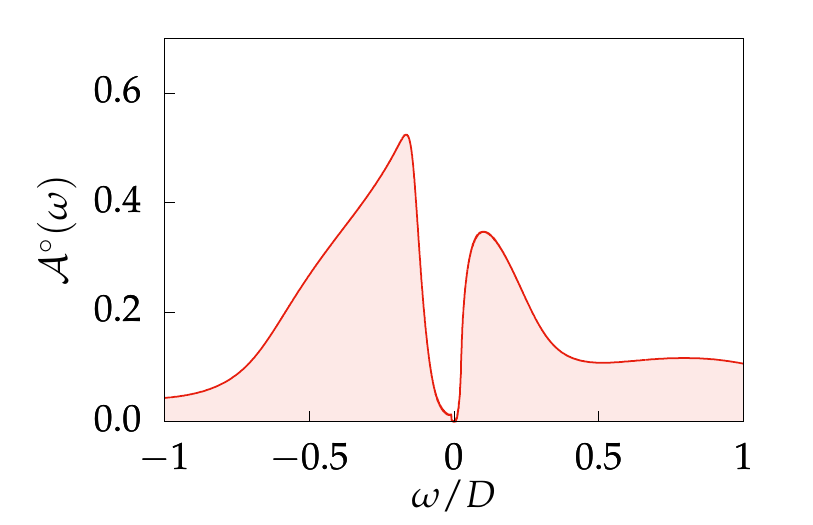}
\vspace{-\baselineskip}
\end{subfigure}
\begin{subfigure}{\linewidth}
\caption*{\vspace{-\baselineskip} \qquad\qquad $\mu/D = 1.3$}
\includegraphics[scale=1]{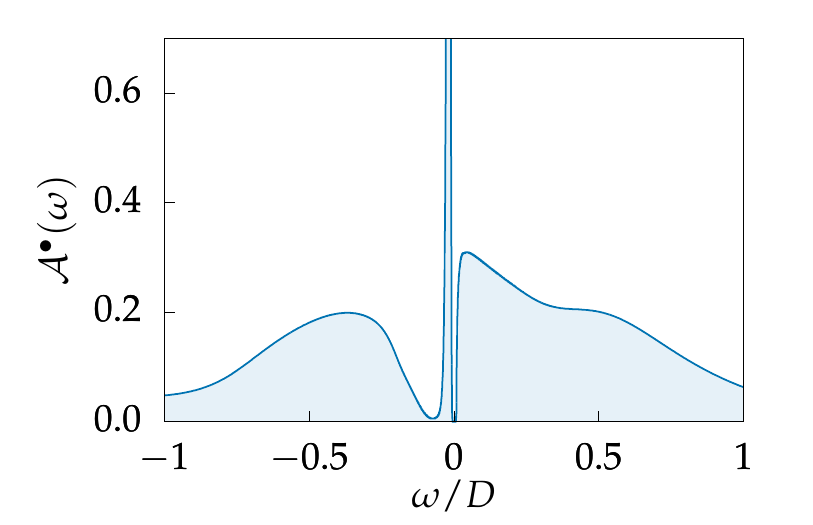}
\includegraphics[scale=1]{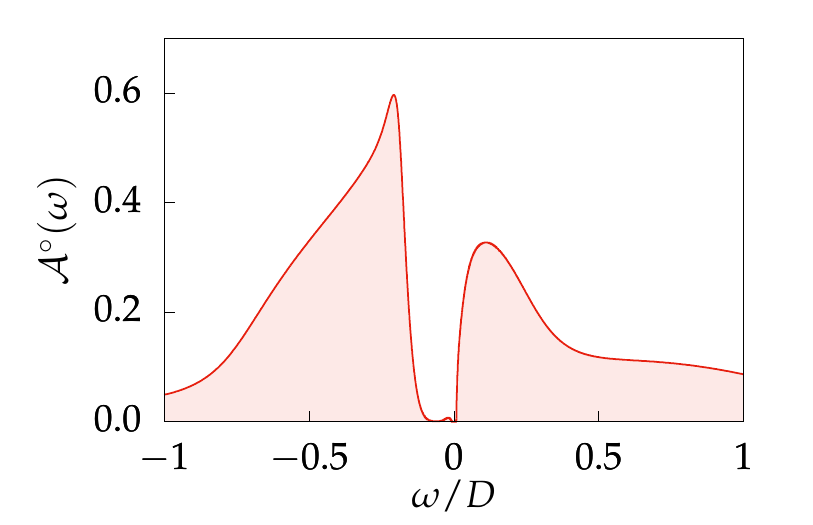}
\vspace{-\baselineskip}
\end{subfigure}
\begin{subfigure}{\linewidth}
\caption*{\vspace{-\baselineskip} \qquad\qquad $\mu/D = 1.5$}
\includegraphics[scale=1]{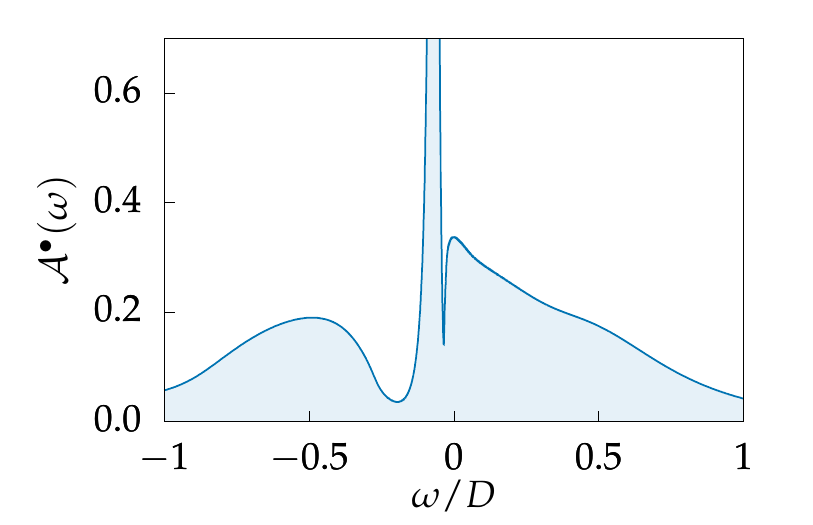}
\includegraphics[scale=1]{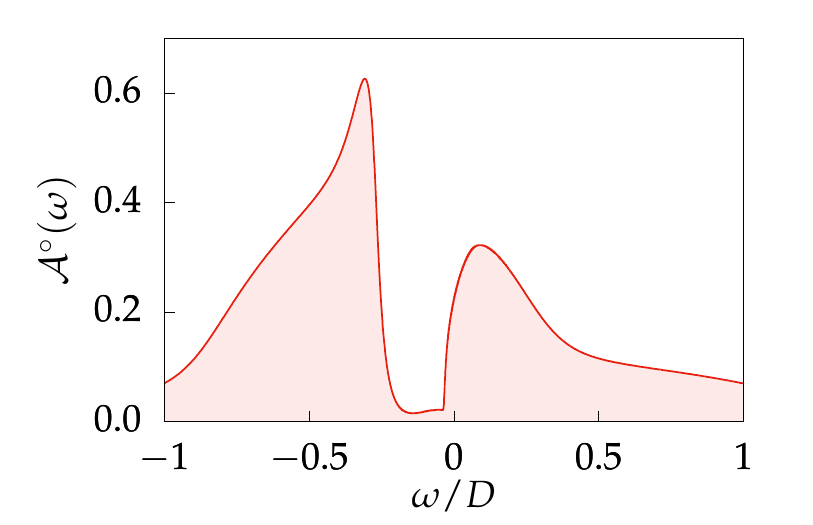}
\vspace{-\baselineskip}
\end{subfigure}
\caption[Spectral functions on the $\bullet$- and $\circ$-sites of the HSSH model at $U/D=3.6$ with doping $\varepsilon = \varepsilon_0 - \mu$ where $\varepsilon_0/D = -1.8$]{Spectral functions on the $\bullet$- and $\circ$-sites of the HSSH model at $U/D=3.6$ with doping $\varepsilon = \varepsilon_0 - \mu$ where $\varepsilon_0/D = -1.8$. Note the doping induced phase transition at $\mu/D\approx 1.3$. Exploded detail plots shown in Fig.~\ref{fig:dopingpt}.\label{fig:phasymmsequence}}
\end{figure}
The calculation presented here is only for a single value of $U$ characteristic of the $U<U_c$ phase. In principle this work could straightforwardly be extended to capture other regions of the parameter space to develop a full phase diagram of the system in the $U$--$\mu$-plane. 
\begin{figure}[ht!]
\includegraphics[scale=1]{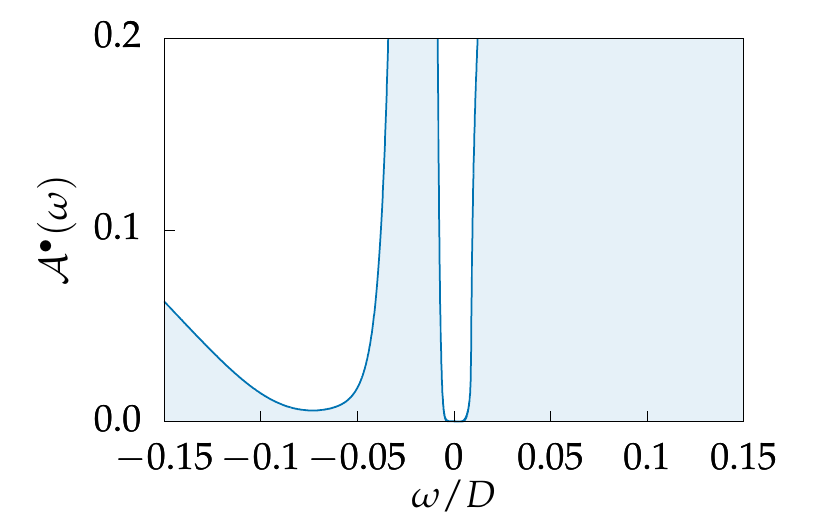}
\includegraphics[scale=1]{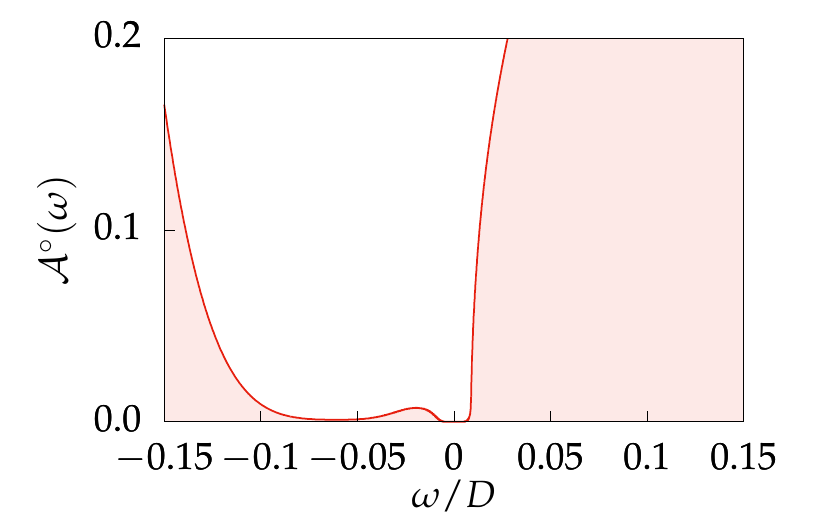}
\caption{Doping induced gap in spectrum on both sites at $\mu/D=1.3$.\label{fig:dopingpt}}
\end{figure}

\section{Outlook}

This chapter developed a scheme for treating the topological SSH model with interactions in the strongly correlated limit. New to the literature here is the reformulation of a topological insulator in infinite dimensions whose characteristics retain the distinction between topological and trivial configurations, and the subsequent exact solution in the non-perturbative strongly correlated case by means of DMFT+NRG. In particular, this calculation shows that a Mott transition may occur in a topological insulator.

The infinite dimensional Bethe SSH model devised for this calculation may be a platform for future experimental work. The standard $1d$ SSH model has been replicated by various quantum simulators, such as in cold atom experiments~\cite{zakobs}. Similar experiments could be engineered to replicate the the Bethe SSH model and confirm the appearance of topological states on the alternating shells of the Bethe lattice.

The SSH model is a prototypical example of a topologically non-trivial system in $1d$. %\index{$0$@\textbf{List of Edits}!400@expanded outlook} 
A second prototypical example of a $1d$ topological system is the Kitaev superconducting wire~\cite{kitaev}\index{Kitaev superconductor}, of class $BD$I which is similar to the $A$III class of the SSH model. A higher dimensional case of this model fitted to the Bethe lattice also exists and has not previously been reported in the literature.\footnote{The name ``Kitaev superconductor'' is employed here rather than the more generic ``Kitaev model'' to preserve the distinction between the superconducting wire and another ``Kitaev model'' which is a spin liquid model on a honeycomb lattice.}
The basic structure of the Kitaev superconductor is that of a $1d$ $p$-wave superconductor described by the Hamiltonian
\begin{equation}
	\hat{H}_{\textsc{k}}
	=
	\sum_{j} \left[ \tensor*{\varepsilon}{_{j}} \opd{c}{j} \op{c}{j} + \tensor*{t}{_{j}} \left( \opd{c}{j+1} \op{c}{j} + \opd{c}{j} \op{c}{j+1} \right) + \tensor*{\Delta}{_{j}} \opd{c}{j+1} \opd{c}{j} + \tensor*{\Delta}{^*_{j}} \op{c}{j} \op{c}{j+1} \right]
\end{equation}
where $\tensor*{\Delta}{_j}$ is the superconducting order parameter.
In the Majorana\index{Majorana} representation, the Hamiltonian for this model reads
\begin{equation}
	H_{\textsc{k}\gamma} = 
	\i \sum_{j} \Big\{ \tensor*{\varepsilon}{_j} \gamma_{[j,1} \gamma_{j,2]} + \tensor*{t}{_j} \gamma_{[j+1,1} \gamma_{j,2]}
	- \tensor*{\Delta}{^*_j} \gamma_{j,(2|} \gamma_{j+1,|1)} + \tensor*{\Delta}{_j} \gamma_{j+1,(2|} \gamma_{j,|1)} \Big\}
\end{equation}
where $c_j = (\gamma_{j,1} - \text{i} \gamma_{j,2})$. 
The conventional index notations~\cite{schutz} $A_{[i} B_{j]} = \frac12 (A_i B_j - A_j B_i )$ and $A_{(i|j} B_{k|l)} = \frac12 (A_{ij} B_{kl} + A_{lj} B_{ki} )$ have been employed in this expression.
The superconducting order parameter $\Delta_j$ can be described by $\Delta_j = \lvert \Delta \rvert e^{\text{i}\varphi(j)}$

On the $1d$ Kitaev superconductor, zero-energy Majorana zero modes can be generated on every site provided that the magnitude of $\Delta$ alternates sign,
\begin{equation}
	\lvert \Delta_j \rvert = \begin{cases} +t_j & j \text{ odd} \\ -t_j & j \text{ even} \end{cases} \,.
\end{equation}
This is tantamount to the phase $\varphi$ rapidly oscillating such that $|\Delta_j| = - |\Delta_{j+1}|$. In the conventional parameterization of the Kitaev superconductor, the Majorana zero modes manifest only on the boundary sites of the wire.
These Majorana modes, however, only manifest themselves at the special point $\varepsilon_j = 0$. This is in contrast to the standard Kitaev superconductor in $1d$ where the Majorana modes manifest for the entire parameter regime $\lvert \varepsilon \rvert < \lvert 2 t \rvert$. In a similar manner to the SSH model discussed above, this configuration for the Kitaev superconductor can be mapped onto a Bethe lattice and the limit taken to infinite coordination number.
A detailed DMFT analysis of this system is left as an outlet for future work. However the Majorana decomposition briefly discussed here will be generalized and taken into a new context in the next chapter.

The approach for treating an interacting topological insulator presented here was based on performing the infinite dimensional counterpart of a $1d$ model in class $A$III.  A complementary approach which has appeared in the literature has treated the infinite dimensional limit of the $2d$ Chern insulator~\cite{interactingchern}, which is a topological insulator in the Cartan class $A$. Due to Bott periodicity systems of this class are topologically non-trivial in dimensions of $d=2\mod2$, or in all even dimensions. 
%\index{$0$@\textbf{List of Edits}!402@added clarification}
This approach makes use of the Clifford algebra valuedness of the Hamiltonian to iteratively construct Hamiltonians in successively larger even dimensions. The limit to infinite dimensions is accomplished by taking the limit through only even dimensions. This strategy allows the application of DMFT to $2d$ Chern insulators.
This scheme is complementary to the approach developed in this chapter. It is well known that the topological phases of matter are dimension dependent~\cite{hk,tenfoldclassification}. The construction of this chapter can be viewed as a method of obtaining a well-defined infinite dimensional limit of a $1d$ topological insulator. The method of~\cite{interactingchern} on the other hand, can be viewed as a systematic way of taking the infinite dimensional limit of a $2d$ topological insulator.

%It is worth noting that a study of the $1d$ SSH model with interactions using DMRG~\cite{manmana} found that the topological state is augmented with a $U$-dependent power-law, which is in contrast with the $U$-independent power-law discovered here. The origin of this discrepancy may involve differences between $1d$ and infinite-$d$, or possibly finite size effects. This point remains open for further consideration.

This chapter studied the effects of strong interactions on a system whose non-interacting counterpart is topological. In contrast, the following chapters investigate strongly correlated systems and reveal how a notion of topology can be found within them.

\chapter{Impurity Effective Models\label{ch:aux}}

The difficulties in dealing with strongly correlated systems motivates the development of effective models which capture the dynamics of the full interacting problem, but are themselves noninteracting. 
%
%The difficulty in solving strongly correlated systems and the necessity of approximations means that 

The DMFT described and used in preceding chapters is itself an example of an effective model for treating strongly correlated systems. Rather than treating the entire interacting system, it treats an approximation of the system where interactions only occur locally at a single point within a larger non-interacting background.
This chapter is devoted to the development of two such similar effective models, the first being based on non-linear canonical transformations and Majorana\index{Majorana} degrees of freedom, and the second based on a novel auxiliary field mapping.

The generalized Majorana decomposition developed here is similar to previous work involving their use in non-linear canonical transformations of strongly correlated fermionic systems \cite{bazzanella,bazzanellakondo,bazzanellanlct,bazzanellamott}. In this previous work, Majoranas were employed to form representations of the symmetry groups of specific models, such as the Hubbard model~\cite{bazzanellamott} and Kondo lattice~\cite{bazzanellakondo}.

The approach to Majorana degrees of freedom taken in this thesis is conceptually similar to the geometric algebra framework of mathematical physics~\cite{hestenes,hestenessobczyk,doranlasenby}. In contrast to standard vector calculus in physics where vectors are taken as elements living in sections of the tangent bundle, in geometric algebra vectors representing physical quantities are taken to be elements of sections of a Clifford bundle. The inspiration here comes from using such a framework to manipulate Majorana degrees of freedom as they behave as Clifford algebra valued vectors, as well as providing a framework for manipulating higher-order tuples of Majoranas. In contrast to the work of~\cite{bazzanella,bazzanellakondo,bazzanellanlct,bazzanellamott}, the transformation defined in this thesis aims to be more general and not dependent on the symmetries of the specific model under consideration.
%\index{$0$@\textbf{List of Edits}!500@emphasized difference with previous literature}
%
%As will be shown below, a key insight is that the Majorana triplet $\i \gamma_a \gamma_b \gamma_c$ behaves algebraically the same as a single Majorana $\gamma_a$.

The second section of this chapter is devoted to the development and basic application of an auxiliary field mapping scheme which systematically maps interacting lattice models onto fully non-interacting lattice models that still reproduce the original dynamics.
The non-interacting equivalent model is not only simpler to treat, but offers insights into the underlying structure and dynamics of such systems.
An application showcasing the utility of this effective model is given for the case of calculating quantum transport through an interacting impurity.

This auxiliary field mapping will play a central role the subsequent chapter where it is applied to the Mott transition in the Hubbard model.

\section{Non-Linear Canonical Transformations\label{sec:majorana}}

The idea of exploiting non-linear canonical transformations as a method of treating strongly correlated systems in~\cite{bazzanella,bazzanellanlct} was inspired by the observation that the local symmetry group of the Hubbard model possesses a non-linear $U(1)$ symmetry~\cite{oestlundmele}. In~\cite{oestlundmele} it was noted that the local symmetry group of the Hubbard model is $G = SU(2)_S \otimes SU(2)_I \otimes U(1)_{NL} \otimes \mathbbm{Z}_2$ where the $SU(2)$ groups correspond to spin and isospin symmetries, and the $\mathbbm{Z}_2$ group is a parity symmetry which exchanges spin and isospin. The electromagnetic charge symmetry $U(1)_Q$ is a subgroup of the isospin group.

The Hubbard model also admits a local non-linear $U(1)_{NL}$ symmetry which acts as
%\begin{subequations}
\begin{align}
	\opd{c}{j,\sigma} &\mapsto \opd{c}{j,\sigma} ( 1 - \op{n}{-\sigma} ) + \e^{2\i\chi} \opd{c}{j,\sigma} \op{n}{-\sigma}
\end{align}
%\end{subequations}
The action of this transformation is to change the phase of the doubly occupied state by $\e^{-2\i\chi}$ while preserving the zero and singly occupied states. 
This transformation can be generated by $\opd{\tilde{c}}{\sigma} = R^\dagger(\chi) \opd{c}{\sigma} R(\chi)$ where
\begin{equation}
	R(\chi) = \e^{2 \i \chi \left( \hat{n}_{\uparrow} - \frac12 \right) \left( \hat{n}_{\downarrow} - \frac12 \right)} \,.
\end{equation}
Under the conventional Majorana decomposition
\begin{align}
	\opd{c}{\uparrow} &= \frac{\gamma_1 + \i \gamma_2}{2}
	&
	\opd{c}{\downarrow} &= \frac{\gamma_3 + \i \gamma_4}{2}
	\,,
\end{align}
the generator of the non-linear transformation becomes
\begin{equation}
	R(\chi) = \e^{- \i \frac{\chi}{2} \gamma_1 \gamma_2 \gamma_3 \gamma_4} \,.
\end{equation}
This generator may be applied directly to the Majorana basis as $\tilde{\gamma}_{i} = R^\dagger(\chi) \gamma_{i} R(\chi)$ yielding Majoranas in the transformed basis as
\begin{equation}
	\tilde{\gamma}_{i} = \gamma_{i} \cos\chi + \sgn(\pi) \i \gamma_{j} \gamma_{k} \gamma_{l} \sin\chi
\end{equation}
where $i \neq j,k,l$ and the sign depends on the permutation $\pi$ of the Majoranas. As evidenced by this transformation, non-linear canonical transformations can be implemented in a Majorana basis in a straightforward manner and such transformations also lead to the introduction of higher-order Majorana tuplets as degrees of freedom.

\subsection{General Majorana Decomposition}

Previous approaches to non-linear canonical transformations using Majorana degrees of freedom~\cite{bazzanella} were based on realizing representations of relevant symmetry groups in terms of Majoranas. For example, the transformations in~\cite{bazzanella} were based on realizing representations of $SU(2)_{S} \otimes SU(2)_{I}$, or more generally, $SU(2^n) \otimes SU(2^n)$.

The approach taken here is a more abstract and less model dependent approach and is conceptually similar to, and motivated by, the geometric algebra\index{geometric algebra} framework of mathematical physics~\cite{hestenes,hestenessobczyk,doranlasenby}. As a primer for the generalized Majorana formalism developed below, presented here is a brief review of geometric algebra. Geometric algebra and the associated geometric calculus is a framework which replaces the standard Gibbs-Heaviside vector algebra, where functions on space(time) manifold $M$ are taken to be sections of the tangent bundle $TM$, with functions which are sections of the real Clifford bundle over $M$, a vector bundle whose fibers are a real Clifford algebra. Formally, a Clifford algebra is defined for a given vector space $V$ and a quadratic form $Q$. The present analysis of Majoranas takes the vector space to be over the field of real numbers $\mathbbm{R}$ and the quadratic form to be the Euclidean metric. An orthonormal basis for the real $k$-dimensional Clifford algebra $\text{Cl}_{k}(\mathbbm{R})$ is given by $\{e_n\}$ such that a Clifford vector $a$ can be written in components as $a = a^1 e_1 + \cdots + a^n e_n$. 
The product of two Clifford vectors $a$ and $b$ is described by~\cite{hestenes,hestenessobczyk,doranlasenby}
\begin{equation}
	a \, b = a \cdot b + a \extp b
\label{eq:geometricproduct}
\end{equation}
where $a \cdot b \vcentcolon= \frac12(a \, b + b \, a)$ is the symmetric part and $a \extp b \vcentcolon= \frac12(a \, b - b \, a)$ is the antisymmetric part. 
%Clifford algebras possess an antisymmetric subalgebra which is 
The antisymmetric subalgebra is an exterior algebra, which allows the construction of higher rank vectors with bases given by exterior products of the vector basis. To illustrate, the graded basis elements of $\text{Cl}_{3}(\mathbbm{R})$ are
\begin{equation}
\begin{gathered}
	1
	\\
	e_1 \qquad e_2 \qquad e_3
	\\
	e_1 \extp e_2 \qquad e_2 \extp e_3 \qquad e_3 \extp e_1
	\\
	e_1 \extp e_2 \extp e_3
\end{gathered}
\end{equation}
which are termed the scalar, vector, bivector, and trivector basis elements respectively. These vectors are bases of contravariant vector spaces. Given a metric, there is exists an isomorphism between these higher rank vectors and exterior forms of the same rank. Since the basis elements are orthogonal to each other, $e_i \cdot e_j = \delta_{ij}$, the higher grade elements can be written directly in terms of the Clifford product rather than the wedge product, \textit{e.g.} $e_1 \extp e_2 \equiv e_1 e_2$. The highest rank vector of a Clifford algebra is also called the pseudoscalar as it functions as a scalar quantity which is odd under parity transformations. The unit pseudoscalar also functions as an operator in $\text{Cl}_k$ as the generator of duality transformations, analogously to the Hodge star operator in the algebra of exterior forms. In $d$ dimensions, the Hodge star operator is a map $\star \vcentcolon \bigwedge_n \to \bigwedge_{d-n}$ which maps $n$-forms to $(d-n)$-forms. This is accomplished in geometric algebra by means of taking the geometric product of an element with the unit pseudoscalar and contracting terms with the inner product.
\begin{align}
	e_1 \, e_1 e_2 e_3 &= e_2 e_3
	&
	e_2 \, e_1 e_2 e_3 &= e_3 e_1
	&
	e_3 \, e_1 e_2 e_3 &= e_1 e_2
\end{align}
This shows that the unit bivectors in $\text{Cl}_{3}(\mathbbm{R})$ are dual to the unit vectors. A historical prototype of the vector algebra of $3d$ space was that of the quaternions $\mathbbm{H}$, the four-dimensional normed division algebra over the reals invented by William Rowan Hamilton in 1843. A quaternion $q \in \mathbbm{H}$ is defined as
\begin{equation}
	q \vcentcolon= \{ a + b \boldsymbol{i} + c \boldsymbol{j} + d \boldsymbol{k} \;\vert\; a,b,c,d \in \mathbbm{R} ;\; \boldsymbol{i}^2 = \boldsymbol{j}^2 = \boldsymbol{k}^2 = \boldsymbol{i} \boldsymbol{j} \boldsymbol{k} = -1 \} \,.
\end{equation}
In terms of representing vector objects in $\mathbbm{R}^3$, the elements $\boldsymbol{i}$, $\boldsymbol{j}$, and $\boldsymbol{k}$ can be taken as an orthonormal basis. Within this vectorial approach, the complex quaternionic units do not actually correspond to the unit vectors $e_a$, but rather the unit bivectors $e_a e_b$. 
Historically the use of quaternions to describe physical objects in space was superseded by the Gibbs-Heaviside algebra as it was computationally simpler and more visually intuitive, however quaternions prove to be useful in the computer graphics industry for implementing rotations in $3d$ graphics where they have advantages over Euler angles, for instance, by avoiding the issue of gimbal lock.

Examples of Clifford algebra known in conventional physics are the algebra of the Pauli matrices and of the Dirac matrices. 
Indeed the Clifford product Eq.~\eqref{eq:geometricproduct} is already known in conventional physics by means of the Pauli matrix identity
\begin{equation}
	{\sigma}_a \; {\sigma}_b = \left(\vec{a} \cdot \vec{b}\right) \mathbbm{1} + \i \, \left(\vec{a} \times \vec{b}\right) \cdot \vec{\sigma}
\end{equation}
where ${\sigma}_a = \vec{a} \cdot \vec{\sigma}$ with Euclidean metric.
The use of Clifford algebras is also well known in the description of relativistic spinors~\cite{pdg}. In this context the internal Clifford metric possesses Lorentzian signature.

Interest in Majorana degrees of freedom in contemporary condensed matter literature is dominated by their potential application as qubits in quantum computation~\cite{kitaev,tqcreview}. 
The conventional decomposition of fermionic degrees of freedom to Majorana degrees of freedom within the context of second-quantized operators is given by
\begin{align}
	\opd{c}{a} &= \frac{\gamma_{1,a} + \i \gamma_{2,a}}{2}
	&
	\op{c}{a} &= \frac{\gamma_{1,a} - \i \gamma_{2,a}}{2}
\label{eq:stdmajorana}
\end{align}
where the Majorana operators obey the Clifford algebra relation
\begin{subequations}
\begin{align}
	\{ \tensor*{\gamma}{_\mu} , \tensor*{\gamma}{_\nu} \} = 2 \tensor*{\delta}{_{\mu,\nu}}
\intertext{and are self-conjugate,}
	\gamma_\mu^\dagger = \gamma_\mu \;.
\end{align}
\end{subequations}
The indices $\mu$ and $\nu$ represent all relevant degrees of freedom, such as lattice site and spin.
Immediate consequences of these relations are that Majorana operators are idempotent, $\gamma_\mu^2 = 1$, and obey antisymmetric statistics, $\tensor*{\gamma}{_\mu} \tensor*{\gamma}{_\nu} = - \tensor*{\gamma}{_\nu} \tensor*{\gamma}{_\mu}$ for $\mu \neq \nu$. Another consequence of this property is that it implies that there does not exist the notion of a Fock space for Majorana operators. Since $\gamma^2 = 1$, it follows that there is no notion of a vacuum state such that $\gamma \lvert \Omega \rangle = 0$.

Majorana degrees of freedom as described here are sometimes referred to as ``Majorana fermions'' in the condensed matter literature due to their similarity with the Majorana fermions that appear in relativistic field theory which are their own anti-particle, a property which manifests as the relation $\gamma^\dagger = \gamma$. However, the Majorana degrees of freedom which appear in condensed matter contexts, such as in the Kitaev wire, do not generally obey fermionic statistics, but rather non-Abelian anyonic statistics. Although this property does not feature in this work, the characterization of Majorana degrees of freedom as ``fermions'' will be eschewed for this reason as well as to more easily distinguish between Majorana degrees of freedom and the complex fermion degrees of freedom.

The Clifford algebraic properties of the Majorana operators also allow for generalizations of this basic decomposition to operators which go beyond the single Majorana operator $\gamma$. Taking inspiration from the geometric algebra formalism discussed above, the following develops a new generalized formalism for decomposing fermions into Majorana degrees of freedom. This generalized decomposition is expressed in the form
%\index{$0$@\textbf{List of Edits}!501@highlight new contribution}
\begin{align}
	\opd{c}{} &= \frac{\Gamma^\dagger_\Re + \Gamma^\dagger_\Im}{2}
	&
	\op{c}{} &= \frac{\Gamma_\Re + \Gamma_\Im}{2}
\label{eq:genmajorana}
\end{align}
where $\Gamma_\Re$, $\Gamma_\Im$ are now arbitrary polynomial functions of the basic Majorana operators $\gamma$ and indices associated to the fermions and internal degrees of freedom have been suppressed for clarity.

In order to preserve fermionic anticommutation relation $\{ \opd{c}{a} , \op{c}{b} \} = \tensor{\delta}{_{a,b}}$, the generalized Majorana operators must obey the conditions
\begin{align}
	\tensor*{\Gamma}{_\Re^\dagger} &= \tensor*{\Gamma}{_\Re} \,,
	&
	\Gamma_\Re^2 &= +1 \,,
	&
	\tensor*{\Gamma}{_\Im^\dagger} &= -\tensor*{\Gamma}{_\Im} \,,
	&
	\Gamma_\Im^2 &= -1 \,,
	&
	&\text{and}
	&
	\{ \tensor*{\Gamma}{_\Re} , \tensor*{\Gamma}{_\Im} \} &= 0 \,.
\label{eq:genmajoranarelations}
\end{align}
The conjugation ($\dagger$) operation involves reversing the order of constituent elementary Majorana operatiors $\gamma$ as well as taking the complex conjugate. For example, for $\Gamma = \i \gamma_{\mu_1} \gamma_{\mu_2} \cdots \gamma_{\mu_n}$, conjugation returns $\Gamma^\dagger = -\i \gamma_{\mu_n} \gamma_{\mu_{n-1}} \cdots \gamma_{\mu_1}$. %Such a compound Majorana operator may also be written in the form of $\Gamma = \tensor{\epsilon}{_{j_1 j_2 \cdots j_n}} \gamma_{j_1} \gamma_{j_2} \cdots \gamma_{j_n}$ where $\tensor{\epsilon}{_{j_1 j_2 \cdots j_n}}$ is the totally antisymmetric Levi-Civita symbol.

Observe that the generalized decomposition Eq.~\eqref{eq:genmajorana} does not exhibit an explicit dependence on the imaginary unit, but rather it appears implicitly through the definition of $\Gamma_\Re$ and $\Gamma_\Im$ due to the algebraic properties of the $\dagger$ operation. In the case of the standard two-fold decomposition Eq.~\eqref{eq:stdmajorana}, $\Gamma_\Re = \gamma_1$ and $\Gamma_\Im = -\i \gamma_2$. In this case the imaginary unit appears within the definition of the object $\Gamma_\Im$. The imaginary unit however is not restricted to lie within the definition of $\Gamma_\Im$, nor must it appear there. 

A first example of a generalized Majorana operator is $\Gamma_\Re = \i \gamma_1 \gamma_2 \gamma_3$. Its properties can be confirmed to obey \eqref{eq:genmajoranarelations}:
\begin{align*}
&\begin{aligned}[t]
	\Gamma_\Re^2
	&= (\i \gamma_1 \gamma_2 \gamma_3)(\i \gamma_1 \gamma_2 \gamma_3)
	\\
	&= - \gamma_1 \gamma_2 \gamma_3 \gamma_1 \gamma_2 \gamma_3
	\\
	&= - (-1)^3
	\\
	&= 1 \,,
\end{aligned}
&
&\begin{aligned}[t]
	\Gamma_\Re^\dagger
	&= -\i \gamma_3 \gamma_2 \gamma_1
	\\
	&= -\i (-1)^3 \gamma_1 \gamma_2 \gamma_3
	\\
	&= \Gamma_\Re \,.
\end{aligned}
\end{align*}
This calculation shows that a triplet of Majorana operators times the imaginary unit is algebraically equivalent to a single Majorana operator. As demonstrated here the imaginary unit may appear in the definition of $\Gamma_\Re$, even though $\Gamma_\Re$ captures the real part of the original complex fermion degree of freedom.
Similarly, a Majorana which obeys the properties of $\Gamma_\Im$ can be obtained from $-\i\Gamma_\Re = \gamma_1 \gamma_2 \gamma_3$.
In accordance with the definition \eqref{eq:genmajorana}, a fermion that can be constructed from these operators may be
\begin{align}
	\opd{c}{} &= \frac{\i \gamma_1 \gamma_2 \gamma_3 + \gamma_4 \gamma_5 \gamma_6}{2} \,,
	&
	\op{c}{} &= \frac{\i \gamma_1 \gamma_2 \gamma_3 - \gamma_4 \gamma_5 \gamma_6}{2} \,.
\end{align}
This example demonstrates the rationale of not explicitly including the imaginary unit in the definition \eqref{eq:genmajorana}. The definition avoids the potentially misleading association between the postfactor of the imaginary unit with the component which has odd parity under Hermitian conjugation.

A pair of generalized Majoranas which lead to an unconventional decomposition are
\begin{align}
	\Gamma_\Re &= \gamma_1 \,,
	&
	\Gamma_\Im &= \gamma_2 \gamma_3 \gamma_4 \,.
\end{align}
This pair represents an unusual representation as it manifests as a purely $\mathbbm{R}$-valued decomposition
\begin{align}
	\opd{c}{} &= \frac{\gamma_1 + \gamma_2 \gamma_3 \gamma_4}{2} \,,
	&
	\op{c}{} &= \frac{\gamma_1 + \gamma_4 \gamma_3 \gamma_2}{2} = \frac{\gamma_1 - \gamma_2 \gamma_3 \gamma_4}{2} \,.
\end{align}
The adjoint property manifests not as complex conjugation, but rather as in the reversal of the Majorana polynomial.

A more elaborate example of this generalized transformation involves
\begin{align}
	\Gamma_\Re &= \frac{\gamma_1 + \gamma_2}{\sqrt{2}} \,,
	&
	\Gamma_\Im &= \frac{\gamma_3 \gamma_4 \gamma_5 - \i \gamma_6}{\sqrt{2}} \,.
\label{eq:majoranapolynomial}
\end{align}
This examples demonstrates that the fermions need not be a monomial, but can in general be polynomials of higher order Majorana terms. To preserve the correct commutation relations, the $\Gamma$ are restricted to be odd degree polynomials.

A key aspect of the Majorana operators that will be exploited in this work is the geometric interpretation of their Clifford algebraic structure. 
In the Majorana representation, the Clifford algebraic properties of the Majorana algebra allow for the expression of canonical transformations in terms of a non-linear $U(1)$ transformation~\cite{bazzanella}. This transformation can be implemented by means of the generalized Euler identity
\begin{equation}
	\e^{\pm \mathcal{I} N \theta} = \mathbbm{1} \cos(N\theta) \pm \mathcal{I} \sin(N\theta)
\label{eq:generaleuler}
\end{equation}
where $N \in \mathbbm{R}$, $\theta \in \mathbbm{R}/2\pi\mathbbm{Z}$, $\mathcal{I}^2 = -\mathbbm{1}$, and $\mathcal{I}$ must be power associative. This transformation amounts to a rotation in the $k$-dimensional Clifford algebra $\mathrm{Cl}_k$. This expression generalizes the conventional Euler identity which appears as a special case for $\mathcal{I} = \pm\i$.

A unitary transformation operator acting on a Majorana operator is the form of a rotor\index{rotor}~\cite{hestenes,hestenessobczyk,doranlasenby}
\begin{equation}
	R(\theta) = \e^{\mathcal{I} \theta / 2}
\label{eq:rotor}
\end{equation}
whose actions are implemented by the transformation
\begin{equation}
	\gamma' = R(\theta) \gamma R^\dagger(\theta)
\end{equation}
with $R^\dagger(\theta) = \e^{\mathcal{I}^\dagger \theta / 2} = \e^{-\mathcal{I} \theta / 2}$.

With a Hamiltonian in its Majorana representation, it is possible to perform a non-linear canonical transformation in Clifford algebra space to a new Majorana representation, which can then be recombined into a new fermion representation.

\subsubsection{Hubbard Atom}\label{nlctha}

An example of the utility of these non-linear canonical transformations is that of application to the Hubbard atom. 
This case serves as an example of how a non-linear canonical transformation can transform an interacting model into a non-interacting one. This transformation relies on mixing the physical degrees of freedom of the Hubbard atom with auxiliary fermion degrees of freedom which are initially decoupled from the physical system. The auxiliary part of the system consists of two species of spinless fermions independently isolated from each other and the physical spinful fermions. 
The Hamiltonian of the total system is
\begin{equation}
	\hat{H}
	=	\underbrace{
		\varepsilon \left( \opd{c}{\uparrow} \op{c}{\uparrow} + \opd{c}{\downarrow} \op{c}{\downarrow} \right) + U \opd{c}{\uparrow} \op{c}{\uparrow} \opd{c}{\downarrow} \op{c}{\downarrow}
		}_{\text{Physical}}
		+
		\underbrace{
		\varepsilon_f \hat{f}^\dagger \hat{f} + \varepsilon_g \hat{g}^\dagger \hat{g}
		}_{\text{Auxiliary}}
\end{equation}
where $\op{c}{\uparrow}$, $\op{c}{\downarrow}$, $\hat{f}$, and $\hat{g}$ obey the usual fermionic commutation relations.
\begin{comment}
\begin{equation}
	H
	=	\underbrace{
		\varepsilon \left( \tensor*{c}{^\dagger_\uparrow} \tensor*{c}{_\uparrow} + \tensor*{c}{^\dagger_\downarrow} \tensor*{c}{_\downarrow} \right) + U \tensor*{c}{^\dagger_\uparrow} \tensor*{c}{_\uparrow} \tensor*{c}{^\dagger_\downarrow} \tensor*{c}{_\downarrow}
		}_{\text{Physical}}
		+
		\underbrace{
		\varepsilon_f f^\dagger f + \varepsilon_g g^\dagger g
		}_{\text{Auxiliary}}
\end{equation}
\end{comment}
The Hilbert space may be broken into two components, one for the physical degrees of freedom and one for the auxiliary degrees of freedom. The total Hilbert space for the combined system is $\mathcal{H} = \mathcal{H}_{\textsc{ha}} \otimes \mathcal{H}_{f} \otimes \mathcal{H}_{g}$ where $\mathcal{H}_{f}$ and $\mathcal{H}_{g}$ are the Hilbert spaces of the auxiliary fermions and $\dim\mathcal{H} = 8$.
These Hilbert subspaces are both spanned by a four-dimensional Majorana representation, represented in the following by $\gamma$ and $\mu$ respectively. The fermionic degrees of freedom of the original model can be transformed into a Majorana representation as
\begin{align}
	\tensor*{c}{^\dagger_\uparrow} &= \frac{\gamma_1 + \i \gamma_2}{2} \,,
	&
	\tensor*{c}{^\dagger_\downarrow} &= \frac{\gamma_3 + \i \gamma_4}{2} \,,
	\\
	f^\dagger &= \frac{\mu_1 + \i \mu_2}{2} \,,
	&
	g^\dagger &= \frac{\mu_3 + \i \mu_4}{2} \,.
\end{align}
These $\gamma$ and $\mu$ degrees of freedom obey the characteristic Clifford algebra relations of Majorana operators
\begin{align}
	\{ \gamma_i , \gamma_j \} &= 2 \delta_{ij}
	=
	\{ \mu_i , \mu_j \} \,,
	&
	\{ \gamma_i , \mu_j \} &= 0 \,.
\end{align}
Under this transformation the original Hamiltonian becomes
\begin{equation}\begin{aligned}[b]
	H
	&=
	\begin{multlined}[t]
	\frac{\varepsilon}{2} \left( 2 - \i \gamma_1 \gamma_2 - \i \gamma_3 \gamma_4 \right) + \frac{U}{4} \left( - \gamma_1 \gamma_2 \gamma_3 \gamma_4 - \i \gamma_1 \gamma_2 - \i \gamma_3 \gamma_4 + 1 \right) \\+ \frac{\varepsilon_f}{2} \left( 1 - \i \mu_1 \mu_2 \right) + \frac{\varepsilon_g}{2} \left( 1 - \i \mu_3 \mu_4 \right)
	\end{multlined}
	\\
	H
	&=	\i \Delta \left( \gamma_1 \gamma_2 + \gamma_3 \gamma_4 \right) - \frac{U}{4} \gamma_1 \gamma_2 \gamma_3 \gamma_4 - \i \frac{\varepsilon_f}{2} \mu_1 \mu_2 - \i \frac{\varepsilon_g}{2} \mu_3 \mu_4
	+ \varepsilon + \frac{U}{4} + \frac{\varepsilon_f}{2} + \frac{\varepsilon_g}{2}
\end{aligned}\end{equation}
with \( \Delta = \frac{\varepsilon}{2} + \frac{U}{4} \) a factor which parameterizes the doping on the Hubbard atom. %$\eta = 1 - 2\varepsilon/U$

The Hamiltonian is now of a form to which a non-linear canonical transformation can be applied based on \eqref{eq:generaleuler}.
To achieve this, it is useful to define the unit pseudoscalar $P$ for each Majorana subspace,
\begin{align}
	P_\gamma &= \gamma_1 \gamma_2 \gamma_3 \gamma_4
	&
	&\text{and}
	&
	P_\mu &= \mu_1 \mu_2 \mu_3 \mu_4 \,,
\end{align}
which have the properties
\begin{align}
	P_\gamma^2 &= +1 = P_\mu^2 \,,
	&
	\{ \gamma_j , P_\gamma \} &= 0 = \{ \mu_j , P_\mu \} \,,
	&
	[ \gamma_j , P_\mu ] &= 0 = [ \mu_j , P_\gamma ] \,,
	&
	[ P_\gamma , P_\mu ] &= 0 \,.
\end{align}
The unit pseudoscalar parameterizes a $U(1)$ symmetry within each subspace with transformation generator $\e^{\i P \theta}$.
The unit pseudoscalars enter into the definition of a rotation kernel which serves as the basis for the element $\mathcal{I}$ in Eq.~\eqref{eq:generaleuler},
\begin{subequations}
\begin{align}
	\mathcal{I}_{\gamma,j} &= \i ( \gamma_j \mu_j ) \gamma_1 \gamma_2 \gamma_3 \gamma_4 = \i (\gamma_j \mu_j ) P_\gamma \,,
	\\
	\mathcal{I}_{\mu,j} &= \i ( \gamma_j \mu_j ) \mu_1 \mu_2 \mu_3 \mu_4 = \i ( \gamma_j \mu_j ) P_\mu \,.
%	\\
%	\mathcal{I}_{\gamma} &= \sum_{j=1}^{4} \mathcal{I}_{\gamma,j}	
%	&
%	\mathcal{I}_{\mu} &= \sum_{j=1}^{4} \mathcal{I}_{\mu,j}
\end{align}
\end{subequations}
A distinct rotation kernel is defined for each Majorana of each subspace. A specific example of one of the kernels is
\begin{equation}
\begin{aligned}
	\mathcal{I}_{\mu,1}
	&= \i (\gamma_1 \mu_1) \mu_1 \mu_2 \mu_3 \mu_4
	\\
	&= \i \gamma_1 \mu_2 \mu_3 \mu_4
\end{aligned}
\end{equation}
The principle behind this definition is to produce an object such that when a Majorana is acted upon by the operator \eqref{eq:generaleuler}, it will contract with elements of the rotation kernel and the resulting expression, for a given $\theta$, will be a properly normalized Majorana in a new basis.
A Majorana of the $\gamma$-basis can for example be transformed to a tuple in the $\mu$-basis by
\begin{equation}
\begin{aligned}[b]
	\gamma_1 \mapsto \gamma^\prime_1(\theta)
	&=	\e^{\mathcal{I}_{\mu,1} \theta/2} \gamma_1 \e^{-\mathcal{I}_{\mu,1} \theta/2}
	\\
	&=	\left( \cos\tfrac\theta2 + \i \gamma_1 \mu_2 \mu_3 \mu_4 \sin\tfrac\theta2 \right) \gamma_1 \left( \cos\tfrac\theta2 - \i \gamma_1 \mu_2 \mu_3 \mu_4 \sin\tfrac\theta2 \right)
	\\
	&=	\gamma_1 \cos\theta - \i \mu_2 \mu_3 \mu_4 \sin\theta
	\\
	\gamma^\prime_1(\tfrac\pi2)
	&= - \i \mu_2 \mu_3 \mu_4 \,.
\end{aligned}
\end{equation}
In this way, a Majorana degree of freedom from the physical subspace may be canonically transformed to a function of Majorana degrees of freedom of the auxiliary subspace.

With this rotation kernel the original Hamiltonian at half-filling, $\varepsilon = - U/2$,
\begin{align}
	H	&=	- \frac{U}{4} \gamma_1 \gamma_2 \gamma_3 \gamma_4 - \i \frac{\varepsilon_f}{2} \mu_1 \mu_2 - \i \frac{\varepsilon_g}{2} \mu_3 \mu_4
\end{align}
can be put into a new form with respect to a new Majorana basis. 
Each Majorana of the phase space can be canonically transformed using the rotor operator\index{rotor} with a specified angle for each Majorana
\begin{subequations}
\begin{align}
	\gamma'_j &= R_{\mu,j}(\theta_{\mu,j}) \, \gamma_j \, R_{\mu,j}^\dagger(\theta_{\mu,j})
	\\
	\mu'_j &= R_{\gamma,j}(\theta_{\gamma,j}) \, \mu_j \, R_{\gamma,j}^\dagger(\theta_{\gamma,j})
\end{align}
\end{subequations}
The angles for each rotor are chosen to be
\begin{align}
	\left. \begin{matrix} \theta_{\gamma,1} \\ \theta_{\mu,1} \end{matrix} \right\}
	&=	\frac\pi2
	&
	\left. \begin{matrix} \theta_{\gamma,2} ,\, \theta_{\gamma,3} ,\, \theta_{\gamma,4} \\ \theta_{\mu,2} ,\, \theta_{\mu,3} ,\, \theta_{\mu,4} \end{matrix} \right\}
	&=	0
\end{align}
Following this parameterization, the Majoranas in the new rotated basis are
\begin{align}
&\begin{aligned}
	\gamma'_1	&=	\gamma_1
	\\
	\gamma'_2	&=	\i \gamma_3 \gamma_4 \mu_1
	\\
	\gamma'_3	&=	-\i \gamma_2 \gamma_4 \mu_1
	\\
	\gamma'_4	&=	\i \gamma_2 \gamma_3 \mu_1
\end{aligned}
&
&\begin{aligned}
	\mu'_1	&=	-\i \gamma_2 \gamma_3 \gamma_4
	\\
	\mu'_2	&=	\mu_2
	\\
	\mu'_3	&=	\mu_3
	\\
	\mu'_4	&=	\mu_4
\end{aligned}
\label{eq:nlctangles}
\end{align}
In the rotated basis, the Hamiltonian at half-filling takes the form
\begin{equation}
	H'	=	-\i\frac{U}{4} \gamma'_1 \mu'_1 + \frac{\epsilon_f}{2} \gamma'_2 \gamma'_3 \gamma'_4 \mu'_2 - \i\frac{\epsilon_g}{2} \mu'_3 \mu'	_4 \,.
\end{equation}
As the $f$ and $g$-sites are not coupled to the original physical degrees of freedom of the Hubbard atom, they can be considered to be ``gauge'' degrees of freedom, meaning that the physics is independent of the parameterization of the $f$ and $g$ orbitals. A convenient gauge choice for the $\epsilon_f$ and $\epsilon_g$ are $\epsilon_f = 0$ and $\epsilon_g = \frac{U}{2}$. This leads to the Hamiltonian taking the form
\begin{equation}
	H' = -\i \frac{U}{4} \left( \gamma'_1 \mu'_1 + \mu'_3 \mu'_4 \right) \,.
\label{eq:dimerinmajoranans}
\end{equation}
In this form, the remaining Majoranas can be recombined into a new set of fermions,
\begin{align}
	a^\dagger &= \frac{\mu'_3 + \i \gamma'_1}{2}
	&
	b^\dagger &= \frac{\mu'_1 + \i \mu'_4}{2} \,,
\end{align}
and the Hamiltonian is then recast into the form
\begin{equation}
	H' = t \left( a^\dagger b + b^\dagger a \right)
\label{eq:nlctha}
\end{equation}
where $\displaystyle t = \frac{U}{2}$. This is the form of a non-interacting dimer. This overall process of transforming the Hubbard atom with gauge fermions to a non-interacting dimer is illustrated in Fig.~\ref{fig:hubbardatom}. 
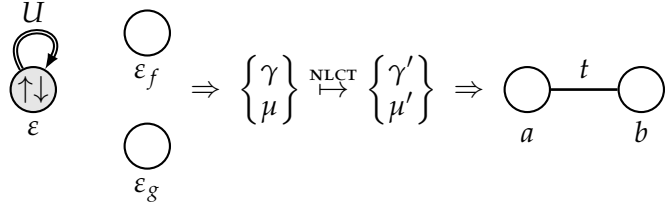
\begin{figure}[htp!]
\centering
\begin{tikzpicture}[every path/.style={in=60,out=120,looseness=6}]
	\node[circle,draw=black,fill=black!10,thick,inner sep=6pt] (H) at (-4,0){};
	\node[circle,draw=black,thick,inner sep=6pt] (a) at (2.5,0){};
	\node[circle,draw=black,thick,inner sep=6pt] (b) at (4,0){};
	\node[circle,draw=black,thick,inner sep=6pt] (f) at (-2.5,0.75){};
	\node[circle,draw=black,thick,inner sep=6pt] (g) at (-2.5,-0.75){};
	\node[below=8pt] at (f) {$\varepsilon_f$};
	\node[below=8pt] at (g) {$\varepsilon_g$};
	\node at (H) {$\uparrow\downarrow$};
	\node[below=8pt] at (H) {$\varepsilon$};
	\node[below=8pt] at (a) {$\phantom{b}a\phantom{b}$};
	\node[below=8pt] at (b) {$b$};
	\path (H) edge[-latex,line width=0.75pt,double distance=0.5pt] node[above] {$U$} (H);
	\node at (0,0) {$\Rightarrow \; \begin{Bmatrix} \gamma \\ \mu \end{Bmatrix} \overset{\textsc{nlct}}{\mapsto} \begin{Bmatrix} \gamma' \\ \mu' \end{Bmatrix} \;\Rightarrow$};
	\draw[-,line width=1pt] (a)--(b) node[midway,above] {$t$};
\end{tikzpicture}
\caption[Non-linear canonical transformation of the interacting Hubbard atom to a non-interacting dimer using gauge degrees of freedom]{Non-linear canonical transformation of the interacting Hubbard atom to a non-interacting dimer using gauge degrees of freedom ($f$ and $g$).\label{fig:hubbardatom}}
\end{figure}
It can be shown that the Green functions for the original physical system and the auxiliary system coincide:
\begin{equation}
	\Green{\op{c}{\uparrow}}{\opd{c}{\uparrow}}_z = \Green{\op{a}{}}{\opd{a}{}}_z \,.
\end{equation}

The above calculation was performed on the Hubbard atom with particle-hole symmetry at half-filling at zero temperature. It is possible to extend this calculation to more general cases, although the necessary algebraic manipulations becomes much more involved. For the particle-hole symmetric case, the choice of angles for the rotors in Eq.~\eqref{eq:nlctangles} could be chosen by inspection of the desired form of the final Hamiltonian \eqref{eq:dimerinmajoranans}. For the particle-hole asymmetric case, the choice of angles in principle involves a system of coupled equations for the eight rotor angles. A solution to this set of equations which transforms the particle-hole asymmetric Hubbard atom to a non-interacting system remains to be found.

The preceding discussion shows how the concept of decomposing fermions into Majorana\index{Majorana} degrees of freedom can be generalized into a more elaborate scheme. 
Within this scheme, nontrivial and non-linear canonical transformations may be applied which manifest as rotations in the Clifford algebra space spanned by the Majorana degrees of freedom, generated by the generalized Euler identity~\eqref{eq:generaleuler}. The generalized Euler identity opens up the possibility of even more elaborate transformations than those considered in the previous discussion. Following the example in Eq.~\eqref{eq:majoranapolynomial}, it is possible to find a representation of the rotation kernel $\mathcal{I}$ which is a Majorana polynomial, such as $\mathcal{I} = \frac{\gamma_3 \gamma_4 \gamma_5 - \i \gamma_6}{\sqrt{2}}$. 

An additional potential aspect which could be generalized is the internal metric within the Majorana algebra. The metric appears in the anticommutation relation $\{ \gamma_a , \gamma_b \} = 2 g_{ab}$ where $g$ is the metric tensor which is usually taken to be Euclidean. This could be modified to be a Riemannian metric. It remains to be shown whether such a scheme would preserve the appropriate Hermitian fermion statistics, or if this generalization would only be applicable to non-Hermitian systems.

A potential continuation of exploiting the Clifford algebraic properties obeyed by the Majorana degrees of freedom is to develop a Majorana calculus constructed along the lines of geometric calculus~\cite{hestenessobczyk,doranlasenby}, where the existence of a geometric object which functions algebraically as the imaginary unit $\sqrt{-1}$ facilitates a generalization of complex analysis to higher dimensions. Within the framework of geometric calculus the notion of a complex number $z = x + \i y$ is generalized to $\mathcal{Z} = x + \mathcal{I} y$ where as in the above $\mathcal{I}$ has some internal structure with a geometric interpretation and has the property that $\mathcal{I}^2 = -1$. The standard complex analysis in geometric calculus can be reproduced using the Clifford algebra space $\text{Cl}_{2}(\mathbbm{R})$ with the ``imaginary'' unit being the unit pseudoscalar $\mathcal{I} = e_1 \extp e_2$. A first generalization which goes beyond this is in the $3d$ case where $\mathcal{I} = e_1 \extp e_2 \extp e_3$. It is in this sense that geometric calculus can extend complex analysis to higher dimensions. Conceptually, this calculus could be interpreted as a variant of the Gra{\ss}mann calculus used in the calculation of fermionic path integrals, although it is not immediately obvious what a physically meaningful integrand might be.

This non-linear canonical transformation is an analytic transformation which transforms interacting degrees of freedom to non-interacting degrees of freedom. This transformation however suffers from algebraic complexity.
As such, it is unwieldy for practical purposes, but serves as an illustration showing that formally an analytic canonical transformation from interacting degrees of freedom to non-interacting degrees of freedom does exist.

%%%%%%%%%%%%%%%%%%%%%%%%%%%%%%%%%%%%%%%%%%%%%%%%%%%%%%%%%%%%%%%%%%%%%%%%%%%%%%%
%%%%%%%%%%%%%%%%%%%%%%%%%%%%%%%%%%%%%%%%%%%%%%%%%%%%%%%%%%%%%%%%%%%%%%%%%%%%%%%
%%%%%%%%%%%%%%%%%%%%%%%%%%%%%%%%%%%%%%%%%%%%%%%%%%%%%%%%%%%%%%%%%%%%%%%%%%%%%%%

\section{Auxiliary Chain Mapping\label{sec:finitemapping}}

The preceding framework presented a method of converting an interacting system to a non-interacting one by means of non-linear canonical transformations and the introduction of gauge degrees of freedom. For simple cases such as the Hubbard atom with particle-hole symmetry an analytic transformation is derivable. However for more general cases, even such as the Hubbard atom without particle-hole symmetry, the method quickly becomes drastically more complicated to implement and at this stage of the method's development serves more as an illustrative proof-of-concept rather than a generally applicable tool to deliver quantitative results.

Presented now in this section is a mapping scheme which also transforms an interacting system to a non-interacting one which can be applied to a much wider range of systems. This method is similar in spirit to the previous scheme, but it is founded on a different technical basis, which is that of Green functions.

This section builds up the formalism of constructing auxiliary systems starting from the simplest cases with finite degrees of freedom and treating systems of gradually increasing complexity, including finite temperature effects, before then describing the method for treating systems of infinite degrees of freedom in the thermodynamic limit.

The form of the effective models which are constructed here are that of $1d$ tight-binding chains with nearest-neighbor dynamics. This form is chosen for its computational simplicity, but in principle the mapping could be performed to other prescribed systems. The auxiliary mapping is not itself a solution strategy for solving strongly correlated systems, but rather enables a reinterpretation of their solutions. A prerequisite for constructing the auxiliary model is a solution to the original physical system. However, the solutions derived below may serve as a starting point for approximate toy-model solutions to certain interacting systems.

\subsection{Exact Mappings for Finite Systems}

The auxiliary chain effective models are first constructed here for finite systems of modest size which can be solved using exact diagonalization.
For a system of finite size, its spectrum consists of a finite number of discrete poles. In terms of the effective chain model, this corresponds to a Green function described by a continued fraction of finite depth.

\subsubsection{Hubbard Atom}

The simplest example of an application of this mapping is the Hubbard model in the atomic limit, $U\to\infty$. This system reduces to a set of isolated Hubbard atoms, each of which consists of a single site of interacting fermions with the Hamiltonian given by
\begin{equation}
	\hat{H}_{\textsc{h.a.}} = \varepsilon \left( \opd{c}{\uparrow} \op{c}{\uparrow} + \opd{c}{\downarrow} \op{c}{\downarrow} \right) + U \opd{c}{\uparrow} \op{c}{\uparrow} \opd{c}{\downarrow} \op{c}{\downarrow} \,.
\end{equation}
As the sites are decoupled, each site can be analyzed independently.
As calculated by the Green function equation-of-motion approach in \S\ref{sec:hubbardatomgf}, the Green function for the Hubbard atom is
\begin{align}
	\Green{\op{c}{\sigma}}{\opd{c}{\sigma}}_z
%	&=	\frac{1}{z - \varepsilon} + \frac{U}{z-\varepsilon} \frac{\langle \op{n}{-\sigma} \rangle}{z - \varepsilon - U}
%	\\
	&=
	\frac{1 - \langle \op{n}{-\sigma} \rangle}{z - \varepsilon} + \frac{\langle \op{n}{-\sigma} \rangle}{z - \varepsilon - U} \,.
\intertext{This Green function can be written in continued fraction form as}
	\Green{\op{c}{\sigma}}{\opd{c}{\sigma}}_z
	&=
	\cfrac{1}{z - \varepsilon - \langle \op{n}{-\sigma} \rangle U - \cfrac{( 1 - \langle \op{n}{-\sigma} \rangle ) \langle \op{n}{-\sigma} \rangle U^2}{z - \varepsilon - ( 1 - \langle \op{n}{-\sigma} \rangle ) U}} \,. \label{eq:hacontfrac}
\end{align}
This Green function can be reinterpreted as the Green function for an auxiliary system as
\begin{equation}
	\Green{\op{a}{1}}{\opd{a}{1}}_z
	=
	\cfrac{1}{z - \epsilon_1 - \cfrac{\tilde{V}^2}{z - \epsilon_2}}
\end{equation}
which is clearly the form of a tight binding dimer with on-site potentials $\epsilon_1$ and $\epsilon_2$, and the hopping amplitude between the two sites $\tilde{V}$.
In terms of the degrees of freedom of the original physical system, the parameters of the effective system are
\begin{align}
	\epsilon_1 &= \varepsilon + \langle \op{n}{-\sigma} \rangle U \,,&
	\epsilon_2 &= \varepsilon + ( 1 - \langle \op{n}{-\sigma} \rangle ) U
	\,,&
	\tilde{V}^2 &= ( 1 - \langle \op{n}{-\sigma} \rangle ) \langle \op{n}{-\sigma} \rangle U^2
	\,,
%\\
\intertext{or in the particle-hole symmetric case for $-U < \varepsilon < 0$ with $T \ll U, |\varepsilon|$ where $\langle \op{n}{-\sigma} \rangle = \frac12 = \langle \op{n}{+\sigma} \rangle$, the parameters are}
	\epsilon_1 &= -\frac{U}{2} \,,&
	\epsilon_2 &= -\frac{U}{2} \,,
	&
	\tilde{V} &= \frac{U}{2} \,.
\end{align}
As in \S\ref{nlctha} (\textit{cf.} the calculation leading to Eq.~\eqref{eq:nlctha}), his shows again that there exists a formal mapping from the single-site interacting Hubbard atom to a non-interacting tight-binding dimer. This mapping is depicted schematically in Fig.~\ref{fig:haauxmap}.
Following again the discussion in \S\ref{sec:hubbardatomgf}, in more general circumstances the expectation value of the fermion number is given by
\begin{equation}
	\langle \op{n}{-\sigma} \rangle = \int \d\omega f(\omega) \mathcal{A}_\sigma(\omega)
\end{equation}
where 
$f(\omega)$ is the Fermi-Dirac distribution. With $-U < \varepsilon < 0$, this leads the expectation value of the number operator to be
\begin{equation}
\begin{aligned}[b]
	\langle \op{n}{\pm\sigma} \rangle
	&=	f(\varepsilon) (1-\langle \op{n}{mp\sigma} \rangle) + f(\varepsilon+U) \langle \op{n}{mp\sigma} \rangle
	\\
	\langle \op{n}{\pm\sigma} \rangle
	&=	\frac{f(\varepsilon)}{1-f(\varepsilon+U)+f(\varepsilon)} \,.
\end{aligned}
\end{equation}
This shows that the functional form of the auxiliary dimer remains the same at finite temperature, but the exact values of the parameters change.
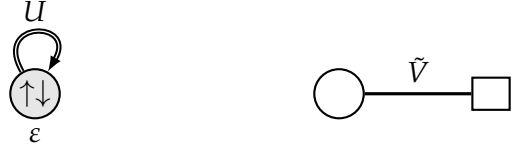
\begin{figure}[htp!]
\centering
\begin{tikzpicture}[every path/.style={in=60,out=120,looseness=6}]
	\node[circle,draw=black,fill=black!10,thick,inner sep=1pt] (h) at (-2,0){$\uparrow\downarrow$};
	\node[circle,draw=black,thick,inner sep=1pt] (a1) at (2,0){$\phantom{\uparrow\downarrow}$};
	\node[rectangle,draw=black,thick,inner sep=1pt] (a2) at (4,0){$\phantom{\uparrow\downarrow}$};
	\node[below=9pt] at (h) {$\varepsilon$};
	\draw[line width=1.2pt](a1)--(a2) node[midway,above] {$\tilde{V}$};
	\path (h) edge[-latex,line width=0.75pt,double distance=0.5pt] node[above] {$U$} (h);
\end{tikzpicture}
\caption{Mapping of the Hubbard atom (left) to a non-interacting auxiliary system (right).\label{fig:haauxmap}}
\end{figure}
While the continued fraction form of the Hubbard atom Green function \eqref{eq:hacontfrac} is well-known in the literature, its equivalent interpretation as the Green function of a non-interacting dimer is new.
%\index{$0$@\textbf{List of Edits}!502@highlighted new contribution}

\subsubsection{Anderson Dimer}
The next simplest system to consider is that of the Anderson dimer, a model which consists of a single interacting site hybridized to a non-interacting bath comprising a single lattice site. For simplicity, the analysis will first be performed at particle-hole symmetry at zero temperature. The Anderson dimer at finite temperature case will be analyzed in a following subsection, as well as Anderson models with larger bath sizes. The Hamiltonian of the Anderson dimer at half-filling, $\varepsilon_d = -U/2$, is given by
\begin{equation}
%	\hat{H} = -\frac{U}{2} \left( \opd{d}{\uparrow} \op{d}{\uparrow} + \opd{d}{\downarrow} \op{d}{\downarrow} \right) + U \opd{d}{\uparrow} \op{d}{\uparrow} \opd{d}{\downarrow} \op{d}{\downarrow} + V \sum_{\sigma} \left( \opd{d}{\sigma} \op{c}{\sigma} + \opd{c}{\sigma} \op{d}{\sigma} \right)
	\op{H}{\text{dimer}} = U \left( \opd{d}{\uparrow} \op{d}{\uparrow} - \tfrac12 \right) \left( \opd{d}{\downarrow} \op{d}{\downarrow} - \tfrac12 \right) + V \smashoperator{\sum_{\sigma\in\{\uparrow,\downarrow\}}}\ \left( \opd{d}{\sigma} \op{c}{\sigma} + \opd{c}{\sigma} \op{d}{\sigma} \right)
\end{equation}
where the $\op{d}{}$ operators act on the impurity site and the $\op{c}{}$ operators on the bath.
\begin{figure}[htp!]
\centering
\begin{subfigure}[b]{0.45\linewidth}
\centering
\begin{tikzpicture}[every path/.style={in=60,out=120,looseness=6},scale=0.75]
	\node[circle,draw=black,fill=black!10,thick,inner sep=1pt] (3) at (0,0){$\uparrow\downarrow$};
	\node[circle,draw=black,thick,inner sep=1pt] (4) at (2,0){$\phantom{\uparrow\downarrow}$};
	\node[below=9pt] at (3) {$\varepsilon_d$};
	\draw[line width=1.2pt](3)--(4) node[midway,above] {$V$};
	\path (3) edge[-latex,line width=0.75pt,double distance=0.5pt] node[above] {$U$} (3);
	\node[draw=none,thick,rectangle,inner sep=6pt] (a1) at (0,-1.5) {};
	\node[draw=none,thick,rectangle,inner sep=6pt] (a2) at (0,-3) {};
	\draw[draw=none,line width=1.2pt](3)--(a1) node[midway,left] {};
	\draw[draw=none,line width=1.2pt](a1)--(a2) node[midway,left] {};
	\node at ($(4)+(2,1)$) {\subref*{fig:andersondimerschematic}};
\end{tikzpicture}
\phantomsubcaption{\label{fig:andersondimerschematic}}
\end{subfigure}
\begin{subfigure}[t]{0.45\linewidth}
\centering
\begin{tikzpicture}[every path/.style={in=60,out=120,looseness=6},scale=0.75]
	\node[circle,draw=black,thick,inner sep=1pt] (3) at (0,0){$\phantom{\uparrow\downarrow}$};
	\node[circle,draw=black,thick,inner sep=1pt] (4) at (2,0){$\phantom{\uparrow\downarrow}$};
	\draw[line width=1.2pt](3)--(4) node[midway,above] {$V$};
	\node[draw=black,thick,rectangle,inner sep=6pt] (a1) at (0,-1.5) {};
	\node[draw=black,thick,rectangle,inner sep=6pt] (a2) at (0,-3) {};
	\draw[line width=1.2pt](3)--(a1) node[midway,left] {$\tilde{V}$};
	\draw[line width=1.2pt](a1)--(a2) node[midway,left] {$\tilde{t}_1$};
	\node at ($(4)+(2,1)$) {\subref*{fig:dimerauxlatt}};
\end{tikzpicture}
\phantomsubcaption{\label{fig:dimerauxlatt}}
\end{subfigure}
\caption{Mapping of the Anderson dimer \subref{fig:andersondimerschematic} to a non-interacting auxiliary system \subref{fig:dimerauxlatt}.}
\end{figure}
The ground state of the Anderson dimer has quantum numbers $Q=2$ and $S_z=0$ with Hamiltonian
\begin{equation}
	\boldsymbol{H}_{2,0} = \begin{pmatrix*}[r] -\frac{U}{2} & 0 & V & V \\ 0 & -\frac{U}{2} & -V & -V \\ V & -V & 0 & 0 \\ V & -V & 0 & 0 \end{pmatrix*}
\end{equation}
in terms of the basis
\begin{equation}
	\left\lvert \phi \right\rangle_{2,0}
	=
	\begin{pmatrix}
	\lvert \uparrow , \downarrow \rangle
	\\
	\lvert \downarrow , \uparrow \rangle
	\\
	\lvert \uparrow\!\downarrow , - \rangle
	\\
	\lvert - , \uparrow\!\downarrow \rangle
	\end{pmatrix} \,.
\end{equation}
The Schr\"odinger equation for the Hamiltonian of this subspace is
\begin{equation}
	\boldsymbol{H}_{2,0} \left\lvert 2 , 0 ; j \right\rangle = E^j_{2,0} \left\lvert 2 , 0 ; j \right\rangle
\end{equation}
with eigenenergies $E^j_{2,0}$ and eigenstates $\left\lvert 2 , 0 ; j \right\rangle$ with $j\in\{1,\ldots,4\}$. The ground state energy of this subspace is $E^1_{2,0} = -\frac14 ( U + \sqrt{U^2 + 64 V^2} )$.

The system also allows excited states with quantum numbers $Q=3$ and $S_z=1/2$ where the Hamiltonian in this subspace is
\begin{equation}
	\boldsymbol{H}_{3,\frac12} = \begin{pmatrix*}[r] 0 & -V \\ -V & -\frac{U}{2} \end{pmatrix*}
\end{equation}
in terms of the basis
\begin{equation}
	\left\lvert \phi \right\rangle_{3,\frac12}
	=
	\begin{pmatrix}
	\lvert \uparrow\!\downarrow , \uparrow \rangle
	\\
	\lvert \uparrow , \uparrow\!\downarrow \rangle
	\end{pmatrix} \,.
\end{equation}
The Schr\"odinger equation for this subspace is
\begin{equation}
	\boldsymbol{H}_{3,\frac12} \left\lvert 3, \tfrac12 ; j \right\rangle = E^j_{3,\frac12} \left\lvert 3, \tfrac12 ; j \right\rangle
\end{equation}
with eigenenergies $E^j_{3,\frac12}$ and eigenstates $\left\lvert 3, \tfrac12 ; j \right\rangle$ with $j\in\{1,2\}$. The eigenenergies of this subspace are $E^1_{3,\frac12} = -\frac14 ( U + \sqrt{U^2 + 16 V^2} )$ and $E^2_{3,\frac12} = -\frac14 ( U - \sqrt{U^2 + 16 V^2} )$.
%where
The Green function on the impurity can be obtained from the Lehmann representation\index{Lehmann representation} as
\begin{equation}
	G(z) \equiv
	\Green{\op{d}{\uparrow}}{\opd{d}{\uparrow}}_z
	=
	\sum_{j=1}^{2} \left\lvert \left\langle 3,\tfrac12 ; j \left\lvert \opd{d}{\uparrow} \right\rvert 2,0 ; 1 \right\rangle \right\rvert^2 \left[ \frac{1}{z - E_{2,0}^{1} + E_{3,\frac12}^{j}} + \frac{1}{z + E_{2,0}^{1} - E_{3,\frac12}^{j}} \right] \,.
\label{eq:dimergreenfunction}
\end{equation}
Accounting for degeneracies, the $T=0$ Green function obtained by Eq.~\eqref{eq:dimergreenfunction} consists of four distinct poles. This implies that the interacting two-site system can be mapped onto a non-interacting four-site system.
The parameters of the effective chain can be obtained from the finite poles of the Lanczos representation by applying the Lanczos algorithm described in \S\ref{sec:lanczos}. The Lehmann representation of the Green function delivers the set of weights $\{ w_p \}$  and positions $\{ \omega_p \}$ of the spectral poles with the spectral function taking the form of $\mathcal{A}(\omega) = \sum_{j} w_j \delta(\omega-\omega_j)$. These define the form of the Hamiltonian in a diagonal representation $\boldsymbol{H}_{\text{D}}$. This Hamiltonian is then processed by Lanczos algorithm to produce the corresponding Hamiltonian in tridiagonal form $\boldsymbol{H}_{\text{T}}$, which defines the set of parameters $\{ \tilde{V} , \tilde{t}_1 \}$ from the off-diagonal terms. This set parameterizes the system as a tight-binding system for the configuration shown in Fig.~\ref{fig:dimerauxlatt} with Hamiltonian
\begin{equation}
	\op{\widetilde{H}}{\text{dimer}} = \sum_{\sigma\in\{\uparrow,\downarrow\}} \left[ V \left( \opd{d}{\sigma} \op{c}{\sigma} + \opd{c}{\sigma} \op{d}{\sigma} \right) + \tilde{V} \left( \opd{d}{\sigma} \op{f}{\sigma;1} + \opd{f}{\sigma;1} \op{d}{\sigma} \right) + \tilde{t} \left( \opd{f}{\sigma;1} \op{f}{\sigma;2} + \opd{f}{\sigma;1} \op{f}{\sigma;2} \right) \right]
\end{equation}
where the operators $\op{f}{\sigma;n}$ act on the auxiliary sites (square sites in Fig.~\ref{fig:dimerauxlatt}). Note that there are technically two copies of the auxiliary lattice Fig.~\ref{fig:dimerauxlatt}, one for each spin. Since there is no coupling between different spins in the auxiliary model, each spin sector can be analyzed independently.
The Green function on the $d$-site of the auxiliary model takes the form
\begin{equation}
	\widetilde{G}(z) \equiv \Green{\op{d}{\sigma}}{\opd{d}{\sigma}}_z = \cfrac{1}{z - \cfrac{\tilde{V}^2}{z - \cfrac{\tilde{t}_1^2}{z}} - \cfrac{V^2}{z}} \,,
\label{eq:auxdimergreenfunction}
\end{equation}
such that the spectral function corresponds to the original physical system:
\begin{equation}
	-\frac1\pi \Im \widetilde{G}(z) \overset{!}{=} -\frac1\pi \Im G(z) \,.
\end{equation}
%with $\widetilde{G}(z) = \Green{\op{d}{\uparrow}}{\opd{d}{\uparrow}}_z$.
Given Eq.~\eqref{eq:auxdimergreenfunction} and Eq.~\eqref{eq:dimergreenfunction}, this is an algebraic expression which relates the auxiliary parameters $\tilde{V}$ and $\tilde{t}$ to the physical parameters $U$ and $V$.
Performing the analysis,\footnote{While it may appear that this is a single algebraic equation for two unknowns, the fractions involved can be rationalized, where the numerators of the rationalized fractions are polynomials in $z$. Matching coefficients for different powers of $z$ yields a set of algebraic equations which can be simultaneously solved for $\tilde{V}$ and $\tilde{t}$ by making use of the fact that monomials of different powers are linearly independent.} it is found that $\tilde{V} = U/2$ and $\tilde{t}_1 = 3 V$.

\subsection{Anderson $\boldsymbol{N}$-mers}
This mapping to an auxiliary system can naturally be generalized to treat $1d$ Anderson models of longer finite length, or Anderson $N$-mers, where $N$ being the length of the non-interacting $1d$ bath.\footnote{The nomenclature is more properly ``$(N+1)$-mer'' as for $N=1$ the system is described as being the Anderson \textit{di}mer and so on. The nomenclature is chosen to preserve the convention that the end of the chain is an impurity coupled to a bath via a hybridization $V$ and the bath parameterized by $t_n$ with $n=1,2,\ldots,N$.}
The setup is an interacting impurity with $U=1$ and $\varepsilon=-U/2$, coupled to a chain of $N=1,3,5,7,9$ bath sites. As a concrete example, the hybridization between impurity and bath is $V=0.2$ and the hopping between the other bath sites is uniformly $t=0.4$. The Green functions for these systems are evaluated at $T=0$. The spectrum is obtained as in the preceding dimer example by means of exact diagonalization to construct the Lehmann representation of the Green function.
%I only do odd N because the ground state is then a spin singlet and you don't have to worry about quantum number degeneracies.
%
%I have provided the numerical output for the aux-chain parameters, which couple on the other side of the impurity. At ph symmetry, it is worth noting that the first coupling is *always* U/2. This is to do with the normalization of the self-energy (although it is not obvious how that comes about via the Lancoz method!).
%Then the original (pole-broadened) spectra for the impurity are A_imp_real_N... and the spectra for the mapped system are A_imp_map_N...
%
%Now: there's a caveat. For the chains N >=3, you get a lot of poles and the aux chain is rather long. However, I have truncated all the poles of very small pole weight <1e-6. This then yields essentially the same aux chain and output spectrum but with a modest number of aux sites. You can see this because the aux chain parameters at some point just fall to zero.
%But, this truncation approximation begins to break down for longer chains because there is a proliferation of many many poles with slightly different energies of very small weights. Of course ultimately these are forming the continuum spectrum in the thermodynamic limit, where the aux chain is infinitely long. Anyway, even truncating at 1e-6 as here introduces small but noticeable errors at N=7 and larger errors near the band edges for N=9 as you'll see.

$N$-mers of odd $N$ are taken such that the ground state is a singlet. For even $N$ the ground state is a doublet which then leads to more complicated transitions between the ground state and the excited states due to quantum number degeneracies. It is not necessary to entertain this complication to obtain results which appropriately illustrate the construction of the auxiliary chain for $N$-mers of various lengths (the results are qualitatively similar).

As a precursor to mapping to the auxiliary system, the spectrum is truncated by discarding all elements of the Lehmann sum whose weight is $< 10^{-6}$. For auxiliary chains associated to $N$-mers of length $N \geq 3$, the auxiliary systems become very long. The truncation cutoff limits the length of the auxiliary chain to a modest number of sites while still recovering the primary features of the desired spectrum. Lowering the truncation threshold significantly increases the dimension of the auxiliary system Hilbert space but the increase in accuracy of reproducing the desired spectrum is minimal. For few bath sites, the parameters of the constructed auxiliary system are not meaningfully affected by the truncation. The reconstructed spectrum from the auxiliary chain obtained from the truncated spectrum is nearly identical to the original exact spectrum. However, this truncation begins to produce errors as the number of bath sites is increased. In the spectrum of the larger $N$-mers there is a proliferation of very many poles with slightly different energies with very small weights. Ultimately these are forming the continuum spectrum in the thermodynamic limit, where the auxiliary chain is infinitely long. The construction of an auxiliary system for an impurity model in the thermodynamic limit will be discussed in the following section.
With the truncation, on systems of a larger number of bath sites the reconstructed spectrum develops deviations from the true spectrum near the outer band edges of the spectrum. These small errors can be seen in the spectral plots for $N=7$ and $N=9$ in Fig.~\ref{fig:Nmers}. These errors are analogous to the errors encountered when modeling a continuum, or system of infinite extent, with a finite number of degrees of freedom.

The Lehmann representation of the spectral functions yields a form of the Hamiltonian in the diagonal basis $\boldsymbol{H}_{\text{D}}$.
The auxiliary chain parameters are obtained by transforming the Hamiltonian into a tridiagonal basis, which results in a Green function of continued fraction form.
The Hamiltonian in its tridiagonal basis is obtained from the Lanczos algorithm as described in \S\ref{sec:lanczos}. The tridiagonal Hamiltonian produced by the Lanczos algorithm consists of a set of parameters $\{ \tilde{V} , \tilde{t}_n \}$ where $n = 1,\ldots,N$ ranges over the number of bath sites in the original $N$-mer.
\begin{figure}[htp!]
\centering
\begin{tikzpicture}[every path/.style={in=60,out=120,looseness=6},scale=0.75]
	\node[circle,draw=black,fill=black!10,thick,inner sep=1pt] (0) at (0,0){$\uparrow\downarrow$};
	\node[circle,draw=black,thick,inner sep=1pt] (1) at (2,0){$\phantom{\uparrow\downarrow}$};
	\node[circle,draw=black,thick,inner sep=1pt] (2) at (4,0){$\phantom{\uparrow\downarrow}$};
	\node[circle,inner sep=3pt] (int) at (6,0){};
	\node[circle,draw=black,thick,inner sep=1pt] (N) at (8,0){$\phantom{\uparrow\downarrow}$};
	\node[below=9pt] at (0) {$\varepsilon$};
	\draw[line width=1.2pt](0)--(1) node[midway,above] {$V$};
	\draw[line width=1.2pt](1)--(2) node[midway,above] {$t$};
	\draw[line width=1.2pt](2)--($(2)+(1,0)$) node[above] {$t$};
	\draw[line width=1.2pt,dashed]($(2)+(1,0)$)--(int);
	\draw[line width=1.2pt](N)--($(N)-(1,0)$) node[above] {$t$};
	\draw[line width=1.2pt,dashed](int)--($(N)-(1,0)$);
	\path (0) edge[-latex,line width=0.75pt,double distance=0.5pt] node[above] {$U$} (0);
\end{tikzpicture}
%\caption{Schematic of the Hubbard model in $1d$.\label{fig:hubbard1d}}
%\end{figure}
\\
%\begin{figure}[htp!]
%\centering
\begin{tikzpicture}[every path/.style={in=60,out=120,looseness=6},scale=0.75]
	\node[circle,draw=black,thick,inner sep=1pt] (0) at (0,0){$\phantom{\uparrow\downarrow}$};
	\node[circle,draw=black,thick,inner sep=1pt] (1) at (2,0){$\phantom{\uparrow\downarrow}$};
	\node[circle,draw=black,thick,inner sep=1pt] (2) at (4,0){$\phantom{\uparrow\downarrow}$};
	\node[circle,inner sep=3pt] (int) at (6,0){};
	\node[circle,draw=black,thick,inner sep=1pt] (N) at (8,0){$\phantom{\uparrow\downarrow}$};
	\draw[line width=1.2pt](0)--(1) node[midway,above] {$V$};
	\draw[line width=1.2pt](1)--(2) node[midway,above] {$t$};
	\draw[line width=1.2pt](2)--($(2)+(1,0)$) node[above] {$t$};
	\draw[line width=1.2pt,dashed]($(2)+(1,0)$)--(int);
	\draw[line width=1.2pt](N)--($(N)-(1,0)$) node[above] {$t$};
	\draw[line width=1.2pt,dashed](int)--($(N)-(1,0)$);
	\node[draw=black,thick,rectangle,inner sep=6pt] (a1) at (0,-1.5) {};
	\node[draw=black,thick,rectangle,inner sep=6pt] (a2) at (0,-3) {};
	\node[inner sep=6pt] (ai) at (0,-4.5) {};
	\node[draw=black,thick,rectangle,inner sep=6pt] (an) at (0,-6) {};
	\draw[line width=1.2pt](0)--(a1) node[midway,left] {$\tilde{V}$};
	\draw[line width=1.2pt](a1)--(a2) node[midway,left] {$\tilde{t}_1$};
	\draw[line width=1.2pt,dashed](ai)--($(a2)-(0,0.75)$);
	\draw[line width=1.2pt](a2)--($(a2)-(0,0.75)$) node[left] {$\tilde{t}_2$};
	\draw[line width=1.2pt,dashed](ai)--($(an)+(0,0.75)$);
	\draw[line width=1.2pt](an)--($(an)+(0,0.75)$) node[left] {$\tilde{t}_N$};
\end{tikzpicture}
\caption[Schematic illustrating the auxiliary field mapping from an interacting Anderson $N$-mer to a non-interacting system]{Schematic illustrating the auxiliary field mapping from an interacting Anderson $N$-mer (top) to a non-interacting system hybridized to an auxiliary system (bottom).\label{fig:Nmerschematic}}
\end{figure}
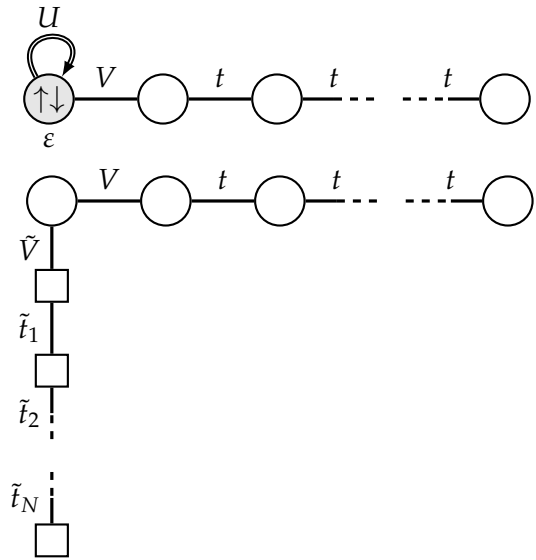
\begin{figure}[htp!]
%\centering
	\includegraphics{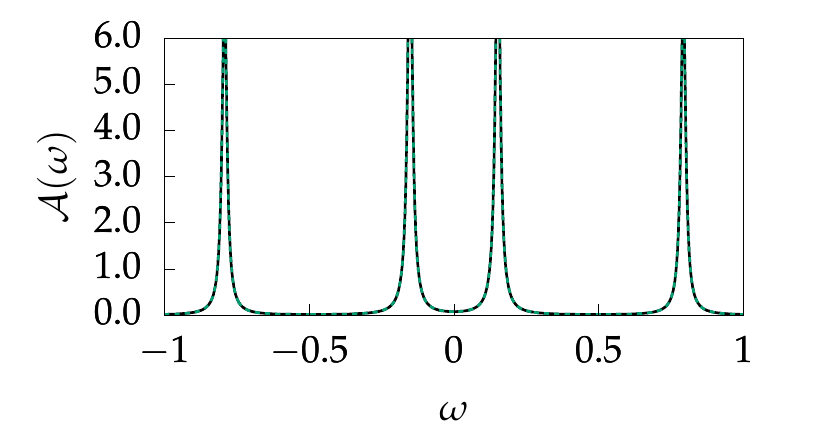}
	\includegraphics{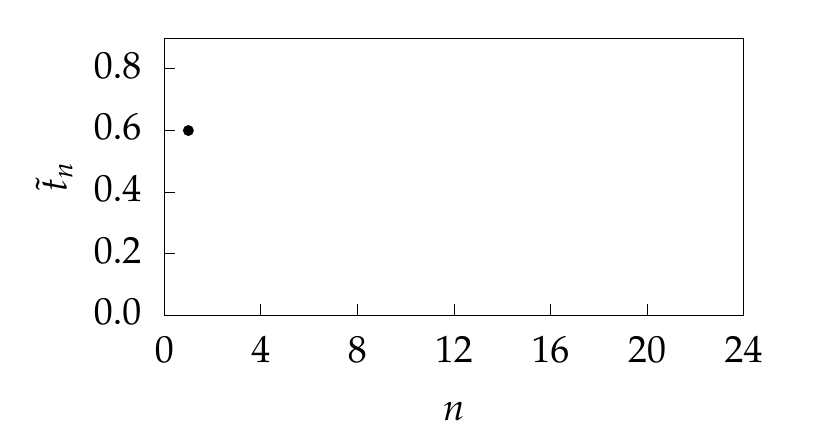}
	\includegraphics{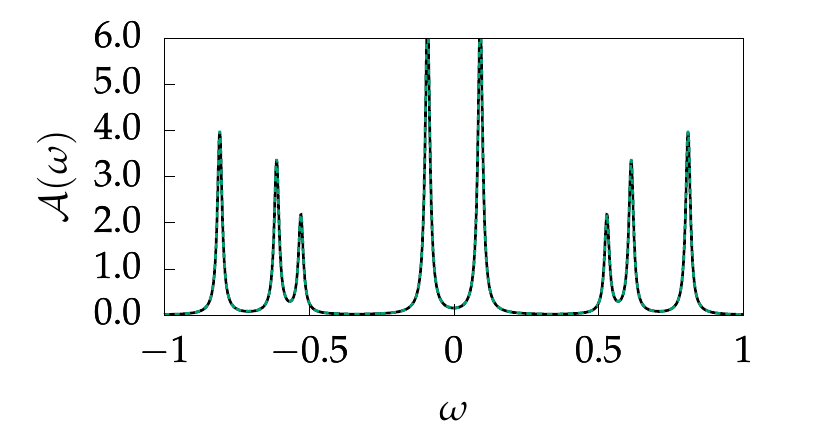}
	\includegraphics{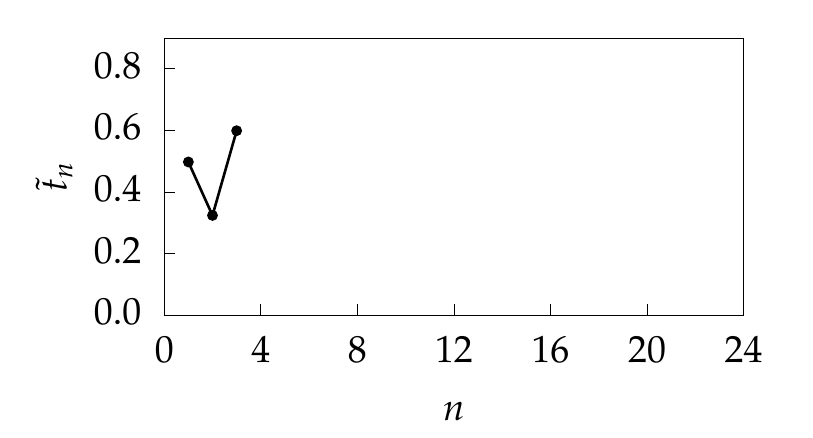}
	\includegraphics{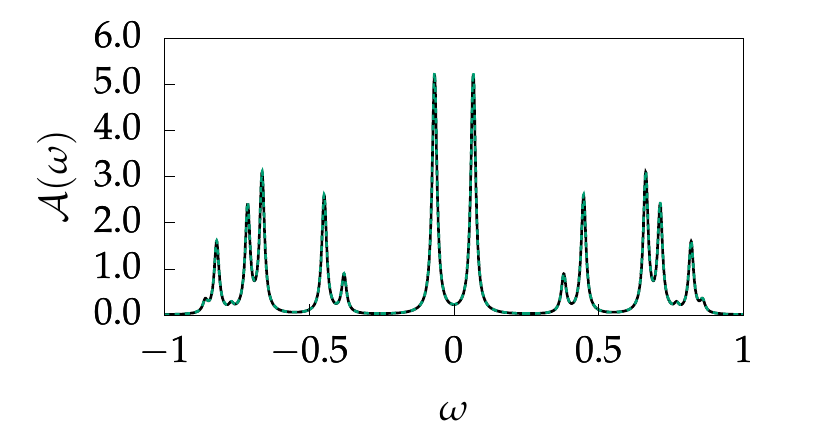}
	\includegraphics{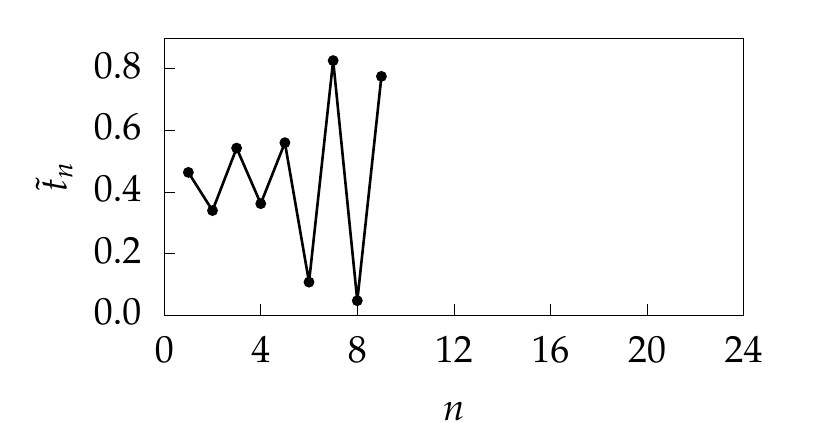}
	\includegraphics{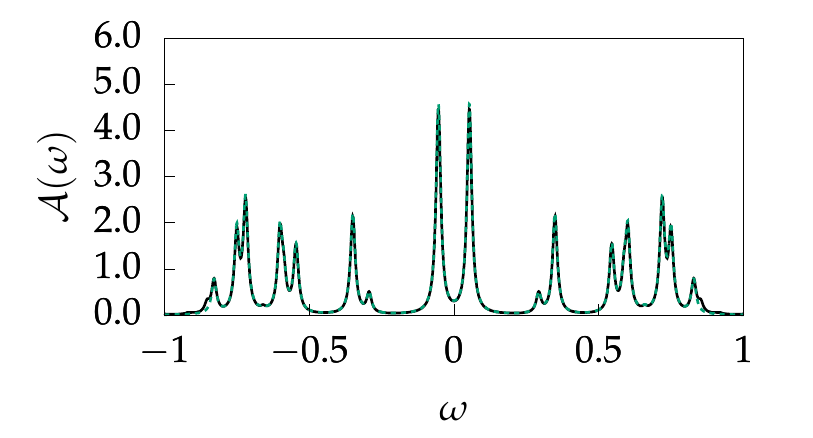}
	\includegraphics{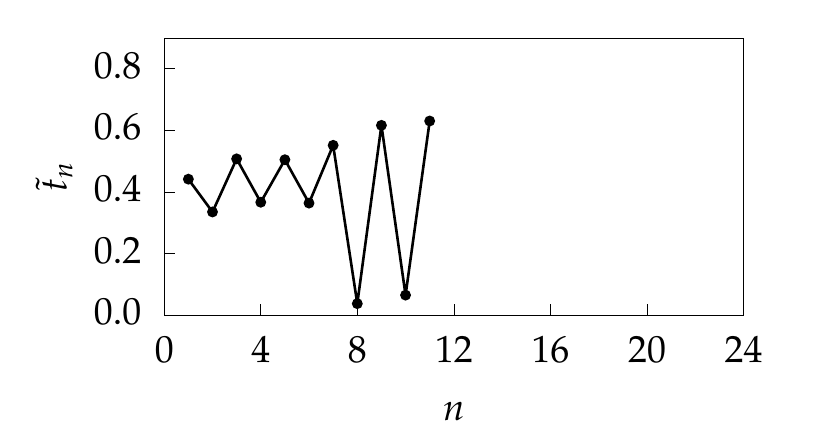}
	\includegraphics{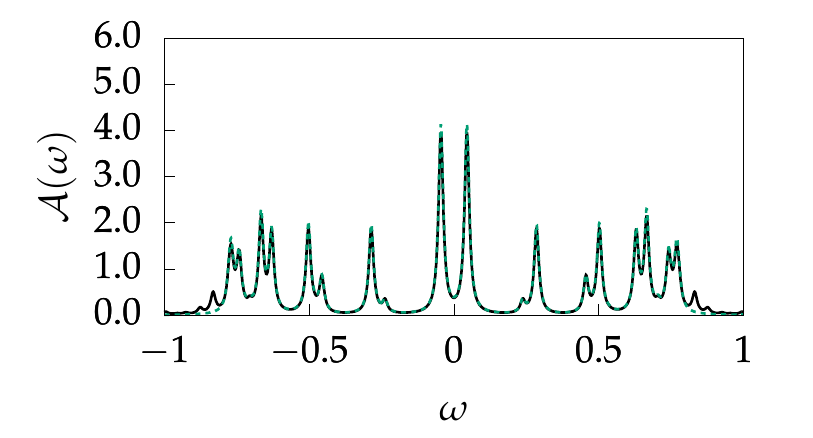}
	\includegraphics{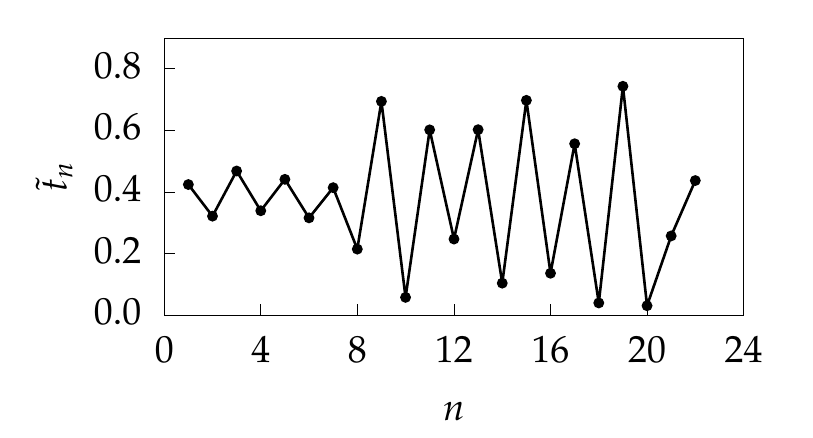}
\caption[Spectral functions and auxiliary chain hopping parameters for Anderson $N$-mers]{Spectral functions and auxiliary chain hopping parameters for Anderson $N$-mers with $N = 1, 3, 5, 7, 9$ at $T=0$. The true spectral function is plotted in solid lines and the reconstructed spectral function from the auxiliary chain is plotted in dashed teal lines.\label{fig:Nmers}}
\end{figure}
The auxiliary chains calculated for $N$-mers of various lengths are shown in Fig.~\ref{fig:Nmers}. The right-hand panels show the values of the parameters $\{ \tilde{t}_n \}$ and the left-hand panels show the reconstructed spectrum $\mathcal{A}(\omega) = -\frac1\pi\Im\widetilde{G}(\omega)$ and their comparison to the original spectrum as calculated exactly from the exact diagonalization of the Anderson model via the Lehmann sum.

Note that $\tilde{V} = U/2$ in all cases. For $\tilde{t}_n$ with $n>2$, the chain parameters develop complexity, embodying the spectral complexity.

\subsubsection{Finite Temperature}

At finite temperature there exist transitions between excited states, which increase the multiplicity of finite matrix elements, and hence poles, in the Green function. This then results in an auxiliary chain of greater number of sites than for the zero temperature case. The system considered in this analysis is the Anderson dimer, although the results generalize to larger systems. Unlike the case for the Hubbard atom, the effective model for the Anderson dimer at finite temperature is not simply a reparameterization of the effective model for the zero temperature case.

In the following calculations the Anderson impurity model is parameterized by $U = 1.0$ and $V = 0.2$.
As in the previous examples of finite chains, the hybridization to the effective auxiliary system is fixed to $\tilde{V} = U/2$. Numerical results are shown in Fig.~\ref{fig:nmertns}.
\begin{figure}[htp!]
\includegraphics{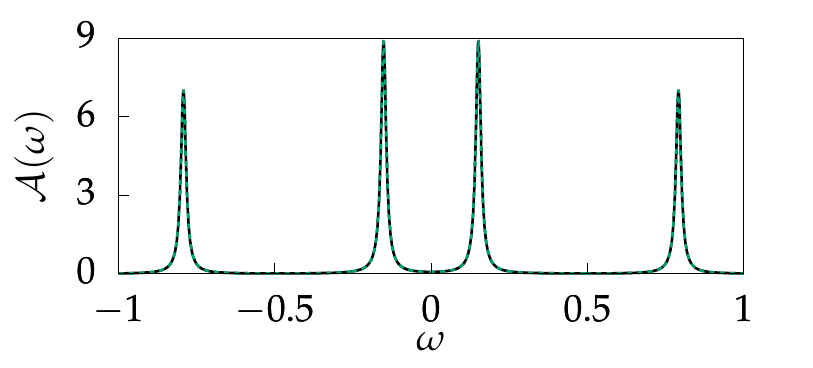}
\includegraphics{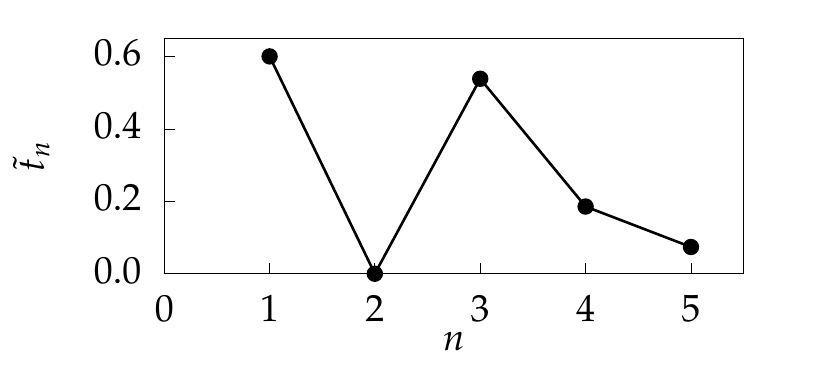}
\vspace{-0.5\baselineskip}
\includegraphics{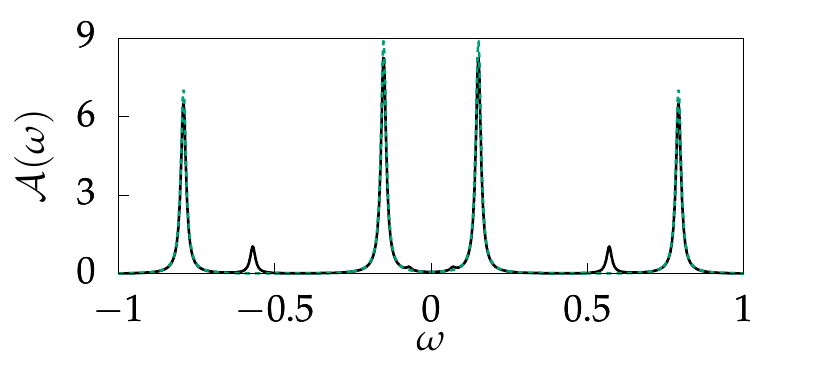}
\includegraphics{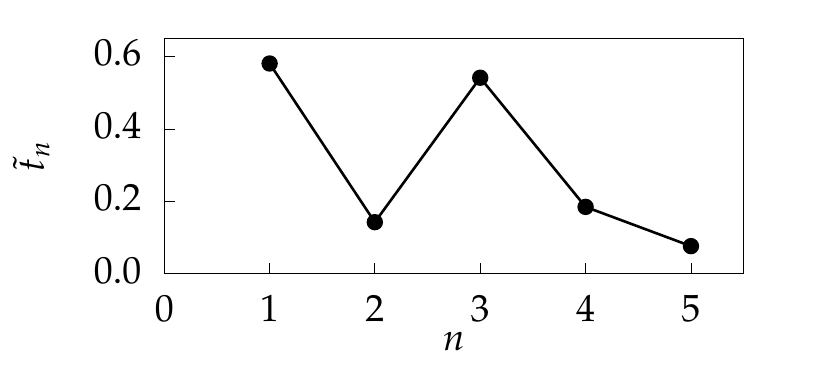}
\vspace{-0.5\baselineskip}
\includegraphics{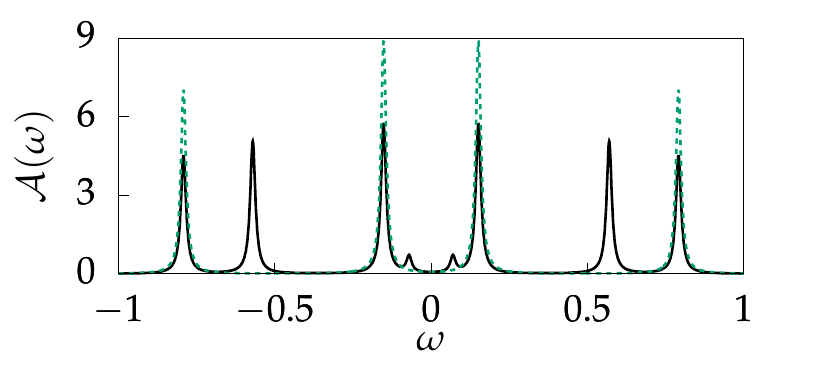}
\includegraphics{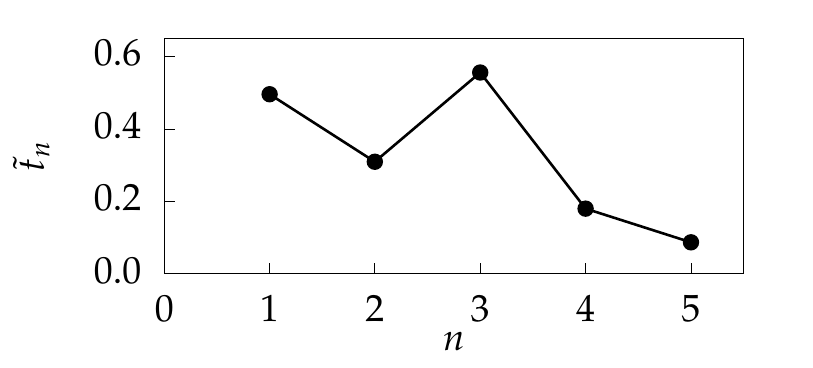}
\vspace{-0.5\baselineskip}
\includegraphics{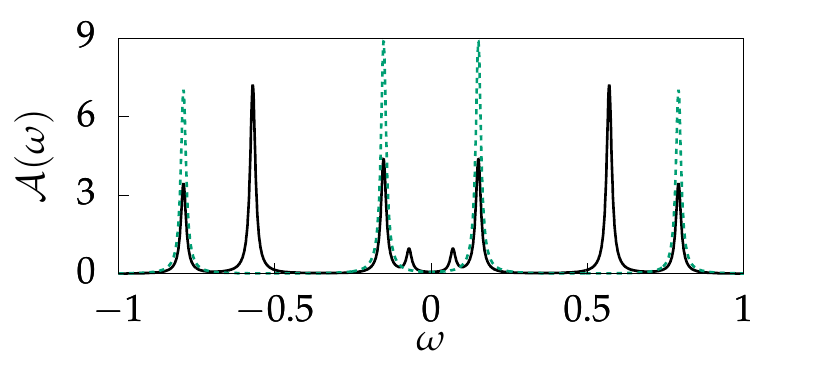}
\includegraphics{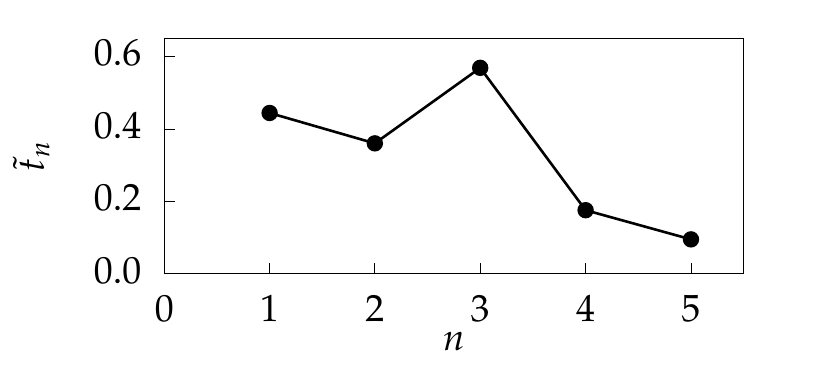}
\vspace{-0.5\baselineskip}
\includegraphics{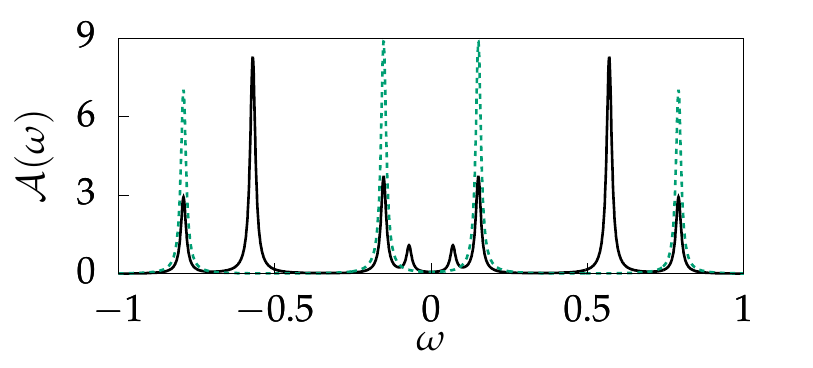}
\includegraphics{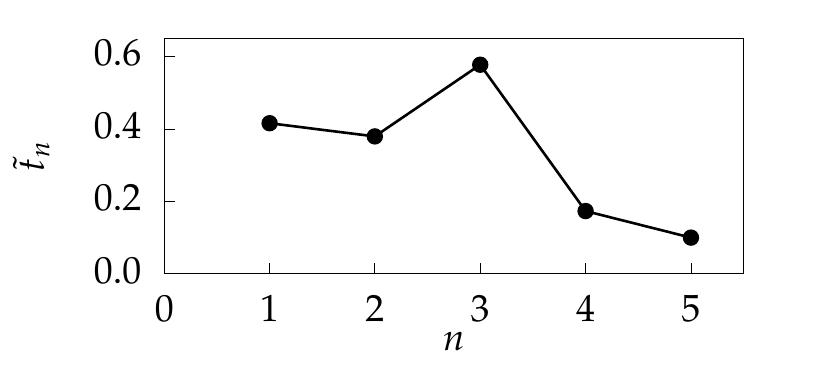}
\vspace{-0.5\baselineskip}
\includegraphics{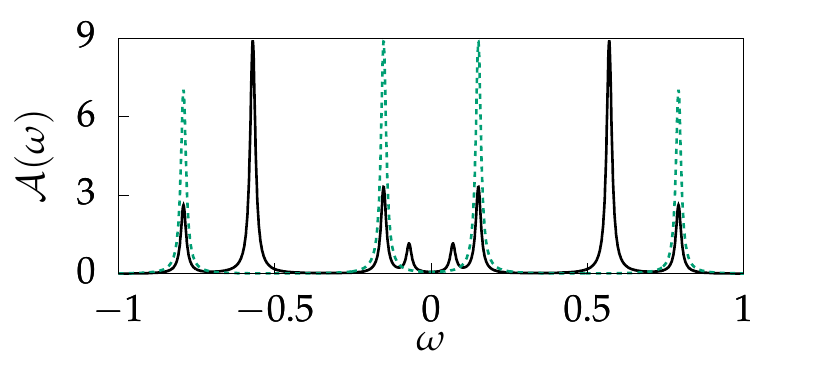}
\includegraphics{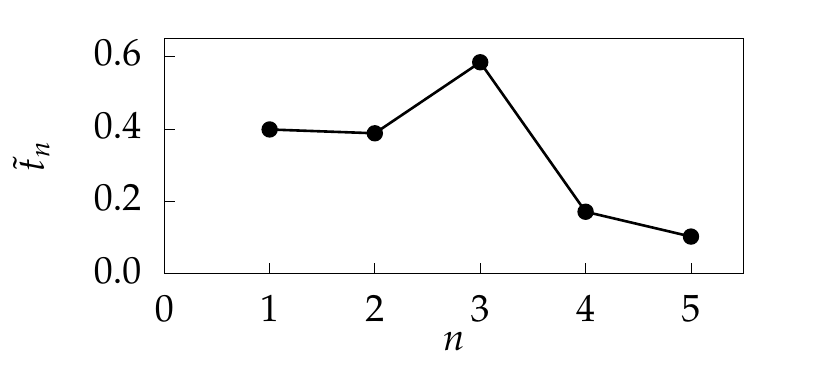}
\caption[Axiliary chain parameters and their associated reconstructed spectrum of the Anderson dimer at finite temperature]{Axiliary chain parameters and their associated reconstructed spectrum of the Anderson dimer at finite temperature for $T = 0.01, 0.04, 0.08, 0.12, 0.16, 0.20$. The $T=0$ spectrum is plotted in teal dashed lines for comparison.\label{fig:nmertns}}
\end{figure}
\begin{figure}[htp!]
\centering
	\includegraphics{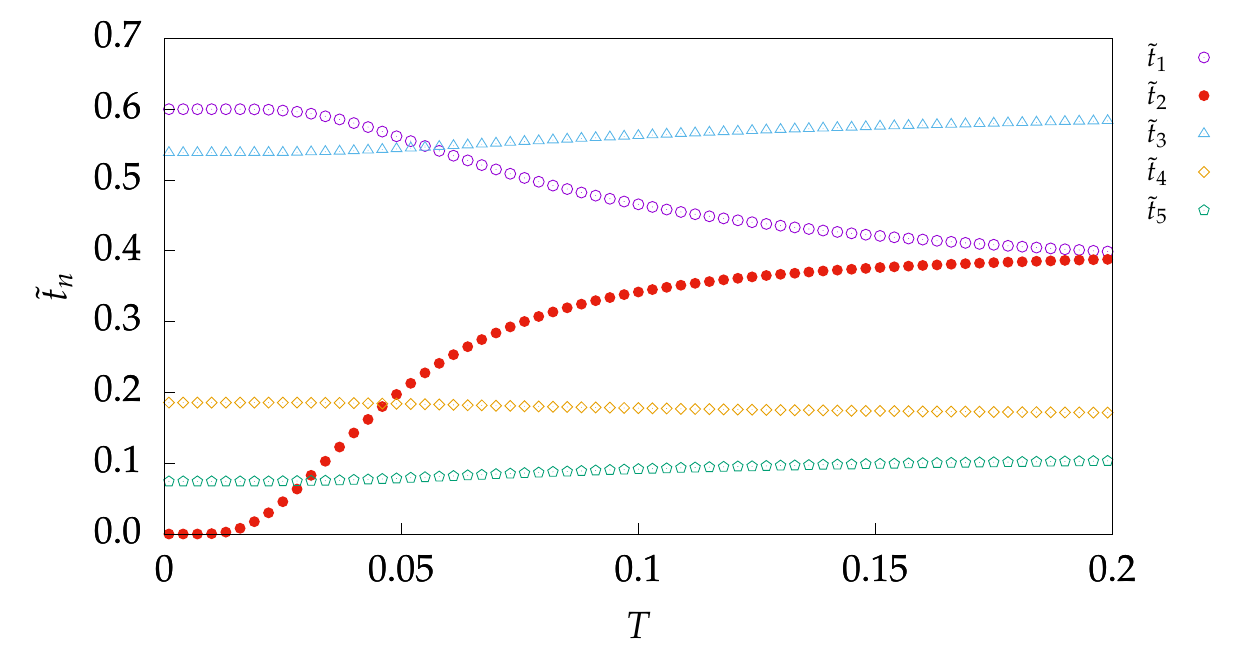}
\caption{Temperature dependence of the auxiliary chain hopping parameters for the Anderson dimer.\label{fig:dimertT}}
\end{figure}

At the low (non-zero) temperature of $T=0.01 > 0$ the value of $\tilde{t}_2$ is small, but finite ($\tilde{t}_2 \approx 4.73\times10^{-4}$). Owing to this finite value the Lanczos algorithm does not truncate at this stage and continues to produce finite values for further hopping amplitudes down the chain. However, at the level of the Green function this small value essentially splits the auxiliary chain into two disjoint parts, so that the spectral function on the end of the chain does not see the entire chain, just the first two auxiliary sites. The small value of $\tilde{t}_2$ remains small with marginally higher temperature, up to a certain threshold. As the temperature increases beyond this threshold, the amplitude of $\tilde{t}_2$ also increases to $\tilde{t}_2 \sim \mathcal{O}(\tilde{V})$, thereby increasing the hybridization to the sites deeper into the auxiliary chain and the Green function is no longer effectively truncated. This is shown by the filled red circles in Fig.~\ref{fig:dimertT} where the value of $\tilde{t}_2$ is negligible until $T \sim 0.02$ where it begins to take on appreciable values. The meaning of this transition stems from the correspondence of the sites of the auxiliary chain to the states which are accessible to the original system. At low temperature the number of accessible states is limited. As temperature is increased, a greater number of excited states become available. This increase in the availability of excited states is captured in the increase in amplitude of $\tilde{t}_2$, which governs the hybridization to the additional sites of the auxiliary chain replicating these excited states. 
%\index{$0$@\textbf{List of Edits}!503@clarification of threshold}
With increasing temperature, the values of the hopping amplitudes $\tilde{t}_3$, $\tilde{t}_4$, and $\tilde{t}_5$ do not significantly change from their values at low temperature. The temperature dependence of the chain parameters is plotted in Fig.~\ref{fig:dimertT}.

Extrapolating this analysis to $N$-mers of longer length, it can be inferred that finite temperature effects of an auxiliary chain can effectively be modeled by a chain partitioned into two parts. The head of the chain captures the properties of the physical system at low temperature. The tail partition of the chain consists of hopping amplitudes that remain essentially fixed over a range of temperature scales. The hybridization between the two partitions is negligible at very low temperatures, but becomes finite when the temperature increases beyond a particular threshold. The parameters of the head of the chain however, do change over different temperature scales. 
Empirical analysis of the auxiliary chain parameters show that the hybridization from the impurity site to the auxiliary chain is fixed to $\tilde{V} = U/2$, but the nature of how the remaining parameters of the head of the chain modulate with respect to temperature remains an open question.

%%%%%%%%%%%%%%%%%%%%%%%%%%%%%%%%%%%%%%%%%%%%%%%%%%%%%%%%%%%%%%%%%%%%%%
%%%%%%%%%%%%%%%%%%%%%%%%%%%%%%%%%%%%%%%%%%%%%%%%%%%%%%%%%%%%%%%%%%%%%%
\section{Auxiliary Field Mapping for Infinite Systems\label{sec:mapping}}
%%%%%%%%%%%%%%%%%%%%%%%%%%%%%%%%%%%%%%%%%%%%%%%%%%%%%%%%%%%%%%%%%%%%%%
%%%%%%%%%%%%%%%%%%%%%%%%%%%%%%%%%%%%%%%%%%%%%%%%%%%%%%%%%%%%%%%%%%%%%%

So far the systems considered for the effective field mapping were of finite size. For these systems, the spectrum consisted of a finite number of poles which could then be mapped exactly to an auxiliary chain, also of a finite depth. For systems of infinite size with a continuous spectrum, a more sophisticated method for generating the parameters of the auxiliary chain is here needed.
Rather than mapping the spectrum of the infinite system onto an auxiliary system, here only the self-energy is mapped. The mapping is performed such that the physical Green function possesses the same dynamics.

\subsection{Recursion Algorithm}

The method of generating the parameters of the auxiliary chain for an infinite system is performed recursively from an input interaction self-energy, rather than exact diagonalization of the physical system. 
For an impurity system with a general bath, the Green function on the impurity is
\begin{equation}
	G(z) \equiv \Green{\op{d}{\sigma}}{\opd{d}{\sigma}}_z = \cfrac{1}{z - \varepsilon - \Gamma(z) - \Sigma(z)}
\end{equation}
where $\Gamma(z)$ represents the hybridization of the impurity to the physical bath and $\Sigma(z)$ is the local interaction self-energy on the impurity. Since the Green function is analytic, it follows that the self-energy is also an analytic function~\cite{analyticse}.
This can been seen from inverting the Dyson equation
\begin{align}
	\Sigma(z) &= z - \varepsilon - \Gamma(z) - \frac{1}{G(z)}
%	\\
%	&= \omega-\varepsilon + \i \delta - \Re\Gamma(z) - \i \Im\Gamma(z) - \frac{\Re G(z)}{\Re G(z)^2 + \Im G(z)^2} + \i \frac{\Im G(z)}{\Re G(z)^2 + \Im G(z)^2}
%	\\
%	\Re\Sigma(z) &= \omega-\varepsilon - \Re\Gamma(z)- \frac{\Re G(z)}{\Re G(z)^2 + \Im G(z)^2}
%	\\
%	\Im\Sigma(z) &= - \Im\Gamma(z) + \frac{\Im G(z)}{\Re G(z)^2 + \Im G(z)^2}
\end{align}
and noting that the self-energy obeys the Kramers-Kronig relations\index{Kramers-Kronig relations} just as the Green function does. This also implies that the self-energy is causal.
Since the self-energy is analytic and causal, it is possible for a hybridization function, which is also analytic and causal, to play the same role as the self-energy in the Green function as the mathematical structure of the Green function would be preserved. The self-energy may then be replaced by a hybridization $\Sigma(\omega) \equiv \Delta_0(\omega)$ to auxiliary degrees of freedom described by some noninteracting Hamiltonian $\hat{H}_{\text{aux}}$. The full single-particle dynamics can therefore be reproduced by replacing $\hat{H}_{\text{int}} \mapsto \hat{H}_{\text{aux}} + \hat{H}_{\text{hyb}}$. Specifically, $\hat{H}_{\text{aux}}$ is taken to be a noninteracting semi-infinite tight-binding chain coupled at one end to the physical lattice degrees of freedom. The general form of such an auxiliary Hamiltonian is
\begin{equation}
	\hat{H}_{\text{aux}} = \sum_{\substack{n=m \\ \sigma \in \{\uparrow,\downarrow\}}}^{\infty} \left[ {\epsilon}_{n} \opd{f}{\sigma;n} \op{f}{\sigma;n} + {t}_{n} \left( \opd{f}{\sigma;n} \op{f}{\sigma;n+1} + \hc \right) \right]
\label{eq:impHaux}
\end{equation}
where the $n$ index labels sites within each auxiliary chain. In $\hat{H}_{\text{aux}}$ the $m$ index is fixed to $m=1$, but will be considered as a dummy index for illustrative purposes in the following explanation.

The coupling of the auxiliary systems to the physical impurity site is given by the hybridization Hamiltonian
\begin{equation}
	\hat{H}_{\text{hyb}} = {t}_0 \sum_{ \sigma \in \{\uparrow,\downarrow\}} \left( \opd{d}{\sigma} \op{f}{\sigma;1} + \opd{f}{\sigma;1} \op{d}{\sigma} \right) \,.
\label{eq:impHhyb}
\end{equation}

The mapping onto this auxiliary chain is analogous to the mapping onto the Wilson chain in NRG, but there are some key differences.
Both schemes involve the mapping of an analytic function onto the parameters of a tight-binding chain. In NRG it is the non-interacting bath of the impurity which is mapped to Wilson chain.
In the recursion algorithm here it is the local self-energy of an interacting site which is mapped to the auxiliary chain.
A key distinguishing difference between the two mappings is that the Wilson chain involves the mapping of a discretized spectrum, but the auxiliary chain here is constructed from the full continuous spectrum of the self-enegy without approximations.

Similarly to the Wilson chain, features at low energy are captured by parameters further down into the chain.

\begin{figure}[htp!]
\centering
\begin{tikzpicture}[every node/.style={line width=1.2pt,inner sep=6pt,scale=1.0}, every path/.style={line width=2pt,scale=1.5}]
\input{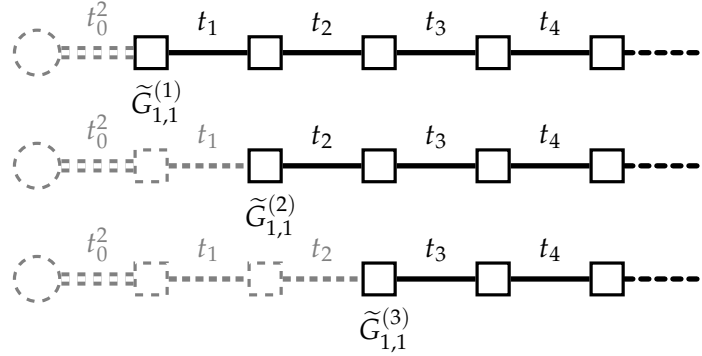}
\end{tikzpicture}
\caption[Hierarchy of the effective hybridization Green functions]{Hierarchy of the effective hybridization Green functions $\tensor*{\widetilde{G}}{^{(m)}_{1,1}}(z)$. The effective Green functions only take into account sites down the chain (shown in black).}
\end{figure}
The Green function $\tensor*{\widetilde{G}}{^{(m)}_{1,1}}(z) = \Greenline{\op{f}{\sigma;m}}{\opd{f}{\sigma;m}}_{z}$ is the Green function on the $m$-th site of the auxiliary chain with all sites $<m$ removed. It is the Green function on the boundary of the truncated chain, hence the $1,1$ subscript.

The effective Green functions $\widetilde{G}$ are obtained from the self-energy of the physical model and the hybridization onto the auxiliary chain $V$,
\begin{equation}
	\Sigma(\omega) \mapsto \Delta_0(\omega) \equiv {t}_0^2 \widetilde{G}^{(1)}_{1,1}(\omega) ,
\end{equation}
 which serves as a normalization factor to satisfy the requirement of a Green function
 \begin{equation}
 	-\frac1\pi \int \Im \widetilde{G}^{(1)}_{1,1}(\omega)\text{d}\omega \overset{!}{=} 1 .
\end{equation}
The self-energy $\Sigma(\omega)$ is not an object which is solved for using the auxiliary system, but it is rather an input. In order for the mapping scheme to be initialized, the solution to the self-energy must be provided as an input. As in the rest of this work, the solution for the self-energy is taken from the NRG calculation of a defined quantum impurity model.

\begin{comment}
\begin{equation}
	\hat{H}_{\text{aux}} = \sum_{\substack{j \in \Gamma \\ \sigma \in \{\uparrow,\downarrow\}}} \sum_{n=m}^{\infty} \left[ \tensor{e}{_{n}} \opd{f}{j,\sigma;n} \op{f}{j,\sigma;n} + \tensor{t}{_{n}} \left( \opd{f}{j,\sigma;n} \op{f}{j,\sigma;n+1} + \hc \right) \right]
\end{equation}
The outer sum sums over all sites $j$ of the physical lattice $\Gamma$ for each spin $\sigma$. The inner sum captures the dynamics of the auxiliary degrees of freedom where the $n$ index labels sites within each auxiliary chain. In $\hat{H}_{\text{aux}}$ the $m$ index is fixed to $m=1$, but will be considered as a dummy index for illustrative purposes in the following explanation.
%
The hybridization from the physical site to the auxiliary system is described by the Hamiltonian
\begin{equation}
	\hat{H}_{\text{hyb}} = V \smashoperator[r]{\sum_{\substack{j \in \Gamma \\ \sigma \in \{\uparrow,\downarrow\}}}} \left( \opd{c}{j,\sigma} \op{f}{j,\sigma;1} + \opd{f}{j,\sigma;1} \op{c}{j,\sigma} \right)
\end{equation}
%
\end{comment}

From the equations of motion for Green functions of a non-interacting chain, the Green function for the edge site of a chain beginning from site $n$ of the effective chain is
\begin{equation}
\begin{aligned}[b]
	{\widetilde{G}}^{(n)}_{1,1}(z) &\equiv \cfrac{1}{z - {\epsilon}_{n} - \cfrac{{t}_{n}^2}{z - {\epsilon}_{n+1} - \cfrac{{t}_{n+1}^2}{z - \ddots}}}
	\\
	&=
	\cfrac{1}{z - {\epsilon}_{n} - {t}_{n}^2 {\widetilde{G}}{^{(n+1)}_{1,1}}(z)}
	\\
	\widetilde{G}^{(n+1)}_{1,1}(z)
	&=
	\frac{1}{{t}_n^2} \left[ z - {\epsilon}_{n} - \frac{1}{\widetilde{G}^{(n)}_{1,1}(z)} \right] \,.
\end{aligned}
\label{eq:greenrecursion}
\end{equation}
Since $\tensor*{\widetilde{G}}{^{(n)}_{1,1}}(z)$ is a Green function for any $n$, it holds that $-\frac1\pi \int \Im \tensor*{\widetilde{G}}{^{(n)}_{1,1}}(\omega)\text{d}\omega = 1$ also for any $n$. Using this fact and the previous equation, an expression for $t_n$ are computed as
\begin{equation}
	{t}_{n}^2 = \frac1\pi \int_{-\infty}^{\infty} \Im \frac{1}{\widetilde{G}^{(n+1)}_{1,1}(\omega)} \text{d}\omega \,.
\end{equation}

The effective Green functions are related to the hybridization functions by $\Delta_{n}(\omega) = {t}_{n}^2 \widetilde{G}^{(n+1)}_{1,1}(\omega)$. %with the identification ${t}_0 \equiv {V}$. 
In terms of the hybridization functions, the auxiliary field mapping may be cast as
\begin{equation}
	\Sigma(z)
	\equiv	\Delta_{0}(z)
	=	\cfrac{{t}_0^2}{z - {\epsilon}_{1} - \cfrac{{t}_1^2}{z - {\epsilon}_{2} - \cfrac{{t}_2^2}{z - \ddots}}}
\label{eq:continuedfraction}
\end{equation}
with the recursion relation
\begin{equation}
\begin{aligned}[b]
%	\Delta_{0}(z) &= \frac{V^2}{z - \Delta_{1}(z)}
%	\\
%	\Delta_{1}(z) &= z - \frac{V^2}{\Delta_{0}(z)}
%	\\
%	&\vdotswithin{=}
%	\\
	\Delta_{n+1}(z) &= z - {\epsilon}_{n} - \frac{{t}_{n}^2}{\Delta_{n}(z)} \,.
\end{aligned}
\label{eq:recursion}
\end{equation}
From the normalization of the Green functions, the $\{\tilde{t}_n\}$ in terms of the hybridization are
\begin{equation}
	{t}_n^2 = -\frac1\pi \Im\!\int \d\omega\ \Delta_n(\omega)
\end{equation}
and the on site potentials can be constructed as
\begin{equation}
	{\epsilon}_n = -\frac{1}{\pi {t}_{n-1}^2} \Im \int \d\omega \, \omega \Delta_{n-1}(\omega) \,.
\end{equation}
At particle-hole symmetry, the hybridization functions $\Delta_{n}(\omega)$ are even functions of $\omega$. This means that the integrand of the formula for the ${\epsilon}_{n}$ is odd. Therefore the auxiliary chain of a particle-hole symmetric system features ${\epsilon}_{n} = 0$ $\forall n$.

A high energy cutoff $D$ for the domain of $\omega$ is enforced such that $\Im \Sigma(\omega) \propto \theta(D - |\omega|)$ as small numerical errors can produce large perturbations in the real component at high energies, which leads to the algorithm delivering inaccurate results due to the recursive nature of the algorithm compounding the errors into both the real and imaginary parts of the iterated hybridization functions. This additionally breaks the Kramers-Kronig relationship between the real and imaginary parts of the auxiliary Green functions, which further contributes to the breakdown of the recursion algorithm.
Introducing the cutoff $D$ stabilizes the algorithm, but is found not to affect physical results for $\lvert \omega \rvert < D$.

\label{sec:cfetechnicalities}
Before proceeding to applications it is worthwhile to first mention some technicalities associated with the mapping in the case where the self-energy exhibits vanishing power-law behavior near the Fermi level~\cite{motttopology}. This is the case which includes the Fermi liquid, a state which arises in such relevant systems as the Anderson impurity model and the Hubbard model~\cite{hewson,dmft}. %The discussion here follows that of~\cite{motttopology}.

In mapping the self-energy of a Fermi liquid there exists a technical complexity in determining the $\{{t}_n\}$ numerically due to the low energy form of the input self-energy.
%It is observed from the recursion algorithm in Eq.~\eqref{eq:recursion} that for an initial hybridization which vanishes at low energy, every other site features a low energy divergence in the next hybridization function $\Delta_{n}(\omega)$. The first iteration of the recursion algorithm is
%\begin{equation}
%	\Delta_{1}(z) = z - {\epsilon}_{0} - \frac{{t}_{0}^2}{\Delta_{0}(z)} \,.
%\end{equation}
%A $\Delta_0(\omega)$ which vanishes at low energy necessitates a $\Delta_{1}(\omega)$ which diverges at low energy. In particular, for a Fermi liquid this results in a pole in $\Delta_{1}(\omega)$ at zero energy. Similarly, continuing with the recursion algorithm this divergence causes $\Delta_{3}(\omega)$ to vanish at zero energy. This pattern then continues with all $\Delta_{n}(\omega)$ with $n$ even exhibiting zero energy poles and all $\Delta_{n}(\omega)$ with $n$ odd vanishing at zero energy. It is therefore crucial to the application of this recursion algorithm to Fermi liquid self-energies that these singularities be appropriately managed numerically.
Following from Eq.~\eqref{eq:lowese}, a Fermi liquid at low energy takes the form of
\begin{equation}
	\Sigma(\omega) \overset{\omega\to0}{\sim} a_0 \omega + \i b_0 \omega^2
\end{equation}
with $a_0 , b_0 < 0$.
%
%
%In the Fermi liquid phase, the technical complexity in determining $\{ t_n\}$ numerically is due to the low energy form of the (input) self-energy. In this case, 
%\begin{align}
%	\Delta_0(\omega)\overset{\omega\to 0}{\sim} a_0\omega+ib_0\omega^2 \qquad (a_0, b_0< 0) \;.
%\end{align}
%
After making the initial association of $\Sigma(\omega) \equiv \Delta_0(\omega) = t_0^2 \tensor*{\widetilde{G}}{^{(1)}_{1,1}}(\omega)$, the first step in the recursion algorithm is the calculation of $\Delta_1(\omega)$ as
\begin{align*}
    \Delta_1(\omega)=t_1^2 \tensor*{\widetilde{G}}{^{(2)}_{1,1}}(\omega) = \omega^+ - \frac{1}{\tensor*{\widetilde{G}}{^{(1)}_{1,1}}(\omega)} \,.
\end{align*}
Since both the real and imaginary parts of $\tensor*{\widetilde{G}}{^{(1)}_{1,1}}(\omega)$ are vanishingly small as $\omega\to 0$ and are equal to zero at $\omega=0$, this leads to a non-analyticity in $\Delta_1(\omega=0)$ and hence a singular part in the corresponding Green function $\tensor*{\widetilde{G}}{^{(2)}_{1,1}}(\omega)$. The Green function can be written as $\tensor*{\widetilde{G}}{^{(2)}_{1,1}}(\omega) = \tensor*{\widetilde{G}}{^{(2)}_{1,1}}^{\text{reg}}(\omega)+ \tensor*{\widetilde{G}}{^{(2)}_{1,1}}^{\text{sing}}(\omega)$, where $\tensor*{\widetilde{G}}{^{(2)}_{1,1}}^{\text{reg}}(\omega)$ denotes the regular (continuum) part, and $\tensor*{\widetilde{G}}{^{(2)}_{1,1}}^{\text{sing}}(\omega)$ denotes the singular part. More precisely,
\begin{align*}
	\Delta_1(\omega\to0)
	&=	\omega^+ - \frac{t_0^2}{a_0 \omega + \i b_0 \omega^2}
	\\
	&=	\omega^+ - \frac{t_0^2}{a_0} \left[ \frac{1}{\omega^+} - \frac{\i b_0}{a_0 + \i b_0\omega}\right] \,.
\end{align*}
Therefore,
\[
\Delta_1^{\text{reg}}(\omega\to 0) = \omega + \frac{t_0^2 b_0^2 \omega}{a_0(a_0^2 + b_0^2 \omega^2)} + \i \frac{t_0^2 b_0}{a_0^2 + b_0^2 \omega^2} \;,
\]
such that
\begin{equation}
	\Delta_1(\omega\to 0) = \Delta_1^{\text{reg}}(\omega\to0) - \frac{t_0^2}{a_0 \omega^+} \,.\label{eq:sigma1}
\end{equation}
The second term on the right-hand side of Eq.~\eqref{eq:sigma1} corresponds to a pole in the imaginary part concomitant with a diverging real part of $\Delta_1(\omega)$ at $\omega=0$. Furthermore, this pole resides on top of a background function, $\Delta_1^{\text{reg}}(\omega)$, such that $\Delta_1^{\text{reg}}(0)=\beta_1$. The residue of this pole is   $\alpha_1=\frac{t_0^2}{|a_0|}$. To fix $t_1^2$ the spectral normalization is used,
\begin{align}
   t_1^2 = \int \mathcal{A}_{1}^{\text{reg}}(\omega) \d\omega +\alpha_1 \,.
    \label{eq:t1sq}
\end{align}
Proceeding to the next step of the recursion, the low energy spectral behavior of $\Delta_2(\omega)$ is
\begin{align}
    \Delta_2(\omega)=\omega-\frac{\omega t_1^2}{\omega \Delta_1^{\text{reg}}(\omega)+\alpha_1}
\end{align}
whose imaginary part is
\begin{align}
\Im\Delta_2(\omega)&=-\omega t_1^2 \Im \frac{1}{\omega\Delta_1^{\text{reg}}(\omega)+\alpha_1}\nonumber \\
    &=\frac{\omega^2 t_1^2 \Im\Delta_1^{\text{reg}}}{(\omega \Re\Delta_1^{\text{reg}} + \alpha_1)^2 + (\omega \Im\Delta_1^{\text{reg}})^2} \,.
\label{eq:sigma_even_site}
\end{align}
Substituting in the respective low energy dependencies of $\Delta_1^{\text{reg}}(\omega)$ it is found that  ${\Im}\Delta_2(\omega)\overset{\omega\to0}{\sim} b_2\omega^2$. Similarly, from the evaluation of $\Re\Delta_2(\omega\to 0)$ it follows from the presence of a non-zero $\alpha_1$ that $\Re\Delta_2(\omega) \overset{\omega\to0}{\sim} a_2\omega$, just as in a Fermi liquid. The advantage of separating the regular and singular parts of the odd-site chain hybridization functions is clear from the structure of Eq.~\eqref{eq:sigma_even_site}, where the information about the underlying pole from the previous iteration is embedded via its weight, and allows the recursion algorithm to deal with a regular function numerically.

Based on the above, it is evident that at every odd recursion step $\Delta_{2n+1}(\omega)$ has a pole structure similar to $\Delta_1(\omega)$ with pole weight $\alpha_{2n+1}=\frac{t_{2n}^2}{|a_{2n}|}$ and subsequently the Fermi liquid character will follow for every even numbered site in the chain, $\Delta_{2n}(\omega)$. In summary, the following recursion relations describe the flow of the low-energy expansion coefficients,
\begin{subequations}
\begin{align}
	\Delta_{2n}(\omega\to 0) &= a_{2n}\omega + \i b_{2n}\omega^2 & (a_{2n},b_{2n}&<0) \,,
	\\
	\Delta_{2n+1}(\omega\to 0) &= \frac{\alpha_{2n+1}}{\omega^+} + \i\beta_{2n+1} \,,
\end{align}
\label{eq:flhybparams}
\end{subequations}
where
\begin{subequations}
\begin{align}
	a_{2n} &= 1-\frac{t_{2n-1}^2}{\alpha_{2n-1}} \,,
	&
	b_{2n} &= \frac{t_{2n-1}^2 \beta_{2n-1}}{\alpha_{2n-1}^2} \,,
	\\
	\alpha_{2n+1} &= \frac{t_{2n}^2}{|a_{2n}|} \,,
	&
	\beta_{2n+1} &= \frac{t_{2n}^2 b_{2n}}{a_{2n}^2} \,.
\end{align}
\end{subequations}
The asymptotic properties of the $\Delta_n(\omega)$ are therefore completely determined by the initialized values $a_0$, $b_0$ and the $\{t_n\}$ determined by the above continued fraction expansion.
Note that the above analytic structure is characteristic of a Fermi liquid: All $\Im\Delta_n(\omega)$ for {even} $n$ have low-energy quadratic behavior, while all {odd} $n$ functions have zero-energy poles.

On the practical level of the numerical calculations, for every odd iteration the singular pole feature is cut from $\mathcal{A}_{2n+1}(\omega)$ and a Hilbert transform is then performed to obtain the correct corresponding regular real part, and hence $\Delta_{2n+1}^{\text{reg}}(\omega)$. The even or odd $t_{n}$ are subsequently evaluated at each iteration from the normalization of $\mathcal{A}_{n}(\omega)$ as above.

This recursion scheme is highly sensitive to the precision of the input. 
In order to capture the appropriate low energy behavior of the imaginary part of the self-energy in the chain parameters it is typically necessary to manually correct some of the numerical artifacts which appear. The types of artifacts which are necessary to correct are described in~\S\ref{sec:seproblems}. The region where the irregularities appear is relatively small, so it is generally easy to extrapolate the correct behavior of $\Im\Sigma(\omega)$.

\subsection{Single Impurity Anderson Model\label{sec:auxsiam}}

As a first case study, this section details how this auxiliary field mapping can be applied to the single impurity Anderson model.
The initial starting point of the auxiliary field mapping is obtaining a solution for the self-energy of the Anderson model. The solution here is obtained from NRG analysis of the single impurity Anderson model.
From \S\ref{ch:methods} the Hamiltonian for the single-impurity Anderson model is
\begin{equation}
	\hat{H}_{\textsc{aim}} = \sum_{k,\sigma} \tensor*{\varepsilon}{_{k}} \opd{c}{k,\sigma} \op{c}{k,\sigma} + \sum_{k,\sigma} \left( \tensor*{V}{_{k,\sigma}} \opd{c}{k,\sigma} \op{d}{\sigma} + \tensor*{V}{^*_{k,\sigma}} \opd{d}{\sigma} \op{c}{k,\sigma} \right) + \tensor*{\varepsilon}{_{d}} \opd{d}{\sigma} \op{d}{\sigma} + U \opd{d}{\uparrow} \op{d}{\uparrow} \opd{d}{\downarrow} \op{d}{\downarrow} \tag{\ref*{eq:siam}}
\end{equation}
The initial condition for the impurity model is taken as a flat band hybridization of bandwidth $D$
\begin{equation}
	\Im\Delta(\omega) = -\frac{V^2 \pi}{2 D} \Theta(D-|\omega|) \,.
\end{equation}
The real part of the hybridization is obtained by a Hilbert transform and is found to be
\begin{equation}
	\Re \Delta(\omega) = \frac{V^2}{D} \ln\left\lvert \frac{\omega+D}{\omega-D} \right\rvert \,.
\end{equation}
This ensures that the hybridization function obeys the Kramers-Kronig relations\index{Kramers-Kronig relations} and possesses the correct analytic structure.
In the following, the parameters are chosen to be $V/D=0.1$ and $D=1.0$.
This flat band hybridization is plotted in Fig.~\ref{fig:inputflathyb}. 
\begin{figure}[htp!]
\centering
\includegraphics{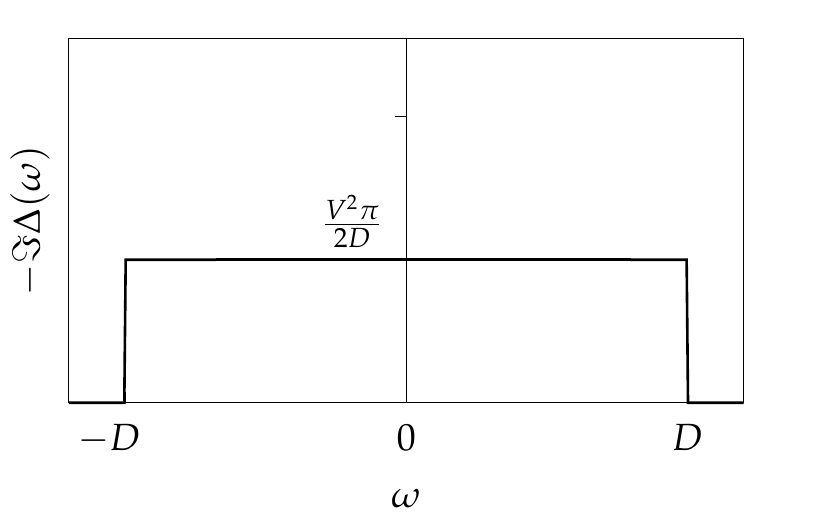}
\caption[Anderson model flat band hybridization function]{Anderson model flat band hybridization function as the input of the NRG self-energy calculation.\label{fig:inputflathyb}}
\end{figure}
The impurity is treated with particle-hole symmetry at half-filling with $U/D = 0.3$ and $\varepsilon_d/D = -0.15$. The resulting self-energy as obtained from NRG at different temperatures is plotted in Fig.~\ref{fig:siamV0.1}.

As will be shown below, the auxiliary chains take on the form of generalized SSH models which exhibit features belonging to the same class of generalizations discussed in \S\ref{ch:genssh}. The present chapter will discuss primarily the qualitative features of these auxiliary chains. A more detailed quantitative study analysis of the auxiliary chains which arise in infinite systems will be postponed until the following chapter, \S\ref{ch:motttopology}, where a more elaborate story will be told.
The primary focus of the analysis here is in the temperature dependence of the auxiliary parameters.

Before analyzing the finite temperature case, it is worth understanding the form of the auxiliary chain for the $T=0$ Fermi liquid self-energy, shown in Fig.~\ref{fig:siamV0.1T0}. The auxiliary chains exhibit a uniform stagger, taking the form of $t_n = \overline{t} \pm \delta t_n$. In this way the auxiliary chains are said to be of generalized SSH form.
Plotted in Fig.~\ref{fig:T1e-5V1tn} are the chain parameters generated by the auxiliary field mapping, and plotted in Fig.~\ref{fig:finiteTchainzoom1e-5} are the odd (blue) and even (red) hopping parameters in the asymptotic regime. Plotted here are the $\delta t_n$'s only, with the normalization $\overline{t} = 0$. The overall form of the chain parameters is that of $1/n$ decay, which is consistent with the pattern uncovered in \S\ref{sec:pseudogapssh}. 
At $T=0$, the long distance behavior of the auxiliary chain hopping parameters takes the form
\begin{equation}
	t_n \sim  \frac{1}{2}\sqrt{1-(-1)^n\frac{2}{n+d}} \,,
\label{eq:amsetn}
\end{equation}
where $d \sim 1/Z$ is related to the quasiparticle weight $Z$,
\begin{equation}
	Z = \left( 1 - \left. \frac{\d \Re \Sigma}{\d \omega} \right\rvert_{\omega=0} \right)^{-1} \,.
\end{equation}
The factor of 2 in Eq.~\eqref{eq:amsetn} is understood to arise from $-\Im\Sigma(\omega) \overset{|\omega| \ll D}{\sim} \omega^2$, which matches the expected behavior from the generalized SSH models constructed from Eq.~\eqref{eq:powerlawtn}.

At zero temperature the imaginary part of the self-energy exhibits Fermi liquid behavior at low energy, $-\Im\Sigma(\omega) \overset{|\omega| \ll D}{\sim} \omega^2$,
while at finite temperatures the self-energy plateaus to a finite value at zero frequency. The high energy ($|\omega|\gg T$) features of $-\Im\Sigma(\omega)$ have no significant temperature dependence.
\begin{figure}[htp!]
\centering
\begin{subfigure}{\linewidth}
\centering
\phantomsubcaption{\label{fig:siamV0.1T0}}
\vspace{-\baselineskip}
\begin{tikzpicture}
\node at (0,0) {\includegraphics[scale=1]{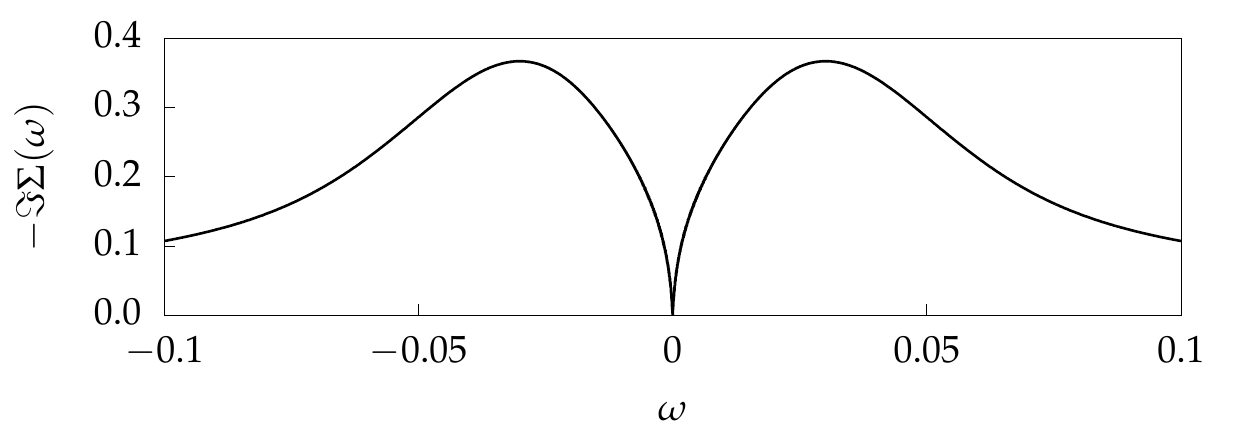}};
\node at (5.25,1.5) {\footnotesize{\subref*{fig:siamV0.1T0}}};
\end{tikzpicture}
\end{subfigure}
\begin{subfigure}{\linewidth}
\centering
\begin{tikzpicture}
\node at (0,0) {\includegraphics[scale=1]{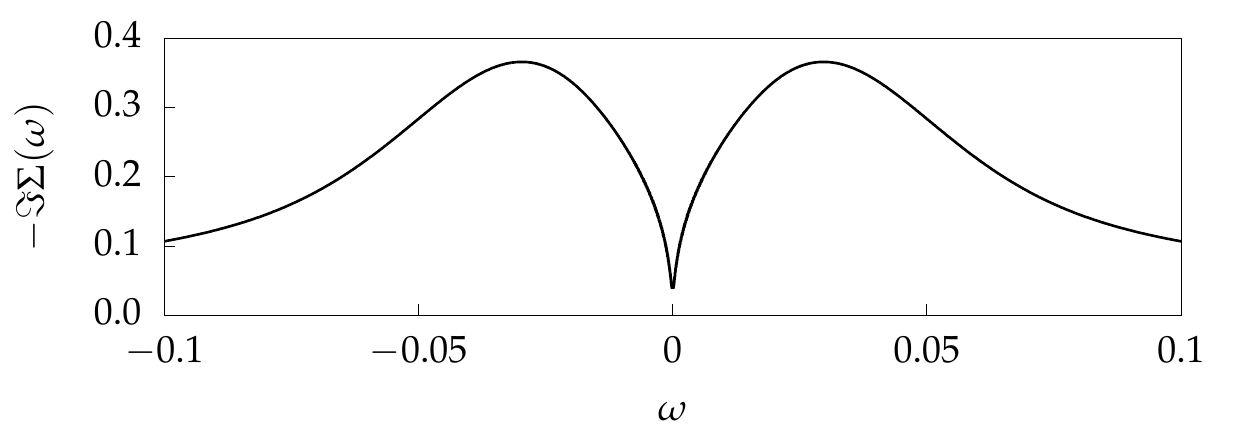}};
\node at (5.25,1.5) {\footnotesize{\subref*{fig:siamV0.1T1e-4}}};
\end{tikzpicture}
\phantomsubcaption{\label{fig:siamV0.1T1e-4}}
\vspace{-\baselineskip}
\end{subfigure}
\begin{subfigure}{\linewidth}
\centering
\begin{tikzpicture}
\node at (0,0) {\includegraphics[scale=1]{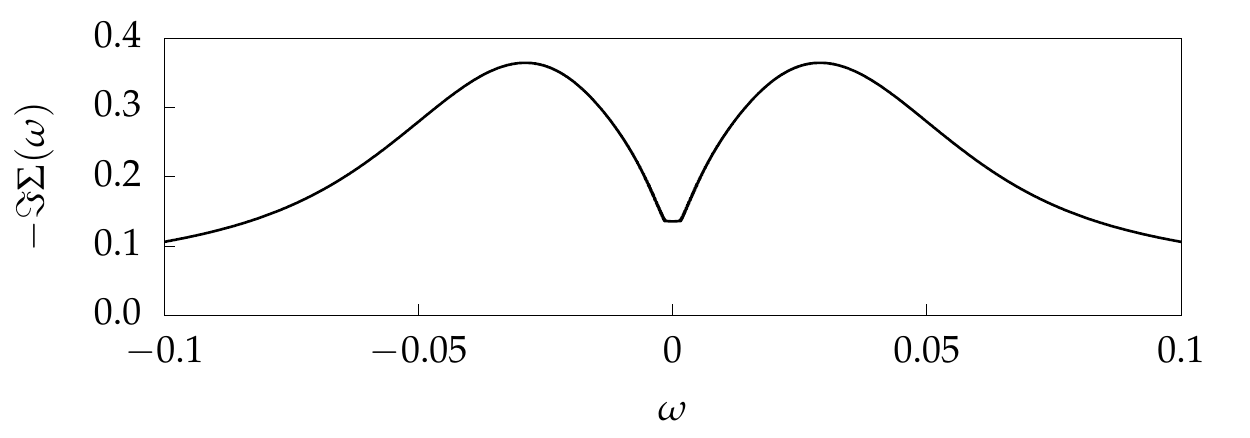}};
\node at (5.25,1.5) {\footnotesize{\subref*{fig:siamV0.1T1e-3}}};
\end{tikzpicture}
\phantomsubcaption{\label{fig:siamV0.1T1e-3}}
\vspace{-\baselineskip}
\end{subfigure}
\begin{subfigure}{\linewidth}
\centering
\begin{tikzpicture}
\node at (0,0) {\includegraphics[scale=1]{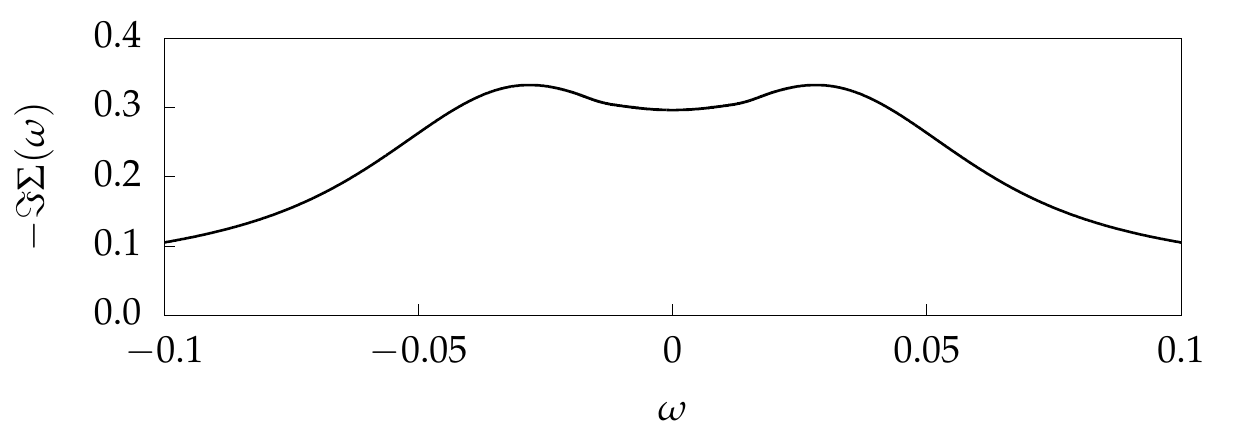}};
\node at (5.25,1.5) {\footnotesize{\subref*{fig:siamV0.1T1e-2}}};
\end{tikzpicture}
\phantomsubcaption{\label{fig:siamV0.1T1e-2}}
\end{subfigure}
\caption[$-\Im\Sigma(\omega)$ for the single impurity Anderson model at finite temperatures]{$-\Im\Sigma(\omega)$ for the single impurity Anderson model with $V=0.1$ and $U=0.3$ at half-filling evaluated at \subref{fig:siamV0.1T0} $T=0$, \subref{fig:siamV0.1T1e-4} $T=10^{-4}$, \subref{fig:siamV0.1T1e-3} $T=10^{-3}$, and \subref{fig:siamV0.1T1e-2} $T=10^{-2}$.
At $T=0$ $-\Im\Sigma(\omega) \overset{|\omega| \ll D}{\sim} \omega^2$, but at finite temperatures $-\Im\Sigma(\omega)$ plateaus to a finite value. Note that only low energy parts of $-\Im\Sigma(\omega)$ are meaningfully affected by temperature; the high energy ($|\omega|\gg T$) features are essentially unchanged across the parameter regime.
\label{fig:siamV0.1}}
\end{figure}

As seen in Fig.~\ref{fig:finiteTchain}, the head of the chain corresponding to various temperatures takes on a similar form, but the envelope which follows changes with the temperature. Since the parameters at the head of the chain determine the high energy features of the spectrum, it is expected that these would be nearly identical for the various temperatures from visual inspection of the self-energies in Fig.~\ref{fig:siamV0.1}. The detailed effects of temperature on the envelope of the chain are illustrated in Fig.~\ref{fig:finiteTchainzoom}.

A prominent characteristic of these auxiliary systems is the appearance of a thermal length scale $\xi_T$ in the auxiliary chains.
\begin{figure}[htp!]
\begin{subfigure}{0.49\linewidth}
\phantomsubcaption{\label{fig:T1e-5V1tn}}
\begin{tikzpicture}
	\node at (0,0) {\includegraphics{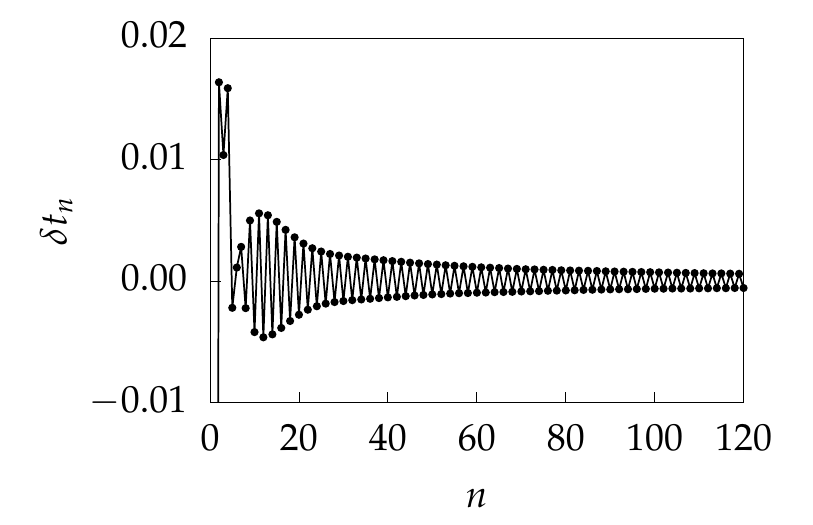}};
	\node at (3.125,2) {\footnotesize \subref*{fig:T1e-5V1tn}};
\end{tikzpicture}
\end{subfigure}
\hfill
\begin{subfigure}{0.49\linewidth}
\phantomsubcaption{\label{fig:T1e-4V1tn}}
\begin{tikzpicture}
	\node at (0,0) {\includegraphics{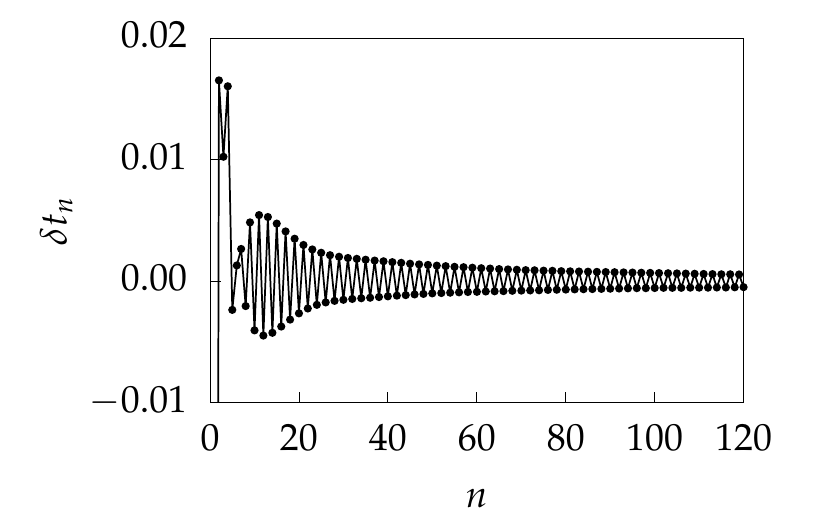}};
	\node at (3.125,2) {\footnotesize \subref*{fig:T1e-4V1tn}};
\end{tikzpicture}
\end{subfigure}
\begin{subfigure}{0.49\linewidth}
\phantomsubcaption{\label{fig:T1e-3V1tn}}
\begin{tikzpicture}
	\node at (0,0) {\includegraphics{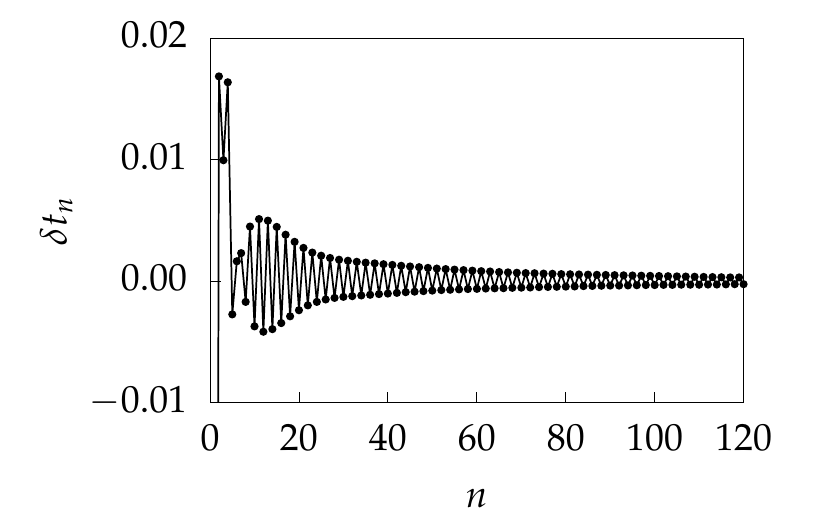}};
	\node at (3.125,2) {\footnotesize \subref*{fig:T1e-3V1tn}};
\end{tikzpicture}
\end{subfigure}
\hfill
\begin{subfigure}{0.49\linewidth}
\phantomsubcaption{\label{fig:T1e-2V1tn}}
\begin{tikzpicture}
	\node at (0,0) {\includegraphics{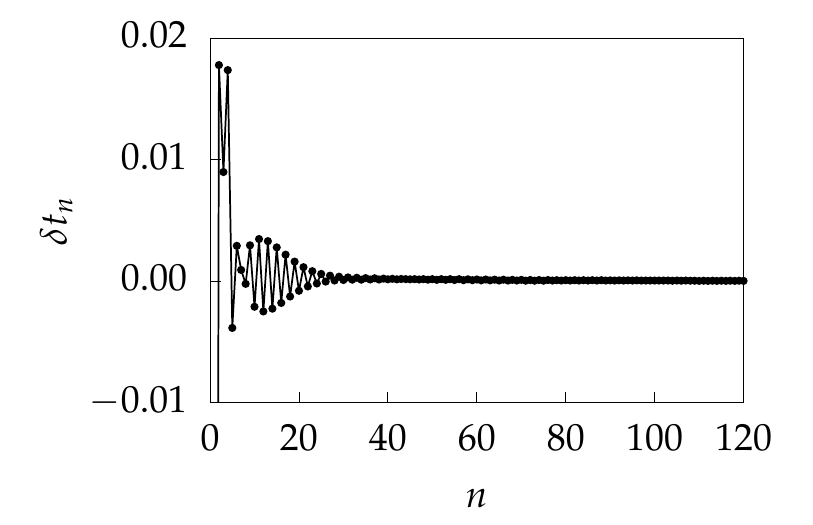}};
	\node at (3.125,2) {\footnotesize \subref*{fig:T1e-2V1tn}};
\end{tikzpicture}
\end{subfigure}
\caption[Auxiliary chain parameters for the self-energy of the single impurity Anderson model at finite temperatures]{Auxiliary chain parameters for the self-energy of the single impurity Anderson model at \subref{fig:T1e-5V1tn} $T=0$, \subref{fig:T1e-4V1tn} $T=10^{-4}$, \subref{fig:T1e-3V1tn} $T=10^{-3}$, and \subref{fig:T1e-2V1tn} $T=10^{-2}$. \label{fig:finiteTchain}}
\end{figure}
$t_n$ at large $n$ correspond to features approximately at $\omega \sim 1/n$. The fact that at finite temperature the $t_n$'s reach a fixed value without oscillations, $t_{2n} \simeq t_{2n-1}$, is expected as at finite-$T$, $-\Im\Sigma(\omega)$ does not tend to zero, but rather plateaus at a finite value.
\begin{figure}[htp!]
\begin{subfigure}{\linewidth}
\begin{tikzpicture}
	\node at (0,0) {\includegraphics{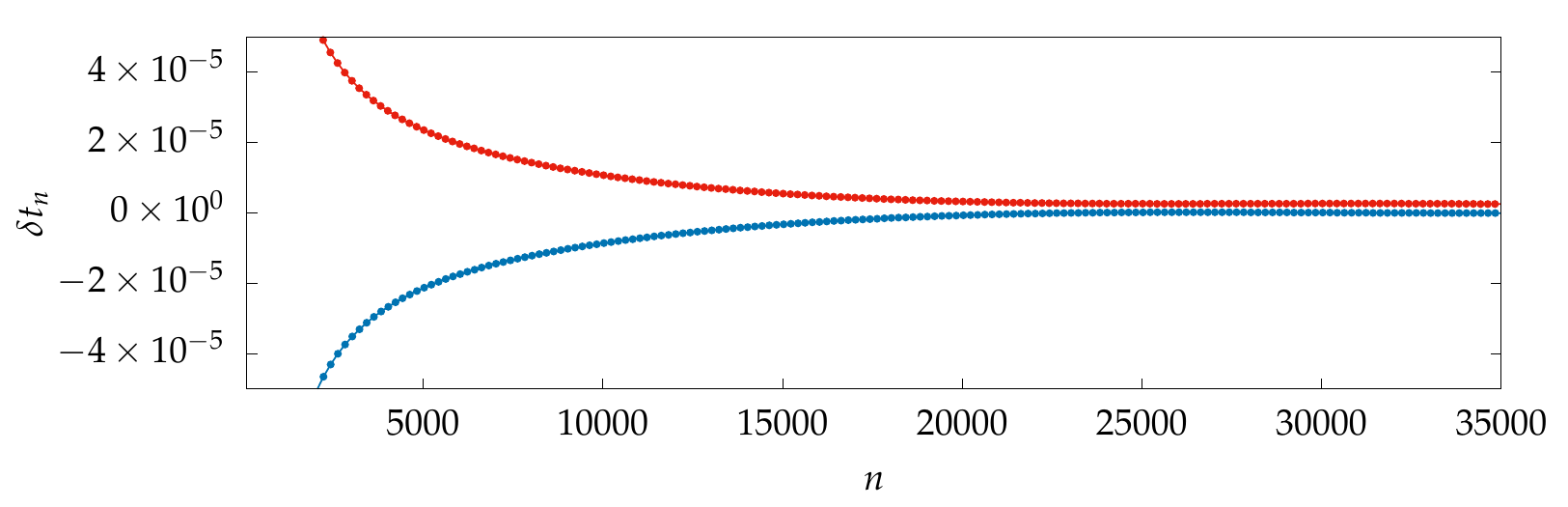}};
	\node at (7.25,2) {\footnotesize\subref*{fig:finiteTchainzoom1e-5}};
\end{tikzpicture}
\phantomsubcaption{\label{fig:finiteTchainzoom1e-5}}
\vspace{-2\baselineskip}
\end{subfigure}
\begin{subfigure}{\linewidth}
\begin{tikzpicture}
	\node at (0,0) {\includegraphics{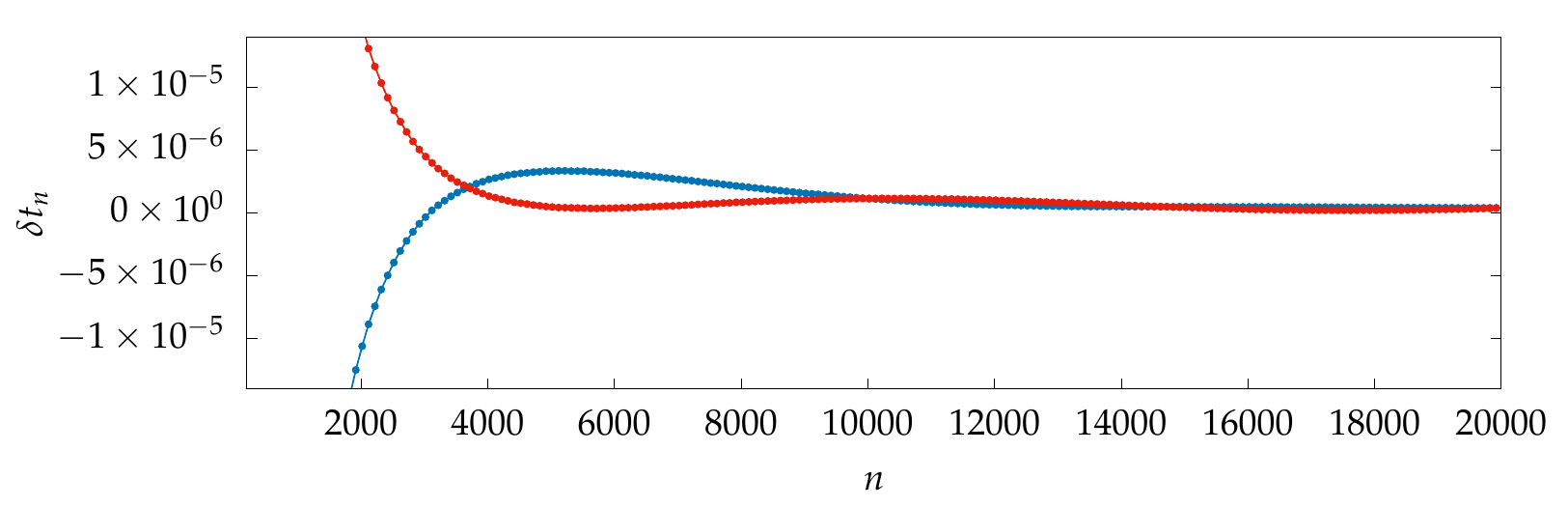}};
	\node at (7.25,2) {\footnotesize\subref*{fig:finiteTchainzoom1e-4}};
\end{tikzpicture}
\phantomsubcaption{\label{fig:finiteTchainzoom1e-4}}
\vspace{-2\baselineskip}
\end{subfigure}
\begin{subfigure}{\linewidth}
\begin{tikzpicture}
	\node at (0,0) {\includegraphics{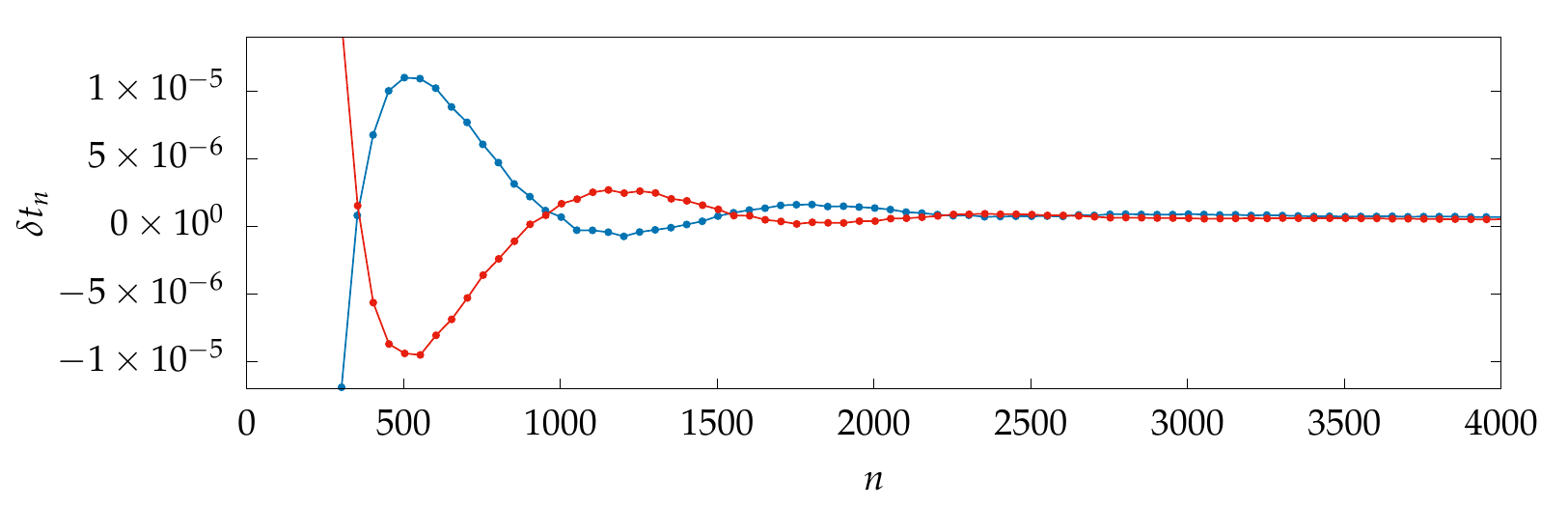}};
	\node at (7.25,2) {\footnotesize\subref*{fig:finiteTchainzoom1e-3}};
\end{tikzpicture}
\phantomsubcaption{\label{fig:finiteTchainzoom1e-3}}
\vspace{-2\baselineskip}
\end{subfigure}
\begin{subfigure}{\linewidth}
\begin{tikzpicture}
	\node at (0,0) {\includegraphics{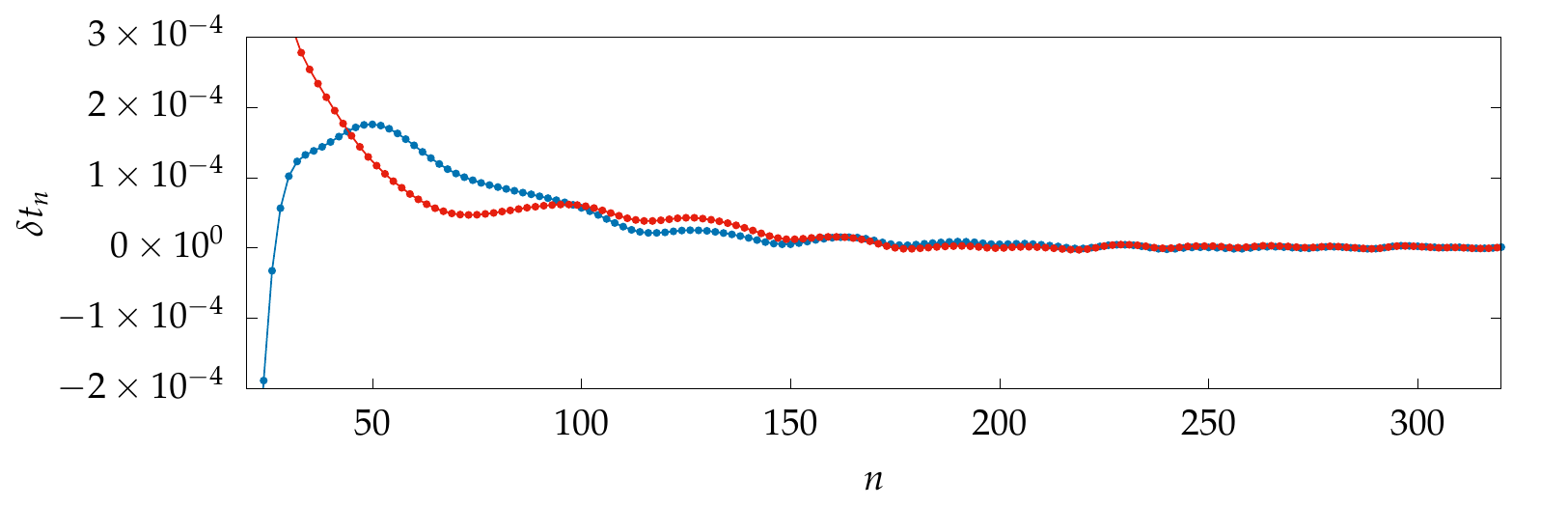}};
	\node at (7.25,2) {\footnotesize\subref*{fig:finiteTchainzoom1e-2}};
\end{tikzpicture}
\phantomsubcaption{\label{fig:finiteTchainzoom1e-2}}
\vspace{-\baselineskip}
\end{subfigure}
\caption[Auxiliary chain parameters for the Anderson model at finite temperatures]{Envelopes of the auxiliary chain hopping parameters $t_n = t_0 + \delta t_n$ for single-impurity Anderson model at temperatures of \subref{fig:finiteTchainzoom1e-5} $T=0$, \subref{fig:finiteTchainzoom1e-4} $T=10^{-4}$, \subref{fig:finiteTchainzoom1e-3} $T=10^{-3}$, and \subref{fig:finiteTchainzoom1e-2} $T=10^{-2}$. The even $t_n$ are plotted in red and the odd $t_n$ are plotted in blue. The chain parameters exhibit an enveloped behavior at the front of the chain before settling down to a constant value at a thermal length scale $\xi_T \sim 1/T$. The $T=0$ case does not, which reflects $\xi_T \to \infty$. Points plotted are a subset of the total data set to best illustrate the envelopes.\label{fig:finiteTchainzoom}}
\end{figure}
At finite temperature, the $\{t_n\}$ exhibit $\sim1/n$ decay until settling at a constant value (within systematic numerical noise) on the order of $\xi_{T}$.
Analysis of the auxiliary chain parameters in Fig.~\ref{fig:finiteTchainzoom} reveals that the order of the thermal length scale is
\begin{equation}
	\xi_{T} \sim \mathcal{O}(T^{-1}) \,.
\end{equation}

%at large $n$, the even and odd $t_n$ exhibit small oscillations of the form $t_{n'} \sim t_{n'}' + (-1)^{n'} \delta t_{n'}$ with $n'$ odd or even and $t_{n'}'$ some mean value. These micro oscillations are much smaller than the overall oscillating behavior of the $t_n$, $\delta t_{n'} \ll \delta t_n$ and are due to numerical errors in the calculation of such small numbers for the $t_n$ from the continued fraction expansion.

A chain parameterized by constant $t_n$'s corresponds to a boundary spectral function which is metallic and has finite value at the Fermi level.

%%%%%%%%%%%%%%%%%%%%%%%%%%%%%%%%
\subsection{Hubbard-SSH Revisited}
%%%%%%%%%%%%%%%%%%%%%%%%%%%%%%%%

The above analysis can also naturally be applied to the interacting Hubbard-SSH on the Bethe lattice seen in \S\ref{ch:bethessh}. Recall from \S\ref{sec:hsshse} that the self-energy on the $\circ$-sites for $U < U_{c}$ is a Fermi liquid with $-\Im\Sigma^{\circ}(\omega) \sim \omega^2$, just as in the Anderson model above. The auxiliary chain parameters for the $\Sigma^{\circ}(\omega)$ is also of the form of Eq.~\eqref{eq:amsetn}.
\begin{figure}[htp!]
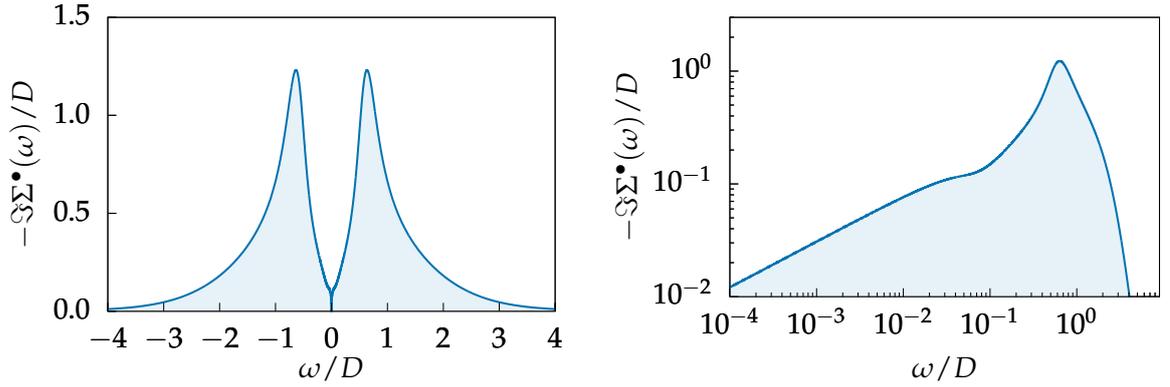

\begin{subfigure}[t]{0.49\linewidth}
\includegraphics{hsshSU3a.pdf}
%\caption{$-\Im\Sigma^\bullet(\omega)$ at $U/t = 3.0$}
\end{subfigure}
\begin{subfigure}[t]{0.49\linewidth}
\includegraphics{SU3loga.pdf}
%\caption{$-\Im\Sigma^\circ(\omega)$ at $U/t = 3.0$. At low energy the self energy scales as $\omega^2$, which is indicative of a Fermi liquid.}
\end{subfigure}
\caption{Self-energy of the HSSH model as obtained by the DMFT-NRG calculation of \S\ref{ch:bethessh}. \label{fig:hsshseredux}}
\end{figure}
As observed in Fig.~\ref{fig:hsshseredux}, the self-energy of the $\bullet$-sites for $U < U_{c}$ has low energy power-law behavior of $ -\Im\Sigma^{\bullet}(\omega) \sim \lvert \omega^{r_\bullet} \rvert$, but it is non-Fermi liquid with $r_\bullet \neq 2$.
Here again it is revealed that the long distance behavior of the auxiliary chains take the form of
\begin{equation}
	t_n \overset{nZ\gg1}{\sim} \frac{D}{2} \sqrt{1 + (-1)^{n} \frac{r}{n+d}}
\end{equation}
where now $r$ no longer takes the value of 2 as characteristic of the Fermi liquid. Instead $r$ takes the value of $r = r_\bullet$, the power of the self-energy at $\lvert\omega\rvert\ll1$. This analysis reaffirms the assumptions leading into the toy model in the previous section, as well as the characteristics determined in the generalized SSH models discussed in \S\ref{ch:genssh}.

%%%%%%%%%%%%%%%%%%%%%%%%%%%%%%%%%%%%%%%%%%%%%%%%%%%%%%%%%%%%%%%%%%%%%%%%%%%%%
%%%%%%%%%%%%%%%%%%%%%%%%%%%%%%%%%%%%%%%%%%%%%%%%%%%%%%%%%%%%%%%%%%%%%%%%%%%%%
\section{Quantum Transport\label{sec:transport}}
%%%%%%%%%%%%%%%%%%%%%%%%%%%%%%%%%%%%%%%%%%%%%%%%%%%%%%%%%%%%%%%%%%%%%%%%%%%%%
%%%%%%%%%%%%%%%%%%%%%%%%%%%%%%%%%%%%%%%%%%%%%%%%%%%%%%%%%%%%%%%%%%%%%%%%%%%%%

The auxiliary field mapping presented in the previous section maps an interacting system to a non-interacting auxiliary system. This mapping is not a solution technique, as a solution for the self-energy is needed to initialize the auxiliary field mapping in the first place. A practical utility of this scheme is in the calculation of transport properties of quantum impurity models.

A standard method of calculating conductance is by means of the Kubo formula~\cite{bruusflensberg,kubotransport}, which is a general formula for calculating the linear response conductance for interacting systems in arbitrary configurations at arbitrary temperature. A difficulty in the application of the Kubo formula is that the current-current correlator is difficult to compute compared to the Green functions.

There exist numerous methods of calculating transport via Green functions, however these formulae are typically of limited scope and can only be applied in certain circumstances. A basic example is the Landauer formula~\cite{landauer,landauer2,greenlandauer}, which is applicable to transport through a non-interacting impurity with two terminals. This can be generalized to the multi-terminal case in the form of the Landauer-B\"{u}ttiker formula~\cite{landauerbuttiker}.

An electric transport equation which incorporates fully interacting impurities in non-equilibrium with two terminals is the Meir-Wingreen formula~\cite{meirwingreen}. However, the formulation in linear response is in terms of equilibrium retarded Green functions and is limited to the proportionate coupling case.
For non-proportionate coupling systems, Meir-Wingreen requires explicit non-equilibrium Green functions, even in linear response. These are notoriously challenging to calculate for interacting systems, and so these calculations typically resort to the Kubo formula in linear response.

However for a Fermi liquid, as in the case of the Anderson impurity model, the linear response at zero temperature in terms of the Green functions can be obtained by the Oguri formula~\cite{oguri}.
%
%Among this plethora of formulae certain ones are easier or harder to calculate and the correct formula needs to be applied in order to capture the correct physics.
In general it would be useful to have here an equilibrium Green function based formulation for finite temperature transport in linear response for interacting models without the proportionate coupling condition. No such formulation is known.

The principal concept of the application here is to map an interacting quantum dot system to an auxiliary system without interactions in order to avoid the incorporation of interactions in the calculation of quantum transport expressions. This potentially leads to the simplification in the calculation of transport properties of quantum impurity models.
With an effective non-interacting model at hand, quantum transport can be calculated using standard Green function methods within the Landauer-B\"uttiker framework.

For physical concreteness, in this section (\ref{sec:transport}) the Planck constant $h$ is included explicitly as well as the unit electric charge $e$. This is done so that physically observable quantities have prefactors of meaningful dimension. The electrical conductance $\mathfrak{G}_C$ for instance has the appropriate dimensionful prefactor of $e^2/h$.

\subsubsection{Single Quantum Dot}

%Recently in Ref.~\cite{motttopology},  topological properties of the Mott metal-insulator transition in the Hubbard model were uncovered by mapping the interaction self-energy of the effective impurity problem within dynamical mean field theory, to auxiliary non-interacting degrees of freedom. 

In this section the consequences of the mapping described above are explored in the context of quantum transport. A first application is to the case of transport through a single quantum dot, where the dot is represented by the paradigmatic single-impurity Anderson model,
\begin{equation}\label{eq:aim}
	\hat{H}_{\text{SQD}} = \hat{H}_{\text{leads}} + \varepsilon_d ( \hat{n}_{\uparrow} + \hat{n}_{\downarrow} ) + \sum_{\alpha,\sigma} \left( \tensor{V}{_{\alpha}} \opd{d}{\sigma} \op{c}{\alpha \sigma} + \hc \right) + \hat{H}_{\text{int}} \;,
\end{equation}
where $\hat{n}_{\sigma} = \opd{d}{\sigma} \op{d}{\sigma}$, $\hat{H}_{\text{int}} = U_d \hat{n}_{\uparrow}\hat{n}_{\downarrow}$, and
\begin{equation}
	\hat{H}_{\text{leads}} =  \sum_{\alpha} \hat{H}_{\text{leads}}^{\alpha} = \sum_{\alpha,k,\sigma} \tensor*{\varepsilon}{_k} \opd{c}{\alpha k \sigma} \op{c}{\alpha k \sigma}
\label{eq:transportleads}
\end{equation}
with $\alpha = \text{s}, \text{d}$ for source and drain. The leads are characterized by their free Green functions, denoted $\mathcal{G}^0_{\alpha\alpha}(\omega) \equiv \Green{\op{c}{\sigma}}{\opd{c}{\sigma}}_{\omega}$, with corresponding free density of states $\varrho(\omega)=-\tfrac{1}{\pi} \Im \mathcal{G}^0_{\alpha\alpha}(\omega)$ which are taken to be the same for both source and drain leads.

As described in the previous section, the auxiliary field mapping is an exact representation of the single-particle dynamics for an interacting system in terms of a completely non-interacting one. The Dyson equation for the Anderson impurity reads as
\begin{equation}\label{eq:Gaim}
	G_{\sigma}(\omega)\equiv \Green{\op{d}{\sigma}}{\opd{d}{\sigma}}_{\omega} = \frac{1}{\omega - \varepsilon_d/\hslash -\sum\limits_{\alpha}\Delta^{\alpha}(\omega)- \Sigma(\omega)} \,,
\end{equation}
where $\Delta^{\alpha}(\omega)=V_{\alpha}^2 \mathcal{G}^0_{\alpha\alpha}(\omega)$ is the hybridization between the impurity and the physical lead $\alpha \in \{ \text{s}, \text{d} \}$ and $\Sigma(\omega)$ is the interaction self-energy. For convenience the static contribution to the self-energy is absorbed into the definition of the renormalized level $\varepsilon_d^* = \varepsilon_d + \hslash~\Re \Sigma(0)$, and work with the dynamical part of the self-energy $\tilde{\Sigma}(\omega) = \Sigma(\omega) - \Re \Sigma(0)$ .

%%%%%%%%%%%%%%%%
\begin{figure}[htp!]
%\centering
\begin{subfigure}[c]{0.49\linewidth}
\begin{tikzpicture}[{thick}]
	\node[circle,draw=black,fill=black!10,inner sep=1pt] (imp) at (0,0) {$\uparrow\downarrow$};
	\def\p{1.25}
	\def\l{2.5}
	\def\w{0.5}
	\coordinate (s) at (-\p,0);
	\coordinate (d) at (\p,0);
	\draw (d) arc(0:-90:-\l cm and \w cm);
	\draw (d) arc(0:-90:-\l cm and -\w cm);
	\draw (s) arc(0:-90:\l cm and \w cm);
	\draw (s) arc(0:-90:\l cm and -\w cm);
	 \node at ($(s)+(-1.5,0)$) {source};
	 \node at ($(d)+(1.5,0)$) {drain};
	 \draw[-,line width=1.5pt] (s)--(imp) node[midway,above] {$V_{\text{s}}$};
	 \draw[-,line width=1.5pt] (d)--(imp) node[midway,above] {$V_{\text{d}}$};
	\path[in=60,out=120,looseness=6] (imp) edge[-latex,line width=0.5pt,double distance=0.5pt] node[above] {$U_d$} (imp);
	\begin{scope}[scale=0.67,draw=none]
	\def\first{1}
	\def\alast{3}
	\def\nalast{2}
	\foreach[evaluate=\s as \sc using (\s+0.5)] \s in {\first,...,\alast}
	{
	\node[rectangle,inner sep=5pt] (d\s) at (0,-\sc) {};
	}
	\draw[draw=none,line width=1.25pt](imp)--(d\first);
	\foreach[evaluate=\s as \n using int(\s+1)] \s in {\first,...,\nalast}
	{
	\draw[draw=none,line width=1.25pt](d\s)--(d\n) node[midway,above] {};
	}
	\draw[draw=none,line width=1.25pt, dashed, line cap=round] (d\alast)--+(0,-1) {};
	\end{scope}
	\node at (3.75,1.4) {\footnotesize\subref*{fig:singlei}};
\end{tikzpicture}
\phantomsubcaption{\label{fig:singlei}}
\end{subfigure}
\begin{subfigure}[c]{0.49\linewidth}
\begin{tikzpicture}[{thick}]
	\node[circle,draw=black,inner sep=6pt] (imp) at (0,0) {};
	\def\p{1.25}
	\def\l{2.5}
	\def\w{0.5}
	\coordinate (s) at (-\p,0);
	\coordinate (d) at (\p,0);
	\draw (d) arc(0:-90:-\l cm and \w cm);
	\draw (d) arc(0:-90:-\l cm and -\w cm);
	\draw (s) arc(0:-90:\l cm and \w cm);
	\draw (s) arc(0:-90:\l cm and -\w cm);
	\node at ($(s)+(-1.5,0)$) {source};
	\node at ($(d)+(1.5,0)$) {drain};
	\draw[-,line width=1.5pt] (s)--(imp) node[midway,above] {$V_{\text{s}}$};
	\draw[-,line width=1.5pt] (d)--(imp) node[midway,above] {$V_{\text{d}}$};
	\begin{scope}[scale=0.67]
	\def\first{1}
	\def\alast{3}
	\def\nalast{2}
	\foreach[evaluate=\s as \sc using (\s+0.5)] \s in {\first,...,\alast}
	{
	\node[rectangle,draw=red,inner sep=5pt] (d\s) at (0,-\sc) {};
	}
	\draw[red,line width=1.25pt](imp)--(d\first);
	\foreach[evaluate=\s as \n using int(\s+1)] \s in {\first,...,\nalast}
	{
	\draw[red,line width=1.25pt](d\s)--(d\n) node[midway,above] {};
	}
	\draw[red,line width=1.25pt, dashed, line cap=round] (d\alast)--+(0,-1) {};
	\end{scope}
	\node at (0.45,-0.6) {$V_{\text{aux}}$};
	\node at (3.75,1.4) {\footnotesize\subref*{fig:singleimpaux}};
\end{tikzpicture}
\phantomsubcaption{\label{fig:singleimpaux}}
\end{subfigure}
  \caption[Auxiliary field mapping of transport through single quantum dot]{Schematic of the auxiliary field mapping. The interaction self-energy $\tilde{\Sigma}(\omega)$ of the single-impurity Anderson model \subref{fig:singlei} is mapped to an auxiliary non-interacting tight-binding chain \subref{fig:singleimpaux}. The current between physical source and drain leads due to a bias voltage in the interacting system is reproduced in the mapped non-interacting system through a zero-current constraint for the auxiliary `lead' in the 3-terminal Landauer-B\"uttiker formula.% Figure reproduced from~\cite{multiorbitaltransport}.
  }
\label{fig:singleimp}
\end{figure}
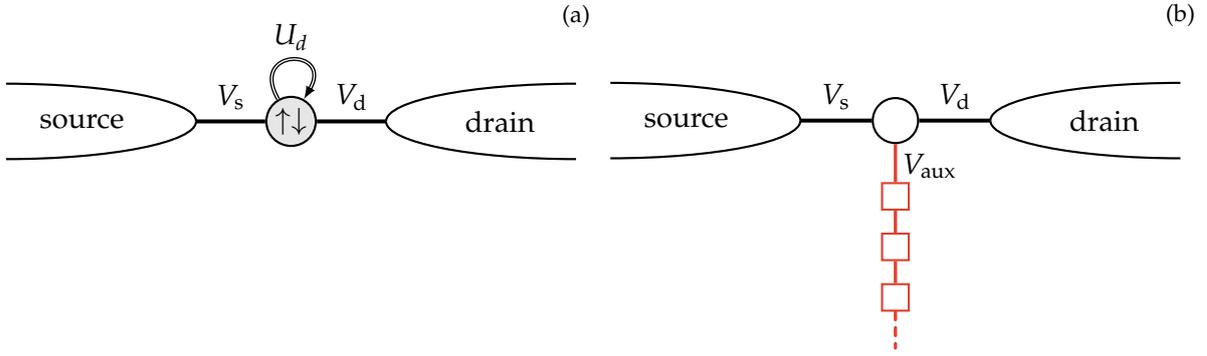
%%%%%%%%%%%%%%%%

Following the mapping described previously in \S\ref{sec:mapping}, the dynamical part of the self-energy is interpreted as a hybridization to a bath of auxiliary non-interacting degrees of freedom $\tilde{\Sigma}(\omega) \mapsto \Delta^{\text{aux}}(\omega)$. 
The effect of the electronic scattering due to the Coulomb interaction on the impurity is reproduced exactly by a proper choice of the auxiliary bath. For simplicity it is assumed that the impurity is at particle-hole symmetry, with $\varepsilon_d = -U_d/2$, so that $\varepsilon_d^*=0$ and Eq.~\eqref{eq:impHaux} does not require inclusion of onsite potentials, $\tensor*{\epsilon}{_{n}} = 0$ $\forall n$. For notational clarity, the hybridization parameter from the impurity site the auxiliary chain in Eq.~\eqref{eq:impHhyb} will be notated $\tensor{V}{_{\text{aux}}}$ to avoid any conflation with hybridization to the leads.
The mapping is illustrated schematically in Fig.~\ref{fig:singleimp}.

\begin{comment}%%%%%%%%%%%%%%%%%%%%%%
Here we describe the auxiliary system as a semi-infinite linear chain,
\begin{eqnarray}\label{eq:Haux}
	\hat{H}_{\text{aux}} = \sum_{n=0}^{\infty}\sum_{\sigma} \left( t_n^{\phantom{\dagger}} \opd{f}{n\sigma} \op{f}{(n+1)\sigma} + \hc \right) \,,
\end{eqnarray}
where for simplicity we have now assumed particle-hole symmetry, $\epsilon_d=-U_d/2$ (such that $\epsilon_d^*=0$ and Eq.~\eqref{eq:impHaux} does not require inclusion of onsite potentials). The auxiliary chain is coupled at one end to the impurity,
\begin{eqnarray}\label{eq:Hhyb_aux}
	\hat{H}_{\text{imp-aux}}= \sum_{\sigma} \left( \tensor{V}{_{\text{aux}}} \opd{d}{\sigma} \op{f}{0\sigma} + \hc \right) \,.
\end{eqnarray}

The impurity-auxiliary chain hybridization function is thus given by $\Delta^{\text{aux}}(\omega)=V_{\text{aux}}^2 \tensor*{\widetilde{G}}{^{(0)}_{1,1}}(\omega)$, where 
$\tensor*{\widetilde{G}}{^{(0)}_{1,1}}(\omega) = \Green{\op{f}{0\sigma}}{\opd{f}{0\sigma}}^0_\omega$ 
is the boundary Green function of the isolated auxiliary system. The latter can be expressed simply as a continued fraction in terms of the tight-binding parameters $\{t_n\}$ in Eq.~\eqref{eq:impHaux} as $\tensor*{\widetilde{G}}{^{(0)}_{1,1}}(\omega)=1/[\omega^+ -\Delta^{\text{aux}}_0(\omega)]$ with $\Delta^{\text{aux}}_n(\omega)=t_n^2/[\omega^+-\Delta^{\text{aux}}_{n+1}(\omega)]$. 
\end{comment}%%%%%%%%%%%%%%%%%%%%%%%%%

As per the auxiliary field mapping, the auxiliary parameters $V_{\text{aux}}$ and $\{t_n\}$ are uniquely determined by setting $\Sigma(\omega) \mapsto \Delta^{\text{aux}}(\omega) =V_{\text{aux}}^2 \tensor*{\widetilde{G}}{^{(1)}_{1,1}}(\omega)$.
%Specifically, $V_{\text{aux}}^2 = -\tfrac{1}{\pi} \Im\int \d\omega \Sigma(\omega)$ initializes the iterative scheme in which successive $t_n$ are obtained by $t_{n}^2 = -\tfrac{1}{\pi} \Im \int \d\omega \Delta_n^{\text{aux}}(\omega)$. 
At half-filling, $V_{\text{aux}} = U_d/2$. The other auxiliary parameters are regular and well-behaved, with the recursion being efficient and numerically stable.
Since the original self-energy $\Sigma(\omega)$ is a continuous function, the recursion does not terminate and the auxiliary chain is semi-infinite; however the $t_n$ settle down to a regular pattern after a finite number of steps.

The key point for the present discussion is not the specific form of these parameters for a given model, but rather the fact that this mapping exists and is unique.
For the mapped non-interacting system (Fig.~\ref{fig:singleimpaux}) there are {three} effective leads, with a resonant level impurity Green function,
\begin{equation}\label{eq:auxG}
G_{\sigma}(\omega)= \frac{1}{\omega -\varepsilon_d^*/\hslash -\sum\limits_{\gamma} \tensor{\Delta}{^{\gamma}}(\omega)} \,,
\end{equation}
where $\gamma \in \{ \text{s}, \text{d}, \text{aux} \}$.

To obtain the conductance through the dot it is now necessary to calculate the current $I^{\text{d}}$ flowing from source lead to drain lead due to a source-drain bias voltage $\Delta V_{\text{b}}$. In the physical interacting system (Fig.~\ref{fig:singlei}), the linear response conductance $\mathfrak{G}_C$ follows from the Meir-Wingreen formula~\cite{meirwingreen},
\begin{equation}\label{eq:MW_cond}
	\mathfrak{G}_C(T) = \frac{e^2}{h} \int \d\omega \left[- \frac{\partial f_{\text{eq}}(\omega)}{\partial \omega} \right] \mathcal{T}(\omega,T) \;, 
\end{equation}
where $f_{\text{eq}}(\omega) =1/(1+\e^{\hslash\omega/k_{\text{B}} T})$ is the equilibrium Fermi function, and $\mathcal{T}$ effective transmission function
\begin{equation}\label{eq:trans}
    \mathcal{T}(\omega,T) = \frac{4\pi V_{\text{s}}^2 V_{\text{d}}^2 \varrho(\omega)}{V_{\text{s}}^2+V_{\text{d}}^2} \sum_{\sigma} \left [ -\Im G_{\sigma}(\omega,T) \right]
\end{equation}
in terms of the lead-coupled interacting impurity Green function Eq.~\eqref{eq:Gaim}.

For the mapped non-interacting system (Fig.~\ref{fig:singleimpaux}), it is instead possible to use the 3-terminal linearlized Landauer-B\"uttiker formula~\cite{landauerbuttiker} for the current into lead $\gamma$,
\begin{equation}\label{eq:LB}
I^{\gamma}(T) = \frac{e}{h} \sum_{\beta\ne\gamma}\int \d\omega \left[ \frac{\partial f_{\text{eq}}(\omega)}{\partial \omega} \right](\mu_{\gamma}-\mu_{\beta})\mathcal{T}_{\gamma\beta}(\omega) \;,
\end{equation}
where $\mu_{\gamma}$ is the chemical potential of lead $\gamma$ and $\mathcal{T}_{\gamma\beta}(\omega)=4\Gamma_{\gamma}(\omega)\Gamma_{\beta}(\omega)\sum_{\sigma}|G_{\sigma}(\omega)|^2$, 
with $G_{\sigma}(\omega)$ the effective non-interacting Green function given in Eq.~\eqref{eq:auxG}, and $\Gamma_{\gamma}(\omega) = - \Im \Delta^{\gamma}(\omega)$. Eq.~\eqref{eq:LB} is a generalization of the usual Landauer-B\"uttiker formula to the case with inequivalent leads with arbitrary density of states. This is important because the auxiliary `lead' has a specific form that must be accounted for. Here it is assumed for simplicity that the source and drain leads are equivalent, such that $\Gamma_{\text{s}}(\omega)=\pi V_{\text{s}}^2\varrho(\omega)$ and $\Gamma_{\text{d}}(\omega)=\pi V_{\text{d}}^2\varrho(\omega)$ with the same, but otherwise arbitrary, density of states $\varrho(\omega)$.

The auxiliary lead is not a physical lead and so there is no voltage applied to it ($\mu_{\text{aux}} = 0$), and therefore no current flows into or out of it, $I^{\text{aux}} = 0$. The latter property is also required by current conservation in the physical system, $I^{\text{s}} = -I^{\text{d}}$. From Eq.~\eqref{eq:LB} these constraints imply that $V^2_{\text{s}} \tensor*{\mu}{_{\text{s}}} + V^2_{\text{d}} \tensor*{\mu}{_{\text{d}}}=0$. The voltage bias $e \Delta V_{\text{b}} \equiv \mu_{\text{s}} - \mu_{\text{d}}$ must be split across source and drain leads in a specific way to satisfy this constraint, with $\mu_{\text{s}} = e \Delta V_{\text{b}} (1+V^2_{\text{s}}/V^2_{\text{d}})^{-1}$ and $\mu_{\text{d}} = - e \Delta V_{\text{b}} (1 + V^2_{\text{d}} / V^2_{\text{s}})^{-1}$. Note however that the $\Delta V_{\text{b}} \to 0$ linear response conductance does not depend on the details of this splitting. 
Substituting these expressions into Eq.~\eqref{eq:LB} results in the current into the drain lead being
\begin{equation}\label{eq:LBaux}
I^{\text{d}}(T) = \Delta V_{\text{b}} \frac{e^2 }{h} ~\frac{4\pi V^2_{\text{d}} V^2_{\text{s}}}{V^2_{\text{d}} + V^2_{\text{s}}} \int \d\omega \left[ -\frac{\partial f_{\text{eq}}(\omega)}{\partial \omega} \right] \varrho(\omega) \sum_{\gamma,\sigma}\Gamma_{\gamma}(\omega)|G_{\sigma}(\omega)|^2 \;.
\end{equation}
This reduces correctly to Eqs.~\eqref{eq:MW_cond} and \eqref{eq:trans} since $\sum\limits_{\gamma}\Gamma_{\gamma}(\omega)|G_{\sigma}(\omega)|^2 = - \Im G_{\sigma}(\omega)$ from the Dyson equation~\eqref{eq:auxG}. 

These arguments generalize trivially to any multi-orbital two-lead system in proportionate coupling. The Green function for the effective orbital $\op{\overline{d}}{\sigma}$ coupling to the leads can always be expressed as
\begin{equation}\label{eq:G_PC_aux}
	\overline{G}_{\sigma}(\omega)\equiv \Green{\op{\overline{d}}{\sigma}}{\opd{\overline{d}}{\sigma}}_{\omega} = \frac{1}{\omega - \varepsilon_d/\hslash - \sum\limits_{\alpha}\Delta^{\alpha}(\omega)- \Sigma'_{\sigma}(\omega)} \,,
\end{equation}
where $\Sigma'_{\sigma}(\omega)$ includes the effect of scattering from coupling of $\op{\overline{d}}{\sigma}$ to the other impurity degrees of freedom, as well as accounting for electronic interactions. Following the same steps as before, $\Sigma'_{\sigma}(\omega)$ is mapped to a single non-interacting auxiliary chain \eqref{eq:impHaux} coupled at one end to a single resonant level \eqref{eq:impHhyb}, which is also coupled to the physical source and drain leads. Schematically, the mapped system is identical to that depicted in Fig.~\ref{fig:singleimpaux}. Such an example will be illustrated for the triple quantum dot in a following subsection.

The non-proportionate coupling case is more subtle, since the equivalent non-interacting form of the transmission function for use in Eq.~\eqref{eq:LB} must be determined. In general this requires mapping the effective self-energies to {two} auxiliary chains. The mapping for the case of the serial two-impurity Anderson model is illustrated in the following subsection.

To summarize this subsection, quantum transport for interacting systems can be understood in terms of the non-interacting Landauer-B\"uttiker formula~\cite{landauerbuttiker}, in which the self-energy plays the role of an additional fictitious lead, subject to a zero-current constraint. 
This formulation provides a simple way of viewing the correction to quantum transport due to interactions. 

Furthermore, the auxiliary chain representation may provide a route to simple approximations, given its convenient structure and well-defined asymptotic form. For example, at $T=0$ in the metallic Kondo screened case,
\begin{equation}
t_n \sim  \frac{D}{2}\sqrt{1-(-1)^n\frac{2}{n+d}} \,,
\end{equation}
for large $n$, where $D$ is the effective bandwidth and $d \sim 1/Z$ is related to the quasiparticle weight $Z$. This is the same asymptotic form as the single impurity Anderson model analyzed previously in \S\ref{sec:auxsiam}. The conductance formulae can be expressed in terms of the auxiliary chain parameters.

\subsubsection{Double Quantum Dot}

Calculating the transport properties can be simplified by using transport equations for non-interacting leads.
This is made possible my mapping the interaction self-energy to an auxiliary non-interacting tight-binding chain that is then regarded as an additional lead.

The next more general system which can be analyzed through this framework is transport through an impurity which consists of two tandem quantum dots that are symmetrically hybridized to the source and drain leads as illustrated in Fig.~\ref{fig:dqdp}.

\begin{figure}[htp!]
\centering
\begin{subfigure}{\linewidth}
\centering
\begin{tikzpicture}[thick]
	\node[circle,draw=black,fill=black!10,inner sep=1pt] (imp1) at (-0.75,0) {$\uparrow\downarrow$};
	\node[circle,draw=black,fill=black!10,inner sep=1pt] (imp2) at (0.75,0) {$\uparrow\downarrow$};
	\node[above=0.5cm] at (imp1) {$\footnotesize1$};
	\node[above=0.5cm] at (imp2) {$\footnotesize2$};
	\draw[-,line width=1.5pt] (imp1)--(imp2) node[midway,above] {$t$};
	\def\p{2}
	\def\l{3}
	\def\w{0.5}
	\coordinate (s) at (-\p,0);
	\coordinate (d) at (\p,0);
	\draw (d) arc(0:-90:-\l cm and \w cm);
	\draw (d) arc(0:-90:-\l cm and -\w cm);
	\draw (s) arc(0:-90:\l cm and \w cm);
	\draw (s) arc(0:-90:\l cm and -\w cm);
	 \node at ($(s)+(-1.5,0)$) {source};
	 \node at ($(d)+(1.5,0)$) {drain};
	 \draw[-,line width=1.5pt] (s)--(imp1) node[midway,below] {$V$};
	 \draw[-,line width=1.5pt] (d)--(imp2) node[midway,below] {$V$};
	\node at ($(d)+(3,1)$) {\footnotesize \subref*{fig:dqdp}};
\end{tikzpicture}
\\
\phantomsubcaption{\label{fig:dqdp}}
\end{subfigure}
\\
\vspace{\baselineskip}
\begin{subfigure}{\linewidth}
\centering
\begin{tikzpicture}[thick]
	\node[circle,draw=black,fill=black!10,inner sep=1pt] (imp1) at (0,0.75) {$\uparrow\downarrow$};
	\node[circle,draw=black,fill=black!10,inner sep=1pt] (imp2) at (0,-0.75) {$\uparrow\downarrow$};
	\node[above=0.5cm] at (imp1) {e};
	\node[below=0.5cm] at (imp2) {o};
	\def\p{2}
	\def\l{3}
	\def\w{0.5}
	\coordinate (s) at (-\p,0);
	\coordinate (d) at (\p,0);
	\draw (d) arc(0:-90:-\l cm and \w cm);
	\draw (d) arc(0:-90:-\l cm and -\w cm);
	\draw (s) arc(0:-90:\l cm and \w cm);
	\draw (s) arc(0:-90:\l cm and -\w cm);
	 \node at ($(s)+(-1.5,0)$) {source};
	 \node at ($(d)+(1.5,0)$) {drain};
	\draw[line width=1.5pt](imp1)--(d) node[midway,above] {$\frac{V}{\sqrt{2}}$};
	\draw[line width=1.5pt](imp1)--(s) node[midway,above] {$\frac{V}{\sqrt{2}}$};
	\draw[line width=1.5pt](imp2)--(d) node[midway,below] {$-\frac{V}{\sqrt{2}}$};
	\draw[line width=1.5pt](imp2)--(s) node[midway,below] {$\frac{V}{\sqrt{2}}$};
	\node at ($(d)+(3,1.5)$) {\footnotesize \subref*{fig:dqdeo}};
\end{tikzpicture}
\\
\phantomsubcaption{\label{fig:dqdeo}}
\end{subfigure}
\caption[Interacting double quantum dot system in the physical and even/odd bases]{Interacting double quantum dot system in the physical~\subref{fig:dqdp} and even/odd~\subref{fig:dqdeo} bases.}
\end{figure}

The impurity Green function is
$\boldsymbol{G}(\omega) = \left[ \left[\boldsymbol{G}^0(\omega)\right]^{-1} - \boldsymbol{\Sigma}(\omega) \right]^{-1}$
where $\boldsymbol{G}^0(\omega)$ is the free Green function is given by
\begin{equation}
	\left[ \boldsymbol{G}^{0}(\omega) \right]^{-1} = \begin{pmatrix} \omega - \varepsilon/\hslash - \Delta(\omega) & t/\hslash \\ t/\hslash & \omega - \varepsilon/\hslash - \Delta(\omega) \end{pmatrix}
\end{equation}
where $\Delta(\omega)$ is the physical hybridization function for the leads.

The auxiliary field mapping cannot be directly implemented in this configuration as the self-energy is not a local scalar function, but rather a matrix function. The impurity can be diagonalized into a form such that the self-energy is a local scalar function by performing a decomposition into an even/odd parity basis as $\op{d}{\text{e}/\text{o}} \vcentcolon= \frac{1}{\sqrt{2}} ( \op{d}{1} \pm \op{d}{2} )$. In this basis the Green function becomes diagonal as
\begin{equation}
	\boldsymbol{G}(\omega)
	=
	\left[ \left[\boldsymbol{G}^0(\omega)\right]^{-1} - \boldsymbol{\Sigma}(\omega) \right]^{-1} = \begin{pmatrix} G_{\text{e}}(\omega) & 0 \\ 0 & G_{\text{o}}(\omega) \end{pmatrix}
\end{equation}
where
\begin{equation}
	G_{\text{e}/\text{o}}(\omega) = \frac{1}{\omega - (\varepsilon \mp t)/\hslash - \Delta(\omega) - \Sigma_{\text{e}/\text{o}}(\omega)} \,.
\end{equation}
With the self-energy now diagonalized in the even/odd basis with $\Sigma_{\text{e}/\text{o}}(\omega)$ scalar functions, it is now possible to apply the auxiliary chain mapping of substituting the self-energies with hybridizations to auxiliary subsystems by making the identification $\Delta_{\text{e}/\text{o}}(\omega) \equiv \Sigma_{\text{e}/\text{o}}(\omega)$. This auxiliary model for the double quantum dot in the even/odd basis is shown in Fig.~\ref{fig:evenodddqd}
\begin{figure}[htp!]
\centering
\begin{tikzpicture}[thick]
	\node[circle,draw=black,inner sep=4pt] (imp1) at (0,0.75) {e};
	\node[circle,draw=black,inner sep=4pt] (imp2) at (0,-0.75) {o};
	\def\p{2}
	\def\l{3}
	\def\w{0.5}
	\coordinate (s) at (-\p,0);
	\coordinate (d) at (\p,0);
	\draw (d) arc(0:-90:-\l cm and \w cm);
	\draw (d) arc(0:-90:-\l cm and -\w cm);
	\draw (s) arc(0:-90:\l cm and \w cm);
	\draw (s) arc(0:-90:\l cm and -\w cm);
	 \node at ($(s)+(-1.5,0)$) {source};
	 \node at ($(d)+(1.5,0)$) {drain};
	\draw[line width=1.5pt](imp1)--(d) node[midway,above] {$\frac{V}{\sqrt{2}}$};
	\draw[line width=1.5pt](imp1)--(s) node[midway,above] {$\frac{V}{\sqrt{2}}$};
	\draw[line width=1.5pt](imp2)--(d) node[midway,below] {$-\frac{V}{\sqrt{2}}$};
	\draw[line width=1.5pt](imp2)--(s) node[midway,below] {$\frac{V}{\sqrt{2}}$};
	\begin{scope}[scale=0.67]
	\def\first{2}
	\def\alast{3}
	\def\nalast{2}
	\foreach[evaluate=\s as \sc using (\s+0.5)] \s in {\first,...,\alast}
	{
	\node[rectangle,draw=red,inner sep=5pt] (u\s) at (0,\sc) {};
	\node[rectangle,draw=red,inner sep=5pt] (d\s) at (0,-\sc) {};
	}
	\draw[red,line width=1pt](imp1)--(u\first);
	\draw[red,line width=1pt](imp2)--(d\first);
	\foreach[evaluate=\s as \n using int(\s+1)] \s in {\first,...,\nalast}
	{
	\draw[red,line width=1pt](u\s)--(u\n) node[midway,above] {};
	\draw[red,line width=1pt](d\s)--(d\n) node[midway,above] {};
	}
	\draw[red,line width=1pt, dashed, line cap=round] (u\alast)--+(0,1) {};
	\draw[red,line width=1pt, dashed, line cap=round] (d\alast)--+(0,-1) {};
	\end{scope}
\end{tikzpicture}
\caption[Auxiliary field representation of a double quantum dot in the even/odd basis]{Auxiliary field representation of a double quantum dot in the even/odd basis. The impurities in the even/odd basis are now non-interacting sites which are hybridized to auxiliary chains (in red), which are also non-interacting. The total system is now a four-lead system without interactions.\label{fig:evenodddqd}}
\end{figure}
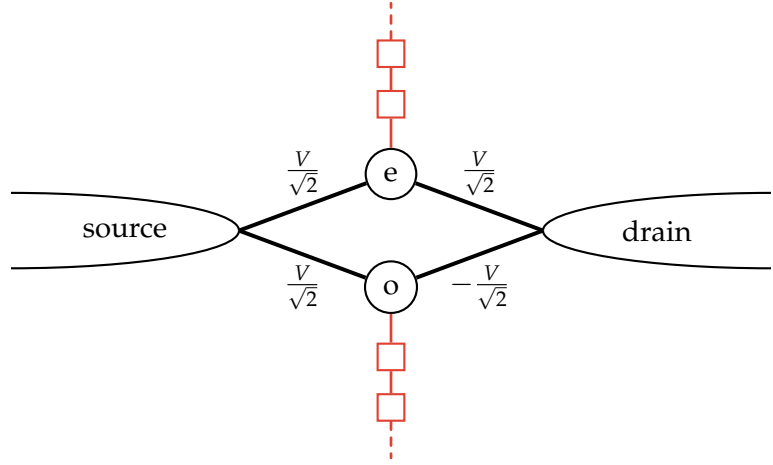

The resulting system is now a non-interacting four-lead system with effective Green functions given by
\begin{equation}
	G_{\text{e}/\text{o}}(\omega) = \cfrac{1}{\omega - (\varepsilon \pm t)/\hslash - \Delta(\omega) - \Delta_{\text{e}/\text{o}}(\omega)} \,.
\end{equation}
Transport through the dots can now be calculated via the Landauer-B\"uttiker formula for non-interacting multi-terminal transport
\begin{equation}
	I^\gamma = \frac{e}{h} \sum_{\beta \neq \gamma} \int \d\omega \left[ -\frac{\partial f(\omega)}{\partial \omega} \right] ( \mu_\beta - \mu_\gamma ) \mathcal{T}_{\gamma\beta}(\omega)
\end{equation}
where the voltage biases are
$\mu_{\text{s}} = +\frac12 e V_{\text{b}}$, $\mu_{\text{d}} = -\frac12 e V_{\text{b}}$, $\mu_{\text{e}/\text{o}} = 0$.
The 
transmission function $\mathcal{T}_{\gamma\beta}(\omega)$
is given by
\begin{equation}
	\mathcal{T}_{\gamma\beta}(\omega) = 4 \sum_{\sigma} \Tr \left[ \boldsymbol{G}_\sigma(\omega) \boldsymbol{\Gamma}_\gamma(\omega) \boldsymbol{G}^*_\sigma(\omega) \boldsymbol{\Gamma}_\beta(\omega) \right]
\end{equation}
with
$\boldsymbol{\Gamma}_{\alpha}(\omega) = -\Im \Delta_{\alpha}(\omega)$ and $\beta,\gamma \in \{\text{s}, \text{d}, \text{e}, \text{o}\}$. In the even/odd basis the
$\boldsymbol{\Gamma}$ are now frequency dependent and leads do not need to be equivalent.

\begin{equation}
\begin{aligned}
	\boldsymbol{\Gamma}_{\text{source}} &= \begin{pmatrix} \frac12 & \frac12 \\[0.25em] \frac12 & \frac12 \end{pmatrix} \Gamma \,,
	&
	\boldsymbol{\Gamma}_{\text{drain}} &= \begin{pmatrix*}[r] \frac12 & -\frac12 \\[0.25em] -\frac12 & \frac12\end{pmatrix*} \Gamma
\end{aligned}
\end{equation}
where $\Gamma = \pi V^2 \varrho_0$.

\begin{equation}
\begin{aligned}
	\boldsymbol{\Gamma}_{\text{even}} &= \begin{pmatrix} -\Im\Sigma_{\text{even}}(\omega) & 0 \\ 0 & 0 \end{pmatrix}
	&
	\boldsymbol{\Gamma}_{\text{odd}} &= \begin{pmatrix} 0 & 0 \\ 0 & -\Im\Sigma_{\text{odd}}(\omega) \end{pmatrix}
\end{aligned}
\end{equation}

%$-\Im\Delta_{\text{e}/\text{o}}(\omega) = -\Im\Sigma_{\text{e}/\text{o}}(\omega)$

The linear response conductance is
\begin{equation}
	\mathfrak{G}_C \equiv \frac{I^{\text{d}}}{V_{\text{b}}} = \frac{2e^2}{h} \int \d\omega \left[ -\frac{\partial f(\omega)}{\partial \omega} \right] \mathcal{T}_{\text{eff}}(\omega)
\end{equation}
with transmission function
\begin{equation}
	\mathcal{T}_{\text{eff}}(\omega) = \Gamma^2 \left\vert G_{\text{e}}(\omega) - G_{\text{o}}(\omega) \right\rvert^2 - \Gamma \, \Im \Sigma_{\text{e}}(\omega) \left\lvert G_{\text{e}}(\omega) \right\rvert^2 - \Gamma \, \Im \Sigma_{\text{o}}(\omega) \left\lvert G_{\text{o}}(\omega) \right\rvert^2 \,.
\end{equation}
The transmission function
reduces to the Oguri formula~\cite{oguri} at $\omega = 0$,
\begin{equation}
\begin{aligned}[b]
	\mathcal{T}_{\text{eff}}(0)
	&=	\Gamma^2 \left\lvert G_{\text{e}}(\omega) - G_{\text{o}}(\omega) \right\rvert^2
	\\
	&\equiv	\Gamma^2 \left\lvert G_{12}(\omega) \right\rvert^2 \,.
\end{aligned}
\end{equation}
In order to match the behavior of the original physical system it is required that
$I^{\text{e}/\text{o}} = 0$ and $I^{\text{s}} = -I^{\text{d}}$.

There does however appear to be a complication with this implementation. In making the transformation into the even/odd basis and decomposing the local even/odd self-energy into the auxiliary chain form, the voltage bias applied to the source and drain leads does not appear to be the same as the voltage bias applied to the physical source and drain leads of the original system. Simply imposing that $\mu_{\text{e}} = 0 = \mu_{\text{o}}$ is apparently insufficient to fully reproduce the exact conductance. Further analysis is left for future investigation.

\subsubsection{Triple Quantum Dot}

\begin{figure}[htp!]
\centering
\begin{tikzpicture}[{thick}]
	\node[circle,draw=black,fill=black!10,inner sep=1pt] (1) at (0,0) {$\uparrow\downarrow$};
	\node[circle,draw=black,fill=black!10,inner sep=1pt] (2) at ($(1)+(-0.707,1.125)$) {$\uparrow\downarrow$};
	\node[circle,draw=black,fill=black!10,inner sep=1pt] (3) at ($(1)+(0.707,1.125)$) {$\uparrow\downarrow$};
	\node[below=0.25cm] at (1) {$1$};
	\node[above left=0.125cm] at (2) {$2$};
	\node[above right=0.125cm] at (3) {$3$};
	\draw[-] (1)--(2) node[midway,left] {$t$};
	\draw[-] (3)--(2) node[midway,above] {$t'$};
	\draw[-] (1)--(3) node[midway,right] {$t$};
	\def\p{1.25}
	\def\l{3}
	\def\w{0.5}
	\coordinate (s) at (-\p,0);
	\coordinate (d) at (\p,0);
	\draw (d) arc(0:-90:-\l cm and \w cm);
	\draw (d) arc(0:-90:-\l cm and -\w cm);
	\draw (s) arc(0:-90:\l cm and \w cm);
	\draw (s) arc(0:-90:\l cm and -\w cm);
	 \node at ($(s)+(-1.5,0)$) {source};
	 \node at ($(d)+(1.5,0)$) {drain};
	 \draw[-,line width=1.5pt] (s)--(1) node[midway,below] {$V$};
	 \draw[-,line width=1.5pt] (d)--(1) node[midway,below] {$V$};
\end{tikzpicture}
\caption{Triple quantum dot in symmetric proportionate coupling.\label{fig:tqd}}
\end{figure}
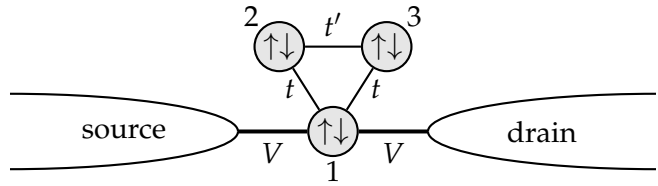
The triple quantum dot consists of three sites with local interactions all of which are hybridized to each other with amplitudes $t_{ij}$. The case taken under consideration here is with the triple quantum dot in the proportionate coupling configuration, where the source and drain leads are symmetrically hybridized to the same element of the dot (site $1$ of the dot). The configuration will also be taken to be the symmetric coupling case where the couplings from site $1$ to the sites $2$ and $3$ are equal, $t_{12} = t = t_{13}$, but the coupling between sites $2$ and $3$ is different, $t_{23} = t' \neq t$. This configuration is diagrammed in Fig.~\ref{fig:tqd}.
This system is described by the Hamiltonian
$\hat{H} = \hat{H}_{\text{TQD}} + \hat{H}_{\text{hyb}} + \hat{H}_{\text{leads}}$
where
\begin{subequations}
\begin{align}
	\hat{H}_{\text{TQD}} &=
	\begin{multlined}[t]
	\sum_{j=1,2,3} \left[ \varepsilon \left( \hat{n}_{j\uparrow} + \hat{n}_{j\downarrow} \right) + U \hat{n}_{j\uparrow} \hat{n}_{j\downarrow} \right]
	\\
	+ t \sum_{\sigma} \left( \opd{d}{1\sigma} \op{d}{2\sigma} + \opd{d}{1\sigma} \op{d}{3\sigma} + \hc \right)
	+ t' \sum_{\sigma} \left( \opd{d}{2\sigma} \op{d}{3\sigma} + \hc \right) \,,
	\end{multlined}
	\\
	\hat{H}_{\text{hyb}} &= V \sum_{\sigma} \left( \opd{d}{1\sigma} \op{c}{\sigma} + \hc \right) \,,
\end{align}
\end{subequations}
with the leads Hamiltonian taking the same form as Eq.~\eqref{eq:transportleads} and $\op{c}{\sigma} = \frac{1}{\sqrt{2\pi}} \sum_k \tensor{V}{_k} \op{c}{k}$.

The matrix Green function on the quantum dot is given by
\begin{align}
	\boldsymbol{G}_{\text{TQD}}(z) &= \left[ z \mathbbm{1} - \boldsymbol{h} - \boldsymbol{\Delta}(z) - \boldsymbol{\Sigma}(z) \right]^{-1}
\intertext{with}
	\boldsymbol{h} &= \frac1\hslash \begin{pmatrix} \varepsilon & t & t \\ t & \varepsilon & t' \\ t & t' & \varepsilon \end{pmatrix}
\intertext{and}
	\boldsymbol{\Delta}(z) &= \begin{pmatrix} V^2 G^0_{\text{bath}}(z) & 0 & 0 \\ 0 & 0 & 0 \\ 0 & 0 & 0 \end{pmatrix} \,,
\end{align}
where $-\frac1\pi \Im G^0_{\text{bath}}(\omega) = \rho(\omega)$ as before.
The impurity problem for this quantum dot is solved using NRG, yielding a $\Sigma(\omega)$ which may then serve as the input for the auxiliary field mapping. The noninteracting auxiliary system for the triple quantum dot is of the same form as that used for the single impurity system as shown schematically in Fig.~\ref{fig:singleimpaux}. Analysis reveals that the triple dot exhibits a local moment phase when $t' < t$, and exhibits a strong coupling phase when $t' > t$~\cite{akmtqd}.

For very low energy scales, corresponding to very long distances down the auxiliary chain, the recursion algorithm no longer returns accurate results. However, important characteristics of the self-energy appear at very low energy scales of $\omega \lesssim 10^{-6}$. At such low energy scales the recursion algorithm is unable to properly capture the resolution of the input data and the reconstructed spectrum no longer matches the self-energy. In order to construct an auxiliary chain which captures these features, it is necessary to perform the recursion calculation on a data set which has a rescaled frequency axis. This is done by rescaling the $\omega$ of the input data by a factor of $10^{4}$, such that the breakdown scale of $10^{-6}$ in the rescaled data corresponds to a scale of $10^{-10}$ in the original unscaled data. The resulting $t_n$ parameters of the auxiliary chain as calculated by the recursion algorithm are then also scaled by a factor of $10^{4}$. This rescaled calculation is shown in the right-hand panel of Fig.~\ref{fig:tqdparameters} with the unscaled calculation for $\omega \gtrsim 10^{-6}$ plotted in the left-hand panel.

The system parameters for the calculation are chosen to be $U = 0.4$, $\epsilon = -U/2$, $V = 0.1$, and $t = 0.0005$ with conduction bandwidth $D=1.0$. The remaining parameter $t'$ is tuned to alter the phase of the triple dot impurity. For the local moment phase, the parameter $t'$ is chosen to be $t' = 0 < t$, and is chosen to be $t' = 0.01 > t$ for the strong coupling regime.
\begin{figure}[htp!]
\centering
	\includegraphics[width=0.67\linewidth]{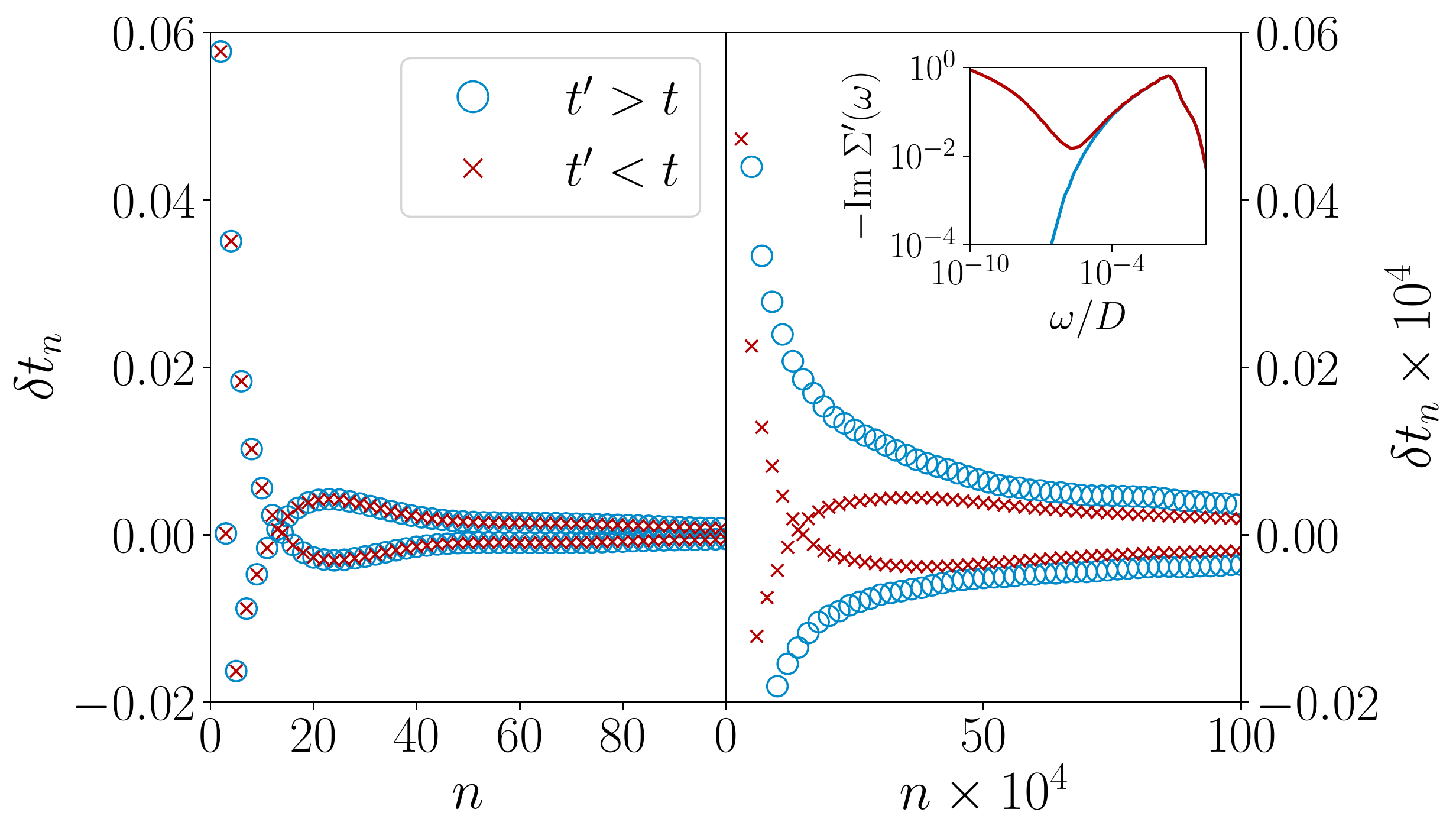}
	\caption[Auxiliary chain parameters for the triple quantum dot]{Auxiliary chain parameters for the triple quantum dot. To clearly show the envelope structure only the values of $\delta t_n$ are plotted, the $t_n$ are still exhibit the characteristic alternating pattern. $\Sigma'_{\sigma}(\omega)$ in the inset denotes the self-energy including the effect of scattering from coupling to the other impurity degrees of freedom. Figure reproduced from~\cite{multiorbitaltransport}.\label{fig:tqdparameters}}
\end{figure}
%Parameters are: U=0.4; eps=-0.2; V=0.1; t=0.0005 and t'=0 (local moment phase) or 0.01 (strong coupling phase)
Fig.~\ref{fig:tqdparameters} shows that at higher energies where the self-energies of the two models agree, the corresponding $t_n$ are also essentially identical for $n<100$. At lower energies, the self-energies become very different, with $-\Im\Sigma(\omega) \sim \omega^2$ in the Kondo strong coupling phase and $-\Im\Sigma(\omega) \sim \frac{1}{\ln\lvert\omega\rvert}$ in the local moment phase. In this regime the $t_n$ also become different at large $n$, $n \gg 100$. In particular there is a crossing point in the envelope of the $\delta t_n$ at $n_c$ in the local moment phase, corresponding to the crossover behavior of the self-energy at $\omega_c \sim 1/n_c$.
Importantly, the rather simple form of the $\delta t_n$ may motivate the development of simple approximate toy models for the self-energy in interacting systems.

\section{Outlook}

Presented in this chapter were the development of two novel methods of mapping interacting systems to non-interacting systems while preserving the original system's dynamics.

The first of these methods involved the introduction of gauge degrees of freedom and implementation of non-linear canonical transformations applied to fermionic systems in their Majorana basis. The Majorana modes exhibited by the standard decomposition spurred a great deal of theoretical and experimental research due to their appearance in the Kitaev superconducting wire. It is possible the generalized decomposition developed here could also lead to physically interesting models where Majorana modes appear nontrivially.

The second scheme presented here was based on generating an effective model from the original system's spectrum or self-energy involving auxiliary non-interacting degrees of freedom.

While the choice of auxiliary system here was taken to be a $1d$ tight-binding model, it is possible that other systems may be chosen as well. Such an example might be a semi-infinite square lattice, where the impurity site is hybridized to the end corner of the square.

It is also plausible to extend the scheme constructed here to the case where the self-energy is not purely local and possesses momentum dependence. For this situation it is envisioned that the auxiliary systems would take the form of interconnected ladders rather than simple disjoint chains.

The difficulty in analyzing strongly correlated systems motivates the development of approaches which aim to reduce the complexity of the calculations involved. The auxiliary field mapping developed in this chapter is one such example which simplifies the system to be analyzed by mapping it to a fully non-interacting system which preserves the dynamics of the original strongly interacting system. As this mapping requires already the solution to the self-energy, it is not a method for solving strongly interacting systems, but serves as a platform for determining characteristics and behavior of such systems, as demonstrated by the example of its application in quantum transport situations above.

Even though it is a powerful tool, DMFT is constrained computationally primarily by the need to solve the local impurity problem. While NRG is the impurity solver employed throughout this thesis, another common solver is exact diagonalization. Exact diagonalization has some advantages in that the impurity problem can be solved exactly in equilibrium, as well as non-equilibrium cases.

In the auxiliary mappings developed in this chapter, additional (non-interacting) degrees of freedom are added to the system to take into account the dynamics of the interactions. A method which takes a converse approach was developed in~\cite{weberembedding}. The central concept of this approach is to emulate a non-interacting bath with many degrees of freedom with an interacting bath of few degrees of freedom.
The context of this work is that of the utilizing exact diagonalization as the impurity solver within DMFT. While this solver allows the production of the exact impurity model spectrum, also in the full non-equilibrium situation, it is computationally limited by the size of the bath and produces a severely discretized spectrum.
The key idea presented in~\cite{weberembedding} is that the computational cost for calculating the spectrum of an Anderson impurity model with a non-interacting bath is the same as for that with an interacting bath. Compared to the non-interacting bath, diagonalization of the interacting bath results in a much higher density of poles in the discrete spectrum. Use of an interacting bath is therefore analogous to using a much larger non-interacting bath, but at the same computational cost.
There is however no guarantee that the system with the interacting bath generates the same physics as the original system with the non-interacting bath. The discrepancy here is due to the fact that in the non-interacting case the self-energy is a scalar function which exists only on the impurity site, whereas for an interacting bath the self-energy is a matrix function existing also on the bath.
In~\cite{weberembedding} this is partially compensated for by tuning the bath interactions such that the self-energy takes a block diagonal structure. This limits the non-local contributions to the self-energy which then leads to the interacting bath producing more accurately the physics of the non-interacting bath.
This strategy provides another method for treating strongly correlated systems with a reduced computational cost.
There are clear conceptual similarities with the auxiliary field approach developed in this chapter.

The auxiliary field mapping constructed in \S\ref{sec:mapping} provides a numerical method of producing an effective model of strongly correlated systems which is fully non-interacting. Up to now analysis of the form of the auxiliary systems has been brief. The next chapter presents a detailed study of these auxiliary systems within the context of the Mott transition in the Hubbard model.

\chapter{The Mott Transition as a Topological Phase Transition\label{ch:motttopology}}
\chaptermark{Mott Topology}

This chapter presents a more elaborate application of the auxiliary field mapping developed in the previous chapter. In doing so, it uncovers a relationship between the Mott metal-insulator transition in the Hubbard model with the topological phase transition in the SSH model. The form of the auxiliary chains belong to the classes of generalized SSH models which were discussed in~\S\ref{ch:genssh}. The case of mid-gap bands with finite width as well as the case of power-law spectral functions which were only briefly discussed in~\S\ref{ch:genssh} are examined here in more detail within the context of the Hubbard model's auxiliary field mapping.

This chapter is based on Ref.~\cite{motttopology}.

To reiterate from \S\ref{ch:methods}, the one band Hubbard model is described the Hamiltonian
\begin{equation}
	\hat{H}_{\textsc{h}} = \varepsilon \sum_{j,\sigma} \opd{c}{j,\sigma} \op{c}{j,\sigma} + {t} \sum_{j,\ell,\sigma} \left( \opd{c}{j,\sigma} \op{c}{j+\ell,\sigma} + \opd{c}{j+\ell,\sigma} \op{c}{j,\sigma} \right) + U \sum_{j} \opd{c}{j,\uparrow} \op{c}{j,\uparrow} \opd{c}{j,\downarrow} \op{c}{j,\downarrow}
\tag{\ref*{eq:hubbard}}
\label{eq:hubbard0}
\end{equation}
where $j\in\Gamma_{\textsc{bl}}$ in the following.

One of the initial motivations for this work lies in the observation that the character of the Hubbard model self-energy as the system undergoes the Mott metal-insulator transition is qualitatively similar to the SSH model as it undergoes its topological phase transition. Compare Fig.~\ref{fig:analogy} for the Hubbard model self-energy with the SSH model spectrum in its topological and trivial phases, Fig.~\ref{fig:sshbands}.
\begin{figure}[h]
\subfiglabel{\includegraphics{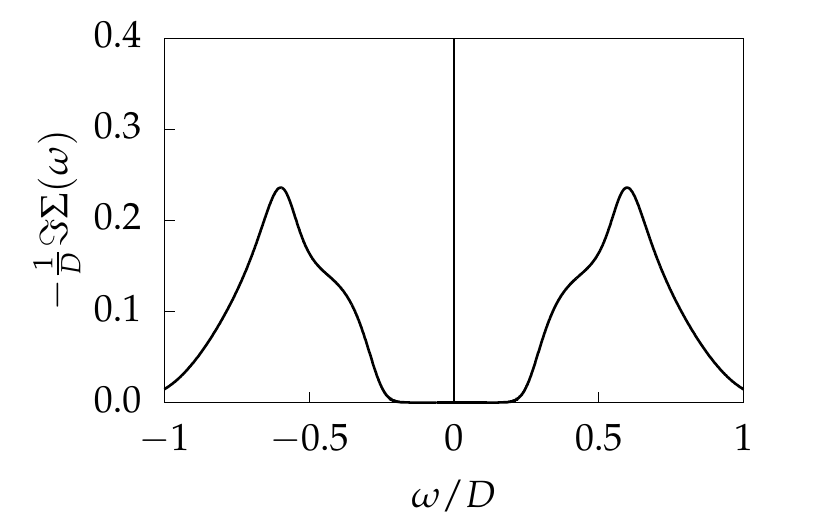}}{3.125,2}{fig:SU9}
\subfiglabel{\includegraphics{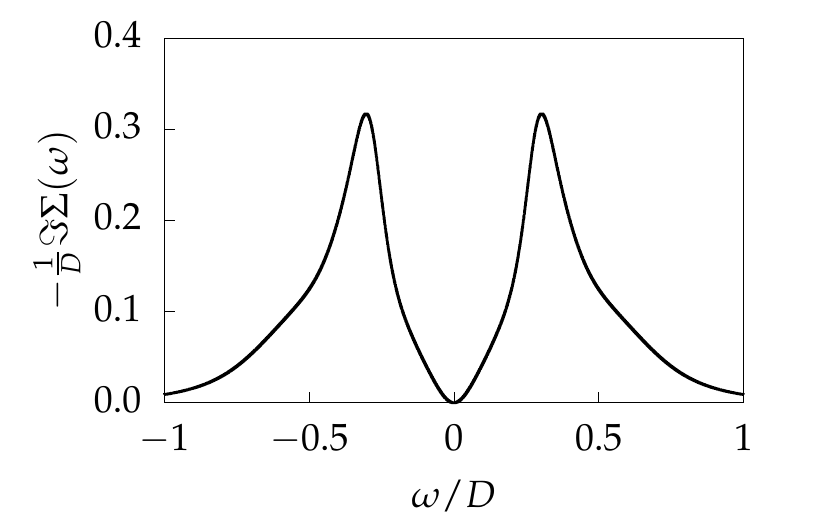}}{3.125,2}{fig:SU3}
\caption{Self-energy of the Hubbard model at half-filling in the Mott insulator ($U/t=9$ with bandwidth $D=4$) \subref{fig:SU9} and metallic phases ($U/t=3$ with bandwidth $D=3$) \subref{fig:SU3}.\label{fig:analogy}}
\end{figure}
%\begin{figure}[h]
%\centering
%\includegraphics{SU5_82zoom.pdf}
%\end{figure}
%The position of the peaks is proportional to $\sqrt{Z}$ where $Z$ is the quasiparticle weight.
As shown in Fig.~\ref{fig:analogy}, the self-energy of the Hubbard model possesses a phase which is (pseudo)gapped, $-\Im\Sigma(\omega) \sim \omega^2$ for $|\omega|\ll D$, and another phase in which there exists a localized spectral pole within a hard gap, $-\Im\Sigma(\omega) \sim \delta(\omega)$ for $|\omega|\ll D$. This is qualitatively similar to the SSH model which features a gapped phase, and a phase involving a pole within a spectral gap.

In this chapter the critical interaction strength for the Mott transition is taken to be $U_{c} = U_{c2}$, and the physics across this transition is explored. %The results are qualitatively identical for $U_{c1}$.

\section{Mapping to Effective Model}
\sectionmark{The Effective Model}

Following the procedure developed in \S\ref{sec:mapping}, the local self-energy of the infinite-dimensional Hubbard model can be mapped to an effective $1d$ tight-binding chain. 
For the case of the Hubbard model on the Bethe lattice, this mapping is shown schematically in Fig.~\ref{fig:hubbardmapping}. 
\begin{figure}[h]
\centering
\begin{tikzpicture}[every node/.style={line width=1pt,scale=1.0}, every path/.style={line width=1pt,scale=1.0}]
\input{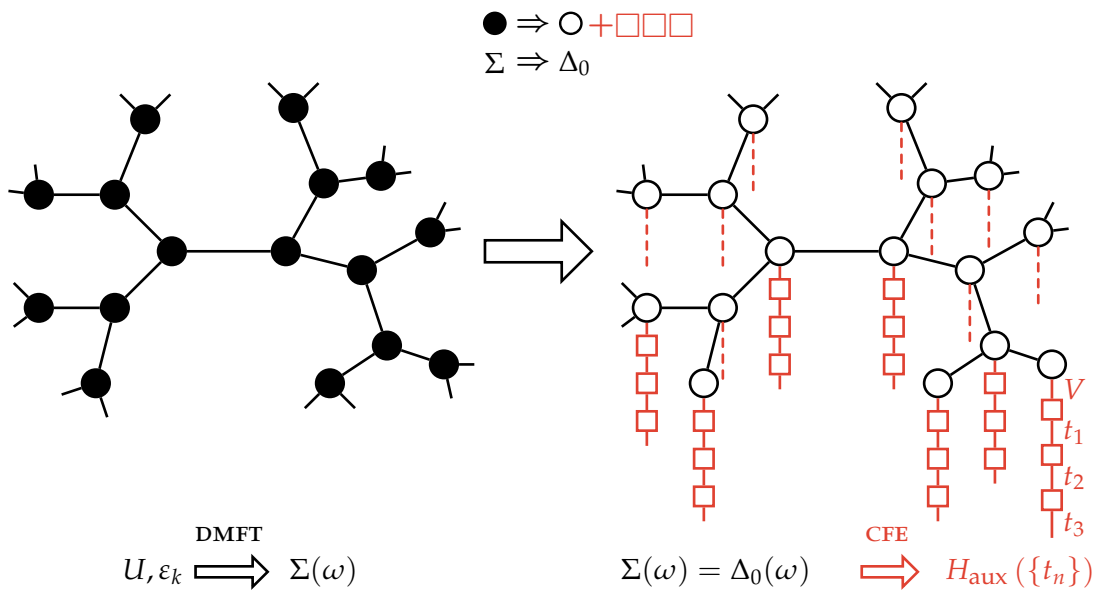}
\end{tikzpicture}
\caption[Auxiliary chain mapping of the Hubbard model]{Auxiliary chain mapping of the Hubbard model on the Bethe lattice. The original fully interacting system (left) with on-site interactions (filled circles) is mapped to a system (right) where the self-energy from the local interactions $\Sigma$ has been substituted by hybridization $\Delta$ to auxiliary degrees of freedom (red boxes), resulting in a fully non-interacting system (open circles). Figure adapted from~\cite{motttopology}.\label{fig:hubbardmapping}}
\end{figure}
As a preliminary step, the input self-energy for the effective mapping is calculated for the infinite-dimensional Hubbard model on the Bethe lattice using DMFT-NRG. The procedure as developed in \S\ref{sec:mapping} was for impurity models where the self-energy is localized on the impurity site. In application to the Hubbard model, it is necessary to perform the mapping within the context of DMFT in the infinite dimensional limit to ensure a momentum independent self-energy. The DMFT equations are solved self-consistently so that the local self-energy $\Sigma(\omega)$ at convergence is the proper lattice self-energy, which is local for the infinite dimensional Bethe lattice. In the case of finite systems with a momentum dependent self-energy, the mapping method needs to be modified. This is left for future investigation

The mapping is performed first for the Hubbard model at $T=0$ and particle-hole symmetry with
the $T=0$ particle-hole asymmetric case discussed later in \S\ref{sec:phasymmmap}.

The complete Hamiltonian of the effective model shown in Fig.~\ref{fig:hubbardmapping} is given by
\begin{subequations}
\begin{align}
	\op{H}{\text{eff}} &= \varepsilon \sum_{j,\sigma} \opd{c}{j,\sigma} \op{c}{j,\sigma} + \tilde{t} \sum_{j,\ell,\sigma} \left( \opd{c}{j,\sigma} \op{c}{j+\ell,\sigma} + \hc \right) + \hat{H}_{\text{hyb}} + \hat{H}_{\text{aux}}
\\
	\op{H}{\text{hyb}} &= V \smashoperator[lr]{\sum_{\substack{j \in \Gamma_{\textsc{bl}} \\ \sigma \in \{\uparrow,\downarrow\}}}} \left( \opd{c}{j,\sigma} \op{f}{j,\sigma;1} + \hc \right) \label{eq:Hhyb}
\\
	\op{H}{\text{aux}} &= \sum_{\substack{j \in \Gamma_{\textsc{bl}} \\ \sigma \in \{\uparrow,\downarrow\}}} \sum_{n=1}^{\infty} \left[ \tensor{e}{_{n}} \opd{f}{j,\sigma;n} \op{f}{j,\sigma;n} + \tensor{t}{_{n}} \left( \opd{f}{j,\sigma;n} \op{f}{j,\sigma;n+1} + \hc \right) \right] \,. \label{eq:Haux}
\end{align}
\end{subequations}
The outer sums sum over all sites $j$ of the physical Bethe lattice $\Gamma_{\textsc{bl}}$ for each spin $\sigma$. $\tilde{t}$ is the regularized hopping amplitude between sites on the Bethe lattice, $\tilde{t} = {t}/\sqrt{\kappa}$, with $t$ the bare hopping amplitude (without index). In $\hat{H}_{\text{aux}}$ the inner sum captures the dynamics of the auxiliary degrees of freedom where the $n$ index labels sites within each auxiliary chain. Within this model there is one auxiliary chain per physical lattice site; auxiliary chains are decoupled due to the local nature of the self-energy in the infinite dimensional limit.

Before preceding to the results, a short technical remark is in order with respect to the treatment of the Mott pole in the self-energy of the Mott insulating phase. These remarks are analogous to those discussed in \S\ref{sec:cfetechnicalities} for Fermi liquid inputs to the auxiliary field mapping. The notation here follows that discussion.

For a Mott insulator, $\Delta_0(\omega)$ is hard-gapped for $\omega\in[-\delta,\delta]$, where $2\delta$ is the size of the Mott gap. Inside the gap resides the zero-energy Mott pole, such that $\Delta_0(\omega\to 0)=\frac{\alpha_0}{\omega^+}$, \textit{i.e.} the only low energy contribution to the hybridization comes from the singular Mott pole. Based on the analysis for the Fermi liquid self-energy in \S\ref{sec:cfetechnicalities}, it is readily observed that in the Mott insulating case, the role of odd and even chain sites is now interchanged. 
Upon iteration of recursion algorithm $\Delta_{n+1}(z) = z - \frac{t_n^2}{\Delta_{n}(z)}$, the next hybridization function is found to be $\Delta_1(\omega\to0) = \omega^+ - \frac{t_0^2}{\alpha_0} \omega^+$. 
%\index{$0$@\textbf{List of Edits}!601@improved $\beta = 0$, $b=0$ explanation}
Note that with reference to the form of the even and odd hybridizations in Eq.~\eqref{eq:flhybparams}, the $\beta_n = 0$ for all (even) $n$ for the Mott insulator because the pole is sitting inside the Mott gap and the $b_n$ coefficients are also zero for all odd $n$, as observed from the form of $\Delta_1(\omega\to0)$. The low-energy asymptotic behavior and coefficient recursion (for $n\ge1$) follows,
\begin{subequations}
\begin{align}
	\Delta_{2n-1}(\omega\to 0) &= a_{2n-1}\omega & (a_{2n-1}&<0) \,,
	\\
	\Delta_{2n}(\omega\to 0) &= \frac{\alpha_{2n}}{\omega^+}\,,
\end{align}
\end{subequations}
where
\begin{subequations}
\begin{align}
	a_{2n-1} &= 1-\frac{t_{2n-2}^2}{|\alpha_{2n-2}|} \,,
	&
	b_{2n-1} &= 0\,,
	\\
	\alpha_{2n} &= \frac{t_{2n-1}^2}{|a_{2n-1}|} \,,
	&
	\beta_{2n} &= 0 \,.
\end{align}
\end{subequations}
Thus the pole structure of the Mott insulator $\Delta_n(\omega)$ is reversed with respect to the Fermi liquid. 
Importantly, the imaginary part of the hybridizations $\Delta_n(\omega)$ is hard gapped for all $n$, but contains a mid-gap zero energy pole for all \textit{even} $n$.

\section{Topological Phases of the Effective Models}
\sectionmark{Topological Phases}

%%%%%%%%%%%%%%%%%%%%%%%%%%%%%%%%%%
\subsection{Mott Insulator Regime}
%%%%%%%%%%%%%%%%%%%%%%%%%%%%%%%%%%

For interaction strength $U>U_c$, the Hubbard model Eq.~\ref{eq:hubbard0} describes a Mott insulator, with two Hubbard bands separated by a hard spectral gap of width $2\delta$. The corresponding self-energy at zero temperature is shown in Fig.~\ref{fig:SU9}, obtained by DMFT-NRG for $U/t=9$. The imaginary part of the self-energy features a mid-gap `Mott pole'\index{Mott pole} throughout the Mott insulating phase, pinned at $\omega_{\textsc{mp}}=0$ (and with finite weight at the transition).  

\begin{comment}%###########################
\begin{figure}[h]
\includegraphics[width=1.0\columnwidth]{fig2.pdf}
  \caption{Lattice self-energy at $T=0$ obtained from DMFT-NRG [panels (a,d)] and corresponding $t_n$ of the auxiliary chain [panels (b,e)]. Left panels show results for the MI ($U/t=9$, $D=4$): the hard gap in $\Im\Sigma(\omega)$ and the Mott pole at $\omega=0$ produce an SSH-type chain in the topological phase, hosting an exponentially-localized boundary zero mode, panel (c). Right panels show the metallic FL ($U/t=3$, $D=3$): the low-energy $\omega^2$ psuedogap in $\Im\Sigma(\omega)$ produces a generalized SSH chain with $1/n$ decay, in the trivial phase.}
  \label{fig:2}
\end{figure}
\end{comment}%###########################

Mapping to the auxiliary non-interacting chain, Eq.~\ref{eq:Haux}, leads to a model of modified SSH type, as exhibited in Fig.~\ref{fig:mit}. In particular, the hard gap in $\Im\Sigma(\omega)$ generates an alternating sequence of $t_n$ in $\hat{H}_{\text{aux}}$ at large distances from the physical degrees of freedom,
\begin{equation}
    \label{eq:tn_MI}
    t_n ~~ \overset{\frac{n \delta}{D} \gg 1}{\sim} ~~ \tfrac{1}{2}[D+(-1)^n\delta] \,.
\end{equation}
In the Mott insulator phase, the auxiliary chain parameters are alternating for all $n$, starting from a \textit{weak} bond ($t_1<t_2$). It is this feature that produces the Mott mid-gap pole at $\omega=0$. Additional structure in the Hubbard bands merely gives rise to transient structure in the $t_n$ for small $n$, but importantly the parity of the alternation, $t_{2n-1}/t_{2n}<1$, is preserved for all $n$.

The SSH model in its topological phase (Eq.~\ref{eq:Haux} with $t_n$ given by Eq.~\ref{eq:tn_MI} for all $n\ge 1$) hosts an exponentially-localized boundary zero-mode that is robust to parity-preserving perturbations~\cite{ssh,shortcourse}. Similarly, the zero-energy Mott pole corresponds to a robust and exponentially-localized state living at the end of the auxiliary chain (on its boundary with the physical degrees of freedom of the original lattice). This can be readily seen from the transfer matrix method, which gives the wavefunction amplitude of the zero-energy state at odd sites $(2n-1)$ of $\hat{H}_{\text{aux}}$ as
\begin{equation}
	|\psi_0(2n-1)|^2\sim \prod_{x=1}^{n} \frac{t_{2x-1}}{t_{2x}} ,
\end{equation}
which at large $n$ decays exponentially as $\exp(-n/\xi)$ with $\xi \approx D/2\delta$ for small $\delta$ 
(while $|\psi_0(2n)|^2=0$ for all $n$)~\cite{ssh,shortcourse}. The boundary-localized nature of this zero-mode state is confirmed by exact diagonalization of $H_{\text{aux}}$, shown in Fig.~\ref{fig:miwavefunction}. 

\begin{figure}[h]
\begin{subfigure}{0.36\linewidth}
\begin{tikzpicture}
	\node at (0,0) {\includegraphics[scale=1]{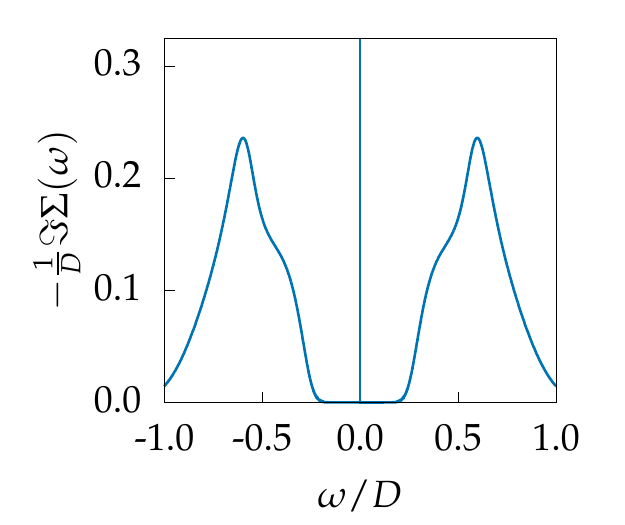}};
	\node at (2.2,2) {\footnotesize\subref*{fig:mis}};
\end{tikzpicture}
\phantomsubcaption{\vspace{-\baselineskip}\label{fig:mis}}
\end{subfigure}
\begin{subfigure}{0.69\linewidth}
\begin{tikzpicture}
	\node at (0,0) {\includegraphics[scale=1]{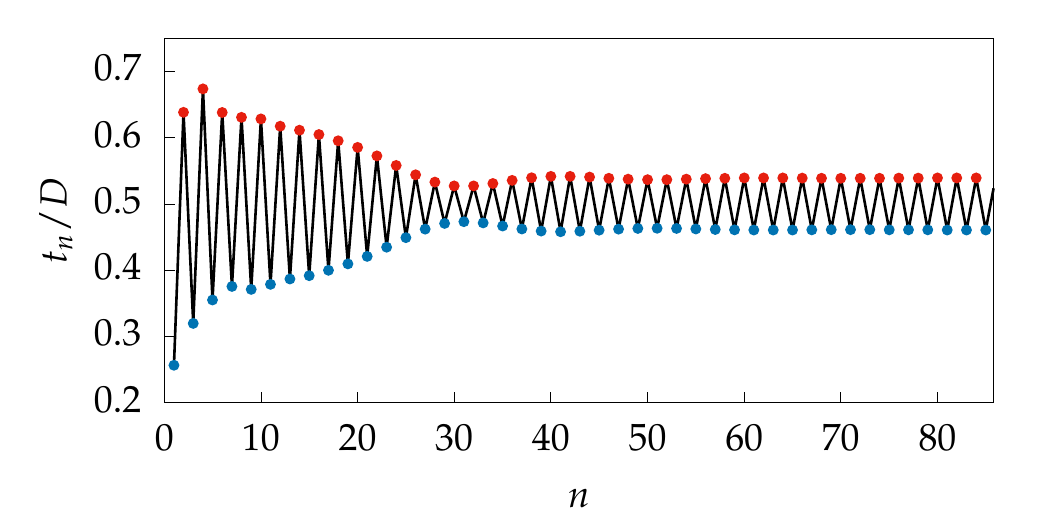}};
	\node at (4.4,2) {\footnotesize\subref*{fig:mit}};
\end{tikzpicture}
\phantomsubcaption{\vspace{-\baselineskip}\label{fig:mit}}
\end{subfigure}
\caption{Self-energy \subref{fig:mis} and effective chain parameters \subref{fig:mit} corresponding to the Mott insulating phase, $U/t = 9$. Analogously to the topological phase of the SSH model, the auxiliary chain is initialized with a weak bond.\label{fig:miresult}}
\end{figure}

\begin{figure}[h]
\subfiglabel{\includegraphics[scale=1]{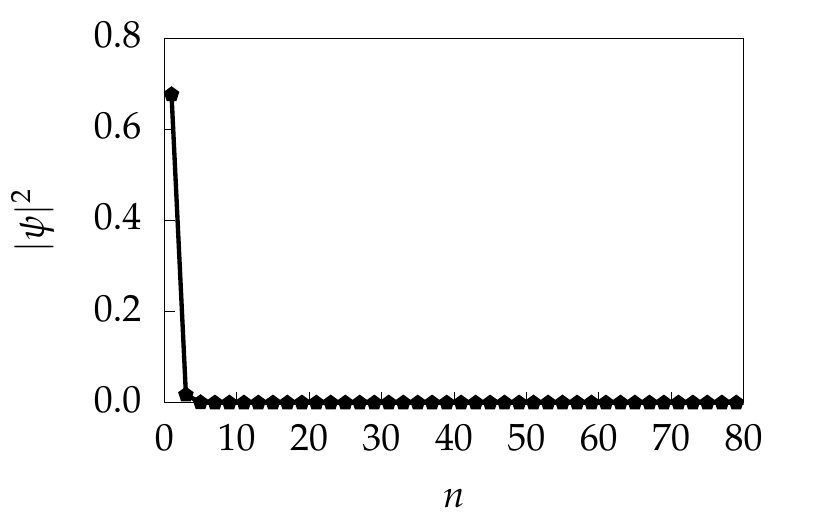}}{3.125,2}{fig:miwf}
\subfiglabel{\includegraphics[scale=1]{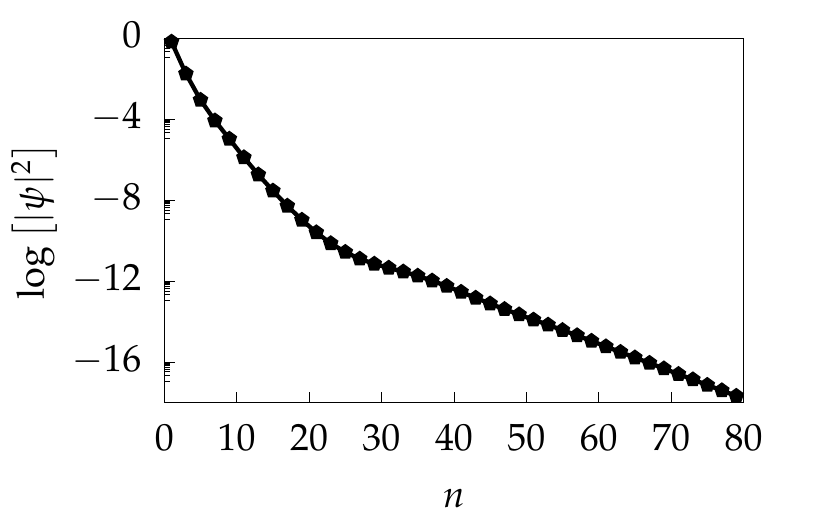}}{3.125,2}{fig:miwflog}
\caption[Wavefunction amplitude]{Zero energy wavefunction amplitude of the effective chain. The wavefunction exhibits exponential localization. Only the wavefunction on odd sites is plotted as the wavefunction has zero support on the even sites.\label{fig:miwavefunction}}
\end{figure}

%%%%%%%%%%%%%%%%%%%%%%%%%%%%%%%%
\subsection{Fermi Liquid Regime\label{sec:flregime}}
%%%%%%%%%%%%%%%%%%%%%%%%%%%%%%%%

For $U<U_c$, the Hubbard Hamiltonian describes a correlated metal, with low-energy Fermi liquid properties characterized by a quadratic dependence of the self-energy, $-t\Im\Sigma(\omega\rightarrow 0) \sim (\omega/Z)^2$, in terms of the quasiparticle weight $Z$~\cite{hewson,bullahubbard}\index{quasiparticle weight}. The quasiparticle weight is obtained from the real part of the self-energy and is given by~\cite{marino,bullahubbard}
\begin{equation}
	Z = \left( 1 - \left. \frac{\d \Re \Sigma}{\d \omega} \right\rvert_{\omega=0} \right)^{-1} \,.
\end{equation}
Fig.~\ref{fig:fls} shows the $T=0$ self-energy deep in the Fermi liquid phase, obtained by DMFT-NRG for $U/t=3$.
A distinctive form for the auxiliary chain hopping parameters is obtained from the continued fraction expansion, arising due to the low-energy pseudogap in $\Im\Sigma(\omega)$, 
\begin{equation}
	t_n \overset{nZ\gg1}{\sim} \frac{D}{2} \sqrt{1 - (-1)^{n} \frac{r}{n+d}}
    \label{eq:tn_FL}
\end{equation}
where $r=2$ is the exponent of the low-energy spectral power-law, and $d\sim 1/Z$. 
This result follows from the scaling behavior observed in \S\ref{sec:pseudogapssh}.

Eq.~\ref{eq:Haux} with hopping parameters $t_n$ given by Eq.~\ref{eq:tn_FL} generalizes the standard hard-gapped SSH model to the pseudogapped case: the alternating sequence of $t_n$ again has a definite parity, but with a decaying $1/n$ envelope. Since $t_{2n-1}/t_{2n}>1$ for all $n$ (the chain starting this time from a \textit{strong} bond), the analogous SSH model would be in its trivial phase; likewise here, the Fermi liquid phase of the Hubbard model may be regarded as trivial. There is no localized boundary state of the auxiliary chain in the Fermi liquid phase.
\begin{figure}[h]
\begin{subfigure}{0.36\linewidth}
\begin{tikzpicture}
	\node at (0,0) {\includegraphics[scale=1]{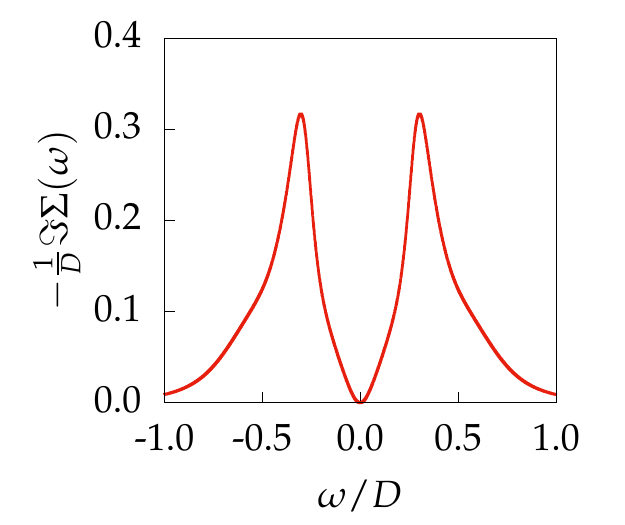}};
	\node at (2.2,2) {\footnotesize\subref*{fig:fls}};
\end{tikzpicture}
\phantomsubcaption{\vspace{-\baselineskip}\label{fig:fls}}
\end{subfigure}
\begin{subfigure}{0.69\linewidth}
\begin{tikzpicture}
	\node at (0,0) {\includegraphics[scale=1]{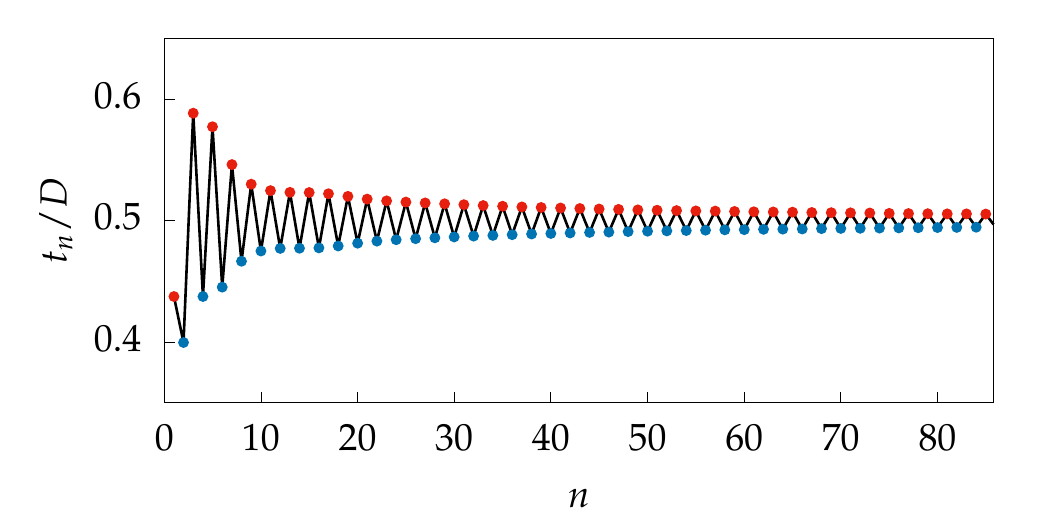}};
	\node at (4.4,2) {\footnotesize\subref*{fig:flt}};
\end{tikzpicture}
\phantomsubcaption{\vspace{-\baselineskip}\label{fig:flt}}
\end{subfigure}
\caption{Self-energy \subref{fig:fls} and corresponding chain parameters \subref{fig:flt} at $U/t = 3$, characteristic of the metallic Fermi liquid phase. Analogously to the trivial phase of the SSH model, the auxiliary chain is initialized with a strong bond.\label{fig:flresult}}
\end{figure}
In both the metallic and Mott insulating regimes, the auxiliary chain takes the form of staggered hopping parameters with period $2\mathbbm{Z}$. This is due to the global appearance of two bands separated by an energy gap. From the analysis shown in \S\ref{ch:genssh}, there is a relationship between the multiplicity of bands and gaps and the period of the stagger in the hopping parameters. 

%Analysis of the qualitative features of self-energies in this regime can be illuminated from considering 
%\S\ref{sec:moments}.

\section{Mid-Gap Peaks and Domain Walls\label{sec:mottbands}}

This section elaborates on the analysis of \S\ref{sec:singledw} on the effect of domain walls within the unit cell of generalized SSH models. There it was seen in \S\ref{sec:singledw} that a single domain wall in a repeated unit cell produces a mid-gap band, as opposed to a mid-gap pole as in a domain wall-free SSH model. The reasoning behind this phenomenon is that the domain wall hosts a localized state which possesses finite overlap with the states localized on neighboring domain walls. 
%we elaborate on the role of the mid-gap peaks in a gapped spectrum using the example of a model system. We extend our analysis represented by Fig~3 of the main text by asking a further question: What change in $\{t_n\}$ results from the consideration of sharp mid-gap peaks centered around $\pm\omega_p$ instead of two poles? Before making connections to the Hubbard model self-energy, we understand this from the perspective of a toy model as discussed in the following.
%
\begin{figure}[h]
\centering
\begin{subfigure}{\linewidth}
\phantomsubcaption{\label{fig:s2a}}
\end{subfigure}
\begin{subfigure}{\linewidth}
\phantomsubcaption{\label{fig:s2b}}
\end{subfigure}
\begin{subfigure}{\linewidth}
\phantomsubcaption{\label{fig:s2c}}
\end{subfigure}
\begin{subfigure}{\linewidth}
\phantomsubcaption{\label{fig:s2d}}
\end{subfigure}
\includegraphics[width=0.625\linewidth]{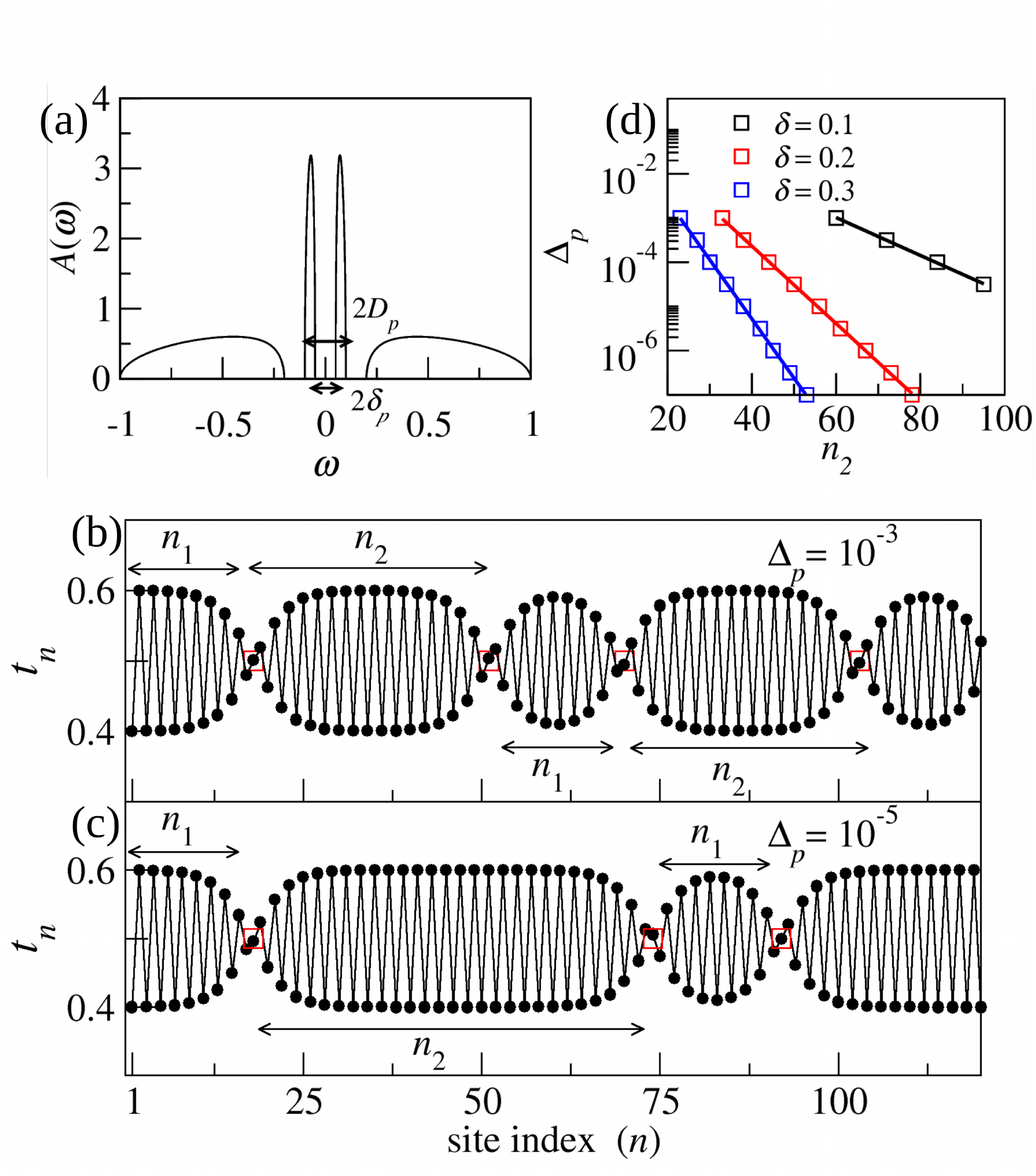}
\caption[Generation of mid-gap bands in SSH models from domain wall superlattice.]{Beating in auxiliary chain hopping parameters, $\{t_n\}$ and their relation to the width of sharp mid-gap peaks. This model is generated first by prescribing the spectrum in \subref{fig:s2a} and then generate the associated $\{t_n\}$ in \subref{fig:s2b},\subref{fig:s2c} from a moment analysis. The width of the mid-gap bands in \subref{fig:s2a} is exaggerated for illustrative purposes to emphasize that these are not poles. On a linear plot the peaks corresponding to the parameters in  \subref{fig:s2b},\subref{fig:s2c} are much narrower than what is illustrated.
%: In panel \subref{fig:s2a}, we design a toy model spectrum of full bandwidth $2D$ consisting of two outer SSH bands separated by a gap $2\delta$. The spectrum also consists of two inner SSH bands of full bandwidth $2D_p$ and gap $2\delta_p$. The width of these mid-gap peaks is defined as $\Delta_p$, such that $\Delta_p=D_p-\delta_p$. In panels \subref{fig:s2b} and \subref{fig:s2c} we plot the respective set of $\{t_n\}$ for $\Delta_p=10^{-3}$ (panel \subref{fig:s2b}) and $\Delta_p=10^{-5}$ (panel \subref{fig:s2c}), at a fixed $\delta=0.2$ and $D=1$. Clearly, the respective set of set of $\{t_n\}$ represents a periodic modulation of two SSH chains of domain lengths $n_1$ and $n_2$, with domain walls indicated as red squares. The width $(\Delta_p)$ of the inner SSH bands mimicking the effect of sharp mid-gap peaks is also a measure of the hybridization energy between these domain walls as shown in panel \subref{fig:s2d}, where we plot $\Delta_p$ as a function of $n_2$. The domain length, $n_2$ (squares) follows $\Delta_p\sim D \exp(-n_2 \delta/D)$ (fit represented as lines). 
Figure reproduced from~\cite{motttopology}.}
\label{fig:s2}
\end{figure}
In Fig~\ref{fig:s2a} such a model spectrum of full bandwidth $2D$ is shown consisting of two outer SSH bands separated by a gap $2\delta$. Inside this spectral gap of $2\delta$, there exists two inner SSH bands of full bandwidth $2D_p$ and gap $2\delta_p$, centered around $\omega_p$, such that the width of the inner SSH band is given by, $\Delta_p = D_p-\delta_p$. 
In order to understand the effect of the mid-gap features, $\omega_p$ is kept fixed and $\Delta_p$ is varied. The moment expansion technique, discussed in \S\ref{sec:moments}, is employed to determine the respective tight binding chain represented by $\{t_n\}$ for $\Delta_p/D = 10^{-3}$ (see Fig~\ref{fig:s2b}) and $\Delta_p/D = 10^{-5}$ (see Fig~\ref{fig:s2c}) for a fixed $\delta/D = 0.2$, and $\omega_p/D = 10^{-2}$. As seen in Figs.~\ref{fig:s2b} and \ref{fig:s2c}, the chain represented by $\{t_n\}$ consists of a periodic beat pattern with multiple domain walls (highlighted with red squares), such that each domain wall contributes to a topological localized state that hybridize amongst themselves. These domain walls form a periodic pattern and thereby the additional domain walls are completely absent in Fig~\ref{fig:dwdistance}, where $\Delta_p=0$ because the spectrum consists of two mid-gap poles at $\pm\omega_p$. 
Recalling the analysis of SSH models with a single domain wall in \S\ref{ch:genssh}, particularly from Fig.~\ref{fig:dwdistance}, it can be inferred that the location of the first domain wall and hence the length $n_1$ is pinned by the value of $\omega_p/D$; $n_1$ grows in magnitude as $\omega_p \to 0$. The additional beating is a manifestation of the presence of a band of topological states, represented as a mid-gap peak of width $\Delta_p$;  $\Delta_p$ determines the length $n_2$ that grows in size as $\Delta_p \to 0$. In other words, the original SSH medium, denoted as $n_2$ is now interrupted by multiple domains of length $n_1$, due to the presence of a band of topological \textit{defects} instead of just two topological excitations, such that the hybridization $\Delta_p$ is determined by $n_2$ which is the real space separation between these \textit{defects}. Indeed, these defects being topological in nature hybridize with each other via an exponentially localized $\Delta_p$ given by $\Delta_p\sim D \exp(-n_2\delta/D)$. This is shown in Fig.~\ref{fig:s2d} where $\Delta_p$ is varied in the calculations and the respective $n_2$ is determined (shown as squares) and it is observed that indeed $\Delta_p$ is exponentially localized in $n_2$ (solid lines).

This is essentially the many-domain wall equivalent to the analysis presented in \S\ref{sec:singledw}. There it was seen that a pair of domain walls produces a pair of mid-gap states (\textit{cf.} Fig.~\ref{fig:tanhdw} where the topological boundary is considered as a domain wall). The distance between the mid-gap poles was seen to be inversely proportional to the distance between the domain walls in the chain (Fig.~\ref{fig:dwdistance}). In the case here, rather than single poles, the mid-gap peaks are bands of finite width. As there are still many domain walls in the system, the states localized on these domain walls are still able to hybridize with each other and maintain the bands. 

In the terminology of a continuum model such as those discussed in \S\ref{sec:continuumfieldtheory}, this configuration of domain walls could be termed a `soliton gas', or `soliton ice' as the solitons are immobile. While the domain walls may informally be described as `propagating' through the chain, the auxiliary chains constructed here have fixed parameters; an auxiliary chain and its set of parameters $\{t_n\}$ are specific to one (self-energy) spectrum of the physical model for a specific set of physical system parameters. The physical system parameters are adjusted adiabatically to produce a new input for the auxiliary field mapping. The auxiliary models are not obtained from dynamical changes in the auxiliary system parameters, and therefore the domain walls do not propagate in a dynamical sense.

%%%%%%%%%%%%%%%%%%%%%%%%%%%%%%%%
\section{The Mott Transition}\label{sec:motttransition}
%%%%%%%%%%%%%%%%%%%%%%%%%%%%%%%%

Deep in either the Mott insulator or Fermi liquid phases of the Hubbard model, the auxiliary chains are of generalized SSH model type, with the Mott insulator phase being topologically non-trivial. A robust and exponentially localized zero-energy state lives on the boundary between the auxiliary and physical systems throughout the Mott insulating phase, corresponding to the Mott pole. However, richer physics is observed on approaching the Mott transition\index{Mott transition} from the Fermi liquid phase. In particular, the Mott transition occurs \textit{without} bulk gap closing of the Hubbard bands. Such a transition is unusual for a topological phase transition.
\begin{figure}[h!]
\centering
\begin{subfigure}{\linewidth}
\centering
\begin{tikzpicture}
	\node at (0,0) {\includegraphics{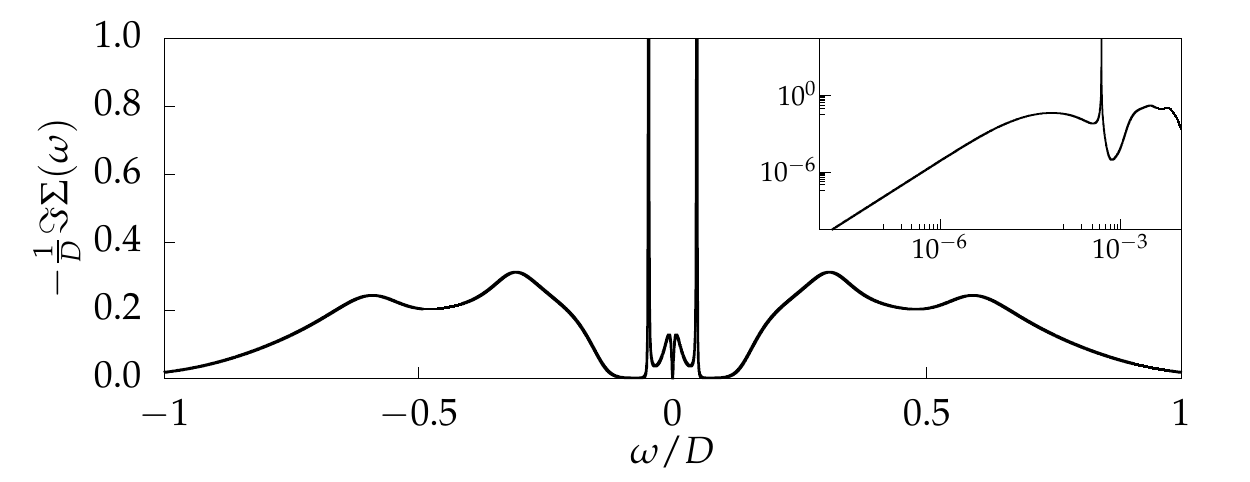}};
	\node at (5.375,1.875) {\footnotesize\subref*{fig:SU5_82zoomlog}};
\end{tikzpicture}
\phantomsubcaption{\vspace{-1.5\baselineskip}\label{fig:SU5_82zoomlog}}
\end{subfigure}
\begin{subfigure}{\linewidth}
\centering
\begin{tikzpicture}
	\node at (0,0) {\includegraphics{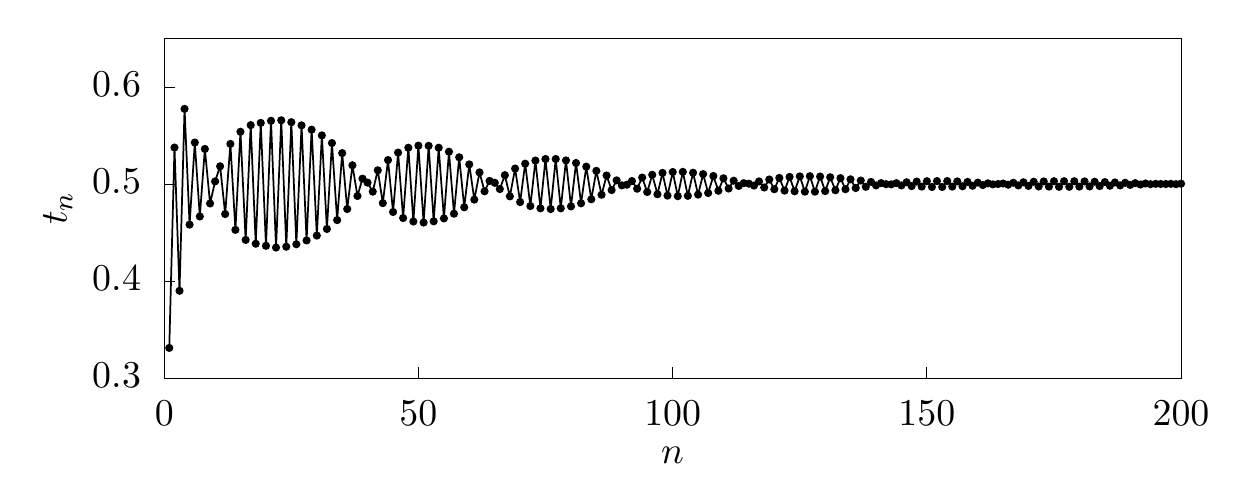}};
	\node at (5.375,1.75) {\footnotesize\subref*{fig:tnU5_82}};
\end{tikzpicture}
\phantomsubcaption{\vspace{-\baselineskip}\label{fig:tnU5_82}}
\end{subfigure}
\caption{%$-\Im\Sigma(\omega)$ \subref{fig:SU5_82zoomlog} and the corresponding set of $t_n$ \subref{fig:tnU5_82} 
Auxiliary field mapping for $U/t = 5.82$. The position of the peaks (inset of \subref{fig:SU5_82zoomlog}) is inversely proportional to the distance between the domain walls in the auxiliary chain, shown in \subref{fig:tnU5_82}. Compare the situation to Fig.~\ref{fig:U5_86}.\label{fig:U5_82}}
\end{figure}
\begin{figure}[h!]
\centering
\begin{subfigure}{\linewidth}
\centering
\begin{tikzpicture}
	\node at (0,0) {\includegraphics{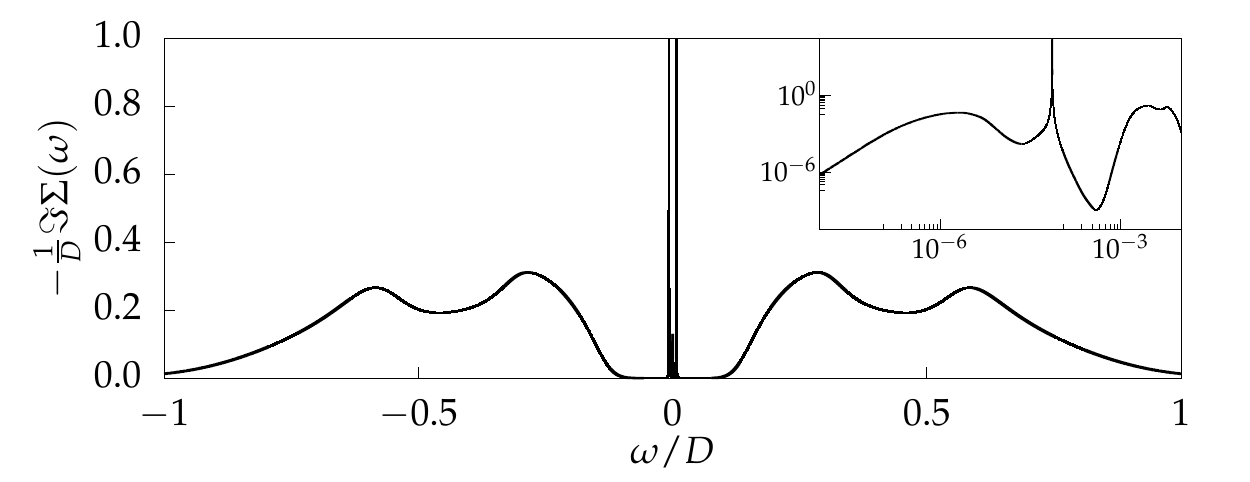}};
	\node at (5.375,1.75) {\footnotesize\subref*{fig:SU5_82zoomlog}};
\end{tikzpicture}
\phantomsubcaption{\vspace{-1.5\baselineskip}\label{fig:SU5_86zoomlog}}
\end{subfigure}
\begin{subfigure}{\linewidth}
\centering
\begin{tikzpicture}
	\node at (0,0) {\includegraphics{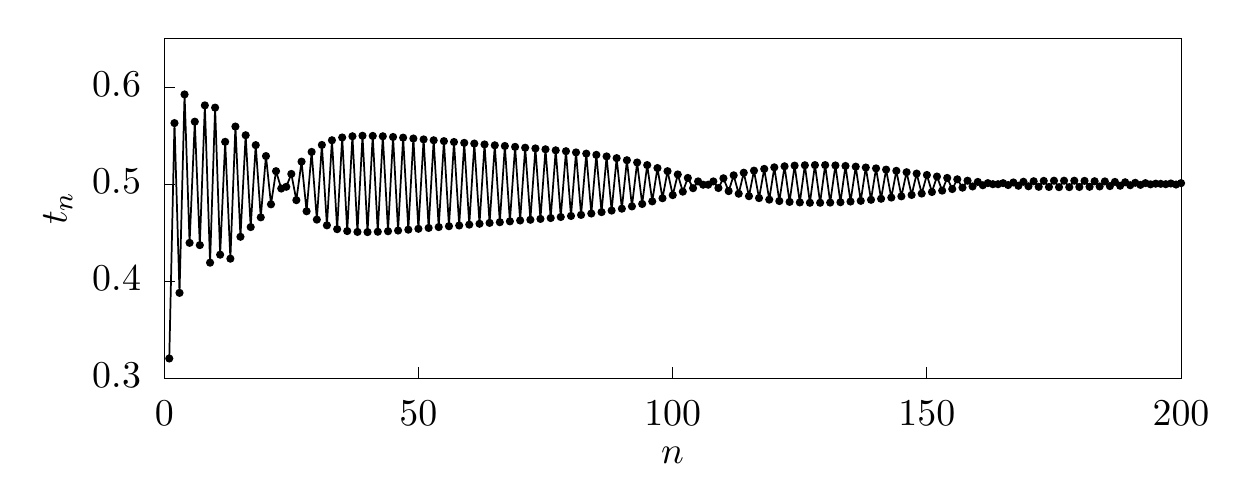}};
	\node at (5.375,1.875) {\footnotesize\subref*{fig:tnU5_86}};
\end{tikzpicture}
\phantomsubcaption{\vspace{-\baselineskip}\label{fig:tnU5_86}}
\end{subfigure}
\caption{%$-\Im\Sigma(\omega)$ \subref{fig:SU5_86zoomlog} and the corresponding set of $t_n$ \subref{fig:tnU5_86} 
Auxiliary field mapping for $U/t = 5.86$. The position of the peaks (inset of \subref{fig:SU5_86zoomlog}) is inversely proportional to the distance between the domain walls in the auxiliary chain, shown in \subref{fig:tnU5_86}. Compare the situation to Fig.~\ref{fig:U5_82}.\label{fig:U5_86}}
\end{figure}
In the vicinity of the transition on the Fermi liquid side, the self-energy develops a preformed gap, inside which are peaks of finite width centered at $\pm\omega_p$ with $\omega_p \propto t \sqrt{Z}$~\cite{bullahubbard}, while quadratic pseudogap behavior sets in on the lowest-energy scales $\lvert \omega \rvert \ll \omega_p$. The transition corresponds to $Z\to0$. 
These characteristics can be interpreted as a composite of features which appeared in the generalized SSH models in \S\ref{ch:genssh}.

%\begin{figure}[h]
%\centering
%\includegraphics{MtransU5_82.pdf}
%\end{figure}
%\begin{figure}[h]
%\centering
%\includegraphics{MtransU5_86.pdf}
%\end{figure}

As shown in the previous section \S\ref{sec:flregime}, deep in the Fermi liquid phase the auxiliary chain begins with a strong bond, however the auxiliary chains in this phase close to the transition feature an initial \textit{weak} bond. In the SSH model such a change in the system parameters necessitates a topological phase transition, but in the Hubbard model there is no such transition within the metallic phase. 
The explanation for this behavior in the auxiliary system is that as $U$ is increased,
a domain wall pair forms at the end of the chain, thereby flipping the parity of the first bond. One domain wall remains fixed to the end of the chain, and the other propagates into the chain with increasing $U$, approaching $U_c$ from below.

Analogously to the case in \S\ref{sec:domwalls}, increasing $U$ produces a superlattice of domain walls in $\hat{H}_{\text{aux}}$, each hosting a localized state which hybridize together to form low energy bands of finite width.

This mechanism is reminiscent of the vortex-pair dissociation in the Berezinskii-Kosterlitz-Thousless transition~\cite{berezinskii1,berezinskii2,kosterlitzthouless1,kosterlitzthouless2}. 
The topological phase transition occurs without bulk gap closing.

%%%%%%%%%%%%%%%%%%%
\subsection{Toy Model}\label{sec:mttoy}
%%%%%%%%%%%%%%%%%%%

The preceding analysis of the auxiliary chains can be checked by engineering a toy model with the intent of reproducing a spectral function with specific features. The toy model to be generated is completely determined by a set of hopping terms $\{t_n\}$ with a given parameterization. The choice of parameterization is informed both by the analysis in the preceding section \S\ref{sec:mottbands} as well as the analysis of extended SSH-type models presented in \S\ref{ch:genssh}.

As shown in the previous section, a self-energy in the metallic phase near $U_{c}$ features two prominent features at low energies: a Fermi liquid pseudogap, and two large spectral peaks outside the pseudogap region, but inside the main gap. A parameterization for a set $\{t_n\}$ for a tight-binding chain whose spectrum would exhibit the features of the Hubbard model self-energy in this regime could be given by
\begin{equation}
	t_n = \frac{D}{2} \sqrt{ 1 - (-1)^{n} \frac{r}{n+d} \left[ 1 - \beta \cos\left( \tfrac{2\pi n}{\lambda} + \phi \right) \right] } \,.
\label{eq:tntoy}
\end{equation}
The structure of this parameterization can be understood as a composition of a part which produces a low energy power-law component 
and a part which generates periodic domain walls. The $1/n$ part imposes an asymptotic decay envelope which results in power-law spectral features at low energy. As in \S\ref{sec:powerlawssh}, $r$ and $d$ parameterize the decay envelope. The domain walls are prescribed by the cosine term, with $\beta$, $\lambda$, and $\phi$ parameters determining the domain wall envelopes and their distribution in the chain.
It may be recalled that the enveloping function for domain walls in \S\ref{sec:domwalls} was a $\tanh$ profile rather than a (co)sine. This turns out not be a significant issue here as the density of domain walls is such that the resulting spectrum is insensitive to this difference in parameterization.
\begin{figure}[h!]
\centering
\includegraphics{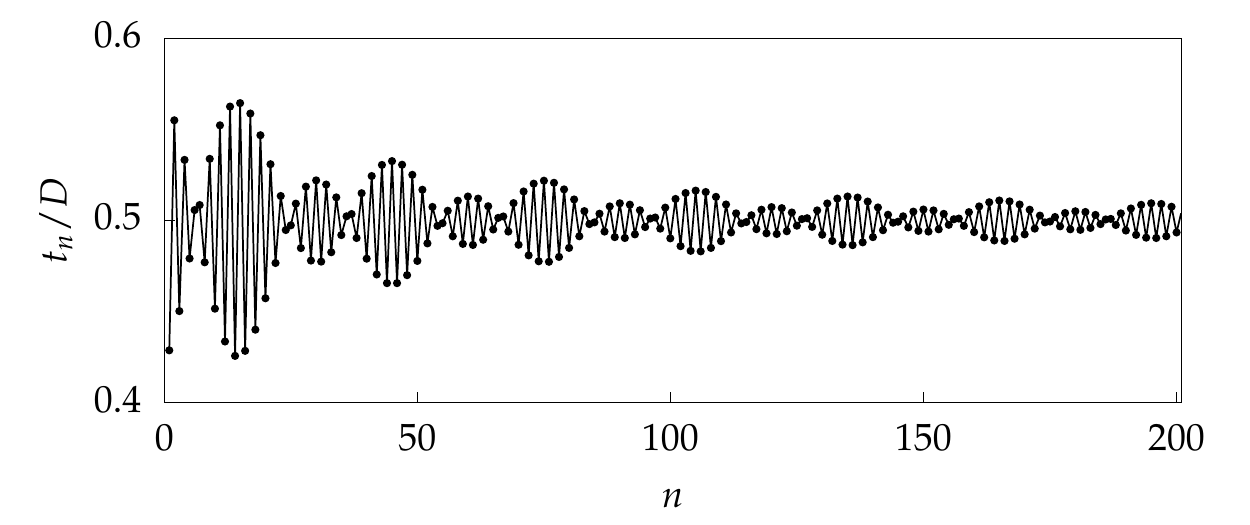}
\caption{Hopping parameters for toy model prescribed by Eq.~\eqref{eq:tntoy} with parameterization $(\beta,d,\phi,\lambda)=(3,15,0.1,30)$. This set of hoppings yields the spectrum plotted in Fig.~\ref{fig:toyspec}.\label{fig:tn_toy}}
\end{figure}
%\begin{figure}[h]
%\centering
%\includegraphics[scale=1]{toys.pdf}
%\caption{\label{fig:toys}}
%\end{figure}
%\begin{figure}[h]
%\centering
%\includegraphics{SU5_6.pdf}
%\end{figure}
\begin{figure}[h!]
\centering
\begin{subfigure}{\linewidth}
\centering
\begin{tikzpicture}
\node at (0,0) {\includegraphics{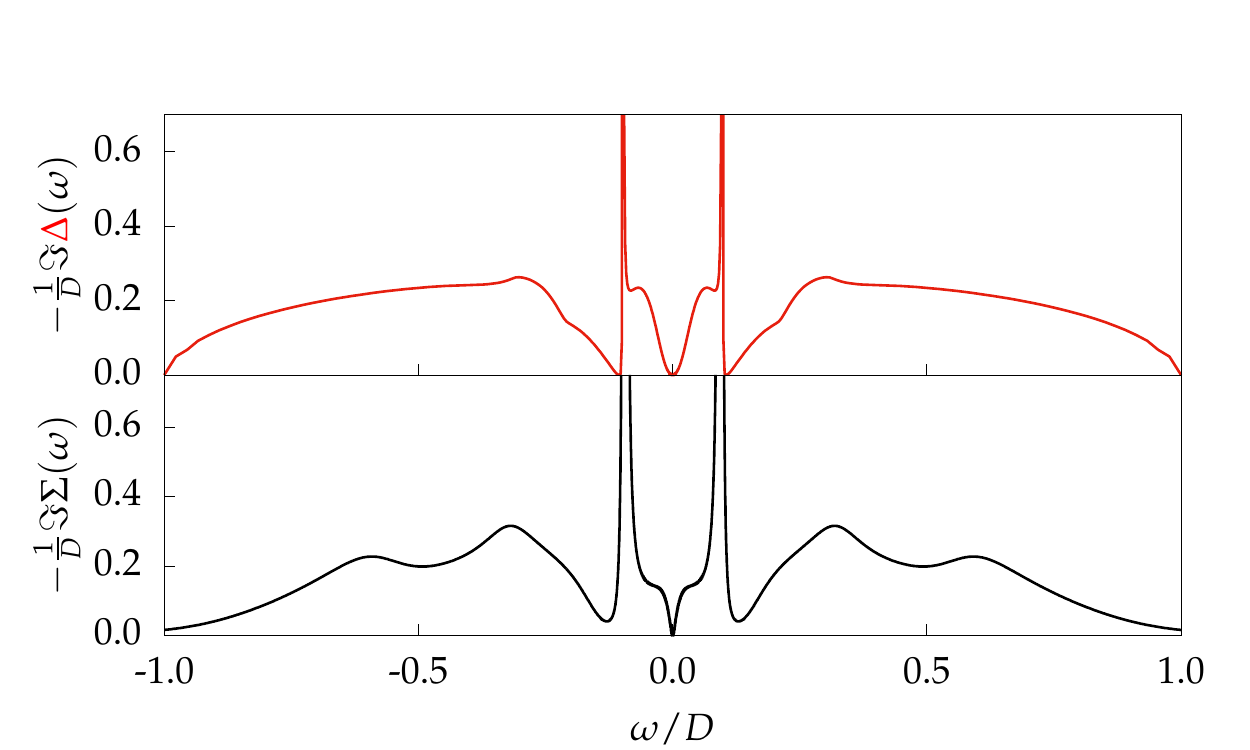}};
\node at (5.375,2.375) {\footnotesize\subref*{fig:toyspec}};
\node at (5.375,-0.25) {\footnotesize\subref*{fig:hsetoy}};
\end{tikzpicture}
\phantomsubcaption{\label{fig:toyspec}}
\phantomsubcaption{\label{fig:hsetoy}}
\end{subfigure}
\caption[Comparison of the toy model with Hubbard data]{Spectrum of the toy model Eq.~\eqref{eq:tntoy} parameterized by $(\beta,d,\phi,\lambda)=(3,15,0.1,30)$ \subref{fig:toyspec} compared to a corresponding self-energy of the Hubbard model at $U/t = 5.6$ with $D=3$ \subref{fig:hsetoy}.}
\end{figure}
In this model, the Mott transition\index{Mott transition} occurs in the limit of $\lambda$, $d \to \infty$. This limit of the parameters leads to the domain walls becoming increasingly more diffuse in the chain. This can be understood in terms of the discussion above in \S\ref{sec:mottbands}. The relation to the Mott transition is in that the positions of the peaks in the self-energy are proportional to the quasiparticle weight $Z$, which vanishes across the transition. As in the above, this can be manifested in the toy model by prescribing that the domain walls infinitely delocalize.

This toy model demonstrates that the lessons learned from \S\ref{ch:genssh} can be applied to generate models whose spectra possess specific features.

%%%%%%%%%%%%%%%%%%%%%%%%%%%%%%%%%%%%%%%%%%%%%%%%%%
\section{Particle-Hole Asymmetry\label{sec:phasymmmap}}
%%%%%%%%%%%%%%%%%%%%%%%%%%%%%%%%%%%%%%%%%%%%%%%%%%

The preceding analysis can also be applied to the more complex situation away from particle-hole (\textit{ph}) symmetry. The \textit{ph}-asymmetry is quantified according to the parameter $\eta \vcentcolon= 1-2\mu/U$, which involves the chemical potential $\mu$ and local Coulomb interaction strength $U$. At \textit{ph}-symmetry, $\mu = U/2$ and $\eta = 0$. For $\eta\neq 0 $,  the system is not \textit{ph}-symmetric, and $\Im\Sigma(\omega)\neq \Im\Sigma(-\omega)$. In the Mott insulating phase in particular, the Mott pole is not located precisely at the Fermi energy. These features lead to some differences in the mapping to the auxiliary chain and the subsequent analysis in terms of an emergent topology. However, the important conclusion, as explained below, is that the topological classification associated to the Fermi liquid phase as trivial and the Mott insulating phase as topologically non-trivial carries over to the \textit{ph}-asymmetric case, and holds for the Mott transition\index{Mott transition} arising at any $\eta$.

First, it is important to distinguish \textit{ph}-asymmetry $\eta$, and the average filling $\langle n \rangle$ which must be determined self-consistently. At \textit{ph}-symmetry $\eta=0$, the system is exactly half-filled, $\langle \op{n} \rangle = 1$. However, note that $\langle \op{n}{} \rangle = 1$ pertains throughout the Mott insulator phase for any $\eta$, and also $\langle \op{n}{} \rangle \rightarrow 1$ as $U \rightarrow U_c^-$ in the metallic phase for any $\eta$~\cite{logangalpin}. The immediate vicinity of the Mott metal-insulator transition is in fact always at half-filling, for any $\eta$. 

In the metallic phase, the lattice self-energy takes the usual Fermi liquid form, $-t\Im\Sigma(\omega)\sim (\omega/Z)^2$ at low energies $|\omega| \ll \omega_c$, where $\omega_c \sim Z t$ is the lattice coherence scale and $Z$ is the quasiparticle weight\index{quasiparticle weight}. This holds for \textit{any} asymmetry $\eta$, and as such there is an emergent low-energy \textit{ph}-symmetry in the precise sense that $\Im\Sigma(\omega) = \Im\Sigma(-\omega)$ for all $\lvert \omega \rvert \ll \omega_c$ independent of $\eta$. Since low energies in $\Sigma(\omega)$ roughly correspond to large $n$ down the auxiliary chain, one therefore expects the `bulk' of the Fermi liquid auxiliary chain to be the same as in the \textit{ph}-symmetric case already studied. At higher energies, \textit{ph}-asymmetry shows up in an asymmetry between the upper and lower Hubbard bands; in general this leads to $\Re\Sigma(\omega \rightarrow 0) \neq 0$ for $\eta \neq 0$. However, also note that $\Re\Sigma(\omega \rightarrow 0) \rightarrow 0$ as $U\rightarrow U_c^-$ approaching the Mott transition from the Fermi liquid side, independent of asymmetry $\eta$ and is as such a stronger definition of emergent \textit{ph}-symmetry in the close vicinity of the Mott transition (see Ref.~\cite{logangalpin} for details). Aspects of the generic Mott transition at $\eta \neq 0$ might therefore be expected to be related to the \textit{ph}-symmetric case at  $\eta = 0$.

%We now discuss the generalization of the mapping to the auxiliary chain away from \textit{ph}-symmetry. 
The auxiliary field mapping of \S\ref{sec:mapping} requires some generalizations to be applicable to the \textit{ph}-asymmetric case~\cite{motttopology}. The auxiliary chain takes the form of Eq.~\eqref{eq:Haux}, with finite on-site potentials, $\{e_n\}$ in general. The continued fraction expansion of the self-energy follows as,  
\begin{align}
    \Sigma(\omega)\equiv \Delta_0(\omega)=\cfrac{t_0^2}{\omega^+-e_1-\cfrac{t_1^2}{\omega^+-e_2-\cfrac{t_2^2}{\phantom{a}\ddots}}} \;.
    \label{eq:G_CFE_ph_asymm}
\end{align}
Similar to the discussion in \S\ref{sec:mapping}, the initial bond of the auxiliary chain %is $V^2=t_0^2$, which 
is obtained from the relation $\int \d\omega \mathcal{A}_0(\omega) \equiv -\tfrac{1}{\pi}\int \d\omega \Im\Delta_0(\omega) = t_0^2$. Subsequently, for all $n>0$ all $t_n$ and $e_n$ can be determined recursively using the relations
\begin{subequations}
\begin{align}
	\int \d\omega \mathcal{A}_n(\omega) \equiv -\frac{1}{\pi}\int \d\omega\, \Im\Delta_n(\omega) &= t_n^2
\intertext{and}
	-\frac{1}{\pi}\int \d\omega\, \frac{\omega{\Im}\Delta_{n-1}(\omega)}{t_{n-1}^2} &= e_n \,,
\end{align}
\end{subequations}
where $e_1$ is the on-site energy of the boundary site in the isolated $\op{H}{\text{aux}}$ coupled to the physical degrees of freedom via $\op{H}{\text{hyb}}$ \eqref{eq:Hhyb}. Note that in the \textit{ph}-symmetric case, $e_n=0$ for all $n$.

\begin{comment}
%##############################
\begin{figure}[h]
%\includegraphics[clip=,scale=0.55]{Fig_MI_FL_asymm.eps}
  \caption{Lattice self-energy at $T=0$ and particle-hole asymmetry, $\eta=1/4$ obtained using DMFT-NRG [panels (a,e)] and corresponding auxiliary chain hopping parameters $\{t_n\}$ and onsite potential energies $\{e_n\}$ [panels (b,c) (MI), panels (f,g) (FL)]. Left panels show results for MI ($U/t=9.0,\,D=4.625$): the hard gap and pole at $E_{MP}\approx 0.12D$ produces a generalized SSH model in the topological phase, hosting a mode $\Psi_{\text{MP}}$ satisfying $H_{\text{aux}}\Psi_{\text{MP}}=E_{MP}\Psi_{\text{MP}}$ which is exponentially-localized on the boundary [panel (d)]. Right panels show results for FL ($U/t=3.0,\,D=4.5$). The low energy $\omega^2$ pseudogap in $\Im\Sigma$ results in an asymptotic $(-1)^n/n$ decay of the respective $t_n$ and $e_n$. Furthermore, the $\{e_n\}$ oscillate around zero (and tend to zero) at large $n$, indicative of a low energy emergent \textit{ph}-symmetric dynamics.}
  \label{fig:MI_FL_ph_asymm}
\end{figure}
\end{comment}

%\subsection{Auxiliary chain representation of a \textit{ph}-asymmetric Mott insulator}
%\label{sec:ph_asymm_MI}
The self-energy $\Im\Sigma(\omega)$ of a \textit{ph}-asymmetric Mott insulator consists of two high energy Hubbard bands centered on some high energy positive $(\omega_+)$ and negative $(\omega_-)$ values respectively, shown in Fig.~\ref{fig:S_eta0_25}. There also exists an insulating charge gap between the Hubbard bands of width $2\delta = | \delta_+ - \delta_- |$ where $\delta_\pm$ denotes the inner band edge on the positive (negative) energy side. Here $|\delta_+| \neq |\delta_-|$ and $|\omega_+ | \neq | \omega_-|$, unlike in the \textit{ph}-symmetric case. Additionally, in the \textit{ph}-asymmetric case the Mott pole inside the insulating gap is located away from the Fermi level at $\omega = e_0 \equiv E_{\text{MP}}$. 
Thus the \textit{ph}-asymmetric Mott insulator self energy is of the form $\Sigma(\omega) \equiv \Delta_0(\omega) = \Delta_0^{\text{reg}}(\omega) + \frac{\alpha_0}{\omega^+ - e_0}$ where $\Im\Delta_0(\omega) = 0$ for $\delta_- < \omega < \delta_+$. Since the Mott pole of weight $\alpha_0$ resides in the gap, $\delta_- < e_0 < \delta_+$ and   
$\alpha_0^{-1}=-\frac{1}{\pi}\int_{-\infty}^{\infty} \d\omega\frac{\Im G(\omega)}{(\omega - e_0)^2}$, where $G(\omega)$ is the local lattice Green function of the Mott insulator. 

For the continued fraction expansion set up, 
\begin{align}
	\Delta_{n-1}(\omega) &= t_{n-1}^2 \tensor*{\widetilde{G}}{^{(n)}_{1,1}}(\omega),\\
	\tensor*{\widetilde{G}}{^{(n)}_{1,1}}(\omega) &= \frac{1}{\omega^+ - e_n - \Delta_n(\omega)}.
\end{align}
Following the same logic based on the analytic structure of the complex $\Delta_n$'s as in the \textit{ph}-symmetric case,
\begin{align}
    \Delta_{2n-1}(\omega)&=\omega^+-e_{2n-1}-\frac{t_{2n-2}^2}{\Delta_{2n-2}^{\text{reg}}(\omega)+\frac{\alpha_{2n-2}}{\omega^+ - e_{2n-2}}},\\
    \Delta_{2n}(\omega)&=\omega^+ - e_{2n}-\frac{t_{2n-1}^2}{\Delta_{2n-1}(\omega)} \notag\\
    &=\Delta_{2n}^{\text{reg}}(\omega)+\frac{\alpha_{2n}}{\omega^+ - e_{2n}},
\end{align}
where $n\ge1$ and every \textit{even} $-\Im\Delta_{2n}(\omega)$ is gapped at low energies with a pole of weight $\alpha_{2n}$ located inside this gap at $\omega = e_{2n}$. Every \textit{odd} $-\Im\Delta_{2n-1}(\omega)$ is gapped and regular and analytic in the complex plane. Since $\Delta_{2n-1}(\omega)$ is purely real for $\delta_-<\omega<\delta_+$ an isolated pole exists inside the gap in $\Delta_{2n}(\omega)$ at $e_{2n}$ where $\Re\Delta_{2n-1} \rvert_{\omega=e_{2n}}=0$. 

The following relations are used to obtain $\{t_n\}$ and $\{e_n\}$ $\forall n\ge1$:
\begin{align}
    &e_n = - \frac{1}{\pi t_{n-1}^2} \int \d\omega \, \omega\Im\Delta_{n-1}^{\text{reg}}(\omega)+\frac{\alpha_{n-1} e_{n-1}}{t_{n-1}^2},\\
    &t_n^2 = - \frac{1}{\pi} \int \d\omega \Im\Delta_{n}^{\text{reg}}(\omega)+\alpha_n,
\end{align}
where, $\Im\Delta_{n}^{\text{reg}} = \Im\Delta_{n}$ for \textit{odd} $n$ and the respective pole weight $\alpha_n=0$ for all \textit{odd} $n$. 
The pole weight, $\alpha_{2n}$ $\forall$ \textit{even} sites is obtained using the relation,
\begin{align}
    \alpha_{2n}^{-1} = -\frac{1}{\pi t_{2n-1}^2} \int_{-\infty}^{\infty} \d\omega\frac{\Im\Delta_{2n-1}(\omega)}{(\omega - e_{2n})^2},
\end{align}
where, $e_{2n}$ is obtained by numerically locating the $\omega$ where $\Re\Delta_{2n-1}=0$ inside the gap. Since $\Delta_0(\omega)$ contains a pole inside the gap it is inevitable that $\Delta_2(\omega)$ will also contain a pole. However, unlike the \textit{ph}-symmetric case the location of the pole will vary along the recursions, albeit remaining inside the gap, until the $\{t_n,e_n\}$ of the auxiliary chain settles down to a staggered alternating form without any attenuation, as shown in Fig.~\ref{fig:phasymmtnen}.

It is important to note that the numerical determination of the $\{\alpha_{2n}\}$ and $\{e_n\}$ is prone to numerical errors due to grid resolution, spectral kinks and/or Hilbert transformation. These errors may propagate down the chain in the initial stages of the recursion leading to spurious features in the auxiliary chain parameters. Therefore care must be taken in the numerical evaluation.

%%%%%%%%%%%%%%%%%%%%%%%%%%%%%%%%%%%%%%%%%%%%%%%%%%%%%%%%%%%%%%

%
\begin{figure}[h]
\centering
\begin{subfigure}{0.49\linewidth}
\begin{tikzpicture}
	\node at (0,0) {\includegraphics{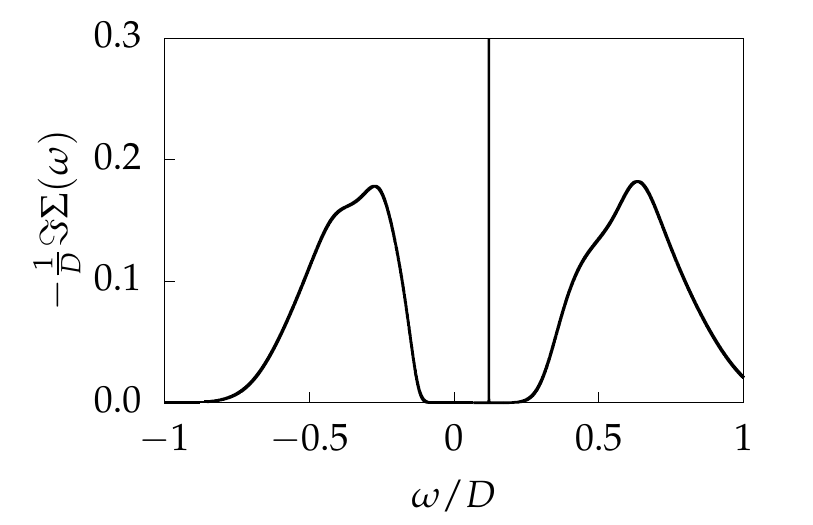}};
	\node at (3.125,2) {\footnotesize \subref*{fig:SU9_eta0_25}};
\end{tikzpicture}
\phantomsubcaption{\vspace{-\baselineskip}\label{fig:SU9_eta0_25}}
\end{subfigure}
\hfill
\begin{subfigure}{0.49\linewidth}
\begin{tikzpicture}
	\node at (0,0) {\includegraphics{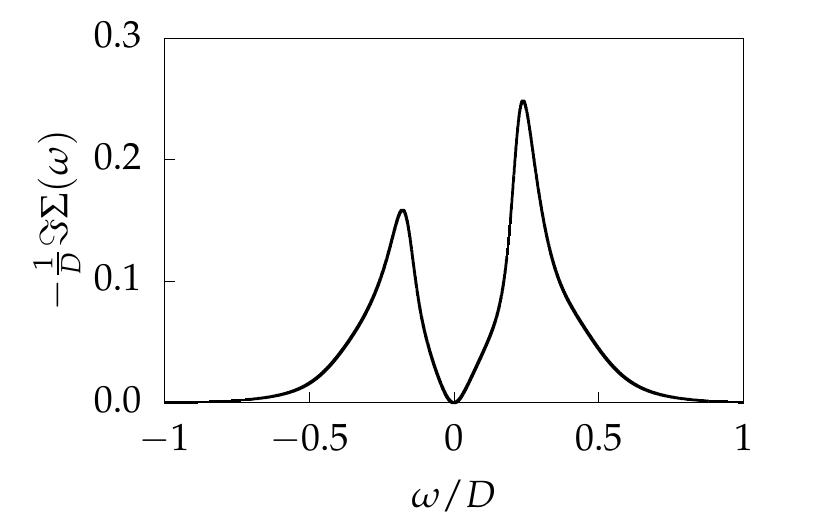}};
	\node at (3.125,2) {\footnotesize \subref*{fig:SU3_eta0_25}};
\end{tikzpicture}
\phantomsubcaption{\vspace{-\baselineskip}\label{fig:SU3_eta0_25}}
\end{subfigure}
\caption{Self-energy of the particle-hole asymmetric Hubbard model for $\eta = \frac14$ at \subref{fig:SU9_eta0_25} $U/t = 9$ with $D = 4.625$, and \subref{fig:SU3_eta0_25} $U/t = 3$ with $D = 4.5$.\label{fig:S_eta0_25}}
\end{figure}
\begin{figure}
\begin{subfigure}{0.49\linewidth}
\begin{tikzpicture}
	\node at (0,0) {\includegraphics{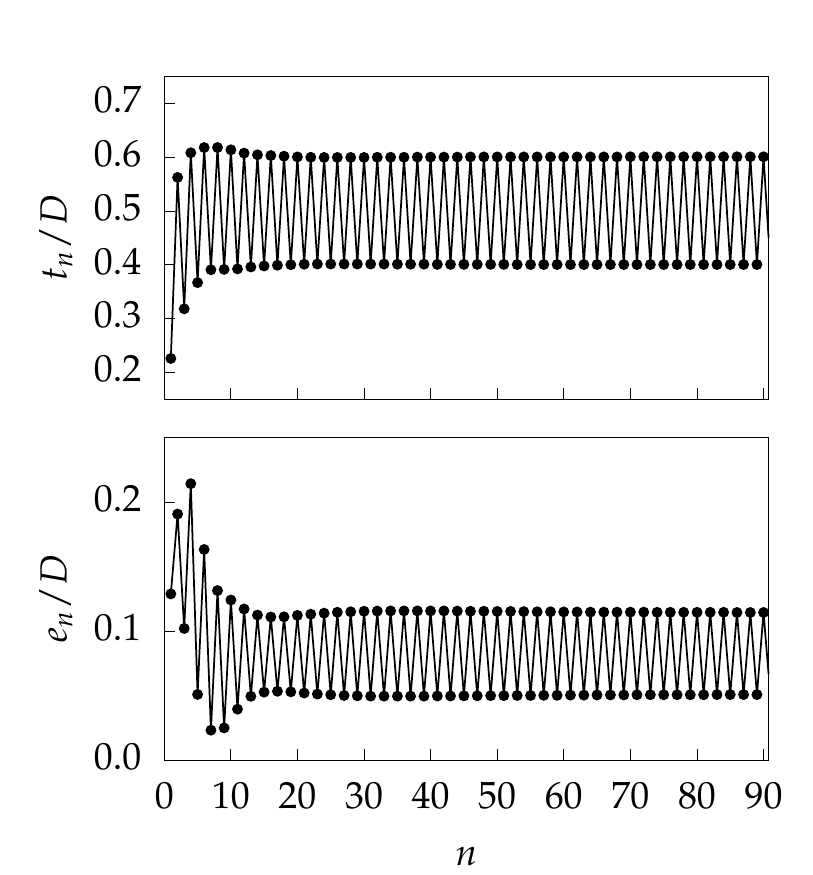}};
	\node at (3.375,3.375) {\footnotesize \subref*{fig:tnU9_eta0_25}};
	\node at (3.375,-0.25) {\footnotesize \subref*{fig:enU9_eta0_25}};
\end{tikzpicture}
\phantomsubcaption{\label{fig:tnU9_eta0_25}}
\phantomsubcaption{\label{fig:enU9_eta0_25}}
\end{subfigure}
\hfill
\begin{subfigure}{0.49\linewidth}
\begin{tikzpicture}
	\node at (0,0) {\includegraphics{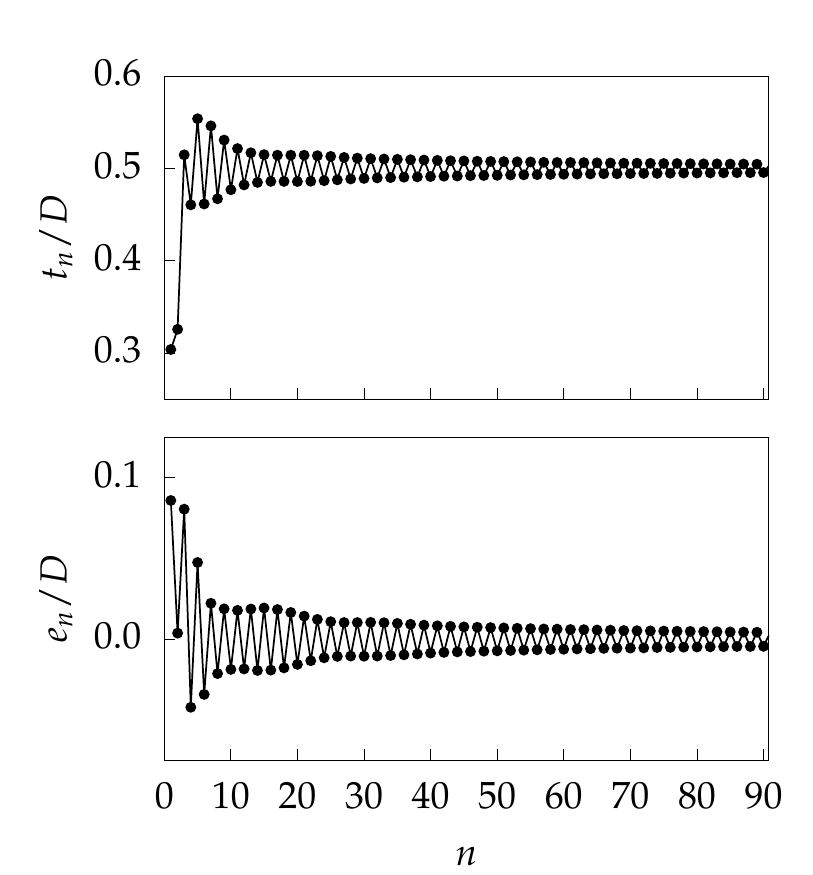}};
	\node at (3.375,3.375) {\footnotesize \subref*{fig:tnU3_eta0_25}};
	\node at (3.375,-0.25) {\footnotesize \subref*{fig:enU3_eta0_25}};
\end{tikzpicture}
\phantomsubcaption{\label{fig:tnU3_eta0_25}}
\phantomsubcaption{\label{fig:enU3_eta0_25}}
\end{subfigure}
\vspace{-\baselineskip}
\caption{Parameters $t_n$ and $e_n$ for the auxiliary chain corresponding to $-\Im\Sigma(\omega)$ of the \textit{ph}-asymmetric Hubbard model with $\eta = \frac14$ at $U/t=9$ \subref{fig:tnU9_eta0_25},\subref{fig:enU9_eta0_25} and at $U/t=3$ \subref{fig:tnU3_eta0_25},\subref{fig:enU3_eta0_25}.\label{fig:phasymmtnen}}
\end{figure}
Characteristic self-energies for the \textit{ph}-asymmetric Hubbard model are shown in Fig.~\ref{fig:S_eta0_25}. The parameters chosen here for analysis are $\eta = \frac14$ with the Mott insulating phase parameters $U/t = 9$ with $D = 4.625$ (self-energy plotted in Fig.~\ref{fig:SU9_eta0_25}), and parameters $U/t = 3$ with $D = 4.5$ for the Fermi liquid (self-energy plotted in Fig.~\ref{fig:SU3_eta0_25}). The notable characteristics of these spectra are that the Mott pole at $\eta \neq 0$ no longer sits at the Fermi level, but rather has been shifted to $\omega_{\textsc{mp}} \approx 0.12 D$. On the other hand, the vertex of the pseudogap in the Fermi liquid still does sit at $\omega = 0$, although it has an asymmetry in the weight of the spectral bands. Employing the modified auxiliary field mapping from above, auxiliary chains can now be constructed for each of these self-energies.
The auxiliary chain parameters for the $\eta = \frac14$ Mott insulator and Fermi liquid are shown in Fig.~\ref{fig:phasymmtnen}. It is seen that
away from particle-hole symmetry, it is no longer true that the $e_n = 0$ $\forall n$ in the effective chain. Instead, these parameters take on an alternating pattern analogous to the $t_n$ parameters. The resulting effective model is then more appropriately a generalized Rice-Mele model~\cite{ricemele} rather than an SSH model. A Rice-Mele model has the same alternating hopping parameters as the SSH model, but additionally has alternating on-site potentials as well.

The auxiliary chain parameters for the Mott insulator are shown in Figs.~\ref{fig:tnU9_eta0_25} and \subref{fig:enU9_eta0_25}. The hoppings exhibit the characteristic strict alternating pattern initialized with a weak bond that is to be expected for the production of a spectral pole within a gap. The on-site potentials also exhibit a strict alternating form, with the chain initialized with $e_0/D \approx 0.12$, giving the position of the spectral pole. The spectral pole is also not centered in the gap. The staggered potentials alternate around a non-zero value, which shifts the center of the gap away from the Fermi level. 

The auxiliary chain parameters for the Fermi liquid are shown in Figs.~\ref{fig:tnU3_eta0_25} and \subref{fig:enU3_eta0_25}. The asymptotic behavior of the hopping parameters again has $1/n$ envelope as in the \textit{ph}-symmetric case for a power-law spectrum. This $1/n$ envelope is also present in the potentials as well.
Unlike in the Mott insulating case, the alternating potentials is about zero, which reflects that the vertex of the pseudogap still lies at the Fermi level. 

In both cases, all the auxiliary chain parameters exhibit a strong asymmetry in the initial head of the chain, reflecting the strong asymmetry present in the higher energy parts of the self-energies.

\subsection{Asymmetric Topology}\label{sec:phasymmtopology}
Based on the analysis in Ref.~\cite{extendedrm}, the following shows that this auxiliary system is indeed topological.
The original model considered in Ref.~\cite{extendedrm} was that of a SSH model with parameterized next-nearest neighbor (NNN) dynamics. The model considered here instead incorporates on site potentials. For systems consisting of a two site unit cell, as in the case here, the two schemes result in the same effective Hamiltonian in momentum space as both NNN and potential terms enter into the diagonal entries of the momentum space Hamiltonian.

As shown in Figs.~\ref{fig:tnU9_eta0_25} and \subref{fig:enU9_eta0_25} the bulk of the auxiliary chain in the Mott insulator has strictly alternating hoppings $t_A$ and $t_B$ as well as alternating
on-site potentials $\epsilon_A$ and $\epsilon_B$ on well-defined $A$ and $B$ sublattices. This implies that the bulk can be taken to be uniformly periodic, meaning that momentum is a good quantum number. Importantly, it allows the analysis to take place in momentum space where traditional methods for studying band topology can be employed, \S\ref{sec:topology}~\cite{fruchartcarpentier,shortcourse,altlandsimons}. Fourier transforming to momentum space yields $\hat{H} = \sum_{k} \tensor*{\vec{f}}{_k^\dagger} \boldsymbol{h}(k) \tensor*{\vec{f}}{_k}$, where $\tensor*{\vec{f}}{_k^\dagger} = \adjvec{\opd{f}{Ak} & \opd{f}{Bk}}\,$, and
\begin{align}
	\boldsymbol{h}(k) = 
	\begin{pmatrix} 
	\epsilon_A & t_A + t_B e^{\i k} \\ 
	t_A + t_B e^{-\i k} & \epsilon_B 
	\end{pmatrix}.
\label{eq:nnn}
\end{align}
Following Ref.~\cite{extendedrm} the coefficients of the Hamiltonian are re-parametrized as
%\begin{subequations}
\begin{equation}
	t_A = t_0 (1 - \delta t \cos\theta) \,,\;
	t_B = t_0 (1 + \delta t \cos\theta) \,,\;
	\epsilon_{A} = q \cos(\theta + \phi) \,,\;
	\epsilon_{B} = q \cos(\theta - \phi) \,. 
\end{equation}
%\end{subequations}
The introduction of a periodically modulated parameter such as $\theta$ as a means of determining the topology of a system is well-known in the literature in the context of the Thouless pump~\cite{thoulesspump,shortcourse}. This concept has also been generalized to the study of topological insulators with synthetic dimensions, with a prominent example being the $4d$ quantum Hall effect~\cite{4dqhe,4dqheuca,syntheticdimensions}.
These methods for endowing systems with synthetic dimensions have also been used to engineer systems with effective magnetic fields or effective gauge fields~\cite{lightgauge,syntheticdimensions}. In addition to being an intriguing theoretical construct, such ideas have been implemented in various experimental setups~\cite{lightgauge,syntheticdimensions,synthetichallribbons,bosegasqhe,4dcircuit}. 

The effective model takes the form of a system in one real spatial dimension and one synthetic dimension. In momentum space the two dimensions appear with equal footing, thereby resulting in a system which is $2d$ in momentum space. The topology of this effective model can then be measured using the Berry curvature and the Chern number.

In the momentum space representation, the cyclic parameter $\theta$ plays the role of the momentum in a synthetic dimension alongside the usual quasimomentum $k$ to produce an effective two-dimensional Brillouin zone. The topological invariant is given by the Chern number\index{Chern number}
\begin{equation}
	\mathrm{Ch} = \frac{1}{2\pi}\oint_{\textsc{bz}} F,
\end{equation}
with the usual notation of exterior forms
\begin{equation}
\begin{aligned}
	F &= \d A
	\\
	&= \left( \partial_k A_\theta - \partial_\theta A_k \right) \d k \extp \d\theta,
\end{aligned}
\end{equation}
where $A_{k}$ is the $k$ component of the Berry connection and $A_\theta$ is the $\theta$ component. For $\theta \in [\frac{\pi}{2},\frac{3\pi}{2}]$, as is the case for the Mott insulator self energy considered here, the Chern number is explicitly given by~\cite{extendedrm,kaufmann}
\begin{equation}
	\mathrm{Ch} = \frac12[\sgn(2 q \sin\phi) - \sgn(-2 q \sin\phi)]=\pm 1 \,,
\end{equation}
meaning the system is topological.
Note that in case of \textit{ph}-symmetry, $t_0=D/2$, however here it is observed that $t_0\approx D/2$ which could be numerical. For the mapping demonstrated here the numerical values are used. 
Furthermore, unlike the \textit{ph}-symmetric case, here the high energy cutoff $D_+(D_-)$ on the positive (negative) side is different and it is chosen to be $D=(D_++|D_-|)/2$.

The respective parameters for this system, as obtained from the continued fraction expansion mapping of the Mott insulator self-energy for $U/t=9$ and $\eta=\frac14$ plotted in Fig.~\ref{fig:SU9_eta0_25}, are $\epsilon_A = 0.24$ and $\epsilon_B = 0.52$, and hopping amplitudes $t_A = 1.85$ and $t_B = 2.78$. In order to cast this set of parameters in terms of $\delta t$, $q$, $\theta$, and $\phi$, $t_0=2.31$ as obtained from the calculation with $\delta t$ chosen to be $\delta t = 0.5$, as it is a free parameter. This yields $\{q,\,\theta,\,\phi\}\approx\{-0.97,\, 1.98,\,-0.16\}$. The $k$ is chosen to be the momentum point where the band gap in $\hat{H}_{\text{aux}}$ is minimum and is equal to the spectral gap in $-\Im\Sigma(\omega)$. This always occurs at $k=\pi$.
%%%
\begin{figure}[h]
\centering
\begin{subfigure}{0.5\linewidth}
\centering
\includegraphics[scale=1]{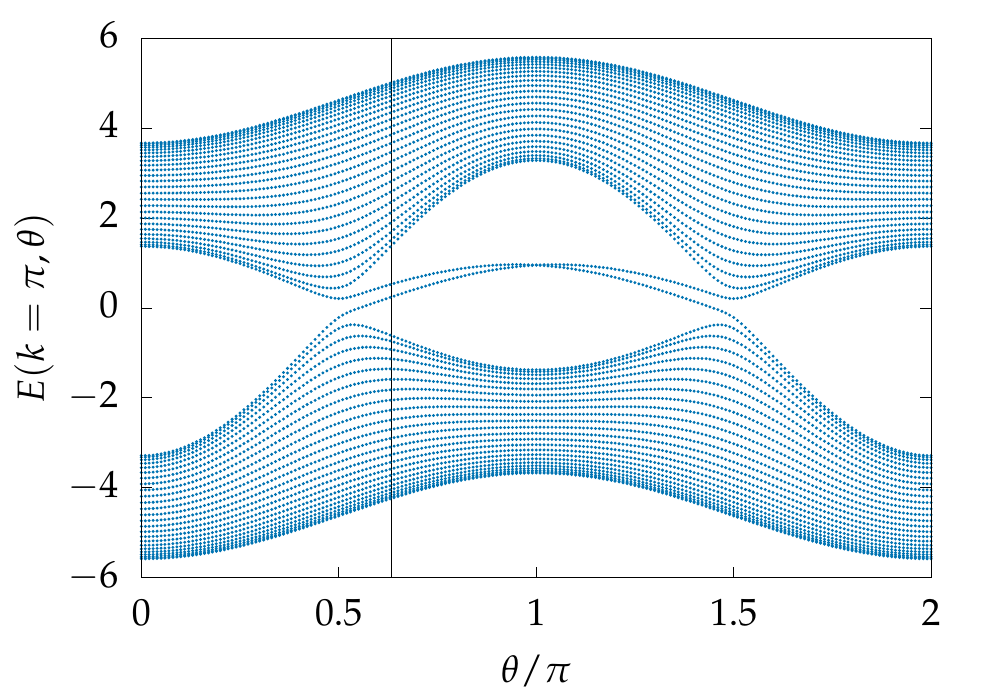}
\end{subfigure}
\hfill
\begin{subfigure}{0.4\linewidth}
\centering
\includegraphics[scale=1]{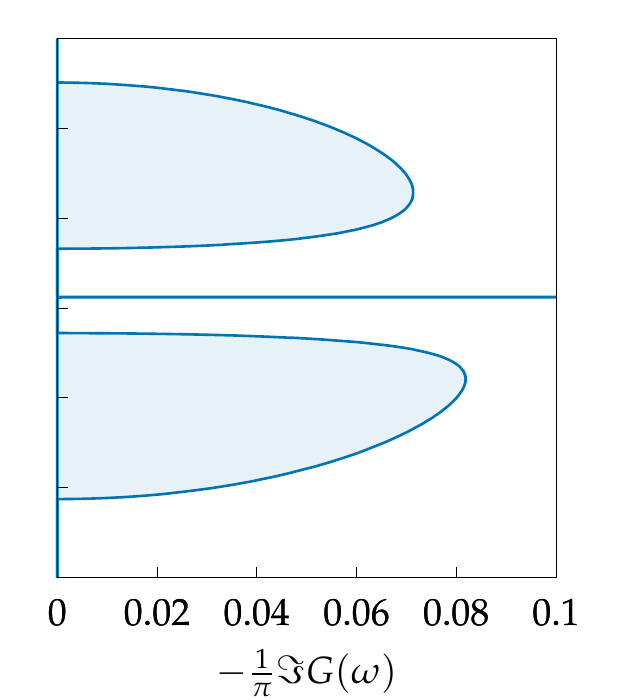}
\end{subfigure}
\caption[Topological band structure of particle-hole asymmetric chain]{Band structure of the extended Rice-Mele model in the $(E,\theta)$-plane at $k=\pi$, evaluated with the parameters $\{q,t_0,\delta t,\phi\} \approx \{-0.97,2.31,0.50,-0.16\}$ corresponding to the Mott insulator. The solid vertical line marks the cut $\theta = 1.99$, which reproduces exactly the appropriate spectral function of the chain boundary (right). The two intragap bands correspond to localized states on the left and right boundaries; only the left boundary is physical in the semi-infinite auxiliary chain (lower intragap band in $\theta<\pi$ region).\label{fig:phasymbands}}
\end{figure}
%%%
Using the above parametrization, it follows that the Chern number $\mathrm{Ch} = 1$ is quantized and the system is in the topological phase. The topological character of the parametrized Mott insulating phase shown here, is further exemplified by the intragap band crossing shown in Fig.~\ref{fig:phasymbands}. Within this auxiliary model, the system is topological only if the intragap bands cross~\cite{extendedrm}. This occurs if there exists a $\theta$ such that $h \sin\theta \sin\phi = 0$. This is always the case for $\theta \in [\frac\pi2,\frac{3\pi}{2}]$, which is when $t_A < t_B$, exactly as in the standard SSH model.  Therefore, it can be concluded that the topological state is robust to perturbations in the hopping amplitudes as well as the on-site potentials.

\begin{figure}[h]
\centering
\begin{subfigure}{0.49\linewidth}
\centering
\begin{tikzpicture}
\node at (0,0) {\includegraphics[scale=1]{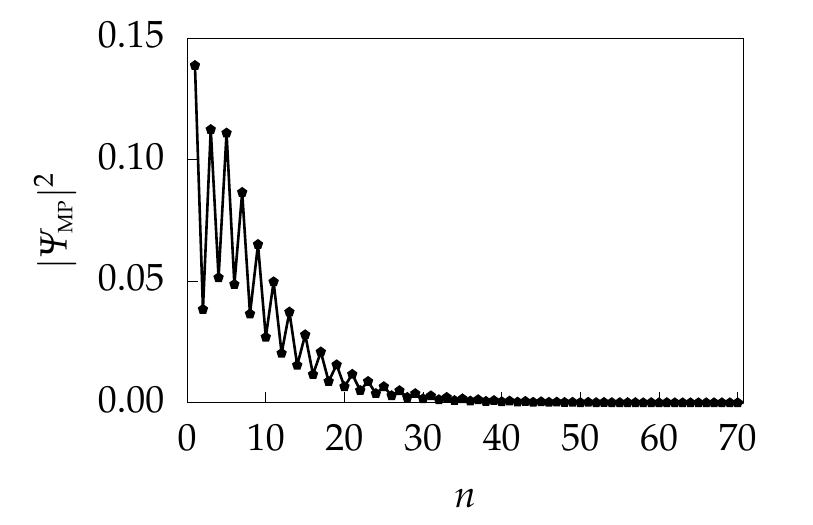}};
\node at (3.125,2) {\footnotesize \subref*{fig:wfnU9_eta0_25}};
\end{tikzpicture}
\phantomsubcaption{\label{fig:wfnU9_eta0_25}}
\end{subfigure}
%
%\hfill
%
\begin{subfigure}{0.49\linewidth}
\centering
\begin{tikzpicture}
\node at (0,0) {\includegraphics[scale=1]{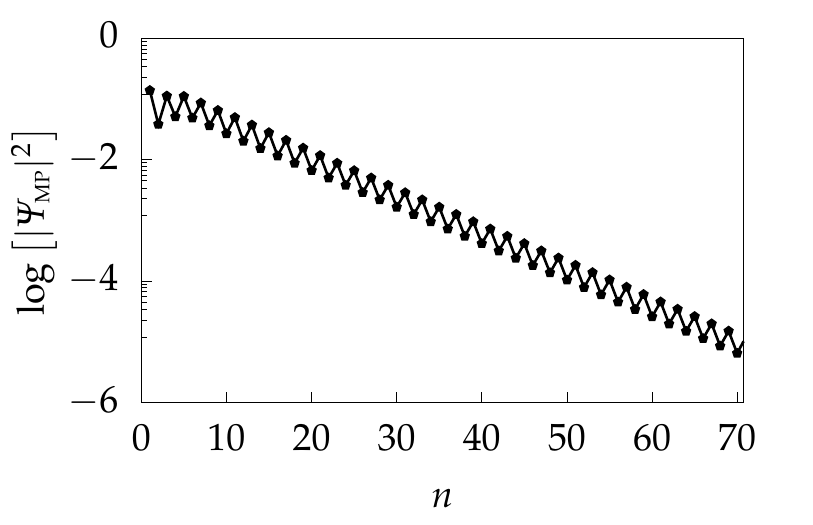}};
\node at (3.125,2) {\footnotesize \subref*{fig:wfnlogU9_eta0_25}};
\end{tikzpicture}
\phantomsubcaption{\label{fig:wfnlogU9_eta0_25}}
\end{subfigure}
\vspace{-2\baselineskip}
\caption[Asymmetric wavefunction amplitude]{Zero energy wavefunction amplitude of ph-asymmetric the effective chain. The wavefunction exhibits exponential localization. Note that unlike the symmetric case, the wavefunction is not zero on all even sites, but rather takes finite value on all sites due to the broken symmetry.\label{fig:asymmwfn}}
\end{figure}

Finally, the mid-gap Mott pole, which arises here at $E_{\textsc{mp}} \vcentcolon= e_0\approx0.12D$, corresponds to a bound state that is exponentially localized on the boundary between the auxiliary system and the physical degrees of freedom of the lattice. This state is denoted the `Mott pole state' 
$\mathnormal{\Psi}_{\textsc{mp}}$, and it satisfies $\hat{H}_{\text{aux}}\mathnormal{\Psi}_{\textsc{mp}}=E_{\textsc{mp}}\mathnormal{\Psi}_{\textsc{mp}}$. 
Numerical diagonalization of $\hat{H}_{\text{aux}}$ allows the wavefunction amplitude of this state to be plotted as a function of chain site index $n$. This is plotted in Fig.~\ref{fig:asymmwfn} and shows that $\psi_{\textsc{mp}}$ is exponentially-localized on the edge of the chain. This is a result of the strict alternation of the chain parameters, with the coupling $t_1$ at the start of the chain being a \textit{weak} bond. Even though the chain parameters near the boundary exhibit variations/perturbations with respect to the bulk of the chain, the strict alternation guarantees that the Mott pole state is localized. 
This is confirmed by the transfer matrix method (described in \S\ref{sec:calcmeth}), which yields explicitly
\begin{equation}
	\lvert \mathnormal\Psi_{\textsc{mp}}(n) \rvert^2 \sim  \prod_{m=1}^{n-1} \frac{E_{\textsc{mp}} - e_{m} - t_{m-1}}{t_{m}} .
\end{equation}
This expression gives a stringent condition connecting the parameters $t_n$, $e_n$ and the pole position $E_{\textsc{mp}}$ for a localized state. The state shows considerable robustness against perturbations to the chain parameters. In particular, if the Mott pole lies inside the gap (as it must in the insulator, by definition) then the corresponding state in the auxiliary chain is exponentially-localized on the boundary. Inverting the parity of the chain oscillations for all sites involves bulk gap closing; while local parity flips down the chain generate additional domain wall states (i.e.~multiple mid-gap poles, which are not seen in the Mott insulator). Although the Mott pole position $E_{\textsc{mp}}$ is affected by the boundary potential $e_1$, removing the boundary state by some boundary potential perturbation is equivalent to shifting the pole out of the gap, in which case the spectrum no longer represents a Mott insulator. 

When analyzing the robustness of the boundary-localized state to perturbations, the perturbations which must be considered are those which are to the original physical degrees of freedom of the Hubbard model, not unphysical perturbations to the auxiliary chain. Perturbations to the Hubbard model constitute variations in the values $U/t$ and $\eta$.  However, provided such perturbations are not so drastic as to cause the system to cross into a different phase, the Mott insulator will always have a Mott pole inside the gap, and hence a corresponding boundary-localized state. It is concluded that a Mott insulator, by definition, has an exponentially-localized state on the boundary of its auxiliary system, which is robust to the physically-relevant perturbations to the underlying Hubbard model.
The auxiliary system is therefore concluded to be topologically non-trivial, even on breaking particle-hole symmetry.

\section{Topological Indicator}

While the auxiliary chains constructed here take the form of generalized topological SSH models, they are spatially inhomogeneous and have a boundary, and therefore momentum is not a good quantum number. This means it is not possible to perform a transform to momentum space and evaluate a standard topological invariant, such as the Zak phase\index{Zak phase}, as in \S\ref{ch:genssh}.

However, it has been shown recently for the Hubbard model at $T=0$ that the Luttinger integral\index{Luttinger integral} takes distinct constant values in the Fermi liquid and Mott insulator phases for any $\eta \neq 0$~\cite{logangalpin} 
\footnote{For a detailed discussion of the Luttinger integral in non-Fermi liquid phases see \cite{logantuckergalpin}.},
\begin{equation}
\label{eq:lutt}
\begin{aligned}
	I_{\textsc{L}}
	&= \frac{2}{\pi} \Im \int_{-\infty}^{0} \d\omega\, G(\omega) \frac{\d \Sigma(\omega)}{\d \omega}
	\\
	&= \begin{cases} 0 & \text{Fermi Liquid} \\ 1 & \text{Mott Insulator} \end{cases}
\end{aligned}
\end{equation}
The finite value of $I_{\textsc{L}}$ for the generic Mott insulator can be traced to the Mott pole, which is identified here as the topological feature of the Mott insulator. $\eta=0$ is a special point at $T=0$ where it is found that $I_{\textsc{L}}=0$~\cite{logangalpin}. This appears to be an order-of-limits issue and $I_{\textsc{L}}=1$ in the Mott insulator phase is expected if the $\eta\rightarrow 0$ limit is taken before the $T\rightarrow 0$ limit~\cite{galpinprivate}. The generic Mott insulator has $I_L = 1$.

Since the evolution of the self-energy with interaction strength drives the Mott transition\index{Mott transition}, the Luttinger integral is a natural quantity to characterize the distinct topologies of the Fermi liquid and Mott insulator phases, and may be regarded as a topological invariant.

\begin{comment}
A further signature of the topological nature of the Mott insulator self-energy, is a quantized fictitious $T=0$ `conductance' through the end of the auxiliary chain 
\begin{equation}
	\mathfrak{G}_{C} = \frac{e^2}{\hslash}\Gamma \mathcal{A}^{\text{aux}}(\omega=0) = 1
\end{equation}
where $\Gamma$ is the hybridization to fictitious electrodes, and $\mathcal{A}^{\text{aux}}(\omega)$ is the spectral function at the end of the electrode-coupled auxiliary chain. By contrast with the Mott insulator, the fictitious $T=0$ conductance through the end of the auxiliary chain precisely vanishes in the Fermi liquid phase,
\begin{equation}
	\mathfrak{G}_{C} = \frac{e^2}{\hslash} \Gamma \mathcal{A}^{\text{aux}}(\omega=0) = 0 .
\end{equation}
\end{comment}

To reiterate, the signatures of topology from the Luttinger integral described above are \textit{not} being employed as a topological invariant of the Hubbard model, but rather they serve as an indicator for the topology of the auxiliary chain constructed for the effective system. The topology here lies in the auxiliary degrees of freedom of the effective system, and not in the physical degrees of freedom of the Hubbard model.

\section{Outlook}\label{sec:motttopologyoutlook}

%\cite{weberembedding}

Presented here was an interpretation of the classic Mott transition in the infinite dimensional one band Hubbard model as a topological phase transition. The lattice self-energy, determined here by DMFT-NRG, is mapped to an auxiliary tight-binding chain, which is found to be of generalized SSH model type. The Mott insulator is the topological phase, with a boundary-localized state corresponding to the Mott pole. The transition from Fermi liquid to Mott insulator involves domain wall dissociation.

It is concluded that any system with such a pole in its local self-energy may be regarded as topological. The analysis could also be extended to multiband models, where the auxiliary chains become multilegged ladders. A speculation is that a superconducting Hubbard model my map to auxiliary Kitaev chains\index{Kitaev superconductor} involving Majorana\index{Majorana} modes. For a fully momentum-dependent self-energy of a $d$-dimensional lattice, the mapping generalizes to an auxiliary lattice in $d+1$ dimensions. For a Mott insulator, such an auxiliary lattice may be a topological insulator with a localized boundary state.

This chapter's detailed study of treating the Hubbard model with the auxiliary field mapping developed in this thesis serves as a robust proof of concept of the method and demonstrates that non-trivial interpretations of conventional results may emerge as a product of the mapping. The auxiliary models for the proposed extensions above may similarly possess interesting features.
%\index{$0$@\textbf{List of Edits}!606@expanded outlook}

\chapter{Conclusion\label{ch:conclusion}}

%novel results

The work of this thesis centered around two paradigms in theoretical physics: models of topologically non-trivial electronic states and the construction of effective models for complex interacting many-body systems.

With regards to topological phenomena in condensed matter, a novel result developed here (\S\ref{ch:bethessh}) was the
adaptation of a prototypical $1d$ topological insulator to infinite dimensions thereby allowing the application of DMFT to produce an exact solution when strong correlations are included.
It was found that the topological state becomes broadened to a power-law diverging spectrum and that the system experiences a metal-insulator transition to a Mott insulator above a critical interaction strength.
The SSH model is also a prototypical, if somewhat tautological, example of a crystalline topological insulator. A potential generalization of the technique developed here is to other crystalline topological insulators to treat interactions with DMFT. This treatment may involve extensions of the base DMFT as well, such as multi-band or cluster DMFT.

Also within the topological paradigm, this thesis developed several classes of generalizations of the SSH model and demonstrated the cases in which these models do or do not exhibit topological features (\S\ref{ch:genssh}). The lessons learned here provide the basis of a toolkit for engineering $1d$ tight-binding chains with nearest-neighbor kinetics which possess a wide variety of desired spectral features. Given the similarities between the SSH model and the Kitaev superconductor, it is possible that generalizations of the Kitaev superconductor similar to the generalizations of the SSH model developed here also exist, with possible phases exhibiting Majorana zero modes. Such generalized models may be of interest to experiments attempting to utilize Majorana modes as qubits for quantum computation.
%\index{$0$@\textbf{List of Edits}!700@added proposed extension}

The second part of this thesis involved the development of auxiliary models for strongly correlated systems which replicate the dynamics of the correlations, but are themselves fully non-interacting (\S\ref{ch:aux}).

An application of this type of effective model was demonstrated for computing quantum transport through quantum dots.
This use of effective model has the potential to simplify transport calculations as non-interacting transport formulas can be used rather than the more complicated transport formulas for fully interacting systems. 
In principle it could be possible to circumvent the requirement of exactly solving the quantum dot impurity problem by instead employing specifically designed auxiliary models. The auxiliary chains as developed in this thesis are well understood in terms of the relation between their parameterization and their spectral output. Estimating the desired features of the dot self-energy would be sufficient to engineer an auxiliary system modeling the desired interactions. For more sophisticated quantum dots, such as multiorbital impurities, multilegged ladders would need to be employed, as referenced in \S\ref{sec:motttopologyoutlook}. Developing these types of auxiliary models and benchmarking them in a manner similar to the development of the toy model in \S\ref{sec:mttoy}, these multilegged models could also be employed to circumvent the necessity of exact solutions.
%\index{$0$@\textbf{List of Edits}!701@added proposed extension}

It was also found that these auxiliary models have features that make them interesting systems themselves, and their properties can lead to interesting interpretations of the original system. This was demonstrated in the construction of the auxiliary models for the Hubbard model, where it was revealed that the auxiliary models take the form of generalized SSH chains and that the Mott metal-insulator transition manifests itself as a topological phase transition in the auxiliary model.
Using features of the self-energy of the Hubbard model near the transition, a toy model was constructed which replicated the relevant qualitative features of this spectrum. The construction of this toy model built upon the understanding of generalized SSH models gathered from \S\ref{ch:genssh}.

Another type of effective model developed here was based on the use of novel Majorana decompositions of fermions to generate non-linear canonical transformations (\S\ref{ch:aux}). The key property exploited here was the observation that various compound Majorana operators can be combined to form objects which obey the usual fermionic commutation relations. 
In addition to forming the basis of non-interacting auxiliary models for strongly correlated systems, these decompositions may lead to more interesting Majorana models and novel appearances of Majorana degrees of freedom in condensed matter systems. In the conventional Majorana framework, the Kitaev superconducting wire is a model on which Majorana degrees of freedom can be identified. The generalized Majorana framework developed here may facilitate the demonstration of Majorana degrees of freedom on other models for which the standard decomposition is not applicable. In particular, the polynomial type decomposition briefly referenced may have interesting applicability to interacting models where quartic terms appear in the Hamiltonian.
%\index{$0$@\textbf{List of Edits}!703@added future work}

The various effective models developed in this thesis have ample opportunity to be developed to an even more sophisticated level and the case study applications investigated here may prove to be only a shallow foray into their full potential.

\appendix

\chapter{Some Mathematics \& Computations}\label{ch:appendix}

\section{Elements of Complex Analysis}

\subsubsection{Dirac delta distribution}
\begin{align}
	\delta(x-x_0) &= \frac1\pi \lim_{\eta\to0^+} \frac{\eta}{(x-x_0)^2 + \eta^2}
\label{eq:lorentzian}
	\\
	\delta(x-x_0) &= \frac{1}{2\pi} \int_{-\infty}^{+\infty} \d s\, \e^{-\i (x-x_0) s}
\end{align}
Being a distribution, the Dirac delta is technically only well-defined when as the argument of a functional.

\subsubsection{Heaviside step function}
\begin{align}
	\theta(t-t') &= \smashoperator{\int_{-\infty}^{t}} \d s\, \delta(s-t')
	\\
	\theta(t-t') &= \frac{\i}{2\pi} \lim_{\eta\to0^+} \int_{-\infty}^{+\infty} \d s \frac{\e^{-\i s (t-t')}}{s + \i \eta}
\end{align}

\subsubsection{Cauchy principal value}\index{Cauchy principal value|see {$\fint$}}\index{$\displaystyle\fint$}
%$f(\xi)$ has pole at $\xi_0$
%\begin{equation}
%	\fint_a^b f(\xi) \d\xi
%	\equiv \lim_{\eta\to0} \left[ \int_a^{\mathrlap{\xi_0-\eta}} f(\xi) \d\xi + \int_{\xi_0+\eta}^b f(\xi) \d\xi \right]
%\end{equation}
\begin{equation}
	\fint_a^b \frac{F(\xi)}{\xi - \xi_0} \d\xi
	\equiv \lim_{\eta\to0} \left[ \int_a^{\xi_0-\eta} \frac{F(\xi)}{\xi - \xi_0} \d\xi + \int_{\xi_0+\eta}^b \frac{F(\xi)}{\xi - \xi_0} \d\xi \right]
\end{equation}

\subsubsection{Sokhotski--Plemelj theorem}\index{Sokhotski-Plemelj theorem} Theorem over the real line
\begin{equation}
	\lim_{\epsilon\to0} \int_{a}^{b} \frac{f(x-x_0)}{x - x_0 \pm \i \epsilon} \d x = \mp \i \pi \int_{a}^{b} f(x) \delta(x-x_0) \d x + \fint_{a}^{b} \frac{f(x-x_0)}{x-x_0} \d x
\label{eq:sokhotskiplemelj}
\end{equation}

\subsubsection{Fourier transform}
\begin{align}
	f(q) &= \mathfrak{F} f(p) = \frac{1}{\sqrt{2\pi}} \int \d p\, f(p) \e^{-\i p q}
	&
	f(p) &= \mathfrak{F}^{-1} f(q) = \frac{1}{\sqrt{2\pi}} \int \d q\, f(q) \e^{\i p q}
\end{align}

\subsubsection{Kramers-Kronig relations}\index{Kramers-Kronig relations}
\begin{align}
	\Re G_k(\omega) &= \phantom{-}\frac1\pi \fint_{-\infty}^{\infty} \frac{\Im G_k(\omega')}{\omega'-\omega} \d\omega'
	=	-\fint_{-\infty}^{\infty} \frac{\mathcal{A}_k(\omega')}{\omega'-\omega}\d\omega'
	=	\mathfrak{F}^{-1}\left[ \mathfrak{F}\mathcal{A}_k(\omega) \cdot \mathfrak{F}\frac1\omega \right]
	\\
	\Im G_k(\omega) &= -\frac1\pi \fint_{-\infty}^{\infty} \frac{\Re G_k(\omega')}{\omega' - \omega} \d\omega'
\end{align}

%%%%%%%%%%%%%%%%%%%%%%%%%%%%%%%%%%%%%%%%%%%%%%%%%%%%%%%%%%%%%%%%%%%%
\section{Equations of Motion Identities}\label{appendixeom}
%%%%%%%%%%%%%%%%%%%%%%%%%%%%%%%%%%%%%%%%%%%%%%%%%%%%%%%%%%%%%%%%%%%%

Here collected are several commutator identities which are useful in the computation of the Green function equations of motion.

\begin{align}
	[ \hat{A} , \hat{B} \hat{C} ]
	&=	\hat{B} [ \hat{A} , \hat{C} ] + [ \hat{A} , \hat{B} ] \hat{C}
\end{align}
\begin{align}
	[ \hat{A} \hat{B} , \hat{C} ]
	&=	\hat{A} [ \hat{B} , \hat{C} ] + [ \hat{A} , \hat{C} ] \hat{B}
\end{align}

\begin{align}
	\{ \hat{A} , \hat{B} \hat{C} \}
%	&=	\hat{A} \hat{B} \hat{C} + \hat{B} \hat{C} \hat{A}
%	\\
%	&=	\hat{A} \hat{B} \hat{C} + \hat{B} \hat{C} \hat{A} + \hat{B} \hat{A} \hat{C} - \hat{B} \hat{A} \hat{C}
%	\\
	&=	\{ \hat{A} , \hat{B} \} \hat{C} + \hat{B} [ \hat{C} , \hat{A} ]
\end{align}
\begin{align}
	\{ \hat{A} \hat{B} , \hat{C} \}
%	&=	\hat{A} \hat{B} \hat{C} + \hat{C} \hat{A} \hat{B}
%	\\
%	&=	\hat{A} \hat{B} \hat{C} + \hat{C} \hat{A} \hat{B} + \hat{A} \hat{C} \hat{B} - \hat{A} \hat{C} \hat{B}
%	\\
	&=	\hat{A} \{ \hat{B} , \hat{C} \} + [ \hat{C} , \hat{A} ] \hat{B}
\end{align}

$[\hat{A},\opd{c}{}] = -[\hat{A},\op{c}{}]$ $\hat{A}$ Hermitian

$\op{T}{ir,s} \vcentcolon= \opd{c}{i,s} \op{c}{i+r,s} + \opd{c}{i+r,s} \op{c}{i,s}$

\begin{equation}
	\{ \op{c}{A} , \opd{c}{B} \} = \delta_{AB}
\end{equation}
\begin{equation}
	\{ \op{c}{A} , \op{c}{B} \} = 0 = \{ \opd{c}{A} , \opd{c}{B} \}
\end{equation}

\begin{equation}
	\{ \op{n}{A} , \op{n}{B} \} = 2 \op{n}{A} \op{n}{B}
\end{equation}

\begin{align}
	[ \op{n}{i,s} , \opd{c}{j,\sigma} ] &= \delta_{ij,s\sigma} \opd{c}{i,s}
	&
	[ \op{n}{i,s} , \op{n}{j,\sigma} ] &= 0
	&
	[ \op{n}{i,\uparrow} \op{n}{i,\downarrow} , \opd{c}{j,\sigma} ] &= \delta_{ij} n_{i,-\sigma} \opd{c}{j,\sigma}
	&
	[ \op{n}{i,\uparrow} \op{n}{i,\downarrow} , \op{n}{k,\sigma'} \opd{c}{j,\sigma} ] &= \delta_{ij} \op{n}{k,\sigma'} \op{n}{i,-\sigma} \opd{c}{j,\sigma}
\end{align}
\begin{align}
	[ \opd{c}{i,s} , \opd{c}{j,\sigma} ] &= 2 \opd{c}{i,s} \opd{c}{j,\sigma}
	&
	[ \op{c}{i,s} , \op{c}{j,\sigma} ] &= 2 \op{c}{i,s} \op{c}{j,\sigma}
	&
	[ \op{c}{i,s} , \opd{c}{j,\sigma} ] &= \delta_{ij,s\sigma} - 2 \opd{c}{j,\sigma} \op{c}{i,s}
\end{align}
\begin{align}
	[ \opd{c}{A,s} , \opd{c}{a,\sigma} \opd{c}{b,\sigma'} \op{c}{c,\sigma''} ] &= \opd{c}{a,\sigma} \opd{c}{b,\sigma'} \left( \delta_{Ac,s\sigma''} - 2 \opd{c}{As} \op{c}{c,\sigma''} \right)
	\\
	[ \op{c}{B,s'} , \opd{c}{a,\sigma} \opd{c}{b,\sigma'} \op{c}{c,\sigma''} ] &= -\opd{c}{a,\sigma} \opd{c}{c,\sigma''} \delta_{Bb,s'\sigma'} + \opd{c}{b,\sigma'} \opd{c}{c,\sigma''} \delta_{Ba,s'\sigma} + 2 \opd{c}{a,\sigma} \opd{c}{b,\sigma'} \op{c}{B,s'} \op{c}{c,\sigma''}
\end{align}
\begin{align}
	[ \opd{c}{A,s} , \opd{c}{a,\sigma} \op{c}{b,\sigma'} \op{c}{c,\sigma''} ]
	&= - \opd{c}{a,\sigma} \op{c}{c,\sigma''} \delta_{Ab,s,\sigma'} + \opd{c}{a,\sigma} \op{c}{b,\sigma'} \delta_{Ac,s\sigma''} + 2 \opd{c}{A,s} \opd{c}{a,\sigma} \op{c}{b,\sigma'} \op{c}{c,\sigma''}
	\\
	[ \op{c}{B,s'} , \opd{c}{a,\sigma} \op{c}{b,\sigma'} \op{c}{c,\sigma''} ] &= \left( \delta_{Ba,s'\sigma} - 2 \opd{c}{a,\sigma} \op{c}{B,s'} \right) \op{c}{b,\sigma'} \op{c}{c,\sigma''}
\end{align}

\begin{align}
	[ \op{T}{ir,s} , \hat{A} ]
	&=	\opd{c}{i,s} [ \op{c}{i+r,s} , \hat{A} ] + [ \opd{c}{i,s} , \hat{A} ] \op{c}{i+r,s} + \opd{c}{i+r,s} [ \op{c}{i,s} , \hat{A} ] + [ \opd{c}{i+r,s} , \hat{A} ] \op{c}{i,s}
\end{align}

\begin{align}
	[ \op{T}{ir,s} , \opd{c}{j,\sigma} ]
	&=	\delta_{i+r,j;s\sigma} \opd{c}{i,s} + \delta_{i,j;s\sigma} \opd{c}{i+r,s}
	\\
	\sum_{i,s} [ \op{T}{ir,s} , \opd{c}{j,\sigma} ]
	&=	\opd{c}{j-r,\sigma} + \opd{c}{j+r,\sigma}
\end{align}

\begin{align}
	[ \tensor*{\hat{T}}{_{ir,s}} , \tensor*{\hat{n}}{_{j,\sigma}} ]
	&=	\delta_{i+r,j;s\sigma} \opd{c}{i,s} \op{c}{j,\sigma} - \delta_{i,j;s\sigma} \opd{c}{j,\sigma} \op{c}{i+r,s} + \delta_{i,j;s\sigma} \opd{c}{i+r,s} \op{c}{j,\sigma} - \delta_{i+r,j;s\sigma} \opd{c}{j,\sigma} \op{c}{i,s}
	\\
	\sum_{i,s} [ \tensor*{\hat{T}}{_{ir,s}} , \tensor*{\hat{n}}{_{j,\sigma}} ]
	&=	\opd{c}{j-r,\sigma} \op{c}{j,\sigma} - \opd{c}{j,\sigma} \op{c}{j+r,\sigma} + \opd{c}{j+r,\sigma} \op{c}{j,\sigma} - \opd{c}{j,\sigma} \op{c}{j-r,\sigma}
\end{align}

%%%%%%%%%%%%%%%%%%%%%%%%%%%%%%%%%%%%%%%%%%%%%%%%%%%%%%%%%%%%%%%%%%%%%%
\section{Non-Linear Canonical Transformations}
%%%%%%%%%%%%%%%%%%%%%%%%%%%%%%%%%%%%%%%%%%%%%%%%%%%%%%%%%%%%%%%%%%%%%%

Collected here are some useful identities for the computations involved in the non-linear canonical transformations performed in the main text.

\begin{align}
	S_{\gamma,i}^2 &= +1 = S_{\mu,j}^2
	&
	( \pm \i S_{\gamma_j} )^2 &= -1 = ( \pm \i S_{\mu,j} )^2
\end{align}
\begin{align}
	\left[ S_{\gamma,i} , S_{\gamma,j} \right] &= 0 = \left[ S_{\mu,i} , S_{\mu,j} \right]
	&
	\left[ S_{\gamma,i} , S_{\mu,j} \right] &= 0
	&
	\left[ S_{\gamma} , S_{\mu} \right] &= 0
\end{align}
\begin{align}
	\{ \gamma_k , S_{\gamma,j} \} &= 2 \delta_{jk} \gamma_j S_{\gamma,j}
	&
	\{ \mu_k , S_{\mu,j} \} &= 2 \delta_{jk} \mu_j S_{\mu,j}
	\\
	[ \gamma_k , S_{\mu,j} ] &= 2 \delta_{jk} \gamma_j S_{\mu,j}
	&
	[ \mu_k , S_{\gamma,j} ] &= 2 \delta_{jk} \mu_j S_{\gamma,j}
\end{align}
\begin{align}
&\begin{aligned}
	P_\gamma S_{\gamma,j}
	&= - P_\gamma \gamma_j \mu_j P_\gamma
	\\
	&= \gamma_j \mu_j P_\gamma P_\gamma
	\\
	&= - S_{\gamma_j} P_\gamma
\end{aligned}
&
&\begin{aligned}
	P_\gamma S_{\mu,j}
	&= - P_\gamma \gamma_j \mu_j P_\mu
	\\
	&= \gamma_j \mu_j P_\gamma P_\mu
	\\
	&= - S_{\mu,j} P_\gamma
\end{aligned}
\end{align}

Baker-Campbell-Hausdorff
\begin{align}
	\e^{X+Y + \sum_n f_n\left([X,Y]^{(n)}\right)} &= \e^{X} \e^{Y}
\end{align}

\begin{align}
	V_{1,j} &= \e^{- \i S_{\gamma,j} \theta_{1,j}/2}
	&
	V_{2,j} &= \e^{- \i S_{\mu,j} \theta_{2,j}/2}
	\\
	&= \cos\left(\tfrac{\theta_{1,j}}{2}\right) - \i S_{\gamma,j} \sin\left(\tfrac{\theta_{1,j}}{2}\right)
	&
	&= \cos\!\left(\tfrac{\theta_{2,j}}{2}\right) - \i S_{\mu,j} \sin\!\left(\tfrac{\theta_{2,j}}{2}\right)
	\\
	V_{1} &\vcentcolon= \e^{- \i \sum_j S_{\gamma,j} \theta_{1,j}/2} = \prod_{j=1}^{4} V_{1,j}
	&
	V_{2} &\vcentcolon= \e^{- \i \sum_j S_{\mu,j} \theta_{2,j}/2} = \prod_{j=1}^{4} V_{2,j}
\end{align}

\begin{align}
	V_{\gamma,j} V_{\mu,k}
	&=	\e^{-\i S_{\gamma,j} \theta_{\gamma,j}/2} \e^{-\i S_{\mu,k} \theta_{\mu,k}/2}
	\\
	&=	\left( \1 \cos\left(\tfrac{\theta_{\gamma,j}}{2}\right) - \i S_{\gamma,j} \sin\left(\tfrac{\theta_{\gamma,j}}{2}\right) \right) 
		\left( \1 \cos\left(\tfrac{\theta_{\mu,k}}{2}\right) - \i S_{\mu,j} \sin\left(\tfrac{\theta_{\mu,k}}{2}\right) \right)
	\\
	&=	\left( \1 \cos\left(\tfrac{\theta_{\mu,k}}{2}\right) - \i S_{\mu,j} \sin\left(\tfrac{\theta_{\mu,k}}{2}\right) \right)
		\left( \1 \cos\left(\tfrac{\theta_{\gamma,j}}{2}\right) - \i S_{\gamma,j} \sin\left(\tfrac{\theta_{\gamma,j}}{2}\right) \right) 
	\\
	&=	V_{\mu,k} V_{\gamma,j}
\end{align}
\begin{align}
	\left[ V_{\gamma,i} , V_{\gamma,j} \right] &= 0 = \left[ V_{\mu,i} , V_{\mu,j} \right]
	&
	\left[ V_{\gamma,i} , V_{\mu,j} \right] &= 0
	&
	\left[ V_{\gamma} , V_{\mu} \right] &= 0
\end{align}
\begin{align}
	\{ \gamma_k , V_{\gamma,j} \} &= 2 \delta_{jk} \gamma_j S_{\gamma,j}
	&
	\{ \mu_k , V_{\mu,j} \} &= 2 \delta_{jk} \mu_j S_{\mu,j}
	\\
	[ \gamma_k , V_{\mu,j} ] &= 2 \delta_{jk} \gamma_j S_{\mu,j}
	&
	[ \mu_k , V_{\gamma,j} ] &= 2 \delta_{jk} \mu_j S_{\gamma,j}
\end{align}

}

\bibliographystyle{custombib}
\bibliography{master}

\printindex

\end{document}